# Barriers Facing E-Service Adopting and Implementation At Local Environment Level in Nigeria

By

**Oseni, Kazeem Oluwakemi**

B. Tech (Hons) FUTA, Nigeria, MSc Staffordshire University, United Kingdom

The thesis is submitted in partial fulfillment of the requirements for the award of the Degree of Doctor of Philosophy (PhD) of the University of Portsmouth, United Kingdom

February, 2017

# Abstract

E-Government services offer a great deal of potential to improve government activities and citizen support. However, there is a lack of research covering E-Government services at the local government level, particularly in developing countries. However, implementing successful E-Service technology in this part of the world will not come without its barriers considering the unstable and fragile economies in most developing countries. The research aim is to identify the barriers facing E-Service adoption and implementation at a local environment level, using Nigeria as a case example.

This thesis adopts an interpretive paradigm and uses action research. It consists of a large field study in Nigeria (interviews), an online survey of government officials, online focus groups, and analyses government documents and E-Service initiatives. A structured literature review method consisted of sifting through 3,245 papers. The main theoretical tools used in this thesis are the diffusion of innovation (DOI) theory and the theory of change.

This thesis makes multiple contributions. The study found that the autonomy issue at the local environment level in Nigeria and corruption are the major barriers facing E-Service development. In mapping the applicability of these theories, this thesis developed a novel diffusion of innovation (DOI) theory-based E-Service framework consisting of an interactive process to investigate the barriers, solutions, and the success factors for successful implementation. In dealing with the corruption issues, this research proposed a theory of change model to investigate the attitudinal behaviours of E-Service development stakeholders towards corruption, including the long-term goals and outcomes. This model also suggests rewarding packages and human development to boost staff morale and necessary sanctions, as well as anti-corruption policies to guide against corruption-hindering E-Service development and implementation.

# Declaration

Whilst registered as a candidate for the above degree, I have not been registered for any other research award. The results and conclusions embodied in this thesis are the work of the named candidate and have not been submitted for any other academic award.

…………………………………................

**OSENI, Kazeem Oluwakemi**

# Acknowledgements

This work would not have been possible without the help of Almighty Allah (SWT), the King of Kings and several people, whom I would like to thank and appreciate forever.

I appreciate my late grandmother, Hajia Sariwiat Ariyehun Oseni (RIP), for the undiluted love she had for me until her death. I also appreciate Chief Idowu Orogbemi (Nigeria), Mr. Wasiu Raji (Canada) and Mr. Hameed Opetubo (USA) for making my journey and the acquisition of a British education a reality. I am thankful for my first supervisor, Dr. Kate Dingley, for her unfailing advice, her support and direction, and her willingness to share her encyclopaedic knowledge. I am also thankful for my second and third supervisors, Dr. Penny Hart and Dr. Carl Adams, the internal and external examiners, for supplying a sense of direction, perspective and reason at critical stages of this research.

I am indebted to all the participants involved in this research for their invaluable help in setting up the field studies, their contributions and for giving up their time and thoughts so generously. All the respondents who filled the online survey, for their willingness to engage with the method, I am indeed very grateful. My colleagues, lecturers, supporting staff in the School of Computing and the staff at the Post-Graduate School, University of Portsmouth, United Kingdom for their support throughout this research work. Also, I appreciate my parents, Tolani and Remilekun Ibrahim-Oseni, my siblings and cousins, Mr. & Mrs. Olushina and Olufunke Shomoye, Dr. & Mrs. Laide and Busola Olugbile, Mr. Ayolaja Seriki and family, Pastor Peter Olorunkosebi, Mr. Lanre Shonubi, Mr. Samson Iyapo, Pastor Lawrence Ogunkoya, Mr. Gbenga Shomoye, Mr. Olumide Disu, and others for their love, prayers, assistance, and interest, expressed in long conversations and insightful enquiries.

Finally, I am grateful to Abimbola, Oluwateniola, and Fiyinfoluwa Oseni, for supporting and tolerating a family member absent in spirit for much of the time. I love you guys.

# Dedication

This Ph.D. thesis dedicated to:

My children, **Kazeem Oluwateniola OSENI** (Jnr) and **Naeema Fiyinfoluwa OSENI**

"May you grow to appreciate that having a quality and equal right to education makes the world a better place"

# List of Publications

1. **Oseni, Kazeem Oluwakemi**, Dingley, Kate and Hart, Penny (2018) Instant Messaging and Social Networks – The Advantages in Online Research Methodology, International Journal of Information and Education technology, Volume 8, Number 1, ISSN 2010-3689, pp.56-62

2. **Oseni, Kazeem Oluwakemi**, Dingley, Kate and Hart, Penny (2017) Instant Messaging and Social Networks – The Advantages in Online Research Methodology, Proceedings of the International Conference on Educational and Information Technology (ICEIT), Cambridge, United Kingdom.

3. **Oseni**, **Kazeem Oluwakemi**; Dingley, Kate; Hart, Penny; Adewole, Ayoade Iyabode and Dawodu, Oluwakayode Abayomi (2016) A Review of E-Service and Mobile Technology in Earthquakes Relief Operations, International Journal of Managing Information Technology (IJMIT). Volume 8, Issue 2. Available in AIRCC Digital Library.

4. **Oseni**, **Kazeem Oluwakemi**; Dingley, Kate and Hart, Penny (2015) E-Service Security: Taking Proactive Measures to Guide Against Theft, Case Study of Developing Countries International Journal for E-Learning Security (IJeLS). Volume 5, Issue 2.

5. **Oseni, Kazeem Oluwakemi** and Dingley, Kate (2015) Roles of E-Service in Economic Development, Case Study of Nigeria, A Lower-Middle Income Country. International Journal of Managing Information Technology (IJMIT). Volume 7, Number 2. Available in AIRCC Digital Library.

6. **Oseni**, **Kazeem Oluwakemi**; Dingley, Kate and Hart, Penny (2015) Barriers Facing E-Service Technology in Developing Countries: A Structured Literature Review with Nigeria as a Case Study, IEEE International Conference on Information Society (i-Society 2015) Proceedings, London. Available in **IEEE** Xplore Digital Library.

7. **Oseni, Kazeem Oluwakemi** and Dingley, Kate (2014) Challenges of E-Service Adoption and Implementation in Nigeria: Lessons from Asia. International Journal of Social, Education, Economics and Management Engineering. Volume 8, Issue 12, Pages 3689-3696.

# Table of Contents

# List of Figures

# List of Tables

# Abbreviations and Acronyms

| | |
|---|---|
| ALGON | Association of Local Government of Nigeria |
| CBN | Central Bank of Nigeria |
| CFRN | Constitution Federal Republic of Nigeria |
| DOI | Diffusions of Innovation Theory |
| E-id | Electronic Identification Card |
| EIU | Economist Intelligence Unit |
| FIRS | Federal Inland Revenue Service of Nigeria |
| FRN | Federal Republic of Nigeria |
| ICT | Information and Communication Technologies |
| ISR | Information Systems Research |
| LGA | Local Government Area |
| LSG | Lagos State Government |
| NBS | National Bureau of Statistics, Nigeria |
| NIMC | National Identity Management Commission |
| NIS | Nigeria Immigration Service |
| NITDA | National Information Technology Development Agency, Nigeria |
| UN | United Nations |
| TAM | Technology Acceptance Model |
| TNN | The Nation Nigeria Newspapers |
| TVN | The Vanguard Newspapers Nigeria |
| V-SAT | Very Small Aperture Terminal |

# Chapter 1 – Research Introduction

## 1.1 Introduction

The research aim is to identify the barriers facing E-Service adopting and implementation at local environment level, using Nigeria as a case example, and to propose a framework (Guidelines) using Action Research Methodology. Moreover, it will review the usefulness of diffusion of innovations (DOI) theory for predicting the successful adoption and implementation. Success factors will be considered in the study. The use of DOI theory in this study is appropriate as it should capture the benefits derived from technology in the delivery of public services to the citizens by the government (see Section 2.6.4). However, the user's intention to accept and use new technology will be looked into, including focusing on the other factors such as trust, access to appropriate technology and, if possible, expectations. An action research focuses mostly on problem-solving in a real scenario or situation as the participants dictate, with minimal contributions from the researcher. This research is strengthened by the use of a participatory action research methodology.

The researcher's interest is to review the overall functions of the Nigerian local government systems in the delivery of efficient and basic services to the citizens. Poor service delivery, ineffective utilisation, and unavailability of ICT components are very common among local governments in Nigeria (Ainabor et al., 2015). The research will assess the effectiveness of the use of ICT for services delivery. This has influenced the research direction to be an action research-based method, where most participants are the stakeholders in the local government. The action research method will allow the inquiry into the barriers facing E-Service adopting and implementation at a local environment level using Nigeria as a case example (see Section 6.3). The participants will study the research domain themselves, to reflect on their observations of various E-Service barriers and frameworks, with minimal influence from the researcher.

Research into the E-Service at the local environment level is very important, as it is the main level of government that is very close to the citizens (see Section 2.3). Mundy and Musa (2010, pp. 158) stated "The local government level is the lowest form of government but also a very important level because of its proximity to the citizens. Therefore, if E-Government is to succeed at all then it must be a success at this level."

Two E-Service frameworks already exist. The researcher plans to adapt these E-Service frameworks, developed by Azenabor (2013), for the Federal Government in Nigeria, which is a unified framework for the country and the framework developed by Mundy and Musa (2010), which is for State level in Nigeria. Both frameworks are considered significant to this research work as there are three tiers of government in Nigeria, namely: Federal, State, and Local Governments. The purpose of this chapter is to outline (see **Figure 1** below) the research fundamentals, which include an introduction, background, E-Government services in Nigeria, research arguments and gaps, research aims, objectives and questions, research design, motivation, thesis structure, and the chapter summary.

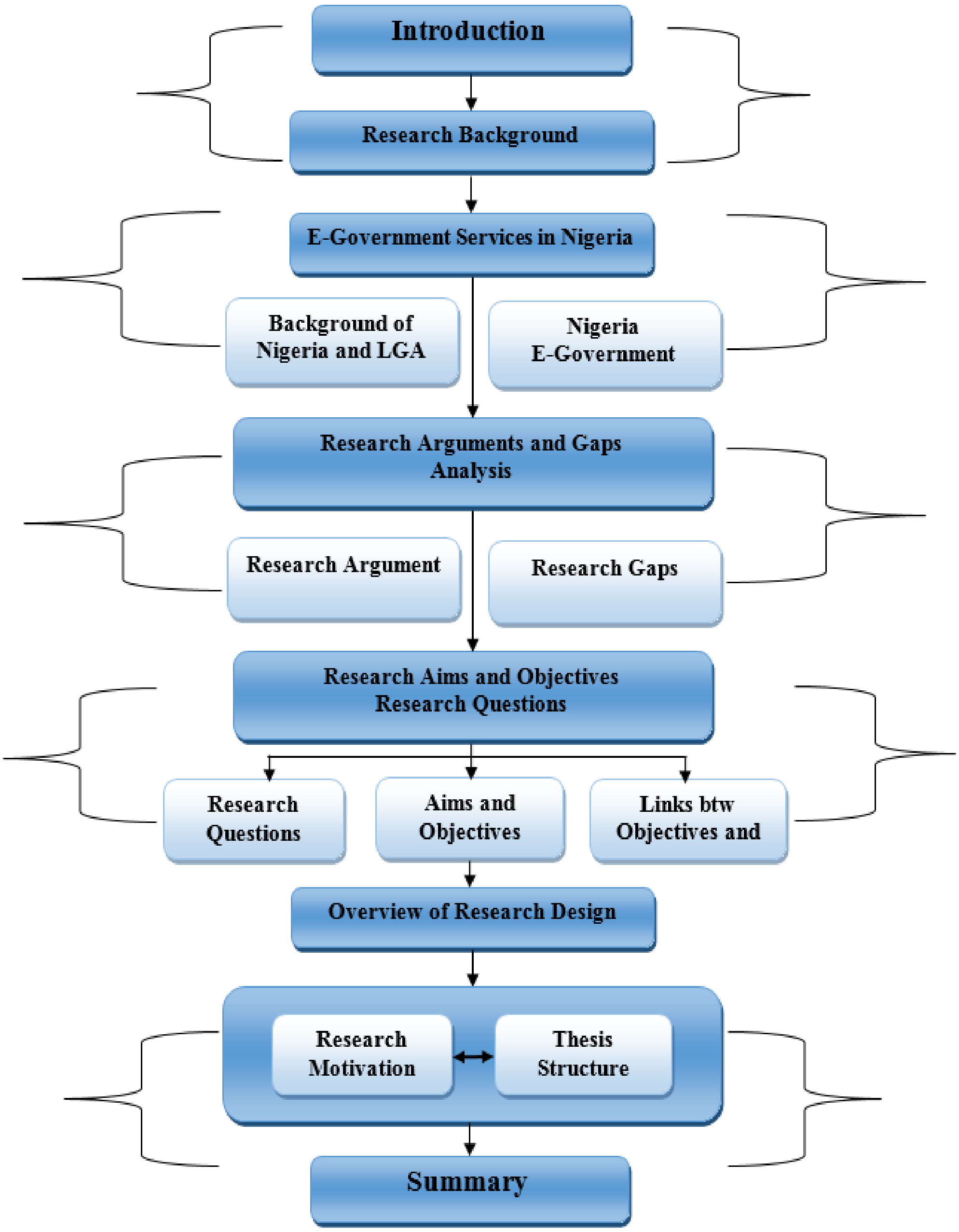

**Figure 1 Chapter One Outline**

## 1.2 Research Background

The recent development in world globalisation has been sustained partly by the advancement in information technology (IT) (Oseni and Dingley, 2014). The internet and telecommunication have brought development to few lower-middle income nations that use E-Government concepts, both at the national and state levels (Ainabor et al., 2015). The successful implementation and adoption of an effective and efficient local government E-Service is a crucial social indicator. The introduction of new technology opens new opportunities for private and public sectors, and E-Services are no exceptions, despite some challenges (Goran and Erik, 2014). Governments in most developed and developing countries wish to implement E-Services, such as e-procurement technology, since, as witnessed in Malaysia, this will enhance transparency and accountability (Arjun et al., 2012). The governments of developed nations have realised that the provision of essential public services cannot be shouldered alone. The complication in demand patterns, coupled with the limited resources, has brought about the need for public-private collaboration (Préfontaine, 2002). In some developed countries, the public-private partnership involves the coming together of the government, private businesses and the society as a whole.

The information system has improved the operations in organisations, both private and public. It has contributed immensely to economic development as the manual-based operation has now shifted to being computer based (Lenk, 2002). E-Service transactions have, to a large extent, phased out face-to-face and telephone-based transactions (Lindgren, 2013). The revolution in the use of the internet in helping organisation communication and business is yielding the business rewards today for those who stake their claims first (Shoniregun, 2007). Apart from reducing the manual cash-handling in the banking sector, Nigeria has overtaken South Africa as the top African economy (Ayoola, 2013). The concept of E-Government emerged in the late 1990s, despite the fact that the history of E-Government as a tool in government establishments could be traced back to the origin of the computer itself; and, just like other e-platform concepts such as E-Commerce, the term E-Government was born out of the internet world (Ake and Horan, 2004). A few decades ago, E-Government as a term and as an identified activity was unknown. Now, however, because of its rapid growth, there is a possible future direction for the research domain (Heeks and Bailur, 2007). It is important to understand that "E-Government initially began as an intra-governmental communication tool", according to Schwester (2009), shortly before government organisations developed their websites with useful information for their citizens. Online transactions began soon after the information of government websites was understood, following the private sector's focus on electronic government (Schwester, 2009). It is nice to know that focusing on customers is the fundamental philosophy of E-Service – that is, to be able to meet customers' needs to enable markets and revenue to grow. Technology has a vital role to play in e-service as it is recognised as an enabler. Thus, businesses can exploit the opportunity provided through technological enhancement to gain a competitive market advantage, as this will open new forms of customer-focused and e-services support services that are more convenient for many users (Roland and Kannan, 2003).

E-Service has moved from the usual manual, and traditional, way of rendering services to electronic service provision for the public. There are several reasons for implementing these services; for example, airline ticketing has gone from the traditional, manual way to an intelligent, web-driven service of purchasing. Many companies have seen their profits doubled through the use of online services in their operation. E-Government service implementation has begun in lower middle-income countries like Nigeria, but the lack of evidence and research has hindered a clear framework for the adoption as expected. In fact, E-Government activities are very low in the country (Mundy and Musa, 2010). From 1960 onwards, the use of information systems, such as E-Services in transforming and improving operations in both public and private organisations, has been a success (Oseni and Dingley, 2014). Also, moving from paper-based operations (manual) to computerised-based ones has been part of the transformation witnessed, as cited by Kolsaker and Lee-Kelley (2009). The face-to-face approach and the use of the telephone in doing business transactions with citizens has been phased out to some extent through the use of online-based services (Oseni and Dingley, 2014).

The establishment of an effective local government in Nigeria is to enable the delivery of efficient services. These services, according to the reasons behind the creation, should be delivered more efficiently than other tiers of the government because of their closeness to the citizens (Ainabor et al., 2015). Unfortunately, the Nigerian local government became an avenue for public civil servants and elected political officers to loot the national treasury (Ukoha, 2006). For example, this included the illegal deduction of local government funds by the ex-military president as mandated by the chairmen, Association of Local Government of Nigeria (ALGON), for the purchase of Toyota land cruiser jeeps (Ukoha, 2006). The adoption and implementation of E-Service at local government level in Nigeria will bring about an open and transparent system of government. The procurement, issuance of contracts, and tenders will be available online. This will reduce corruption in the public service as obtained in the research findings on the success factors for the adoption and implementation of E-Services. For instance, Mr. Adebiyi Mabadeje (TNN, 2015), the immediate past Commissioner for Science and Technology, Lagos

State, Nigeria, during the 2015 ministry's account of stewardship commented on the situation:

> "The administration of former Governor Babatunde Raji Fashola of Lagos State invested in E-Governance through the automation and re-engineering of government business processes, as well as implementing and deploying a Citizens Relationship Management system which delivers efficient services to the citizens more transparently."

Professor Charles Ayo corroborated the statement, during the 2014 Conference on E-Governance in Nigeria (TVN, 2014) hosted by Covenant University, Nigeria. He observed that:

> "E-Governance is an attempt to improve the administration of government through the use of information and communication technology, reducing the bureaucracy of government's businesses."

Professor Ayo also stated that the E-Governance would bring about accountability in managing the affairs of the country. However, in Nigeria's 15 years of unbroken democracy, little or no application of E-Governance has hindered the grassroots from feeling the true dividends of democracy. Ayo further stated that Nigeria is not there yet, but he recognised the efforts of government in the past to institutionalize E-Governance through the installation of V-SAT facilities across the 774 local government areas in Nigeria. Hence, the issue is how long the local governments in Nigeria will keep avoiding the adoption and implementation of E-Services, despite the fact that the same E-Services have been adopted and implemented successfully to some extent in Asia and other parts of the world. There are benefits associated with E-Services. Despite the rapid growth of E-readiness in most countries of the world, from the study conducted by the Economist Intelligence Unit (2010), the Middle East and Africa currently serve approximately 1m internet broadband subscribers. This is a meagre sum compared with the 53m in Asia and 42m in the Americas. Low levels of investment and limited sources of financing constitute the primary reasons for the slow development. With public and private funds for infrastructure development lacking, even broadly available technologies remain too costly for widespread adoption. The EIU study shows that Nigerian E-readiness ranking is 3.46 out of E-readiness maximum points of 10 (Economist Intelligence Unit, 2010).

A participatory action research (see Section 4.4.5) methodology (Bryman, 2016) will help the participants at the local government level in Nigeria to have a thorough understanding of the research domain, and to share knowledge among themselves and with the researcher. The approach focuses on developing a shared understanding within a particular situation, human experience interpretations, and allowing the participants to conduct their investigation while taking ownership with minimum monitoring by the researcher. The research approach also permits the use of different research tools such as document collection and analysis, participant observation recordings, questionnaire surveys, structured and unstructured interviews, and case studies.

## 1.3 E-Government Services in Nigeria

### 1.3.1 Background of Nigeria and Local Governments

According to Azenabor (2013), "Nigeria is situated on the Gulf of Guinea in West Africa with a land mass of 923,768 square kilometres and is divided into 36 states, the capital being Abuja with a federal system of Government. Nigeria is the biggest market in Africa, with a reputation for a huge return on successful investment despite the high cost of doing business." The country gained independence from the United Kingdom (UK) on 1 October 1960, and its formal name is the Federal Republic of Nigeria. It consists of more than 250 ethnic groups, predominantly Yorubas, Hausas, and Ibos.

Nigeria as a lower-middle income country according to the World Bank classification for this current 2017 fiscal year. Its Gross National Income (GNI) per capita is between $1,026 and $4,035 (World Bank, 2017). The GNI is the total domestic and foreign output claimed in any country by the residents. This includes Gross Domestic Product (GDP) added to the factor incomes earned by the foreign residents, while the income earned in the domestic economy by non-residents is removed from the total (World Bank, 2017). The local government in Nigeria is an intermediary player between the federal government and the grassroots in any society (Ahmad, 2013) as it is designed to bring meaningful development to the rural areas. During the colonial era in Nigeria shortly before the independence, the precursor of the local government was the native administration responsible for functions such as the collection of taxes, construction and the maintenance of roads, sanitary inspection and maintaining of law and order (Ukoha, 2006). The idea of local government in Nigeria was a major issue during the colonial regime with its indirect rule, which was designed to avoid physical engagement with the locals who had a well-embedded system of government, most especially in the northern part of Nigeria. Therefore, the system predicted that the traditional rulers should be empowered to carry out the local government administration (Ahmad, 2013). The system flourished well in the north under the leadership of the Sultan, but the reverse is the case in the southern part of Nigeria, which led to the initiation of reform in the 1930s and 1940s. The colonial regime made an attempt to improve the local government system, which was almost destroyed by the Nigeria first leaders after independence in 1960. However, a significant change in the local government system took place in 1976 under the military regime, with 301 local government areas created (Ahmad, 2013). This was also confirmed by Fajobi (2010). Local government areas increased from 301 in 1976 to 774 in 1996 (Fajobi, 2010).

**Table 1 Distribution of LGAs in Nigeria**
(CFRN, 2016)

| Region | Number of LGAs | Percentage of LGAs |
|---|---|---|
| **Northcentral** | 120 | 15.5 |
| **Northeast** | 111 | 14.3 |
| **Northwest** | 186 | 24 |
| **Southeast** | 95 | 12.2 |
| **Southsouth** | 123 | 16 |
| **Southwest** | 139 | 18 |
| **Total** | **774** | **100** |

### 1.3.2 Nigeria E-Government Service Initiatives

In line with the Nigeria initiative and global approach towards E-Government adoption and implementation, the National Information Technology Development Agency (NITDA) was established in 2001 with the aim of implementing the Nigeria Information Technology Policy and co-ordinate overall IT development in the country. The agency has a serious mission to develop and regulate IT for sustainable national development and vision to be the prime catalyst in transforming Nigeria into an "IT Driven Economy." The agency's vision prompted various E-Government service initiatives, which have not only boosted the Nigeria economy but have also created jobs for the residents. In the light of NITDA dynamics of IT and the effects of its application, many NITDA initiatives, which are still ongoing, will enable Nigeria to become a key player in the global IT-driven and knowledge-based economy (NITDA, 2016).

According to the former NITDA Director General, Mr. Peter Jack, “Nigeria must proactively invest in Information Technology to consolidate the gains of the past investments in the sector.” As a result of this, NITDA has been given the mandate to foster the development and growth of IT in Nigeria. NITDA will ensure internet governance, establish and develop IT infrastructures, and empower Nigerians to participate in software and IT system developments. It will improve access to public information for all citizens and bring transparency to government processes (NITDA, 2016). All these mandates could be achieved within a particular period as obtained in developed countries, with an active commitment towards E-Government initiative by all stakeholders.

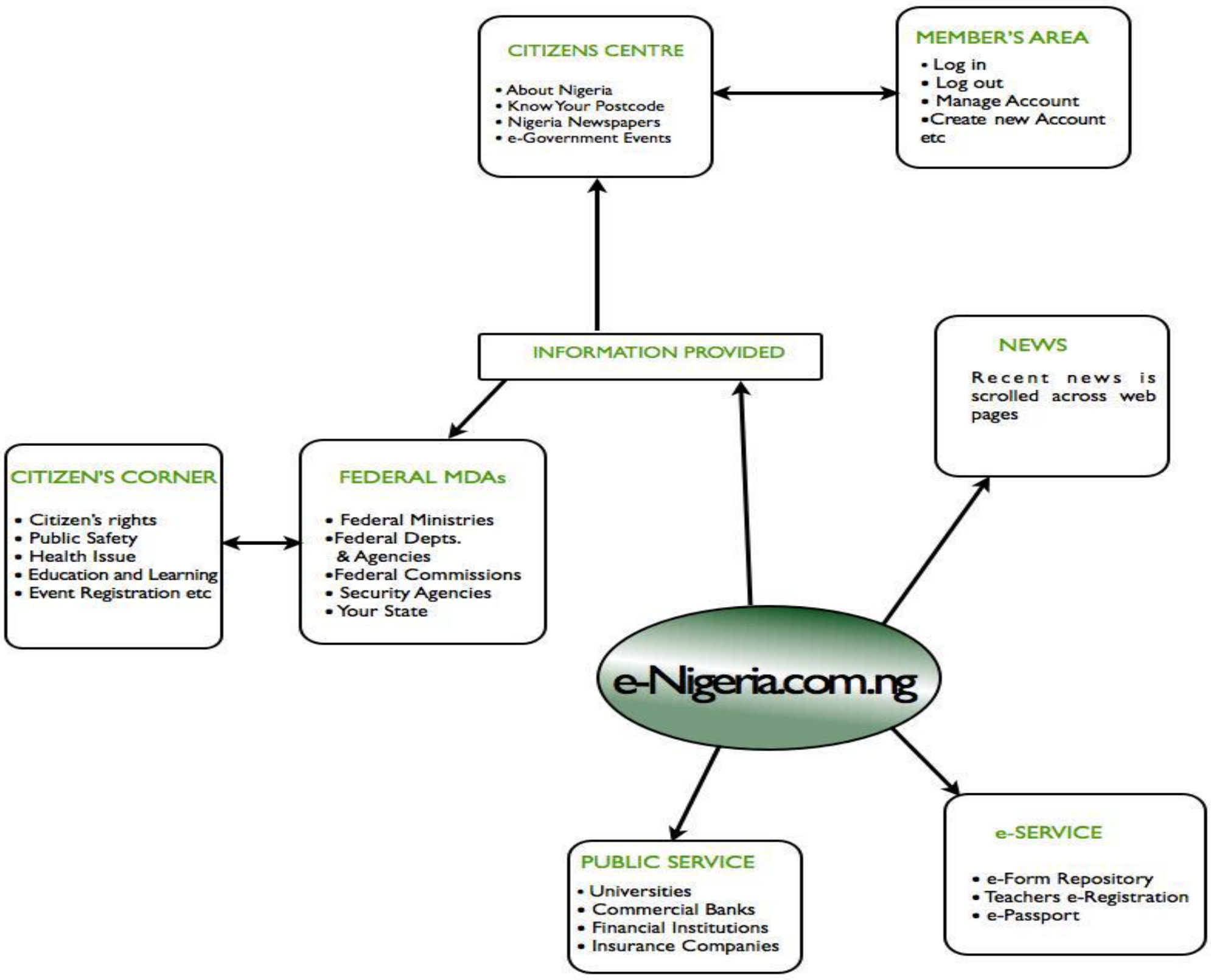

**Figure 2 Nigeria E-Government Model**
**(Azenabor, 2013)**

Currently, there are various laudable E-Government projects initiated in Nigeria, but none of these were undertaken by the local government in Nigeria, even though all citizens and foreigners living legally in the country could benefit from the initiative. Some of these E-Government service projects are given below:

A. **E-Library**: This E-Government project was initiated by the National Information Technology Development Agency (NITDA), which is an agency set up under the Federal Ministry of Communication with the aim of ensuring that IT resources are readily available to promote efficient national development among others. The E-Library project came with standards and guidelines that define the minimum standard for e-libraries in Nigeria. It also mandated all other institutions, private or public, that receive any form of funding from the federal government to make e-library available to the public. The e-library will include all library resources that are accessible in the digital form (NITDA, 2016). Just recently, NITDA has expressed its willingness to partner with a prestigious hospital in Abuja, the capital city of Nigeria, with the aim of improving healthcare delivery in the country through the deployment of E-Health. Hence, the research will look into the barriers affecting the adoption and implementation of E-Service. The strategic positioning of local government in Nigeria, and being the intermediary player between the federal government and the grassroots in the society (Ahmad, 2013), makes the local government significant. Currently, there are limited local E-Service initiatives in Nigeria. Mundy and Musa (2010) articulated for the effective plans towards the development of an E-Government service initiative and model at the local government level in Nigeria.

B. **E-Identification (e-id)**: This is another E-Government initiative where all citizens and residents enroll for electronic identification (e-id). The e-id uses the "National Identification Number (NIN) which is the unique number that identifies you for life and is issued by the National Identity Management Commission (NIMC) after your enrollment. It is also used to match you with your biometric data and other details in the National Identity Database during verification and authentication" (NIMC, 2017). This project will protect individuals not only from identity theft, but also from social vices such as corruption and fraud.

C. **E-Learning**: This E-Government service platform was established by the Lagos State Government, Nigeria, with the aim of introducing the pupils of various schools in the state to E-Learning. The project is achieved through the use of the internet via social media (e.g. Twitter and Facebook) and mobile phones to enhance learning. The project will deploy mobile learning and electronic learning protocols in delivering quality education to the students in Lagos. The students will have to register on its online portal to have access to the facilities, and it is deployed and accessible to all free.

D. **E-Banking**: This excellent E-Government/banking service was implemented by the Central Bank of Nigeria (CBN) in 2012. According to Ayoola (2013), the project is expected to drive development and have a modernised payment system in the country that will position Nigeria as one of the 20 top economies in the world. In supporting the argument, Bayero (2015) noted that the cashless economy policy introduced by CBN would reduce cash management costs, and increase the payment system efficiency. The successful implementation will depend on the infrastructure, among other factors that this study has been able to identify, as barriers facing the E-Service adoption and implementation. Some aspects of the functioning of the cashless economy, as canvassed by Ayoola (2013), are enhanced by e-finance, e-money, e-brokering and e-exchanges. These all refer to how transactions and payments are affected in a cashless economy, and are the laudable projects derived from the E-Service adoption. They will reduce cost and drive financial inclusion to a greater height.

## 1.4 Research Argument and Gaps Analysis

### 1.4.1 Research Argument

The majority of the E-Service Adopting and Implementation research is based on both the state and national levels (Mundy and Musa, 2010; Azenabor, 2013). However, E-Service research at the local government level in Nigeria is not well covered (see Section 3.2). The argument here is whether it's possible to propose a technology based E-Service framework for the successful adopting and implementation that will be suitable for the local environment level in Nigeria as compared to developed countries.

### 1.4.2 Research Gaps

This study is very important as it will produce a solution and a better understanding of the gaps found in previous studies, as explained in the literature review. These gaps are as follows:

- Limited framework covering E-Service Adopting and Implementation at Local Environment Level in Nigeria. Mundy and Musa (2010) proposed an E-Government service framework for the state level, while Azenabor (2013) proposed an E-Government service framework at the federal level.

- Current E-Service frameworks do not capture the full context of the barriers facing E-Service adopting and implementation at Local Environment Level using Nigeria as a case example.
- The roles of humans in accepting the technology and the use of the E-Service is not fully covered at the Local Environment Level using Nigeria as a case example.

## 1.5 Aims and Objectives, Research Questions

### 1.5.1 Research Questions

The main research questions are as follows:

1. What are the barriers facing E-Service Adopting and Implementation at Local Environment Level using Nigeria as a case example?
2. Is it possible to develop an E-Service framework that will be suitable for the Local Environment Level in Nigeria as compared to the developed countries?

### 1.5.2 Aims and Objectives

By answering these questions, the **research aims** are:

1. Identify the barriers facing E-Service adopting and implementation at the Local Environment Level using Nigeria as a case example.
2. Propose a DOI theory-based framework for predicting the successful E-Service adopting and implementation, including the success factors.

The overall **Objectives** of the research will be:

1 To identify the barriers facing E-Service adopting and implementation at local environment level using Nigeria as a case example. Success factors will also be taken into consideration.

2 To learn from the E-Government services in developed and developing countries.

3 To propose a DOI theory-based framework for predicting the successful E-Service adopting and implementation.

4 To evaluate and validate the proposed E-Service adopting and implementation framework for the local environment level in Nigeria.

5 To make recommendations on the study outcomes to the policy makers at local environment level in Nigeria, especially the Local Government Service Commission.

## 1.6 Overview of Research Design and Process

There is a need for the research to have a clear and detailed outline of how the research investigation will take place. This outline typically includes how to collect data and what tools are used for the data collection. It will also state the mode of the data analysis. In the light of this, and in working towards achieving the objectives mentioned above, the researcher carried out the following steps, as shown in **Figure 3** below.

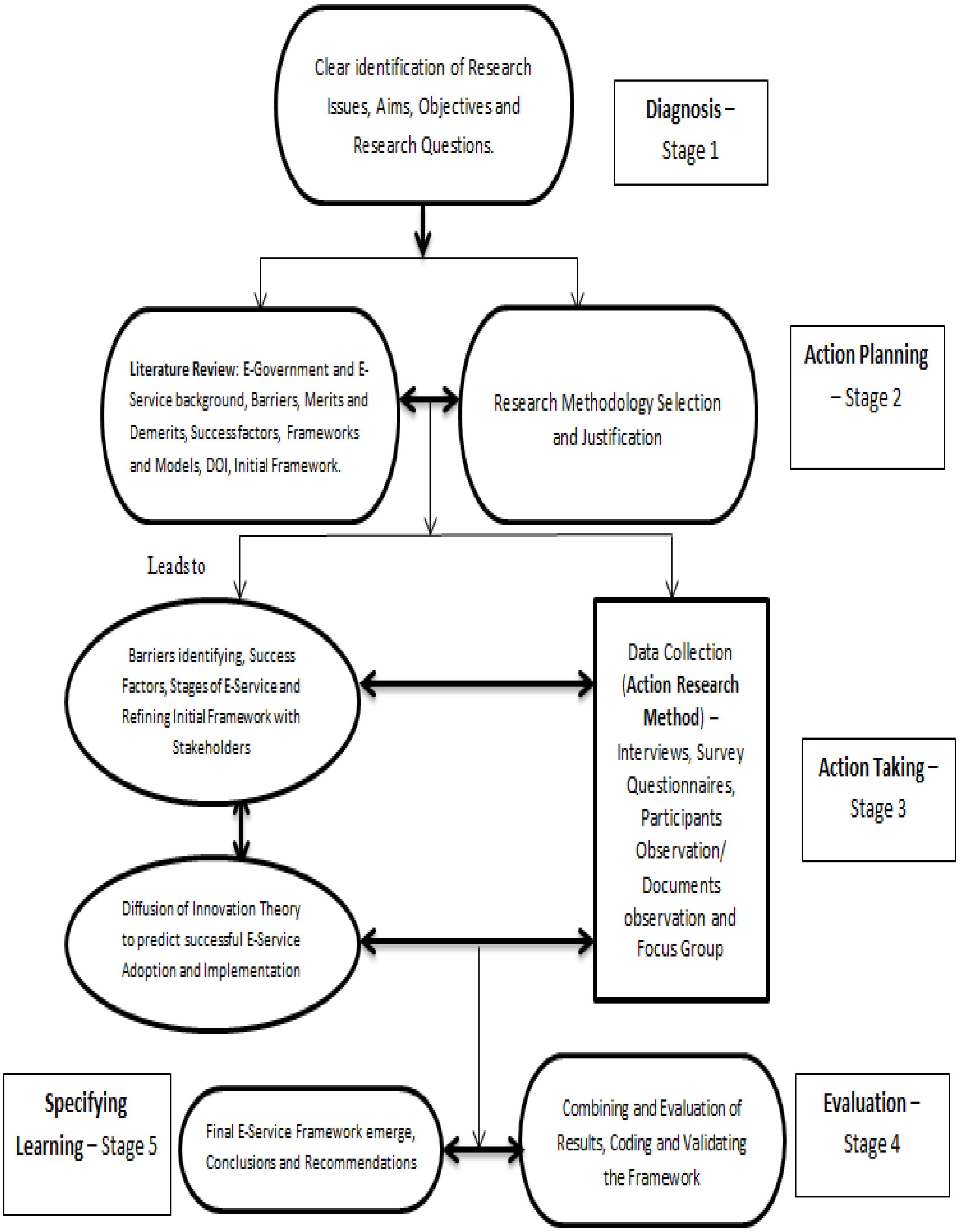

**Figure 3 Research Design**

## 1.7 Research Motivation

The rationale behind this research is the change and advancement in the field of information technology. The government at the local level in Nigeria has not been able to cope effectively with the rate of IT development in the 21st century, as many societies are now moving from a manual/paper-based administration to a paperless administration (Oseni and Dingley, 2014). Hence, there is an urgent need to accelerate and harness the resources in order to meet the growing challenges posed by these developments, especially as this relates to the high level of corruption in government, the storage of data, security, and privacy. Gbadamosi (2007), in one of the most intense reactions on how the adoption and implementation of E-Services are saving the Nigeria Federal Inland Revenue Service (FIRS) a huge amount of money while quoting the FIRS Chairman, stated that:

> "The automation of collection has ensured that tax collected daily by the FIRS from all parts of the country, is swept automatically and electronically into the CBN through our collecting banks. E-tax has stopped the incident of trapped funds in banks, eliminated diversion of cheques by some bad staff and reduced fraud in the collection system. The deployment of Information Communications Technology saved the country $92 million that traditionally vanishes into secret accounts."

Despite the above, the Nigerian economic performance has been very slow (Oseni and Dingley, 2015) compared with many Asian countries that were behind a few decades ago. E-Services contribute substantial revenues to the economic development of any country, and have transformed the economies of many Asian countries, making them major players in the global financial arena (Oseni and Dingley, 2015). However, identifying the research gaps in the early stage has further opened up the need for this research investigation, which firmly corroborates with the research domain. The researcher concludes that the barriers facing the E-Service adoption at the local environment level in Nigeria might have contributed to the following:

- Africa E-Government Ranking 2016, Nigeria in 21st position (UN, 2016).
- 774 local governments in Nigeria, few local governments with a basic website - Publish Stage of E-Government Development (FRN and LSG, 2016).
- The EIU study shows that Nigerian E-readiness ranking is 3.88 out of an E-readiness maximum point of 10, which is very low (EIU, 2010).
- E-Government adopting is low, despite the fact that Nigeria is Africa's largest economy with a GDP calculation of more than $500 billion (Aljazeera News, 2014).

## 1.8 Thesis Structure

This study presents a detailed discussion related to the research introduction, research background, aims and objectives, identified gaps, research motivation, research questions and argument, research design and process, literature review, framework development, and research methodology. The study will also discuss the data collection and analysis, research findings and limitations, and recommendations for the successful adoption and implementation of E-Service projects at the local government level in Nigeria. The thesis is divided into eight chapters, and an illustration is shown in **Figure 4** below:

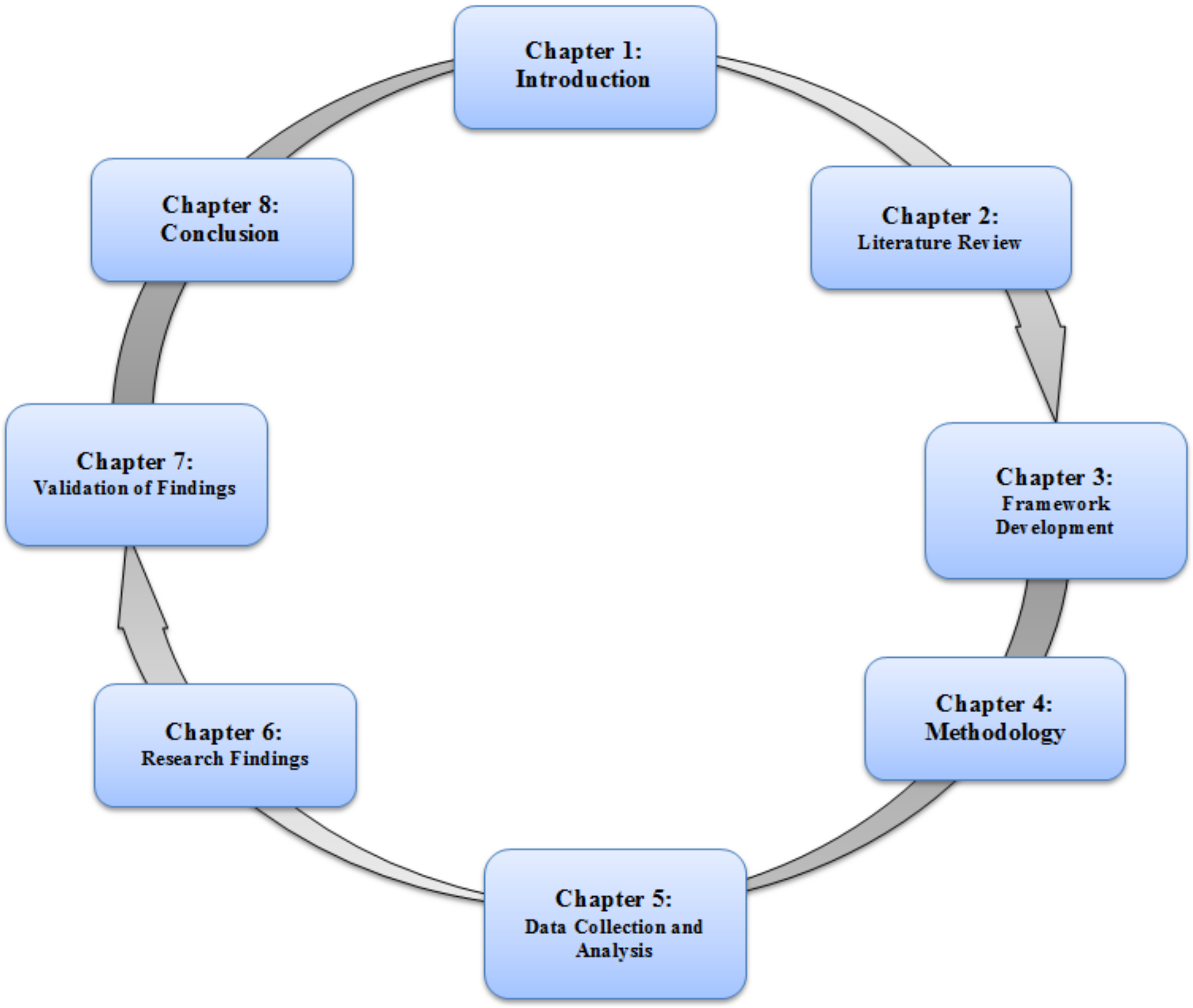

**Figure 4 Thesis Structure**

Chapter One outlined the important research issues. The chapter includes the introduction, the research background, research questions, research argument, research gaps, the research motivations, Nigeria E-Government service initiatives, aims, and objectives. Chapter Two is the literature review and presents a structured and critical examination of the various academic literature on E-Government and E-Services, both in developed and developing countries, including Nigeria. Chapter Two has the clear aim of identifying the research gaps in the existing pool of knowledge, and will enable a better understanding of the investigated and studied areas. Chapter Three comprises a full discussion on the framework development, with an introduction to various existing E-Government frameworks and models. The concluding aspect of Chapter Three will be the introduction of the proposed framework of the study, along with a description of the main parts and success factors. Chapter Four is the methodology. It provides a description of the research strategy, a justification for the adopted method, and the outline of the research methodology that has been followed to ensure that its design is appropriate to provide the answer to the research questions and achieve its aim and objectives.

Chapter Five describes the details of the procedures undertaken for the data collection, in addition to the techniques and plan applied for the data analysis. Next is Chapter Six, which describes and discusses the research findings from this study. Chapter Seven discusses the validation of the research findings and framework, as it is considered to be an important chapter about the observed evidence and findings. Lastly, Chapter Eight is the conclusion chapter and describes the research conclusions. Furthermore, the contributions, achievements of this research and limitations of this study are also discussed in this chapter, and it includes the future directions for this research.

## 1.9 Summary

This chapter briefly discusses the important and usefulness of E-Services adoption and implementation at the local environment level in Nigeria. The chapter looked at various E-Government service initiatives in the country, such as the National Information Technology Development Agency (NITDA) E-Library, and the Nigeria Immigration Service (NIS) E-passport. Other initiatives include the E-id introduced by the National Identity Management Commission (NIMC), and Cashless Economy by the CBN. The research aims and objectives, research questions, identified research gaps, research argument, research design and process, research motivations were also discussed. The introductory chapter also discussed briefly the research background, including trends in IT, ranges from the manual/traditional government services to the online services. The chapter identified various E-Government initiatives in developed countries, as well as Nigeria. Further discussion on the literature review would be revised in the next chapter.

# Chapter 2 - Literature Review

## 2.1 Introduction

In the modern global economy, the need for e-services by the government, public and private bodies has become a central issue because of the critical roles in information systems innovations (Zhang et al., 2014). Despite change and advancement in the field of IT, the government at the local environment level in Nigeria has not been able to cope effectively with the rate of IT development in the 21st century. According to Munda and Musa (2010), many societies are now moving from a manual/paper-based administration into a paperless administration. Hence, there is an urgent need to accelerate and harness the resources in order to meet the growing challenges posed by these developments, especially as they relate to the high level of corruption in government, the storage of data, security, and privacy. There have been concerns over the last few years in the public sector regarding the quality of services, and the e-local government is professed as a significant trend in the development of e-government (Sá et al., 2015). Over the past century, there has been a dramatic increase in reaching out to the citizens and stakeholders in order to encourage them to adopt e-government (Savoldelli et al., 2014). The involvement of the people will improve the policy-making process. Sá et al. (2015) argued that the citizens should appraise the quality of e-Services provided to the public. Heeks and Bailur (2007) added that culture or social context might influence the adoption of E-Government, but that it was dependent mostly on the people.

The governments of both developed and lower-middle-income nations should be determined to ‘take the bull by the horn’ to solve these problems such as corruption, which hinders the non-implementation of E-Services with the support of public and private collaboration. However, the fact is that, no matter the scale or level of cooperation, there is still the need for a sustained and efficient partnering between the principal actors, namely the local government, the private businesses and the society that needs the services. An efficient flow of information between these players, driven by reliable internet and telecommunication infrastructures, may well be the key to a successful e-service implementation. The last two decades have witnessed a dramatic yet remarkable advancement in computer networking technology. The World Wide Web, for instance, has helped to sustain a wide geographical range of users across the world via its global village concept. Consequently, it has reduced the costs associated with maintaining relationships via effective and efficient communication. It is, therefore, no wonder that both private and public sectors have embraced such technology in their quest in reducing administrative costs.

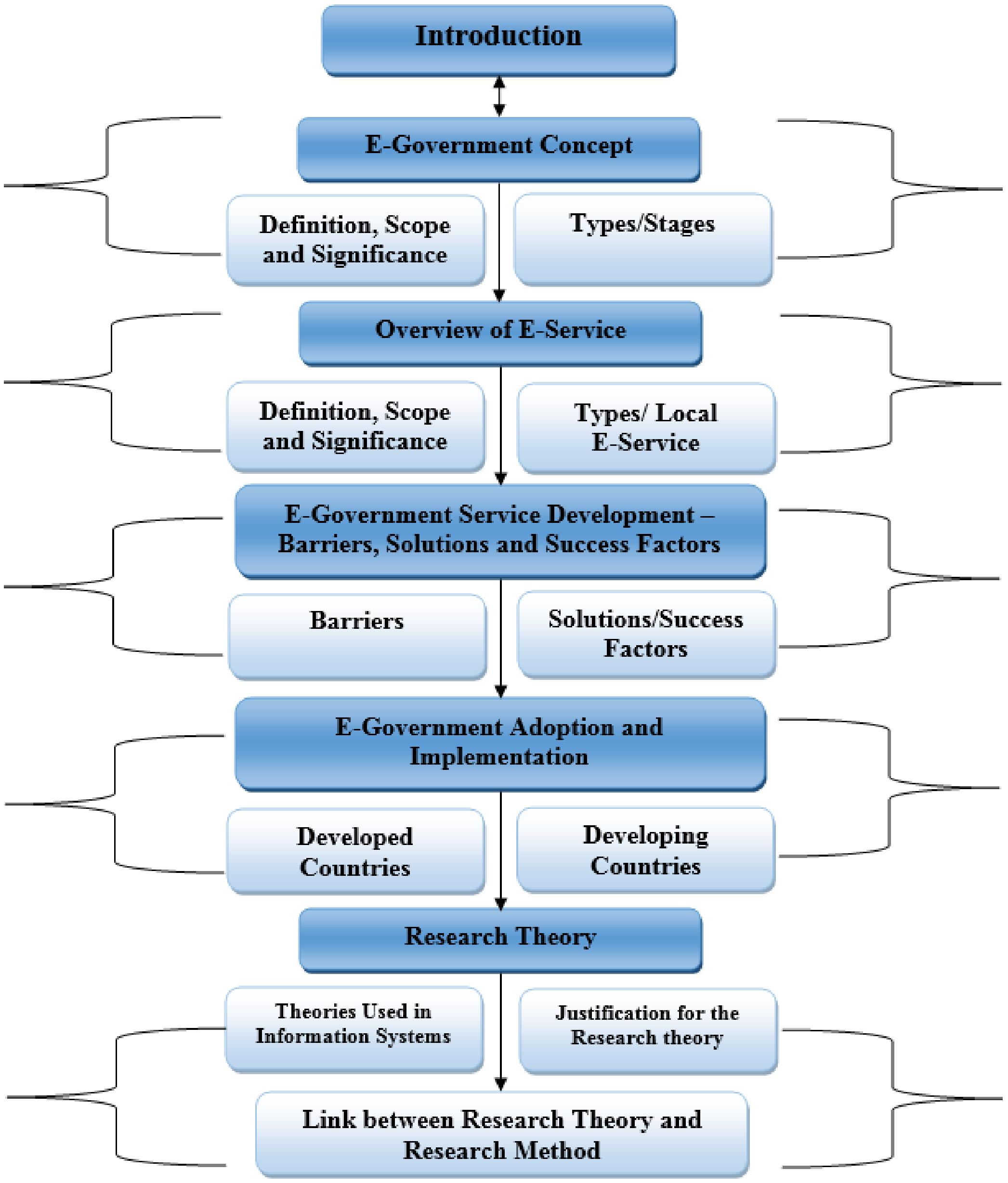

**Figure 5 Chapter Two Outline**

This chapter consists of six sections, as shown in **Figure 5**. The discussion in the introduction will include the literature search, as this research needs to use a robust search method in reviewing past literature on the E-Government services domain to identify possible research gaps. Also, the context and research areas related to E-Government, barriers facing E-Service adopting and implementation, and the use of theory in information systems are thoroughly examined in the literature review, as this will create a foundation for knowledge advancement (Webster and Watson, 2002). Literature review also helps with theory development and early research gaps discovery. Webster and Watson (2002) are also of the opinion that a prior review of relevant literature is imperative in establishing a strong research foundation to build on the established knowledge.

### 2.1.1 Literature Search

The need to use a robust search method for this study meant the researcher had to use a systematic literature review, as suggested by Webster and Watson (2002). There are two parts to the literature search for this study; the first section, as implemented by the researcher, was to search European Conference on E-Government (ECEG) papers from (2009-2013) and E-Government Conference papers from (2009-2014). There are many conference materials, journals, books, online articles on e-government services. It is a Herculean task to read all the references. Therefore, the researcher adopted the method of investigating relevant conference papers from European Conference on E-Government (ECEG) and other E-Government related conferences, as indicated in **Figure 6** below.

A review of previous and relevant literature is a significant custom in academic research (Easterby-Smith and Thorpe, 2002), as cited by Hassan et al. (2011), and gives a better and scientific understanding of the research domain. The literature search (for reviews see Heeks and Bailur (2007); Schwester (2009); Alshehri and Drew (2010); Matavire et al., 2010); Munda and Musa (2010); Bhuiyan (2011); Hassan et al. (2011); Jouzbarkand et al. (2011); Weerakkody et al. (2011); Alateyah et al. (2012); Alshehri et al. (2012); Khan et al. (2012); Rehman et al. (2012); Zafiropoulos et al. (2012); Nkohkwo and Islam (2013); Azenabor (2013); Alateyah et al. (2013); Ashaye and Irani (2013); Zhang et al. (2014); Al-Shboul et al. (2014); Abdelkader (2015)). The search focused on the themes, abstracts, and keywords, as there are many papers on e-government services. The search will allow a view of the main themes in e-government services research. This research uses the European Conference on E-Government (ECEG) as a focus because it is concerned with relevant e-service topics, and it is one of the better-established conferences in the e-government domain. The researcher uses Google Scholar in searching other E-Government conferences because of its credibility and simple way of conducting a broad search for materials. It is also inter-disciplinary, and it allows the finding of the relevant references in journals that are not apparently connected to E-Government or its allied topics. The researcher acknowledges that there is a considerable amount of literature published on E-Service adoption and implementation in developed and developing countries. Unfortunately, there has been no comprehensive research exploring the barriers facing E-Service adopting and implementation at the local environment level in Nigeria, as most literature addressed E-Service adoption and implementation at both federal and state levels.

Therefore, the journals that the researcher came across during the literature search and review are listed below:

- **ECEG** - European Conference on e-Government
- **EJEG** - Electronic Journal of e-Government
- **ICEG** - International Conference on e-Government
- **WASET** - World Academy of Science, Engineering, and Technology
- **BPMJ** - Business Process Management Journal
- **EJSR** - European Journal of Scientific Research
- **JETCIS** - Journal of Emerging Trends in Computing and Information Sciences
- **JEIM** - Journal of Enterprise Information Management
- **IJPSM** - International Journal of Public Sector Management
- **EJISDC** - Electronic Journal of Information Systems in Developing Countries

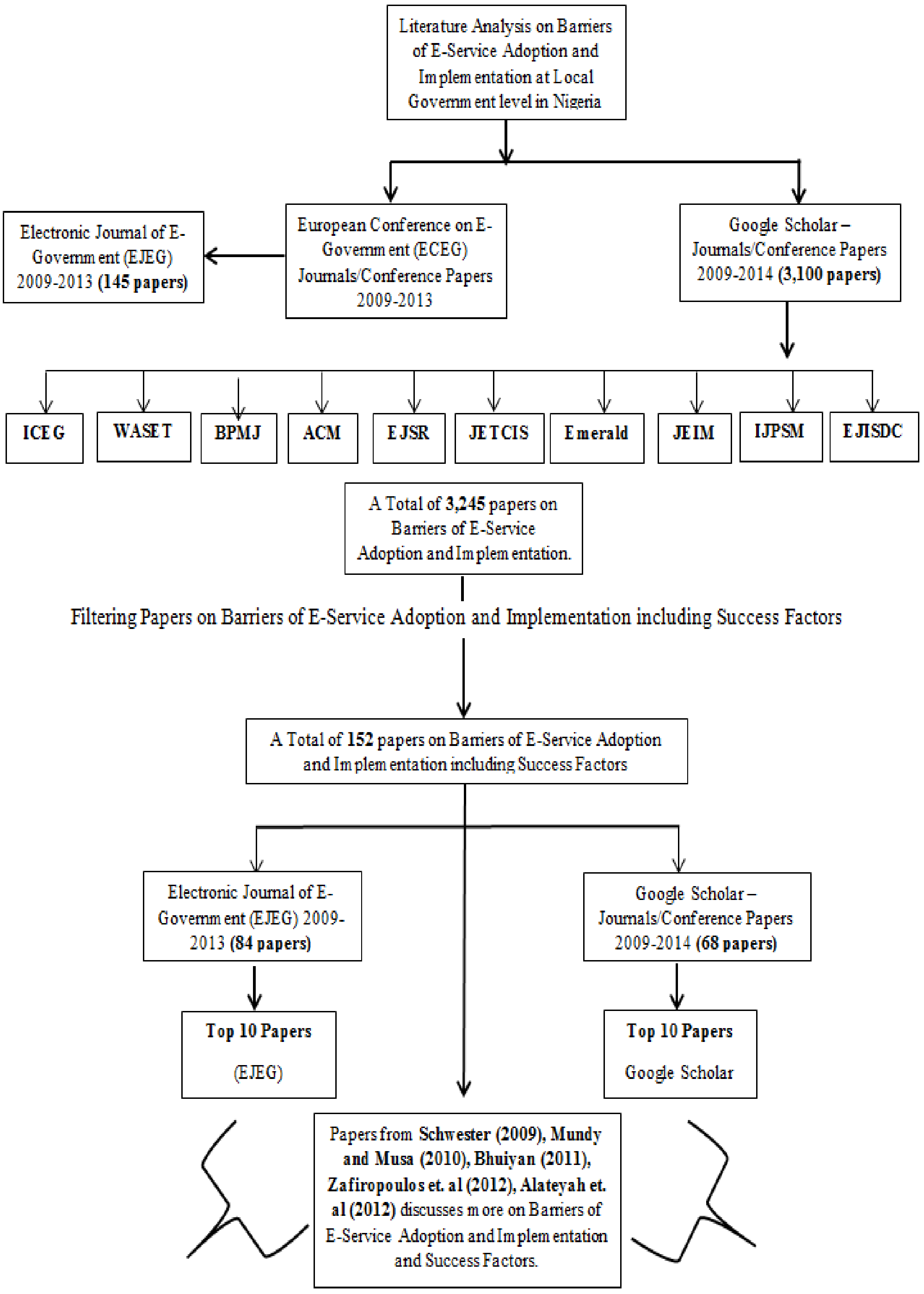

**Figure 6 Literature Analysis**

The search results, as presented in **Figure 6** above, indicated that there are a total of 3,245 papers on E-Government, and barriers facing E-Service adopting and implementations, including success factors.

**Table 2 Previous Literature Search on E-Government Services**

| No | Previous Work | Author(s) (Year) | Method | Success | Gap/ Limitation |
|---|---|---|---|---|---|
| 1. | Examining the Barriers to e-Government Adoption. | Schwester, Richard (2009) | Survey | Barriers successfully examined. | Barrier factors not captured. |
| 2. | Towards a Framework for e-Government Development in Nigeria. | Mundy and Musa (2010) | Framework | Framework developed for e-government at national level. | No Framework developed for Local Government. |
| 3. | Public Sector e-Service Development in Bangladesh: Status, Prospects and Challenges. | Bhuiyan (2011) | Interview | Study identifies the prospect of e-service delivery in Bangladesh. | Lack of Information about E-Service from user's perspective. |
| 4. | Recent advances in e-service in the public sector: State-of-the-art and future trends. | Hassan et al. (2011) | Literature Review | E-service in the public sector necessitates closer working relationships. | Models from literatures reviewed did not capture barriers of e-government adoption. |
| 5. | Towards an integrated model for citizen adoption of E-government services. | Alateyah et al. (2012) | Model | Model developed. | Un-validated model |
| 6. | E-government service use intentions in Afghanistan: Technology adoption and the digital divide in a war-torn country. | Khan et al. (2012) | Survey | Integrated model developed to predict intention to use e-government services. | Early stage of E-Government development. |
| 7. | Factors influencing e-government adoption in Pakistan. | Rehman et al. (2012) | Model | Model proposed to overcome adoption issues. | Limited Targets |
| 8. | Assessing the Adoption of e-Government Services by | Zafiropoulos et al. (2012) | Model | Model developed for teacher's | Reliable Model Needed |
| | Teachers in Greece. | | | e-service adoption. | |
| **9.** | Challenges to the Successful Implementation of e-Government Initiatives in Sub-Saharan Africa. | Nkohkwo and Islam (2013) | Literature Review | Study suggested that sub-Sahara Africa can benefit from the advantages of e-Govt. | Limited research on challenges to the successful implementati on of e-Government initiatives. |
| **10.** | Diffusion of e-Government: A literature review and directions for future. | Zhang et al. (2014) | Framework | Strengtheni ng the diffusion of e-Government | Scattered study on diffusion of e-Government. |

The second part of the literature search by the researcher was to exploit other search engines such as Ethos Library for previous theses and papers on e-government services. Nkohkwo and Islam (2013) added that the use of Google and Google Scholar might not be sufficient enough to find relevant articles. Hence, there is a need to identify further relevant papers considered to be necessary for this study. The Science Direct and the University's Library search engines were also used to capture past research on e-government services adopting adequately, barriers, acceptance, success and implementation. The search method above was utilised by Titah and Barki (2006).

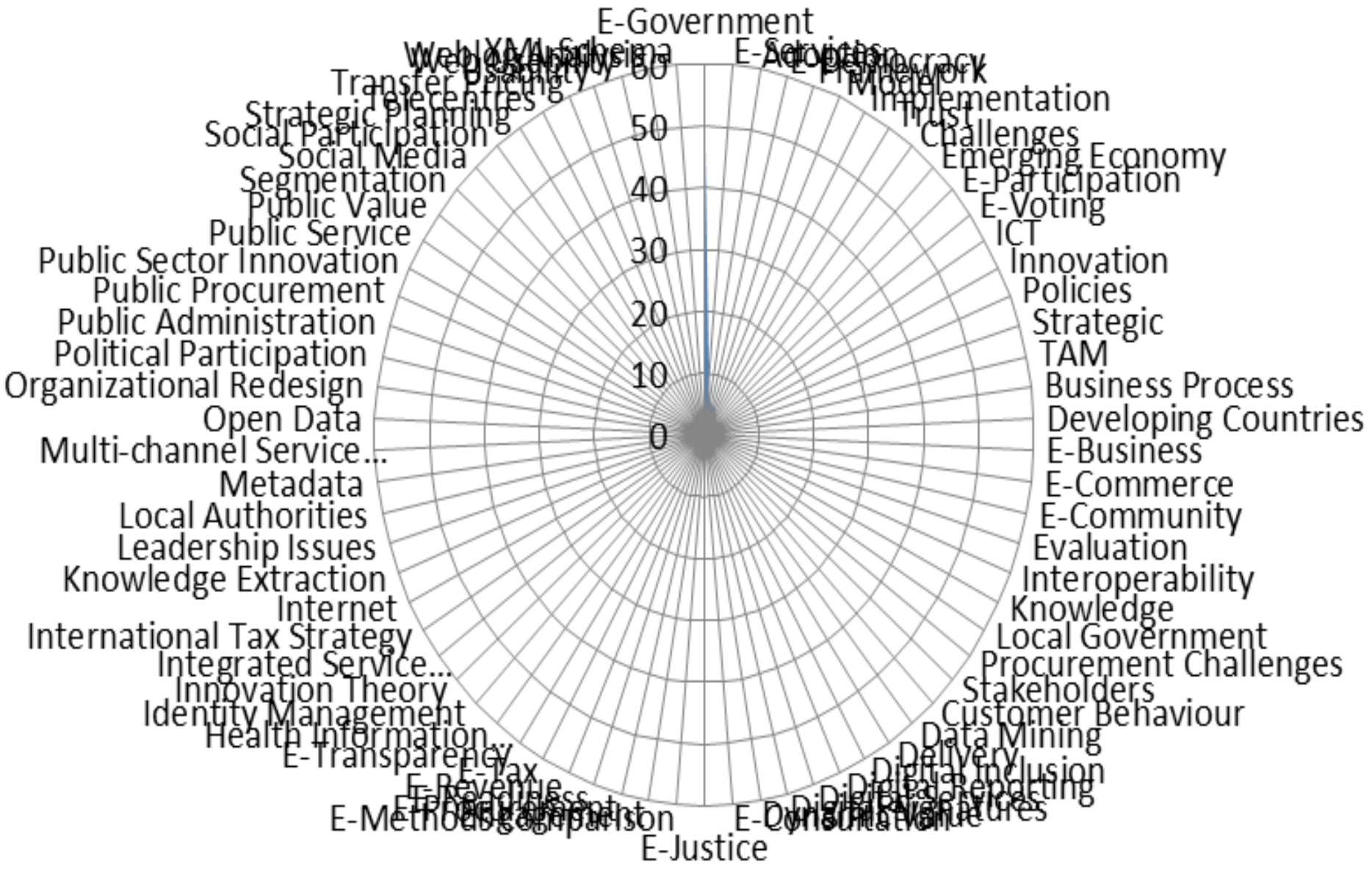

**Figure 7 Search Terms**

## 2.2 E-Government Concept

The concept of E-Government emerged in the late 1990s (Heeks and Bailur, 2007). The history of computing as a tool in government establishments could be traced back to the origin of the computer itself (for example, E-Commerce). The term E-Government was born out of the internet world (Ake and Horan, 2004). A few decades ago, E-Government as a term and as an identified activity was unknown; now, however, because of rapid growth, there is a possible future direction for the research domain (Heeks and Bailur, 2007).

It is important to understand that "E-Government initially began as an intra-governmental communication tool" (Schwester, 2009). This was before government organisations developed their websites with various useful information (Oseni and Dingley, 2014). Online transactions began to emerge after the information of government unit websites had been mastered, followed by the private sector's focus on electronic commerce (Schwester, 2009). Moreover, the term E-Government is more than just a website where emails or the processing of a transaction are carried out using the internet. E-Government has become part of technological advancement in today's society. New concepts have been added to e-Government, which include accountability, transparency and evaluation of government performance through citizen participation (Mohammad et al., 2009).

### 2.2.1 E-Government Definition, Significant, Scope, and Nature

There are many definitions of E-Government. According to the European Union (EU) (2004) in its E-Government initiatives and effort, it defined E-Government as "the use of information and communication technology (ICT) in public administrations. E-Government combined with organisational change and new skills to improve public services and democratic processes." This E-Government definition was also confirmed by the World Bank (2006). Gil-Garcıa and Pardo (2005), in their research journal, described e-Government as the generalised use of information and communication technologies (ICTS) by the government. The ICT usage in E-Government is to provide public services to improve the effectiveness and promotion of democratic values in government. Gil-Garcıa and Pardo (2005) added that the use of ICT would transform government structures and further improve the quality of public services.

E-Government, according to this researcher and for the purpose of this study, is defined as the use of ICTs to advance and enhance the public sector's activities (Hassan, 2011), including the local government. E-Service will provide online services to all the stakeholders at the local government level, including citizens and businesses. **Table 3** below further summarises various E-Government definitions found in the literature. According to Hassan (2011), illustrating these definitions is to explore the E-Government characteristics in each definition, as it is beyond the scope of this thesis to investigate all the definitions of E-Government.

**Table 3 Definitions of E-Government**

| Authors | E-Government Definition |
|---|---|
| Backu (2001) | E-Government is a form of e-business in governance and refers to the processes and structures needed to deliver electronic services to the public (citizens and businesses), collaborate with business partners, and to conduct electronic transactions within an organisational entity. |
| | |
| Gregory et al. (2003) | E-Government as using ICTs by governments and agencies in areas like operational enhancement, public information and service delivery, citizen engagement and public participation in the process of governance. |
| | |
| Yildiz (2007) | E-Government as utilising the ICTs for the implementation of government-related information and services to the public. |
| | |
| Schwester (2009) | E-Government offers internet applications that connect citizens with public administrators, decision-makers and, perhaps, elected officials. |
| | |
| Hassan (2011) | E-Government as the use of ICTs in the government organisation for many purposes: improve efficiency and effectiveness in government administration; enhance coordination and collaboration among governmental organisations; and to provide electronic services to other stakeholders. |
| | |
| Rehman et al. (2012) | E-Government is "the use of ICT and internet to enhance the access to and delivery of all facets of government services and operations for the benefits of its stakeholder groups which includes citizens, businesses, and government itself". |
| | |
| Zafiropoulos et al. (2012) | E-Government involves the use of IT and the internet to improve the delivery of government services to citizens, businesses, and other government agencies. |
| Nkohkwo and Islam (2013) | E-Government can be defined as “use of information and communication technologies to offer citizens and businesses the opportunity to interact and conduct business with government”. |
| | |
| Ziemba et al. (2013) | E-Government is defined as the application of ICTs to transform the efficiency, effectiveness, transparency and accountability of informational and transactional exchanges within government units, between government units at state and local levels, citizens and businesses; and to empower citizens through access and use of public information and public services. |
| | |
| Al-Shboul et al. (2014) | E-Government in its simplest form means using ICT tools to provide services to citizens. |
| | |
| Savoldelli et al. (2014) | E-Government is the process of innovation of public administration in order to achieve innovative forms of government and governance through the use of ICTs. |
| | |
| Abu-Shanab and Bataineh (2014) | E-Government is defined as the utilisation of the internet and World Wide Web for providing government information and services to citizens. |

A major breakthrough in adopting E-Government was witnessed, especially with the collaboration between governments, and public and private companies to achieve a common objective. In New Zealand, for example, the provision of adequate and efficient local E-Services to the citizens (UN, 2004) has enhanced development tremendously. The local government continues to work together with private organisations and consultancies in meeting its sector objectives. The same change has also been witnessed in the area of telecommunications, where private IT companies such as Ericsson have helped to lower the information and communication systems (Préfontaine, 2002). The E-Readiness 2005 ranking, according to EIU (2005), put the Middle East and Africa at the bottom ranking in the world, with a score of 4.42 out of 10 available. Therefore, Schware and Deane (2003) classify the impact and significant of E-Government in **Table 4** below.

**Table 4 E-Government Impact**
**(Schware and Deane, 2003)**

| Impact | Definition | Project Example |
|---|---|---|
| **Direct Citizen Value** | **E-Government Adoption benefits:**<br>-Increased Access<br>-Reduced Delays<br>-Improved Service Delivery<br>-Less Interaction with intermediaries | **India**: Land registration and payment of property taxes by the citizens at Andhra Pradesh only takes five minutes instead of 15 days.<br><br>**Nigeria**: Lagos State online tax payment only takes three minutes instead of spending a whole day at the bank queuing. |
| **Social Value** | -Improved Trust in Government<br>-Increased Sharing of Information<br>-Monitoring of Regulatory Compliance | **Estonia**: Citizens participate in government decision-making via E-Democracy Portal. They review and comment on draft legislation online and send proposal online to the government. |
| **Government Operational Value** | -Performance improvement<br>-Preparation for Future: on-time, network congestion flexibility, completion rate, redundancy | **Philippines**: Quick clearance via e-customs of major transactions has brought down the cost of trade drastically. Cargo is released between two days instead of eight days in the previous system. |
| **Strategic/Political Value** | -Improved Public Image<br>-meeting Legislative Guidelines | **Mexico**: Declaranet requires civil servants to declare their assets online. |
| **Government Financial Value** | -Reduced Cost<br>-Decrease in Cost of Materials<br>-Reduction in Cost of Errors | **Chile**: The e-procurement system estimated to provide government with efficiency gains of $200 million per year. |

From **Table 4** above, and according to Schware and Deane (2003), citizen's benefits from the use of E-Government, as services are now being delivered effectively online, access has increased, and it has reduced the intermediaries between the government and the citizens. In most cases, the citizens can now communicate directly to the government via online portals, review and comment on draft legislation online, and send proposals online to the government. The scope involved in E-Government according to Rana (2014) is about the flow and connection of information between the Government and Businesses (G2B), Government to Employees (G2E), Government and Citizens (G2C), Government and Government (G2G) as it facilitates smooth access to information across the board. This researcher, however, argued that an active E-Government service could witness an improved communication between the Citizens and the Government (C2G) at the local government level in Nigeria (Jouzbarkand et al., 2011) as opposed to Rana (2014) above. In understanding the value of E-Government, it is imperative to clearly define the scope of E-Government projects as it encompasses a broad collection of services, people, products and procedures (Hassan, 2011).

### 2.2.2 Types of E-Government

E-Government usage is rising drastically worldwide through both public and private organisations. Governments are now using ICT to provide services to government agencies, employees, businesses, and citizens (Fang, 2002). The following are the major types of E-Government, although they sometimes overlap each other (Huang and Bwoma, 2003; El-Sofany et al., 2012; Gajendra et al., 2012; Haque and Pathrannarakul, 2013):

1. **Government to Citizen (G2C):** This form of E-Government is described as the communication link between government and private individuals. It brings about the relationship between government and citizens (Haque and Pathrannarakul, 2013; Gajendra et al., 2012). This communication link open doors for citizens to learn much about governance as feedback could be given in respect of public services, accountability and democracy, as cited by Haque and Pathrannarakul (2013). In another direction, G2C is described by Huang and Bwoma (2003) as keeping a one-on-one relationship with citizens by the government in order to provide public services like social security information.

2. **Government to Business (G2B):** This is a type of E-Government interaction between government and a business community which is important to economic development (Huang and Bwoma, 2003; Haque and Pathrannarakul, 2013). G2B involves various service exchanges between government and the business community, including registration of business, obtaining and renewal of licences, obtaining a permit and so on. E-procurement is a typical example of G2B communication as this makes the "bidding process transparent and enables the smaller business to bid for large government procurement projects" (Haque and Pathrannarakul, 2013). According to Mahadeo (2009), the Mauritius Revenue Authority (MRA) online tax filing and payments service to the citizens is another interesting case in G2B transactions, where there is an electronic business relationship between MRA and stakeholders in business.

3. **Government to Employees (G2E):** This form of E-Government explains the relationship between the government and its employees. According to Alateyah et al. (2013), the services provided by government would be available online for the employees, such as comprehensive insurance, e-payslip, housing benefits, credit tax and so on. Fang (2002) pointed out that the exchange of information between Government and employees is mostly related to work and performance issues, personnel policies, career management, and development.

4. **Government to Government (G2G):** This is another type of E-Government where all tiers of governments, namely federal, state and local, are represented and integrate all their internal systems and procedures to form a central system (Alsaghier et al., 2009). The objective here is to facilitate the collaborations and processes among the inter-government organisations, as also stated by Seifert (2008). Government to Government (G2G) e-government is mainly about sharing and doing transactions electronically among governmental actors, with the aim of improving consistency and reducing the number of personnel needed to complete a given task system. G2G activities include e-identity and e-security services (Alsaghier et al., 2009).

### 2.2.3 Stages of E-Government Development

There are different stages involved in E-Government development. According to Hassan (2011), the implementation of E-Government means a transformation process at different levels. Hence, the factors that hinder the change is to recognise the various stages of E-Government development (Hassan, 2011). Researchers in the literature attempted to describe different stages of E-Government development in terms of web interface and the degree of technological complexity (Mundy and Musa, 2010; Alfarraj et al., 2011;

Hassan, 2011). The following are the studies as presented in the literature regarding E-Government development:

1. **<u>Nigeria Five-Stage E-Government Development (Mundy and Musa, 2010)</u>**

The five-stage Nigeria E-Government development model (Mundy and Musa, 2010) consists of five different levels: Publish, Interact, Transact, Portal and Participate level. According to Mundy and Musa (2010), 70% of the state websites were still very much in the publish stage of E-Government, as indicated in **Figure 8** below.

a. **<u>Publish Stage</u>**: This is the basic form of an E-Government stage, as governments publish very simple and limited information on their websites. The information may include the vision and mission of the agency, contact information, official hours, and documents. Most states in Nigeria are still at this stage, as displayed in **Figure 8**.

b. **<u>Interact Stage</u>**: Also known as the interaction stage, simple communication is achieved at this stage between the government and public. Griffin et al. (2007) further stated that it is the stage at which basic search capabilities are provided, including downloaded forms and email addresses of offices. Few states in Nigeria are at this stage of E-Government development.

c. **<u>Transact Stage</u>**: The transaction stage focuses on building self-service applications for the public to access online (Griffin et al., 2007). Application for ID cards and renewals of licences online could be made at this stage as the government provides online services for citizens' access and usage (Alfarraj et al., 2011). No state is at this level of E-Government development stage in Nigeria according to Munda and Musa (2010). However, the researcher observed that there is a new shift in the transaction stage of E-Government development in Nigeria. For citizens residing in the states such as Lagos, Rivers, and Abuja, the federal capital territory could now pay their taxes online. There is an online resources platform now developed by Lagos State Government Electronic Banking System of Revenue Cycle Management for citizens to register and make payments online for taxes, land-use charge, and even make complaints or comments that will assist in enhancing governance in the state.

d. **<u>Portal Stage</u>**: At this stage of E-Government development, there is a presence of a single portal for the government, including contact. It provide links to other government sites and serves as a point of coordination for several organisations to develop their applications (Hassan, 2011). The E-Government adoption in Nigeria today has evolved past what we have had during the last five years. In many state government portals such as Lagos, Ogun, Rivers, and Abuja, the federal capital territory now provides links to many other government parastatals and agencies.

e. **<u>Participate Stage</u>**: The is the E-Government development stage where states achieve the purpose and long-term aims of E-Government. This stage assists government parastatals to improve their performance. There is an availability of tools such as online voting, pollings, and surveys; these increase the level of social participation and citizen engagement. Unfortunately, no state in Nigeria today is at this stage of E-Government development (Mundy and Musa, 2010).

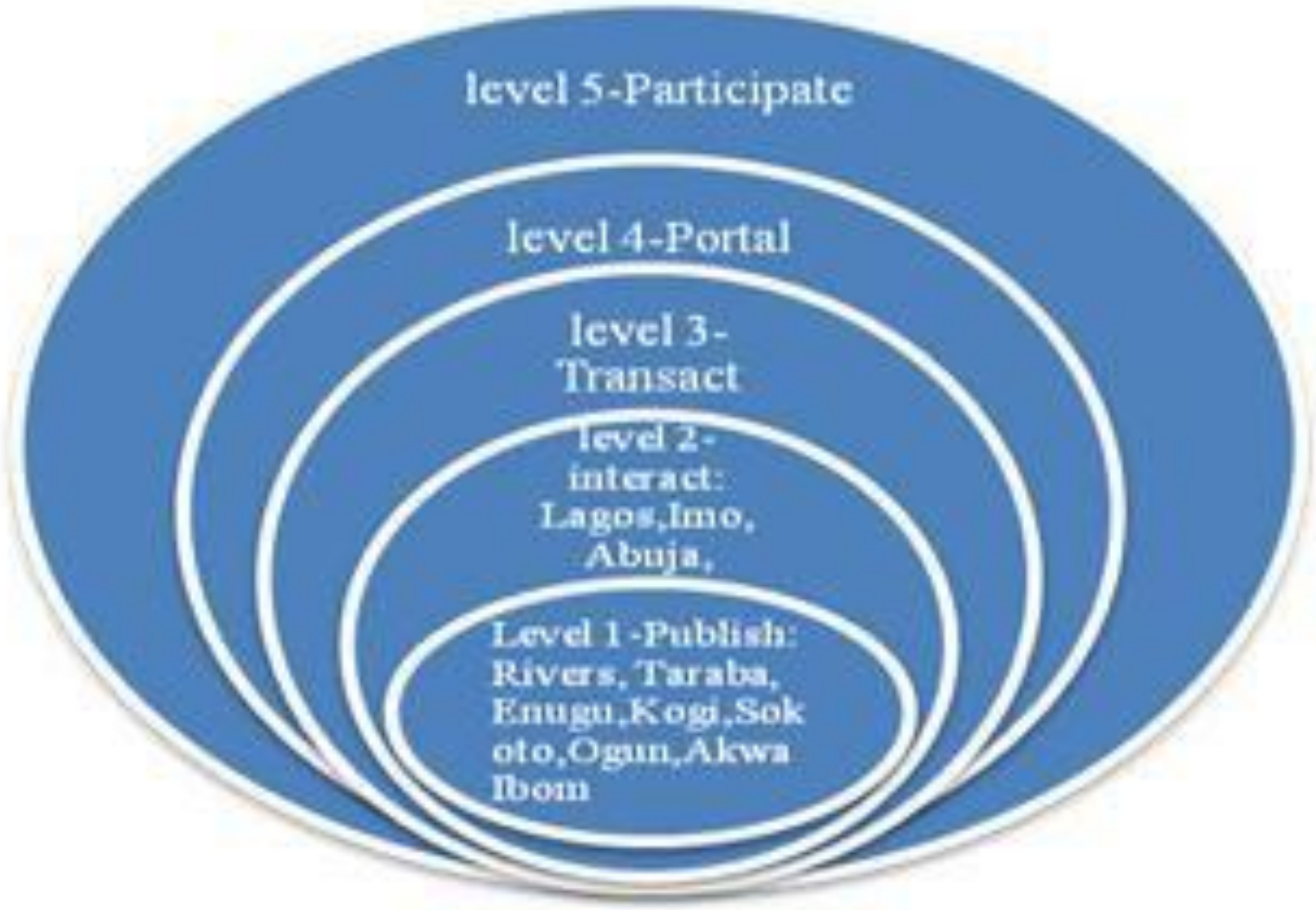

**Figure 8 Nigeria Five-Stage E-Government Development (Mundy and Musa, 2010)**

2. **<u>Five Stages of E-Government Development (Alfarraj et al., 2011)</u>**

Another five-stage E-Government development model has also been proposed. This model comprised of five different levels, namely: Basic Site, E-Publishing, Interactive, Transactional and Holistic E-Government. This model took another dimensional definition compared to the first five-stage model discussed earlier. For more clarification, there is a need to discuss each stage of this model further.

a. **Basic Site**: This level of E-Government stage is very similar to the publish stage of the model developed by Mundy and Musa (2010), as governments publish very basic and limited information on their websites. According to Alfarraj et al. (2011), the primary aim at this stage is to display the government agency's mission, addresses, opening hours and other things relevant to the public. Fath-Allah et al. (2014) added that this stage only serves as a static method by which information is provided to the public.

b. **E-Publishing**: Information on various government activities at this stage of E-Government development is available online. The information may include the vision and mission of the agency, contact information, official hours, and documents.

c. **Interactive**: This stage of E-Government development is similar to the third level of the United Nation's (UN) five-stage. The interactive level, according to Fath-Allah et al. (2014), includes a more sophisticated level of formal communication between citizen and government. Furthermore, this stage provides simple communication between the public and government agency that includes email system, basic search engine, and officially downloaded forms.

d. **Transactional**: The transactional stage of E-Government development enables the exchange of information among government agencies. Activities such as online payment are possible at this stage, and the citizens can conduct other online transactions such as e-passport and e-licence. According to Hassan (2011), governments begin to transform themselves by engaging the citizens through two-way communication on websites.

e. **Holistic E-Government**: In the final stage of E-Government development model, Fath-Allah et al. (2014) stated that a Government presence at this stage includes the availability of online activities/software for the public and agencies to identify if an E-Government project will be successful or not. An example is the Capacity Assessment Toolkit which is used to examine project capabilities.

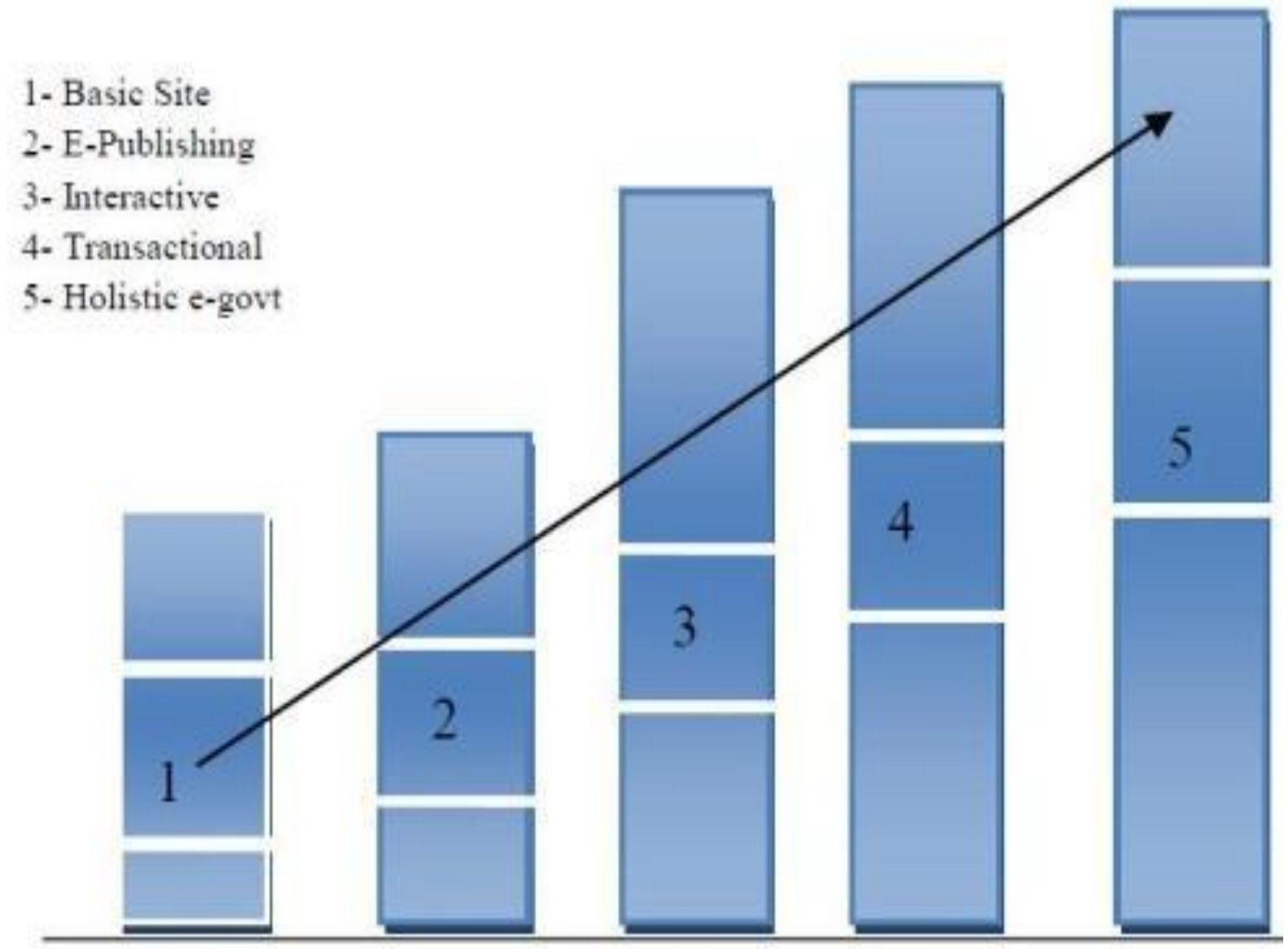

**Figure 9 Five stages of E-Government Development (Alfarraj et al., 2011)**

3. <u>**Five-Stage of E-Government Service Development (Hassan, 2011)**</u>

The UN five-stage E-Government development index attempts to evaluate the performance level of countries through the model of maturity, as shown in **Figure 10**. According to Hassan (2011), the model makes assumptions about countries typically starting an emerging online presence (basic websites) based on general observation and reflection among experts. This further progresses to a developed state and through a transactional level with many online services provided and, finally, to an integrated web-connected state. Below are detailed analyses by Hassan (2011) on the UN five-stage E-Government development level.

a. **Emerging**: At this stage of E-Government development, there is an online presence of a website, and it provides basic information on governance, public policy, laws, and regulations. It has links to other government departments and ministries, but the level of interaction is very low at this stage as much of the information presence on the websites is static (Adeyemo, 2011).

b. **Enhanced**: There is more information available online at this stage of E-Government development from government websites compared to the earlier emerging stage. Government websites are delivering one-way or simple two-way enhanced e-communication between government and the citizen, according to Hassan (2011). Services such as downloaded forms, email addresses, and newsletters are provided.

c. **Interactive**: This stage of E-Government development includes a more sophisticated level of formal interaction between citizen and government. Also, Fath-Allah et al. (2014) concluded that this stage provides simple communication between the public and government agency that includes the basic search engine, email system, and officially downloaded forms.

d. **Transactional**: As earlier discussed, the transaction stage focuses on building self-service applications for the public to access online (Griffin et al., 2007). This stage, according to Fath-Allah et al. (2014), allows the public to perform online transactions such as personal information updates, course registrations, and licence applications. Financial transactions are also handled at this level.

e. **Connected**: The connected stage of the UN five-stage E-Government development represents the most advanced level of government inventiveness. According to Alfarraj et al. (2011), a government at this level has an online presence to meet the demand of the public at a comfortable and modernised way. Hassan (2011) added that a government at this level creates a robust environment in which to empower the public to be more familiar with government activities, and for the citizens to have an input in decision-making activities.

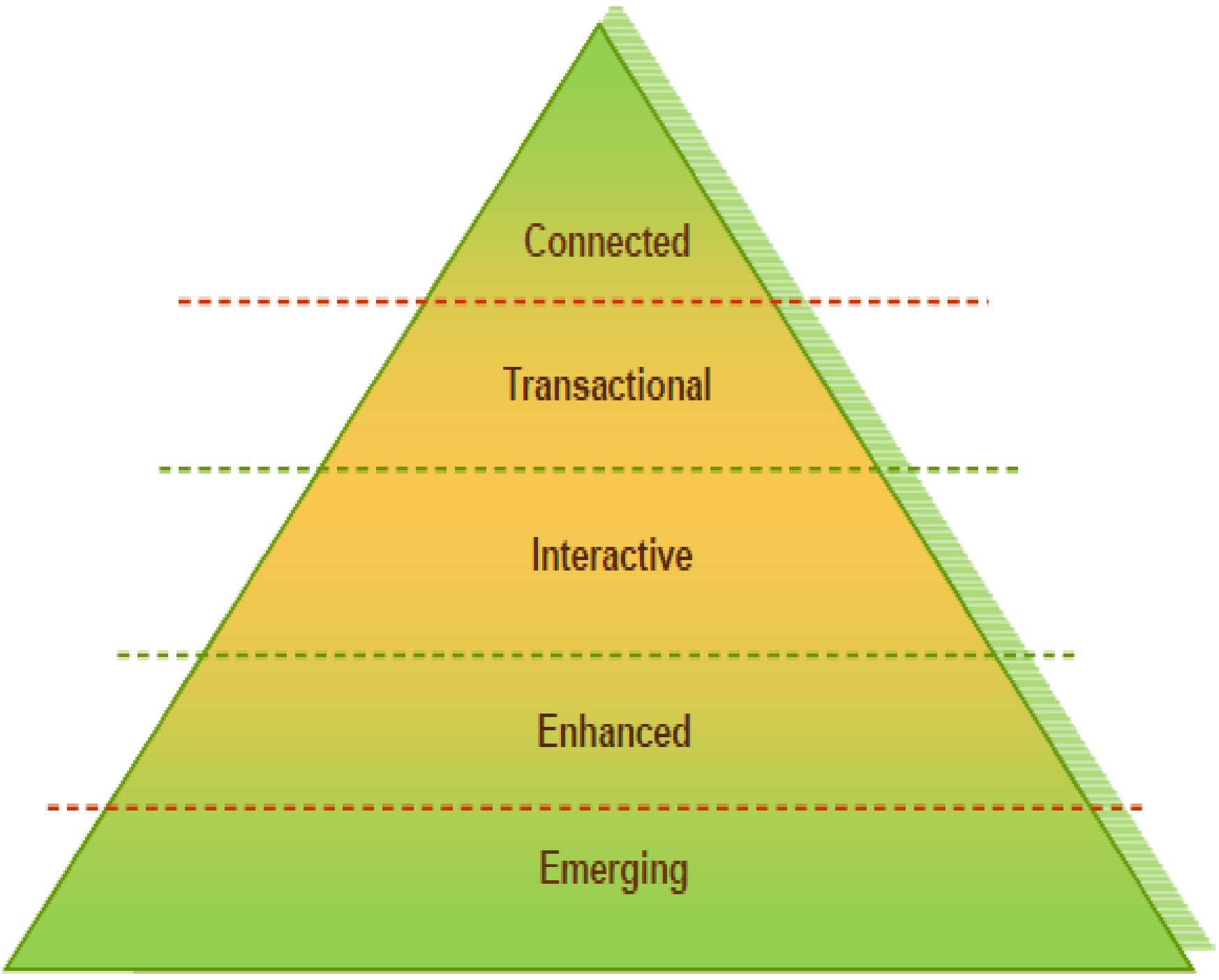

**Figure 10 The 5-Stage of E-Government Service Development (Hassan, 2011)**

In addition to the various E-Government Service Development stages shown above, **Table 5** shows other E-Government Service Development stages. Layne and Lee's four-stage model comprises of cataloguing, transaction, vertical and horizontal integration. The cataloguing level according to Alfarraj et al. (2011) includes an online presence and availability of information for the public. It is also possible for the citizens to download forms from the websites. The transaction level, as presented in other models, allows the direct dealings between government and the citizens through online interfaces. The vertical integration level of the Layne and Lee four-stage model allows all the tiers of government, namely local, state and federal, to be connected to various government functions and services to effect permanent changes in the concepts and processes of the government.

In the horizontal integration according to Hassan (2011) and Alfarraj et al. (2011), there is direct communication between the databases across different government functionalities with the purpose of sharing information. Hiller and Belanger also identified four stages of E-Government Service Development, which is very similar to the Layne and Lee model. The information stage is similar to the Cataloguing stage of the Layne and Lee model, while the second stage, which is the transaction, is also similar to that of Layne and Lee's model. However, Hiller and Belanger's model added a two-way communication to their second stage and the last level, which is an integration that is similar to both the vertical and horizontal integration of Layne and Lee's model.

The World Bank three-stage E-Government development model comprises of publishing, interact and transact. As discussed above, the publishing stage is the basic level where governments publish little information on their websites, such as the vision and mission of the agency. The interact level involves the communication between the government and its citizens. The stage includes the availability of services such as downloaded forms on government websites for the citizen's usage. The transact stage of E-Government development allows the public to perform online transactions such as e-payment, updating personal information, course registrations and licence applications.

**Table 5 Summary of Other E-Government Development Stages (Griffin et al., 2007)**

<table>
<tr><th>Layne and Lee</th><th>Hiller and Belanger</th><th>World Bank</th><th>Gartner</th></tr>
<tr><td>Cataloguing</td><td>Information</td><td>Publish</td><td>Presence</td></tr>
<tr><td rowspan="2">Transaction</td><td>Two-way Communication</td><td>Interact</td><td>Interaction</td></tr>
<tr><td>Transaction</td><td rowspan="3">Transact</td><td>Transaction</td></tr>
<tr><td>Vertical Intergration</td><td rowspan="2">Intergration</td><td rowspan="2">Transformation</td></tr>
<tr><td>Horizontal Intergration</td></tr>
</table>

Lastly, the Gartner four-stage E-Government Development model, according to Griffin et al. (2007), has the presence stage. The primary function at this first level of the model is to have basic information about the government available to the citizens online. Such information that is relevant to the public includes the address, mission, and opening hours. The second level of this model is the interaction, basic search capabilities of which are provided (Griffin et al., 2007; Alfarraj et al., 2011), as well as links to other agencies of the government. The transaction stage, which is the third level, focuses on providing self-service applications available online for the citizens. The fourth level refers to the transformation stage, and this stage redefines the delivery of government services. Griffin et al. (2007) added that this stage provides a single transparent medium to the citizens by the government.

In light of the above, it has been argued traditionally that most of the reviewed E-Government development models are already implemented in many developed countries (Hassan, 2011). The researcher believes that these models are not able to represent the progress of E-Government development in a developing country such as Nigeria, where the cost involved in providing the latest technology is still very high. Moreover, Zarei et al. (2008) argued that, with different technological infrastructures in the developing countries, there is a need for more modified E-Government development models.

## 2.3 E-Service: An Overview

Electronic Service, shortened as 'E-Service', refers to any service provided by any electronic means, for example services provided via the internet/website, mobile devices or kiosk (Bhuiyan, 2011). Ruyter et al. (2001) examined E-Service as a term that represents content-centred and interactive internet-based customer service "driven by the customer and integrated with related organisational customer support processes and technologies with the goal of strengthening the customer-provider relationship." Moreover, Sukasame (2004) argued that E-Service is not only about electronic and service, but that the real e-service operation may comprise part, if not all, the interaction between service provider and the customer carried out via the internet. The E-Service definition was also substantiated by Surjadjaja et al. (2003). E-Service has moved from the usual manual and traditional way of rendering a service, to an electronic service provision for the public. There are several reasons for implementing these services; for example, airline ticketing has gone from its traditional manual way of purchasing, to an intelligent web-driven service. Many companies have seen their profits doubled through the use of online services in their operation (Oseni and Dingley, 2015). A typical example, as highlighted in the previous chapter, is Hewlett Packard (HP), which is rapidly transforming its after-sales business into a profit-generating E-Service business unit (Ruyter et al., 2001).

### 2.3.1 E-Service Definition, Significance, Scope, and Nature

Recently, considerable literature has grown up around the theme of E-Government service as it has been described as a positive change driven by the use of technology (Hassan, 2011). Recent developments in the field of E-Services for the public at the local government level have led to a renewed interest in looking at the core value of E-Government services and the ability of the stakeholders to have access to government services at any time (Hassan, 2011). Kumar et al. (2007) suggest that the quality of services provided to businesses and public could be improved drastically with E-Service, and that this can enhance the E-Service general acceptance. In this regard, **Table 6** below further reviews various E-Service definitions found in the literature.

**Table 6 Definitions of E-Service**

| Authors | E-Service Definition |
|---|---|
| **Ruyter et al. (2001)** | E-Service represents content-centred and interactive internet-based customer service, driven by the customer and integrated with related organisational customer support processes and technologies with the goal of strengthening the customer-provider relationship. |
| | |
| **Sukasame (2004)** | E-Service as a term is not only about "electronic" and "service", but the true E-Service operation may be that where part, if not all, interaction between service provider and customer is done via the internet. |
| | |
| **Kelleher and Peppard (2009)** | E-Service can be broadly viewed as an interactive information service where services are provided or consumed using internet-based or electronic systems, and where the service organisation and customer(s) use the information gathered about each other to co-create a better service experience. |
| | |
| **Bhuiyan (2011)** | E-Service in a simple term refers to any service that is provided by any electronic means. |
| | |
| **Hassan et al. (2011)** | E-Service is defined as the provision of service over electronic networks such as the internet. |

The scope and significance of E-Service involved in providing the opportunities for individuals, and private and public organisations for deploying and developing online services will enhance customer/citizen interactions and experience. According to Kelleher and Peppard (2009), there is a need for organisations to envisage deploying E-Service delivery strategy, as this will not only give them the competitive advantage, but will propel them to success. However, the implementation of E-Service comes with an unavoidable cost, and it is useful to obtain important feedback from the customers to improve the service. A dedicated online customer service agent in any organisation benefits from E-Services. Furthermore, the deployment and delivery of an E-Service system has been associated with an increase in the ways businesses and their customers interact. Rowley (2006) concluded that E-Service delivery contributes to the total service experience, and not just the consumer's experience and evaluation of E-Service. Schware and Deane (2003) discuss major public E-Service priorities, and they are of the opinion that form processing, public complaints, information access, procurement, customer response and polling are the most valuable services and the reasons why the general public goes online. More details of these public services are explained in **Table 7**.

**Table 7 Summary of Public E-Services Priorities**
**(Schware and Deane, 2003)**

| Public E-Services | Definition |
|---|---|
| **Form Processing** | To enable the public to submit and process various applications electronically. |
| **Public Complaints** | To provide a single interface for the public to communicate complaints effectively and conveniently. |
| **Information** | To provide public access to information, statistics, budgets, legislations and laws. |
| **Payments** | To enable electronic transfer of funds between the public and government for the payment of services fees and fines. |
| **Procurement** | To obtain information, purchase orders and requisition forms electronically. |
| **Customer Response** | To provide the public with help information. |
| **Polling** | To provide secured channels for citizens' opinion polls, voting and government decision-making participation. |

### 2.3.2 Types of E-Service

There are many E-Services available today at local government level in most developed and developing countries. E-complaint, E-assessment, E-licencing, E-forum, E-submission, E-rental are different types of Local Government E-Services according to Khadaroo et al. (2013).

a. **E-Complaint**: This is an online service which provides citizens/public, including government contractors, the facility to lodge complaints against any issue related to the services rendered. The complaints may even be against staff behaviours or suspected fraud perpetrated against the government. This facility enables quick and efficient communication between citizens and government. Also, it provides a faster feedback to improve the services. In the UK, this facility is very common and is found on the Royal Court of Justice website; there is always a procedure to go about it online. When the e-complaint is used in conjunction with the emailing and telephoning systems, citizens can make a complaint about a judge, tribunal members, or a magistrate, and every complaint received is treated in confidence by a complaint advisory committee (Khadaroo et al., 2013).

b. **E-Assessment**: This online service uses technology for prompt delivery and managing assessment. E-Assessment is used in an educational setting to assess students, and private organisations make use of this facility when recruiting new staff or conducting promotional exercises. Recently, the Federal Civil Service in Nigeria introduced an E-Assessment exercise for top civil servants during annual promotion examinations (FRN, 2016). There has been a high level of E-Service adoption at the Federal level in Nigeria (Azenabor, 2013). Although the E-Assessment exercise has been a big success, there may be some setbacks, especially in the use of the technology by some staff. The federal government has set up IT training sessions for the staff through the federal civil service commission. Currently, the E-Assessment facility is still lacking at the local government level in Nigeria due to barriers that this study has been able to address. The use of E-Assessment facility offers a method for immediate feedback, and it could help students or other users to improve their knowledge and performance. It could be accessed from any location in the world, and it saves time with exclusive functionality.

c. **E-Licencing**: The E-Licencing is an online facility that allows the issuance or renewal of licences online. Drivers could make use of this service as it provides expected processing and it saves time. This was introduced to remove the bottleneck faced with paper applications, and it provides an interactive interface. The users would be allowed to register for the facility online with the creation of the username and a password; this is a secured online facility, as it welcomes payment transactions either through card payments or by the use of PayPal. Users could check the status of their permit application when logged on, while other services under the E-Licencing facility include the issuance of birth and death certificates, renewal of council permits, business licences and so on (Khadaroo et al., 2013).

d. **E-Forum**: This is an electronic discussion forum used in sharing information and knowledge. This facility could be set up by the individual or a group for businesses or research purposes. A typical example of the E-Forum is an e-forum for Chinese teachers developed to improve knowledge sharing and networking (Khadaroo et al., 2013).

### 2.3.3 Local Government Structure in Nigeria and E-Government Service

In this study, the researcher has taken into consideration Nigeria's 1999 Constitution on the functions of the local government, as this will give us more information about the types of E-Government services. **Figure 11** below provides more insight into various E-Services that should be available at local environment level in Nigeria.

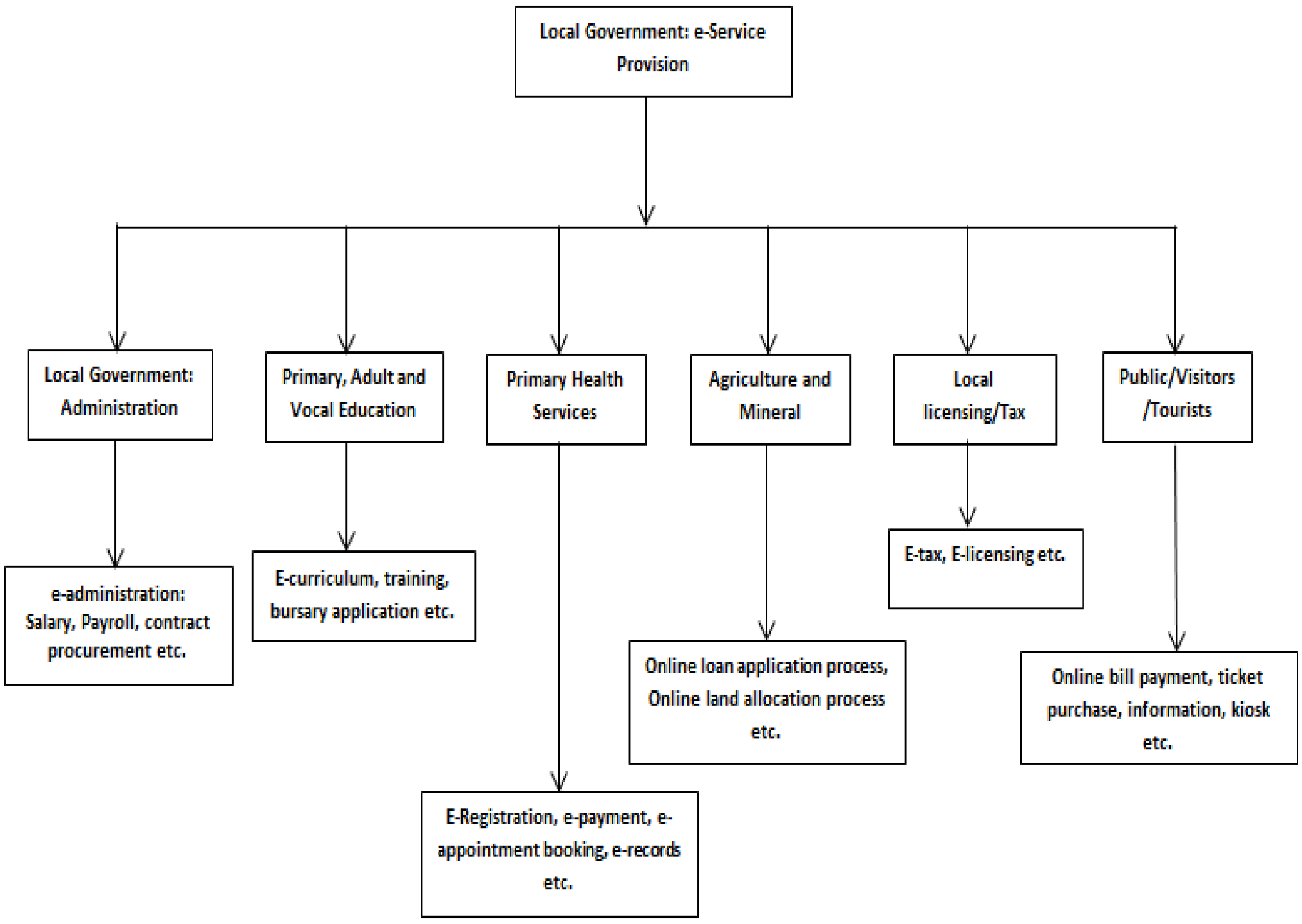

**Figure 11 Nigeria Local Government E-Services Provision (CFRN, 2016)**

The Nigerian Constitution (1999) is very clear on the functions of the local government to the citizens. As shown in **Figure 11**, it is pertinent to note that there should be the following E-Government services among others at the local government level in Nigeria:

- **E-Licencing**
- **E-Tax**
- **E-Curriculum**
- **E-Payroll**
- **E-Payment**
- **E-Registration**
- **E-Health**
- **E-Procurement, etc.**

Recent evidence suggests (Mundy and Musa, 2010; Azenabor, 2013) that the low adopting of E-Service and limited availability of a functional website for the local government has contributed to the present predicament, as these E-Services mentioned above are not currently available at the local environment level in Nigeria. Local governments with websites in Nigeria are still in the early stages of E-Government development, as mentioned by Mundy and Musa (2010). These websites are at E-Government stage, known as the Publish Stage, where governments publish very basic and limited information on their websites (Oseni and Dingley, 2015). The researcher believes the situation would be improved if the suggested solutions in this research study to the barriers facing E-Service Adopting at the Local Government Level in Nigeria are taken seriously and acted upon by the government in collaboration with public/private sectors.

## 2.4 Barriers, Solutions and Success Factors of E-Government Service Development

The low level of E-Service adoption in Nigeria (Mundy and Musa, 2010; Azenabor, 2013) is due to various barriers, as reviewed in the literature. Moreover, despite the increased rate of adoption in developed countries (Hassan, 2011), the implementation level varies from one country to another. The adoption of E-Service in Nigeria according to Adeyemo (2011) will result in excellent and efficient public service delivery by making information about the Government available promptly to the citizens. E-Services reduce the level of corruption cases due to accountability and transparency (Oseni and Dingley, 2014).

The barriers, solutions and the success factors to the E-Government service adoption and implementation show that the reviewed literature has taken into consideration the gaps between the developing and developed countries (see section 2.1.1 for the reviewed literature). The implementation of E-Government faces many problems and obstacles, including the literacy of the users, and the ability to use a computer without any assistance (Azenabor, 2013). As illustrated in **Figure 12** below, leadership, finance, and legislation issues are other barriers that are facing the implementation of E-Government in the West Cape of the South Africa (Matavire et al., 2010). The progress of adopting and implementation is slower in the developing countries, and developed countries could overcome these barriers more quickly than the developing countries (Hassan, 2011).

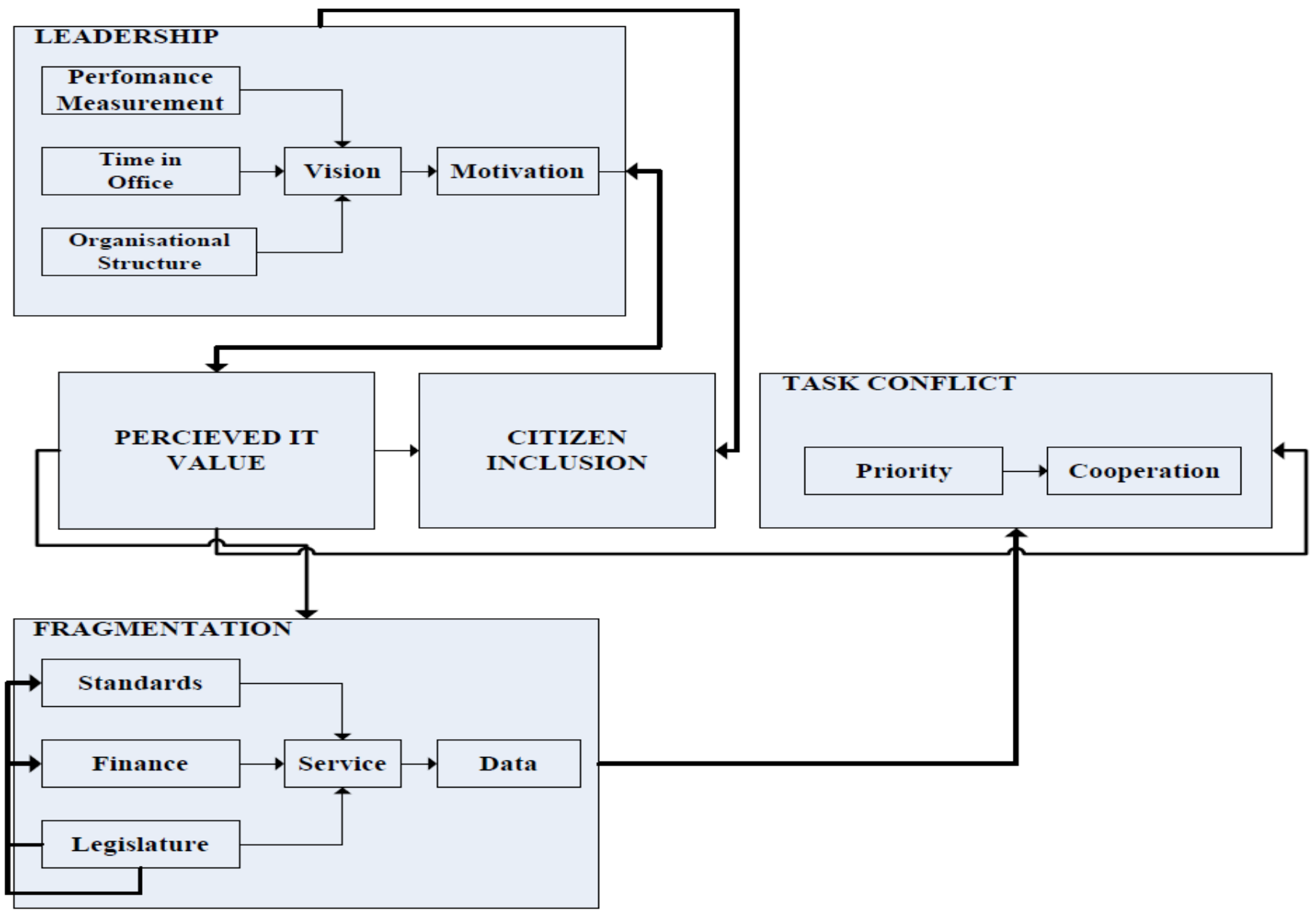

**Figure 12 Barriers Affecting the Implementation of E-Government in the West Cape, South Africa**
**(Matavire et al., 2010)**

### 2.4.1 Barriers Identified from the Literature

In reviewing the literature on the barriers facing E-Government Service adopting and implementation, there is a need to consider the following:

1. The barriers are classified in many different ways, such as barriers to adopting or diffusion (Hassan, 2011), based on diverse perspectives on the level and position of E-Government in various countries, either local, state or federal, and the involvement of stakeholders (Munda and Musa, 2010; Azenabor, 2013).

2. Previous studies conducted within the E-Government research domain to identify and classify the barriers facing the adopting and diffusion had different objectives in their barriers classification. For instance, Alateyah et al. (2013) classified the barriers as Technical Infrastructure, Lack of Awareness, Privacy Issues, and Culture. Moreover, Hassan (2011) added that the barriers are also classified according to the research context in which the research is applied (for example, the Barriers or Challenges facing E-Government in Developing Countries, Africa or Asia). The barriers are also classified into other major categories, according to Alshehri and Drew (2010); Mundy and Musa (2010); Hassan (2011); Alshehri et al. (2012); Khan et al. (2012); Alateyah et al. (2013); Al-Shboul et al. (2014); Abdelkader (2015). These barriers are:

- **Political Barriers**
- **Administrative/Organisational Barriers**
- **Cost, Economic and Resource Barriers**
- **Technological Barriers**
- **Legislative Barriers**
- **Cultural Barriers**

3. The research evidently recognises that there are contextual issues in identifying the barriers facing the E-Government Service Adopting and Implementation. Therefore, the results from the literature may not offer practical and novel guidelines for researchers and practitioners' future plans concerning E-Government adopting in developing countries.

However, the literature does not take into account corruption in the full context as the major barrier facing E-Government Service adoption or diffusion in developing countries. Neither does the literature examine the few other barriers in-depth, as shown in **Table 8** below, while critical barriers were merged. The researcher is of the opinion that, for a better and fair result, there is a need to conduct an in-depth study. The in-depth study would include interviews, an online survey, and an online focus group. The findings on some major barriers could help in future adopting and implementation plans. The barriers identified in the literature are presented in **Table 8** below.

**Table 8 Barriers Identified from the Literature for E-Government Services Adopting**

| No | Barriers | Authors and Year |
|---|---|---|
| **1.** | Organisational or staff resistance/ Resistance to change to electronic ways | Schwester (2009), Alshehri and Drew (2010), Matavire et al. (2010); Mundy and Musa (2010), Jouzbarkand et al. (2011), Bhuiyan (2011), Alshehri et al. (2012), El-Sofany et al. (2012), Azenabor (2013), Nkohkwo and Islam (2013), Al-Shboul et al. (2014), Abu-Shanab and Bataineh (2014). |
| **2.** | Budget/Finance | Schwester (2009), Alshehri and Drew (2010), Matavire et al. (2010); Mundy and Musa (2010), Bhuiyan (2011), Jouzbarkand et al. (2011), Hassan et al. (2011), Alshehri et al. (2012), El-Sofany et al. (2012), Ashaye and Irani (2013), Azenabor (2013), Nkohkwo and Islam (2013), Al-Shboul et al. (2014), Abu-Shanab and Bataineh (2014), Abdelkader (2015). |
| **3.** | Privacy/Security and Trust Issues | Schwester (2009), Mundy and Musa (2010), Hassan et al. (2011), Alshehri et al. (2012), Rehman e. al. (2012), Alateyah et al. (2013), Al-Shboul et al. (2014), Ashaye and Irani (2013), Abdelkader (2015). |
| **4.** | Cultural | Carter and Weerakkody (2008), Hassan et al. (2011), El-Sofany et al. (2012), Alateyah et al. (2013), Azenabor (2013), Abdelkader (2015). |
| **5.** | Lack of IT Skills/Literacy/Training | Carter and Weerakkody (2008), Alshehri and Drew (2010), Mundy and Musa (2010), Alshehri et al. (2012), El-Sofany et al. (2012), Alateyah et al. (2013), Al-Shboul et al. (2014). |
| **6.** | Lack of ICT Infrastructure | Alshehri and Drew (2010), Mundy and Musa (2010), Bhuiyan (2011), Hassan et al. (2011), Alshehri et al. (2012), Azenabor (2013), Nkohkwo and Islam (2013), Al-Shboul et al. (2014), Abu-Shanab and Bataineh (2014). |
| **7.** | Lack of partnership and collaboration | Alshehri and Drew (2010), Matavire et al. (2010); Bhuiyan (2011), Alshehri et al. (2012), Al-Shboul et al. (2014). |
| **8.** | Lack of policy and regulation/ Inadequate policies on freedom of information | Alshehri and Drew (2010), Matavire et al. (2010); Bhuiyan (2011), Hassan et al. (2011), Jouzbarkand et al. (2011), El-Sofany et al. (2012), Alshehri et al. (2012), Al-Shboul et al. (2014), Abdelkader (2015). |
| **9.** | Lack of strategic planning | Alshehri and Drew (2010), El-Sofany et al. (2012), Azenabor (2013), Al-Shboul et al. (2014). |
| **10.** | Lack of Leadership and Management Support | Matavire et al. (2010); Mundy and Musa (2010), Alshehri and Drew (2010), El-Sofany et al. (2012), Azenabor (2013), Zhang et al. (2014). |
| **11.** | Internet Connection/ Cost of Internet usage | Mundy and Musa (2010), Bhuiyan (2011), Alshehri et al. (2012), Alateyah et al. (2013), Azenabor (2013). |
| **12.** | Accessibility | Mundy and Musa (2010), Alateyah et al. (2013). |
| **13.** | Unstable power supply | Mundy and Musa (2010), Bhuiyan (2011), Ashaye and Irani (2013), Azenabor (2013). |
| **14.** | Lack of interoperability between IT systems | Bhuiyan (2011), Jouzbarkand et al. (2011). |
| **15.** | Implementation and maintenance of e-service systems | Mundy and Musa (2010), Jouzbarkand et al. (2011), Azenabor (2013), Abdelkader (2015). |

### 2.4.2 Solutions Identified from the Literature

There is a growing body of literature that recognises the importance of E-Government in public administration (Mundy and Musa, 2010; Hassan, 2011; Fatile, 2012; Nkohkwo and Islam, 2013; Oseni and Dingley, 2014) and this will improve the system in terms of efficiency. However, the potential of E-Government to bring much-needed transformation to the public service has been hampered by many barriers. The fact that E-Government Service development progress has been hindered by barriers, the implementing of solutions encourages and motivates the E-Service development. Literature has identified some solutions to the barriers facing E-Service adoption and implementation, which are the motivating influences to successful E-Government initiatives.

From the study conducted by Oseni and Dingley (2014), the availability of funds as identified in **Table 9** could serve as an enabler to E-Government Service development barriers. The authors are of the opinion that more money should be allocated to E-Government service initiatives in the country's yearly budget, as presented to the legislative arm of government by the executive arm. If any country is to achieve expected and desirable results from E-Government initiatives, plans and strategies are needed to be mapped out to consolidate the initiatives, officials should eschew corruption, and dishonest IT project contractors should be phased out.

Another significant analysis and discussion on the solutions to the barriers facing E-Service adoption and implementation were presented by Tarus et al. (2015), whereby they elaborated on the need for collaborations and partnerships with other successful stakeholders in E-Service provision to battle the barriers facing the implementation of E-Learning for teachers in Kenya. Collaborations and partnerships, according to the authors, will encourage best practices in the sharing of useful resources, and will further strengthen technical and human capacity building for teaching, learning, and research. The solutions to the barriers identified in the literature are presented in **Table 9** below.

**Table 9 Solutions to the Barriers Identified from the Literature for E-Government Services Adopting**

| No | Solutions | Authors and Year |
|---|---|---|
| 1. | Provision of appropriate legislation and policy for e-government | Mundy and Musa (2010), Jouzbarkand et al. (2011), Abdelkader (2015), Al-Shboul et al. (2014). |
| 2. | Increase broadband internet lines/Low cost of internet usage | Mundy and Musa (2010), Bhuiyan (2011), Jouzbarkand et al. (2011), Alshehri et al. (2012), Alateyah et al. (2013), Azenabor (2013). |
| 3. | Funding | Alshehri and Drew (2010), Mundy and Musa (2010), Bhuiyan (2011), Hassan et al. (2011), Alshehri et al. (2012), El-Sofany et al. (2012), Azenabor (2013), Ashaye and Irani (2013), Nkohkwo and Islam (2013), Al-Shboul et al. (2014), Abu-Shanab and Bataineh (2014), Abdelkader (2015). |
| 4. | Cultural integration | Carter and Weerakkody (2008), Hassan et al. (2011), Alateyah et al. (2013), Azenabor (2013), Al-Shboul et al. (2014), Abdelkader (2015). |
| 5. | Improve awareness level | Alshehri and Drew (2010), Khan et al. (2012), El-Sofany et al. (2012), Al-Shboul et al. (2014), Abu-Shanab and Bataineh (2014). |
| 6. | IT Skills/Literacy/Training of employees | Carter and Weerakkody (2008), Alshehri and Drew (2010), Mundy and Musa (2010), Jouzbarkand et al. (2011), Alshehri et al. (2012), El-Sofany et al. (2012), Alateyah et al. (2013). |
| 7. | Partnership and collaboration/investment between government, private and public. | Alshehri and Drew (2010), Jouzbarkand et al. (2011), El-Sofany et al. (2012), Abdelkader (2015), Tarus et al. (2015). |
| 8. | Provision of ICT infrastructure | Alshehri and Drew (2010), Mundy and Musa (2010), Bhuiyan (2011), Hassan et al. (2011), Alshehri et al. (2012), El-Sofany et al. (2012), Azenabor (2013), Ashaye and Irani (2013), Nkohkwo and Islam (2013), Al-Shboul et al. (2014), Abu-Shanab and Bataineh (2014). |
| 9. | Improve on Security, Trust and Privacy issues. | Alshehri and Drew (2010), Mundy and Musa (2010), Bhuiyan (2011), Hassan et al. (2011), El-Sofany et al. (2012), Alshehri et al. (2012), Azenabor (2013), Nkohkwo and Islam (2013). |
| 10. | Leadership/clear strategic plans for e-government | Alshehri and Drew (2010), Azenabor (2013), Al-Shboul et al. (2014), Ashaye and Irani (2013), Nkohkwo and Islam (2013), Al-Shboul et al. (2014), Abu-Shanab and Bataineh (2014). |

### 2.4.3 Success Factors Identified from the Literature

Recent advances in the field of E-Government development have led to a renewed interest in the research domain. However, despite the importance in public service delivery, the implementation of E-Service at the local government level in developing countries is highly problematic (Nabafu and Maiga, 2012; Elkadi, 2013). Success factors identified from the literature as shown in **Table 10** are the benefits derived from the successful adopting and implementation of the E-Service. Ziemba et al. (2013) describe the success factors as the "social, political and economic conditions and a correlation between country's conditions with its level of E-Government maturity". A matured and successful E-Government enables government services to be more accessible, convenient, are cost reductive, improve the quality of decision-making and open the door for an accountable and transparent government. Successful E-Government services implementation creates robust information sharing and dissemination in the public service, while improving record management. The success factors for the E-Government Services adopting and implementation identified in the literature are presented in **Table 10**.

**Table 10 Success Factors Identified from the Literature for E-Government Services Adopting**

| No | Success Factors | Authors and Year |
|---|---|---|
| 1. | Quality of service delivery | Kumar et al. (2007), El-Sofany et al. (2012), Fatile (2012), Abu-Shanab and Bataineh (2014). |
| 2. | Cost reduction and efficiency gains | Ruyter et al. (2001), Kumar et al. (2007), El-Sofany et al. (2012), Fatile (2012), Nabafu and Maiga (2012), Abu-Shanab and Bataineh (2014). |
| 3. | Network and community creation /Bridging the digital divide | Kumar et al. (2007), Hassan (2011), Nabafu and Maiga (2012), Abu-Shanab and Bataineh (2014) |
| 4. | Transparency, anti-corruption, and accountability | Mohammad et al. (2009), Fatile (2012), Nabafu and Maiga (2012), Adeyemo (2013), Ziemba et al. (2013), Abu-Shanab and Bataineh (2014). |
| 5. | Improve the quality of decision-making | Kumar et al. (2007), Abu-Shanab and Bataineh (2014) |
| 6. | Internet usage increase | Hassan (2011), Gil-Gracia (2013), Ziemba et al. (2013) |
| 7. | ICT infrastructure development | Hassan (2011), Ziemba et al. (2013) |
| 8. | Improved records management | Hassan (2011), Ziemba et al. (2013) |

The literature has raised important questions about the desire of the researchers and stakeholders in the E-Government services deployment to have full digitisation of the public sector (Hassan, 2011). However, the transformation has been confronted by significant barriers, especially in developing countries. There are no precise IT plans and strategies, poor ICT infrastructures, and lack of both human and financial resources, a high level of corruption, as well as the absence of IT policy and regulations. The belief is that the barriers, solutions and the success factors for E-Government services development that are mentioned in the literature will help in the next stage of this research.

## 2.5 Electronic Government Adopting and Implementation

Different attempts have been made to implement E-Government models and initiatives to facilitate organisation change (Nabafu and Maiga, 2012). E-Government, according to Safeena and Kammani (2013), represents an essential change in the structure of the public sector, and the culture and value of conducting business through the use of technology by the government agency. Information and communication technology (ICT) has been a major factor in the reform of the public sector, as it reduces the cost of governance and improves performance. Azenabor (2013) emphasised that the major purpose of E-Government adoption is to phase out paper and manual administration via the use of information and communication technology facilities, and to reach out to the public through the utilisation of the internet for better information sharing and dissemination. In constructing the E-Service adoption model, there is a need for practitioners to thoroughly investigate the barriers hindering the initiatives, as the literature has indicated that the implementation success varies in different environments (Nkohkwo and Islam, 2013).

The challenges facing E-Government adoption and implementation, especially in the developing countries, tend to be more pronounced, as the stakeholders see the innovation as mere computerisation of government operations. According to Al-Shboul et al. (2014), the transformation in E-Government services is seen as a reform that needs a restructuring process. There should be a clear vision for the E-Service process, while the government, in collaboration with private sectors, needs to devote time and resources to the successful implementation. Above all, the adoption and implementation of E-Government initiatives should be targeted at facilitating a stress-free government service delivery and improving citizen participation. A further discussion on E-Government adoption and implementation in developed and developing countries will be included in the next section, as well as the current status of E-Government adoption and implementation in Nigeria.

### 2.5.1 Electronic Government Adopting and Implementation in Developed Countries

The development and implementation of E-Government in developed countries have been on a high and encouraging level. This cannot be compared to the developing countries due to the unavailability of efficient ICT facilities. Other obstacles are the high levels of literacy and lack of government political will to support E-Government initiatives (Hassan, 2011; Anazebor, 2013; Nkohkwo and Islam, 2013). Rokhman (2011) as cited by Anazebor (2013) argued that, in order to have successful E-Government initiatives, the citizens' desire and willingness to adopt E-Government services will complement the efforts and support of the government.

Bidgoli (2004) reported that the United States E-Government groundwork laid out in the 1980s the introduction of the Paperwork Reduction Act, which mandated an information resources management approach to the federal data. This established, for the very first time, a single policy E-Government framework at the national level. Furthermore, Bidgoli (2004) confirmed that the use of the internet by the government in the United States (US) is on the increase. Schlæger (2013) argued that E-Government development in China is now moving towards local E-Government and there has been an increase in the local level E-Government policy and implementation. The author added that, despite the new interest in the local E-Government in China, there are barriers confronting the successful implementation such as the lack of relevant laws and regulations. Hence, there is no clear internal procedures and rules regarding automated processing in China. However, despite the critical issue, Schlæger (2013) acknowledged that there is an effective method to solving the barriers facing E-Government Services adoption (Schlæger, 2013).

| Country or region | Examples of e-government practices by category | | | | |
|---|---|---|---|---|---|
| | G2C | G2B | G2G | IEE | Cross-cutting |
| The U.S. | GovBenefits.gov: providing a single point of access for citizens to locate and determine potential eligibility for government benefits and services. | Federal Asset Sales: creating a single, one-stop access point for businesses to find and buy government assets. | e-Grants: providing a single, online portal for all federal grant customers to access and apply for grants. | Government Human Resource Integration: streamlining and automating the exchange of federal employee human resources information. | The e-Authentication project: providing a secure infrastructure for online transactions. |
| The Europe Union | Single Point of Access for Citizens of Europe (an EU-project): supporting citizens' travel within Europe. | The Net-Enterprises Project (France): allowing enterprises, through Internet, to send standardized notifications to government agencies. | Interchange of data between administrations (the IDA program): networking of public administrative units. | Government Secure Intranet (UK): a governmentwide communications infrastructure for joined-up government. | IDA e-Link: a communication middleware solution to enable reliable and secure information exchanges among administrative units across Europe. |
| Singapore | eCitizen Portal: providing a single access point to government information and services, which are organized and integrated in intuitive categories. | G2B Portal: the entry point for all local and international businesses to access a full suite of aggregated and integrated G2B information and services. | GeBIZ Enterprise: coordinating the purchasing needs of the public sector procurement officers. | InfoComm Education Programme (IEP): facilitating learning and enabling public officers to appreciate and work toward the objective of a "Networked Government." | Singapore Personal Access (SingPass): a nationwide personal authentication framework for e-services. |
| South Korea | Home Tax Service (HTS) via the Internet: providing 24/7 online service such as tax declaration and payment. | Integrated e-Procurement System: a single procurement window, allowing all procurement related processes electronic. | Integrated National Finance Management System: a system for information sharing and linkage for finance related institutions. | Integrated Administration Information System in Local Government: promoting the application of information systems for all administrative affairs. | Government e-Signature & e-Seal System: securing reliability for information distribution and e-administration such as private information protection and security. |
| Taiwan | Online motor vehicle services system: providing 21 applications and payment services to individual citizens. | Government Procurement Information Center: enabling government procurement with businesses much more transparent and efficient. | Interdepartmental E-mail Delivery infrastructure: delivering official messages via electronic delivery systems not bound by time and geographical constraints. | Online Central Personnel Administration: improving the administrative efficiency in government human resource management. | Government Root Certification Authority (GRCA): providing the public, businesses, and government agencies with secure and error-free means of making online applications and transmitting data. |

**Figure 13 E-Government Practices among the Selected Leading Countries (Lee et al., 2005)**

It is interesting to note that the Chinese Government Online Project (GOP), among other projects, which will enable the public to acquire information and other public services through the internet. This process will encourage electronic procurement and provide more efficient communications across government agencies and the public. It will also accelerate the use and acceptance of IT in China. **Figure 13** above shows the E-Government practices among the selected developed leading countries with their regional ranking. Furthermore, the EU has been a major administrative and political priority since the 1990s (Lee et al., 2005) by using the E-Government services. As shown in **Figure 13**, the EU countries were ranked second according to a UN report in the geographic regions. The European Commission (EC) had been the primary motivator of E-Government services among the EU countries to spread the benefits of information society across EU member states in the 1990s (Lee et al., 2005). As part of the E-Government initiative in the UK, and to boost the cyber security for UK businesses, the government through the Department for Business Innovation and Skills announced a new scheme to help businesses stay safe online. This system is also applicable to other organisations, such as universities, as it will encourage a more online presence in the UK, boost confidence and profitability (Nkohkwo and Islam, 2013).

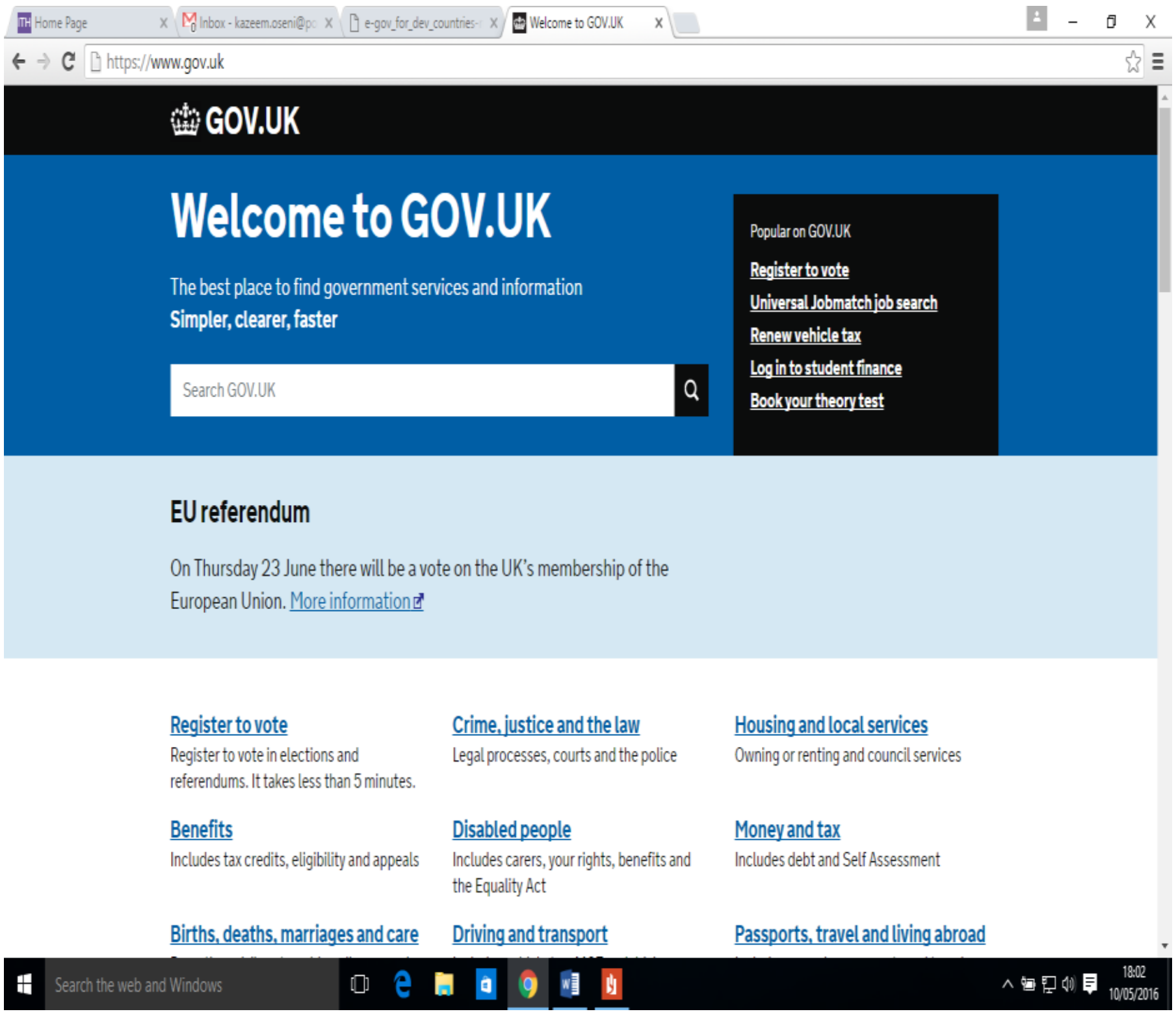

**Figure 14 United Kingdom E-Government Portal**

The UK E-Government service portal (**Figure 14**) displayed information on various UK online government services. Relevant information to the public such as voters' registration, vehicle tax, student finance, booking a driving theory test, marriage, birth, and death certificate bookings, and passport and immigration services can be obtained from the portal. There are E-forms available to the public to download ranging from newly established right to rent, complaints, benefits, housing, pension issues, council tax and many others. The public could use the search engine to search other services, including job vacancies, and these services are easier and faster to use. **Figure 15** below shows the world E-Government leaders with a Very High E-Government Development Index (EGDI) according to the 2016 United Nations Survey.

| Country | Region | EGDI | 2016 Ranking |
|---|---|---|---|
| United Kingdom | Europe | 0.9193 | 1 |
| Australia | Oceania | 0.9143 | 2 |
| Republic of Korea | Asia | 0.8915 | 3 |
| Singapore | Asia | 0.8828 | 4 |
| Finland | Europe | 0.8817 | 5 |
| Sweden | Europe | 0.8704 | 6 |
| Netherlands | Europe | 0.8659 | 7 |
| New Zealand | Oceania | 0.8653 | 8 |
| Denmark | Europe | 0.8510 | 9 |
| France | Europe | 0.8456 | 10 |
| Japan | Asia | 0.8440 | 11 |
| United States of America | Americas | 0.8420 | 12 |
| Estonia | Europe | 0.8334 | 13 |
| Canada | Americas | 0.8285 | 14 |
| Germany | Europe | 0.8210 | 15 |

**Figure 15 World E-Government Leaders: Top 15 (United Nations, 2016)**

### 2.5.2 Electronic Government Adopting and Implementation in Developing Countries

In order to gain a better understanding of the barriers facing E-Government services adoption and implementation at the local environment level in Nigeria, there is a need for the researcher to review the E-Government initiatives in some other developed and developing countries. Most of the reviewed and current literature on E-Government adoption and implementation pay particular attention to the increasing rate of adoption of E-Services in the developing countries, but the level of implementation varies from one country to another (Abdelkader, 2015). However, the ability of the developing countries according to Matavire et al. (2010) to fully reap the benefits of E-Government services adoption is hindered by countless factors, such as economic, political, infrastructural and social. Nabafu and Maiga (2012) argued that most of the E-Government models used in developed countries have been unable to give desirable results in developing countries when implemented because of infrastructural issues, and social, political and economic factors. Gant (2008) points out that many governments in the developing countries struggle to actively deploy ICT to service citizens due to uncertainty such as the complexity of technology. Also, the author argued that most of the developing countries have a very lower E-Government readiness level, which affects the citizens' willingness to adopt the technology.

Ashaye and Irani (2013) posit that the governments in developing countries should take the lead, in partnership with private organisations, to ensure the successful implementation of E-Government services. This could be achieved through monitoring and evaluation after the post-implementation stage. The national E-Government strategy (YESSER Program) in Saudi Arabia was a bold step taken by the government in creating a user-centric E-Service initiative in the delivery of government services electronically to the public sector, as it enhances transparency and accountability (Alateyah et al., 2013). The government of Saudi Arabia firmly believes in the enormous benefits of the E-Government concepts, and it has attached high importance to the E-Government initiatives and the transformation process to make it a reality and help the national economy. Apart from E-Forms that are available on the website, there are other relevant E-Services available to the public, while information relating to the strategies, regulations, practices, and the media centre are available for general usage.

Moreover, in a move to further encourage E-Governance, the government of Saudi Arabia endeavours to promote the collaboration and innovation of culture. The government, through the E-Government program (Yesser) seeks to conduct more advanced research work on E-Government. The initiative document practices of E-Government, lessons learnt from various E-Government initiatives, and identifies the prospects of the E-Government regionally or internationally. The move will drive innovative ideas and contribute to national development (Alateyah et al., 2013).

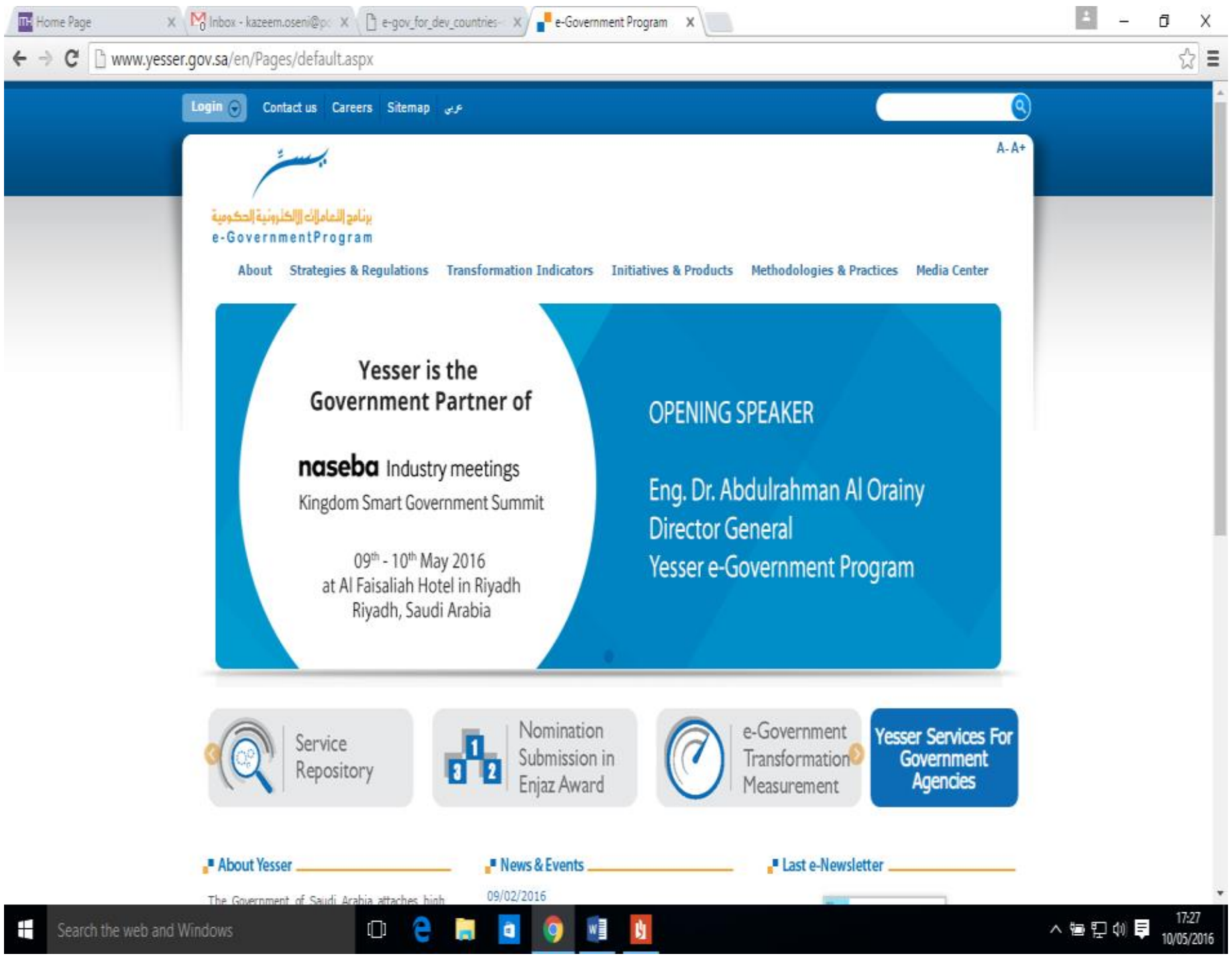

**Figure 16 Saudi Arabia National E-Government Strategy (YESSER Program) (Alateyah et al., 2013)**

However, the E-Government adopting and implementation in developing countries, as witnessed in African countries, including Nigeria, are moderately slow and irregular (UN, 2016). The low rate of E-Government adopting and implementation in Nigeria, which is currently sitting in 21st position in the African E-Government ranking, does not commensurate with the country's new position as the largest economy in Africa (Oseni and Dingley, 2014). According to the World Bank (2017), Nigeria has a GDP calculation of more than $500 billion. The researcher agrees with Azenabor (2013), who posits that the E-Government development in Nigeria is still in its developmental stages, which require a lot of work, including the provision of infrastructures. There is a digital divide between most developing countries and developed countries (Hassan, 2011). This gap could be bridged by raising the level of literacy among citizens of the developing countries (Azenabor, 2013). More awareness needed on E-Government initiatives and the government should be encouraged to massively fund these initiatives, as it helps in raising the economy standard (Azenabor, 2013). **Figure 17** shows the top E-Government ranking countries in Africa, with Mauritius and Tunisia being the two highest ranked countries.

| Country | Region | Sub-Region | OSI | HCI | TII | EGDI |
|---|---|---|---|---|---|---|
| Mauritius | Africa | Eastern Africa | 0.7029 | 0.7067 | 0.4596 | 0.6231 |
| Tunisia | Africa | Northern Africa | 0.7174 | 0.6397 | 0.3476 | 0.5682 |
| South Africa | Africa | Southern Africa | 0.5580 | 0.7253 | 0.3807 | 0.5546 |
| Morocco | Africa | Northern Africa | 0.7391 | 0.4737 | 0.3429 | 0.5186 |
| Seychelles | Africa | Eastern Africa | 0.4058 | 0.6861 | 0.4624 | 0.5181 |
| Cape Verde | Africa | West Africa | 0.4565 | 0.6031 | 0.3629 | 0.4742 |
| Egypt | Africa | Northern Africa | 0.4710 | 0.6048 | 0.3025 | 0.4594 |
| Botswana | Africa | Southern Africa | 0.2826 | 0.6553 | 0.4215 | 0.4531 |
| Libyan Arab Jamahiriya | Africa | Northern Africa | 0.1087 | 0.7588 | 0.4291 | 0.4322 |
| Kenya | Africa | Eastern Africa | 0.5580 | 0.5169 | 0.1808 | 0.4186 |

**Figure 17 Top 10 E-Government Countries in Africa (United Nations, 2016)**

The ranking in **Figure 17** above shows that Nigeria is not in the Africa Top 10 E-Government Index ranking. One could not have imagined how Nigeria, with the biggest economy in Africa (Aljazeera News, 2014), could be far behind in E-Government development, despite the enormous benefits of E-Government in economic development (Azenabor, 2013). Gant (2008) suggests that, as part of various challenges facing E-Government adoption and implementation in developing countries, there is a need for government to fully understand the citizen in meeting up with their E-Government services demand. There is the need to know the statistics related to the citizen behaviour or reaction to the use of E-Government services, the barriers that affect the citizens from going online, the factors that will encourage people to go online, and if the citizens that have been online will help others to do so. All these factors assist the government in the delivery of cost-effective E-Government services as the user population grows, and the E-Government emerging trend is to have services that focus on the citizens' needs. Remarkably, some authors from the literature (Gant, 2008; Adeyemo, 2011; Al-Shboul et al., 2014) have identified that there is growing interest in how to assist developing countries in E-Government implementation, especially in the monitoring of some pilot projects.

### 2.5.3 Electronic Government in Developed Countries versus Developing Countries

The implementation of various E-Government initiatives will no doubt improve the lives of citizens in the developing countries (Hassan, 2011). The primary benefits of different E-Government projects are being gained by the developing countries, as they continue to implement many E-Government initiatives to create friendly, easier and citizen-focused web-driven services (Hassan, 2011). According to Nawafleh et al. (2012), some international reports and studies show that E-Government has become very important in providing communities with the chance to improve the quality of government services to the citizens.

Chen et al. (2006) and Hassan (2011) posit that the real gap between developed and developing countries concerning ICT infrastructures, practices, and usage has become wider over the years. The lack of sufficient capital to build E-Government infrastructures in developing countries has contributed to the lack of sufficient knowledge and skills required to build or develop effective strategies in developing countries to promote E-Government. Similarly, the education indicators in Mozambique, according to Chen et al. (2006), show that only 7% of the total population is enrolled in secondary schools. This will have an effect on E-Government growth in the countries as the level of literacy contributes immensely to E-Government initiatives adoption. Hence, the researcher believes that greater effort should be made in improving the infrastructure and ICT education.

The case is slightly different in Australia, where E-Government strategies have had a tremendous impact on government-citizen interaction. Chen et al. (2006) explained that more than 75% of Australians now file income taxes online. This makes Australia, as well as countries like the US, Canada, Singapore, and the EU, the early leaders in the march towards E-Government.

**Table 11 Difference between Developed and Developing Countries (Chen et al., 2006; Hassan, 2011; Nawafleh et al., 2012)**

| | Developed Countries | Developing Countries |
|---|---|---|
| **History and Culture** | • Government and economy developed early, immediately after independence.<br>• Economy growing at a constant rate, productivity increasing, high standard of living.<br>• Relatively long history of democracy and more transparent government policy and rule. | • Government usually not specifically defined; economy not increasing in productivity.<br>• Economy not growing or increasing productivity; low standard of living.<br>• Relatively short history of democracy and less transparent government policy and rule. |
| **Technical Staff** | • E-Government projects have current staff, needs to increase technical abilities and hire younger professionals.<br>• E-Government projects have outsourcing abilities and financial resources to outsource; current staff would be able to define requirements for development. | • Does not have a staff, or has very limited in-house staff.<br>• Does not have local outsourcing abilities and rarely has the financial ability to outsource; current staff may be unable to define specific requirements. |
| **Infrastructure** | • Good current infrastructure<br>• High internet access for employees and citizens. | • Bad current infrastructure<br>• Low internet access for employees and citizens. |
| **Citizens** | • High internet access and computer literacy; still has digital divide and privacy issues.<br>• Relatively more experienced in democratic system and more likely to participate in governmental policy-making process. | • Low internet access and citizens are reluctant to trust online services; few citizens know how to operate computers.<br>• Relatively less experienced in democratic system and less active participation in governmental policy-making process. |
| **Government Officers** | • Decent computer literacy and dedication of resources; many do not place e-government at a high priority. | • Low computer literacy and dedication of resources; many do not place e-government at a high priority due to lack of knowledge on the issue. |

The differences between developed and developing countries are summarised in **Table 11** above using various E-Government aspects, as adopted from Chen et al. (2006); Hassan (2011); Nawafleh et al. (2012). The comparison was made based on five main factors: History and Culture, Technical staff, Infrastructure, Citizens, and Government Officers. The history and culture between developed and developing countries regarding E-Government initiatives are different in many aspects. The developed countries compared to developing countries are known for their early economic and government growth (Chen et al., 2006) and are most often categorised as early adopters in the diffusion of innovation process due to the high level of literacy, the availability of infrastructural, and the strategic plans for E-Government. Culture difference is another factor in E-Government development in both developed and developing countries, as differences in religious beliefs often prevent certain activities (Chen et al., 2006).

The definition of culture according to Umeoji (2011) has given the impression that it is concerned more with national issues and that the diffusion and adoption of E-Government should be studied along the cultural axis. The Cultural background does contribute to the adoption and usage of ICT, and this could have an effect on the adoption of E-Government. However, another school of thought states that culture should not be examined along a national axis; instead, it should be examined on cultural units that correctly determine ICT diffusion and adoption (Umeoji, 2011). Conclusively, factors such as History and Culture, Technical staff, Infrastructure, Citizens, and Government Officers, as listed in **Table 11**, have contributed to the reasons why major E-Government initiatives failed. In developing countries, there are significant gaps in the strategies and experiences in the developed countries that are not practically applicable to E-Government development in the developing countries (Hassan, 2011).

### 2.5.4 Comparison of E-Government Services between Nigeria and South Africa

The latest and newly released United Nations E-Government Ranking survey 2016, launched in July 2016, ranked the UK as the highest (Very High E-Government Development Index – EGDI) as shown in **Figure 18** for E-Government in Support of Sustainable Development compared to the 2014 ranking, which placed the country in $8^{th}$ position. This is a laudable achievement for the E-Government services in the UK, as the government continues to implement electronic government services and information. The E-Government services in the UK have been explained in Chapter Two of this thesis.

However, it would be interesting to make a further comparison of E-Government in Nigeria and another top E-Government ranked country in Africa. The 2016 E-Government ranking, as released by the UN, revealed South Africa to be the 3rd ranked E-Government country in Africa, while Nigeria was ranked further down at 21st position in Africa, as compared to the 19th position the country maintained in 2014. According to Mutula and Mostert (2010), the E-Government in South Africa is being implemented alongside several poverty alleviation programmes to improve the living standards of its citizens. The investments in ICT infrastructures have been growing steadily as the use of ICT is seen to be crucial in fighting poverty and uplifting the socio-economic and standard living of the citizens. For example, the Gauteng province, as shown in **Figure 19** below, and as part of E-Governance initiatives plan for the citizens, rolled out a new mandate of a core network infrastructure that will connect together all the government buildings, urban renewal zones, and targeted economic zones. However, it is also worth noting that Nigeria has overtaken South Africa in the economic area (Aljazeera News, 2014). Although South Africa is much more aware and is, in fact, adopting innovation in e-government, Nigeria is in a better economic position to bring about change.

| Rank | Country | Region | Sub-Region | EGDI 2016 |
|---|---|---|---|---|
| 147 | Togo | Africa | West Africa | 0.3096 |
| 105 | Tonga | Oceania | Oceania | 0.4700 |
| 70 | Trinidad and Tobago | Americas | Caribbean | 0.5780 |
| 72 | Tunisia | Africa | Northern Africa | 0.5682 |
| 68 | Turkey | Asia | Western Asia | 0.5900 |
| 140 | Turkmenistan | Asia | Central Asia | 0.3337 |
| 151 | Tuvalu | Oceania | Oceania | 0.2950 |
| 128 | Uganda | Africa | Eastern Africa | 0.3599 |
| 62 | Ukraine | Europe | Eastern Europe | 0.6076 |
| 29 | United Arab Emirates | Asia | Western Asia | 0.7515 |
| 1 | United Kingdom of Great Britain and Northern Ireland | Europe | Northern Europe | 0.9193 |
| 130 | United Republic of Tanzania | Africa | Eastern Africa | 0.3533 |
| 12 | United States of America | Americas | North America | 0.8420 |
| 34 | Uruguay | Americas | South America | 0.7237 |
| 80 | Uzbekistan | Asia | Central Asia | 0.5434 |
| 149 | Vanuatu | Oceania | Oceania | 0.3078 |
| 90 | Venezuela | Americas | South America | 0.5128 |
| 89 | Viet Nam | Asia | South-Eastern Asia | 0.5143 |
| 174 | Yemen | Asia | Western Asia | 0.2248 |
| 132 | Zambia | Africa | Eastern Africa | 0.3507 |
| 134 | Zimbabwe | Africa | Eastern Africa | 0.3472 |

| | **EGDI 2016** |
|---|---|
| **World Average** | 0.4922 |

**Figure 18 World E-Government Ranking**
(United Nations, 2016)

Furthermore, several initiatives in policy, regulatory framework, and ICT infrastructure are being undertaken by the South Africa government as part of its efforts to enhance service delivery to its citizens. Gauteng province in South Africa has the vision to be a City region that uses ICT to support the delivery of quality services and equitable and inclusive social economic development of its citizens. However, the E-Government in Nigeria is still in its developmental stages, which require a lot of work including the provision of infrastructures (Azenabor, 2013). Nigeria made a bold step like other developing countries to join the global train of ICT as a consumer of the technology, which has improved drastically over some time with technological innovations, which now make ICT relevant in the country's daily operation needs.

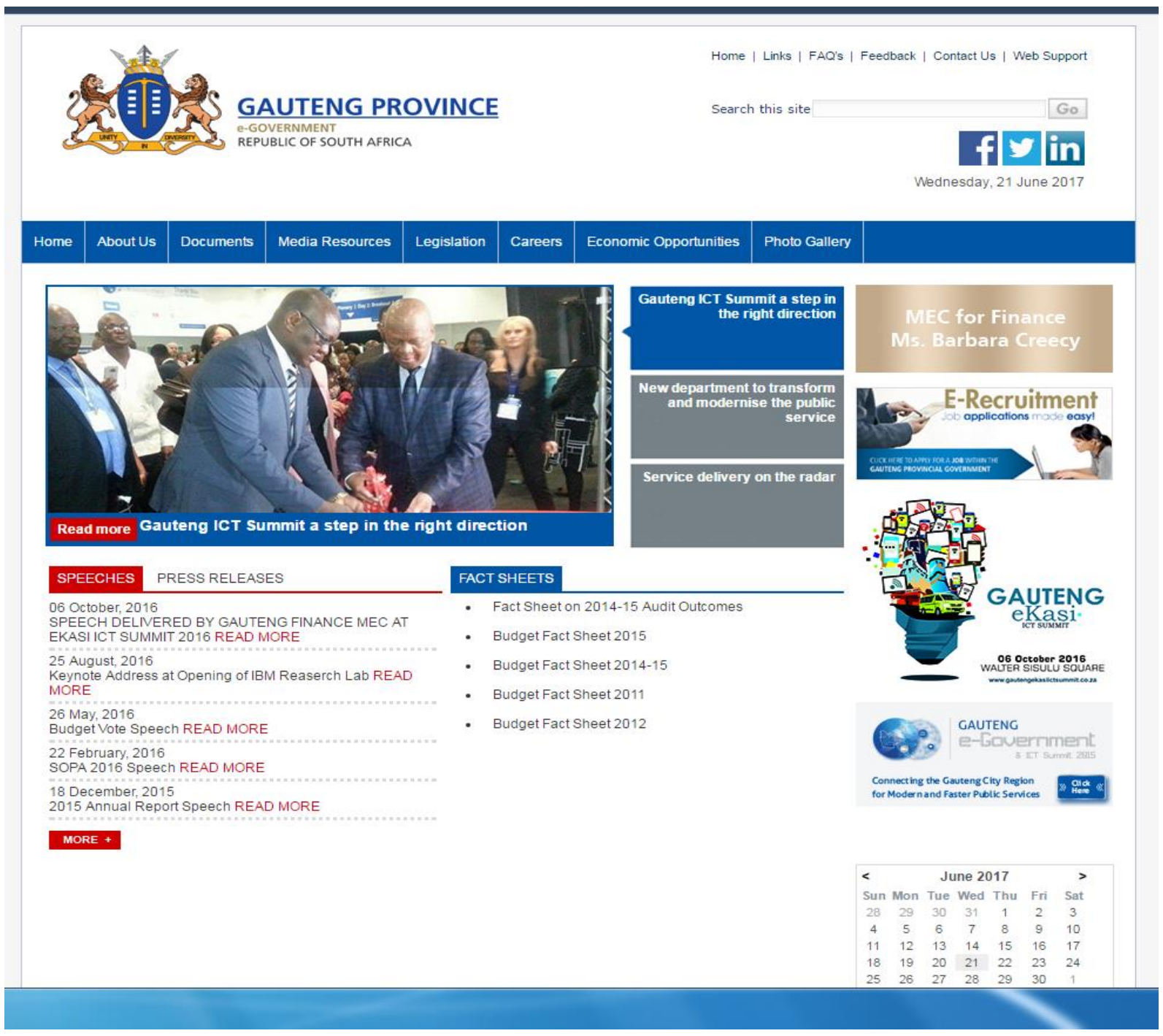

**Figure 19 South Africa Gauteng Province E-Government Portal**

## 2.6 Research Theory

### 2.6.1 Introduction

According to Pettigrew and McKechnie (2001), the use of theory in information science research helps to organise and present large data in a simplified form, especially the intricacies of the social world. In simple terms, the *Oxford English Dictionary* defines "theory" as a system of ideas explaining something. In a related development, Neuman (1997) suggests that the use of theory in research enables the researchers to connect a single study to the large base of knowledge, which then increases awareness of interconnections and broadens the significance of data. Pettigrew and McKechnie (2001) argued that information science has lacked good theories in the past and, according to the philosophy of science, the use of theory by researchers is a hallmark of their academic maturity. Consequently, this allows researchers to make necessary links between the abstract and other statements, either empirical or observational. As a result of the above, the researcher will review in the next section the various theories used in information system research, the justification for the research theory, and the building of theory from available data.

### 2.6.2 Theories Used in Information Systems Research

There has been a limited research in the theoretical foundations of information systems (Lim et al., 2009). Hence, the use of theory is very critical in information systems research (ISR) as researchers strive hard for a strong theoretical contribution (Lim et al., 2009). Although the differences in views of theory depend, to some extent, on philosophical and disciplinary angles, philosophers in physical or natural science will generally like to see theory as the provision of explanations and predictions which can be tested (Gregor, 2006).

**Table 12 Classification of Theory in Information Systems (Gregor, 2006)**

| Theory Classification | | Primary Goals |
|---|---|---|
| **1** | **Analysis and Description** | The theory provides a description of the phenomena of interest, analysis of relationships among those constructs, the degree of generalisability in constructs and relationships and the boundaries within which relationships and observations hold. |
| **2** | **Explanation** | The theory provides an explanation of how, why, and when things happened, relying on varying views of causality and methods for argumentation. This explanation will usually be intended to promote greater understanding or insights by others into the phenomena of interest. |
| **3** | **Prediction** | The theory states what will happen in the future if certain preconditions hold. The degree of certainty in the prediction is expected to be only approximate or probabilistic in information systems. |
| **4** | **Prescription** | A special case of prediction exists where the theory provides a description of the method or structure, or both, for the construction of an artefact (similar to a recipe). The provision of the recipe implies that the recipe, if acted upon, will cause an artefact of a certain type to come into being. |

Gregor (2006) posits that theory in information systems is classified according to the primary goals of the theory, as indicated in **Table 12** above. As obtained in action research methodology, research begins with a problem that needs a solution or some question of interest, and the theory is then developed based on this problem. In the light of the above, the next section will provide and establish the various theories used in ISR as explained by Lim et al. (2009).

The theories used in information systems research have been published and analysed in the literature (see Gregory, 2006; Larsen et al., 2015), but the dominant theories used over time in the ISR are discussed in this study. Most ISR is built upon the use of theories, while some is drawn from other disciplines (Gregory, 2006; Lim et al., 2009). The top ten theory pages visited in 2014 (Larsen et al., 2015), such as DOI theory used in this study, are displayed in **Table 13** below.

**Table 13 The Top 10 Information System Theory (Larsen et al., 2015)**

| | Information Systems Theory | Usage Times (Percentage) in 2014 |
|---|---|---|
| 1 | Institutional Theory | 9.4% |
| 2 | Social Network Theory | 6.7% |
| 3 | Contingency Theory | 6.6% |
| 4 | Organisational Culture Theory | 5.8% |
| 5 | Transaction Cost Economics | 5.6% |
| 6 | DeLone and McLean IS Success Model | 5.1% |
| 7 | Technology Acceptance Model (TAM) | 5.1% |
| 8 | Socio Technical Theory | 4.8% |
| 9 | Garbage Can Theory | 4.0% |
| 10 | Diffusion of Innovation Theory (DOI) | 3.7% |

Lim et al. (2009) further categorised the theories used in ISR based on their dominance over a period between 1998 and 2006 through the review of these two papers, namely Information Systems Research (ISR), and MIS Quarterly (MISQ) published from 1998 to 2006. The nine years capture of these papers, according to Lim et al. (2009), was considered to be comprehensive enough to serve as a representative sample of IS research theory used, including the dominant ones. A total of 386 research articles were identified, being 202 for MISQ and 184 for ISR. The authors arrived at 10 top dominant theories, as shown in **Figure 20** below, used between 1998 and 2006 out of the 386 in the literature through Search, Filter, Peruse, and Confirm. Despite the rejection of TAM by the participants for this study, the TAM model remained in the top 10 of the information system theories frequently used between 1998 and 2006. The results show that Technology Acceptance Model (TAM), Resource Based View, Game Theory, Theory of Reasoned Action, Institutional Theory, Theory of Planned Behaviour, Transaction Cost Theory, Dynamic Capability Theory, Diffusion Innovation Theory, and Agency Theory are the most dominant theories used between the periods.

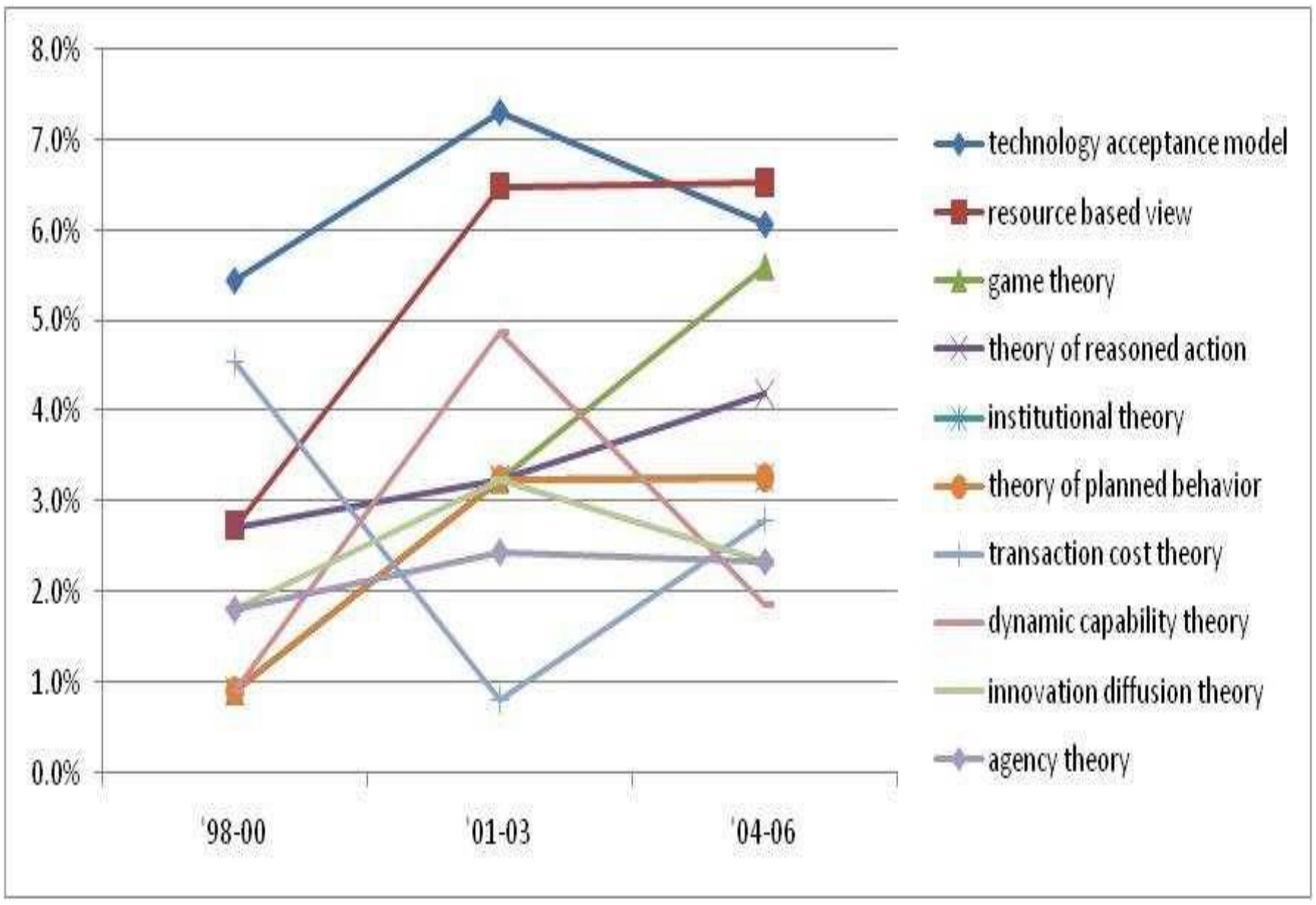

**Figure 20 Top 10 Dominant Theories Used in IS Research between 1998 and 2006 (Lim et al., 2009)**

### 2.6.3 Justification for Research Theory - Diffusion of Innovations theory (DOI)

At the early stage, this research study was set to use the Technology Acceptance Model (TAM) in E-Service adoption at the local government level in Nigeria. As enumerated in the research objectives, the critics of the TAM were in respect of areas such as absence of theory in the framework, lack of cultural relativism (Bagozzi, 2007), the questionable heuristic value, limited explanatory and predictive power (Chuttur, 2009). Notwithstanding such criticism, TAM's popularity remains largely undiminished. Even with these limitations, and from the literature reviewed by the researcher, it is a remarkable model, and it would have been a more beneficial and efficient model to use.

The researcher was privileged to have interacted with participants during the interviews; however, the model does not stand up to scrutiny considering the opinions of the members interviewed, as they were opposed to the use of TAM in E-Service adoption. The participants interviewed pointed out that the use of TAM in technology acceptance is subjective in nature because of self-reported data by the model instead of the actual use data, as explained by Chuttur (2009). The use of students as participants in the application of TAM in the technology acceptance is more prominent in many studies according to Chuttur (2009). Also, the results obtained from those studies could not be generalised to the real world. Chuttur (2009) claimed that students might have different motivations, such as achieving good grades, which may affect their honest participation and commitment to the study. This was also substantiated by Yousafzai et al. (2007). In the action research approach, the view of the participants is considered to play a vital role in determining the outcome of the study. According to McCartan et al. (2012), the relationship between the researcher and the participant, despite them being a partner in the study, is for the researcher to maintain neutral or invisible, as anything else may falsify or threaten the validity of the research findings.

The researcher agreed with McCartan et al. (2012), who found it logical and reasonable to look at other models. The situation had lent itself for a better theory or model for this research. Some authors have considered for more than three decades the process of adopting innovations with one of the most widespread adoption models described in *Diffusion of Innovations*, a book written by Roger Everett (Sherry and Gibson, 2002; Sahin, 2006). The model is used as a framework in many research studies from a broad and variety of multi-disciplines such as education, technology, economics, communications, history and political science, and public health, as mentioned by Dooley (1999). Rogers' DOI theory has been widely used as a theoretical framework in the areas of new technology adoption and diffusion (Dooley, 1999). Many case studies have supported this theory. In their submission, Parisot (1995) and Medlin (2001) argued that Rogers' DOI theory is the most appropriate theory used in investigating technology adoption in a higher educational environment. One unique consequence of DOI theory is the presence of Cultural relativism, as every culture has a different set of values, beliefs, attitudes, and norms that work for their particular situation (Rogers, 1995).

This research will make use of DOI theory to support the investigation of barriers facing the E-Service, adopting at the local environment level using Nigeria as a case example, through action research. Researchers have successfully used the theory through various research methods. Using an online survey, and in-depth interviews, Carter (1998) examined computer-based technology used by faculty members and the factors that may affect the technologies (Carter, 1998). He also reviewed the attitudes towards usage, and he came up with factors such as training, and the resources needed to use these technologies efficiently, and found that the most frequently used computer-based technologies were email, word processing software and internet resources (Carter, 1998). DOI theory was also tested through the use of quantitative research methods by Surendra (2001). Surendra (2001) examined the diffusion factors proposed by Rogers in 1995 to predict web technology acceptance by scholars and administrators of a college. He found that the diffusion factors in Rogers' DOI theory are very useful in the prediction of technology and innovation adoption (Surendra, 2001). The stages of innovation adoption process are shown in **Figure 21**.

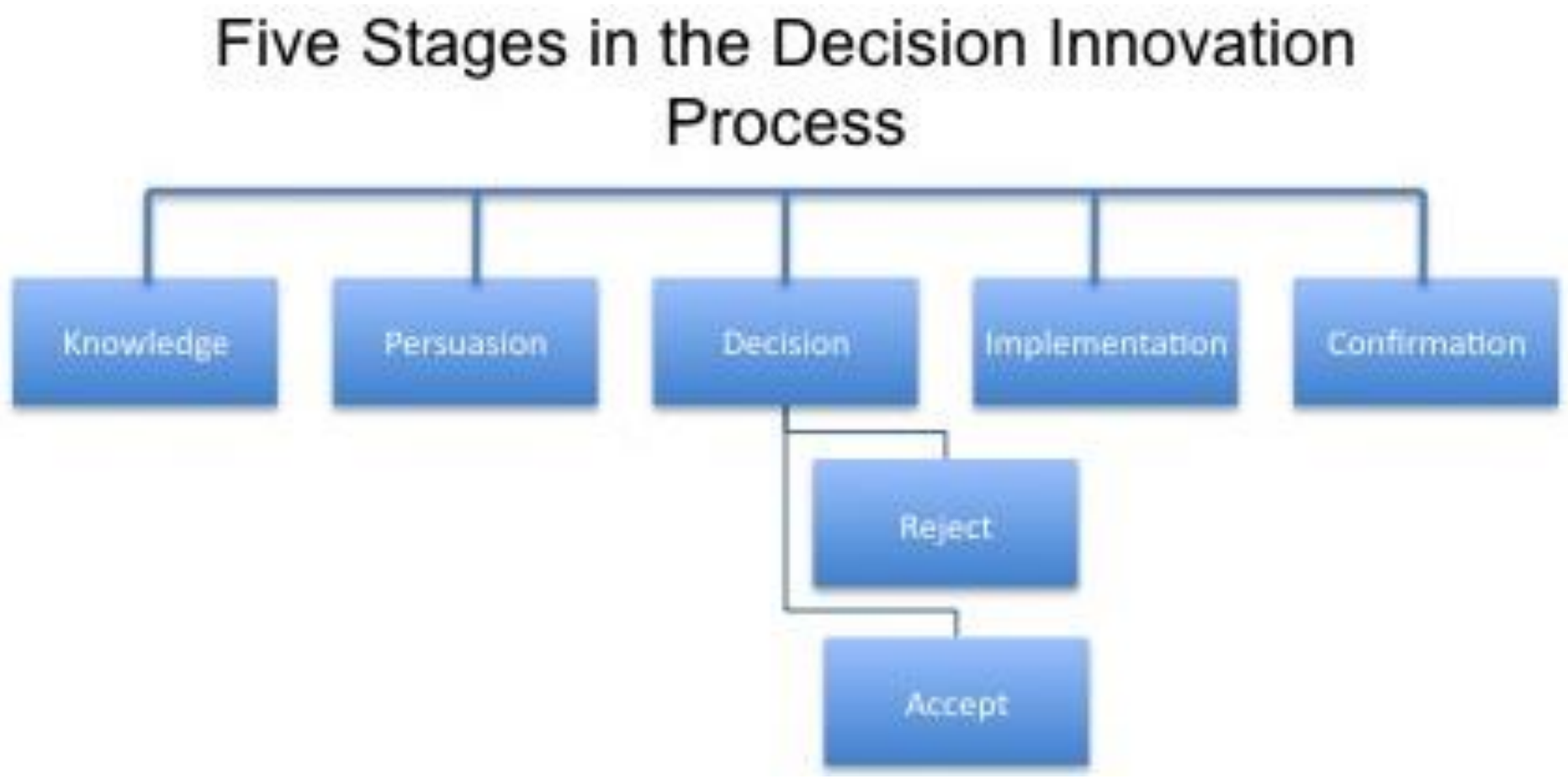

**Figure 21 Five Stages of the Innovation Adoption Process (Rogers, 1962; 1995; 2003)**

Moreover, the DOI according to Robinson (2009) seeks to provide an explanation on how innovations are taken up in a population, and it offers three valuable insights into the process of social change (Robinson, 2009). These are as follows:

- What qualities make an innovation spread?
- The importance of Peer-to-Peer networks (Conversations)
- Understanding the needs of the different users

Furthermore, to answer the questions above, Robinson (2009) looked at why certain innovations spread more quickly than others and why others fail. The author is of the opinion that the following are the five qualities/attributes that determine the success of innovation, as shown in **Table 14** below:

**Table 14 Innovation Success Quality/Attributes (Robinson, 2009)**

| Innovation Success Quality/Attributes | Function |
|---|---|
| **Relative advantage** | This is the degree to which an innovation is perceived as better than the idea it supersedes by a particular group of users, measured in terms that matter to those users, like economic advantage, social prestige, convenience, or satisfaction. The greater the perceived relative advantage of an innovation, the more rapid its rate of adoption is likely to be. |
| **Compatibility with existing values and practices** | This is the degree to which an innovation is perceived as being consistent with the values, past experiences, and needs of potential adopters. |
| **Simplicity and ease of use** | This is the degree to which an innovation is perceived as difficult to understand and use. New ideas that are simpler to understand are adopted more rapidly than innovations that require the adopter to develop new skills and understandings. |
| **Trialability** | This is the degree to which an innovation can be experimented with on a limited basis. An innovation that is trialable represents less risk to the individual who is considering it. |
| **Observable results** | The easier it is for individuals to see the results of an innovation, the more likely they are to adopt it. Visible results lower uncertainty and also stimulate peer discussion of a new idea, as friends and neighbours of an adopter often request information about it. |

Robinson (2009) believes that the study of DOI theory should provide more insight into understanding the needs of different user segments. The adopters (population) are broken down into five different segments based on their willingness to adopt an innovation (Robinson, 2009). As shown in **Figure 22**, the five parts of the adopters are innovators, early adopters, early majorities, later majorities, and laggards. Each group of the adopters has its personality in line with attitude to a particular innovation, and it will spread when innovation progresses in meeting the need of the adopters (Robinson, 2009). From **Figure 22**, innovators are 2.5%, Early Adopters 13.5%, Early Majority 34%, Late Majority 34% and Laggards 16%.

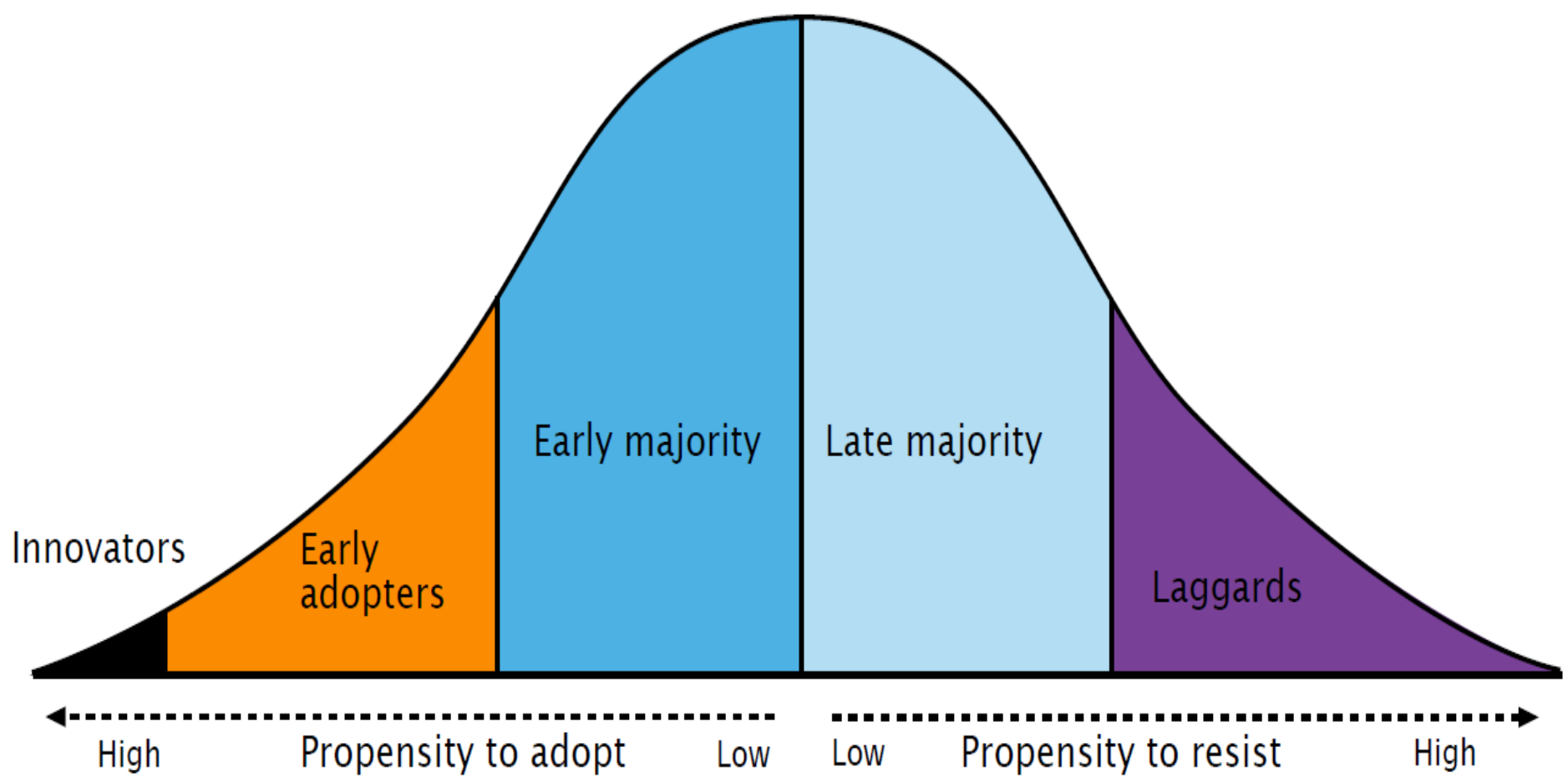

**Figure 22 Innovation Adopters Category (Robinson, 2009)**

Robinson (2009) further explained, as shown in **Table 15** below, the five segments of the diffusion innovation adopters and how researchers could successfully work with different user segments.

**Table 15 Working with Diffusion Innovation Adopters (Robinson, 2009)**

| Diffusion Innovation Adopters | Definition | How to Work with Adopters |
|---|---|---|
| **Innovators** | The adoption process begins with a tiny number of visionary, imaginative innovators. They often lavish great time, energy and creativity on developing new ideas and gadgets. They also love to talk about them. | •Track them down and become their "first followers", providing support and publicity for their ideas.<br>•Invite keen innovators to be partners in designing your project. |
| | | |
| **Early Adopters** | Early adopters leap in once the benefits start to become obvious. They are on the lookout for a | •Offer strong face-to-face support for a limited number |
| | strategic leap forward in their lives or businesses and are quick to make connections between clever innovations and their personal needs. | of early adopters to trial the new idea.<br>•Study the trials carefully to discover how to make the idea more convenient, low cost and marketable.<br>•Reward their egos, e.g. with media coverage.<br>•Promote them as fashion leaders (beginning with the cultish end of the media market).<br>•Recruit and train some as peer educators.<br>•Maintain relationships with regular feedback. |
| | | |
| **Early Majority** | Early majorities are pragmatists, comfortable with moderately progressive ideas, but won't act without solid proof of benefits. They are followers who are influenced by mainstream fashions and wary of fads. | •Offer give-aways or competitions to stimulate buzz.<br>•Use mainstream advertising and media stories featuring endorsements from credible, respected, similar folks.<br>•Lower the entry cost and guarantees performance.<br>•Redesign to maximise ease and simplicity.<br>•Cut the red tape: simplify application forms and instructions.<br>•Provide strong customer service and support. |
| | | |
| **Late Majority** | Late majority are the conservative pragmatists who hate risk and are uncomfortable with your new idea. Practically, their only driver is the fear of not fitting in; hence, they will follow mainstream fashions and established standards. They are often influenced by the fears and opinions of laggards. | •Focus on promoting social norms rather than just product benefits: they will want to hear that plenty of other conservative folks like themselves think it's normal or indispensable.<br>•Keep refining the product to increase convenience and reduce costs.<br>•Emphasise the risks of being left behind.<br>•Respond to criticisms from laggards. |
| | | |
| **Laggards** | Laggards hold out to the bitter end. They are people who see a high risk in adopting a particular product or behaviour. Some of them are so worried that they stay awake all night, tossing and turning, thinking up arguments against it. | •Give them high levels of personal control over when, where, how and whether they do the new behaviour. •Maximise their familiarity with new products or behaviours. Let them see exactly how other laggards have successfully adopted the innovation. |

### 2.6.4 Merit and Demerit of Diffusion of Innovation (DOI)

The definition of diffusion of innovation (Rogers 1983; 1995) takes into consideration the need to explain the elements involved in the DOI. Rogers defined the DOI as the process by which innovation "is communicated through certain channels over time among the members of a social system" (Rogers 1983; 1995). If the diffusion focuses on interpersonal communication, as posited by Rogers (1983; 1995), within a social group, this could contribute to the adoption of innovation and technology. As a result, it will be necessary to expatiate further on the four elements involved in the DOI, as explained in **Table 16** below.

**Table 16 Elements of Diffusion of an Innovation (Rogers, 1995)**

| Elements of Diffusion of an Innovation | Definition |
|---|---|
| **1. Innovation** | Any idea, object, or practice that is perceived as new by members of the social system. |
| **2. Channels of Communication** | The means by which information is transmitted to or within the social system. For example, the media in 2001 played a vital role in the emergency of Global System of Mobile Communication (GSM) in Nigeria through various announcements on likely price and operations. |
| **3. Time** | This is the rate at which innovation is diffused or the relative speed with which it is adopted by the members of the social system. |
| **4. Social System** | This is referring to individuals, organisations, or agencies that share a common "culture" and are potential adopters of the innovation. Examples are opinion leaders and agents of change. |

Furthermore, the relevance of DOI theory in the world today will continue to be strong, as innovations and ideas occur daily and would continue to diffuse in order for the society to adopt them. One could imagine the diffusion process involved in the Global System for Mobile (GSM) communications in Nigeria a few years ago. The introduction in 2001 was viewed with uncertainty according to Tella et al. (2009); the media made the announcement first with a number of adverts on the telecommunication operations, and the innovators, together with early adopters, swiftly grasped the chance when the operation started. The diffusion process continues by encouraging the opinion leaders and the change agents, who made sure the innovation got to the late adopters (Tella et al., 2009). Before the introduction of GSM communications in 2001, the number of connected phone lines in Nigeria was 450,000 in a population of about 120 million citizens (Tella et al., 2009). However, with the level of investment in the technology and the diffusion of the innovation, the statistics released by the Nigerian Telecommunication Commission in 2014 indicated that a total of 94.5 million citizens were connected to mobile communications in a population of about 177 million citizens. To this end, this researcher will, therefore, review the merits and demerits of DOI theory in the next session. The **Merits** are as follows:

- **Spreading of Technological Innovations and New Ideas**: The theory provides the practical guide for information campaigns. For instance, the United States Agency for International Development (UNSAID), according to Anaeto et al. (2008), used the DOI theory strategy to spread agricultural innovations in the third world.
- **Improves Empirical Research**: Rogers (1962; 1995; 2003) successfully integrates a vast amount of empirical research, as the theory drew from existing empirical generalisations, and synthesised them into a coherent, insightful perspective. The theory is consistently used in most finding-involving surveys and persuasion experiments (Anaeto et al., 2008).
- **Improves Communication and Marketing Theories**: DOI theory laid the background for many promotional communication and marketing theories and the campaigns they support, even to this today (Anaeto et al., 2008).

In the light of the above, DOI theory has some limitations. Perhaps the most serious disadvantages of this theory are:

- **Individual Blame Bias**: DOI theory has the tendency to hold individuals responsible for problems instead of the system which an individual belongs to (Anaeto et al., 2008).
- **Lack of Zero Tolerance**: Researchers believed zero tolerance should be integrated into DOI theory adopter's categories to take care of people who are innovators, but who may not be ready to adopt innovations (Anaeto et al., 2008).
- **Pro-innovation Bias**: This is when innovation is assumed to be diffused and adopted by all members of a social system. It anticipated that innovation should not be rejected and that it should be diffused rapidly (Anaeto et al., 2008).
- **Equality Issues**: This is the problem of equality in the diffusion of innovations, as there are socio-economic gaps among the members of a social system (Anaeto et al., 2008).
- **Recall Problems**: DOI research may lead to inaccuracies, especially when the respondents were asked to remember the new ideas adopted over time (Anaeto et al., 2008).

### 2.6.5 Link between Research Theory and Research Method

Theory, method, and analysis in research are closely interconnected, and the decisions about one affect the others, according to Smith (2005). In reality, Smith (2005) argued that doing a good research entails blending the theory with our choice of methodology. By drawing on the concept of Smith (2005), the link between research methodology and theory for this study is presented in **Figure 23**, with a further analysis represented in **Table 17**, showing the research progress and plan.

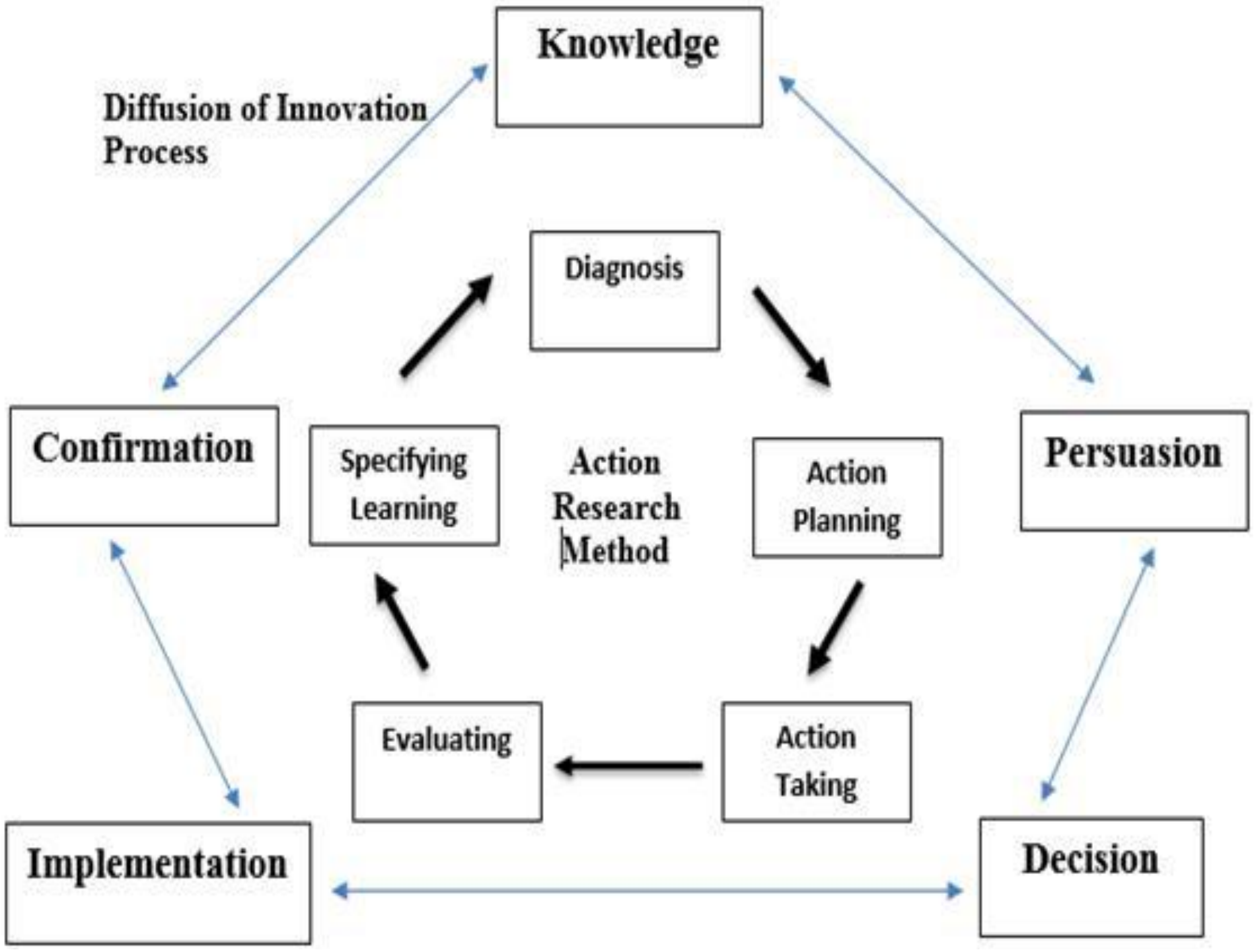

**Figure 23 Link between Research Methodology and Theory (Oseni, 2016)**

As illustrated in **Figure 23**, having a clear, detailed research plan, as enumerated previously, in the research design will give a better understanding of the activities. The link between the research methodology and theory is displayed in **Table 17** below.

**Table 17 Research Plan versus Research Methodology and Theory**

| Research Phase | Action Research Method/Diffusion of Innovation (DOI) Theory | Research Activities |
|---|---|---|
| **Phase 1** | **Diagnosis/Knowledge** | Clear identification of research issues, aims, objectives and research questions. Understand clearly what the research background is. |
| **Phase 2** | **Action Planning/Persuasion** | Literature review, research methodology selection and justification, development of the initial framework.<br>• Seeking information about the new ideas, innovations.<br><br>**<u>Theory</u>**<br>• Discussion on the theory - The use of Technology Acceptance Model (TAM) for the study. |
| **Phase 3** | **Action Taking/Decision** | Data Collection – Interviews and online survey, documents observation, research theory, refining the initial framework. Decision on mode and structure of the interviews, theory.<br><br>**Theory Decision**<br>• Participants engaged in activities that lead to a choice to adopt or reject the theory, advantages and disadvantages.<br>• The use of Technology Acceptance Model (TAM) was rejected.<br>• More deliberations on other theories used in ISR.<br>• Diffusion of Innovation Theory (DOI) was considered. |
| **Phase 4** | **Evaluation/Implementation** | Data analysis, evaluating results, coding and validation of the framework, implementing the changes from the data collection on the initial framework, feedback.<br><br>• **Online Focus Group**: Blind focus group session with anonymised group typing. Participants using a pseudonym of their own choosing by answering E-Service development questions based on fictional scenario, using a fictional country in developing countries. |
| **Phase 5** | **Specifying Learning/Confirmation** | Confirming the final framework, writing up, specifying achievement/contributions and making recommendations. |

## 2.7 Summary

The detailed analysis of the previous literature on E-Government services development was explained in this chapter. It has identified and provided a broad and better understanding of the barriers facing E-Service adoption and implementation at the local government level in developing countries, including Nigeria. Also, the chapter has been able to identify and review the solutions to the barriers facing E-Service adoption and implementation at the local government level in developing countries, including the success factors.

Further analysis was carried out on electronic government adopting and implementation in developed and developing countries, with significance, scopes, and the nature of E-Government discussed. Categorisation of the major stakeholders and different levels of interactions in E-Government service development at the local level have been identified. Lastly, the various theories used in ISR, justification of the choice of research theory for this study, the history, together with merit and demerit of research theory (Diffusion of Innovation), and the link between the research theory and research method, were thoroughly discussed in this chapter. The literature search, and the use of structured literature search, were outlined, as were the stages of E-Government development, barriers to the E-Services adopting and implementation, research theories, and E-Government in developed and developing countries. There is, however, always more research than any researcher can read and assimilate, but this chapter has collated some key concepts and evaluated the spread of e-services to developing countries.

# Chapter 3 – Framework Development and E-Government Models

## 3.1 Introduction

A critical review of the literature presented in the previous chapter on E-Government services has also provided a deep and better understanding of the investigated barriers, and solutions to the barriers, including the success factors for the adoption of E-Services at the local government level in Nigeria. Consequently, the main purpose of this chapter is to:

*Propose a conceptual E-Service framework to identify the barriers facing E-Service adopting and implementation at the environment government level using Nigeria as a case example. The development of the framework for this study is based on the review of the relevant models and frameworks and the combination of previous literature on the E-Government services. The main elements are collated together and will be used in developing the E-Service framework for this study.*

In the light of the above, there are six sections in this chapter. The first section is the introduction, which has stated clearly the main purpose of this chapter. The second section of this chapter briefly reviews the various E-Government models and frameworks in the literature. Section three discusses the proposed framework for this study, and will review the current, transformation and desired phases of the framework. Section four will evaluate the current E-Service projects in Nigeria, including the present stages of E-Government service at the local environment level. Section five will look into the various E-Service initiatives covered in Nigeria by the press/media, as shown in **Figure 24** below. Lastly, the summary section will recapitulate all the points discussed in this chapter.

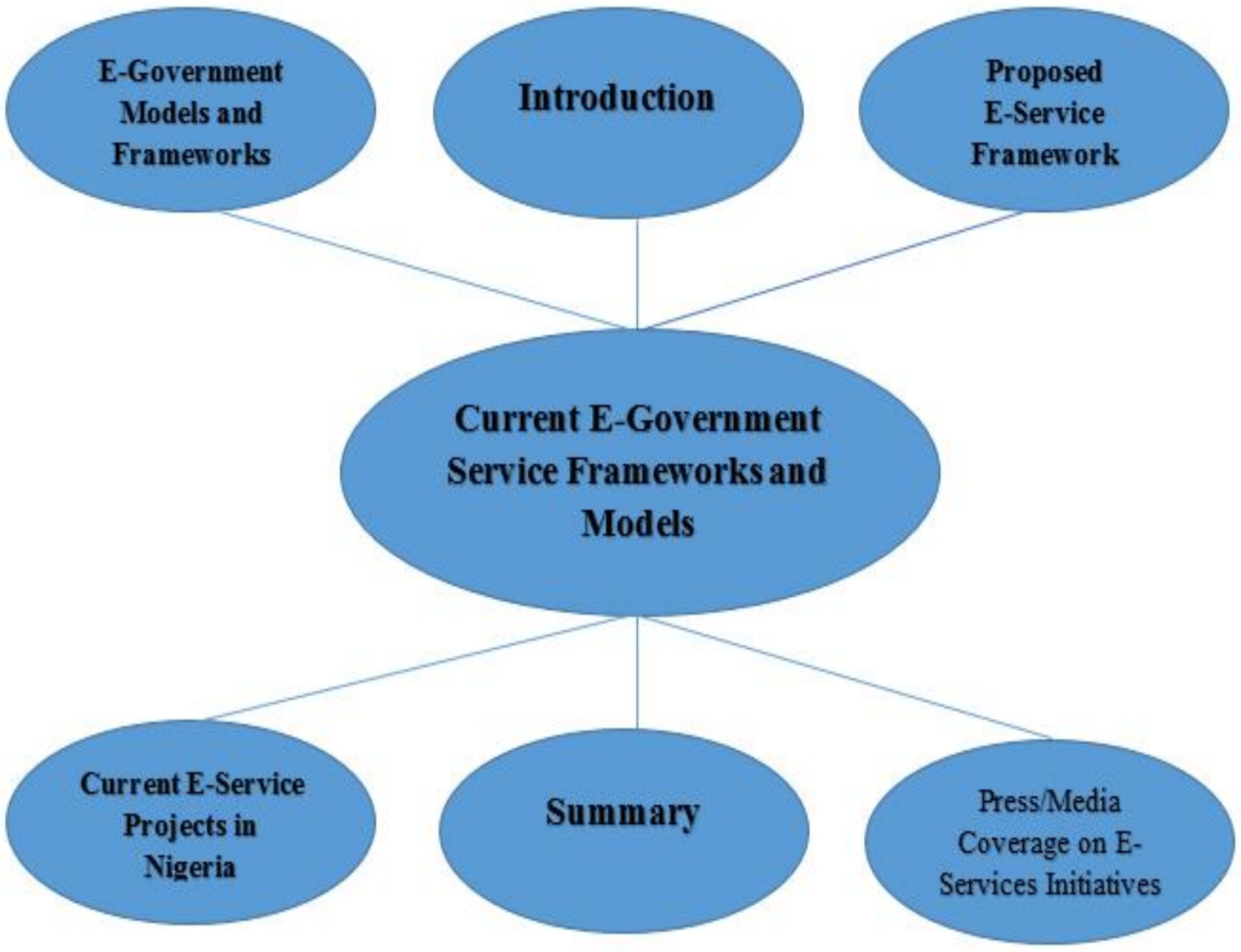

**Figure 24 Chapter Three Summary**

## 3.2 Overview of Various E-Government Models and Frameworks

The E-Government model and framework according to Azenabor (2013) is a representation of the pattern involved in adopting and implementation. The advancement in the actual implementation of E-Government projects and initiatives will fully depend on the availability of an effective E-Government assessment framework (Azab et al., 2009). Therefore, it is pertinent for the researcher to review various E-Government models and frameworks in the literature, as they will serve as a driver and enabler towards the development of a suitable E-Service framework for this study. The models and frameworks are as follows:

1. **E-Service Development Framework (eSDF) - Figure 25** below was developed for the Office of the e-Envoy, Cabinet Office, UK. It includes the process for developing E-Service specifications, including functional requirements, technology-neutral design, and implementation using XML schema (Benson, 2002). The high-level information architecture is the top level which provides a single set of top-level specifications and standards for E-Government service development. This level is followed by the reusable elements that provide the framework necessary for reusable patterns, components and resources, and these will improve consistency and reduce costs. The last level of the framework, which is the lowest level, is the E-Service development level. This involves the development of specific E-Services. This ESDF framework consists of three main phases, namely the requirements phase, the design phase and the implementation phase (Benson, 2002).

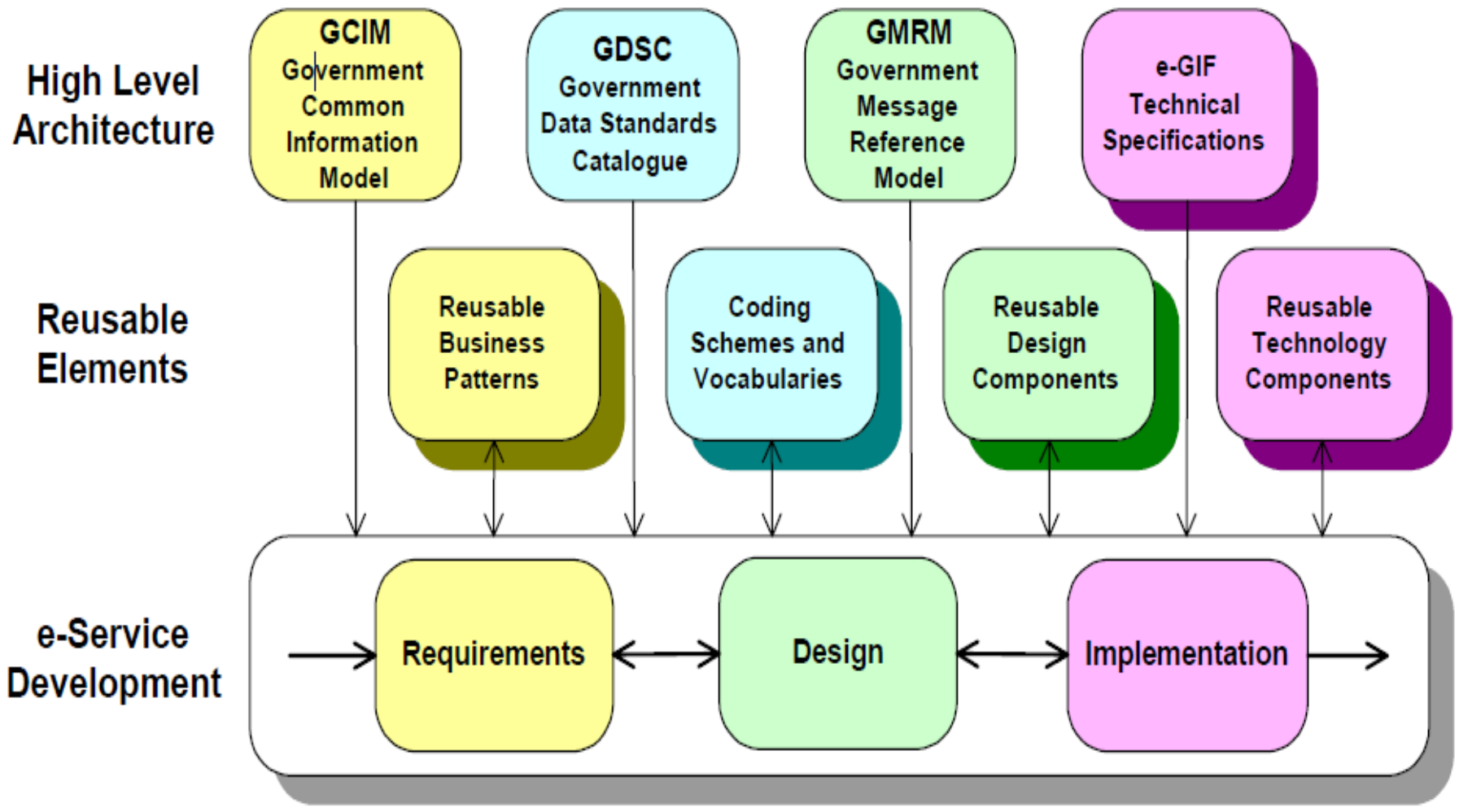

**Figure 25 The E-Service Development Framework (eSDF) (Benson 2002)**

2. **The framework developed by Macintosh and Fraser (2003) for SmartGov Project**. **Figure 26** below was developed with the aim of the government to meet the service needs of the citizens and businesses. The project also targets the need to minimise the reliance on IT skills to develop E-Government services. The framework will benefit public authority in the delivery of electronic transaction services (Macintosh and Fraser, 2003).

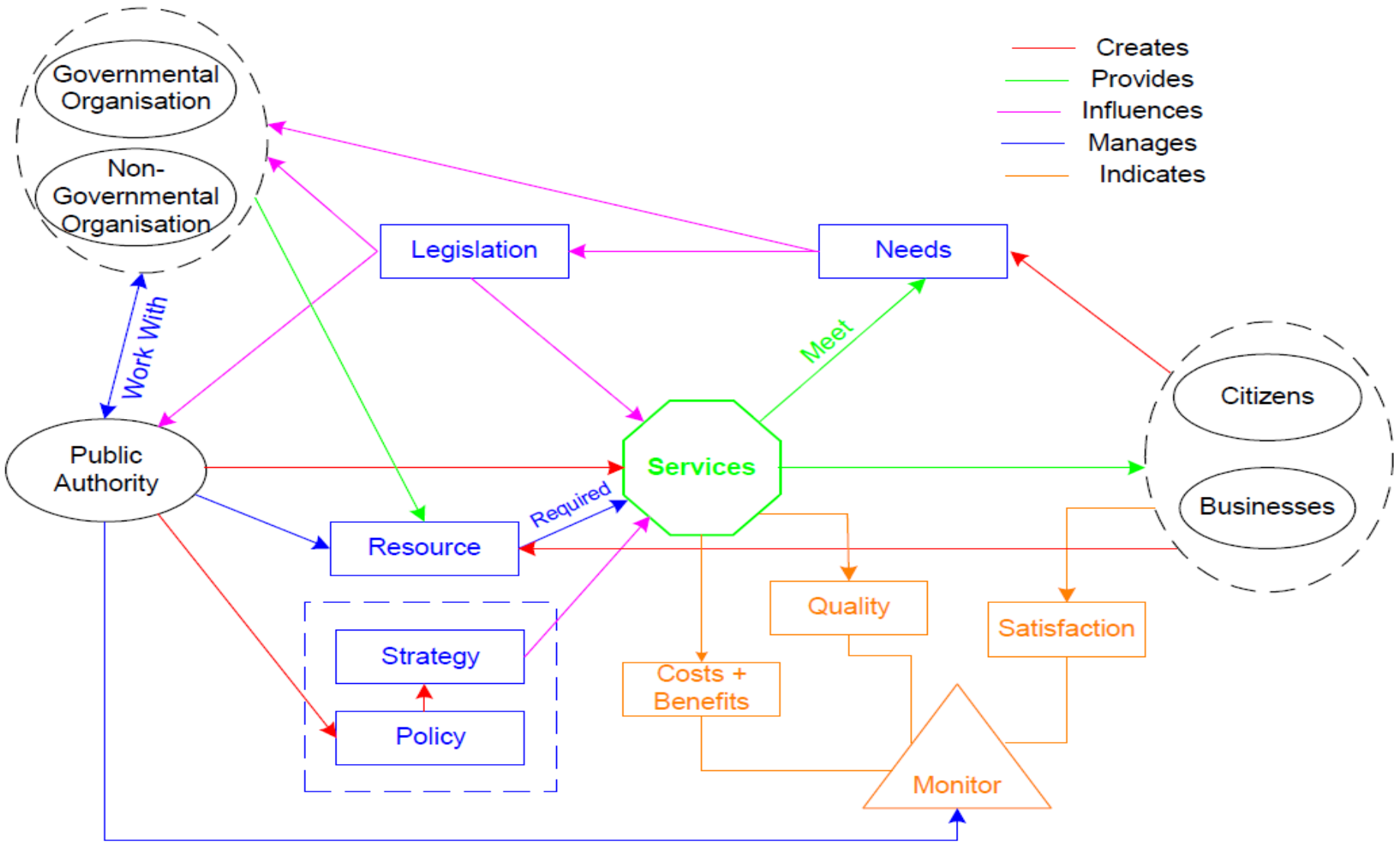

**Figure 26 SmartGov Framework for E-Government Services (Macintosh and Fraser 2003)**

3. **ISRUP E-Service framework** according to Hashemi et al. (2006) was developed to solve the problems associated with business domain specifications. The model is for the agile enterprise architecting and employs Unified Modelling Language (UML) and Rational Unified Process (RUP) terminology to improve the enterprise architecture of business operations. **ISRUP**, as shown in **Figure 27** below, is based on a process that describes what to produce, skills required for the production, and the necessary steps needed to achieve the goals. Some parts of government and today's businesses use ISRUP E-Service framework to make informative and reusable documents as their assets. This framework described the merging of E-Commerce, E-Business, and E-Government, and the service provider or service requester can take advantage of this model for a successful business transaction (Hashemi et al., 2006).

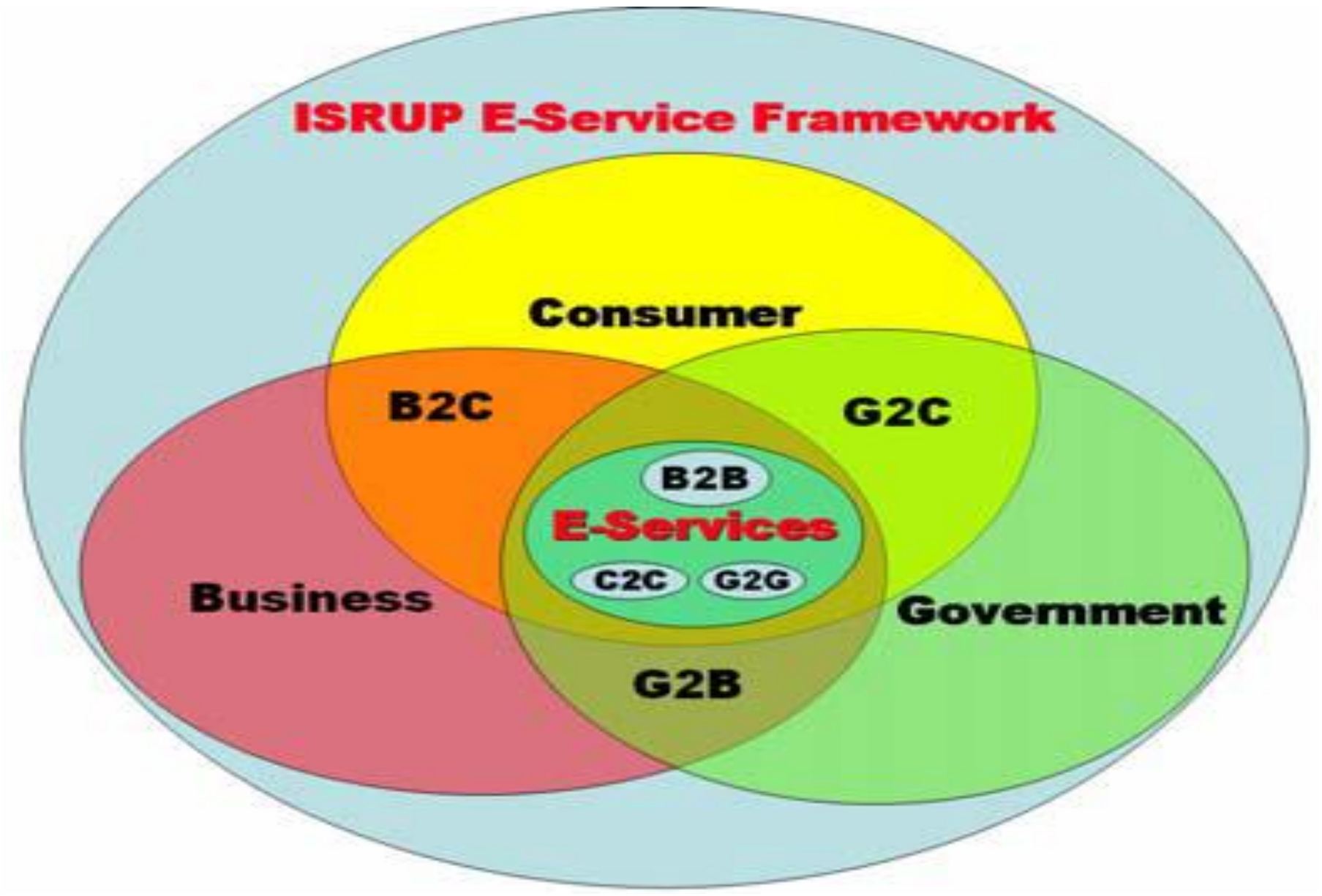

**Figure 27 ISRUP E-Service Framework**
**(Hashemi et al., 2006)**

4. **Three-Level E-Service Approach** towards process and data management. This framework relates the two process specification levels, the external and internal, through a third, being the conceptual level. This approach, inspired by the American National Standards Institute (ANSI) and Standards Planning, and Requirement Committee (SPARC), is termed the ANSI-SPARC model for data management (Grefen et al., 2002). In the first level, which is the external level as shown in **Figure 28** below, the users view the database relevant to the particular user. At the conceptual level, the community views the database as in what data are stored in the database and what their relationships are. The last level, which is the internal level, shows the physical representation of the database on the system. That is how data are stored in the database (Grefen et al., 2002).

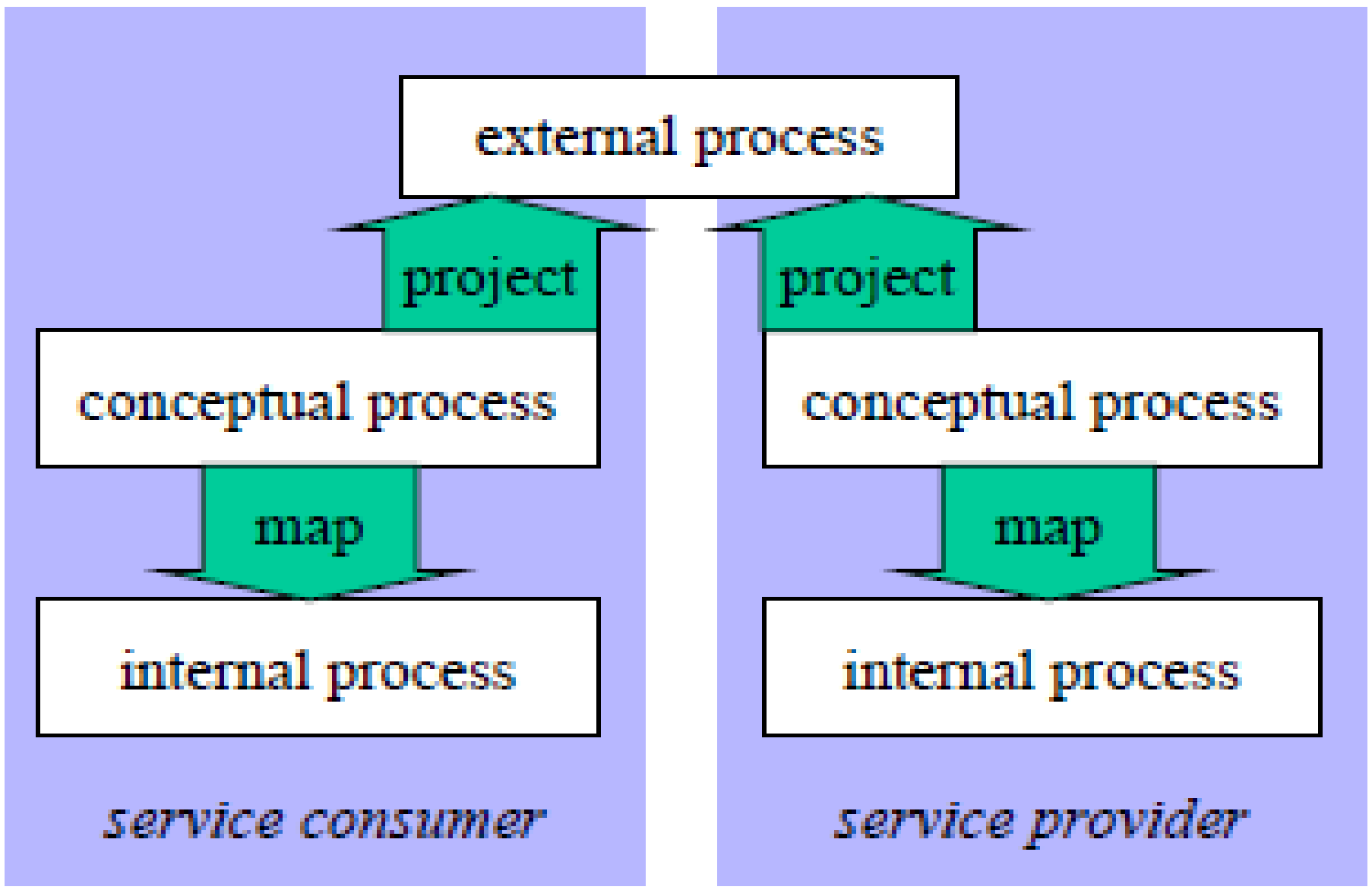

**Figure 28 Three-level E-Service framework (Grefen et al., 2002)**

5. **The E-Government Framework** by Keng and Yuan (2005), as shown in **Figure 29** below, is the overall framework for E-Government. The framework illustrates the interaction between Government to Government (G2G) and Government to Employee (G2E), which are classified as the internal level, while Government to Customers (Citizens) G2C, Government to Business (suppliers) G2B are classified as the external level in the framework. The primary objective of Government to G2G is to enhance cooperation and collaboration between governments (such as federal and state governments) at different levels, irrespective of their location. The activities include the sharing of information and data about the government. The G2E relationship has as its objective to improve the internal efficiency and effectiveness of government administrations. Activities include the provision of training, payroll, and travel expenses to the employee. The G2B relationship is to provide services to businesses by providing a single portal and an integrated database. This eliminates redundant collections of data and reduces cost (Keng and Yuan, 2005). Lastly, the G2C relationship has as its objective to provide satisfactory services to citizens. The activities include access to information, such as policies, loans, benefits, and educational materials (Keng and Yuan, 2005).

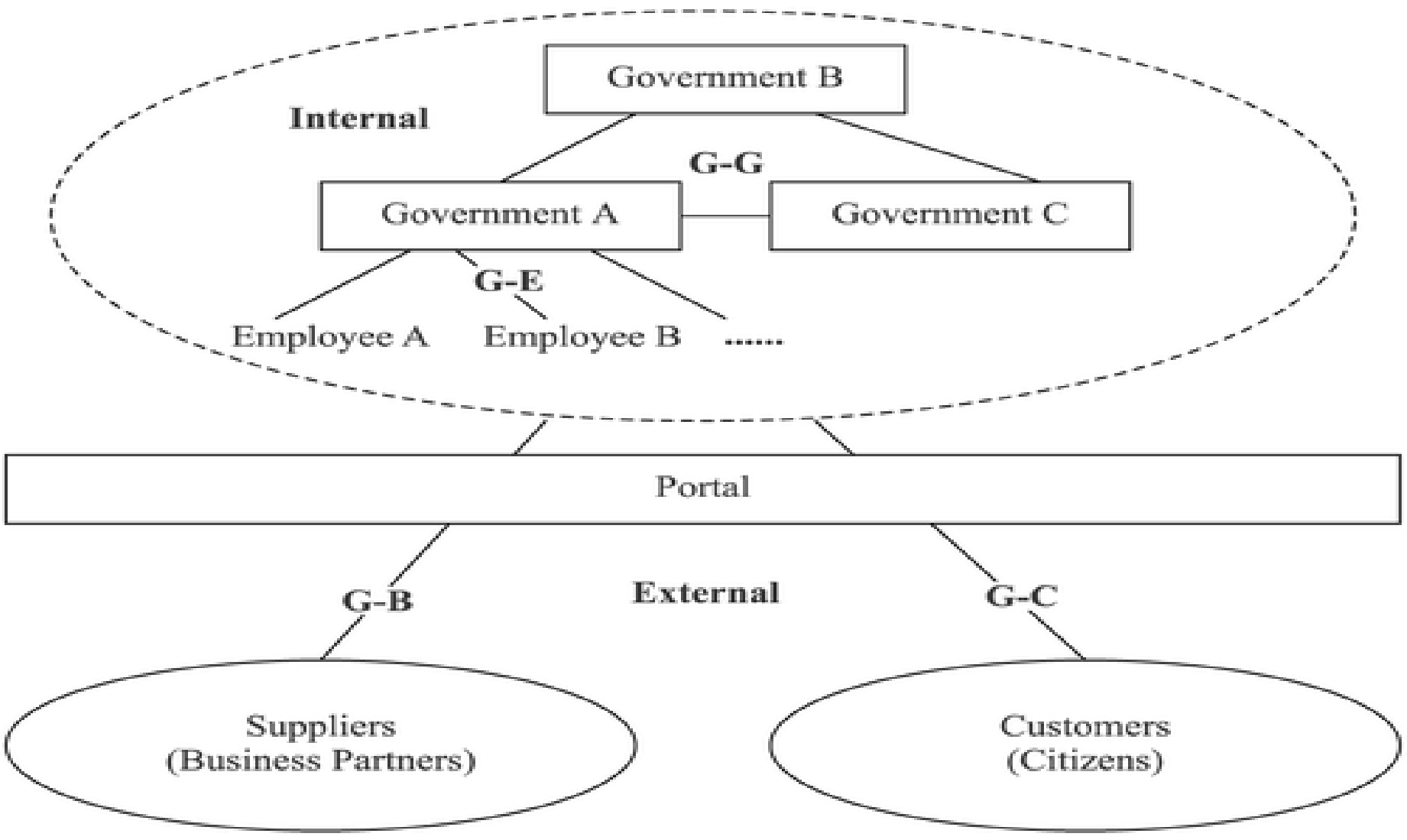

**Figure 29 E-Government Framework**
**(Keng and Yuan, 2005)**

6. **E-Services Adoption Model** was developed by Othman et al. (2012) on the success factors (adoption factors) of the implementation of E-Services in Malaysia. The model as shown in **Figure 30** below consists of 11 E-Service adoption elements, the incorporation of which would make E-Service initiatives more efficient and fruitful (Othman et al., 2012).

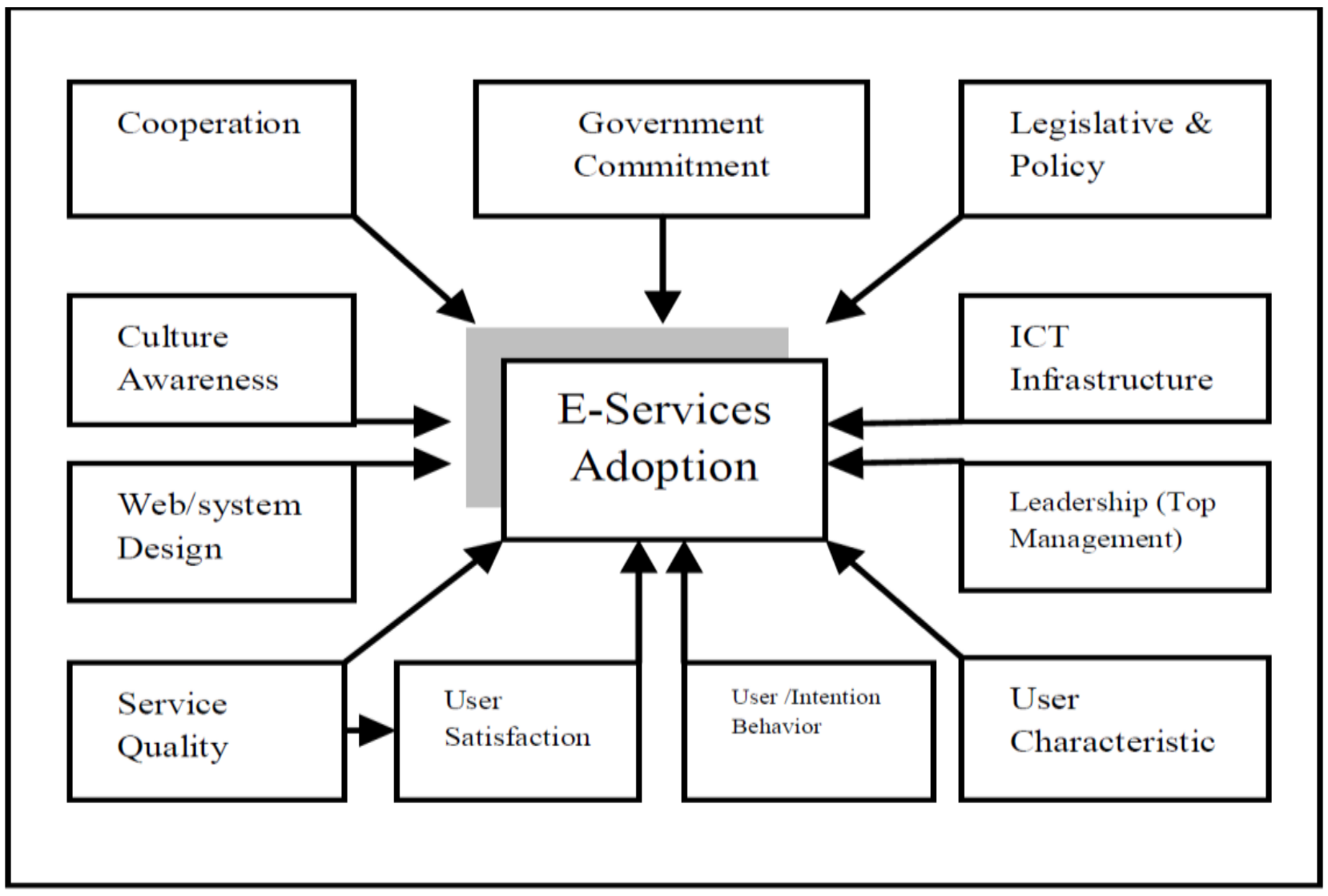

**Figure 30 E-Services Adoption Model for Malaysia**
**(Othman et al., 2012)**

7. **The Strategic Framework for E-Government** adoption, shown in **Figure 31** below, was proposed by Ebrahim et al. (2004). The framework consists of three different parts, namely: the Stage of Growth Model, the Technology-Organisation-Environment Framework and the Benefit and Barrier factors. The framework also identified six strategic objectives needed for E-Government framework, as follows:

- The evolutionary process applied to the E-Government initiatives. This will allow some stages that organisations need to pass through before the E-Government predicted benefits can be fully exploited.
- An implementation strategy that outlines the action plans and the required capabilities. Also, it includes a roadmap on E-Government evolution.
- Provision of guidance towards E-Government initiative potentials realisation.
- Identifying key factors and stages for action in framework integration.
- Developing and maintaining IT capabilities, organisation structure and process, consumer readiness and external partnership relationship.
- The incorporating website features for framework maturity.

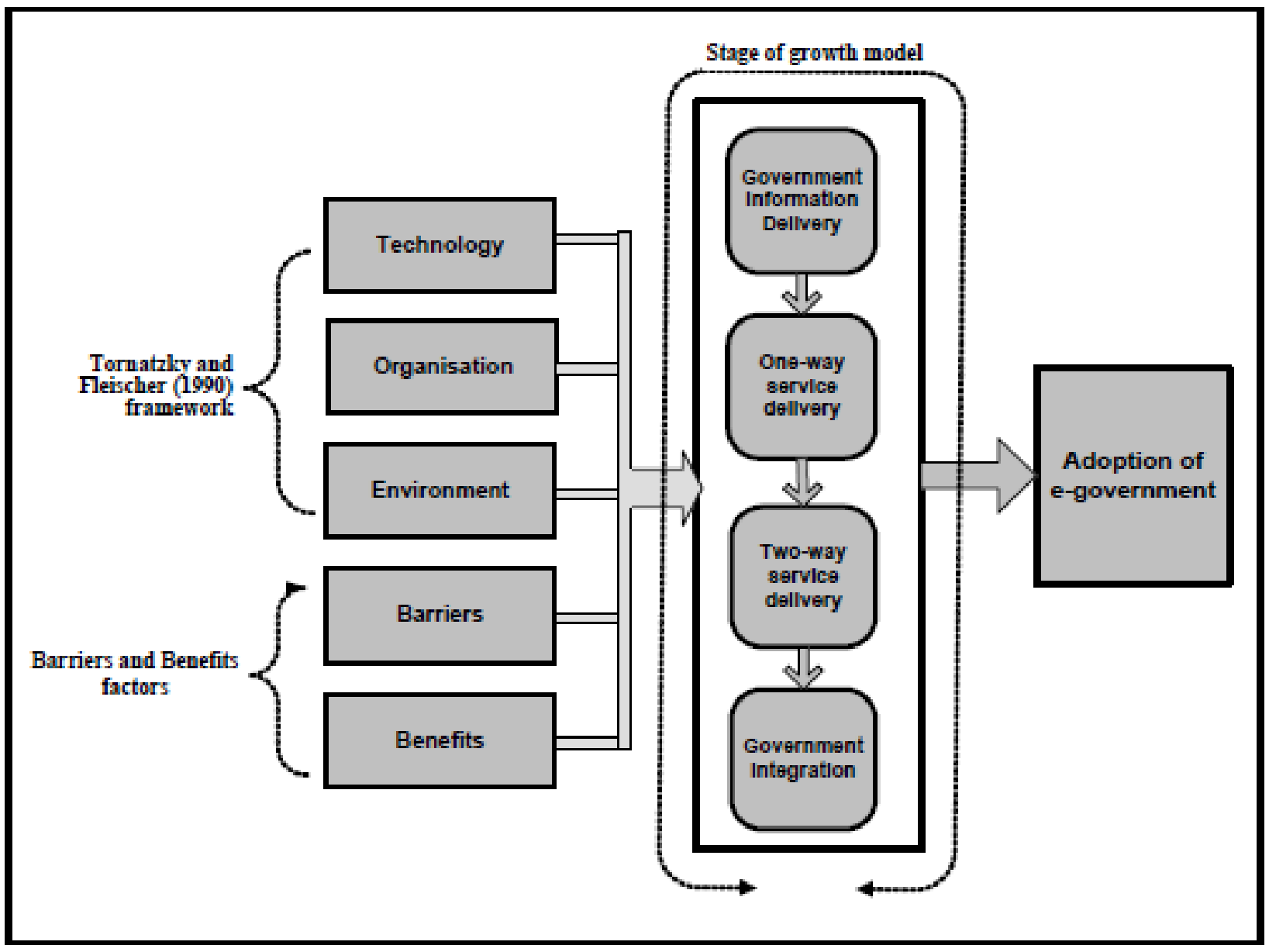

**Figure 31 Strategic Framework to E-Government Adoption (Ebrahim et al., 2004)**

8. **The Unified E-Government Model for Nigeria** was proposed by Azenabor (2013), and it combined the functionalities of other E-Government models examined in Azenabor's work. The unified E-Government model for Nigeria consists of other models in areas such as education, information management, justice, public online services, e-Health, security, and online voter's registration (Azenabor, 2013). The aims of the unified E-Government model for Nigeria, as shown in **Figure 32** below, are to encourage accountability through corruption prevention by offering transparent governance by the government. The structures of the E-Government model, if implemented by the government, will save costs, eliminate geographical boundaries and stop ghost workers, as monthly salaries to the employees will be paid electronically into their bank accounts (Azenabor, 2013).

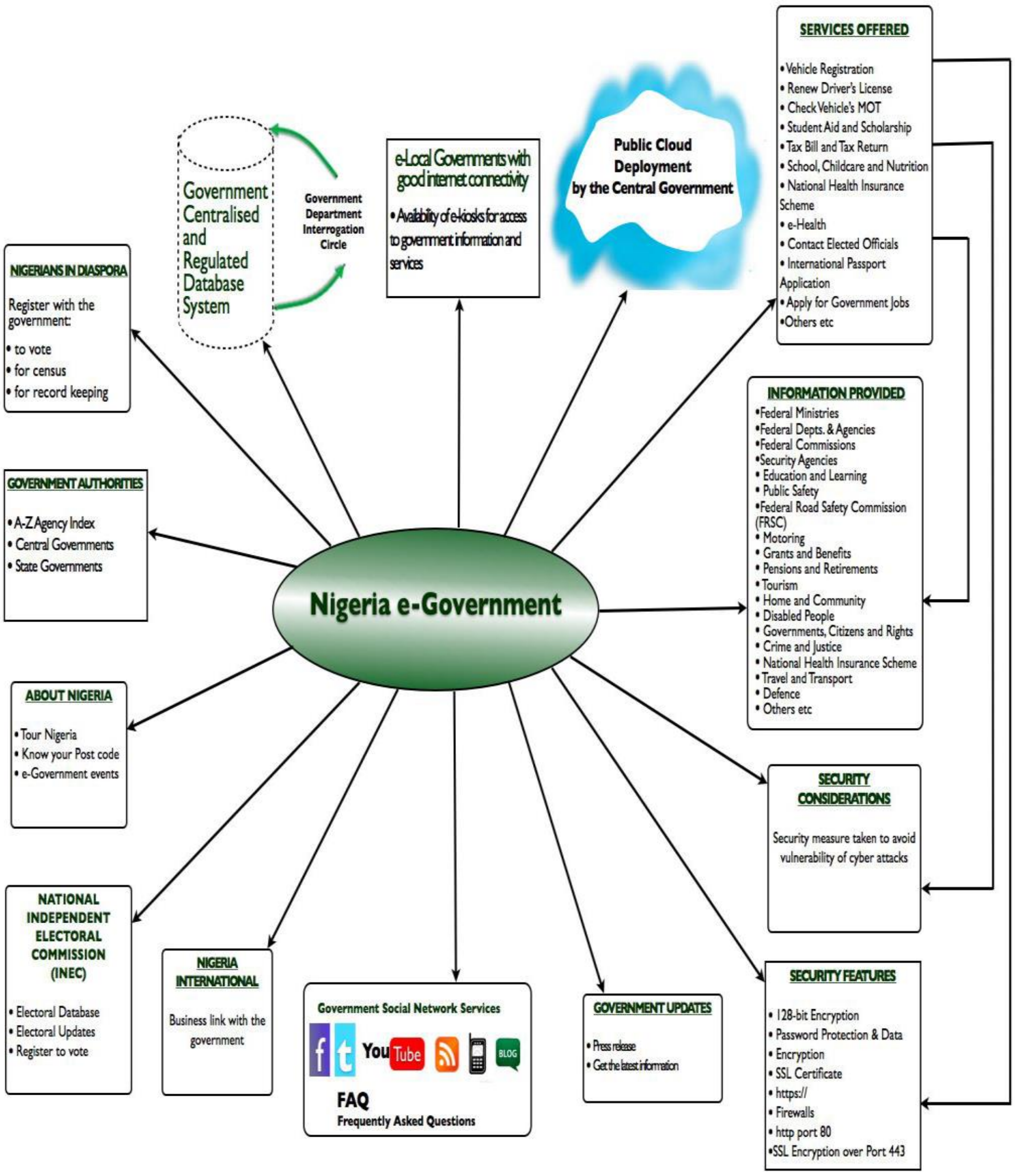

**Figure 32 Unified E-Government Model for Nigeria (Azenabor, 2013)**

9. **The E-Government Framework for States in Nigeria** was proposed by Mundy and Musa (2010). The framework, as captured in **Figure 33** below, is to improve and facilitate the adopting of E-Government at the state level in Nigeria. The framework will adopt part of the policies formulated at the federal level in Nigeria, and it is the responsibilities of the appointed commissioner to ensure the implementation of E-Government initiatives at the state level. Improving the E-Government adopting at the state level in Nigeria requires increasing E-Service adoption awareness, IT education and training, which should be spread among the citizens to improve the standard of illiteracy (Mundy and Musa, 2010).

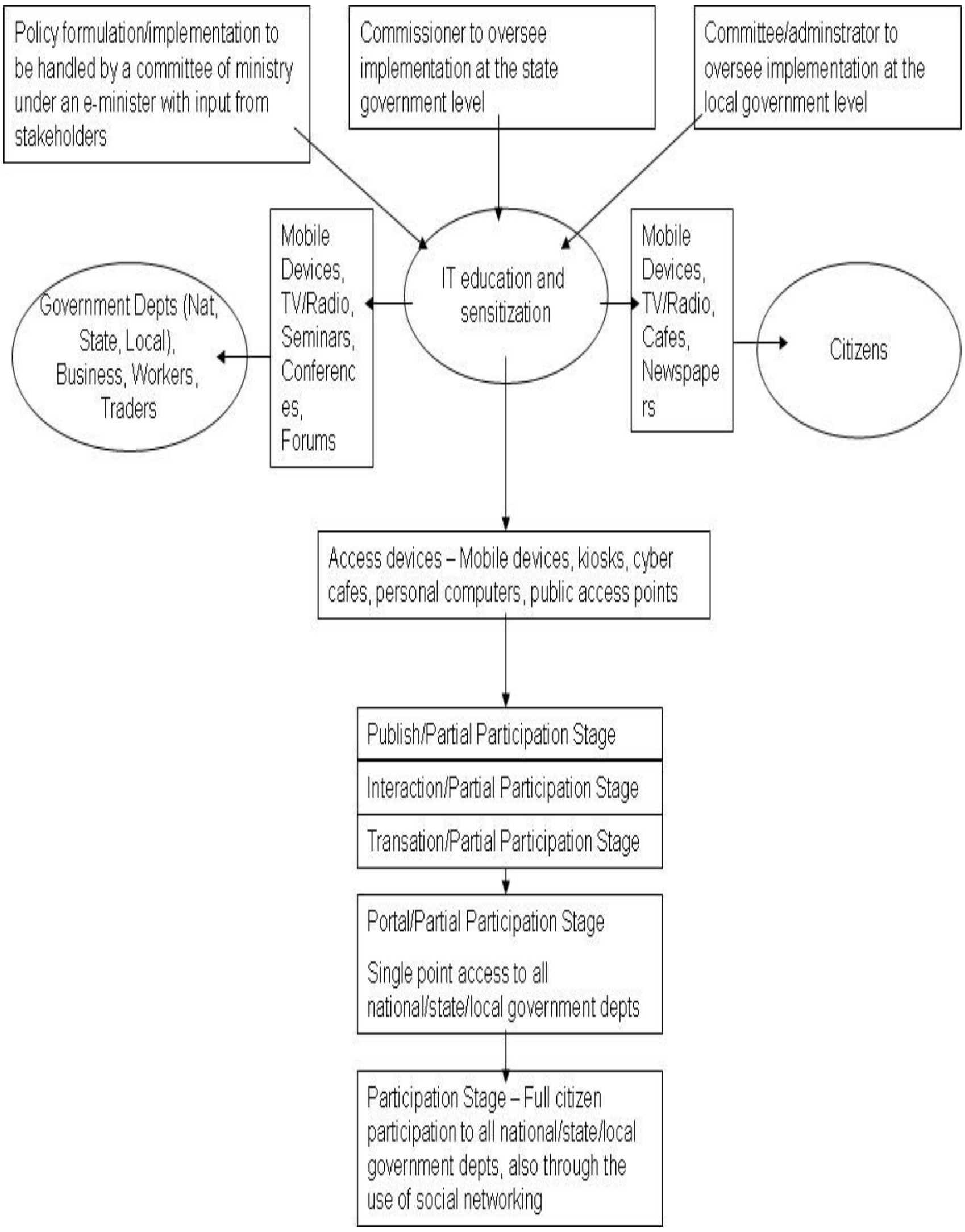

**Figure 33 State Government E-Government Framework in Nigeria (Mundy and Musa 2010)**

10. **The Four-tier Architectural Model** for E-Government applications according to Azenabor (2013) was proposed by the Federal Ministry of the Interior in 2003 for E-Government adopting in Germany. The model was developed based on the public agencies' individual E-Government initiatives with the following aims:

- To facilitate communications, modern IT architectures, technologies, and E-Government structures.

- To provide uniform standards and structure for implementing E-Government projects.
- To identify the available technologies for E-Government applications (Azenabor, 2013).

The four-tier architectural model as shown in **Figure 34** below is a multi-layer architecture with four different tiers. The multi-tier architecture consists of the client tier, the presentation tier, the middle tier, and the persistence tier/back-end (Azenabor, 2013). The functionalities of each tier are as follows:

a. The **Client Tier**: Users and application software interacts here.
b. The **Presentation Tier**: The presentation of application data happens here (for example, website).
c. The **Middle Tier**: This is referred to as the application tier. According to Azenabor (2013), this tier accommodates the component used in implementing the application logic, regardless of their presentation. The control of the program sequence is achieved here.
d. The **Persistence Tier**: The storage of data objects is done in this tier.

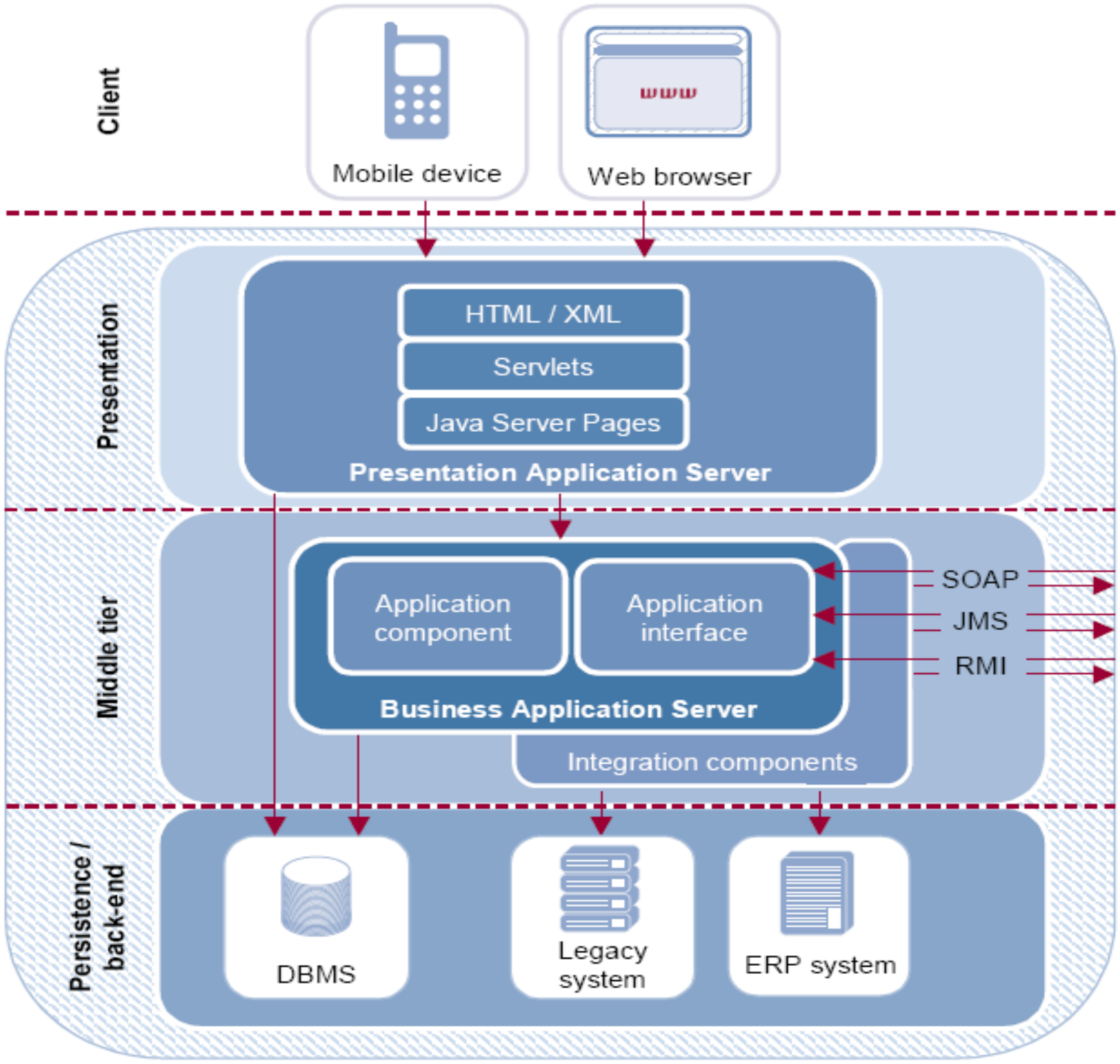

**Figure 34 Four-tier Architectural Model for E-Government Applications (Azenabor 2013)**

## 3.3 Proposed Framework of E-Service Adopting and Implementation

A conceptual framework proposed by the researcher to identify the barriers facing the adopting and implementation of E-Service at the local environment level in Nigeria. This is encouraged by the aspiration to have a successful E-Service project at the local environment level in Nigeria. The proposed framework is based on literature and desk research, and also by collating together E-Services frameworks derived from the combination of comparative study. The framework builds on prior E-Government adopting framework developed by Azenabor (2013) for the Federal Government in Nigeria, which is a unified framework for the country and the framework developed by Mundy and Musa (2010) for State level in Nigeria. Also, the researcher reviewed the features of the frameworks discussed above in **section 3.2**, including the models from Abdelkader (2015) and Hassan (2011) for this proposed framework to emerge.

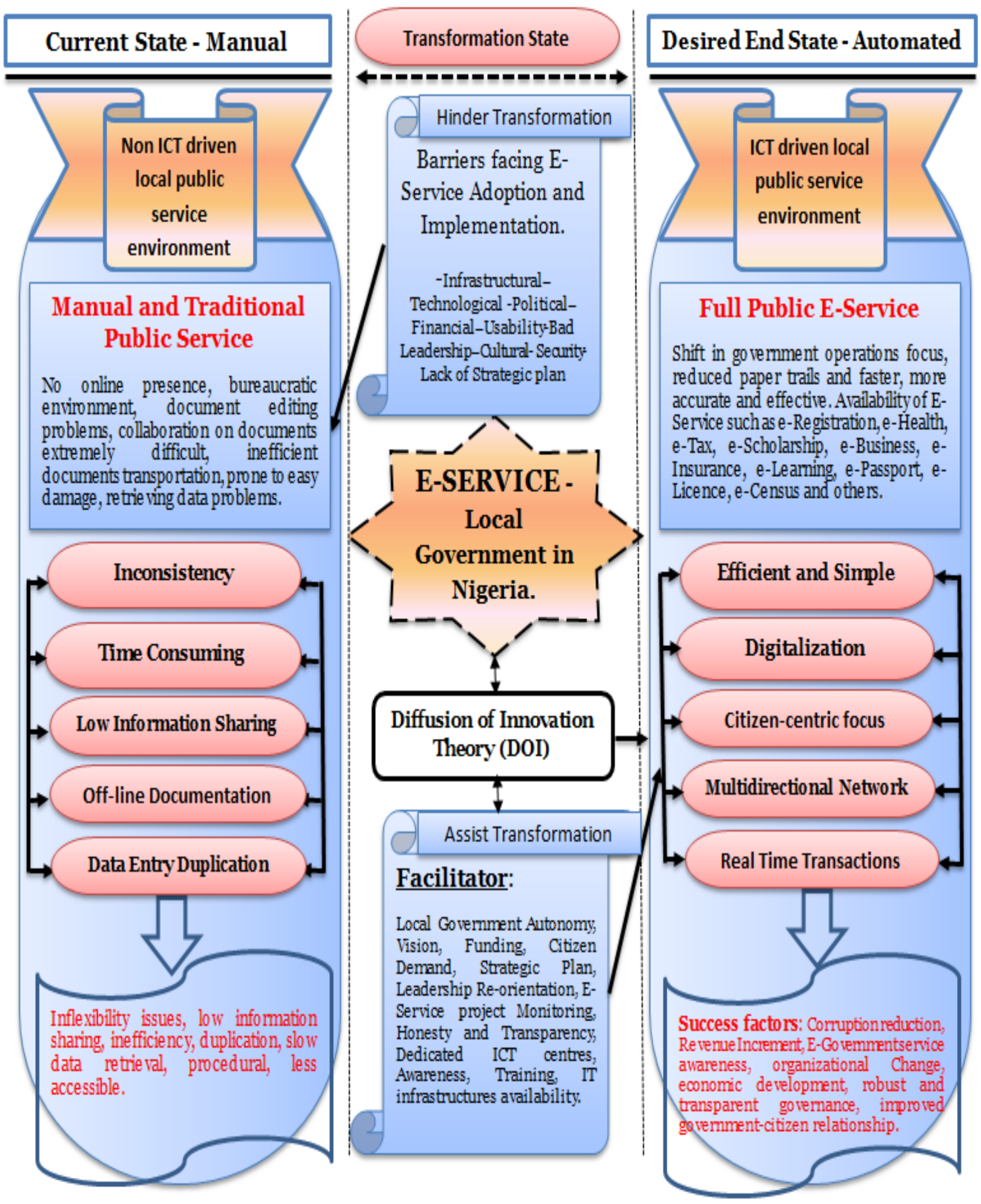

**Figure 35 Proposed Framework for Identifying the Barriers facing E-Service Adopting and Implementation at the Local Government in Nigeria**

The use of a framework is crucial in E-Service development. According to Hassan (2011), it will allow the systematic integration of various elements of a problem in a very consistent and straightforward manner. The framework will help to attain the pursued outcomes, and it allows holding a collective work discipline. One might imagine the stress experienced by the citizens at the local government level in Nigeria searching for government information. It is difficult at this level for the citizens to obtain useful information from the government. Citizens will have to travel from one government department to another searching for information that they could have obtained easily online (Azenabor, 2013).

The proposed framework as shown in **Figure 35** above summarises the barriers facing the adopting and implementation of E-Services based on the literature review in the government context (Oseni, 2016). It also reviews what E-Government services should be available and what actions the government needs to take to reach the designated target (Desired End State). The E-Service development process is considered in the framework as the transformation from the manual (traditional rigid context), as currently experienced at the local government level in Nigeria and many developing countries (Hassan, 2011; Azenabor, 2013), to an automated full public service environment (full public E-Service). However, the transformation as expected from the manual state through the desired automated state is hindered by many barriers. There are three phases involved in the proposed E-Service framework, namely the **Current State**, the **Transformation State,** and the **Desired End State**.

### 3.3.1 Current Phase

The current phase of the framework is known as the manual or traditional state of E-Service development and it is also referred to as the AS-IS Phase (Hassan, 2011; Abdelkader, 2015). This traditional phase, which is mainly a non-ICT driven public service environment, is described as hierarchical bureaucracy (Hassan, 2011). The traditional phase focuses on internal and managerial concerns and is characterised by weak communication. The features of the traditional system are based on behaviour that is standardised by the government officials, enforcing discipline, and by respecting the rules (Petrescu et al., 2010). Similarly, organisation and not the individuals determine career advancement, which depends on technical or other relevant qualifications (Hassan, 2011).

However, the traditional bureaucratic paradigm according to Persson and Goldkuhl (2010) has been criticised by the New Public Management (NPM) movement for underperformance and poor legitimacy in meeting the customer needs. Petrescu et al. (2010) also added that the traditional system is characterised by principles such as the bureaucratic hierarchy, and criticised for its rigidity, slow decision-making and inactive procedures (Hassan, 2011).

### 3.3.2 Transformation Phase

In the transformation phase, there are enablers such as strategic plans, funding, vision, and others that will facilitate the transformation and speed up the E-Service development process to the desired end state. The need for the online services and the focus of the government is an inward-looking approach to the outward one to meet the needs of the end users (Hassan, 2011). As described in Gartner's four-stage E-Government development model, the current operational processes transformed to the automated one to provide more efficient, integrated and unified public service (Baum and Di Maio, 2000). Keng and Yuan (2005) explained that transformation is involved in the way governments provide services and that it could be either vertical or horizontal transformation. In their synthesising E-Government stage model, the vertical transformation could be within governments at different levels, or horizontal in various locations.

### 3.3.3 Desired End Phase

The desired end phase of the proposed E-Service framework is also known as **TO-BE Environment** (Hassan, 2011; Abdelkader, 2015) and is the full ICT-driven public environment where more effective and efficient government services are provided to the public. The government services are delivered through ICT platforms such as the internet, wireless devices, telephone and other communication devices (Hassan, 2011). The failure to provide government services to the public according to Bertot et al. (2016) in many developing countries is due to inducement and accountability problems. The only way to overcome these challenges is through innovative and ICT-driven public service delivery, namely E-Government (Bertot et al., 2016). The desired end phase, which is a full digitalisation state, will identify innovations in government services delivery, and it will focus on the following:

- **Transparency**: Citizens will have bright ideas and an understanding of public policies and services.
- **Anticipatory**: The provision of services to the public is initiated by the government (Bertot et al., 2016).
- **Participatory**: Citizens will have the opportunity to engage the government and be part of the decision-making process.
- **Personalisation**: Citizens can personalise their mode of service delivery (Bertot et al., 2016).

Moreover, the ICT-driven phase model encourages the community ownership process as this will allow the public to take on the ownership of the problems. The partnership between the citizens and the government officials to find the solution to the problems is encouraged. Information sharing, improved communications, and availability of multi-directional network is promoted (Hassan, 2011).

## 3.4 Current E-Service Projects in Nigeria and E-Government Service Stages at the Local Environment Level

This section discusses current E-Service projects in Nigeria, including the stages of the E-Service development at the local government levels. This study was started in February 2014 and the stage of E-Services at the local government level at that time was not on a par with other developing countries, as Nigeria was ranked 19th in the E-Government Index ranking for Africa in 2014 (Oseni and Dingley, 2014). Also, concerning the gaps identified in **section 1.4.2** (Chapter One), there were limited frameworks covering E-Service adopting and implementation at local environment level in Nigeria. Also, the current E-Service frameworks do not capture the full context of the barriers facing E-Service adopting and implementation at the local environment level in Nigeria. In the light of the above, the next sections will discuss current E-Service projects and E-Government service stages at the local environment level in Nigeria.

### 3.4.1 Current E-Service Projects in Nigeria

This section discusses the current E-Government services in Nigeria to complement the E-Service initiatives mentioned in **section 1.3.2** (Chapter One). Most of these services is being sponsored by both the federal and state governments, as explained in **Table 18** below. The distance between the government and the citizens continues to widen in Nigeria (TVN, 2014), and only the laudable E-Services are able to close these communication gaps. There is an urgent need for the government to heavily invest in ICT infrastructures, E-Service project monitoring, and the availability of funds for E-Services, particularly at the local administration levels, to curb the menace posed by the high rate of corruption among the government officials and stakeholders involved in various government service implementations.

**Table 18 Current E-Service Projects in Nigeria**

| S/N | E-Service Initiative | Advantages | Class | Level |
|---|---|---|---|---|
| **1** | **E-Passport**: The electronic passport (EP) service was introduced by the Nigeria government in 2007 to replace the old machine readable passport (MRP). The enforcement started on 30 April 2011, when the citizens with the old machine readable passport were unable to travel without their electronic passports.<br><br>The new e-passport conforms to the global standards and enjoys the confidence of the international communities. The new electronic passport has an additional small integrated circuit known as a “Chip” embedded at the back cover to prevent misuse. The e-passport application is applied for online via the Nigeria Immigration Service website, including the application fees using various means of payment such as a credit or debit card. | The new Nigeria electronic passport (EP) stops reckless abuses from hackers and identity theft experts. | G2C | Federal |
| | | | | |
| **2** | **E-National Database**: This service was initiated in 2014 to combat crimes, as Nigeria is among those countries where inadequate social facilities have made it difficult for crime control. The national identity database empowers security agents to track and apprehend the criminals. Serial offenders are kept under surveillance with the help of the electronic national database. This E-Service project was initiated by the National Identity Management | It prevents crimes committed as the criminals could be identified easily. Examples of such crimes are drug trafficking, internet and credit card frauds. | G2C | Federal |
| | Commission (NIMC) on the behalf of Nigeria government. | | | |
| | | | | |
| **3** | **E-Learning**: This service was introduced by the Lagos State Government in 2015 to enable the students in public schools to learn through the internet. It allows the use of mobile phones and social media platforms. The project is deployed to deliver quality education to students through mobile and electronic learning protocols. Presently, few private and public universities in the country have introduced e-learning to enhance students' learning experiences such as those in developed countries like the UK. | This E-Service project was initiated to boost the students' performance in public examinations and to aid human capital development. | G2C | State |
| | | | | |
| **4** | **E-Banking**: This service enables the citizens to use the computer and the internet to carry out banking instructions. This E-Service was introduced in 2012 by the Central Bank of Nigeria (CBN) with Lagos as a pilot state. The cashless economy through the use of IT helps to reduce the transaction cost for banks. This service is now available in many other states in Nigeria after Lagos, which serves as the pilot state. | According to the CBN and the banker's association, e-banking is more convenient, cheaper and faster, and it reduces the risk associated with cash-related crimes. It also reduces revenue leakages and increases economic development. | G2B, G2C | Federal /Private |
| | | | | |
| **5** | **E-Payment**: There are significant efforts for the governments at both the federal and state to digitise government payments. The e-payment system was introduced by the government on 1 January 2009. Payment of taxes and custom duties are now being paid online. For economic growth, there is a need to accelerate financial inclusion and this could be achieved through e-payment. Governments at both | The main advantage of the e-payment in Nigeria is the introduction of the cashless economy, and this has eliminated the fear of the unknown. Though the system is faced with | G2G, G2B, G2C | Federal /State/ Private |
| | federal and state levels also engage private electronic payment solutions like InterSwitch to collect taxes on their behalf. | challenges such as acceptability issues, it has made drastic efforts to address corruption issues in Nigeria. | | |
| | | | | |
| **6** | **<u>E-Driver Licence</u>**: This E-Service enables the citizens to obtain a new electronic driver licence and to renew their old driving licence in the country. It is embedded with a small integrated circuit known as a "Chip". The information stored in the chip could be used to trace an individual in the case of an accident or crime, as all information is connected to a central national database maintained by the Federal Road Safety Corporation. This new initiative conforms to the international standards as the citizens could process the application online anytime, irrespective of their location. | The new e-driving licence will help to combat fraud and increase trust. It helps to streamline the application process and produce more cost effective licences. Lastly, it ensures compliance and conforms to the international standard. | G2G, G2B, G2C | Federal |

### 3.4.2 E-Government Service Stages at the Local Environment Level

This section discusses the stages of E-Services at the local environment levels in Nigeria. According to the 1999 Constitution of the Federal Republic of Nigeria, there are 774 local governments as shown in **Table 1** (see section 1.3). However, few of these local governments have an online presence in 2014, when the researcher began this study. In fact, the local government's online presence at that time was through the state governments, or private and multi-national companies' executing projects in the local governments, as indicated in **Figures 36** and **37** below.

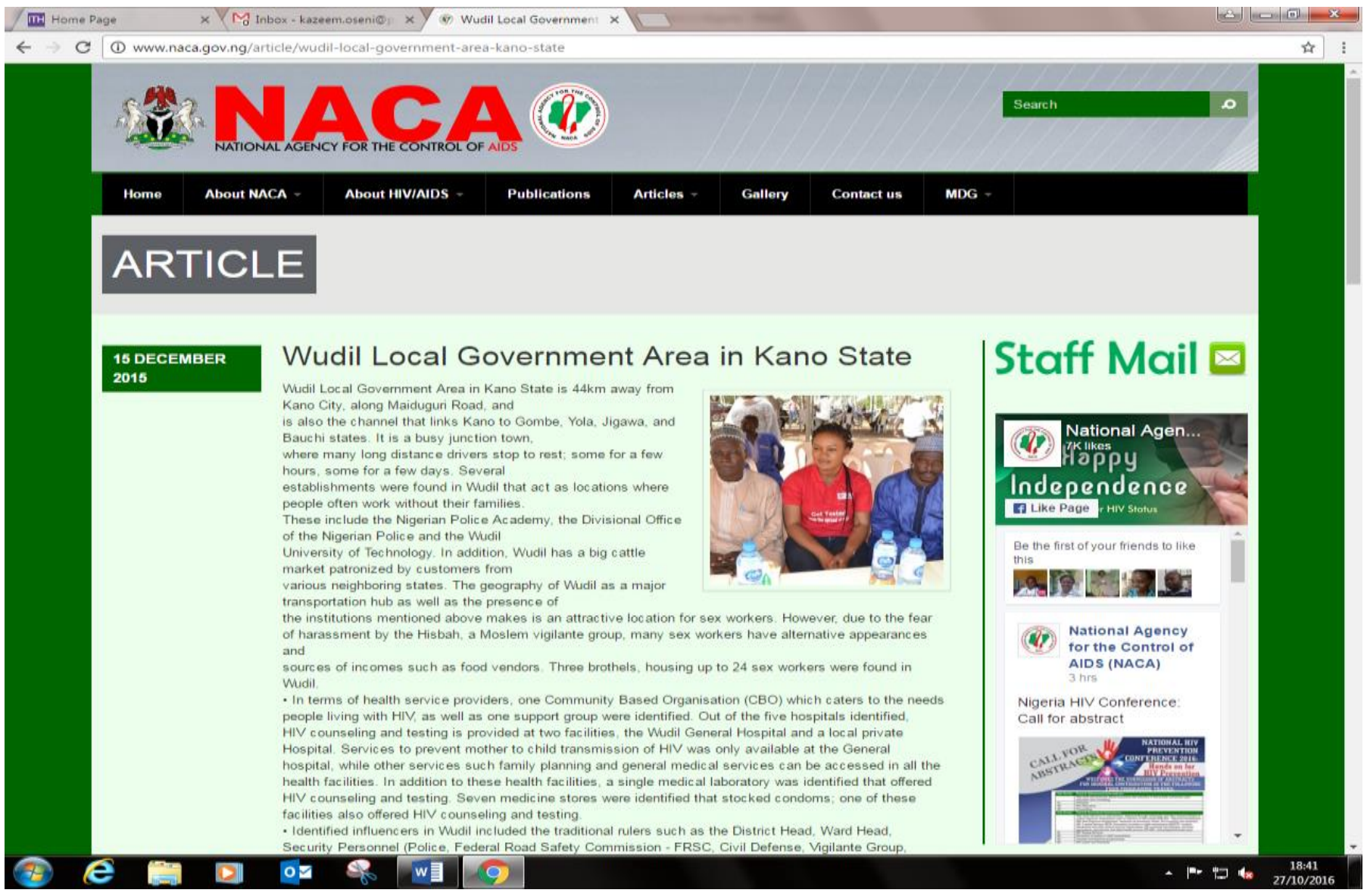

**Figure 36 Wudil Local Government Kano State Nigeria**

As shown in **Figure 36**, in Wudil local government area in Kano state, Nigeria had an online presence in 2015 through the National Agency for the Control of AIDS (NACA), which was carrying out an awareness project for AIDS control at the local government. This researcher made another check, both in 2016 and 2017, on the Wudil local government website, and the results show that the local government was offline as there was no permanent and functioning website for the local government to deploy E-Service initiatives for the local citizens. Secondly, a check on the local governments in Ekiti State by this researcher at the beginning of this study showed that all the 16 local government areas had no functioning websites, but maintained an online presence through the state government website, as illustrated in **Figure 37** below.

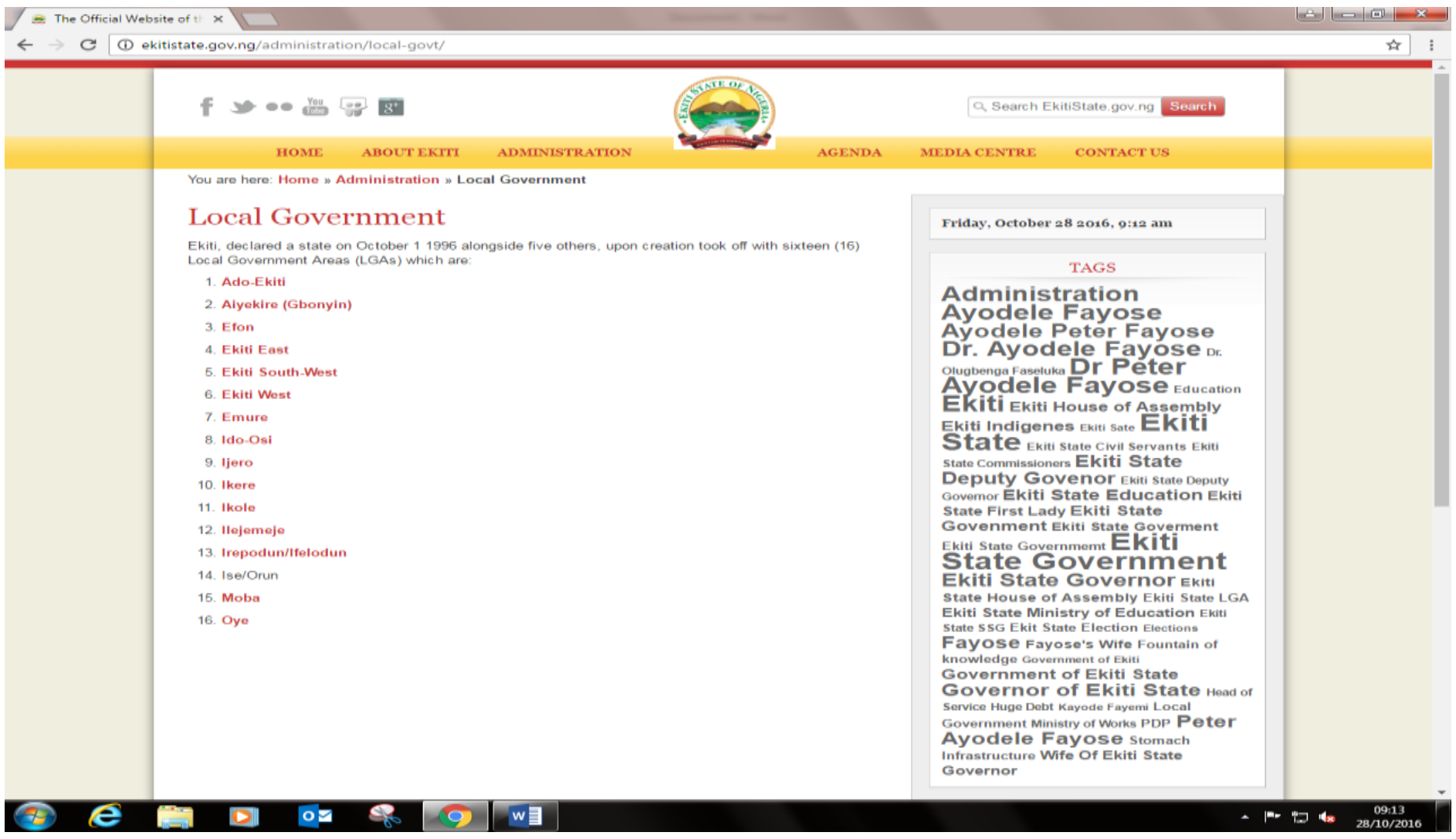

**Figure 37 Ekiti State Government Nigeria Website**

Similarly, as the study progresses, this research made efforts to check the Nigeria local government's online presence because it might have an effect on the research outcomes, and to confirm the reasons for conducting this research work, as indicated in the research gaps. Few of the Nigerian local government websites checked by the researcher showed that they still maintained their online presence through their respective states, but not as an independent body, as expected considering the technology revolution in the world today. Interestingly, some local governments in Lagos State, which is the economic power of Nigeria (Azenabor, 2013), now have an online presence as shown in **Figure 38** below.

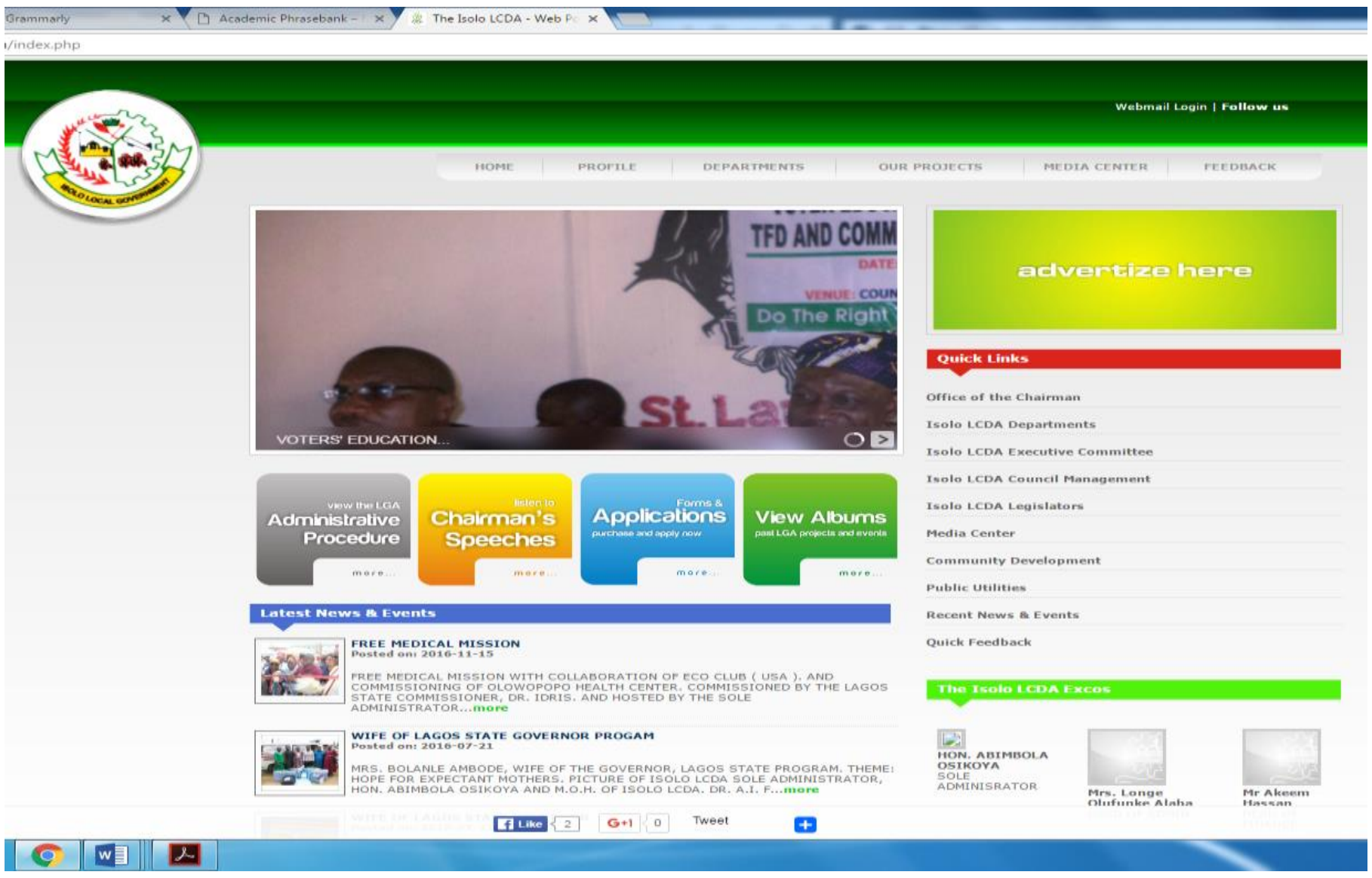

**Figure 38 Isolo Local Government Lagos State Nigeria Website**

Furthermore, the online presence achieved by the Isolo local government, as shown in **Figure 38** above, is commendable. This might be connected with the focus of the council leadership in providing E-Services to the citizens to aid access to the government information. However, the website is still in the publishing stage of E-Government development, as discussed in **section 2.2.3** (Chapter Two). The publish stage is the basic form of E-Government stages, as governments publish very simple and limited information on their websites. The information may include the vision and mission of the agency, contact information, official hours, and documents. Efforts made to download forms proved abortive, and there were no online payment facilities. Hence, these findings support the researcher's decision to conduct this study in E-Government domain. It is important to unveil the barriers facing the E-Service implementation and adopting at the local environment level in Nigeria, and to understand why few local governments could only maintain a basic (publish stage) of E-Government development.

## 3.5 The Press/Media Coverage on E-Services Initiatives in Nigeria

There is no doubt that the press and media, including social media (Twitter, WhatsApp, Facebook, and Instagram), play a vital role in the growth of any country. Mass media convey information to the citizens on economic priorities, cultural, and, political agenda by forcing their attention to certain issues (Olayiwola, 2013). Unfortunately, there is limited press and media coverage for E-Service initiatives in Nigeria. Nonetheless, the media and press covered most of the E-Service initiatives discussed in **Table 18** above. Recent research shows that the media and press in Nigeria are focusing more on reporting on the insurgency in the northern part of the country (Omotosho, 2015). The fact is the media, and the press, remain a major communication medium. The government in Nigeria needs to raise its awareness level about E-Service initiatives through a full implementation and hoped that, with the evolution of modern technology, activities on E-Service projects would be a priority for the media and press to broadcast.

## 3.6 Summary

This chapter has reviewed the relevant E-Government service models and frameworks. It has also presented a proposed E-Service framework that could be adopted at the local government level in Nigeria. The framework is derived from the E-Government Service model from both Abdelkader (2015) and Hassan (2011), with a modification to make it more appropriate to the Nigeria local government level context. The framework builds on previous E-Government adoption frameworks developed by Azenabor (2013) for the Federal Government in Nigeria, which is a unified framework for the country, and by Mundy and Musa (2010), which targets the State level in Nigeria. The chapter also revised current E-Service projects in Nigeria and looked into the current E-Service initiatives covered by the press/media in the country.

However, the chapter has not explored in depth the immediate benefit of the framework developed for this study on E-Service adoption at the local environment level in Nigeria. Also from this chapter, at least a couple of the other frameworks reviewed will not work because Nigeria has unique needs, so a specific framework will be more useful. Hence, these are limitations of the chapter and future research should be able to cover more on the limitations to improve research in E-Government services.

# Chapter 4 - Methodology

## 4.1 Introduction

The research methodology approach in information systems has been given a lot of attention recently, as making a good approach by the researcher will have a significant impact on the study.

The purpose of this chapter is:

*To provide and justify the rationale in selecting a suitable research strategy and outline, and to ensure that the research methodology that has been chosen will be able to answer the research questions, aims, and objectives of this study.*

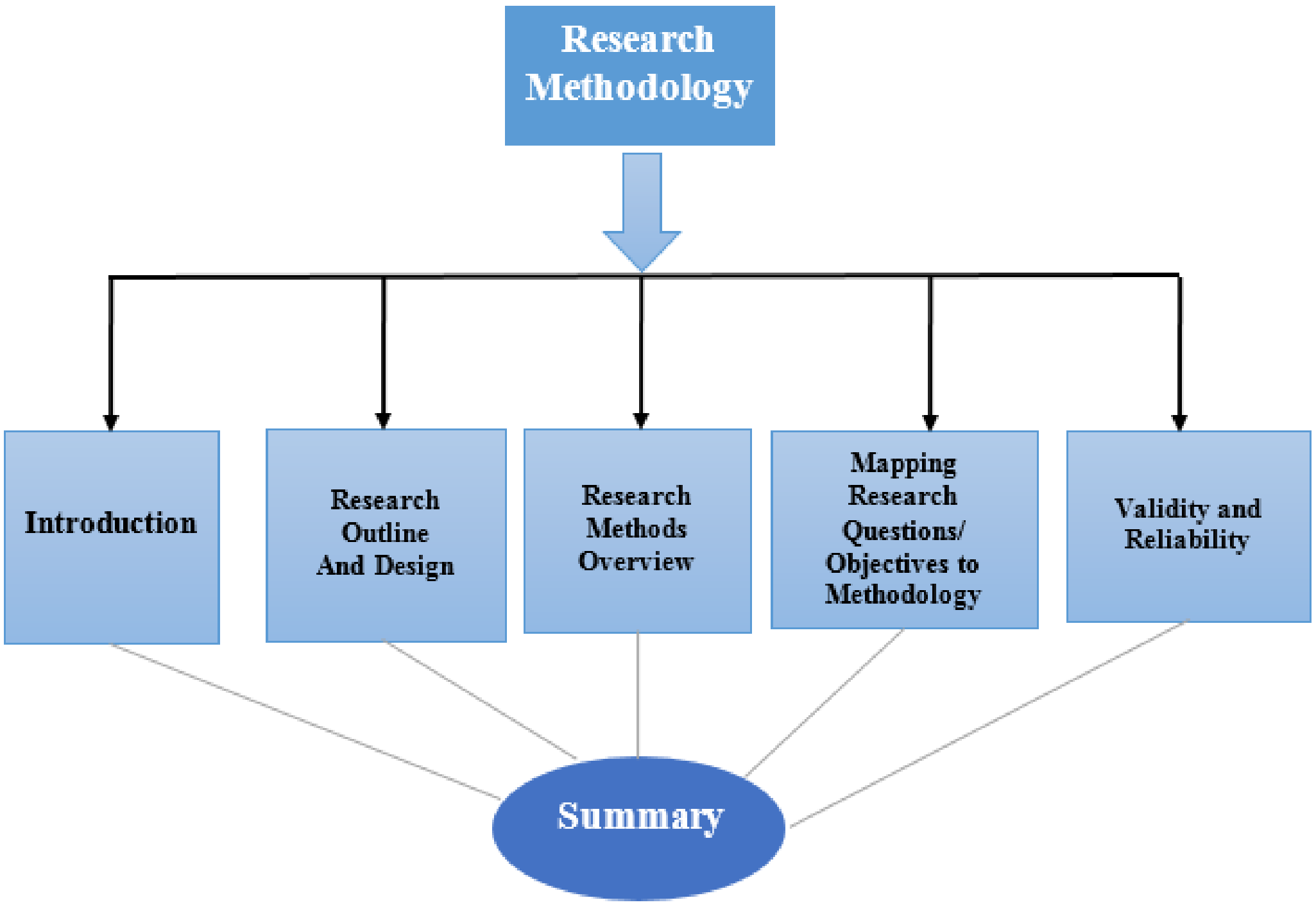

**Figure 39 Research Methodology Summary**

Therefore, this chapter consists of five parts as shown in **Figure 39**. The first part is the introduction, followed by the research outline and design. The third part is the overview of research methods, which includes the various research methods and approaches, philosophical paradigms and research philosophy, the adopted research methodology, and justification for the adoption. Reliability and validity will be discussed in the fourth part, as this is very significant in evaluating any academic research. This chapter will end with a summary.

## 4.2 Mapping Research Questions and Research Objectives to the Methodology

The whole emphasis of this section is to explain the links between the research questions, objectives, and the research methodology as shown in **Table 19** below.

**Table 19 Mapping Research Questions and Research Objectives to the Methodology**

| Research Questions | Research Objectives | Methodology |
|---|---|---|
| What are the barriers facing E-Service Adopting and Implementation at Local Environment Level using Nigeria as a case example? | To identify the barriers facing E-Service adopting and implementation at Local Environment Level using Nigeria as a case example. Success factors will also be taken into consideration. | Action Research Method:<br>- Document collection and analysis (Literature Review)<br>- Keeping a research journal<br>- Search conference<br>- Interviews with the participants<br>- Case Studies<br>- Online Survey<br>- Focus group discussion |
| | To learn from the E-Government Services in developed and developing countries. | Action Research Method:<br>- Document collection and analysis (Literature Review)<br>- Keeping a research journal<br>- Interviews with the participants<br>- Online Survey<br>- Focus group discussion |
| Is it possible to develop an E-Service framework that will be suitable for the Local Government in Nigeria as compared to the developed countries? | To propose a Diffusion of Innovations (DOI) theory based framework for predicting the successful E-Service adopting and implementation using action research methodology by looking at existing frameworks used in advanced countries. | Action Research Method:<br>- Interview results with the participants<br>- Online Survey results<br>- Focus group discussion |
| | To evaluate and validate the proposed E-Service adopting and implementation framework for the Local Environment Level using Nigeria as a case example. | Action Research Method:<br>- Testing: using triangulation, survey questionnaires and interview results to evaluate the usefulness of technology in E-Service adoption and implementation at Local Environment level using Nigeria as a case example.<br>- Focus group discussion |
| | To make recommendations on the study outcomes to the policy makers at Local Environment level in Nigeria, especially the Local Government Service Commission. | Action Research Method:<br>- Using results from the study: interviews and survey results<br>- Focus group discussion<br>- Participants' observations and suggestions |

Dapper (2004) stated that the relationship between the research questions, objectives and methodology is critical to the entire research process, as applying them inappropriately might weaken the results of the research study. Therefore, it is essential that there should be a correlation between them in order for the research to achieve the study goals and with the research objectives, by using the appropriate methodology, which should be able to answer the research questions. In **Table 19** above, the researcher highlighted how the research objectives and research questions would be answered using appropriate research methods.

## 4.3 Research Outline and Design

The research needs to have a detailed outline of how the study investigation will take place. The outline typically includes how data are collected and what tools are used for the data collection. It will also describe the mode of the data analysis. Research design justifies the need to have a framework that will control the collection of the required data for this study, the process, and specification of methods engaged in the research. In light of the above, and working towards achieving the objectives as mentioned above, the researcher carried out the research data collection steps, as shown in **Figure 3** in Chapter One.

### 4.3.1 Description of Data

Data on E-Government services collected for the research include the barriers facing the E-Service adoption and implementation, the perceived or real success factors, and the use of theory. Participants at a local government in Nigeria were contacted to arrange the interview sessions. The participants included members of the local government staff and other stakeholders in E-Government service initiatives. The action research method used included document collection and analysis, participant observation recordings, online questionnaire surveys, structured and unstructured interviews, online focus groups, and case studies.

### 4.3.2 Confidentiality and Ethical Consideration

This research went through an ethical review. The main ethical issue was to ensure anonymous responses were received from the participants. Participants' consent was obtained, and the normal condition of the participants being able to leave at any point should they feel uncomfortable was emphasised. The consent form, participant information sheet, and invitation letter were sent individually to the members. Personal information was treated in confidence and no personal identification was used so that the participants remained anonymous for the online focus group sessions. All data collected will be transcribed where necessary, analysed quickly and original sources removed from servers.

## 4.4 Research Methods Overview

The researcher's main secondary source of data collection is the literature review. It was examined to discover the current issues on the E-Government services adoption and implementation. Reacting to Gill et al. (2008), data collected through any method in the research study will help to explore views, beliefs, and experiences of participants on the topic domain. However, exploring different research methods will assist in gaining an in-depth insight into the research study (Azenabor, 2013). The essence of the prior literature review for this study and as part of the methodology is to see how the previous work on E-Service adoption and implementation in other countries could be incorporated into the Nigerian local government system. More discussions are included in the next section on the research methodology, approaches, philosophy, and research method adopted for this study.

### 4.4.1 Philosophical Paradigms and Research Philosophy

In the research study, it is imperative to review different philosophical views. Research philosophy relates to the nature and development of knowledge, and it is a belief in the way data should be gathered, analysed and used. Hassan (2011) defined a paradigm as a set of shared assumptions or ways of thinking about some aspects of the world. This definition of the term paradigm by Hassan (2011) substantiates the earlier understanding of paradigm by Kuhn (1962), which he defined as a "set of common beliefs and agreements shared between scientists about how problems should be understood and addressed". Accordingly, different philosophical paradigms hold divergent views about assumptions on the nature of the world or reality (Ontology – the ways of constructing reality, how things work) and the set of assumptions on the best means to acquire knowledge (Epistemology) (Easterby-Smith et al., 2002; Hassan, 2011).

Philosophical paradigms as used in this research study are interrelated with the research methodology (Easterby-Smith et al., 2002). As shown in **Figure 40** below, ***ontology*** is the way of constructing the reality, how things are and how things work. In a related development, ***epistemology*** is about the different forms of knowledge of that reality. That is, how do we know the relationship that exists between the inquirer and the inquired? To solve the issue of the reality (truth) in any research study, there is a need to understand the appropriate tools necessary for this, and that is where ***methodology*** comes in (Easterby-Smith et al., 2002). The methodology could be quantitative, qualitative, mixed methods, designed-based research or action research, and the researcher can make use of any dependence on the research study. Every paradigm according to Scotland (2012) is based upon its own ontological and epistemological assumptions. Each has different assumptions about the reality and knowledge, which strengthen their particular research approaches. This is reflected in their methodology and methods (Scotland, 2012).

In the research study, there are many arguments and discussions on the progress of research philosophical paradigms. Nevertheless, it is imperative for the researcher to understand each side of the argument, since research problems often require compromise designs which may be drawn from more than one tradition (Easterby-Smith et al., 2002). In general, there are two major research philosophical paradigms in social sciences, namely ***positivism*** and ***social constructionism (Interpretive)***. However, Oates (2006), Hassan (2011) and Scotland (2012) confirmed that some authors added a third paradigm (**Critical**) and a fourth paradigm (**Pragmatic**) for the research in Information Systems, as shown in **Table 20**.

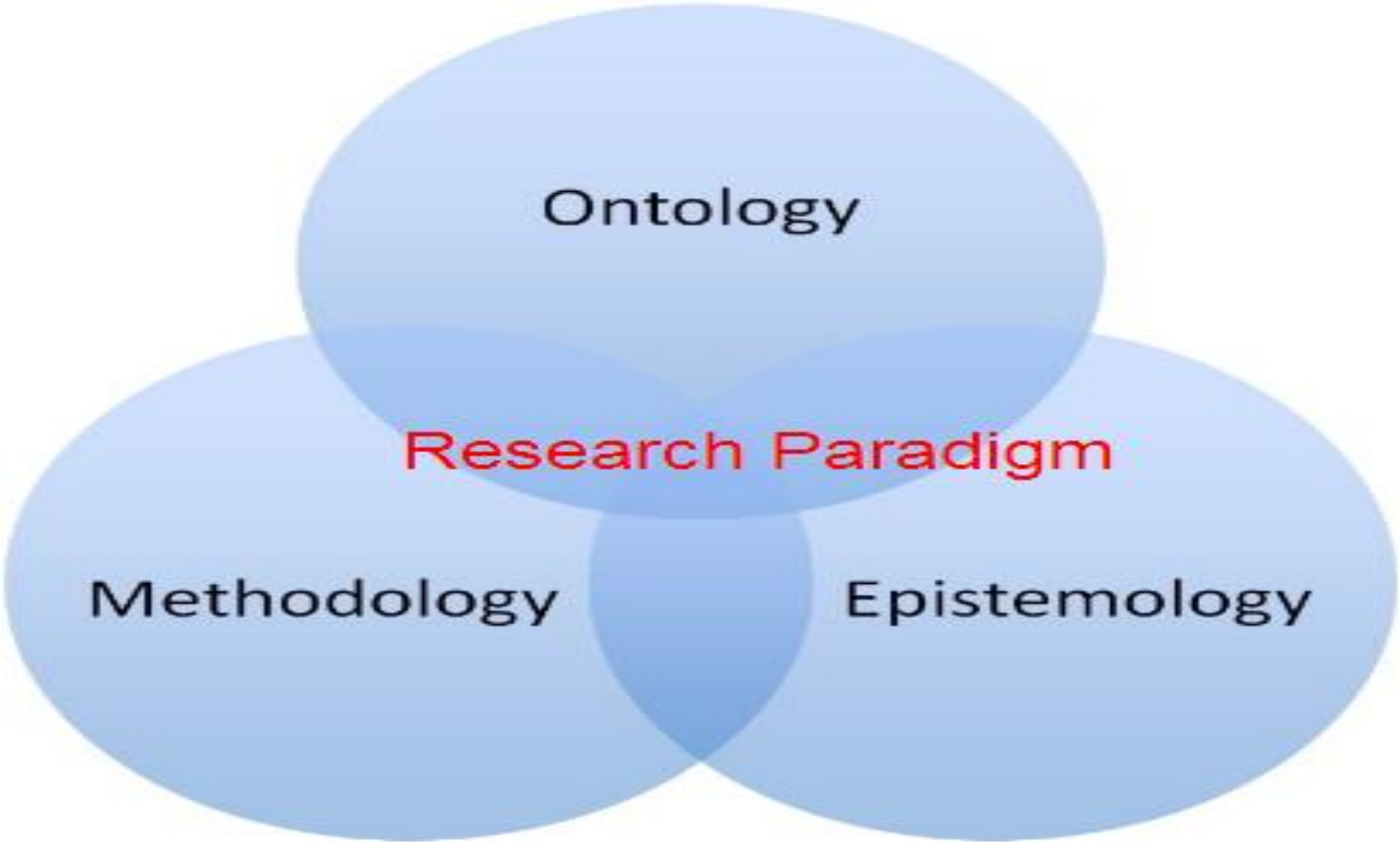

**Figure 40 Research Paradigm**
**(Anderson, 2005)**

**Positivism:** This research paradigm is based on the idea that regards science as the only medium of learning about the truth (Collins, 2010). The positivism paradigm as explained by Aliyu et al. (2014) indicates that both real and factual events can be observed and studied both scientifically and empirically through investigation and analysis. Positivism's fundamental idea according to Hassan (2011) is that the social world exists externally, and it will be realistic if its properties are measured through the objective methods instead of being conditionally subjective through reflection (Hassan, 2011). This approach also limits the researcher's activities regarding data collection and analysis, which are usually observable and quantifiable. Hence, the positivism paradigm in philosophy followed the suggestion that only genuine knowledge gained through observation is trustworthy (Collins, 2010). Furthermore, the position of the positivism paradigm, as explained above, appears to have a contrary opinion according to Aliyu et al. (2014), who argued that positivism lacks the necessary foundation for research and seems to be weak. Positivism as a research paradigm has the following disadvantages according to Collins (2010):

1. Positivism paradigm relies on the experience as a valid source of knowledge.
2. It came with the assumption that all types of processes are perceived as a variation of actions between individuals.

3. Research findings in positivism research are found to be only descriptive, lacking insight into in-depth research issues (Collins, 2010).

**Social Constructionism (Interpretive**): This research paradigm explains how knowledge is acquired through interaction with other people. McKinley (2015) explained that human development is socially situated. Hence, social constructionism endeavours to understand how individuals create and maintain their social worlds (Hassan, 2011). Korsgaard (2007) argued that social constructionism is based on reality being decisively formed by individual perception, as shown in **Table 20** below.

**Table 20 Summary of Research Philosophical Paradigms (Korsgaard, 2007)**

| Paradigm | Ontology | Epistemology | Question | Method |
|---|---|---|---|---|
| **Positivism** | Hidden rules govern teaching and learning process | Focus on reliable and valid tools to undercover rules | What works? | Quantitative |
| **Interpretive/ Constructivist** | Reality is created by individuals in the group | Discover the underlying meaning of events and activities | Why do you act this way? | Qualitative |
| **Critical** | Society is rife with inequalities and injustice | Helping uncover injustice and empowering citizens | How can I change this situation? | Ideological review, civil actions |
| **Pragmatic** | Truth is what is useful | The best method is one that solves problems | Will this intervention improve learning? | Mixed Methods, Design-based |

**Critical**: The critical paradigm explains the belief that research is conducted for the individuals and groups liberation in a democratic society (Mack, 2010). This paradigm involves the study of people’s behaviours within the society, while a critical researcher is seen as someone who works towards making a change and not only to give the account of behaviours in those societies (Mack, 2010). In a related development, Hassan (2011) added that research could be categorised as critical if the task is seen as social critique. Therefore, critical research is seen as liberation and it open opportunities for human development and engagement (Hassan, 2011).

A major comparison, which explained the key features of the positivism and social constructionism philosophical approaches, is shown in **Table 21**.

**Table 21 Differences between Positivism and Interpretivism (Ramanathan, 2008)**

| | Positivism | Social Constructionism |
|---|---|---|
| **The observer** | Must be independent | Is part of what is being observed |
| **Human interests** | Should be irrelevant | Are the main drivers of science |
| **Explanations** | Must demonstrate causality | Aim to increase general understanding of the situation |
| **Research Progresses through** | Hypotheses and deductions | Gather rich data from which ideas are induced |
| **Concepts** | Need to be operationalised so that they can be measured | Should incorporate stakeholder perspectives |
| **Units of Analysis** | Should be reduced to simplest terms | May include the complexity of 'whole' situations |
| **Generalisation through** | Statistical probability | Theoretical abstraction |
| **Sampling Requires** | Large numbers selected randomly | Small numbers of cases chosen for specific reasons |

In the light of the above, this research study is more of an ***interpretive paradigm*** as it involved the acquiring of knowledge through interaction with other people on how a change could occur in their organisation, and the reason for making a change (McKinley, 2015). This is more of qualitative research method. However, it should be noted that making a change within the society also relates to the ***critical paradigm*** as explained in **Table 21** above. The critical paradigm helps to eliminate inequalities in the society, to make a change, and shows how to make a change either through ideological review or civic action. Therefore, this study combines the features of both ***interpretive and critical paradigms***. This justifies the adopted research method used in this study, namely the ***Action Research Method***, which identifies problems and helps to find answers to the problems (Buijs et al., 2012).

### 4.4.2 Research Methods and Approaches

There are many research methods and approaches available in conducting a research study. The researcher needs to make a choice which has to be justified. These options range from the paradigm, methodology, and method. Dick (1993) noted that, within each research paradigm, there are several methodologies available and these methodologies are drawn on some methods used in data collection and analysis. The relationship between paradigm, methodology, and method is shown in **Figure 41** below.

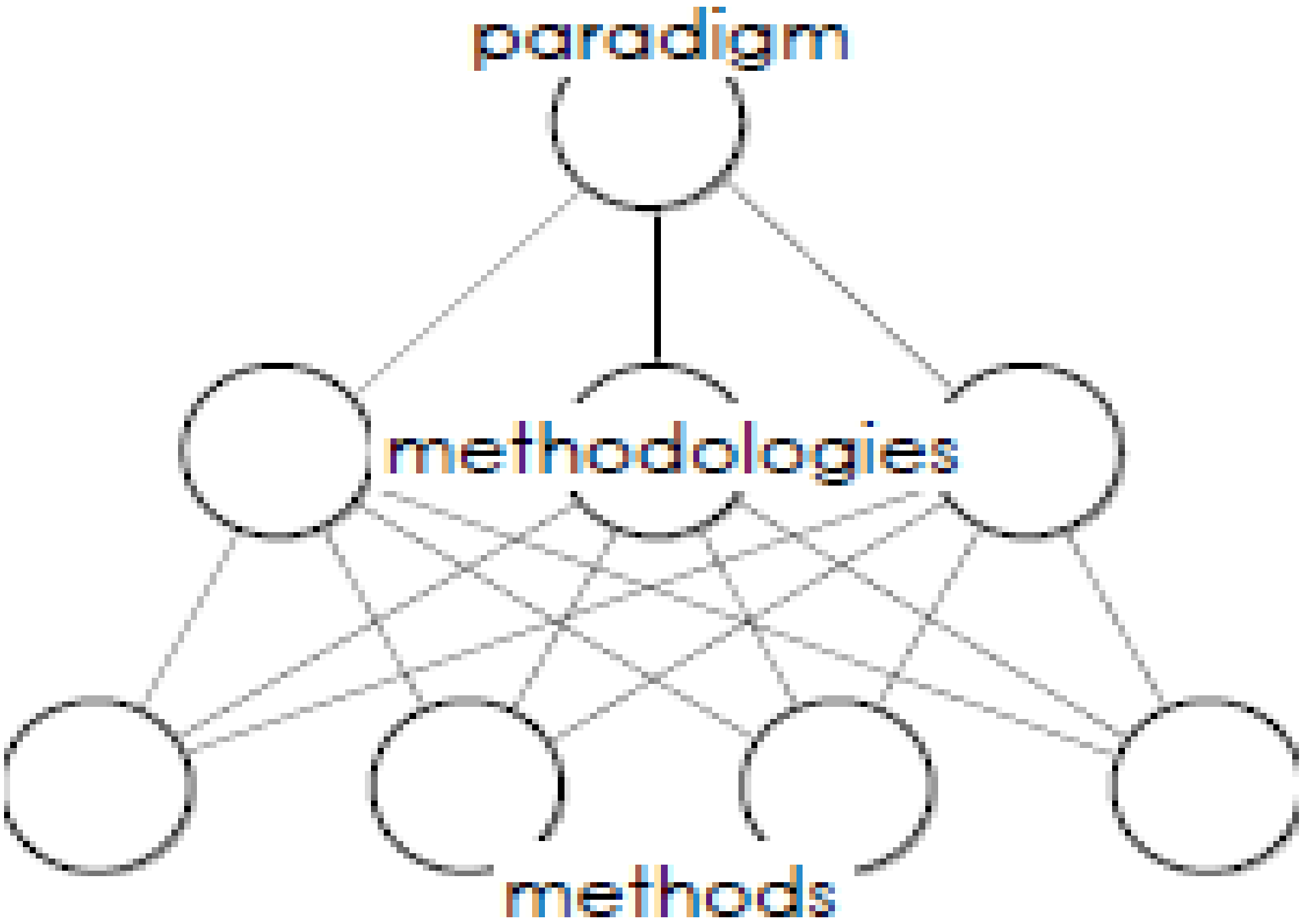

**Figure 41 Relationship between Paradigm, Methodology, and Methods (Dick, 1993)**

Many researchers use both research methods and approaches as one entity by making reference to quantitative and qualitative methods as the two commonest approaches in research design (Hassan, 2011). The finding shows that the **research approach** is known to be a discipline where one can acquire knowledge via different research methods (Bryman and Bell, 2015). Hence, there are three most common research approaches, namely ***Deductive***, ***Inductive*** and ***Abductive***. For the researcher to meet up with the aims and objectives of this study, he made use of ***inductive research approach*** which addresses the research questions, aims, and objectives of this study. Inductive research approach, as a matter of fact, is not involved in the formulation of the hypothesis (Bryman and Bell, 2015).

**Research Methods**: This could be better explained according to Cohen et al. (2007) as a range of approaches in research to gather data used as the basis for inference, interpretation, prediction, and explanation. This is also referred to as the techniques and procedures used in data the gathering process (Bryman, 2016). The research method is categorised generally as either a **quantitative** or **qualitative** research technique (Hassan, 2011). However, many researchers combine both techniques, known as Mixed-Methods (Bryman, 2016; Saunders et al., 2016). Meanwhile, a researcher could focus more on the quantitative approach if the research study is more of a realistic perspective (positivist). Also, the researcher could use the qualitative technique if the study aims to identify and discuss themes (interpretivist) after identifying the research philosophy (Bryman and Bell, 2015).

**Quantitative Research Methods**: This research method, as described in broad terms by Bryman (2016), involves the collection of numerical data. A quantitative research method examines the relationship between variables that are measured numerically and analysed through the use of statistical and graphical techniques (Saunders et al., 2016). Examples of quantitative methods include survey techniques, laboratory experiments, and numerical methods such as mathematical modelling (Hassan, 2011).

**Qualitative Research Methods:** This research method is associated with interpretive philosophy (Saunders et al., 2016). A qualitative research method focuses on the meanings and relationships between the participants through the use of various data collection to develop a conceptual framework (Saunders et al., 2016). Qualitative research sources include interviews and questionnaires, observation and participant observation (fieldwork), documents and texts, and the researcher's impressions and reactions (Hassan, 2011; Bryman, 2016).

### 4.4.3 Ethnographic Research and its Usage in this Study

Ethnography is the study of social interactions, behaviours, and perceptions that occur within groups, teams, organisations, and communities (Reeves et al., 2008). The main aim of ethnography research is to provide holistic insights into the views and actions of the people, including their environment (Reeves et al., 2008). Ethnography as a term overlaps with several other things. According to Hammersley (2016), ethnography as a qualitative method is known as interpretative research and, as a case study, it is termed as participant observation. However, many ethnographic studies try to control the researcher's involvement.

As posited by Dingley (2001), ethnographers should be included in the research space, and not as mere observers. In this study, ethnography as a method involved the researcher in presenting the problem to the participants in an informative and explanatory manner, to find holistic solutions to the barriers facing the E-Service adopting and implementation at local environment level in Nigeria as identified. Also, as part of the features of ethnographers, the researcher studied the behaviours of the participants during the data collection through relatively informal conversations and engagements.

However, despite many advantages of ethnography in research such as the use of participant observation, which helps the ethnographer to generate a rich understanding of social actions in different contexts (Reeves et al., 2008), there is still criticism of it not being scientific enough. Ethnography requires empirical study commitment and finding exceptional behaviours (Dingley, 2001), to move the research forward. This research is ethnographic in nature and it undertakes to achieve the aims and objectives through reflective dimension, interpretative, and informative interaction with participants to find patterns and exceptions. The ethnographer involved in this thesis is the research facilitator/consultant that gives advice, feedback, provides questionnaires and manages the time component. Hence, the information gained by the researcher is given back to the stakeholders who are also the participants, who will bring further improvement. The high return rate from the online survey used in this study affirms the robust relationship between the ethnographer involved in this study and the participants.

### 4.4.4 Boundaries Between Ideographic Versus Nomothetic

The research aim is to identify the barriers facing E-Service adopting and implementation at local environment level. In order to achieve the research aim, it was necessary to examine research methodologies that might best serve the aims. In categorising the methodology types, the two generic models involved are Ideographic and Nomothetic.

**Ideographic Model**: An ideographic study explores a single person, situation or event in detail (Dingley, 2001). The advantages of an ideographic approach include rich qualitative data and further information can be provided about the study. However, an ideographic approach lack scientific rigour and is high in bias. Other disadvantages of an ideographic approach are that it is time consuming and it cannot be generalised when samples collected are small. There are various examples of ideographic research in this research study, including ethnographic studies, case studies, task observations, and usability observations. This research study is more ideographic, as explained in the ethnographic section, where the researcher presents the problem to the participants in an informative and explanatory manner in order to find holistic solutions. The ethnography research aims to provide holistic insights into the views and actions of the people, including their environment (Reeves et al., 2008). Data gathered from this study is given and feedback given to the stakeholders, who are also the participants, to bring further improvement to the local E-Service initiatives in Nigeria.

**Nomothetic Model**: The nomothetic research model attempts to discover laws and principles that govern reality, according to Dingley (2001). The advantages of the nomothetic approach are that it is scientific in nature and it avoids subjective data, it is very reliable, and can help to control or prevent behaviour. However, the nomothetic approach ignores subjective experience and lacks ecological validity. This means that the nomothetic approach cannot be applicable to real-life situations. The studies involved in the nomothetic environment need to be tightly controlled as there is a need to examine sufficient representative occurrences of the phenomenon in question. The required characteristics of the nomothetic research model are: a defined experimental design, independent and dependent variables, a disprovable hypothesis, a participant (subject) sample initially matched to the target population, and subsequently randomly allocated to the experimental and control conditions (Dingley, 2001). Hence, the experimental methodology fits well into this model, especially the studies which are examined within a laboratory, or a closely controlled field setting. Findings are generalised to a target population in the nomothetic model when experiments are used appropriately.

### 4.4.5 Research Methodology Adopted

The methodological approach taken in this study is the **Action Research Method**, which is a mixed methodology based on both qualitative and quantitative features. Furthermore, action research as described by Reason and Bradbury (2001) brings “together action and reflection, theory and practice, in participation with others, in the pursuit of practical solutions to issues of pressing concern to people, and more generally the flourishing of individual persons and their communities."

The action research method has been successfully used and validated by researchers (Phythian et al., 2009; Buijs et al., 2012). However, unlike more controlled experiments, it is hard to generalise findings to other circumstances. Hence, the researcher will also adopt the action research cycle by Baskerville (1999), as shown in **Figure 42** below. Action research is not about imposing knowledge on participants and stakeholders; rather, it is about creating a collaborating environment for the exchange of ideas in problems analysis and the designing of actions to solve the problems (Johnson et al., 2014).

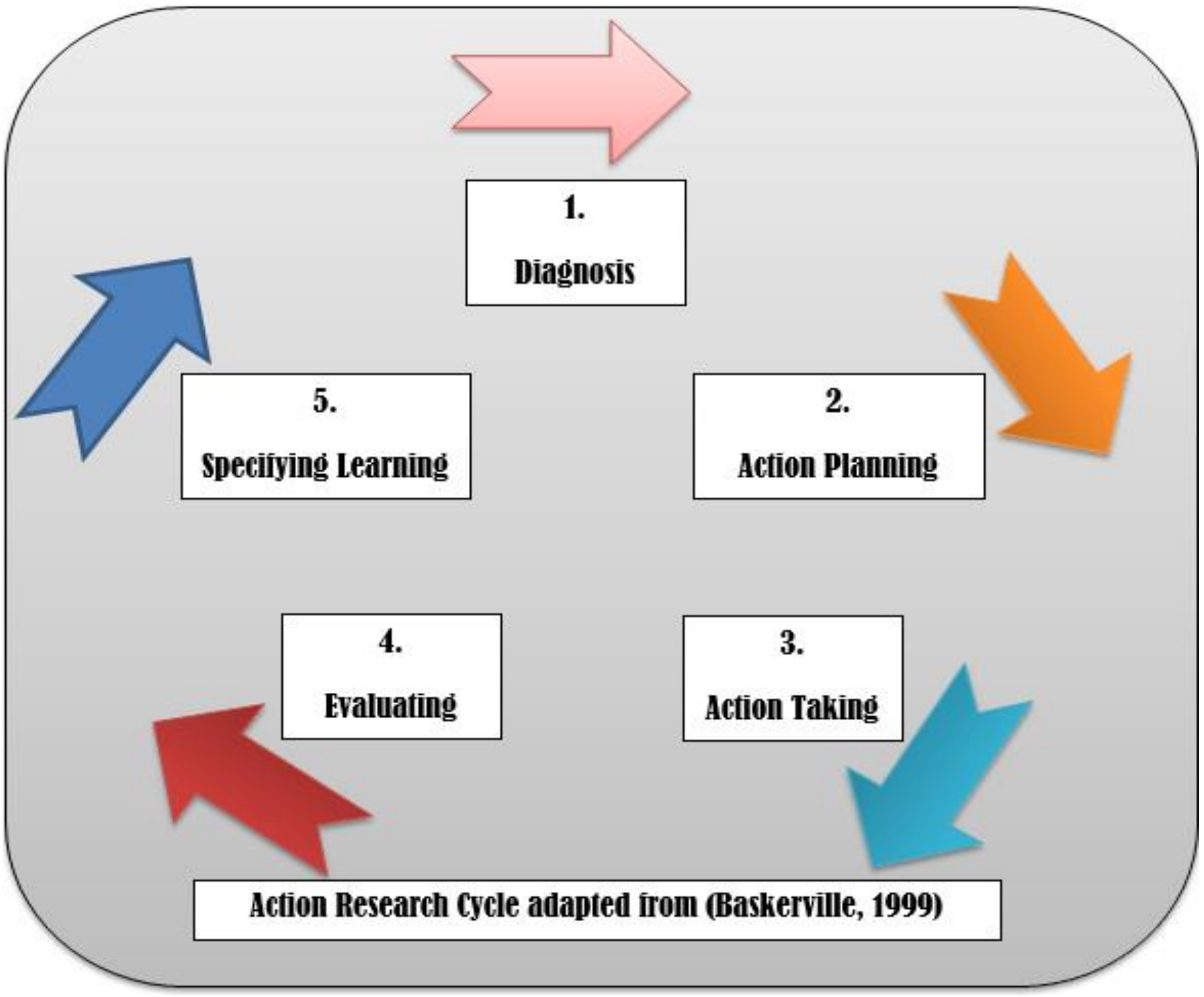

**Figure 42 Research Methodology Adopted - Action Research Method**

The researcher chooses action research because it allows participation with others, either in the academic world or in industries, to suggest a change in the pursuit of practical solutions to issues of pressing concern to people, and more generally the flourishing of individual persons and their communities. Low adoption of E-Services (Azenabor, 2013), particularly at the local government level in Nigeria, is a persistent issue that the government and stakeholders should be concerned about. Hence, the data collection techniques of the Action Research could also be classified as "The 3 Es" as shown in **Figure 43** below. The Action Research methods of collecting data in any research study (Buijs et. al., 2012) are listed below:

- Document collection and analysis
- Participant observation recordings
- Questionnaires
- Online surveys
- Focus Group discussions
- Structured and unstructured interviews and case studies.

For this research, the action research methods of collecting data used in this study are structured and unstructured interviews, questionnaire, online surveys, online Focus Group discussions, document collection, and analysis.

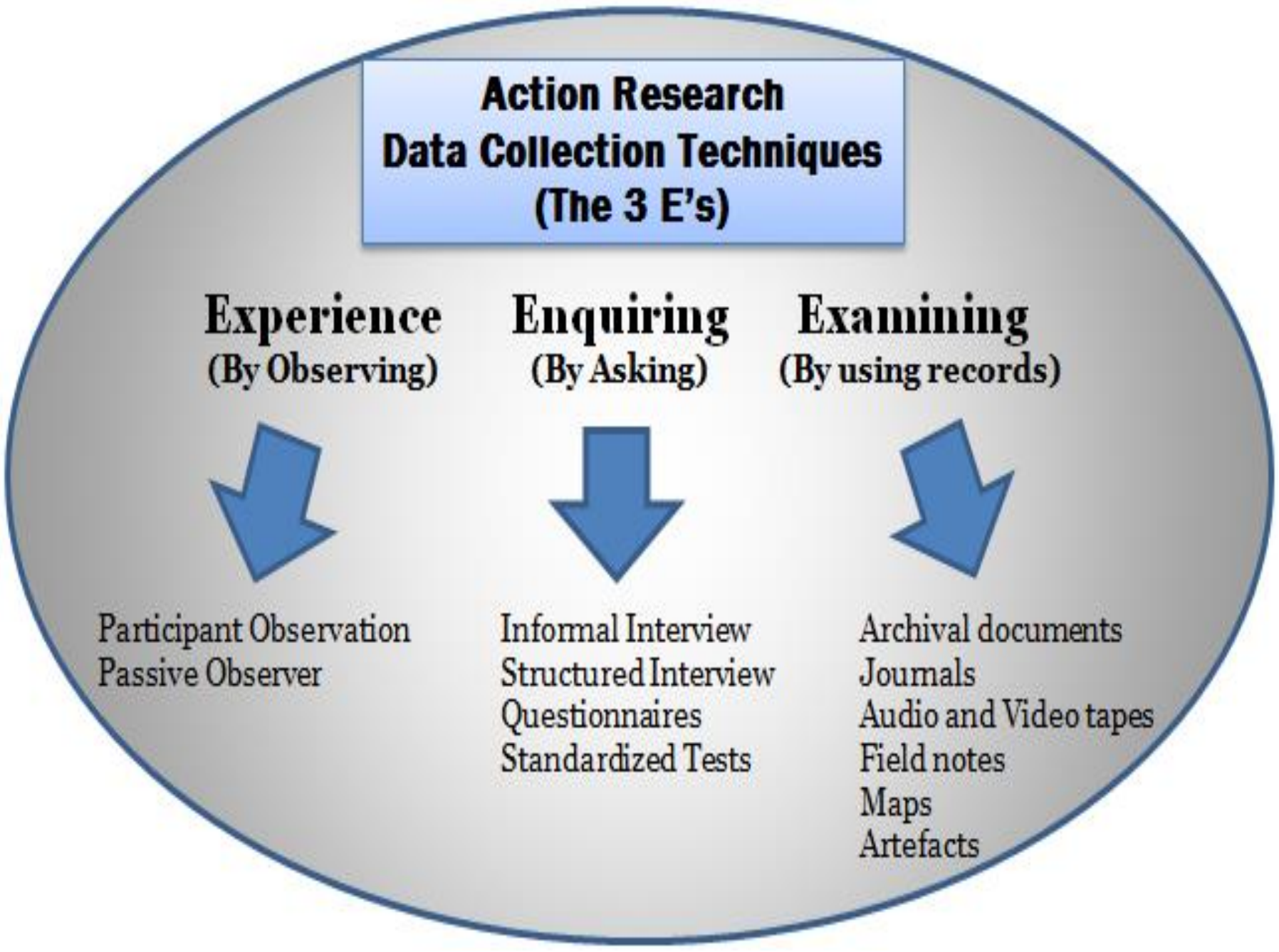

**Figure 43 Action Research Data Collection Techniques (The 3 Es) (Baskerville, 1999)**

### 4.4.6 Justification and Rationale for Adopted Research Methodology

The decision of the researcher to choose action research methodology is based on the fact that it focuses on solving problems. The low adopting of E-Services at the local environment level in Nigeria (Oseni et al., 2015) is because of barriers that are investigated through action research by engaging the stakeholders involved in the implementation of E-Service initiatives in Nigeria. Moreover, action research uses a holistic approach towards problem-solving, rather than a single method for collecting and analysing data (Phythan et al., 2009). Hence, as explained by Adams and McNicholas (2007), action research is a family of research methodologies that pursue action (or change) and research (or understanding) at the same time. This assertion is in line with the researcher orientation and belief in the introduction of the required changes necessary for an organisation's growth. The researcher, who is indeed involved in the system, was formerly a partner in the local government administration, and has confidence that action research will facilitate the process of reflection and inquiry in introducing change. Action research encourages both the participants and the researcher to reflect on the actions concerned to see the organisational change, compared to the conventional research where the researcher just takes a snap of what is happening in the organisation without any actions being initiated (Beard et al., 2007). Action research methodology, according to the researcher and as justified by Dawson et al. (2013), is not just about focusing on the search for the truth or development of the theory; rather, it is designed to make a change from the current situation for a better one.

### 4.4.7 Action Research and Approaches

Action research differentiates itself from the general professional practices, as the researchers systematically study the problem and ensure the involvement informed by the theoretical reflections (O'Brien, 1998). A simple action research model of the cyclical nature was developed by Stephen Kemmis with a four-step cycle, namely ***Plan, Act, Observe and Reflect***, as shown in **Figure 44**. This model is, however, short of the model adapted from Susman (1983), which elaborates more on the phases to be conducted in each action research cycle, as shown in **Figure 45**.

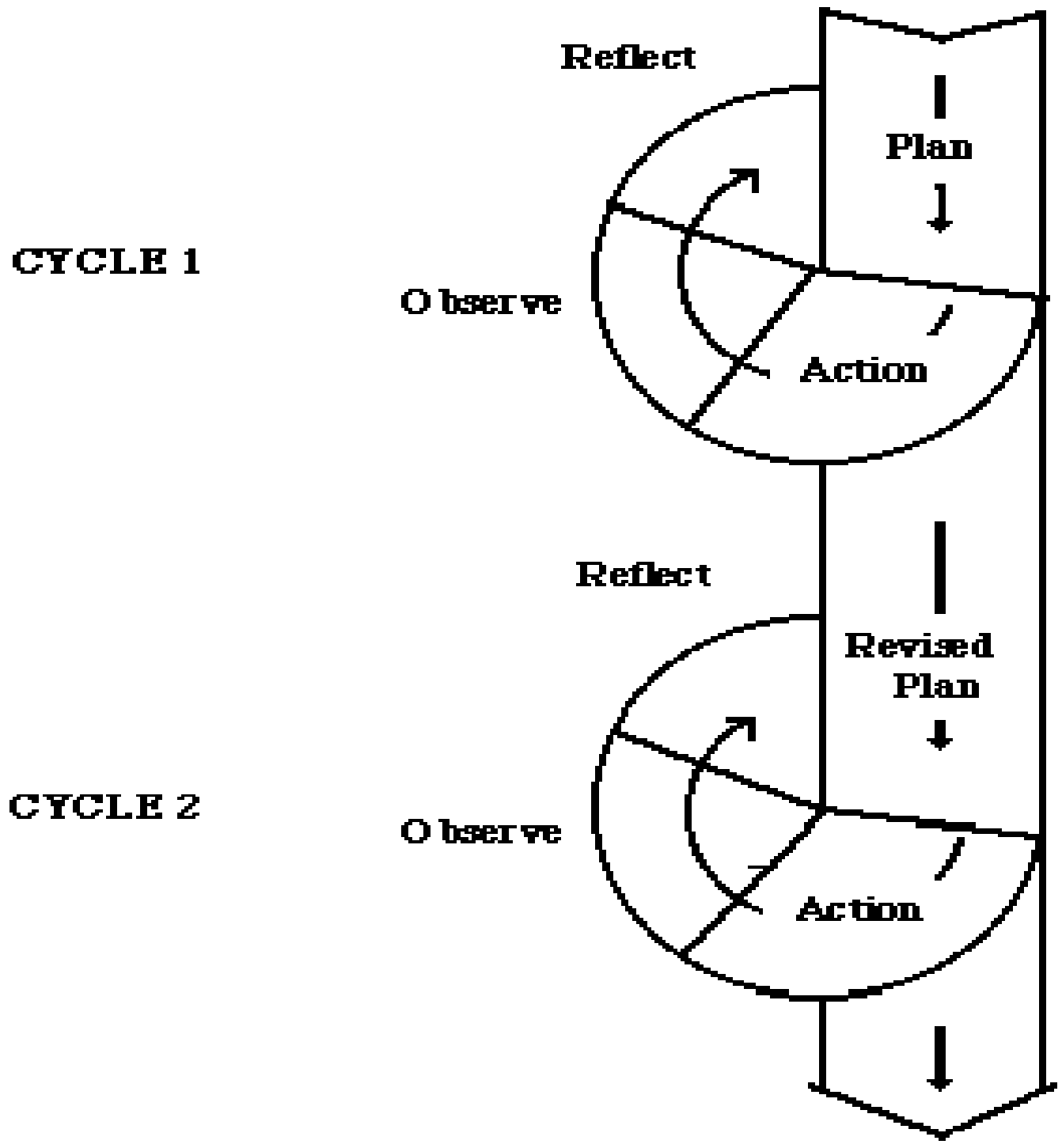

**Figure 44 Simple Action Research Model Developed by Stephen Kemmis**
(O'Brien, 1998)

The action research model adapted from Susman (1983) explains that a problem identified and data collection carried out for a more detailed analysis. This is followed by a collective assumption or belief of several possible solutions, where a single plan of action will emerge and be implemented. The data on the results of the intervention are collected, analysed, and the findings interpreted in line with how successful the action has been. Hence, the problem is re-assessed, and the process begins another cycle. This is a continuous process until the problem is resolved. Action research trend is becoming more popular (Calabrese, 2006), alongside other qualitative research methods, as the researchers and practitioners now benefit from collecting research data through distinct ways and contributing to practice through theory development.

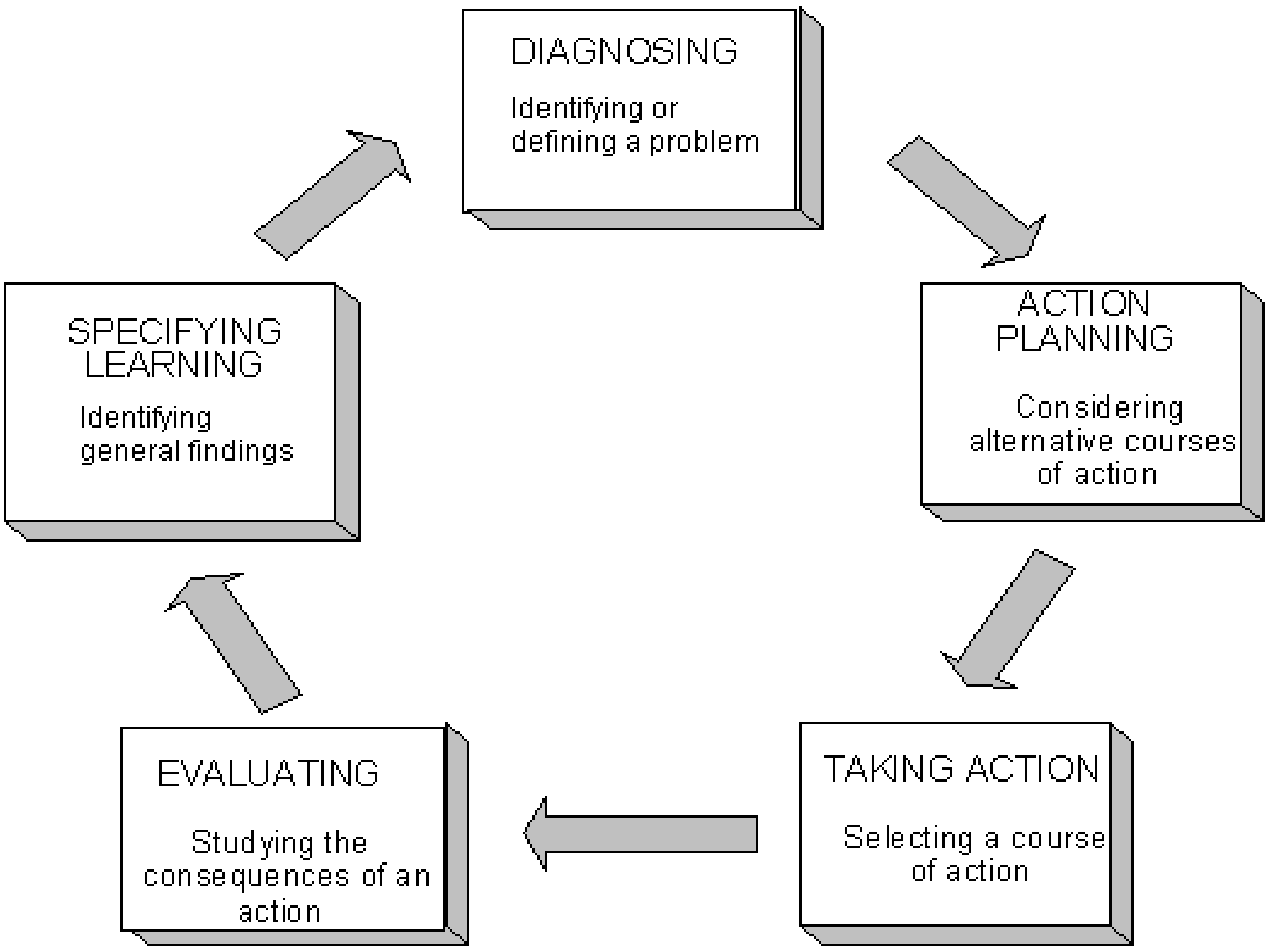

**Figure 45 Detailed Action Research Model**
**(Susman, 1983)**

Hart (2013) added that, when a situation is identified by the participants involved in an action research study, it is examined for the participants to develop ideas on the approach to solving the problem. Action will take place based on these ideas, and the new situation will be investigated in a further cycle of learning, the whole constituting planning, action, and fact-finding as obtained in the action research model (O'Brien, 1998; Hart, 2013). However, as the cycle progresses, the participants take the "ownership" of the process and learn about the situation, including what they can do to change it. The researcher is facilitating the process and, by reflecting on the participants' contributions, learns both about the situation and the research method (Hart, 2013).

**Action Research Approaches:** There are many action research approaches, but the most common ones are Classical Action Research, Action Learning, Participatory Action Research (PAR), Participatory Rural Appraisal (PRA), Community-Based Participatory Research, Living Theory and Appreciative Inquiry (Popplewell and Hayman, 2012). This research will elaborate further on the three types of Action Research Approaches that are widely used for the research study. These are:

1. **Classical Action Research**: This involves the researchers and participants working together to identify and solve problems mainly within an organisational context, with new knowledge generated through collaborative cycles of planning engagement (Popplewell and Hayman, 2012).

2. **Action Learning**: This type of action research approach facilitates either organisational or individual learning. McGill and Beaty (2001) described action learning as a process of learning and reflection to support a group or colleagues occupied with real problems at work with the aim of solving the same. This approach enables people to learn from each other through working on real problems and reflecting on their experiences (McGill and Beaty, 2001). For example, this might be staff development in an organisation achieved through action learning sets, which support the in-house development.

3. **Participatory Action Research (PAR)**: This approach of action research involved participation and action in community research. Hence, the approach strives to understand the world by trying to change it, whether collaboratively or by following reflection (Reason and Bradbury, 2008). PAR highlights collective inquiry and experimentation grounded in experience and social history. In this approach, important questions are addressed and responses issued through communities of inquiry and actions for the participants. It should be noted that three basic features are integrated into the Participatory Action Research by the practitioners. They include **Participation** (Life in society and democracy), **Action** (Engagement with experience and history), and **Research** (Soundness in thought and the growth of knowledge), as explained by Chevalier and Buckles (2013). Participatory Action Research suggests experiment-based research with both evidence and people-based inquiry (Phillips and Kristiansen, 2012). This is the orientation on which this research is established, as it will create robust debates while participation is on and actions being taken.

## 4.5 Validity and Reliability

There are many threats to both validity and reliability in research (Gravetter and Forzano, 2012). It is important that researchers are aware of these threats, which could disrupt the validity of their studies. Validity is defined according to Gravetter and Forzano (2012) as the degree to which a research study measures what it intends to measure. Hassan (2011) argued that validity deals with the issue of whether the research works are accurate, true, or correct. Also, Zohrabi (2013) explained that validity is concerned with whether our research is believable and true, or whether it is evaluating what it is supposed to evaluate. Hence, with the various definitions of the validity as obtained in the literature, this researcher concludes that, for this research study to be accepted as valid, the results from the research finding should be able to address the research objectives and be able to solve the research questions. There are two major types of validity, namely ***internal*** and ***external*** validity. Internal validity deals with the validity of measurement and the test itself. External validity, on the other hand, deals with the ability to generalise the findings to the target population.

Nevertheless, in the analysis of the appropriateness, meaningfulness, and usefulness of research, both the ***internal*** and ***external*** validity are important (Gravetter and Forzano, 2012). However, this study will focus more on the validity of the measurement technique (i.e. **internal validity**) as it deals with the connection of the research findings with the reality (Zohrabi, 2013). The issues of validity and reliability in action research were examined by Turnock and Gibson (2001), and they both argued that validity in action research is not entirely about the methodology, but also the personal and interpersonal issues. Hence, the validity of the action research, according to Turnock and Gibson (2001), rely mainly on the decisions made during the data collection, as it will assist in the judgement of the study validity. However, in assessing and dealing with internal validity, Zohrabi (2013) discussed the four main types of validity based on different perspectives and the relationship between measurements. These are:

1. **Face Validity**: This is when a technique appears as if it should measure the variable it intends to measure (Zohrabi, 2013). Also, Venkitachalam (2015) explained that face validity is the extent to which a measuring instrument appears to be valid on its surface, and there should be a link between the research objectives and the question or item on the research instrument.

2. **Construct Validity**: In this type of internal validity, research findings are tested to determine whether the measurements of a variable in a study behave accurately (Zohrabi, 2013). Part of the construct validity is to examine the past research in a different aspect based on the same variable. Venkitachalam (2015) argued that construct validity could also be measured by correlating performance on the test for which construct validity has been determined.

3. **Concurrent Validity**: The comparison of the results from a new measurement technique to the ones of the more established techniques is done with concurrent validity. The concurrent validity will check whether the claims of the same variables measured are related (Zohrabi, 2013). The validity here must be thoroughly examined as two measurements often behave in the same way, but are not measuring the same variable.

4. **Predictive Validity**: This validity is used to predict future performance. For example, the validity test is used to determine whether the results obtained from measuring a construct could accurately be used to predict behaviours (Zohrabi, 2013).

In the light of the above, face validity is more relevant to this study as there is a link between the research objectives and the research questions. The measuring instrument (questionnaires) appears to be valid as a result obtained from the interviews, and an online focus group is similar to the results obtained from the online survey with a larger instrument. The researcher ensures validity in this study by maintaining anonymity with the participants in the interviews conducted, as well as the online focus group sessions.

**Reliability**: In any research, reliability deals with consistency dependability and replicability of the results obtained. Zohrabi (2013) argued that, to get the expected results, the major requirement is the reliability of the data. In Babbie (2010), reliability is deemed to be the extent by which a researcher could obtain the same answers using the same instruments during a number of times. In essence, research with high levels of reliability will generate the same results under similar conditions and by using the same research methods. For this study, in order to maintain a high level of reliability, the researcher made use of the instruments for both the interview and online survey that were previously used and validated by Susanto and Goodwin (2010), Alshehri and Drew (2010), and Almarabeh and Abu Ali (2010). A slight modification has been carried out to these instruments to accommodate changes made by the participants involved in the action research. The source of data has a significant reliability influence on the research study (Babbie, 2010). It is expected that a very high reliability is maintained in this study. Moreover, this researcher tested the reliability of the questionnaires with the use of Cronbach's alpha to measure the internal consistency of items in the scale through Statistical Package for the Social Sciences (SPSS).

**Cronbach's alpha**: This is a statistic used to measure the consistency of responses to a set of scale items in questionnaires. It consists of the alpha coefficient with values that range between 0 and 1 (Saunders et al., 2016). Data gathered for this study were analysed and coded using SPSS and Nvivo software. Meanwhile, a reliability coefficient instrument required level for high reliability in any research is 0.70 for Cronbach's Alpha according to Carcary (2008). This represents a commonly acceptable level of reliability for any research. However, the ***reliability*** may either be ***higher*** or ***lower*** than the accepted level, especially if the factors (questionnaires) have only a few items (instruments) with different scales (Hair et al., 2006). The Cronbach's alpha reliability test for this study is 0.508, which is low, as shown in **Table 22**. The reason for a lower Cronbach's alpha might be connected with the different measurements used in the questionnaires.

### Scale: ALL VARIABLES

**Table 22 Reliability Statistics**

| Cronbach's Alpha | Cronbach's Alpha Based on Standardised Items | No of Items |
|---|---|---|
| .508 | .568 | 22 |

### 4.5.1 Relationship between Validity and Reliability

For researchers, it is important not to get confused with validity and reliability. There is an established relationship between both of them in a research study (Zohrabi, 2013). While reliability deals with the consistency of results when the experiment is replicated under the same conditions, validity, on the other hand, deals with the degree to which a research study measures what it intends to measure (Gravetter and Forzano, 2012). Hence, as these two evaluations of research studies are independent factors, a study can be reliable and not be valid, and vice versa. Thus, an excellent research study will always be both reliable and valid (Zohrabi, 2013).

**Threats to Validity and Reliability**: The main threats to both validity and reliability in a research study are explained by Hassan (2011). These threats could be eliminated or minimised if addressed early by the researcher. They are:

- **Reactivity**: “This refers to the way in which the researcher’s presence may interfere with the case setting” (Hassan, 2011).

- **Researcher Bias**: “This refers to the assumptions and pre-conceptions that the researcher may bring and lead to a selection of certain people for interview who are likely to generate the desired results” (Hassan, 2011).

- **Respondent Bias**: “This refers to the cases where the respondents treat the researcher as a threat; thus, try to hide information from him/her.” In other words, this refers to instances where the respondent gives an answer which would please the researcher (Hassan, 2011).

In this study, and to deal with the main threats to both validity and reliability as listed above, the research proposed the following strategies, as supported by Hassan (2011):

- **Triangulation**: This involved the use of different methods of data sources to improve the research reliability. This substantiates the action research methodology where the researcher could source for data through many means, such as interviews, field study, observation, and online survey.

- **Member Checking**: This involved getting feedback from the participants as crucial for the research credibility.

- **Prolonged Involvement**: This is when the researcher spends time within the research setting, trying to create relationships with the participants and understanding the culture of the setting. Prolonged involvement could increase the researcher bias. This is covered with the action research methodology, as the researcher is always in contact with the participants to complete the cycles involved in research phases.

- **Purposive Sampling**: This offers the researcher a degree of control rather than being at the mercy of any selection bias inherent in pre-existing groups.

- **Peer Debriefing and Support**: The debriefing sessions after an extended period within the research setting can aid in reducing researcher bias.

- **Audit Trail**: This is when the researcher is keeping a full track of the activities taking place during the research.

Finally, the researcher has adopted four strategies to discuss the reliability and validity of the analysis and results for this study. These strategies are triangulation, purposive sampling, multiple coding, and respondent validation. Validation strategies adopted and the overall validation process in this research are discussed in detail in Chapter Seven, which is the validation of research findings chapter.

## 4.6 Summary

This methodology chapter mostly summaries the research approaches, research philosophy, validity, reliability, and methodology that has been adopted to guarantee that its design is suitable to provide the answers to the research questions and achieve the aim and objectives of this study. The chapter summarised the philosophical positions that encouraged the designs of research management. The chapter also made a comparison between positivist, social constructionism, and critical paradigm. Furthermore, a summary of the different research approaches and disciplines to research design (qualitative, quantitative, action research method) has been carried out, within which knowledge is acquired.

The chapter further provided the rationale for selecting a suitable research strategy (action research method). The chapter went on to describe the adopted data collection procedures and techniques, and the validity and reliability issues in the research study and how to deal with the main threats to both validity and reliability as a researcher. Ethnographic research would always have a bias, but this did not pose a critical problem to this research, because anonymity in the online session with the focus group members was used. The research was able to critically justify the adopted research methodology. The action research method gives room for engaging participants, and a robust engagement brings about solutions to the research problems.

# Chapter 5 – Data Collection and Analysis

## 5.1 Introduction

Peersman (2014) stressed the need for a researcher to use a robust and well-implemented data collection approach as this will help in data evaluation and analysis. Data collection is, therefore, defined by Shamoo and Resnik (2003) as the systematic approach to gathering information from various sources. The completeness, together with the accuracy of the data collected, is stressed by the authors because data collected without clear objectives might put the future data analysis and its findings in jeopardy. This means that the collection of data is likely to be pointless. In a related development, data analysis is defined as the systematic application of statistical or logical techniques to describe and illustrate, summarise, and evaluate data (Shamoo and Resnik, 2003).

The purpose of this chapter is thus:

*Describe in detail the data collection procedures for this study, techniques adopted and data analysis plan, as shown in* **Figure 46** *below.*

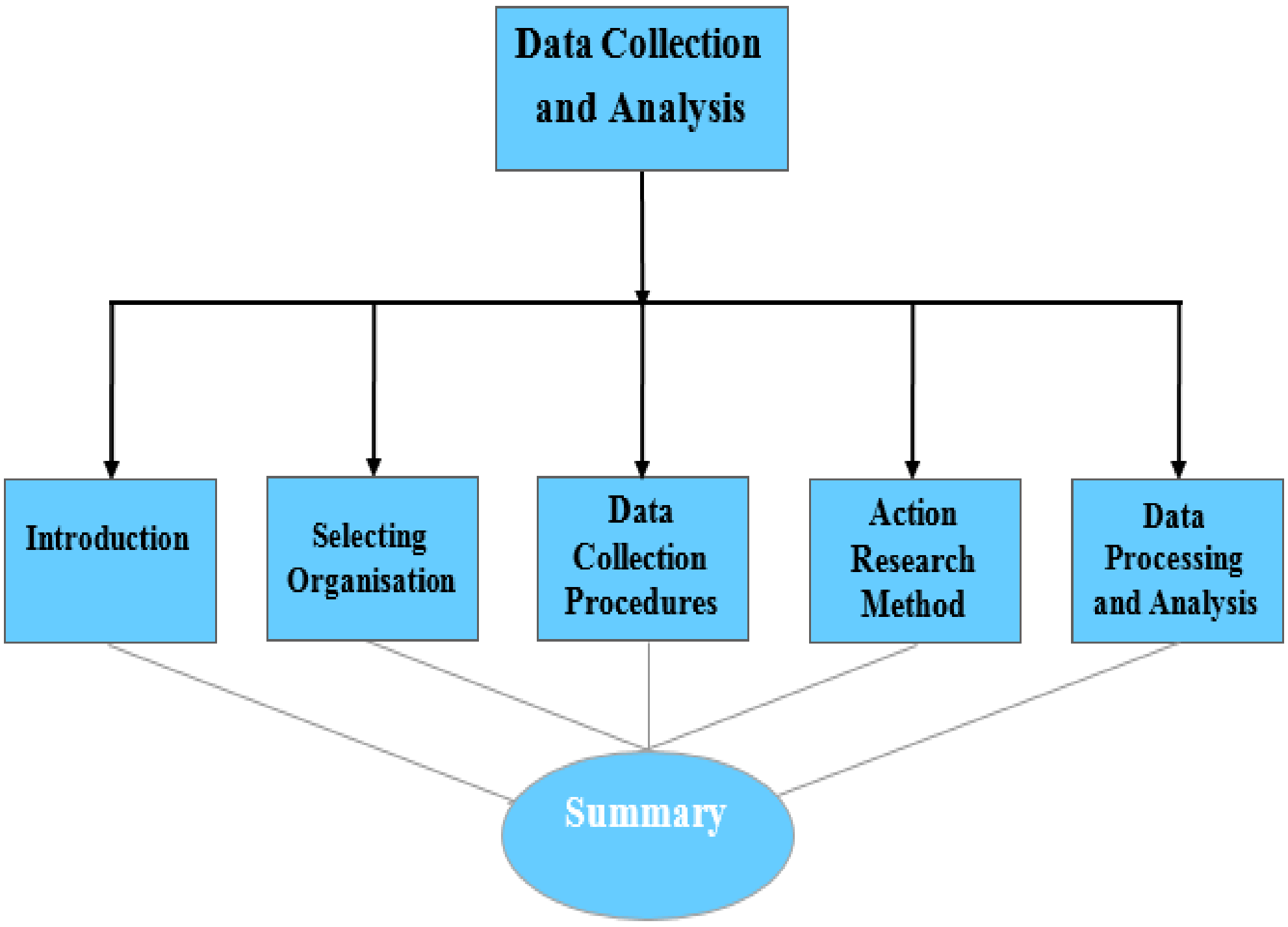

**Figure 46 Chapter Five Summary**

In the researcher's effort to fulfill the purpose, this chapter is divided into six different sections, as shown in **Figure 46**. The sections are the introduction, organisation and participant's selection, data collection procedures, action research method, phases and cycles, data processing and analysis, and summary.

## 5.2 Selecting the Organisation and Participants

### 5.2.1 Purpose

The choice of which organisation to select, and whom to interview for the purpose of this specific research, as well as the location, are important and, of course, not an easy decision to make for the researcher. The participants were selected because they were members of the stakeholders involved in the E-Services implementation at the local government level in Nigeria. The 30 participants interviewed have a combined 491 years of working experience. Moreover, the selection of cases and interviewees for this research tended to be purposive instead of being random. Purposive sampling is very common in qualitative research, which formed part of the action research method and data collection approach. The reason is that the definition of the research case is limited (the research is interested in assessing the barriers, with possible solutions affecting the schemes providing E-Services, targeted to citizens within the Nigeria local government council E-Services context).

### 5.2.2 Selected Organisation in Lagos Nigeria

The governmental organisation was chosen for the purpose of this study based on certain criteria. Firstly, the research study aims to look into the barriers facing the E-Service adoption and implementation at the local environment level using Nigeria as a case example. Hence, it is appropriate for this researcher to select a local government council and the local government service commission in Lagos, Nigeria, as the case example. Secondly, the case example selected is based on the size of the local E-Services potential of beneficiaries and recipients. Lagos is considered to be the most populous city in Africa (African Economist, 2016).

**Reasons for Choosing Lagos as a Case Study**: The study was conducted in Lagos as the state has high information technology penetration in Nigeria. The researcher chose Lagos, which is an advanced state, because the citizens are more keen to be involved in E-Government activities. The data obtained from the National Bureau of Statistics,

Nigeria, as indicated in **Figure 47** below, shows the percentage distribution of households' access to the internet services in Nigeria between 2007 and 2009. This data was released in 2014 and rated Lagos as a State with the highest internet services access in Nigeria. Therefore, it was appropriate for the researcher to consider Lagos because of its widespread internet access, as this would help in data collection and easy access to the online focus group participants. It also means that the citizens of Lagos and the local government staff are likely to be more familiar with the internet than citizens and staff in other states. Having internet experience supports Rogers' (1995, 2003) awareness stage of adoption, as the people involved will have been exposed to the key ideas of socialising and, perhaps, working online. This is an important first stage in the diffusion of innovation (Rogers, 2003).

Moreover, the insurgency witnessed recently in some parts of Nigeria, especially the northern part (Omotosho, 2015), poses a threat to conducting the research study. More importantly, the Boko Haram are trying to stem the spread of the World Wide Web and the internet, in order to prevent Muslim citizens taking part in activities that are associated with the Western world.

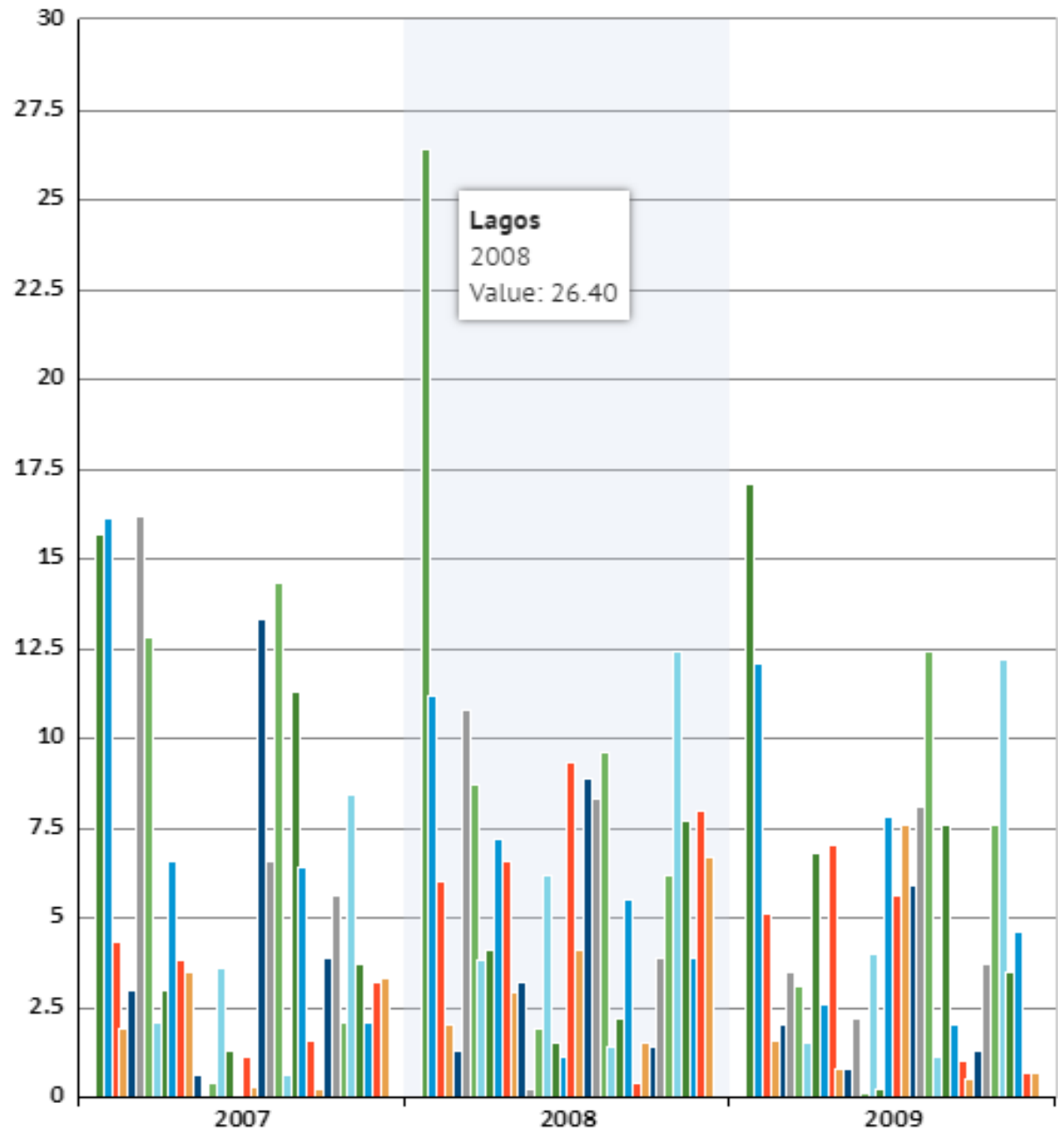

**Figure 47 Percentage Distribution of Households Access to the Internet Services in Nigeria (NBS, 2017)**

**Structure**: The local government area is administered by elected officers consisting of an elected Chairman, who is the Chief Executive of the Local Government Area, and other elected members who are referred to as Councillors. The term Chairman, as referred to in the Nigeria 1999 constitution, is the equivalent of a Mayor in the UK. The local government organisational structure also consists of some operating departments, reflecting the government departments on whose behalf they conduct daily economic activities and for whom they provide expert advice and analysis. Each operating department has a head who administers the activities in his or her department and reports back to the Chairman for further action. It is, therefore, pertinent to note here that, in the absence of the elected Chairmen and Councillors (expired terms before another general election), the state government headed by the elected governor is empowered, according to the Nigeria 1999 constitution, to name an administrator for the local government councils. They will manage the affairs pending a general election (Olasupo, 2013). Both Olasupo (2013) and Olaiya (2016) in their contributions quoted the Nigeria 1999 constitution, section 14, subsection 2A, in the area of "sovereignty", which belongs to the people of Nigeria. This implies that the government should always allow the people to determine who leads them at government levels, including the local government. In recent times, many state governors have failed in conducting the election at the local government levels after the expiration of the incumbent's tenure, as they prefer to compensate their political associates through the appointment of local government administrators, thereby truncating the desires of the citizens.

**Business Model**: The local government councils in Nigeria are constituted and empowered according to the Nigeria 1999 constitution. According to Shamsuddin & SiddigBalal (2014), they perform the following:

- Economic recommendations to the State; including the collection of taxes and fees.
- Establishment and maintenance of cemeteries, burial grounds, and homes for the destitute or infirm. Provision and maintenance of public transportation and refuse disposal.
- Licencing of bicycles, trucks (other than mechanically propelled trucks), canoes, wheel barrows, and carts.
- Establishment, maintenance, and regulation of markets, motor parks, and public conveniences. Also, registration of births, deaths, and marriages.
- Construction and maintenance of roads, streets, drains and other public highways, parks, and open spaces. The naming of roads and streets and numbering of houses.

- Assessment of privately owned houses or tenements for the purpose of levying such rates as may be prescribed by the House of Assembly of a State.
- Control and regulation of outdoor advertising, movement, and keeping of pets of all descriptions, shops and kiosks, restaurants and other places for the sale of food to the public, and laundries.

### 5.2.3 Participant Selection and Sample Size

**Participant Selection**: In identifying the interviewees, the researcher approached them based on their interest as stakeholders and their connection to the E-Service initiatives at the local government council level. Also, in accordance with Hassan (2011), interviewees were chosen based on their policy-making level, interest in the topic, their determination to translate political vision into action plans, and years of working experience. However, it is proper to mention here that the interviewees with fewer years of experience – that is, both the Chairman and Vice Chairman of the council – were elected to their various offices in 2012, and they were exactly three years in their positions by the time the interviews took place in 2015. The Chairman appointed the Secretary of the Local Government Council shortly after his election. Ethical issues were considered, as there was a concern about the confidentiality. The interviewees also declined audio recording; therefore, the use of questionnaires for the interview was adopted.

**Table 23 List of the Interviewees**

| S/N | Job Role | Years of Experience |
|---|---|---|
| 1 | Chairman of the Local Government Council | 3 |
| 2 | Vice Chairman of the Local Government | 3 |
| 3 | The Secretary to the Local Government | 3 |
| 4 | Council Manager/Head, Personnel Management | 25 |
| 5 | Principal Accountant | 23 |
| 6 | Council Treasurer/Head of Finance | 22 |
| 7 | Head, Budget Planning Department | 20 |
| 8 | Head of Accounts | 18 |
| 9 | Accountant | 8 |
| 10 | Deputy Head, Budget Planning Department | 14 |
| 11 | Head, Waste and Environmental Department | 18 |
| **12** | **Senior Information Officer** | **18** |
| **13** | **Head, Works and Infrastructure Department** | **22** |
| **14** | **Senior Administrative Officer** | **8** |
| **15** | **Senior IT Officer** | **19** |
| **16** | **Senior Accountant** | **21** |
| **17** | **Head, Human Resource, and Personnel Management** | **18** |
| **18** | **Principal Executive Officer** | **8** |
| **19** | **Head, Agriculture, and Social Development** | **18** |
| **20** | **Programme Analyst** | **13** |
| **21** | **Senior Programme Analyst** | **18** |
| **22** | **Senior Administrative Officer** | **18** |
| **23** | **Head, Information Technology Department** | **19** |
| **24** | **Head, Primary Health Care** | **23** |
| **25** | **Commissioner, Local Government Service Commission** | **22** |
| **26** | **Head, Education Department** | **24** |
| **27** | **IT Manager** | **13** |
| **28** | **Human Resource Officer** | **9** |
| **29** | **Commissioner, Local Government Service Commission** | **25** |
| **30** | **Head, Legal Department** | **18** |

**Sample Size**: Determining the sample size in a research study is important as it is practically impossible for the researcher to collect data from the total population in the research case study. Therefore, this researcher will take a sample of the population of interest, and it is pertinent to note that the choosing of sample size depends both on the statistical and non-statistical considerations (Marshall et al., 2013). The statistical consideration includes the desired accuracy in the sample numbers while the non-statistical considerations include budget constraints, available resources, ethics, and manpower. The appropriate sample size is also determined by identifying these three criteria, namely the level of precision (sampling error), the confidence level and the degree of variability.

In determining the sample size used for the in-depth interview for the purpose of this study, this researcher considered the Rules of Thumb Based on Data Collection Method and Approach as described and used by Patton (2001) for the number of participants to be interviewed. On the other hand, the sample size calculator and required sample table by Boyd (2006) are considered to determine the expected survey sample for the online survey. The Rules of Thumb Based on Data Collection method, according to Patton (2001), stated that approximately 30 people could be considered for in-depth interviews as represented in **Table 24**, Rules of Thumb Based on Approach using Action Research as mentioned in **Table 25**, and the Length of Interviews as indicated in **Table 26** below:

**Table 24 Rules of Thumb Based on Data Collection Method (Patton, 2001)**

| Data Collection Method | Rule of Thumb |
|---|---|
| 1. Interviewing key informants | Interview approximately 5 people. |
| 2. In-depth interviews | Interview approximately 30 people. |
| 3. Focus groups | Create groups that average 5-10 people each. In addition, consider the number of focus groups you need based on "groupings" represented in the research question. That is, when studying males and females of three different age groupings, plan for six focus groups, giving you one for each gender and three age groups for each gender. |
| 4. Ethnographic surveys | Select a large and representative sample (purposeful or random based on purpose) with numbers similar to those in a quantitative study. |

**Table 25 Rules of Thumb Based on Approach (Patton, 2001)**

| Research Approach | Rule of Thumb |
|---|---|
| 1. Biography/Case Study | Select one case or one person. |
| 2. Phenomenology | Assess 10 people. If you reach saturation prior to assessing 10 people you may use fewer. |
| 3. Grounded theory/ethnography/ action research | Assess 20-30 people, which typically is enough to reach saturation. |

Therefore, the sample size for the in-depth interview for this study and as explained above will be **30**.

**Table 26 Guidelines for Length of Interviews (Patton, 2001)**

| Number of Interviews | Length of each interview |
|---|---|
| 10 | 1-2 hours |
| 20 | 30 minutes-1 hour |
| 30 | 20-40 minutes |

### 5.2.4 Conducting the Study

The necessary permissions were obtained from the interviewees after identifying them, and before proceeding with the interviews. This was necessary for the purpose of ethical consideration. Assistance was, therefore, requested from a contact, an accountant from the local government council. The first meeting with the accountant was informal, and then with the Chairman of the local government council. The meeting afforded the researcher the opportunity to explain them the purpose of the research, aims, and objectives of the research study. Many documents to support the research study were shown to them. These documents are:

- A formal participant information sheet inviting the participant to take part in the research study with study summary, purpose, ethical issues, and data retention clearly explained.
- A formal consent form to seek the participant's permission to participate in the interviews, and confirming their right to withdraw at any time without giving any reason.
- A formal invitation letter requesting permission to conduct interviews and confirming the topics to be discussed.
- An outline of the research, its aim, and objectives.
- A copy of the interview questions (in English).
- The initial list of the possible interviewees that the researcher wished to meet.

As required by the research ethics guiding the research study, it was certified before the beginning of the interviews that the researcher would treat the data collected from the interviewees in a confidential manner and that the anonymity of all participants would be guaranteed at all times. However, tape-recording was not used as questionnaires had been agreed by the researcher and the interviewees. All interviews began with a short description of the research, which included the aim, objectives, estimated time for conducting the interview, and emphasis on the key role of the interviewee's views. The duration of the interviews ranged from 20 to 40 minutes each and notes were taken to record the observations about the meetings. At the beginning of each session, the interviewees were asked to read and sign the consent letter. Also, they were asked to fill out an individual information sheet. The sample invitation letter, consent form, and participant information sheet are included in the appendix.

However, the possibility of manager bias (e.g. the Chairman or any other head of department at the local government council) was considered but ignored, because they were not in a position of influence the other participants (participants were met over a period). The researcher's perception was that the council heads and managers were honest officials, who were interested in what E-Service initiatives could add to the local governments economically and how they could improve the services to citizens.

### 5.2.5 Study Timetable

Although the research study was expected to span over three years (from 1 February 2014 to 31 January 2017), it was later extended until 28 February 2017. The interviews with the participants at the local government council in Lagos, Nigeria, took three months between the months of June and September 2015. The participants were met over the three months on the date earlier agreed between them and the researcher. Hence, to complete the phases and stages required in the action research method (Diagnosis, Action Planning, Action taking, Evaluating and Specifying Learning), the participants later agreed with the researcher the need to constitute a five-person focus group (online) among the 30 interviewees in order to meet with the researcher online when required. The focus group gave another option of meeting with the participants online due to their busy work schedules. The online meetings also prevent some logistic problems, such as frequent travelling to Nigeria.

**Table 27 Field Study**

| | | Venue | Time |
|---|---|---|---|
| 1. | **Interviews** (One-to-one Questionnaires)<br><br>Participants (30) | Site Visit | 29/06/2015 – 11/09/2015 (Meeting days as agreed with each participant during the period above). |
| 2. | **Online Survey** | Online | 06/10/2015 – 29/01/2016 |
| 3. | **Online Focus Group**<br><br>Participants (5) | Online | **Meeting 1**: 29/07/16<br><br>**Meeting 2**: 25/10/16<br><br>**Meeting 3**: 15/12/16 |

### 5.2.6 Challenges

There are many problems this researcher encountered which range from the recruitment of the participants, to ensuring the interviewees were available for the plenary meetings during the interviews. The researcher had prior ideas on this issue, as experienced in the pilot study before the formal interviews for this study. In the pilot study (Oseni and Dingley, 2014), the researchers had to send emails regularly to the participants as a reminder to participate in the survey, despite a prior notice. Therefore, lessons are already drawn from the pilot study concerning meeting arrangements with participants. The researcher always sent a reminder email until the meeting times were eventually agreed. Office environments at the local government council in Nigeria are always very busy because it is one of the tiers of government that is very close to the citizens (Oseni and Dingley, 2014).

Hence, it is always difficult to ensure that the participants will arrange a meeting and turn up for it, while organising a conducive office for the meeting is another challenge. Despite this, a balance has to be set between the promise of a short time commitment, but sufficient time to allow the richness of the ideas to develop during the eventually agreed meetings. Eventually meeting up with the participants came from perseverance, as the researcher had no authority to insist on meeting times. Other problems, such as a refusal for some interviews to be audio recorded, as explained above, were resolved by using questionnaires in the interviews, which were completed face-to-face with the participants. The issue of meeting the participants later to complete the cycle involved in the action research method was also resolved by constituting a five-member online focus group among the interviewees, to meet online when required with the researcher to discuss feedback, research theory, validation, and the testing of the framework as agreed.

## 5.3 Data Collection Procedures

### 5.3.1 Interview - Questionnaires

Interviews, as a method of data collection, help to create a perception of how the participants understand the research domain, and this are an essential method of all qualitative methods. A semi-structured interview is an appropriate method of carrying out a one-on-one meeting between the researcher and the participants (interviewees), as it will give the researcher the needed avenue by which to probe deeply to discover new hints and scopes within the research domain (Hassan, 2011). As discussed in the previous section, in order to maintain a high level of reliability for this study the researcher made use of the instruments (questionnaires) for both the interview and online survey. The interview and survey questions were previously used and validated by Susanto and Goodwin (2010), Alshehri and Drew (2010), and Almarabeh and Abu Ali (2010), with a slight modification to accommodate changes made by the participants involved in the action research. The questionnaires were also tested again by this researcher in the pilot study used for the published paper by Oseni and Dingley (2014) on the “challenges of E-Service Adoption and Implementation in Nigeria: Lessons from Asia.”

Prior notice was given to the interviewees with an outline of the interview structure to be adopted during the session, as required by the research ethics. The action research method allows recording, document observation and note taking during the interview session. The choice of whom to interview first for this particular research purpose was crucial, and it was not an easy decision to make. The participants declined audio recording during the meetings for confidentiality reasons. This is a serious drawback, as recording during the research interviews is an enhanced way to capture qualitative data as it ensures descriptive validity. An audio recording of interviews allows researchers to go back to the interviews in order to retrieve missed data. In the light of the above, there is a need for this researcher to adopt a new interview method. The face-to-face questionnaire was used for the interview as suggested by Harris and Brown (2010) on mixing interview and questionnaires method. Constantinos et al. (2011) further stated that a questionnaire during the interviews could be of great advantage, especially if the interviewees feel socially uncomfortable about discussing some questions on a one-on-one basis with the interviewer. However, the presence of an interviewer allows for complex issues to be clarified, if necessary, to the interviewees.

### 5.3.2 Online Survey

Some authors have considered the use of sample size in the required sample size table (Xu and Yu, 2013; Boyd, 2006) as displayed in the appendix. Therefore, the researcher adopted the required sample as presented by Boyd (2006). Whenever the population size reached 25,000, with a confidence level of 95% and margin error of 5.0%, the required sample for the online survey is **379** (Boyd, 2006). The population size for all the local government councils in Lagos State, Nigeria, being the case study, is 23,123 (LSG, 2016).

Also, it is important for this researcher to exploit other academic means when concluding the number of samples needed for this study. The use of the sample size calculator is considered, and the analysis is presented in **Figure 48** below:

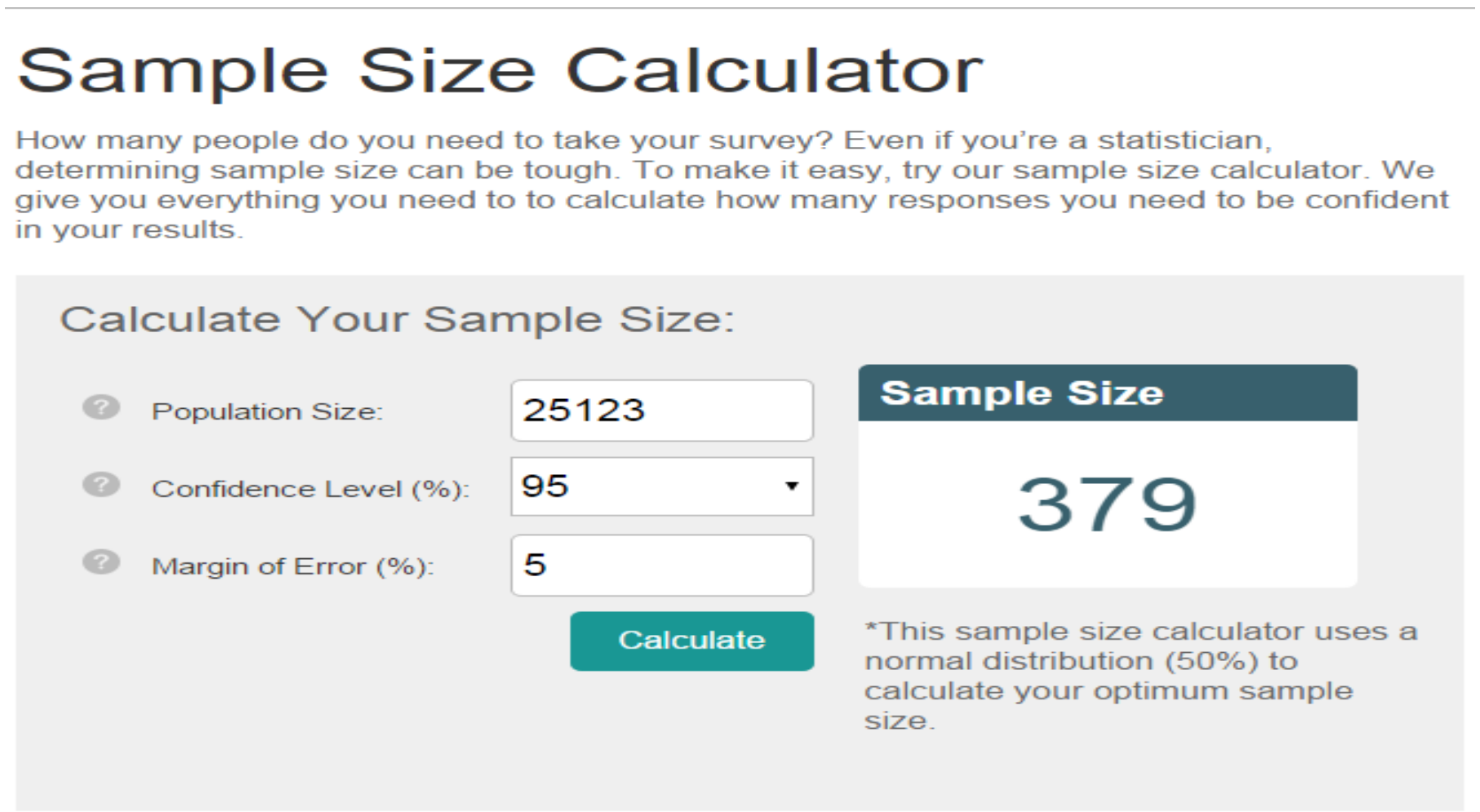

**Figure 48 Questionnaire Based Survey Sample Size Calculator**

**Confidence level:** A measure of how certain you are that your sample accurately reflects the population, within its margin of error. Common standards used by researchers are 90%, 95%, and 99% (Boyd, 2006).

**The margin of error**: A percentage that describes how closely the answer your sample gave is to the "true value" in your population. The smaller the margin of error, the closer you are to having the exact answers at a given confidence level (Boyd, 2006).

Interestingly, there is little significant difference between the sample size obtained by Boyd (2006) and the sample size calculator. Using the average of the two sample sizes, the total number of the sample is 378.5 and approximately equal to 379. Therefore, the sample size needed for the online survey for this study is 379.

### 5.3.3 Pilot Study

The main purpose of a pilot study in a research study is to examine the feasibility of an approach/instrument that is intended to be used in a further larger-scale study. According to Porta (2008), a pilot study is defined as the small-scale test of the methods and procedures to be used on a larger scale. The main importance of the pilot study is to improve the quality of the main study. The pilot study could expose some logistics issues prior to the main study (Porta, 2008). There are many reasons why a researcher should engage the assistance of a pilot study before the main study. These reasons are as follows:

- Pilot study can greatly reduce the number of unanticipated problems; this is because the researcher has an opportunity to redesign parts of the study to overcome difficulties that the pilot study reveals.
- Pilot study often provides the researcher with ideas, approaches, and clues that they may not have foreseen before conducting the pilot study. Such ideas and clues increase the chances of obtaining clearer findings in the main study.
- The pilot study almost always provides enough data for the researcher to decide whether to go ahead with the main study. This may save a lot of time and money.
- Pilot study permits a thorough check of the planned statistical and analytical procedures, giving the researchers a chance to evaluate their usefulness for the data. Researchers may then be able to make the required alterations in the data collecting methods, and therefore analyse data in the main study more efficiently.

Hence, for the purpose of this research study, a prior pilot study was conducted by Oseni and Dingley (2014) to test the instruments (questionnaires) deployed for data collection. The questionnaires were purposely designed to help in collecting data for this study. The questionnaires were divided into two categories. The first category was the demography part, which was used and validated by Alshehri and Drew (2010) and Susanto and Goodwin (2010). The second part of the questionnaires was about E-Service adoption and implementation. These questions have been used and validated by Almarabeh and Abu Ali (2010). Data was collected between 23 September 2014 and 30 October 2014 and a total of 120 responses were received. The responses led to changes, which were replacing the research theory and the use of the online focus group.

### 5.3.4 Theoretical Saturation

As stated in grounded theory, a researcher could carry on collecting data through interviewing, observation or document reviewing until they reached data saturation or theoretical saturation (Bryman, 2012). In other words, means when there are no new relevant data or ideas emerging regarding the research topic. In their detailed analysis of data saturation, Francis et al. (2010) maintained that this emerged from the term "theoretical saturation", which is the process of data gathering and analysis until the point when no new insights are being observed. The issue of data saturation is important in this research, in order to demonstrate the adequate sample used for validity purposes. Francis et al. (2010) argued that data saturation addresses whether a research study is based on an adequate sample that will be good enough for content validity. They also suggested that data saturation occurs after 13 to 15 interviews, and this was consistent with findings by Guest et al. (2006), where 12 interviews conducted in their study resulted in saturation.

For this study and, as explained in the sample size section above (see **Table 25**), in the rules of a thumb-based on approach (Patton, 2001), this researcher interviewed 30 people. This is number is considered to be enough to reach saturation. This researcher discovered that no new data appeared to be occurring concerning the barriers facing the adoption and implementation of E-Services at the local government level in Nigeria. The saturation level also reached for the online survey after this researcher obtained 425 responses against the expected 379, which is the sample size as explained in **section 5.3.2** above. The online survey was opened on 06/10/2015 as agreed by the participants (action research) during the meetings for further data collection to support the interview outcomes. The participants, which comprised the local government staff and other stakeholders in E-Service implementation at the local government Service Commission in Lagos, were invited through emails, text messages, and social media. The online survey was closed for further responses on 29/01/2016 after reaching a saturation level with 425 responses.

## 5.4 Action Research Method Advantages and Disadvantage, Phases and Cycles

### 5.4.1 Advantages and Disadvantage of Action Research Method

One thing that this researcher needs to correct is the statement about the action research method, as presented by Beard et al. (2007). In Beard et al. (2007), in their comprehensive analysis of the action research method, argued that doing action research is easier than conventional research. For the researcher, action research is interesting but difficult to do, as taking responsibility for change is not an easy task. The researcher often puts more effort into the research without getting useful results. Hence, the following are the advantages and disadvantages of the action research method.

**The Advantages of Action Research Method:** There are many reasons for this researcher to choose the action research method for this study, as follows:

- **It allows the researcher to work on a problem**, make a change, and yield answers to the problem, as well as informing theory. Making use of the DOI theory in this study has assisted in predicting the success of E-Service adoption and implementation through action research methodology.

- **It produces rich data from multiple sources**. This is achieved with the researcher, who exploited data collection methods such as interviews, online survey, focus group, document observation, document studies, and so on (Gapp and Fisher, 2006).

- **It empowers participants, enables change, and creates opportunities for organisational learning**. In this research, the honest contributions from the participants go a long way to improving and promoting the future of E-Service initiatives. The E-Service awareness is an alert for the people in authority to take action in delivering the desired technology required at this level (Beard et al., 2007).

- **It creates a solid theory about practice and, hence, leads to improvements**. The theory is expected to emerge from the rich data obtained during the field studies such as interviews, focus group discussions, and observations (Gapp and Fisher, 2006).

- **The use of action research contributes to learning for the participants and this researcher**. The experience gained on how to effect a change during the meetings and deliberations is an active medium for action researcher and practitioners to learn. It is also useful for the participants as they find out more about the topic. This may help in the diffusion of innovation, as 30 local government people will be better informed and engaged with the transformations that E-Government could bring to their services (Gapp and Fisher, 2006).

However, despite these advantages, action research is not without its critics and limitations. Dick (1993) argued that action research may not be a common research method, as psychology tends to have ignored the action research. The following are the **disadvantages of the action research method**:

- **Lack of Impartiality of the Researcher**: This disadvantage, according to Germonprez and Mathiassen (2004), weakens the research as the researcher will not be able to maintain neutrality. In this study, this researcher argues that maintaining neutrality negates the participatory principle of the action research, where solutions to the identified problems are collaboratively mapped out by this researcher and the participants. Ethnomethodology encourages participatory research methods, such as action research.

- **Validity and Reliability Issues**: This is discussed in the previous chapter, where this researcher looked into the relationship between both validity and reliability. While **reliability** deals with the consistency of results, when an experiment is replicated under the same conditions, **validity** deals with the degree to which a research study measures what it intends to measure (Gravetter and Forzano, 2012). There are significant concerns about the effect of validity and reliability on the action research method. Adams and McNicholas (2007) argued that the results obtained from an action research method may not be repeatable, as it is a one-off research and should not be regarded as a conventional method. For this reason, and as stated by Adams and McNicholas (2007), many young researchers avoid using the action research method. This argument was supported by Germonprez and Mathiassen (2004), who claimed that many information system researchers often warned against doing action research since it might slow down their career. However, this researcher was not in total support of the argument as stated above, as he believes that action research is associated with a change that brings about an efficient system. Hence, this should not threaten the validity and reliability of the previous study.

- **Selections of the Participants**: The researcher agreed with the fact that action research is associated with making a change, and it is fundamentally one of the action research method features that enable the participants to have a voice and input. The disadvantage, as experienced by the researcher, is getting the right participants and for them to give their honest contributions. Many participants were approached but, out of fear of the unknown, they preferred not to participate or to remain anonymous because the change in question was related to their organisation.

- **Strange Relationship/Lack of Discipline between Research and Participants**: The participatory and organisational nature of the action research which identified it with the "change" term (Adams and McNicholas, 2007) requires a strong relationship between the researcher and participants. The researcher, however, acknowledged that there might be different opinions in any debate related to the organisational change or improvement. Nevertheless, there is a need for discipline and strong relationships between the researcher and the participants.

### 5.4.2 Action Research Phase and Cycle

There are five phases involved in the **action research** methodology, as shown both in Action Research Cycle (see **Figures 42** and **45**) in Chapter Four (Baskerville, 1999; Susman, 1983) and **Figure 3 (**Research Design) in Chapter One. They are **Diagnosis**, **Action Planning**, **Action Taking**, **Evaluation** and **Specifying Learning** (Susman, 1983; Baskerville, 1999). However, the researchers using an action research method are expected to complete research cycles in the action research phases (Rose et al., 2015). In this approach, the research proceeds as a cycle of joint **Planning**, **Action**, **Observation** and **Reflection**, where the **reflection stage** opens the opportunities for further cycles of planning, acting, observing and reflecting in a spiral of learning, as displayed in **Figure 49** below, depending on the results or actions obtained. Moreover, it is on record that different researchers have used different terminologies for the steps involved in the action research cycle. In a follow-up study, Rose et al. (2015) found that the steps in action research used by Coghlan and Brannick (2010) are labelled as '**Constructing**', '**Planning action**', '**Taking action**', and '**Evaluating action**'.

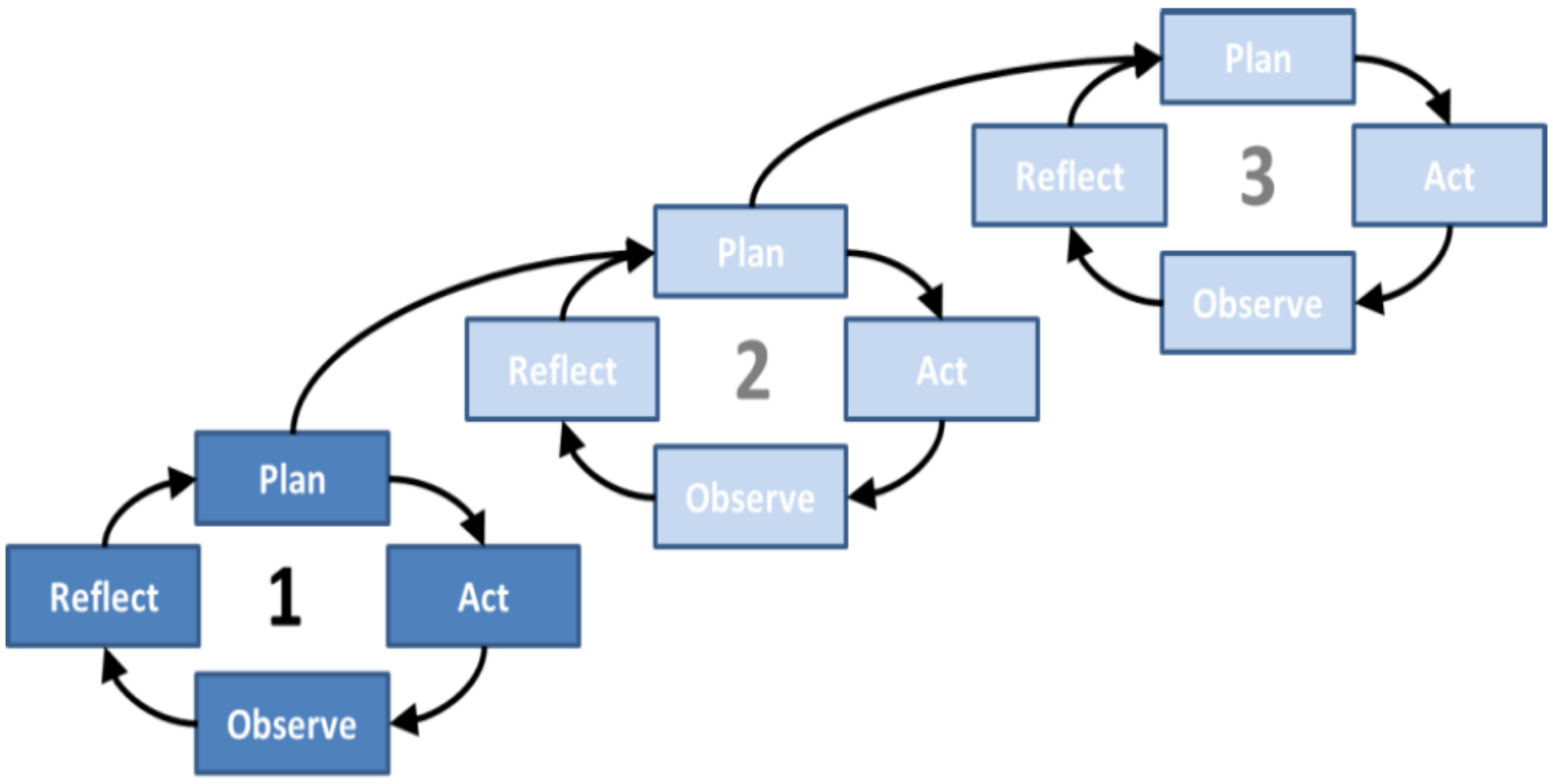

**Figure 49 Action Research and the Cycles (Rose et al., 2015)**

Hence, this research proceeds as a cycle of joint planning, action, observation, and reflection (Coghlan and Brannick, 2010). The reflection phase paved the way for further cycles of planning, acting, observing, and reflecting (Coghlan and Brannick, 2010) as shown in **Figure 49**. In the light of the above, this research went through three cycles of **Interviews**, **Online Survey**, and **Online Focus Group** as illustrated in **Figure 54** (section 6.3.3).

## The Five Action Research Phases

It is essential for the researcher to precisely map out the activities involved in each phase of the action research for this study. These activities are in line with the cycles (see **Figures 49 and 50)** needed to complete the action research phases. Hence, the five action research phases are:

## **Phase 1 – Diagnosis** (Problem Identification)

- Why do you want to do it? Is it an important and practical problem, something worth the researcher's time and effort, something that could be beneficial to the researcher, other researchers, academia, the local government, and others?
- Is the problem stated clearly and in the form of a question? Is it broad enough to allow for a range of insights and findings? Is it narrow enough to be manageable within the researcher's timeframe and daily work?

**Phase 2 –** Action Planning

- Will the researcher develop and implement a new strategy or approach to address the research questions? If so, what will it be?
- Will the researcher focus study on existing practices?  If so, which particular ones?
- What is an appropriate timeline for what the researcher is trying to accomplish?

**Phase 3 – Action Taking** (Data Collection)

- What types of data should the researcher try to collect to answer research questions?
- How will the researcher ensure multiple perspectives?
- What resources exist, and what information from others might be useful in helping the researcher to frame questions, decide on types of data to collect, or to help in interpreting your findings?

**Phase 4 – Evaluation** (Analysis of Data)

- Learning from the data, what patterns, insights and new understandings can the researcher find?
- What meanings do these patterns, insights, and new understandings have for the research and practice?

**Phase 5 – Specifying Learning** (Future Action Plan)

- What can the researcher do differently as a result of this study?
- What will the researcher recommend to others?
- How will the researcher write about what he has learnt so that the findings will be useful to the practice and other stakeholders?

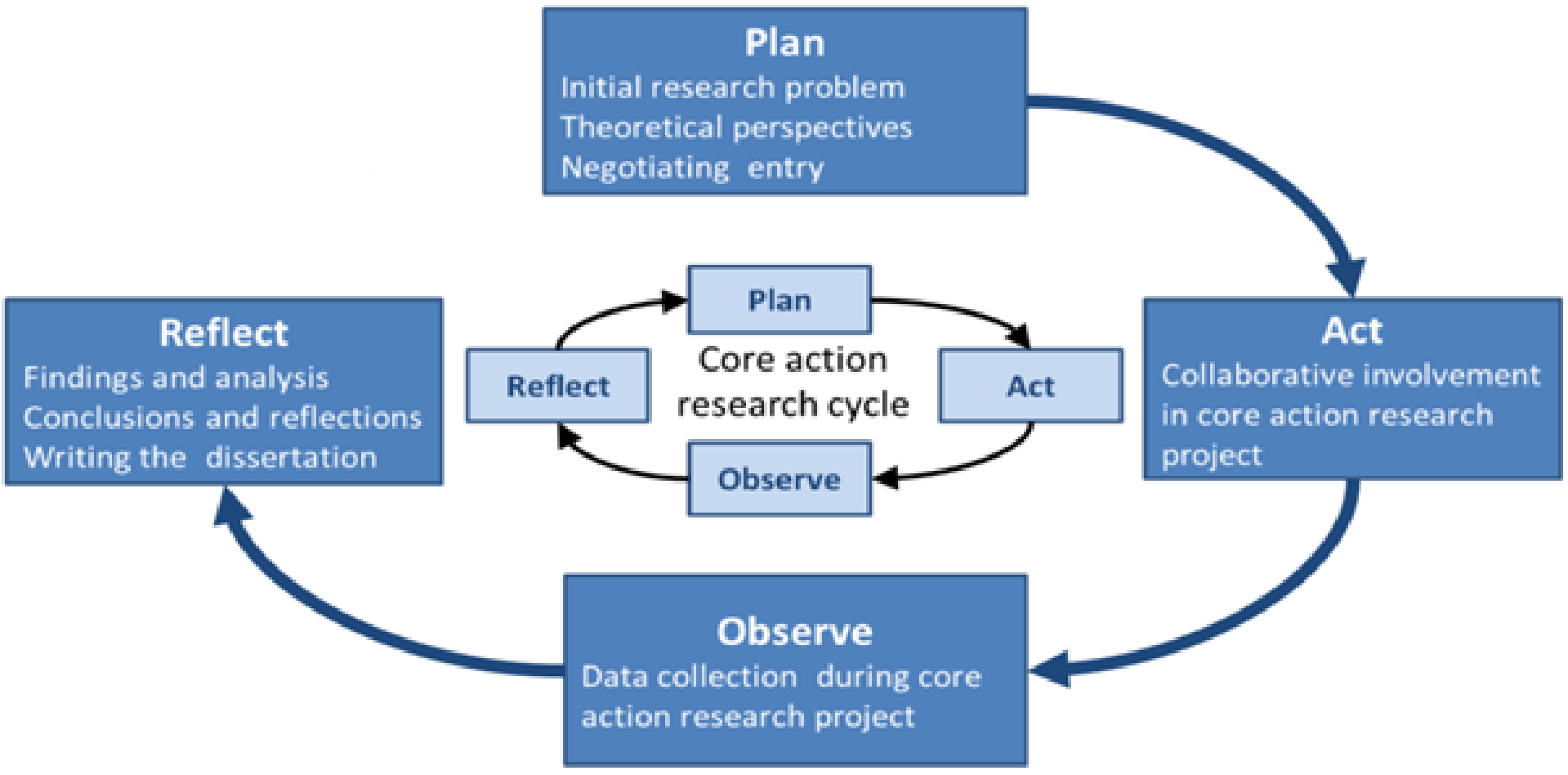

**Figure 50 Action Research Cycle Activities**
**(Rose et al., 2015)**

The action research activities involved in each phase were carried out in conjunction with the research theory adopted in the **Diffusion of Innovation** (DOI) theory (see **Table 28)**. Hence, this will help to evaluate the role of technology and humans in accepting the E-Service initiatives at the local government level in Nigeria.

**Table 28 Action Research Phases and DOI Theory Activities**

| Research Phase | Action Research Method | Research Theory (Diffusion of Innovation Theory (DOI) | Action Research/DOI Activities |
|---|---|---|---|
| Phase 1 | **Diagnosis** | **Knowledge** | **Diagnosis and Knowledge**: Clear identification of research issues, aims, objectives, research questions and gaps found. Understand clearly the research background, the time frame for the research mapped out. |
| Phase 2 | **Action Planning** | **Persuasion** | **Action Planning**: Literature review, selecting research methodology and justification, development of the initial framework. Selecting the participants, giving prior notice and making necessary arrangement for meetings. Timeline concluded also. |
| | | | **Persuasion**: Were the participants influenced by this researcher to participate or accept the technology (making change). |
| Phase 3 | **Action Taking** | **Decision** | **Action Taking**: Data collection via interviews (face-to-face questionnaires) and an online survey, documents observation, refining the initial framework.<br><br>**Decision**: Decision on mode and structure of the interviews. Change of research theory, the use of DOI theory instead of Technology Accepted Model (TAM). The decision was also reached with the participants to further use a 5-member online focus group among the participants to complete the research cycle. |
| Phase 4 | **Evaluation** | **Implementation** | **Evaluation**: Data analysis using SPSS and Nvivo, evaluating results, coding, and validation of the framework. The revelation of any new understanding and contribution to knowledge.<br><br>**Implementation**: Applying necessary changes from the data collection and feedbacks on the framework. Testing and validating the framework. |
| Phase 5 | **Specifying Learning** | **Confirmation** | **Specifying Learning**: Writing up, specifying findings, achievement/contributions, and making recommendations.<br><br>**Confirmation**: Confirming the final framework. |

### 5.4.3 The Use of Online Focus Group in Action Research

Focus group methods involve a group of participants discussing specified aspects. In using focus groups, there is an emphasis on the questioning as it centred on a defined topic (Bryman, 2016). Focus groups, according to Bryman (2016), consist of two methods, namely ***Group interview*** and ***Focused interview***. The **Group Interview** involved the coming together of several people to discuss a particular topic. The **Focused Interview** of the focus group included the selection of participants who had been known to a particular situation (Bryman, 2016). The researcher agreed to use the online focus group with the participants during one of the interviews to have a more robust session, where constructive arguments between the participants and the researcher could be reviewed, and also to complete research cycles. This was not the case in the conventional one-to-one interviews, where the participants rarely challenge the researcher's opinions. Hence, this study involved the use of focused interview as the participants already had the experience of the research.

The focus group has been created by the researcher who is also the moderator. Once the participants register their interest to be a member of the focus group by creating a username and password, they will be able to participate in the focus group discussions and respond to the questions posted by the researchers on the use of internet messaging and social networking in online research methodology. The focus group was created through Focus Group IT online tools. Moreover, the online focus group engaged in this study accommodated both synchronous and asynchronous types. The researcher gave prior notice of online meetings, and the participants who were able to make the session time would engage in the synchronous type, where the discussion is in real time, and contributions and responses made as immediately as the bandwidth allows. The other participants, who are unable to make the online meeting schedule, will engage in the asynchronous type, where the responses are not in real time. Participants are permitted to join the session later and drop in their responses offline.

The online action research with blind (anonymous) focus group meetings was chosen specifically because of the sensitive nature of the topic. Getting responses from the government officials on issues relating to corruption in government is difficult. Therefore, anonymity is key to receiving honest responses, as well as protecting respondents. Since the researcher has developed this novel online research method based on blind focus group sessions (anonymised typing in the group session), a further ethical review was done to ensure that the necessary areas were covered from an ethical point of view. However, the use of anonymised online focus group typing in group sessions was necessitated due to the participants' refusal for the interviews to be audio recorded during the action research interview sessions conducted earlier. Bryman (2016) argued that interviews conducted through focus groups do not have to be audio recorded as this will eliminate an interviewee's apprehension about speaking and being recorded. The online blind focus group sessions were used at the final stage of research data collection and analysis to implement the feedback from the participants, which include the framework validation following the online survey, face-to-face interviews and field studies in Nigeria.

The online action research focus group will also contribute to the study in the following ways:

- It will assist in exploring some of the key topics in greater depth than when they emerged from the surveys.
- It will generate new ideas and knowledge as participants get the chance to discuss and elaborate on each other's ideas (Kapenieks, 2013).
- It will assist this researcher's understanding of the fear that "whistle-blowers" have in a climate of corruption, even when the government headline is to reduce or eliminate corruption.
- It will contribute to advances in theory and practice (Kapenieks & Salīte, 2013).

Furthermore, another benefit of the online focus group to this study is that it is easier for the participants, as they can arrange a time that fits their schedules (Bryman, 2016; Oseni et al., 2017).

### 5.4.4 Recruitment and Approach to the Focus Group

The five participants for the online focus group, as shown in **Table 32**, were selected from those who volunteered their contact details during the one-on-one interview stage of the research. The selection is in line with the rule of thumb based on the data collection method (Patton, 2001), which recommended the numbers of a focus group session, with an average of 5-10 people each. There is no inducement, as all the participants will choose their pseudonym, which will be known to only this researcher. The anonymous online focus group session was deployed to ensure high and reliable responses as well as the safety of the participants and this researcher.

The approach for this study is to provide anonymity by allowing the five focus group participants to use a pseudonym of their choosing. They will use typed text in the session so that the participants can see other's responses, but will not be able to identify whose responses they are (Oseni et al., 2017). The participants could be sitting in the same room as others in the focus group and not know they were participating in the same focus group. The researcher will know the identities and pseudonyms initially for contact purposes, and ensure that anyone that needs the follow-up information can get it. This approach is necessary, as the researchers are dealing with five senior officers at the local government to hear their views on the use of internet messaging and social networking in online research methodology. The online focus group meetings used here prevent conflict of interest and can obtain honest answers (Oseni et al., 2017).

The five participants using the pseudonym **Participant 1**, **Participant 2**, **Participant 3**, **Participant 4** and, **Participant 5** will be required to register to be a member of the focus group ***(http://www.focusgroupit.com/groups/fc236ae6)***. The focus group has been created by the researcher, who is also the moderator. The participants register their interest in being a member of the focus group by setting up a ***username*** and ***password***. They will be able to participate in the focus group discussions and respond to the questions posted by the researcher on the use of internet messaging and social networking in online research methodology, as seen in Oseni et al. (2017). Seven questions asked were which took around 45 minutes to answer by the participants, and the online focus group discussion took place on 29 July 2016 and other arranged dates, as shown in **Table 27**.

## 5.5 Data Processing and Analysis

This section involves a series of actions and methods performed on data collected for this research to organise, describe, transform, integrate, extract for subsequent usage or testing hypotheses. In this study, there is a need to convert data collected to produce information or knowledge needed by this researcher to answer the research questions. The data processing method includes editing, tabulating, coding, charting, and classifying. The processing of data will lead to the manipulation of collected data, as well as irrelevant data removed from the relevant data.

### 5.5.1 Data Processing

The data collected for this study (questionnaires, online survey, field notes, online focus group meetings and official documents) were processed before they were available for further analysis. The researcher, during the one-on-one interaction with the participants during the interviews and online focus group meetings, made notes, selected quotes and wrote his comments. However, it is pertinent to mention here that all the interviews and online surveys were conducted in English. Luckily, English is the official language in Nigeria. The researcher took extraordinary care with facial expressions, word accents, and explanatory gestures to reach a smooth and clear summary of the main thoughts and ideas presented. The raw field notes taken during the interviews were converted into reports as recommended by Hassan (2011). These reports were reviewed and amended for accuracy by the researcher, prepared to be coded and analysed along with the rest of other types of data. Then, all records were organised in a database, as suggested by Hassan (2011), along with primary data (interview questionnaires, online survey, field notes, and online focus group meetings) and secondary data (existing literature, official documents collected from the state and local governments in Nigeria).

### 5.5.2 Quantitative Data Analysis with SPSS

Quantitative data is normally analysed through the use of statistical methods which can be divided into two major categories, namely ***Descriptive*** and ***Inferential*** statistical techniques. The **Descriptive** statistical technique includes the measurement of central tendencies such as average, tables, graphs and charts. The **Inferential** statistical techniques enable inferences to be made about the data collected, and they include t-test, chi-square, and correlation. The usage of these techniques largely depends on what the research has set out to achieve. The researcher made use of the descriptive statistical technique as there is a need to infer from the dataset what the participants might think, and to describe what is going on in the data collected for this study. Also, the descriptive statistics help to simply present a large amount of data collected in a sensible and presentable way.

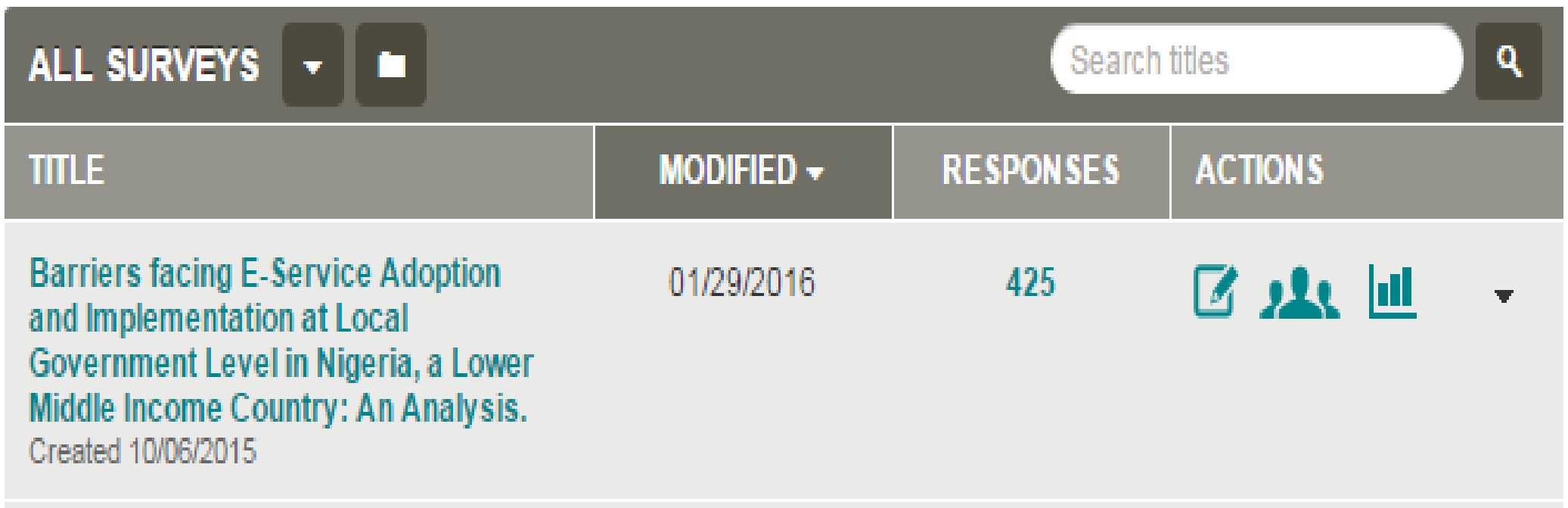

**Figure 51 Survey Monkey Online Survey Data Collector**

Statistical Package for the Social Scientists (SPSS) is a data management and statistical analysis tool with a multipurpose data processing competency. The researcher made use of IBM SPSS Statistics 22 as available at the University of Portsmouth. SPSS data were stored in a spreadsheet-like table, which is very similar to Microsoft Excel. Hence, data can easily be imported into SPSS from both Excel or .csv files. Even though there is a limitation in SPSS about imported .csv file variables, which may not be in the appropriate numerical format, SPSS still offer a more suitable approach to in-depth data analysis (Bryman, 2016). SPSS is used in this study to analyse the data collected by the researcher from both the ***interviews*** with the participants and the ***online survey***. As explained previously, and shown in **Figure 51** above, the online survey data were collected from the respondents during the period **06/10/2015** to **29/01/2016** when the researcher decided to close the survey after reaching a saturation of collected responses.

The data view window displays the research data values (cells), arranged as variables (columns) and cases (rows). The data shown in the data view window could be edited if necessary. The SPSS variable view window contains the description of the attributes of each variable in the data file where rows are the variables, and the columns are the variable attributes. Variables can be added or deleted, and the attributes of variables could also be modified (Bryman, 2016).

### 5.5.3 Qualitative Data Analysis with Nvivo

Nvivo is a qualitative data analysis (QDA) software used by many researchers to analyse collected qualitative research data. The software package is produced by QSR and works better for a deep level of analysis on either a small or large volume of data. The use of Nvivo in the data analysis for this study is borne out of the fact that the software is simply used to manage data and allows continued focus on how to examine the meaning of data collected either by recording or otherwise. Moreover, Nvivo is also used when the data collected by the researchers is presented in a non-numeric form – that is, text or visual (Maxwell, 2013). Using Nvivo in analysing qualitative data obtained by the researcher during the online focus group meeting will help this researcher to manage, query, report, and visualize data. The data from the online focus group for this study were imported into Nvivo for the coding process. The researcher made use of the latest QSR Nvivo 11.3 version, as available in the University library. Coding data in Nvivo is a major phase in the whole qualitative data analysis process (Bryman, 2016). In the Nvivo's main screen, we have the **navigation view**, the **list view,** and the **detail view**. The navigation view consists of components such as sources, nodes, and classifications of the project the researchers intend to access. The list view, on the other hand, provides a list of the contents of imported folders or documents. The items such as interview transcripts and focus group transcripts could all be viewed from the list view. Finally, the detail view shows the content of an opened document or item. Hence, it will enable the researchers to work on the item either by coding, examining or linking it (Maxwell, 2013).

In Nvivo, there are terms that a researcher using Nvivo needs to be familiar with to enable an efficient analysis of the dataset. For example, the **source** is where all the interview questions with the participants are kept. All research materials such as documents, audio, video, memo, framework matrices are available under source. **Auto-coding** is used to organise data before analysis. It could be used to organise each question and response from the participants. The **Query** is used to run the word frequency, text search query, etc. This researcher will be able to know the number of times each word has been used during the interview meetings using text search query.

## 5.6 Summary

In the action research method, the integration of both the quantitative and qualitative research techniques offers an in-depth opportunity towards rich data collection. This combination of approaches is necessary because of the wide range of data needed to obtaining the expected results that will answer both the research questions and objectives for this study. The use of this multiple approach can be time consuming and labour-intensive, but the results this researcher intends to achieve in this study justify its usage.

In this chapter, data collection procedures are described in detail, in addition with the approaches and techniques used for the data analysis. **Figure 46** above shows all the sections discussed in this chapter. The researcher has been able to explain the detailed procedures and why questionnaires are used in the interviews, the use of online focus group, and the online survey for this study. A list of the interviewees' positions in the organisation is provided, although they remain anonymous, as agreed with the participants for this research. Lastly, the reasons and the use of the software packages SPSS and Nvivo are explained as tools to assist with the data analysis.

Furthermore, this chapter also discussed the selection of the organisation and participants. There is no way to judge how honest the answers from the participants were, as the chosen organisation is a political workplace. The researcher was able to address the sincerity issue by providing anonymity in all the online sessions with the focus group members. The advantage of the online anonymity is that the participants feel protected. They would be able to express freely as they are not physically engaging with other people, while the loss of personal connections with other people might be considered as a disadvantage of the online anonymity.

The organisation (a local government) was selected because the study examined the barrier facing the E-Service adopting and implementation at the local environment level, and the organisation is one of the stakeholders in the provision of E-Services at the local level in Nigeria. Limitations in this chapter include the decline of audio recording by the participants, and the unstable political environment, meaning that the participants could not speak their mind for fear of losing their jobs.

# Chapter 6 – Research Findings (Lessons and Reflections)

## 6.1 Introduction

This chapter discusses the research findings (learning and reflections) for this study based on the action research methodology applied by this researcher. To understand the main findings of the research, it is important to commence this chapter with the investigated barriers facing the adopting and implementation of E-Service at the local environment level in Nigeria, including the solutions to these barriers and success factors for the adopting and implementation of E-Service. As illustrated in **Figure 52** below, in the chapter summary, the sections of this chapter include the introduction, investigated/identified barriers, solutions and the success factors. Other points include lessons and reflections (research findings), the respondents' responses from the online survey and the theory of change, the E-Service implementation process and, finally, the chapter summary, which concludes this chapter

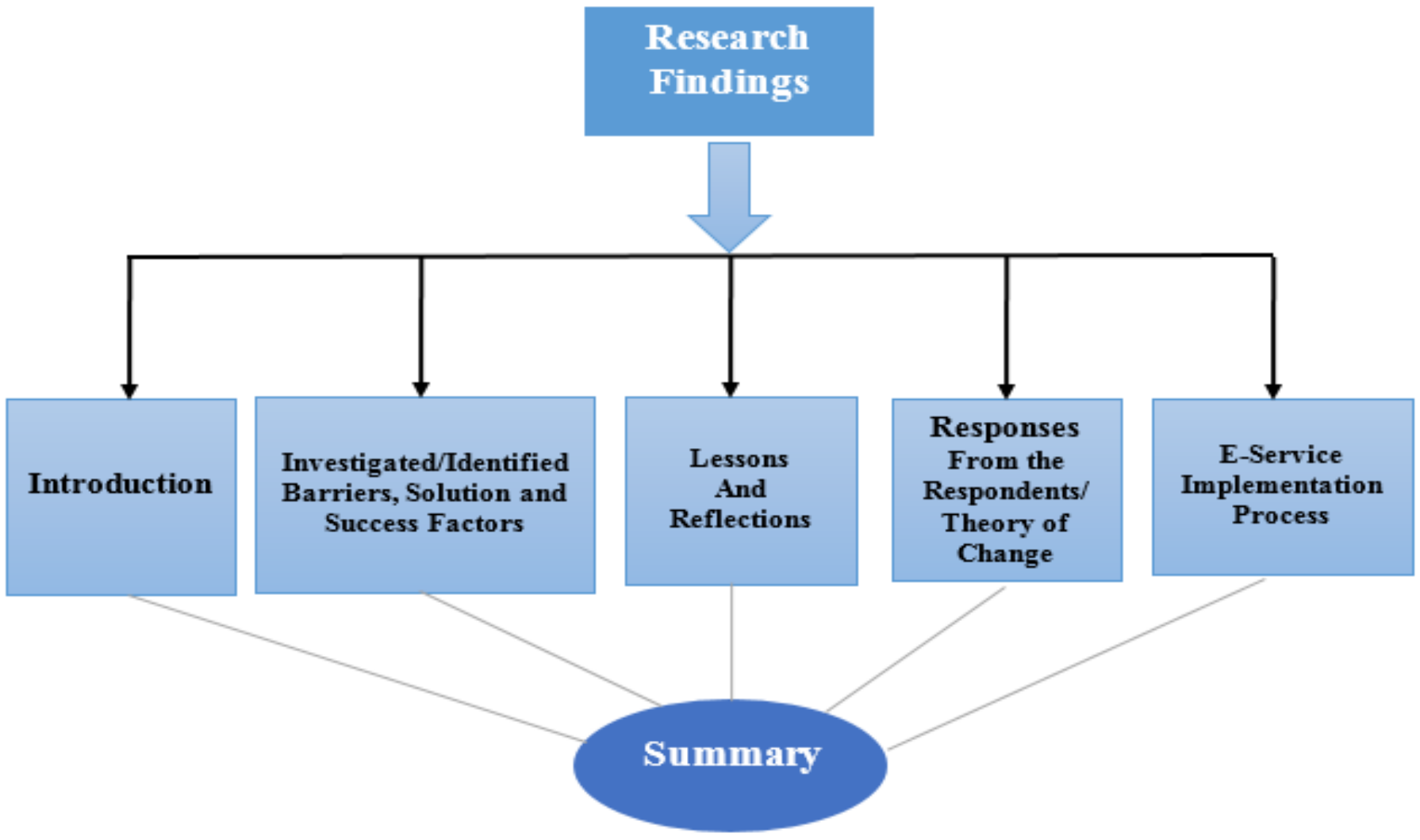

**Figure 52 Chapter Six Summary**

## 6.2 Investigated/Identified Barriers, Solutions, and Success Factors Facing E-Government Services Adopting and Implementation

Reviewing the literature, as explained in Chapter Two, has shed more light on the barriers that impede E-Service development, the solutions to these barriers and the success factors of E-Services adopting and implementation. Barriers have been classified in the E-Government domain on adoption from the diffusion perspective (Hassan, 2011). These barriers have also been classified as Technical Infrastructure, lack of Awareness, Privacy

Issues, and Culture (Alateyah et al., 2013). Nevertheless, barriers can be grouped into major categories such as **Administrative/Organisational**, **Economic**, **Political**, **Legislative, Technological,** and **Cultural Barriers** (Hassan, 2011).

### 6.2.1 Investigated/Identified Barriers

There are various barriers facing the adopting and implementation of E-Service as investigated in this study. The following are the investigated barriers according to the following categorisation, namely: **Administrative/Organisational**, **Economic**, **Political**, **Legislative, Technological**, and **Cultural Barriers** (Alshehri and Drew (2010); Munda and Musa (2010); Hassan (2011); Alshehri et al. (2012); Khan et al. (2012); Alateyah et al. (2013); Al-Shboul et al. (2014); Abdelkader (2015).

1. **Administrative/Organisational Barrier**: The corruption in government, especially in the developing countries, has prevented the management of government agencies from introducing E-Service (Oseni and Dingley, 2014). Other government agencies have reluctantly refused to change to the automated method, their main reason being the cost involved in the deployment of E-Services to the public (Hassan, 2011; Abu-Shanab and Bataineh, 2014). They believe that, if the project is not big enough and has a small target audience, it will be difficult to justify the cost involved in the E-Service provision to the citizens (Hassan, 2011). The leadership aspect of the organisational barrier poses another threat to the successful deployment of the E-Service initiative as there is not enough sufficient skill and communication (Nkohkwo and Islam, 2013). There is a need for effective communication to promote the advantages of E-Services for the public, thereby creating an expected awareness among the citizens. Lack of transparency in management, decision-making and administration (Nurdin et al., 2011) has been a major administrative barrier to the E-Service development, as the goals to implement E-Service by some organisations are unclear. Some barriers under the **Administrative/Organisational barriers** category are:
   - **Corruption/Lack of Transparency**
   - **Low Commitment from both Government and Employees**
   - **Lack of Partnership and Collaboration**
   - **Lack of Participation**
   - **Unclear Mission and Vision for E-Service initiatives**

2. **Economic Barrier**: The high cost involved in ICT equipment procurement, including the setting up and maintenance, has considerable effects on the E-Service development (Oseni and Dingley, 2014). E-Government initiatives are often hindered by finance and a shortage of resources (Hassan, 2011). Unavailability of funds could affect the full implementation of E-Service initiatives, as financial security helps the E-Service project growth and development. Unfortunately, due to a high level of corruption in developing countries and among top managements, the funds allocated for many projects cannot be utilised as expected, and this hinders the successful implementation (Oseni and Dingley, 2014). In a related development**,** a more comprehensive study by Bwalya and Zulu (2012) examined the economic barriers in E-Government development. They concluded that, if citizens are unable to afford a basic access to ICT, then this is a huge setback for E-Government strategic plans for the public, as this hinders the successful implementation of access to ICT. Economic barriers in E-Government service development are:

   - **Lack of Funds**
   - **High Cost of ICT Infrastructures**
   - **High Service User Cost**
   - **High Technology Set-up Cost**

3. **Political Barrier**: This research identified the political barrier as the main barrier facing E-Service adoption in Nigeria. E-Government empowers the citizens to be involved in the political process and contributes to the government decision-making that affects those citizens' lives. It also makes use of ICT to implement reforms to bridge the digital divide and promote transparency (Angelopoulos et al., 2010). Unfortunately, there are many concerns about the unwillingness and inability of political leaders to make a decision that affects E-Government development due to lack of political will and support, over-ambitious agendas, and lack of vision and strategy (Hassan, 2011). There is no doubt that funding and implementing E-Government initiatives will need strong political support and awareness, as these are the foundation of successful E-Services projects (Hassan, 2011). Moreover, E-Government development is being hindered on a daily basis due to political crises, especially in developing countries where democracy is still growing and is on trial (Oseni and Dingley, 2014). For example, there have been politically motivated crises in some countries in Africa that are linked to a particular terrorist group. The issue of "Boko Haram" (who state that Western education is forbidden) in the northern part of Nigeria is worrisome, as many citizens are being killed. This has brought ICT infrastructures down in the area and interrupted the political will (Oseni and Dingley, 2014). According to the authors, the now forbidden Western education equips the average citizens with knowledge and a familiarity with E-Service usage. The continuation of the political crisis will prevent future involvement in information sharing and decision-making, and this is a significant threat to the E-Service development. The political crisis is also a threat to innovation and advancement of that area of the country, which will quickly become non-progressive. Above all, this research has brought to light the significance of political barriers in the E-Services implementation in Nigeria.

4. **Legislative Barrier**: In order to have a successful implementation of E-Government initiatives and projects, it requires the government to put in place an appropriate regulatory environment (Bwalya and Zulu, 2012). These are more related to the security measures, suitable laws and regulations that will protect personal data from misuse by using E-Government services by the public (Hassan, 2011). Also, most security breaches during online services arise due to the lack of a suitable legal framework that involves electronic documents submission (Abdelkader, 2015). The transition to electronic services is threatened as a result of legal issues involving physical appearances, audits, and examinations. An electronic signature is accepted in some countries but not in other countries. The non-acceptability of electronic signatures will hinder E-Service development (Hassan, 2011; Abdelkader, 2015) and globalisation. There is a need for a legal framework for E-Services implementation to prevent issues with electronic document integrity and proof-of-identity. According to Bwalya and Zulu (2012), the following legislations are required in an electronic environment:

- **Electronic Crime Act**
- **Protection of Intellectual Property**
- **Issuance of Electronic Commerce Act**
- **Issuance of E-Signature Law and**
- **Electronic Authentication and Digital Certification Law**

5. **Technological Barrier**: The major bottleneck, as seen in the literature, to the full implementation of E-Services is technology (Hassan, 2011; Abdelkader, 2015). The use of technology in E-Government development includes the availability of networks and servers, which enable government agencies to interact, share information and co-operate together in performing their daily work (El-Sofany et al., 2012). The internet has become an essential tool in E-Government and is used by the authorities to facilitate communication and interaction between the government, businesses, and the citizens. According to Rakhmanov (2009), technological barriers are related to the non-availability of necessary and suitable tools, and infrastructures to develop and deploy electronic services. Other technological challenges are:

   - **Unstable Power Supply**
   - **Lack of Standard for Quality**
   - **Poor infrastructures**
   - **Poor Internet Connections**
   - **Poor Authentication Systems**

6. **Cultural Barrier**: E-Government development is challenged by cultural issues, especially in a country with many groups of ethnic minorities or a huge number of immigrants (Hassan, 2011). Overcoming this barrier in E-Government implementation requires a deep understanding among stakeholders (El-Sofany et al., 2012). This barrier is linked to social or organisational cultures where beliefs, norms, attitudes and behaviours have been passed on to individuals by members of their social environment and which influence the E-Government services implementation (Hassan, 2011).

   Cultural issues have a substantial impact on any community or organisation, and the case of E-Government initiatives is not an exception (Ali et al., 2009). For example, the practice of isolating women from men in Saudi Arabia is derived from the impact of culture (Al-Sowayegh, 2012), especially in regard to females working in the labour market. E-Government initiatives adopted here might suffer a setback, as the goals to transform them into electronic public services will be jeopardised if a section of society is marginalised through isolation, especially if that society does not have the required awareness and access to basic information about E-Government services deployment (Al-Sowayegh, 2012).

### 6.2.2 Investigated/Identified Solutions

As barriers delay the E-Services adopting, implementation, and progress, other elements will serve as facilitators or enablers to motivate the initiatives. As discussed in the literature, implementing the solutions encourages and motivates the E-Service development. Some solutions to the barriers facing E-Service adoption and implementation, which are the motivating influence to successful E-Government initiatives, have been identified. The most important solutions identified in the literature are as follows:

1. **Leadership/Clear Strategic Plans for E-Government**: It is necessary to consider leadership support and a clear vision for E-Government services deployment. Leadership is needed at every level of the government towards successful E-Service implementation and it should ignore corruption at every stage of the project (Oseni and Dingley, 2014). Effective leadership will play a significant role in E-Service initiatives adoption and implementation (El-Sofany et al., 2012).

2. **Funding/Anti-Corruption Measures**: This is very important in achieving the aspiration of any leader towards E-Government services deployment. More money should be allocated to E-Government services development in the country's yearly budget, as presented to the legislative arm of government by the executive arm (Oseni and Dingley, 2014). It is advisable that a committee should be set up by government agencies and ministries to map out a strategy to ensure full implementation and allocation of the funds that are meant for the E-Service projects. These measures will minimise corruption and bring accountability at every level of local government in Nigeria. The availability of sufficient funding is important for government drives towards the implementation of full E-Services for the stakeholders and the citizens (Hassan, 2011).

3. **Provision of Appropriate Legislation and Policy for E-Government**: Appropriate legislation and laws are needed to guide E-Government initiatives, as personal identities of the users need to be protected against theft (Bwalya and Zulu, 2012). For E-Government Services to be fully adopted, users should be able to trust the system. As users will be using their information online, there must be adequate security in place to protect their data and privacy (Oseni and Dingley, 2014), and to ensure that their choices are recorded accurately and not manipulated. As mentioned in the previous section, there is a need for a legal framework for E-Services implementation to prevent issues with electronic document integrity, corruption, and proof of identity (Bwalya and Zulu, 2012).

4. **Provision of Affordable ICT Infrastructures**: The availability of cheaper and accessible ICT infrastructures plays a significant role in E-Government services development. If the cost of the equipment is reduced, it will be easier for the government to acquire it. Hence, there is a need for collaboration between the government, private sectors, insurance and ICT companies in manufacturing affordable equipment. ICT infrastructures are under the technological barrier in E-Government implementation, and this is seen as a major bottleneck in the implementation of E-Services (Hassan, 2011; Abdelkader, 2015).

5. **Improve on Security, Trust, and Privacy**: When using the E-Services by public means, there must be some means of keeping data secure. Citizens want to be assured that their personal information is safe secure using online government services (El-Sofany et al., 2012). This element is a major concern for the successful adoption of E-Government services by the citizens. As much as we are aware of cultural influence on E-Services adoption (Ali et al., 2009), a community where trust and data security is a problem when using these services might discourage their members from adopting the services. Suitable security measures, laws, and regulations that will protect the personal data of the citizens from misuse should be initiated by the government (Hassan, 2011).

### 6.2.3 Investigated/Identified Success Factors

The significant challenge for this study is to identify the barriers facing E-Service adopting and implementation at local government level. The literature identified these barriers and their solutions; it is equally important to establish what the success factors are for the E-Service adopting and implementation. Ziemba et al. (2013) argued that the critical success factors (CSFs) theory laid the foundation and basis on what the E-Service adopters would follow during the adopting process. Hence, the following four important success factors for the E-Service adopting and implementation are identified:

1. **Bridging the digital divide/Network and Community Creation**: Bridging the digital divide is one of the success factors of E-Service adopting and implementation. E-Service development has helped in bridging the digital divide gap and narrowed the disparity between the widespread internet access in developed and developing nations (Helton, 2012). E-Service adoption has accelerated development, as internet access is now available for those countries with limited resources. Saheb (2005) claimed that, in the global world of today, access to the new technology would equip citizens with the vital knowledge needed for the input into the productive measures of the developing countries. Hence, the access to E-Services through the use of ICTs is crucial to a sustainable agenda of socio-economic development (Saheb, 2005). Above all, networking between the government, businesses, and the citizens through the adoption of E-Services and sharing of information will help in bridging the digital divide and facilitate robust community creation.

2. **Transparency, Anti-corruption, and Accountability**: Other success factors of E-Service adopting and implementation, as obtained from the literature, are transparency, anti-corruption, and accountability. The use of ICTs by the government is seen to be more cost-effective and convenient (Bertot et. al., 2010). It helps to promote openness and transparency in the process, thus reducing corruption. The anti-corruption drive through E-Service adoption means that the necessary government procurement documents will be available and accessible online. There are three approaches to the anti-corruption drive. According to Bertot et al. (2010), the first approach is the administrative reform, where government bureaucracies are enhanced in ensuring the effective monitoring of government officials' behaviours in E-Service implementation. Secondly, there is a need for a legal framework to curb corruption in government, while appropriate reform needs to be in place to punish corrupt officials. Lastly, Bertot et al. (2010) concluded that the third approach towards the anti-corruption drive and transparency in government is social change. This social change means that citizens are empowered to participate in institutional reforms by the cultivation of a civil and law-based society, which checks corruption and supports accountability in the system.

3. **ICT Infrastructure Development**: The literature has identified the success factors for adopting E-Government, including ICT infrastructure development. Hence, there are different categories of the success factors, as explained by researchers. For example, Ziemba et al. (2013) categorised the success factors for adopting E-Government services in Poland as Economic, Socio-cultural, Technological, and Organisational. However, the ICT infrastructure development is one of the factors considered under the technological factor. ICT infrastructure development is seen as a critical element that determines the E-Government development and success (Ziemba et. al., 2013). Adoption of E-Services in many ways is dependent on the development and availability of innovative ICTs, such as excellent broadband service, next generation access (NGA) network, mobile devices and government process management systems (Ziemba et al., 2013).

4. **The Quality of Service Delivery:** The issue of the service delivery quality has been the focus of attention in the IT domain for both the practitioners and the researchers (Saha, 2008). Service quality means that the customer's expectation of services is measurable against the service provider's performance. Service delivery quality is an important success factor in E-Service adopting. However, easy delivery of online government services depends on the availability of reliable, accessible, and excellent ICT infrastructures.

## 6.3 Lessons and Reflections

As suggested by Leitch and Day (2000), reflection is considered as a process or activity that is significant to the development of practices. This is connected to what is learnt, and it is adjudged to be linked to cognitive processes of finding and solving problems (Leitch and Day, 2000). Hence, this section will evaluate the results of this study by looking at each phase of the action research method. In doing this, the section will present the reflections from the actions that took place during the study and the recommendations from this researcher to improve the practice and E-Service adopting and implementation at the local environment level in Nigeria.

### 6.3.1 Action Research Findings – Phase One

Phase one of the action research method, as shown in **Table 28** in the previous chapter, is the ***Diagnosis*** phase. In the diagnosis phase, the researcher's achievement was to identify and examine the study background and problems, the aims, objectives, research questions and gaps in the previous research works in E-Government services adopting and implementation at the local environment level in Nigeria. The researcher reviewed literature and, in conjunction with the supervision team, agreed on the research topic. Although the domain chosen as the circumstances of interest was familiar with the researcher, the E-Government interest aids the high level of meaningful communications with the participants.

### 6.3.2 Action Research Findings – Phase Two

This action research phase is as the ***Action planning*** phase. Action planning is needed to clarify the resources necessary to reach the research goals. This researcher achieved major tasks in this phase, which include a further review of the literature, selecting a research methodology and justification, and development of the initial framework. It also includes the selection of the participants, giving prior notice and making necessary arrangement for meetings; formulation of a timeline for when specific research actions need to be completed, and determining what resources are required to complete these actions. Above all, the principles guiding this research work were accepted in advance by the participants before the interviews.

### 6.3.3 Action Research Findings – Phase Three

Phase three in the action research method is the ***Action taking*** phase. The achievements in this phase include data collection via interviews (face-to-face questionnaires). The decision reached with the participants is that this researcher should also use an online survey as a complementary investigation to the interviews for collecting further independent responses on the barriers facing E-Service at the local government level, and will cover similar questions. Furthermore, various government documents on administration were reviewed and observed. The additional decision was agreed on refining the initial framework to welcome contributions from participants. The decision was also reached between this researcher and the participants to further use a 5-member online focus group among the 30 participants interviewed for subsequent meetings to complete the validation processes and for feedback deliberations. However, the use of the TAM as the theory for this research was rejected by the interviewees, and this prompted the need to go through another research cycle, as shown in **Figure 53** below, to unveil more appropriate theory to support this study.

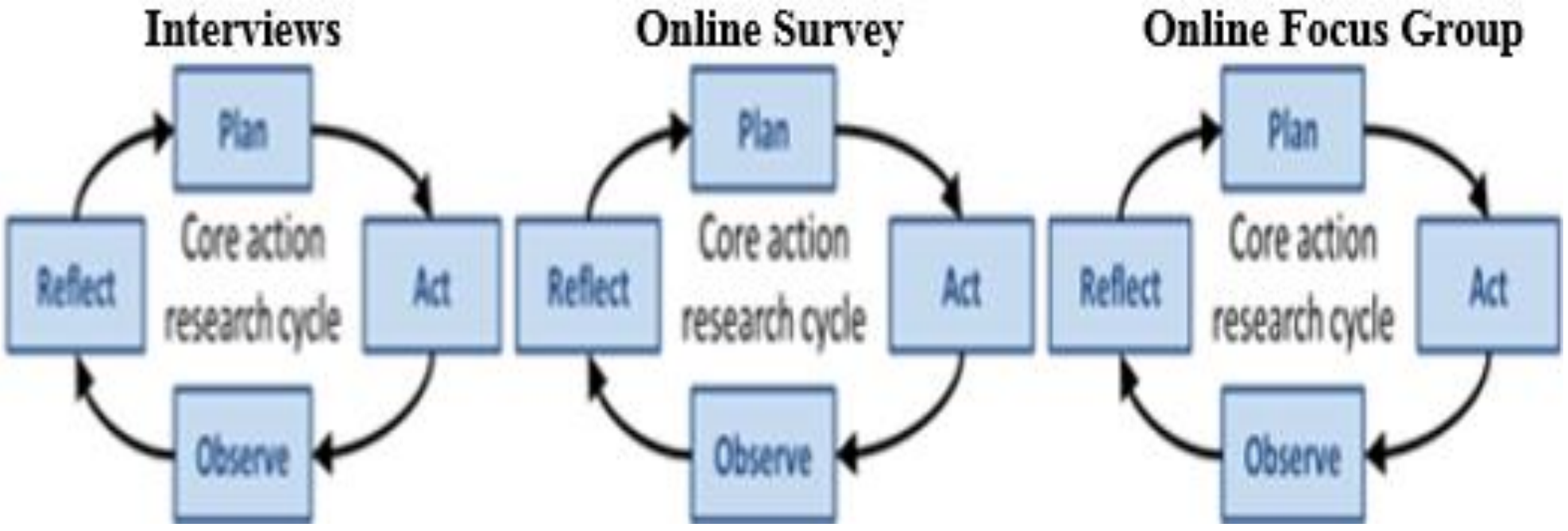

**Figure 53 Cycles for this Study Action Research**

The decision to use the TAM was rejected, as this research theory was initially seen as a setback. However, going forward, the action research model allows new theories to be brought into the research, and many researchers have tested the newly adopted DOI theory through the use of quantitative research methods (Surendra, 2001). Surendra (2001) used the diffusion factors as proposed by Rogers (1995) to predict the web technology acceptance by scholars and administrators of a college, and found that the diffusion factors in Rogers' DOI theory were very useful in the prediction of technology and innovation adoption.

### 6.3.4 Action Research Findings – Phase Four

This action research phase is the ***Evaluation*** phase. The achievement in this phase by this researcher is to evaluate and analyse the data collected during the previous phase, namely the **action taking phase**. In this phase, the validation and coding processes, the emergence of new knowledge and research framework testing were all conducted. Software packages such as SPSS and Nvivo were used as the tools to assist this researcher in the data analysis.

The new understandings from the findings of this study on the **Barriers facing the E-Service Adoption and Implementation** at the local environment level in Nigeria are somehow different to what many previous studies suggested, although they had few barriers in common, as itemised and explained in section 6.2.1 above. One interesting finding is the local government **Autonomy Issues** being a barrier facing the E-Service adopting and implementation at the local environment level in Nigeria. **Autonomy** is defined as the right or condition of self-government, independence, and freedom from external control or influence. In Nigeria, the local government's power to make an independent and bold decision on E-Government deployment may run into difficulties as a result of the State governments due to State-Local Joint Account issues (Okafor, 2010). This has made the local governments in Nigeria rely on external funding (Nabafu and Maiga, 2012) to implement major projects. According to Okafor (2010), there have been repeated demands for the local governments' financial autonomy from the states. The local governments are unable to carry out major projects as the disbursement of funds from the federal accounts goes into the state government's account before the state releases a percentage to the local governments. Nonetheless, the State-Local Governments Joint Account under Nigeria Constitution (1999), Section 162, Schedule 6, empowers the state to maintain a joint account with local governments where fund allocations to the local government councils in Nigeria from the Federal account are paid directly (Okafor, 2010).

This is a major setback to the local governments as the state slashed large amounts from the local government funds, and the small amount of money released to them by the state government can only cater for the payment of staff salaries (Okafor, 2010). A project such as E-Service initiatives has been a big problem. The only solution to this issue is for the local governments in Nigeria to be independent of the state governments. The monthly funds for the local governments from the federal government in Nigeria should go directly into the local government account and not through the state-local government's joint account. For this to be effective, a change in the Nigeria 1999 Constitution (Section 162, Schedule 6) will be required, but this can only be achieved by a majority of two-thirds of the member of the Nigeria Senate. As shown in **Figure 54** and **Table 29** below, from the SPSS data analysis, 83 **(19.5%**) of the total 425 online survey respondents agreed that the local government autonomy issue is a major barrier facing the E-Service adoption and implementation. The respondents believed that having financial independence will aid development and good initiatives such as E-Government services at the local level in Nigeria.

Another significant finding on the barriers facing the E-Service adopting and implementation at the local environment level in Nigeria from this study is the ***Corruption issue***. This result is, however, not surprising considering the level of corruption within the government officials in Nigeria. The results obtained from the Transparency International website show that Nigeria was ranked 136 out of 167 of the most corrupt countries in the 2015 ranking (Ugaz, 2015). The level of impunity is high in the country, as the money allocated for various projects is diverted into private pockets with the help not only from the government officials, but also the banking and judicial officials, who take their lion shares (Okafor, 2010). Ugaz (2015) suggested that corruption could be stamped out if the government officials stop the abuse of power and bribery, and there is a need for the officials to operate a transparent governance and make an attitudinal change towards corruption (see more in **section 6.4.2**, the theory of change). However, since the current government came to power in May 2015, there has been an improved attitude among the government officials towards stealing, as the elected president declared zero tolerance in regards to corruption.

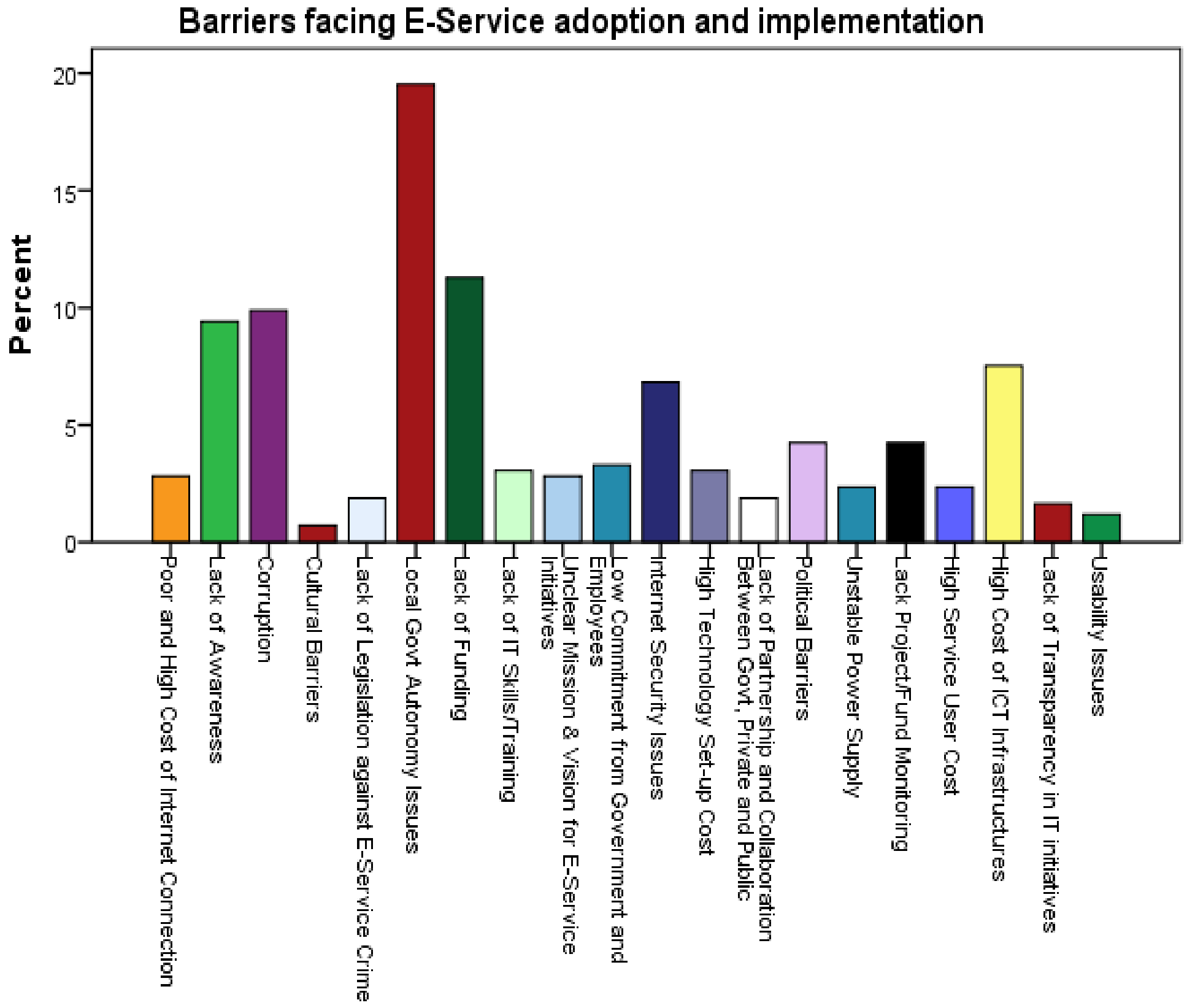

**Figure 54 Barriers facing E-Service Adopting and Implementation (Online Survey)**

The result obtained from this study, and as shown in **Figure 54** and **Table 29**, revealed that 42 (**9.9%**) respondents agreed that corruption is one of the barriers facing the E-Service adopting and implementation at the local environment level in Nigeria. From this study, corruption is ranked third behind the ***local government autonomy*** and ***funding*** issues. More findings are available in **section 6.4**.

Table 29 Barriers facing E-Service Adopting and Implementation (Online Survey)

| | | Frequency | Percent | Valid Percent | Cumulative Percent |
|---|---|---|---|---|---|
| Valid | Poor and High Cost of Internet Connection | 12 | 2.8 | 2.8 | 2.8 |
| | Lack of Awareness | 40 | 9.4 | 9.4 | 12.2 |
| | Corruption | 42 | 9.9 | 9.9 | 22.1 |
| | Cultural Barriers | 3 | .7 | .7 | 22.8 |
| | Lack of Legislation against E-Service Crime | 8 | 1.9 | 1.9 | 24.7 |
| | Local Government Autonomy Issues | 83 | 19.5 | 19.5 | 44.2 |
| | Lack of Funding | 48 | 11.3 | 11.3 | 55.5 |
| | Lack of IT Skills/Training | 13 | 3.1 | 3.1 | 58.6 |
| | Unclear Mission & Vision for E-Service Initiatives | 12 | 2.8 | 2.8 | 61.4 |
| | Low Commitment from Government and Employees | 14 | 3.3 | 3.3 | 64.7 |
| | Internet Security Issues | 29 | 6.8 | 6.8 | 71.5 |
| | High Technology Set-up Cost | 13 | 3.1 | 3.1 | 74.6 |
| | Lack of Partnership and Collaboration Between Government, Private and Public | 8 | 1.9 | 1.9 | 76.5 |
| | Political Barriers | 18 | 4.2 | 4.2 | 80.7 |
| | Unstable Power Supply | 10 | 2.4 | 2.4 | 83.1 |
| | Lack of Project/Fund Monitoring | 18 | 4.2 | 4.2 | 87.3 |
| | High Service User Cost | 10 | 2.4 | 2.4 | 89.6 |
| | High Cost of ICT Infrastructures | 32 | 7.5 | 7.5 | 97.2 |
| | Lack of Transparency in IT initiatives | 7 | 1.6 | 1.6 | 98.8 |
| | Usability Issues | 5 | 1.2 | 1.2 | 100.0 |
| | **Total** | **425** | **100.0** | **100.0** | |

### 6.3.5 Action Research Findings – Phase Five

The last phase of the action research method is ***Specifying Learning***. This researcher is expected to report the findings that will inform the practice and stakeholders, with information about what can be done differently as a result of this study.

## 6.4 Respondents' Responses from the Online Survey and Theory of Change

This section will further review the findings from this study as explained in **section 6.3.4** above (the lessons and reflection section) and the theory of change. Action research typically chooses a theory on which to hang on the research results. The use of the theory of change model will help further in analysing the corruption issue, which is a major barrier that emerged from this study. This researcher believed that the reduction or total removal of corruption at the local environment level in Nigeria would have a positive effect on the full E-Service projects implementation. Hence, the research findings and the responses from the participants and the respondents are divided into three categories, namely **Responses from the Online Survey, Responses from the Interviews**, and **Responses from the Online Focus Group**. The next **section 6.4.1** discusses the respondents' responses from the online survey, while the other results from the interviews and online Focus group are available in Appendix B.

### 6.4.1 Respondents Responses from the Online Survey

The use of the online survey suggested by the action research method allows for multiple data collection methods. The online survey results will complement the results already obtained during the interviews with the participants. The use of the online survey was discussed in both Chapter Five (**section 5.3.2**) and Chapter Six (**section 6.3.3**), respectively. The survey was sent out to the respondents by the researcher, friends, and colleague networks. The results from the respondents who participated in the online survey will help the researcher to learn of the age group and gender of the population, as well as the geographical regions in Nigeria. Other results will show the state of origin, educational background, occupation, E-Government services knowledge and usage, barriers facing E-Service adopting and implementation, knowledge about the diffusion of innovation, willingness to adopt E-Government services, and the solutions and success factors for adopting and implementation of E-Government services. Hence, asking the respondents these questions will enable this researcher to find answers to the research questions and to help the policy makers at the local government level in Nigeria to know the barriers, solutions and the success factors for E-Government services adopting and implementation at the local environment level. IBM SPSS Statistics version 22 is used for this data analysis.

### 6.4.2 Demographic features in the Survey

This research considered the demographic responses, especially gender, from the online survey, to show that the study's result is bridging the internet gap between women and men. In a report tagged "Women and the Web" by Intel Corporation in 2012, it was found that 23 per cent fewer women than men had internet access across the developing countries (Kakar et al., 2012). The report also indicated that women are at risk of being left behind as they are 14 per cent less likely to own a mobile phone than men. According to Kakar et al. (2012), the internet is considered to be a mainstream tool in an increased globalised world and lack of access or awareness might discriminate women from an increasing number of opportunities.

However, the results obtained from this study correlate with the report from the Intel Corporation as it shows more women are now aware of internet usage. From the total 425 responses, 237 respondents, representing 55.8%, were female compared to 188 male, representing 44.2%. By comparison, the result shows that an increasing number of women are involved in the use of internet and this is, to some extent, bridging the internet gap between the genders. According to Melinda Gates of the Bill Gates Foundation, as reported by Kakar et al. (2012), many women who are willing to gain access to financial services and break the cycles of poverty will have a high success rate with access to internet-equipped facilities (Kakar et al., 2012). Other benefits of women using the internet include a potential income boost, connectivity to the national and global news, increased sense of empowerment, increased sense of equity, improved self-esteemed, expression, and participation in the E-Government decision-making process (Kakar et al., 2012). However, these very advantages may cause some men to prevent the women that they are related to from participating in activities that will further emancipate them.

## 1. Age Group

The demographic questions, such as age group, in the online survey will help this researcher to know which respondent is completing the survey. Although there is no specific respondent target for this research, demographic questions will help in determining the factors that influence responses from the respondents. As shown in **Figure 55** below, 64 respondents within the age group 16-25 representing 15.1% of the total participated in the online survey. The results that emerged from the analysis are from respondents between the age groups 26-35 and 36-45, and a total of 146 respondents each from both categories representing 34.4% respectively completed the online survey. This represents the highest respondents out of the total. The researcher believes that there is a correlation between the figures obtained from these age groups as obtained in (Whyte and Lamprecht, 2013). Whyte and Lamprecht (2013) looked into the categories of social network users and found that over 80% of the respondents between the ages of 26 and 45 are present on the social networks. Other results show 42 respondents within the age group 46-55, which represent 9.9% of the total, also completed the survey. Lastly, the age group of 56 and above, with only 27 respondents and representing 6.4%, took part in the online survey. The researcher suggests that the low figure from this age group could be attributed to many factors such as ***lack of participation*** **(E-Service adopting barrier)** in online activities due to old age, security and trust issues.

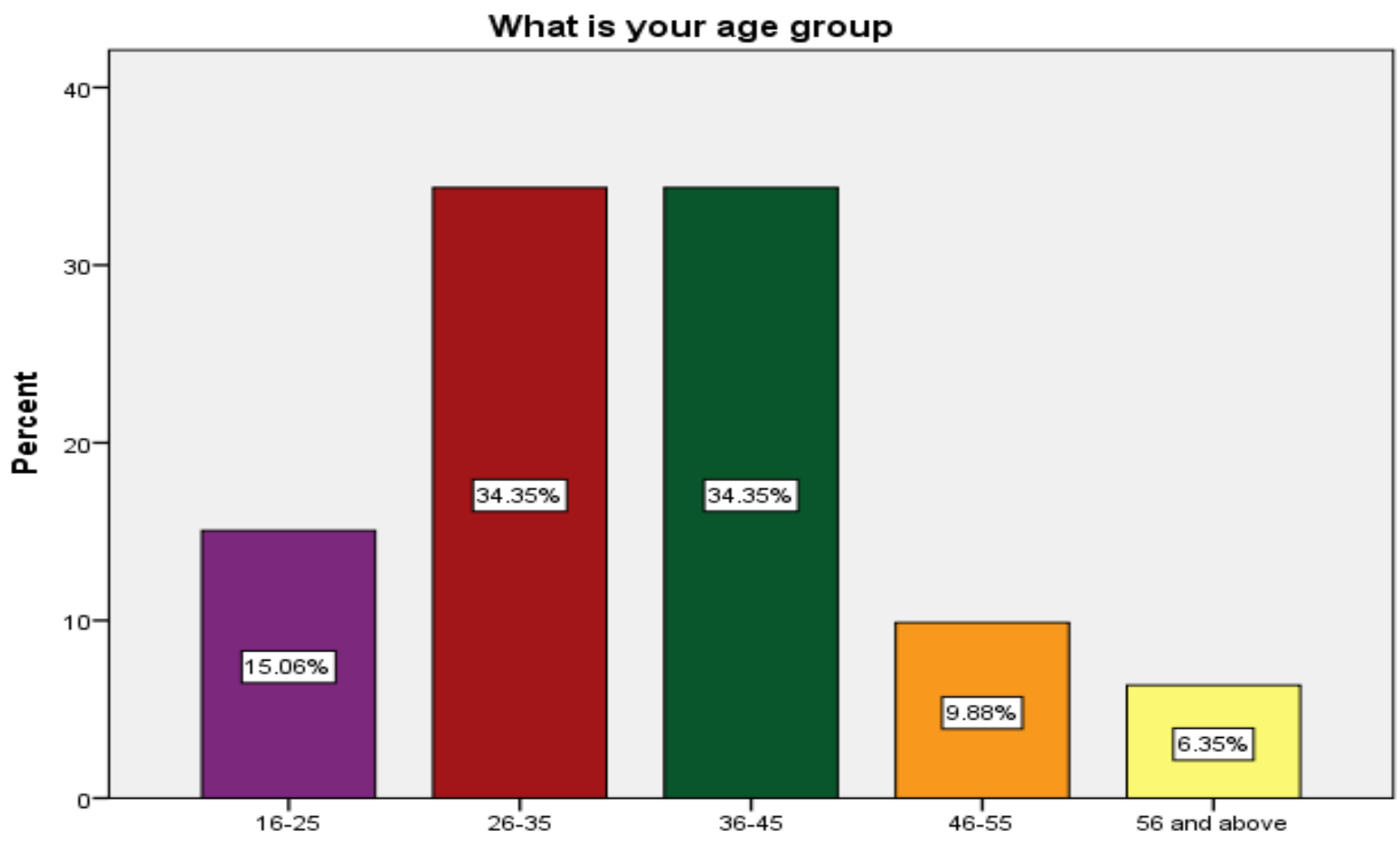

**Figure 55 SPSS Online Survey Analysis – Age Group**

**2. Gender**

This is an impressive result, as more females participated in the online survey than male counterparts. There is a similarity between this result and the ***Gender*** result obtained from the participants during the interviews, as explained in **section 6.4.1** above. In this analysis, and as shown in **Figure 56** below, 237 respondents representing 55.8%, who were female, took part, while 188 respondents representing 44.2%, who were male, participated in the online survey. In **section 6.4.1**, 53.3% of the total were females who took part in the interviews, and the remaining 46.7% were male participants. Interestingly, the differences in the two results are few. This researcher believes that many factors might have contributed to more females taking part in both the online survey and interviews.

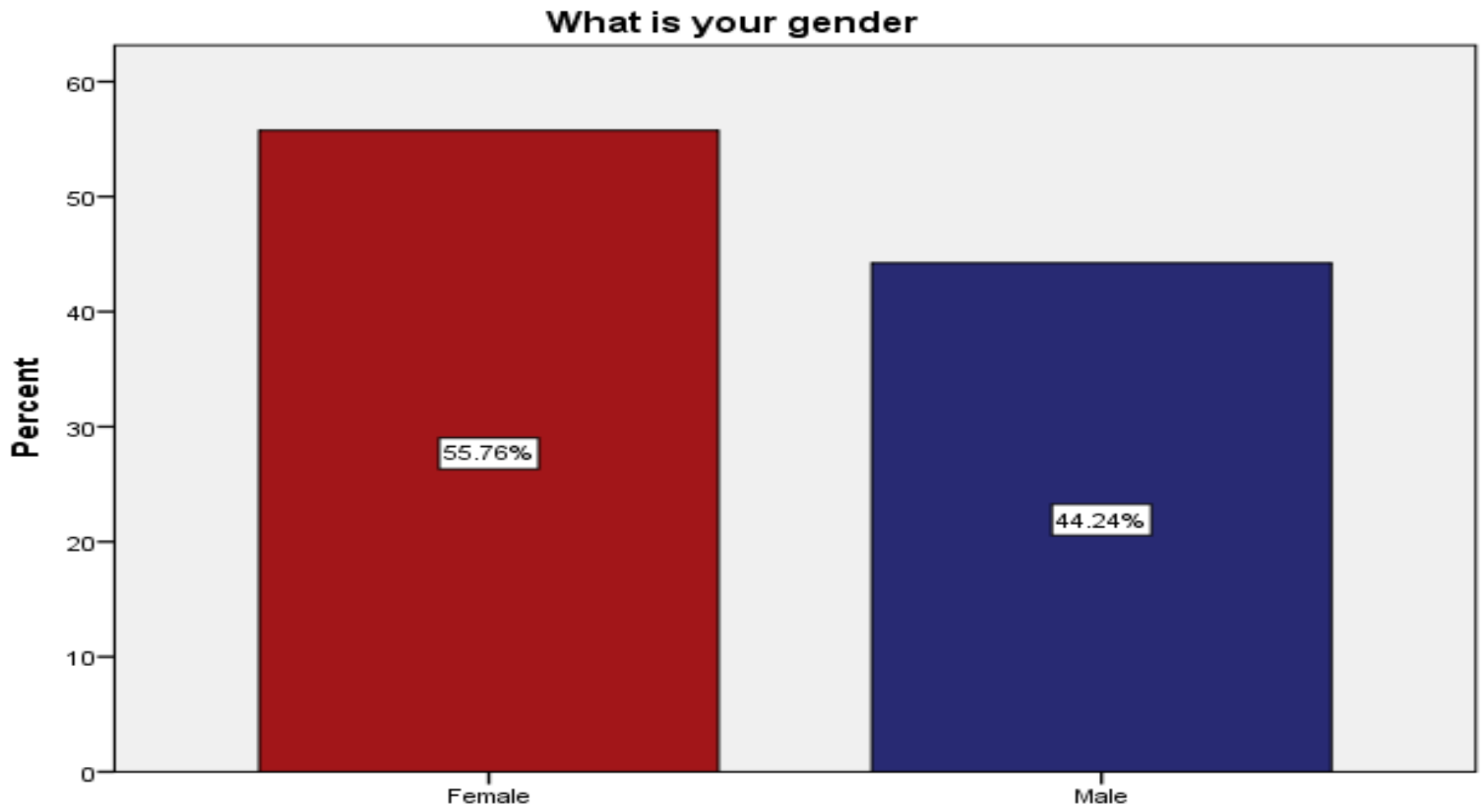

**Figure 56 SPSS Online Survey Analysis – Gender**

**3. Geographical Regions**

There is a need for further information on the geographical regions of the respondents. Nigeria is divided into six (6) geographical regions, namely Northcentral, Northeast, Northwest, Southeast, Southsouth, and Southwest (Oviasuyi and Uwadiae, 2010). Some 22 respondents, representing 5.2% of the total, were from the Northcentral, while 15 respondents from the Northeast took part in the survey, and this represents 3.5% of the total. Twenty-one respondents from the northwest of Nigeria, which represents 4.9%, participated in the survey, while another 37 respondents from Southeast, representing 8.7% of the total, took part in the online survey. Some 40 respondents from the Southsouth part of Nigeria also participated in the online survey, and this represents 9.4% of the total. The most amazing result in this analysis, as shown in **Figure 57**, revealed that 290 respondents from the southwest part of Nigeria participated in the online survey, representing 68.2% of the total.

There are many scenarios the researcher could infer from this result, as obtained from the online survey, regarding the respondents' geographical region. First and foremost, the low respondent responses from the northern regions (Northcentral, Northeast, and Northwest) of Nigeria might be connected to a few investigated barriers facing the E-Service adoption and implementation. For example, the barriers such as ***low literary level***, ***bad infrastructures,*** and ***lack of participation*** from the respondents in the region might be due to the current insurgency (Boko Haram Militancy). The Boko Haram had killed thousands of the citizens in the last six years, and millions of residents have been displaced (Aghedo and Osumah, 2014). Secondly, the large responses from the respondents in the southern regions (Southeast, Southsouth, and Southwest) of Nigeria, as shown in the results, especially from the southwest region, might be facilitated by many factors such as a high literacy level, availability of infrastructure, cheap internet connection, and citizens' readiness towards E-Governance.

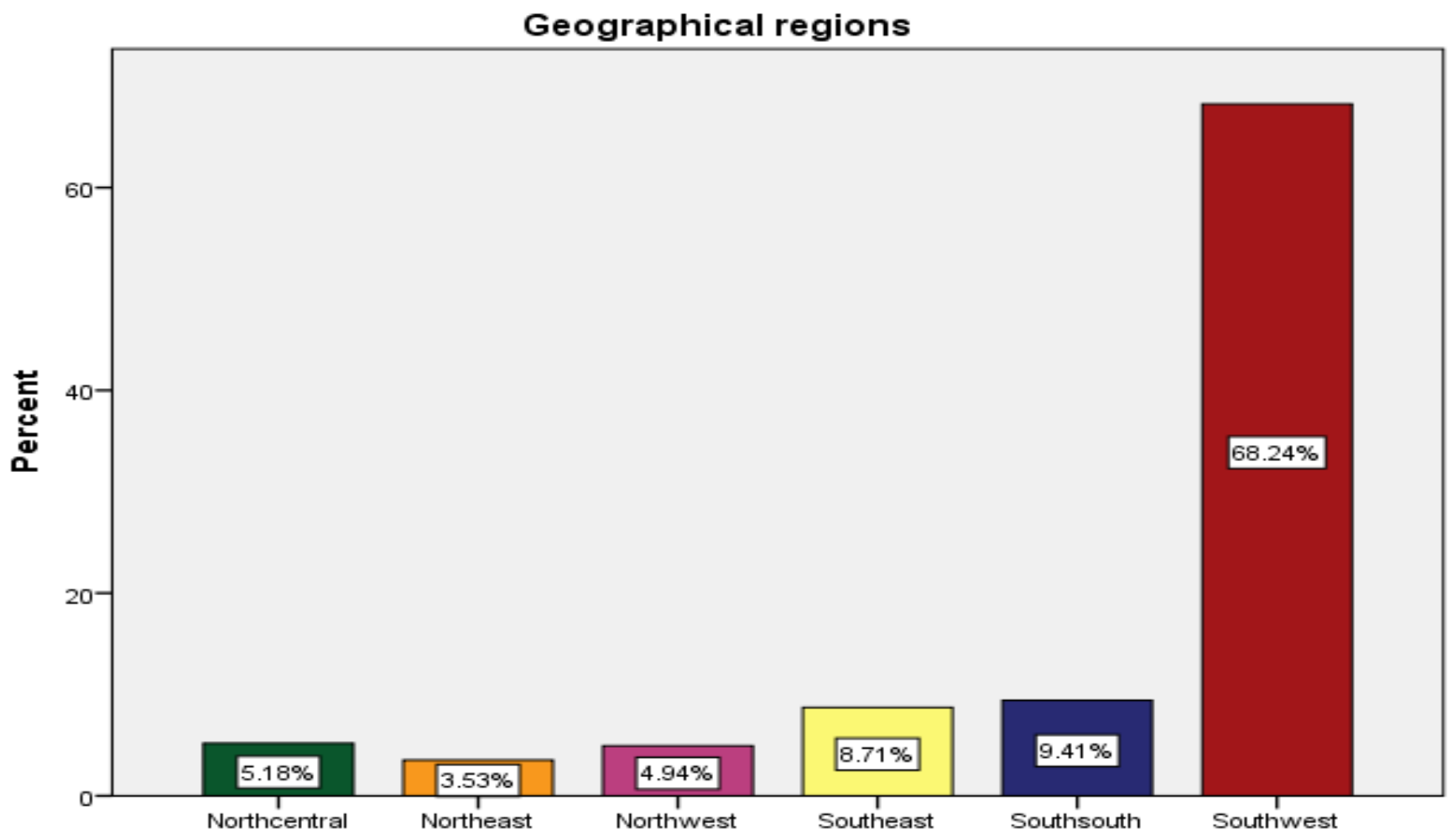

**Figure 57 SPSS Online Survey Analysis – Geographical Regions**

### 4. Education

The level of a respondent's responses might be influenced by their level of education, as respondents with a high level of education might understand better the research domain than other respondents with a lower level of education, and this may have influenced their participation in the survey. In the analysis shown in **Figure 58**, 19 respondents with a Ph.D. or equivalent representing 4.5% participated in the survey, in addition to 145 respondents at Masters level or equivalent, which represent 34.1%, completed the online survey. The analysis also shows that another 151 respondents with a Bachelor degree level or equivalent submitted the online survey, and this represents 35.5% of the total. Sixty-five respondents representing 15.3% are educated to Diploma level or equivalent, and 45 respondents, which represent 10.6% of the total, are educated to High School level or equivalent. Conclusively, having the two highest responses from respondents with Masters and Bachelors or the equivalent signal the respondents' readiness to adopt the E-Service at the local government level in Nigeria. It may be that those with higher levels of education have higher incomes, and can afford technology and infrastructure to support the use of the internet for work, games, and social activities. Whereas, for the respondents with lower qualifications, this may mean lower income, and less money and less time for leisure technology use.

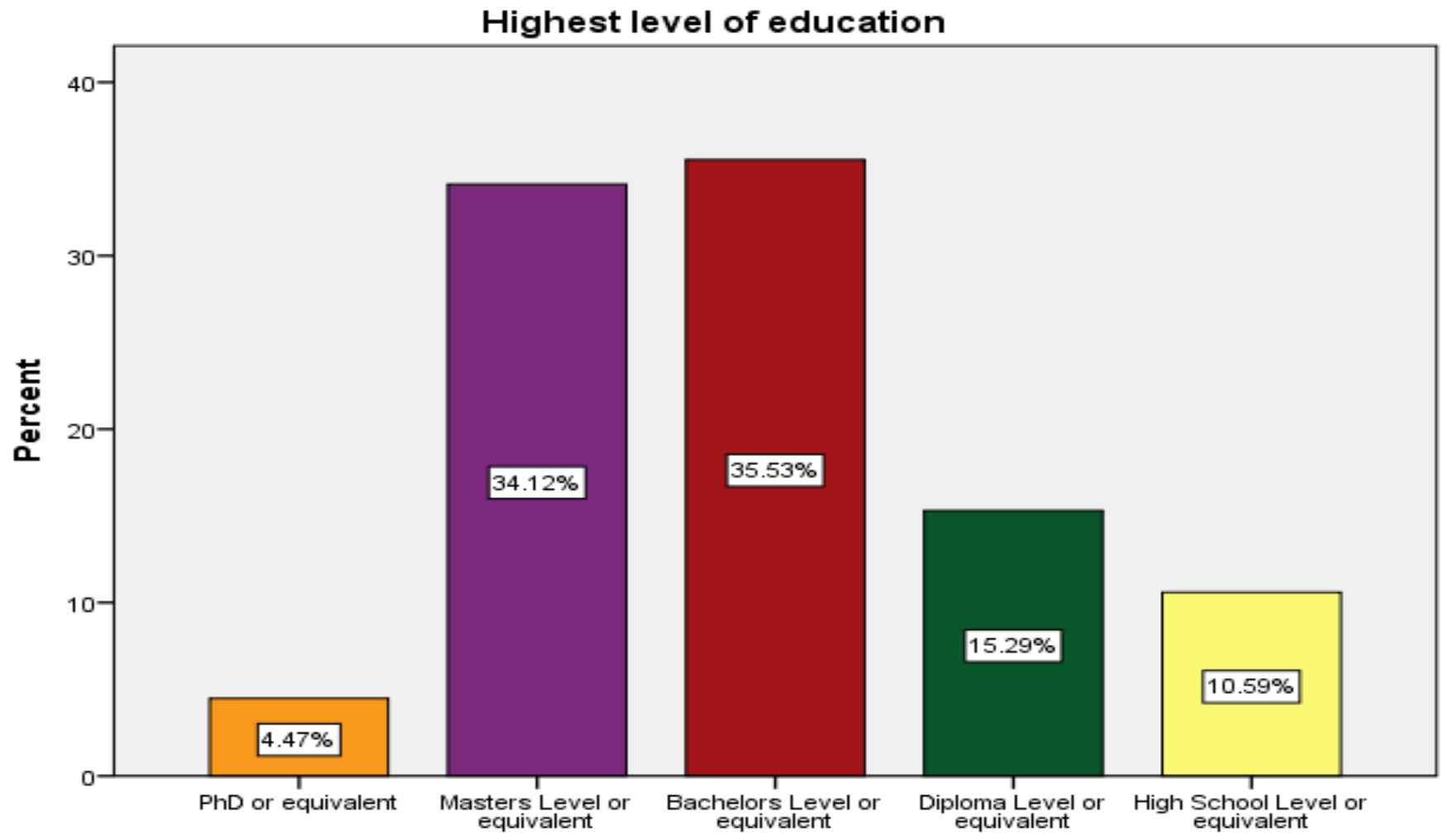

**Figure 58 SPSS Online Survey Analysis – Level of Education**

### 5. E-Government Services Usage

This study investigates the barriers facing E-Service adopting and implementation at the local environment level using Nigeria as a case example and, by asking respondents about their prior usage of the E-Government services, this will help the researcher to understand whether the respondents have opinions on the barriers. This could also confirm the importance of this study, especially in the local environment context in Nigeria. From the analysis shown in **Figure 59**, 181 respondents, representing 42.6% of the total, have used E-Government services, and 237 respondents representing 55.8% of the total have not used E-Government services before this survey was conducted. Another seven respondents, representing 1.6% of the total, confirmed that they are not interested in online developments.

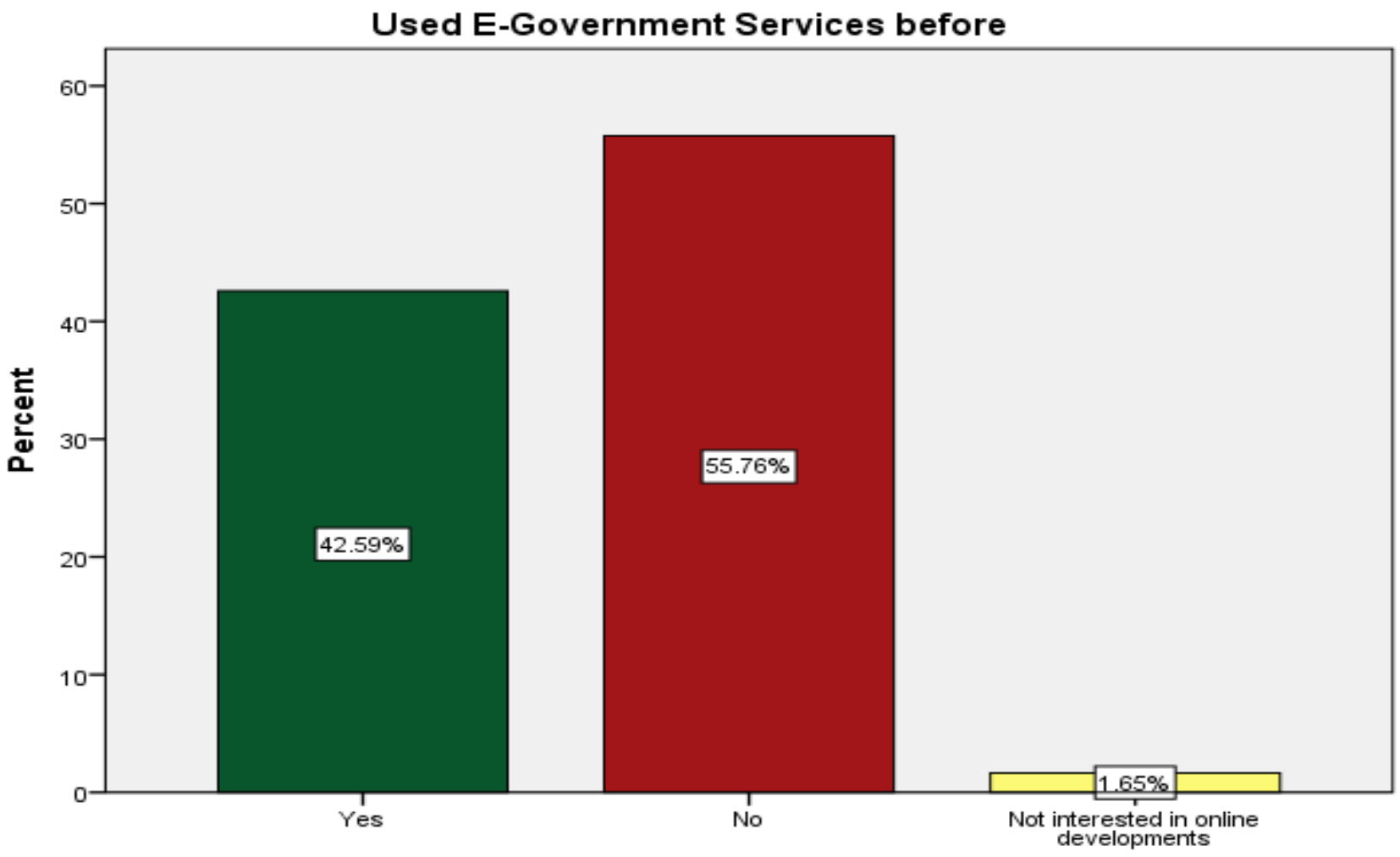

**Figure 59 SPSS Online Survey Analysis: E-Government Services Usage**

With the analysis above, and as obtained during the interviews with the participants, the majority of the interview participants and survey respondents have not used E-Government services before. These results, according to the researcher, were very positive to the topic as there are indeed barriers facing the adopting and implementation of the E-Government services at the local environment level in Nigeria. This might have contributed to the lower adoption rate from the citizens at the local level.

### 6. E-Government Services Adoption

Further to the results obtained in **Figure 59** on E-Government services usage, the respondents were further asked whether they will be ready to adopt and use E-Government service at the local government in Nigeria. Surprisingly, only a minority of respondents confirmed that they are not ready or not interested in the E-Services at the local government level in Nigeria. Some 380 respondents representing 89.4% of the total agreed that they are willing to adopt and use the E-Government services. This is a positive result to this study. Some 35 respondents, which represent 8.2% of the total, confirmed they are not ready for E-Service adoption, while 10 respondents representing 2.4% of the total also confirmed they are not interested in the E-Services at the local government level due to reasons such as internet security and trust, as shown in E-Government Services Adopting results below.

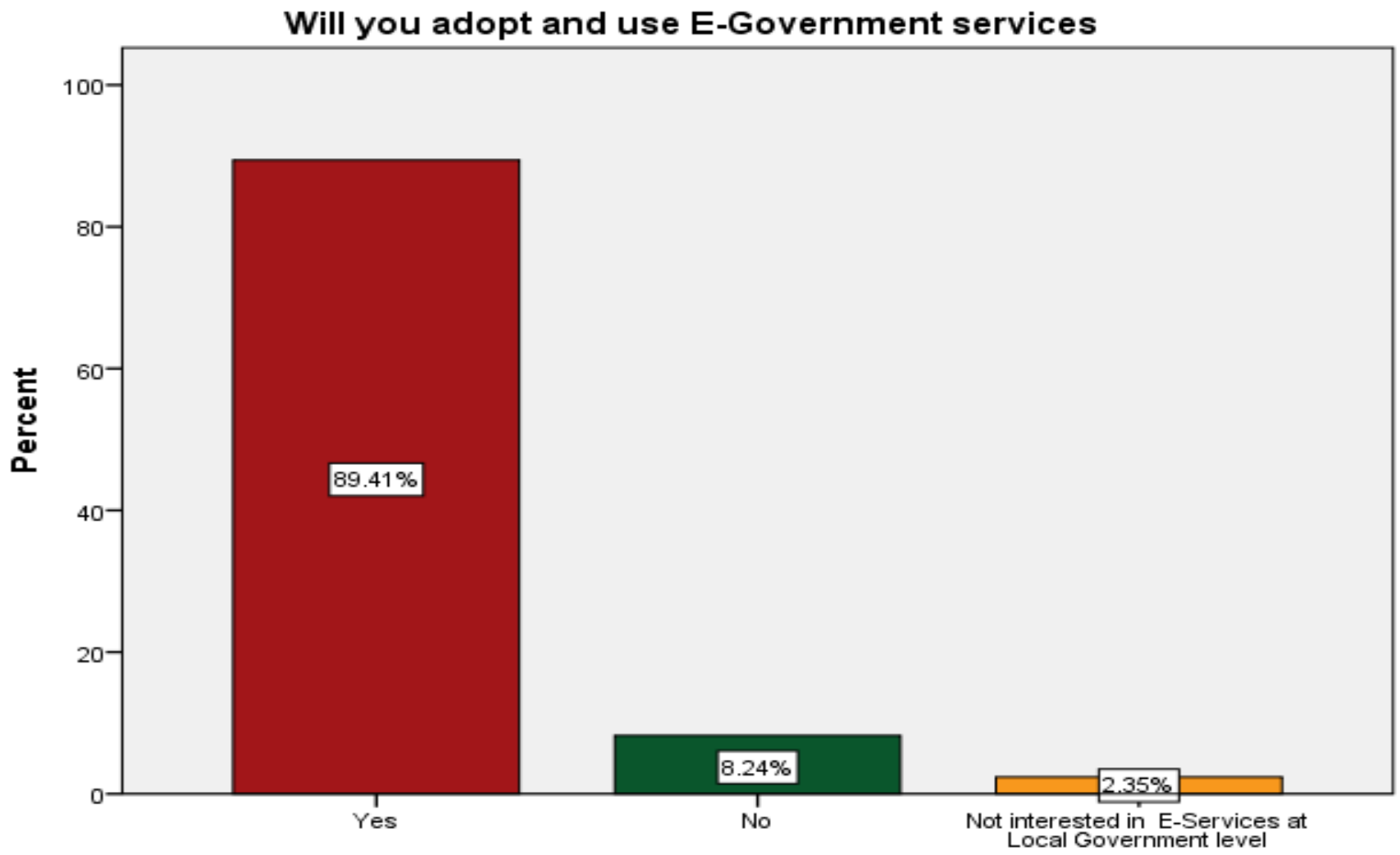

**Figure 60 SPSS Online Survey Analysis: E-Government Services Adoption**

## 7. Diffusion of Innovation Theory (DOI)

The use of Diffusion of Innovation Theory (DOI) for this study was agreed with the participants during the interview stages, and a validation of the theory was carried out after that as the research progressed. The interview participants and survey respondents were asked about their background and knowledge of the theories used in IT, especially DOI theory. The result, however, was not surprising to the researcher, as he is aware of the fact that most participants and respondents might not have a prior IT educational background. Their participation in this study has now opened up an opportunity for them to know more about IT theories. From the results shown in **Figure 61**, 90 respondents (21.2%) had heard about DOI, which is a good development to this research, as having participants with some interest could contribute to obtaining a rich data. More than two-thirds of the respondents (330, being 77.6%) said that they had not heard about DOI before, and a minority of the respondents (5, being 1.2%) indicated that they are not interested in technology deployment.

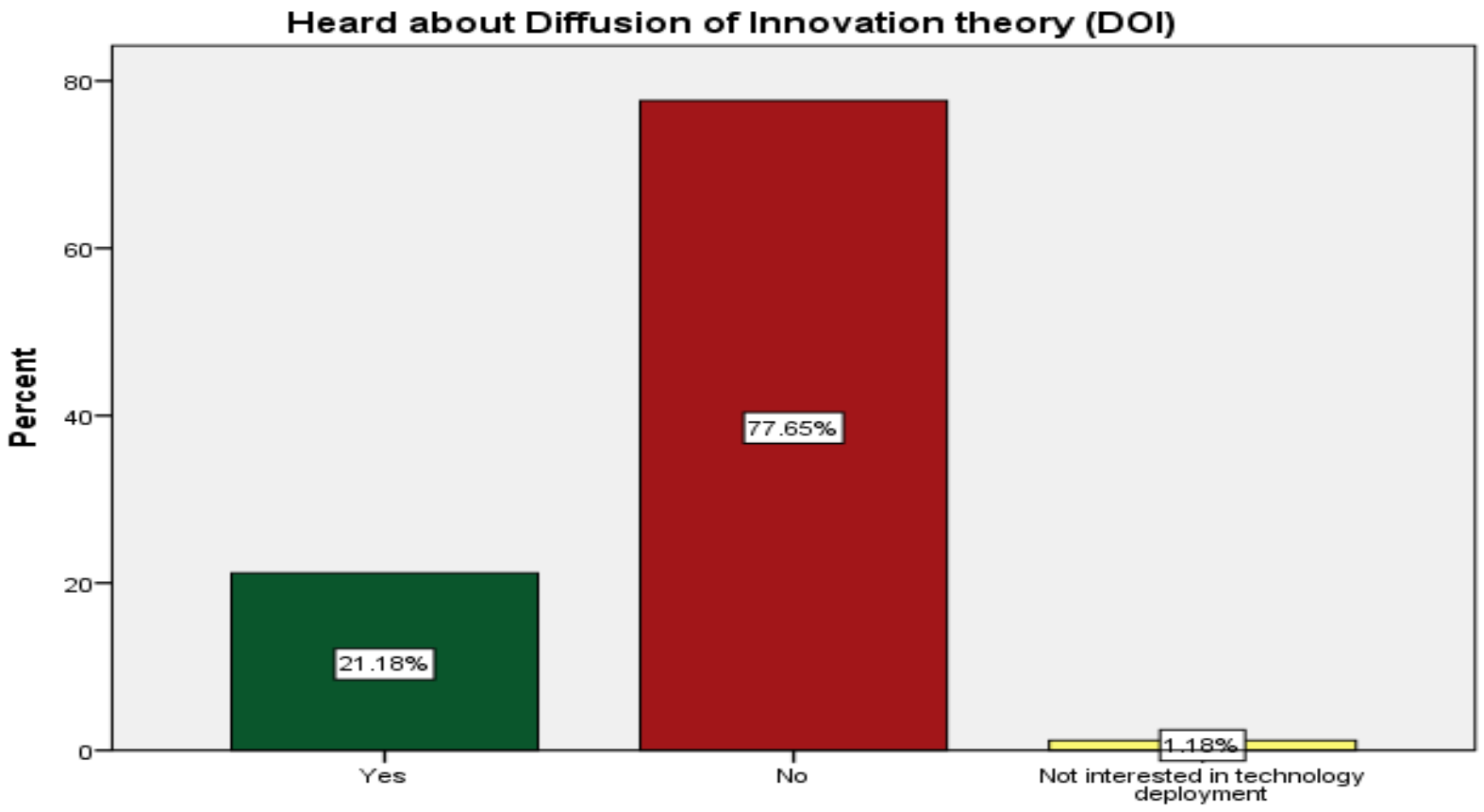

**Figure 61 SPSS Online Survey Analysis - Diffusion of Innovation Theory (DOI)**

## 8. Solutions to the Barriers Facing E-Service Adopting and Implementation

The investigated barriers facing the E-Service adopting and implementation results obtained from the online survey are explained in **section 6.3.4** above and are diagrammatically represented in **Figure 54**. This section of the online survey analysis will further explain the investigated solutions to those barriers, as obtained from the online survey results. It is, therefore, pertinent to conclude here that there is a relationship between the solutions investigated here and the solutions obtained earlier from the participants during the interview stages, as shown in **Figure 62**. Some 12 respondents (2.8%) agreed that ***affordable and cheaper internet connection*** for citizens would help to solve the barriers facing E-Service adoption and implementation. Unfortunately, over 40 million Nigerians have no access to the internet network according to Danbatta (2016), and a country should always aim to have a quality service and cheaper internet service for the citizens, as this would help to improve the adoption rate of the E-Government services. Another 40 respondents (9.4%) of the total concluded that an improved ***awareness*** of E-Service initiatives from the government at the local level would help to increase the adoption rate. Furthermore, ***corruption eradication*** in government will solve the barriers facing the E-Service adoption and implementation according to 42 respondents, which represent 9.9% of the total.

A minority of respondents (0.7%) indicated that ***cultural integration*** would solve these barriers. Some respondents (1.9%) expressed the belief that having a strong ***E-Service usage law*** will help in addressing the barriers facing the E-Service adopting and implementation at the local environment level in Nigeria, as this will reduce the fear of unauthorised accessed to information by hackers. The law should have a severe punishment for people caught taking part in this act. The researcher believes that, if the fear of information theft is minimised, the adoption rate for the E-Services will improve. Furthermore, the majority of the respondents, when answering the question on the solutions to the barriers facing E-Service adoption and implementation, felt that adequate funding will help in E-Service adoption and implementation. Greater funding might improve E-Service initiatives, as the high cost of ICT equipment, setting up and the maintenance of telecommunication equipment need huge allocated funds (Oseni and Dingley, 2014). As shown in **Table 30**, 83 (19.5%) respondents agreed that sufficient ***funding*** is one of the solutions to the barriers facing E-service adoption and implementation at the local government level in Nigeria. Another 48 respondents representing 11.3% of the total responded that ***internet security and trust*** would boost the adoption of E-Services in Nigeria. If the citizens can trust online transactions at the local level, especially when they believe their data are saved anytime online, there will be an increase in the rate of E-Service adopting, as people in various homes and offices do communicate with one another on their experiences about online shopping.

Furthermore, other results from **Table 30** show that 13 (3.1%) respondents agreed on periodic ***IT Skill/Training*** for the local government staff and citizens, with less IT knowledge. Some 12 (2.8%) respondents agreed on solid ***IT strategic plans*** for E-Service initiatives by the stakeholders, which include the local government management. Fourteen (3.3%) respondents agreed on the need for ***leadership trust/focus***, and this will have a positive impact on E-Service projects if the stakeholders show a high level of trust and focus in E-Service project executions.

**Table 30 Solving the Barriers facing E-Service Adopting and Implementation – Online Survey**

| | | Frequency | Percent | Valid Percent | Cumulative Percent |
|---|---|---|---|---|---|
| **Valid** | Affordable Internet Connection | 12 | 2.8 | 2.8 | 2.8 |
| | Awareness | 40 | 9.4 | 9.4 | 12.2 |
| | Corruption Eradication | 42 | 9.9 | 9.9 | 22.1 |
| | Cultural Integration | 3 | 0.7 | 0.7 | 22.8 |
| | E-Service Usage Law | 8 | 1.9 | 1.9 | 24.7 |
| | Funding | 83 | 19.5 | 19.5 | 44.2 |
| | Internet Security/Trust | 48 | 11.3 | 11.3 | 55.5 |
| | IT Skills/Training | 13 | 3.1 | 3.1 | 58.6 |
| | IT Strategic Plans | 12 | 2.8 | 2.8 | 61.4 |
| | Leadership Trust/Focus | 14 | 3.3 | 3.3 | 64.7 |
| | Local Government Autonomy | 29 | 6.8 | 6.8 | 71.5 |
| | Management Re-Orientation | 13 | 3.1 | 3.1 | 74.6 |
| | Partnership Between Government, Private and Public | 8 | 1.9 | 1.9 | 76.5 |
| | Political Stability | 18 | 4.2 | 4.2 | 80.7 |
| | Power Supply | 10 | 2.4 | 2.4 | 83.1 |
| | Project/Fund Monitoring | 18 | 4.2 | 4.2 | 87.3 |
| | Provision of ICT Centres | 10 | 2.4 | 2.4 | 89.6 |
| | Provision of IT Facilities | 32 | 7.5 | 7.5 | 97.2 |
| | Transparency - IT Project | 7 | 1.6 | 1.6 | 98.8 |
| | Usability Improvement | 5 | 1.2 | 1.2 | 100.0 |
| | **Total** | **425** | **100.0** | **100.0** | |

Having an independent local government regarding the spending and control appeared to be the latest research discovery in unveiling the barriers facing E-Service adopting and implementation at the local environment level in Nigeria. Hence, as obtained during the interviews and online focus group meetings, 29 (6.8%) respondents who participated in the online survey also agreed on the need for ***local government autonomy***. Another 13 respondents, which represent 3.1% of the total, agreed that ***management re-orientation*** is one of the solutions to the E-Service adoption and implementation. Eight respondents, representing 1.9%, agreed that a mutual and fruitful ***partnership between the Government, Private, and Public*** could help in the E-Service adopting and implementation.

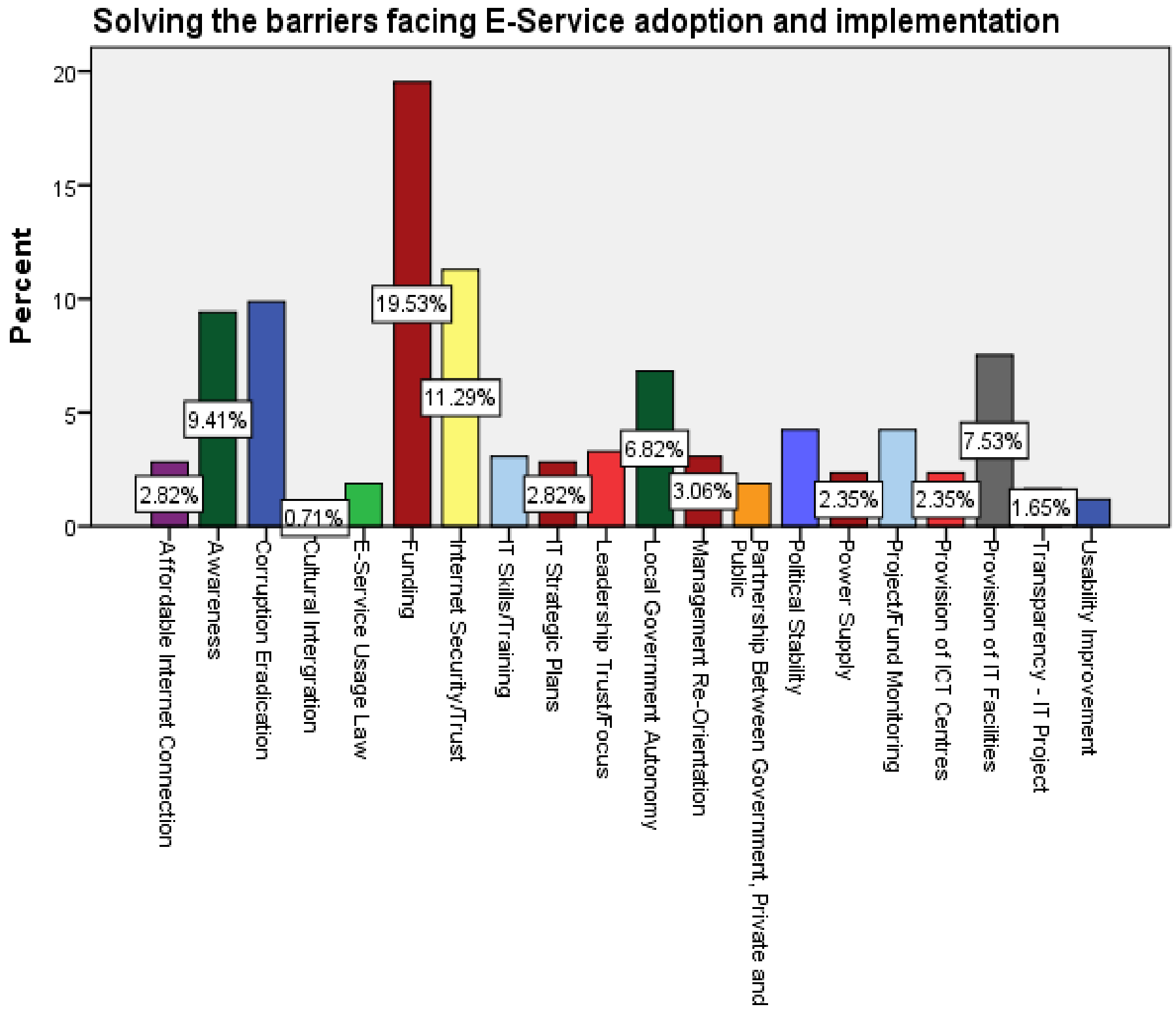

**Figure 62 SPSS Online Survey Analysis – Solutions to the Barriers Facing E-Service Adopting and Implementation**

Other important findings on the solutions to the barriers facing E-Service adopting and implementation at the local environment level in Nigeria, as obtained from **Table 30**, shows that 18 respondents, representing 4.2% of the total, agreed on ***political stability*** as a solution. During the recent sad events in the northern part of Nigeria, where insurgency (Boko Haram) had displaced millions of residents from their various homes, more than 25 telecommunication masts and base stations were destroyed in the process, as reported by Awortu (2015). E-Service adopting and implementation will be difficult when there is political instability, and the massive destruction of telecommunication infrastructures is a big setback. Another ten (10) respondents (2.4%) felt that a ***constant power supply*** will improve the E-Service adoption, as it will be cheaper for the citizens rather than having to buy fuel to run private power generators in the absence of power supplies from the power regulatory authority in the country. Eighteen (4.2%) respondents believed that proper E-Service project/fund monitoring would help to solve the corruption barrier facing E-Service implementation. Provision of ***ICT centres*** at the local government councils will help average citizens with no access to IT infrastructures to upgrade/improve IT skills, or E-Service initiatives adoption rate, as they could visit the centres regularly, such as the public libraries in developing countries. Ten respondents, representing 2.4% of the total, agreed on the need for free ICT centres located at the local government councils.

It can also be seen from the data in **Table 30** that the ***provision of the IT facilities*** is another solution to the barriers facing the E-Service adoption and implementation at the local government level in Nigeria, and 32 respondents representing 7.5% of the total agreed to this. According to the respondents, availability of IT facilities at the local government councils, public places such as community libraries, with trained and skilled IT staff, will improve the adoption rate of E-Government services. Another seven respondents, which represent 16% of the total, agreed that transparency in IT Projects would solve some barriers facing the E-Service adoption and implementation. Lastly, five respondents, which represent 1.2% of the total, decided that ***usability improvement*** would improve the adoption rate of the E-Service at the local government level in Nigeria. In web development, usability is important, as it is a user-centred design which involves elements such as ***clarity*** and ***credibility***. The development of E-Government service is focused on the users (citizens) and, to achieve the aims of the initiatives, there is a need to have a clear, user-friendly website for the E-Services that are efficient and easy to use. Credibility is another important factor in usability, as users of the E-Services internet site (the citizens) should be able to trust the contents, and it is always important to have enough and consistent information about the local E-Service, especially in the tab section "about us".

## 9. Success Factors for E-Service Adopting and Implementation

The results obtained from the investigated success factors for E-Service adopting and implementation (online survey) are represented in **Figure 63** below. This section will discuss the investigated success factors for E-Service adopting and implementation. The key research findings show that 6.4% of respondents agreed that the adopting of E-Service would improve access to government information.

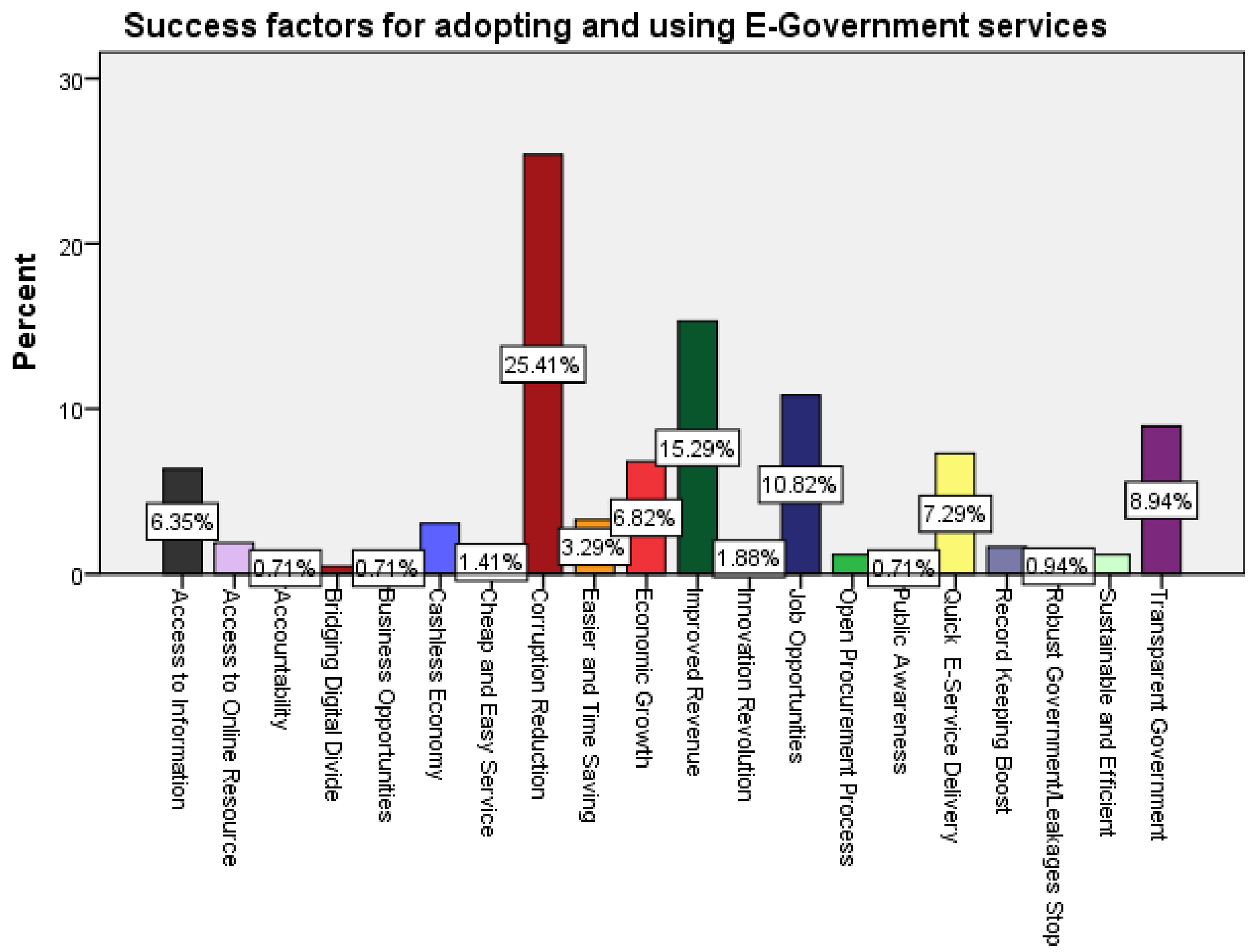

**Figure 63 SPSS Online Survey Analysis – Success Factor for the E-Service Adopting and Implementation**

Furthermore, eight (8) respondents (1.9%) agreed that adopting of E-Services would advance access to the online resources. A minority of respondents (0.7%) indicated that the success of E-Service adopting is accountability. Another minority of those surveyed (0.5%) specified that the success for E-Service adopting is bridging the digital divide. Also, 3.1% respondents agreed that adopting of E-Service would lead to the cashless economy, while a small number of those surveyed (1.4%) concluded that the E-Service adopting would give way for a cheap and easy service. More than one-quarter of the respondents (25.4%) said that E-Service adopting would reduce corruption in government, and 14 respondents (3.3%) agreed that it would save time and be easier to use. A considerable number of the respondents (6.8%) said that the adopting of the E-Service at the local government level in Nigeria would contribute to the economic growth. Some respondents (15.3%) expressed the belief that the adoption of E-Service will improve the revenue of the government, eight respondents (1.9%) agreed that adoption of E-Service would enhance innovation, and 46 respondents 10.8% agreed that it would create job opportunities. The key research findings also show that 1.2% respondents agreed that the adopting and implementation of E-Service at the local government level in Nigeria would open the procurement process (transparency).

Also, another minority of respondents (0.7%) specified that the success for E-Service adoption and implementation is public awareness creation, while a considerable number of the respondents (7.3%) indicated that it creates a quick E-Service delivery. Another 1.6% respondents agreed that adopting of E-Service would boost the record-keeping and 0.9% respondents indicated that the adoption would stop revenue leakages in government. In addition, five (5) respondents 1.2% agreed that the adoption of E-Service at the local government level in Nigeria would bring about sustainability and efficiency. Lastly, a considerable number of the respondents indicated that it enhances transparency in government.

### 6.4.3 Action Research Findings – Application of the Theory of Change

This section describes the core and important research findings obtained from the interview sessions on the barriers facing the adopting of the E-Services and implementation at the local environment level in Nigeria. The local government Autonomy and Corruption issues were the major barriers that emerged from this study. However, past research has rarely discussed how corruption has influenced the non-implementation of E-Service initiatives and why the adoption is very low, especially in developing countries (Aladwani, 2016). These results corroborate with the online survey findings as discussed in **Section 6.3.4** and Appendix B.

As explained in the research methodology, action research involves the acquiring of knowledge through interaction with other people on how a change could occur in their organisation and the reason for making a change. Therefore, action research will typically choose a theory on which to hang these results and helps more with the analysis of the research data. Hence, this researcher used the interviews and the online survey result to assess the various ways on how corruption as the barrier influenced the low adopting and non-implementation of E-Service at the local environment level in Nigeria. The key findings indicated that the solution to the corruption issues as a factor hindering the E-Service full implementation is for the stakeholders in the E-Services implementation at the local government level in Nigeria, which includes the government officials, is to have ***attitudinal change*** towards this menace. Also, the participants agreed that it is important to put in place the anti-corruption structure that will repel any form of corruption. Bertot et al. (2010) added that the main approach towards anti-corruption drive and transparency in government is social change. This means that citizens are empowered to participate in institutional reforms by cultivating a civil and law-based society that checks corruption and supports accountability in the system.

In light of the above, this researcher concluded that, in order to take this research forward, and to complement the anti-corruption drive by the government, this research would look into the ***Theory of Change*** based on the attitudinal change advocated by the participants during the interviews, and the results obtained from the online survey as shown in **Figure 64** below, where 363 respondents out of the total of 425 (85.41%) strongly agreed that deploying E-Service at the local government level in Nigeria will reduce corruption.

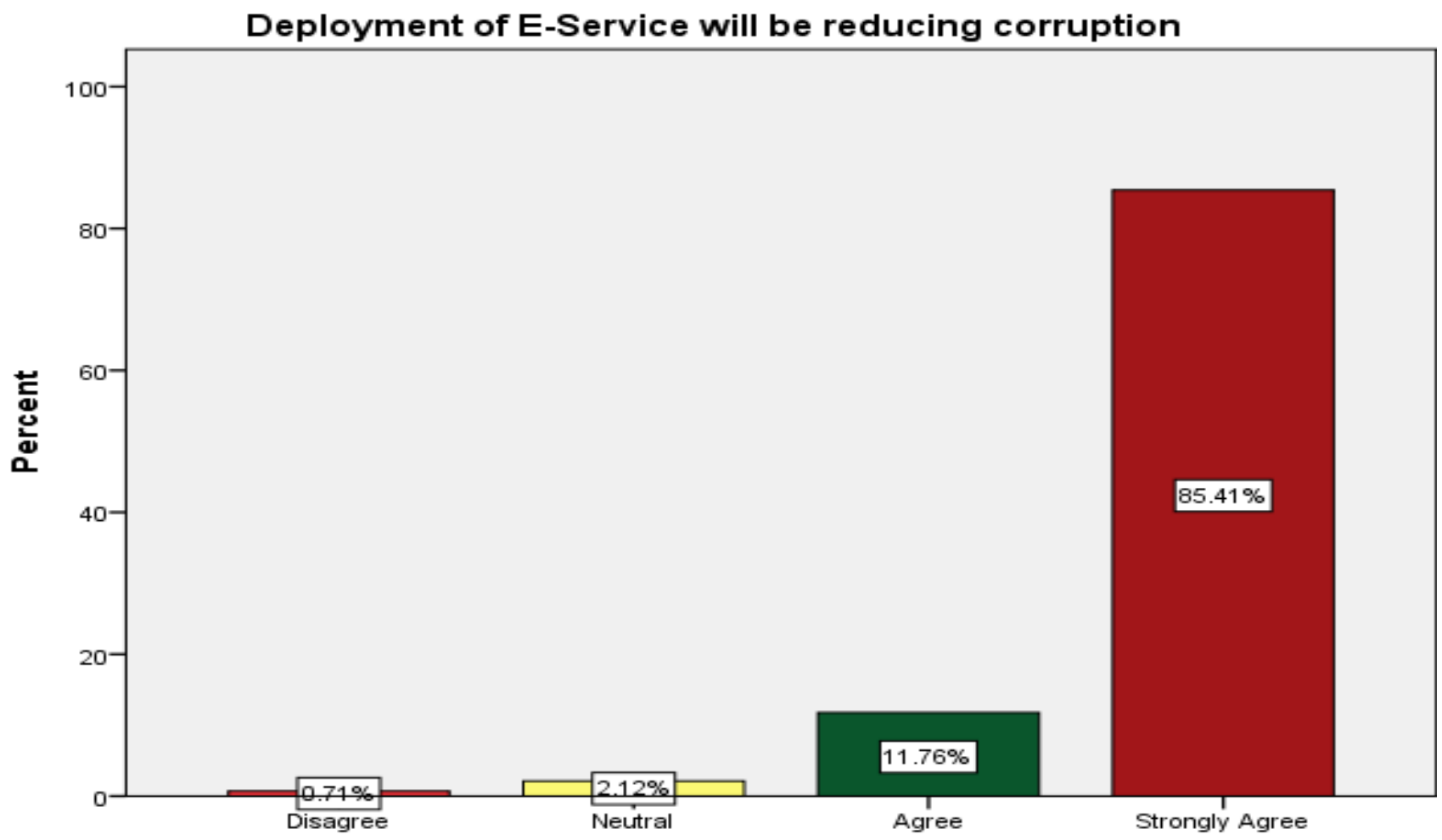

**Figure 64 SPSS Online Survey Analysis on Deployment of E-Service Will be Reducing Corruption**

**The Theory of Change**: The theory of change was developed in order to improve the processes of project design, according to Johnson (2012). It was developed to explore behaviours and outcomes that cannot be measured easily. Hence, the applicability is best suited to the governance and anti-corruption sectors (Johnson, 2012). The theory of change definition by Johnson (2012) is close to those of Taplin and Clark (2012), who describe the theory of change as interventions that bring about the outcomes portrayed in any research work or framework. The main rationale behind this theory is the connection in the research outcomes and why one result is needed to achieve another in future (Taplin and Clark, 2012). Furthermore, Rogers (2014) also described the theory of change as being activities that produce a series of results that make significant contributions to achieving the expected impacts.

Therefore, regarding corruption as a major barrier (research findings) facing the adopting and implementation at the local environment level in Nigeria, the researcher will exploit the theory of change attributes to influence behavioural attitudes. In doing this, it is necessary for this research to explore and identify workable long-term goals and outcomes that will influence the government's attitudinal change towards reducing – if not total eradication of – the corruption involved in the implementation of E-Service at the local environment level in Nigeria. **Figure 65** below is the Theory of Change Model for the Malawi Anti-Corruption Enforcement. The analysis could also be used to design a theory of change model to enforce an anti-corruption drive at the local environment level in Nigeria, and it is hoped that the stakeholders will be committed to the process.

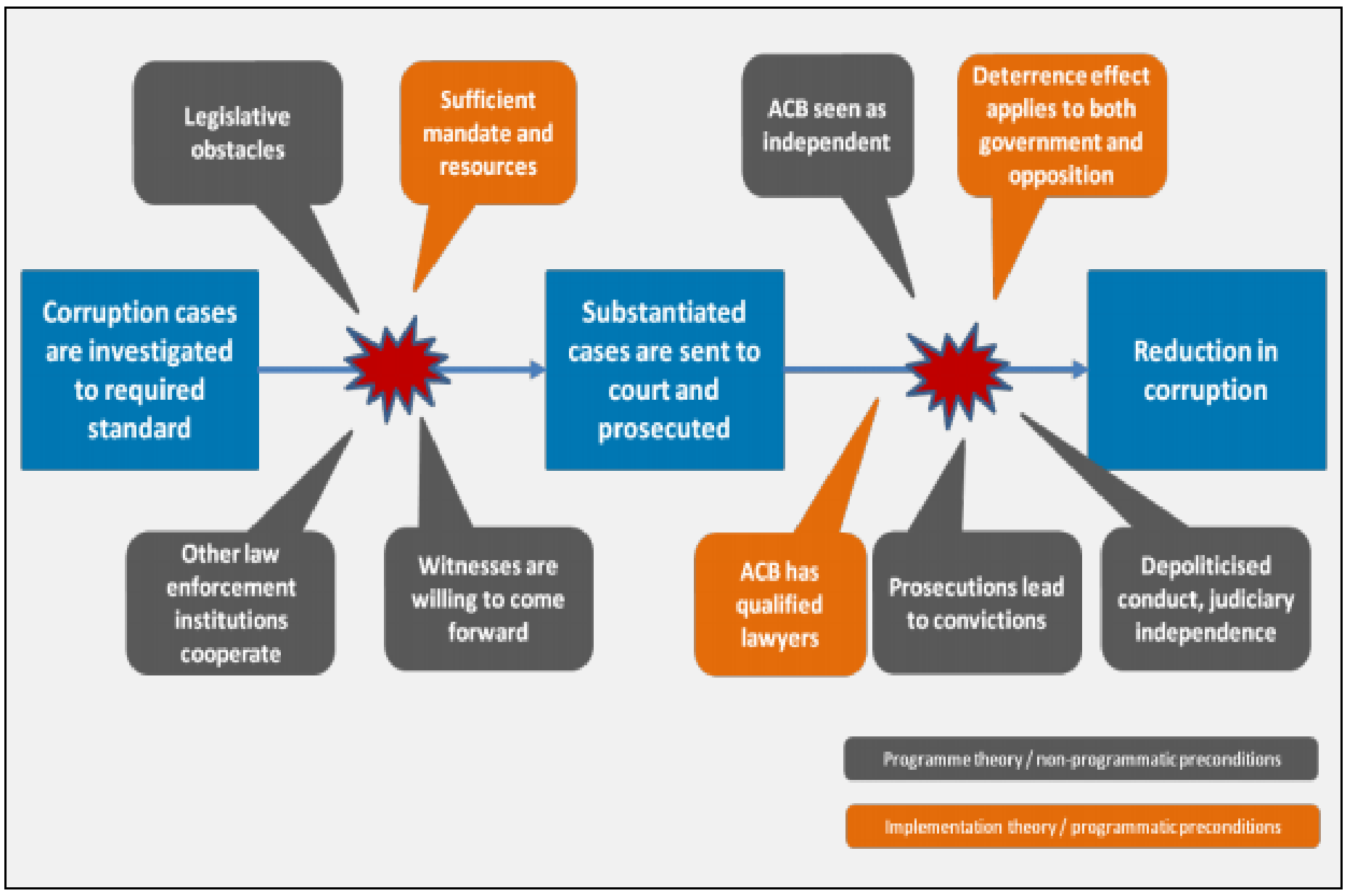

**Figure 65 Theory of Change Model for the Malawi Anti-Corruption Enforcement (Johnson, 2012)**

Johnson (2012) argued that, if the communities (government officials) are better informed about the consequences of corruption, then the system could change. Hence, this researcher proposed a theory of change model, as shown in **Figure 66** below, with anti-corruption guidelines. The model is designed to influence the officials at the local government level, and their attitudinal change to corruption, including the long-term goals and outcomes, and this will boost the anti-corruption drive. It would also influence the full implementation of E-Service projects.

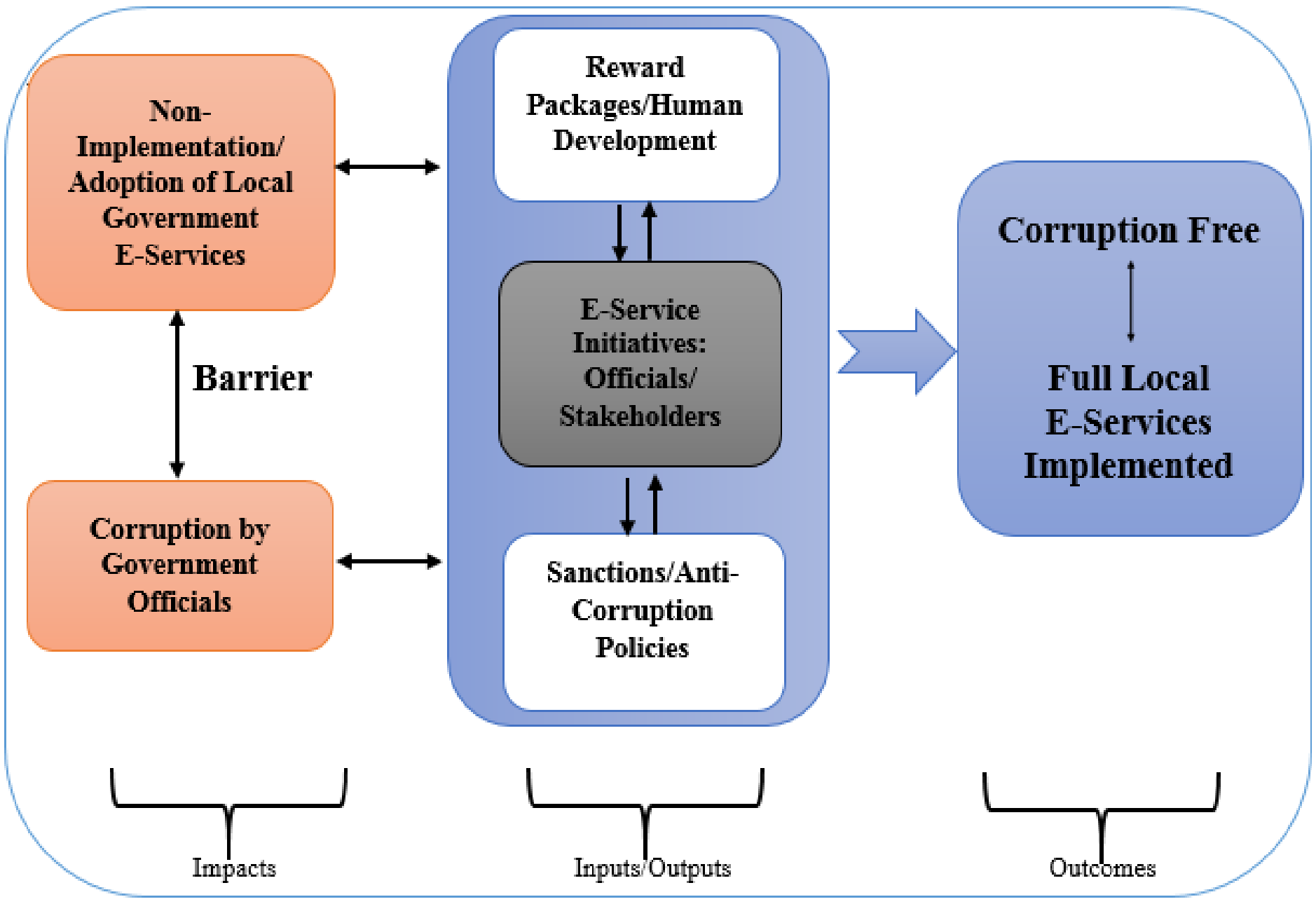

**Figure 66 Proposed Theory of Change Model**

**The Non-Implementation/Adopting of Local Environment E-Services**: This research seeks to discover how the low adoption of the E-Services at the local government level in Nigeria is due to non-implementation of the initiatives as a result of the identified barriers, including corruption.

**Corruption by Government Officials**: This is a serious threat and a major barrier hindering the E-Government services development at the local government levels in Nigeria. Both Okafor (2010) and Ayoola (2013) argued that corruption is a serious menace with the government officials in Nigeria, as the funds appropriated for special projects such as E-Service development are diverted into private pockets.

**Reward Packages/Human development**: Incentives and reward packages should be designed to motivate government officials regularly, high performance needs to be appreciated, and frequent human capacity development through various training and seminars should be introduced for the officials at the local government levels in Nigeria. It is hoped that this can contribute to the expected positive attitudinal change towards corruption.

**Sanctions/Anti-Corruption Policies**: Anti-corruption policies and guidelines should be made available and clearly shown to all the government officials, both old and new ones.

Special courts and tribunals can be set up with the necessary regulations to try the corrupt officials and award needed sanctions, such as suspension or outright dismissal from duties.

In light of the above, the future research includes full testing and validation of the proposed theory of change model within the Nigeria local environment context. Moreover, the future research can also validate the anti-corruption policies, as explained above, with more participants to authenticate its workability, and if it can achieve the expected long-term goals and outcomes in making the necessary change. This has opened more opportunities for further research to be conducted on the theory of change as a model in anti-corruption drive at the local government level in Nigeria.

## 6.5 E-Service Implementation Process

The proposed framework, as discussed in **section 3.3**, itemised the barriers facing the E-Service adoption and implementation. These barriers must be taken into consideration when initiating the E-Service projects implementation. The E-Service implementation process, from the initiation stage until when a fully functional E-Government service is provided, will be discussed in the section. The process explains the different steps that E-Service project managers at the local government level in Nigeria need to follow in implementing future projects, and these include various stages, strategies, and actions (Hassan, 2011). The implementation of E-Government services, according to Alshehri and Drew (2011), not only saves resources, money, and energy, but also reduces the time spent on many government manual transactions, thereby increasing the service quality level comprehensively. The E-Service implementation process is illustrated in **Figure 67** below.

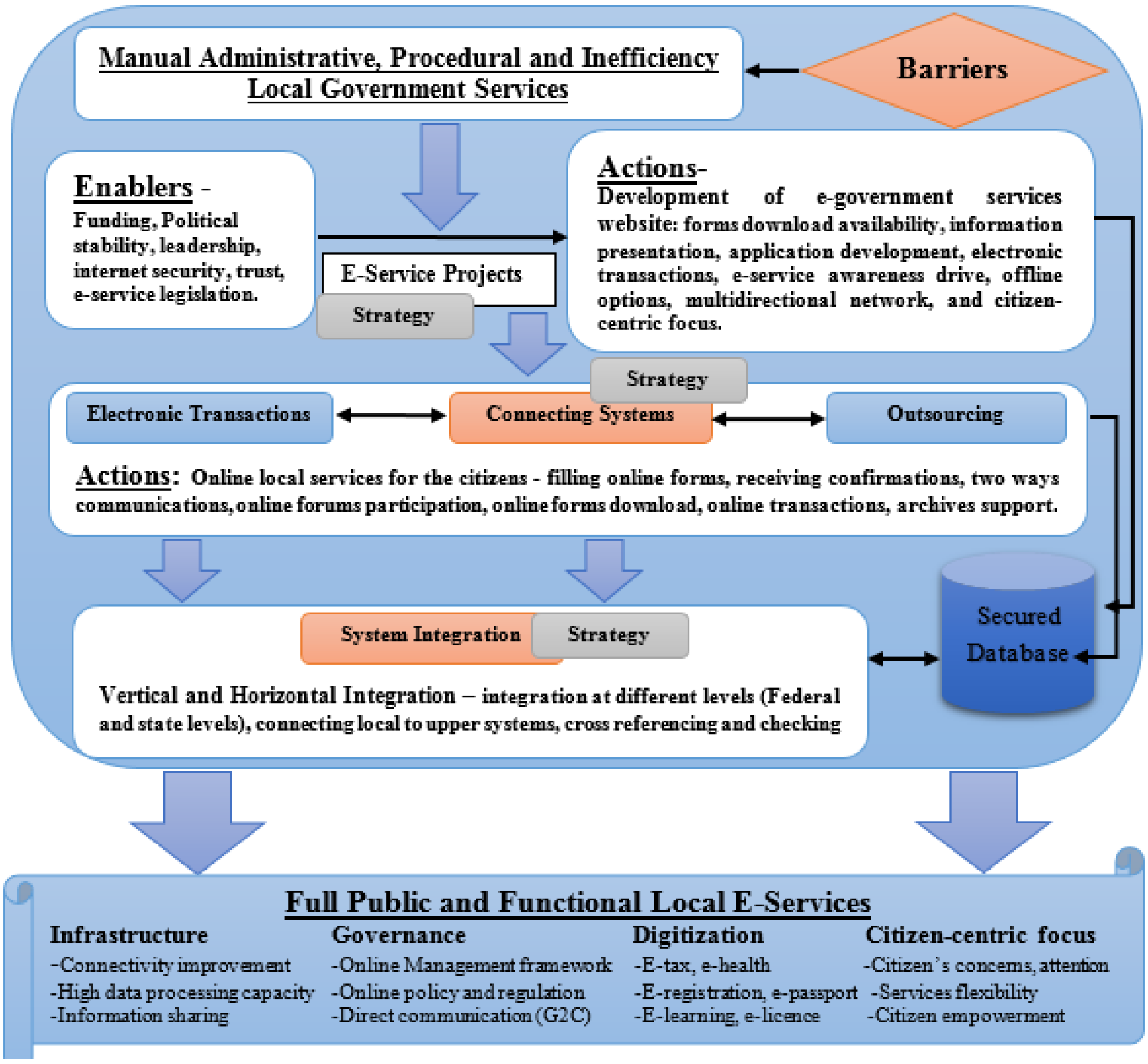

**Figure 67 E-Service Implementation Process**

The E-Service implementation process as shown in **Figure 67** indicates the need for E-Government services initiatives to overcome the identified barriers before they become full and functional public E-Services. These barriers include lack of funding, political instability, corruption, the high cost of internet access and ICT infrastructures, internet security and trust, and lack of E-Service legislation. The solutions to these barriers, as obtained in this study among others, is to have functional E-government services at the local environment level in Nigeria. Government should provide an enabling environment for E-Services projects to flourish. An example of an enabling environment would be availability of funds, a partnership with both public and private organisations, subsidising ICT equipment, provision of affordable internet connection, and provision of a secured online platform among others. The Vice Chancellor of Covenant University, Ota, Nigeria (TVN, 2014), recently confirmed:

> "The distance between the citizens and their government continues to widen. He argued that the efficiency of the current system has become questionable as people are isolated from the governance. He stressed the need to strengthening the means of delivering public services, promoting transparency and enhancing interactions between the government and the citizens." (TVN, 2014).

While the citizens at the local government level in Nigeria do not wield much pressure on the local government administrators on E-Service initiatives, the availability of the necessary enablers (solutions to the barriers) will be a driving force for the E-Service project initiations. As pointed out by the Vice Chancellor at Covenant University, Ota, e-governance, particularly at the local government level, is an attractive choice towards national development. Also, it will help in closing the gap between the citizens and the government through an effective online interaction (TVN, 2014).

Furthermore, after overcoming the significant barriers through the availability of the enablers, E-Service projects will commence with the development of a functional website for the local E-Government services. Citizens will be able to perform many functions using the site such as forms download, information archiving, electronic transactions, E-Service awareness drive information, offline chat options, multi-directional network, and so on. The next stage of applying the necessary strategy is a further connection to the system, where secure online transactions can take place. Outsourcing of some functions is possible here. Depending on the resources and the capacity available on the website, local government administrators could outsource some functionalities on the website to local contractors for awareness and fund-raising purposes. System integration is another stage in the local E-Services implementation process. This involves both vertical and horizontal integration, where local functions on the website could be aligned to the functionalities at both federal and state government level. For example, on the issuance of passports to the citizens, there can be a link to the local government website on how to apply for their passports. Once the users click on this link, it will open the Nigeria immigration website, which is a department under the Federal Government.

In conclusion, there is a need to have a secured system for all the transactions taking place on the local E-Service website. A major barrier to the adoption and implementation of the E-Service at the local government level in Nigeria, as obtained from the study results, is internet security and trust. Therefore, a secured system with the necessary backup is required, and some of this could be outsourced locally or internationally to a responsible database/storage company with an excellent track record of integrity and high performance. This will prevent hackers from tampering with the citizens' details whenever they use the local E-Government services. The E-Government services implementation will improve internet connectivity and data processing capacity, re-engineer business processes, make available the policy and regulatory framework for E-Government, and improve the standard of government operations.

## 6.6 Summary

This chapter has reviewed the research findings from this study, as shown in **Figure 54** above. Investigated/identified barriers facing the E-Service adopting at the local government in Nigeria discussed the solutions to these barriers and the success factors. Discussions from the lessons and reflections section presented detailed actions that took place in each action research method phase. The responses from the online survey respondents were analysed. This chapter concluded with a detailed **Theory of Change Model** and **E-Service Implementation** process, as illustrated in **Figures 66 and 67**, respectively.

Moreover, the results including the barriers, solutions, and success factors for adopting E-Services at the local environment level in Nigeria, were analysed. There is likely to be some bias towards the impact on E-Services adoption rate because it was focused on Lagos, which is the most internet-ready state in Nigeria. As mentioned earlier (section 5.2.2), Lagos was chosen because of its internet readiness, and moving E-Services beyond Lagos will require the infrastructure of the internet, networking and devices to be well established before E-Services can be implemented. The limitation of this chapter is the inability of the researcher to extend the data collection to other states in Nigeria due to such reasons as insurgency and lack of internet penetration. Future research can exploit the limitation and extend the data collection to a few more states in Nigeria as the internet reaches them and internet-ready devices proliferate.

# Chapter 7 – Validation of Research Findings

## 7.1 Introduction

The aim of this study was to identify the barriers facing E-Service adopting and implementation at local environment level, using Nigeria as a case example. The study will also propose a Diffusion of Innovations (DOI) theory-based framework for predicting the successful E-Service adoption and implementation, including the success factors. To achieve the aims of this study, the research used the action research methodology where 30 participants were interviewed on their notion about the E-Services development at the local government level; 425 respondents were surveyed, and five members of an online focus group were involved in the validation processes of the research findings, including the proposed E-Service framework. The online focus group members had an interest in E-Service development and understood this study well, as they were earlier involved in the interview sessions. The five online focus group members were chosen from 30 participants interviewed earlier during the one-on-one interviews stage of research, and their consent was further sought in their participation in the online interview.

In the previous chapter, the findings from the analysis of action research method (interviews, online survey, and online focus group) were presented and discussed. The initial framework was modified to produce the final framework based on the examined and identified barriers facing the E-Service adopting and implementation, the solutions and the success factors. Hence, this chapter discusses various validation techniques adopted to ensure the rigour and quality of this research's findings. This chapter is divided into six different sections, as shown in **Figure 68**, namely, the introduction, validation techniques, final framework validation, the validation of the identified barriers, the general findings feedback sessions and, lastly, the summary section.

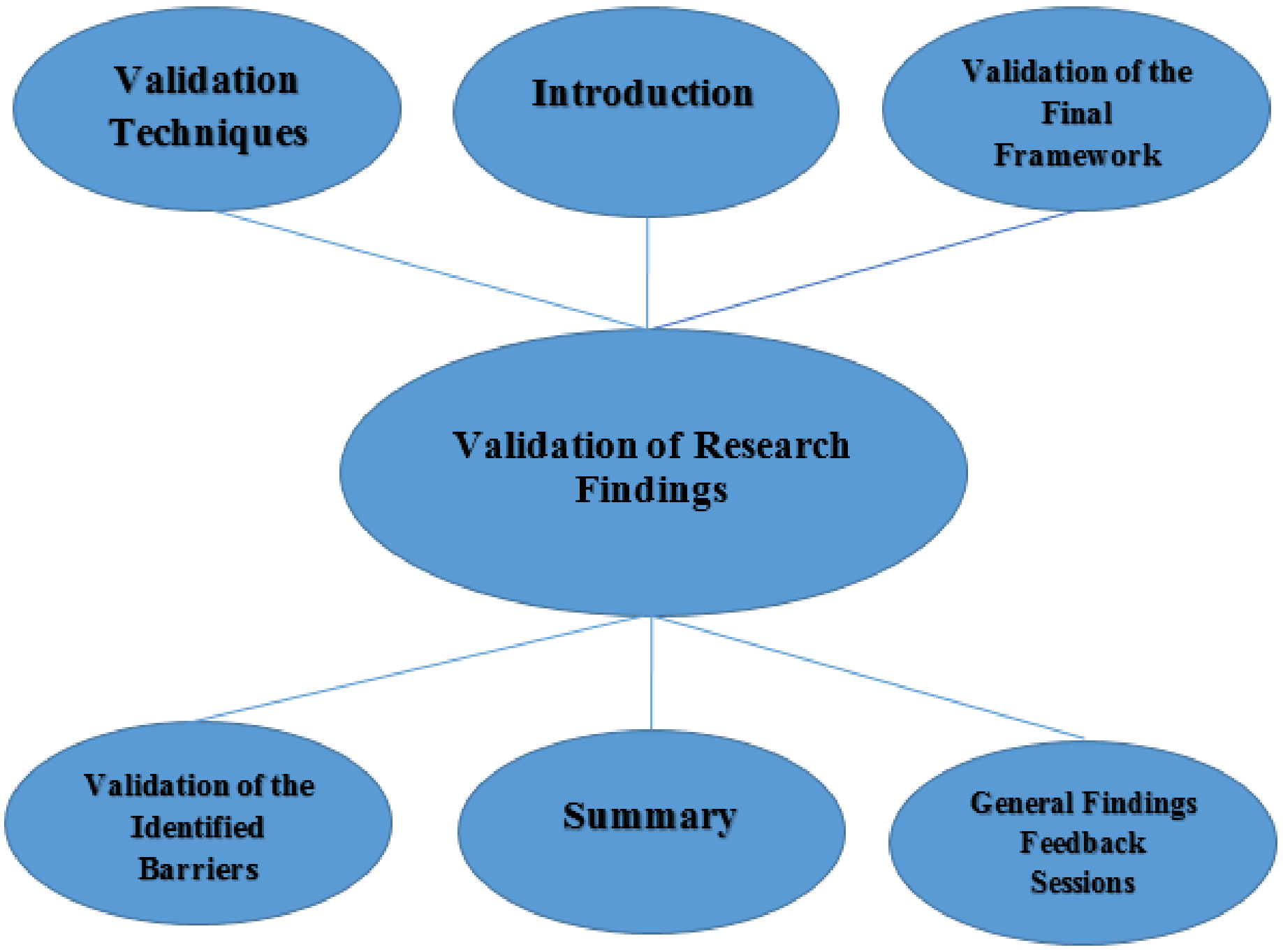

**Figure 68 Chapter Seven Summary**

## 7.2 Validation Techniques

Validating the results obtained from this study will boost the research credibility and will enable this researcher to test the efficiency of the framework. To achieve this, the validation process is performed through the various techniques (strategies) such as triangulation, multiple coding, purposive sampling, and respondents' validation (Hassan, 2011; Torrance, 2012; Schoonenboom, 2015). **Figure 69** below shows the strategies used in assessing the qualitative research as obtained from Hassan (2011).

| Strategy | Concerns Addressed | Realistic Potential |
|---|---|---|
| Purposive Sampling | Bias | Enhancing sample coverage and providing a framework for analysis |
| Multiple Coding | Inter-rater reliability | Refining interpretations or coding frameworks |
| Triangulation | Confirmation or refutation of internal validity | Corroborating or, more often, refining findings |
| Respondent Validation | Confirmation or refutation of interpretations | Corroborating or, more often, refining findings |

**Figure 69 Qualitative Research Assessing Strategies**
(Hassan, 2011)

### 7.2.1 Triangulation

Triangulation, according to Archibald (2016), is seen as the observation of the research issue from two or more angles. Advocating for multiple data collection techniques (action research methods) for this research affirms the esteem conferred on triangulation. Both Hassan (2011) and Torrance (2012) outlined the four crucial aspects of triangulation, which are:

- Triangulation of data (Different sources)
- Triangulation of investigator (Different evaluators)
- Triangulation of methods (Different methods, for examples, interviews, online survey, and focus groups)
- Triangulation of theory (Different theoretical perspectives on the data)

In this research, the two forms of triangulation employed by the researcher are data triangulation, where data are collected from different sources, and methodological triangulation, in which different techniques were used to produce data. Triangulation was used in this study to solve the issue of internal validity through multiple data collection methods, as discussed in Chapters Five and Six, respectively. However, triangulation must be carefully planned to keep the validity of data intact. As observed by Hassan (2011), some problems might occur during analysis, as data obtained from different sources (as shown in **Figure 70**) using different forms were difficult to compare directly with each other. Nevertheless, the most astonishing aspect of this study is that similar findings were obtained despite using different data collection methods. However, the researcher concluded that the research findings produced were accurate and convincing.

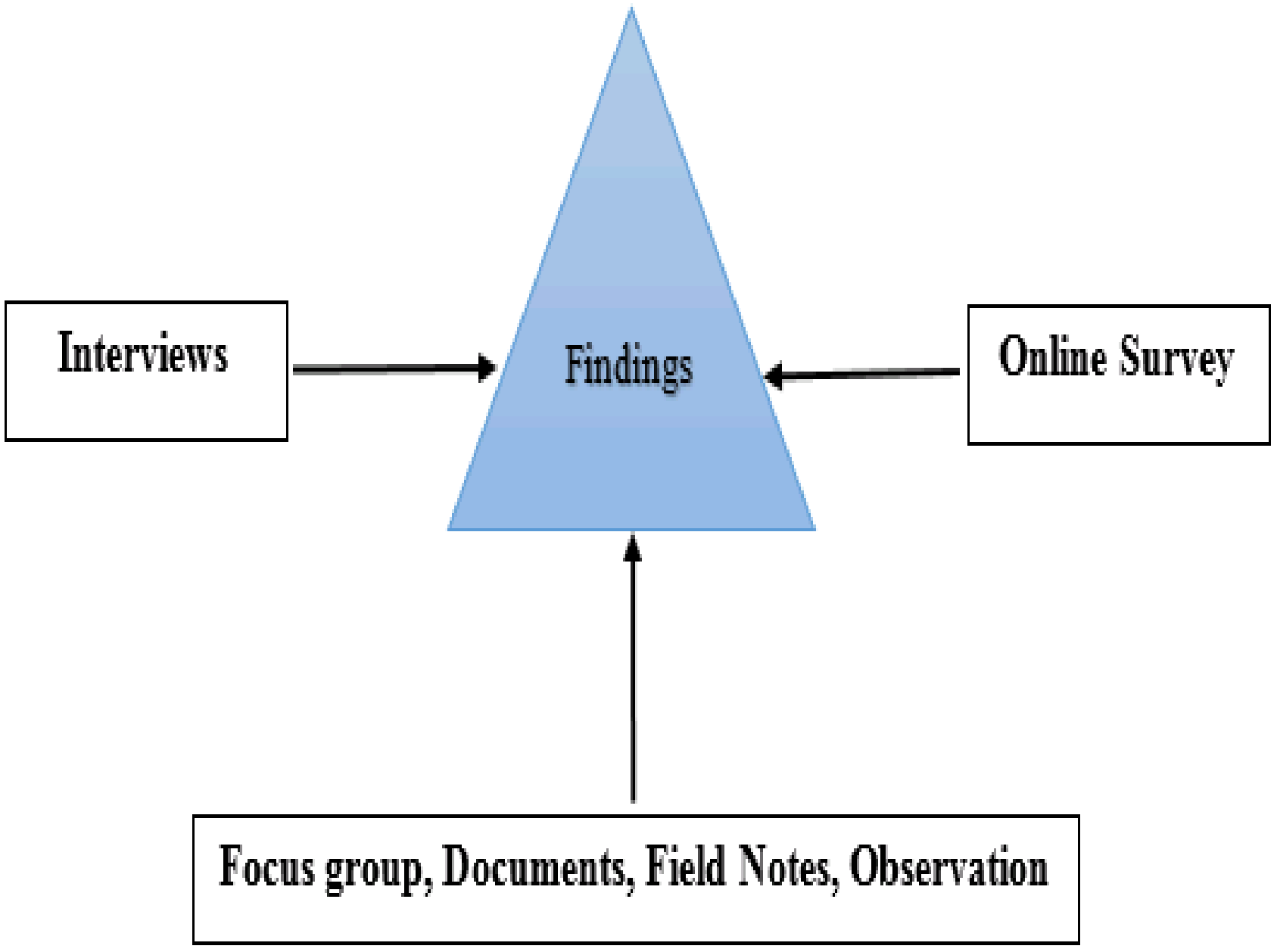

**Figure 70 Evidence of Triangulation - Multiple Data Collection Sources (Hassan, 2011)**

### 7.2.2 Multiple Coding

This is another strategy to validate the research findings in qualitative research (Hassan, 2011). Multiple coding is usually when two or more people work on something or observe something and make independent notes. The researcher then compares and collates the notes, and comes up with the best fit between them. Hassan (2011) posits that multiple coding processes involve the researcher asking colleagues, independent researchers, and supervisors to cross-check the coding and interpretation of data adopted during data analysis. Hassan (2011) used multiple coding, as most of her major supervision meetings with the supervisors formed part of the coding systems. This method enables the researcher to know the opinions of others in relation to the coding, framework validation and other findings during the data analysis. During the cross-checking with other researchers and colleagues, there might be some disagreement; however, the level of the disparity is considered to be irrelevant, as the vision in relation to the discussion is reflected in refining the coding or framework. This process, according to Hassan (2011), supports the diligence of the data gathered and how the analysis was carried out.

### 7.2.3 Purposive Sampling

This validation strategy, according to Hassan (2011), depends on the judgement of selecting research cases which are very close to the research questions. Bryman (2016) added that purposive sampling is conducted with reference to the research questions, so that the selected unit of analysis will be able to answer the research questions. For this research, the organisation was specifically selected instead of using random sampling, as the initial contacts in this organisation have a comprehensive understanding of E-Service initiatives at the local government levels in Nigeria. Moreover, the data obtained from the participants/interviewees in this organisation are rich enough to answer the research questions. However, it is accepted that employees of one organisation may have less varied opinions or organisational culture. This is a difficult decision, as the knowledge and organisational culture of the employees would be hard to obtain elsewhere. Hence, this research adopted Purposive Sampling (or theoretical) as it deals with selection control.

### 7.2.4 Respondent Validation

Respondent validation is another strategy used to validate research findings. The researchers will provide the respondents/participants with the research findings for validation purposes (Bryman, 2016). The process of checking the results with the respondents ensures stability, and this will meet the requirements of reliability according to Hassan (2011). Respondent validation is very common in qualitative research, as this researcher always making sure the correspondence between the respondents and the research findings, and it helps this researcher to improve the accuracy, validity, and credibility of this study (Bryman, 2016).

## 7.3 Validation of the Final Framework

This section discusses the validation of the final framework for this study. It is important to ensure that this framework will work as expected. Initially, the framework validation was carried out with both the interview and focus group participants, as explained in the previous chapter. From the results available in Appendix B (**E-Service Framework Acceptability and Validity)**, 29 participants out of the total 30 participants (96.7%) agreed that the ***framework is appropriate*** and accepted one for this study. Moreover, the online focus group participants indicated that the ***final and refined E-Service framework*** sent to them represented their notion of E-Service at the local environment level in Nigeria (see **Appendix B)**. However, there is a need for the researcher to validate the E-Service final framework further. The focus group members were chosen for the validation due to their interest in the E-Services, and they are part of the stakeholders in the implementation of E-Service initiatives at the local government level in Nigeria. The focus group members, as shown in **Table 32**, have a combined 110 years of working experience.

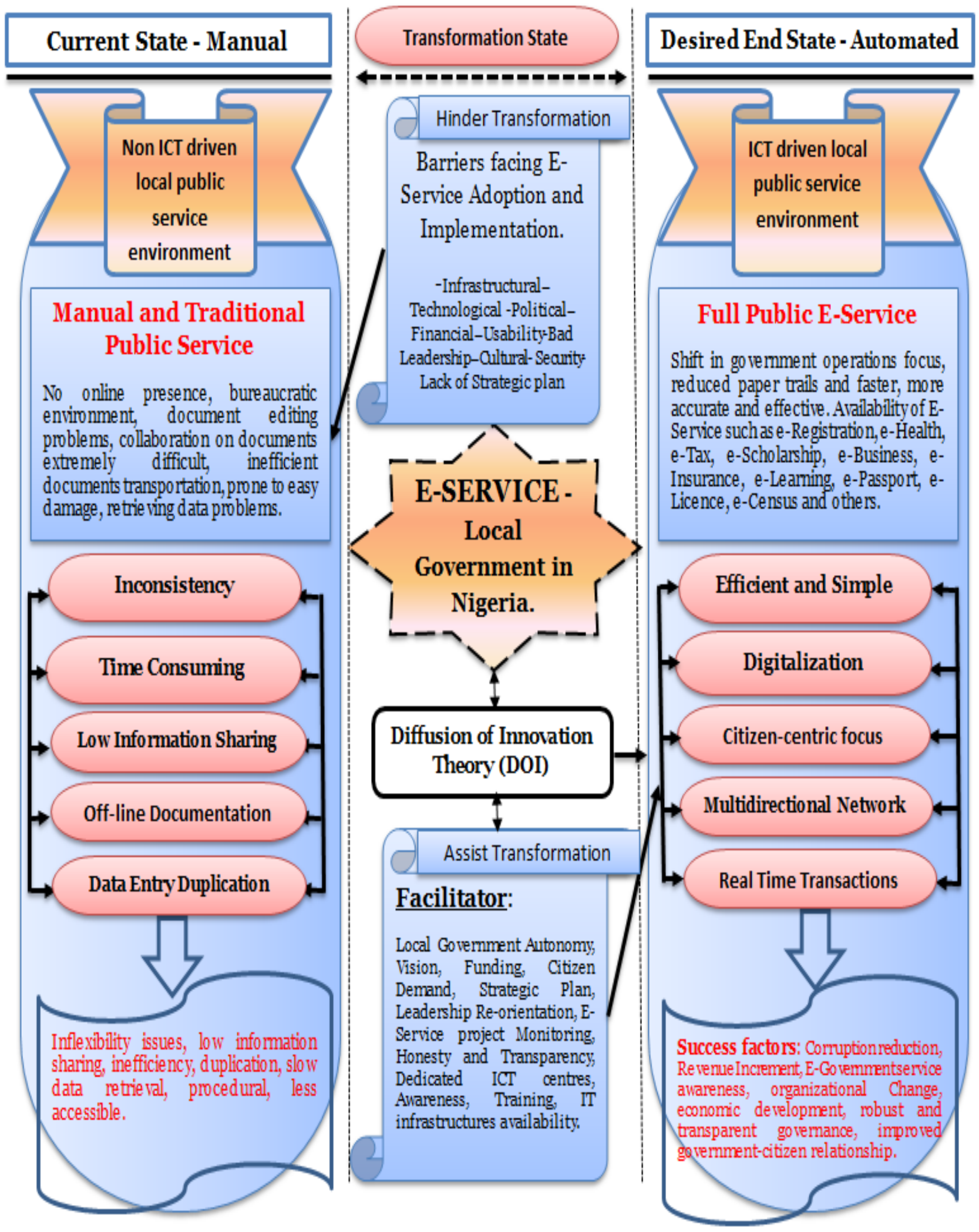

**Figure 71 Final E-Service Framework**

The framework validation meeting agreed with the focus group participants during the last meeting. Invitation and consent letters were sent to the participants. Copies of the letters are available in the **Appendix A**. The validation was carried out on 15 December 2016, and the results are shown in **Figure 73** below. The questions asked are also shown in **Figure 75** and Appendix A. **Figures 72** and **74** below provided the citizens at the local environment level with the necessary local E-Government services such as e-tax, e-payment, e-voting, e-business, e-registration, e-licence, and so on. The E-Service framework can be assessed at http://osekaz.wixsite.com/e-governmentservices. The E-Services are available through the Local Government Service Commission.

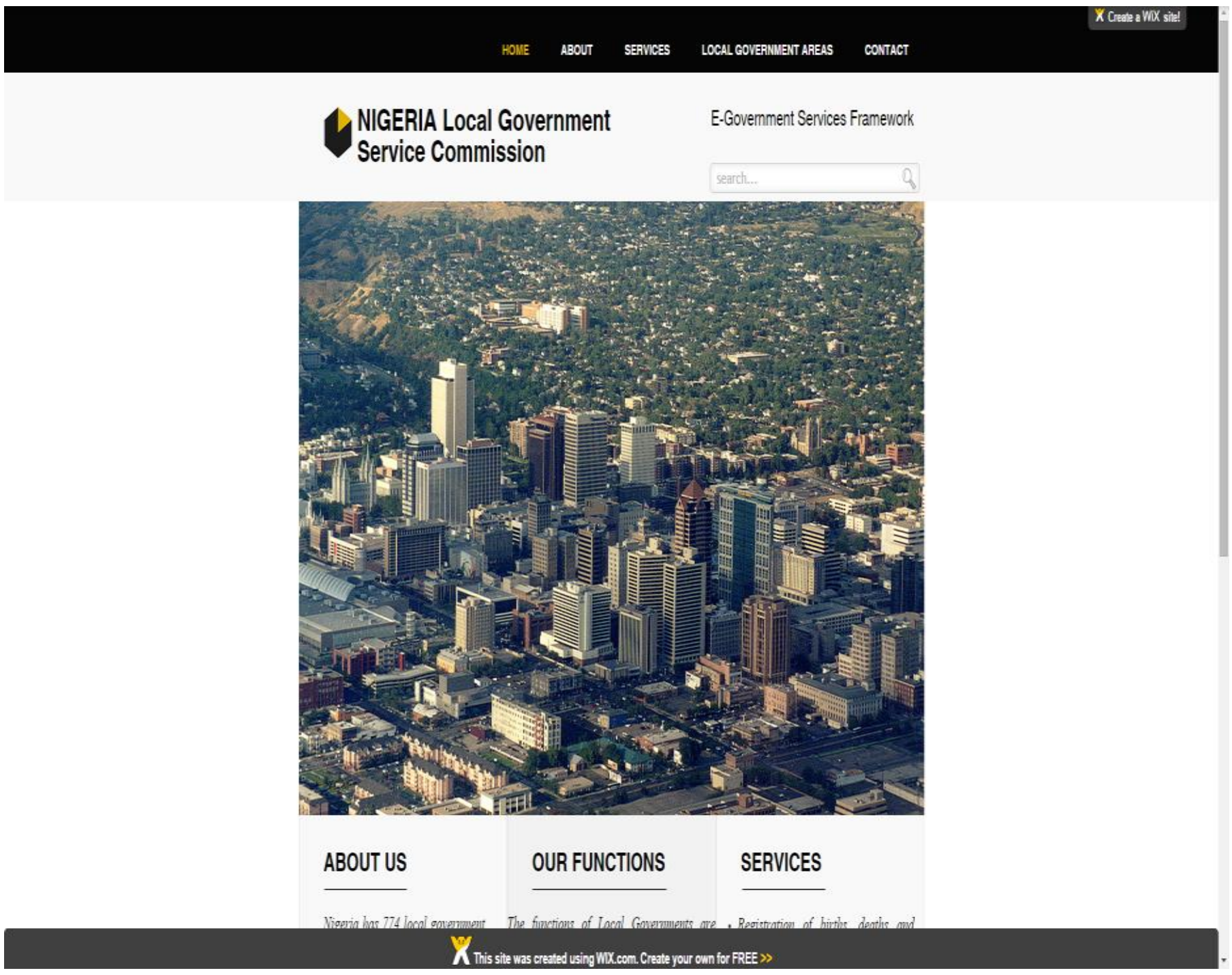

**Figure 72 Local E-Government Services Framework**

However, the study theory has a role to play in the validation process, as there is a need to look at why certain innovations spread more quickly than others and why others fail (Robinson, 2009), as discussed in **Chapter 2**. The researcher agreed with Robinson (2009), Aizstrauta et al. (2015) and Dibra (2015) in that the five qualities/attributes of DOI, as displayed in **Table 31** below, determine the success of innovation. Hence, the local E-Government services framework would be validated based on these five attributes of DOI. The attributes of DOI and innovations questions used by the researcher for the validation purpose have been previously used and validated by Hubbard and Sandmann (2007). The questions are available in **Appendix A**. The results shown in **Figure 73** on the final framework indicated that the participants agreed that the framework is the appropriate one for this study, as it addresses the five attributes of DOI theory. Also, the framework has met with the goals and objectives set out in the E-Service framework at the local environment level in Nigeria.

**Table 31 Attributes of DOI for Framework Validation**

| | Attributes of DOI | Criteria | Description |
|---|---|---|---|
| 1 | **Relative advantage** | **i**. Economic profitability | Economic profitability is an advantage of using this technology. |
| | | **ii**. Low initial cost | Low initial cost is an advantage of using this technology. |
| | | **iii**. Decrease in discomfort | Decrease in some kind of discomfort is an advantage of using this technology. |
| | | **iv**. Social prestige | Use of this technology advances the social prestige of the user. |
| | | **v**. Savings of time/effort | Saving of time and/or effort is an advantage of using this technology. |
| | | **vi**. Immediacy of the reward | The benefits of using technology are immediate and that is an advantage of using this technology. |
| | | **vii**. Social/Cultural values and beliefs | The use of technology is positioned as compatible with social/cultural values and beliefs. |
| 2 | **Compatibility** | **i**. Previously introduced ideas | The use of technology is positioned as compatible with previously introduced ideas. |
| | | **ii**. The client needs | The use of technology is positioned as compatible with client needs. |
| 3 | **Complexity** | **i**. Complexity of technology | The technology is positioned and should be perceived by potential users as easy. |
| 4 | **Trialability** | **i**. Trial availability | There are mechanisms (free search, downloads, trial versions, prototypes), that enable the users to easily try the technology. |
| 5 | **Observability** | **i**. Observability of technology | The results and benefits of technology is easily visible by potential users. |

The results further show that the E-Service framework represents the participant's notion of E-Government services. As shown in **Figure 73** below, the participants agreed that E-Service adopting would increase the local government revenue base, reduce corruption, block revenue leakages, and it would save time/effort used in manual file administration at the local government level. This is the **Relative Advantage feature of Diffusion of Innovation** (**DOI**) in having the E-service framework at the local environment level in Nigeria. **Relative Advantage** shows the degree to which innovation is perceived as better than the existing standard.

| | Does the E-Service framework represent your notion of E-Service | Were the goals and objectives of the framework adequately addressed through the interactive E-Service website | Do you feel this website will increase the revenues at the local government level and saving of time/effort | Do you feel the website is easy to understanding and is interactive enough for the users | Do you feel the website is compatible with the ideas introduced in the framework and that it will meet the user's needs | Do you feel the results and benefits of E-Service adoption are visible by the users through the use of the website | Do you feel that you can test this website reasonably through the mechanisms such as free search and so on | Do you feel the identified barriers from this study represent your notion of the barriers facing E-Service adoption and implementation at the local government levels in Nigeria |
|---|---|---|---|---|---|---|---|---|
| Participant 1 | Yes | Yes | Yes | Yes | Yes | Yes | Yes | Yes |
| Participant 2 | Yes | Yes | Yes | Yes | Yes | Yes | Yes | Yes |
| Participant 3 | Yes | Yes | Yes | Yes | Yes | Yes | Yes | Yes |
| Participant 4 | Yes | Yes | Yes | Yes | Yes | Yes | Yes | Yes |
| Participant 5 | Yes | Yes | Yes | Yes | Yes | Yes | Yes | Yes |

**Figure 73 Results of the Validation of the Final Framework and Barriers**

Moreover, the framework validation results also indicated that the participants agreed that the framework is easy to understand (**Complexity feature of DOI**) and interactive enough for the users. The **Complexity** features explain how easy it is for the E-Service users to understand the new ideas, or how to use them. The participants also concluded that the framework passed the compatibility test (**Compatibility feature of DOI**) with features introduced in the framework, and it would meet the user's needs. The **Compatibility** features explain how easy the E-Service users could use their previous experience to understand the new technology. The Compatibility factors improve the chances of technology adoption (Dibra, 2015). Furthermore, the participants decided that the framework can be reasonably tested through the mechanisms, such as free search (**Trialability feature of DOI**). The **Trialability** features show how effortless the E-Service users can interact with the new technology. The more that potential users can test the technology, the more likely users will adopt it. The participants also agreed that the results and benefits of having E-Services at the local government levels in Nigeria are very visible to the users (**Observability feature of DOI**) during the framework validation. The more visible or noticeable a technology is, and the more likely users will adopt it.

### 7.3.1 E-Service Framework Evaluation

The E-Service DOI-based framework developed for this study would be evaluated using the five attributes of DOI theory. The use of DOI attributes to evaluate the DOI-based framework developed for this study is in line with Robinson (2009), Aizstrauta et al. (2015) and Dibra (2015), as the agreed DOI attributes determine the success of technological innovations. Hence, the approaches to evaluating the framework are described below:

A. **Relative Advantage of the Framework**: The relative advantage evaluation of the E-Service framework achieves by sending out surveys to the users and E-Service stakeholders at the local level to check the benefits of E-Service deployment, as seen in the framework. This includes checking whether technology usage saves time and effort compared to the traditional/manual services. Other framework evaluation includes checking if the technology is a cost advantage, brings economic profitability, and the use of technology is promoting social and cultural values among the citizens.

B. **Compatibility of Framework**: Surveys were sent out to the stakeholders at the local government to understand what they need to change in the system and when to carry out the changes. There is no immediate reward for online services in Nigeria as E-Government is still at the transformation stage (Mundy and Musa, 2010). Hence, moving to online services from the traditional/manual services requires the stakeholders to critically evaluate the framework to check for consistency, and to petition decision-makers for change.

C. **Complexity of the Framework**: The researcher sent out surveys termed a Human Computer Interaction (HCI) survey and engaged the online focus group members on the usability of the framework. This evaluation is essential to the successful implementation of E-Services at the local level in Nigeria. The responses would suggest the framework is perceived as easy to use by the stakeholders/users so that improvement could be made where necessary.

D. **Trialability of the Framework**: This part of the framework's evaluation is done by setting up the online focus group members' webpage (Local E-Government Services). Different software would be uploaded by the researcher for the stakeholders and users to test. The testing includes downloading and updating the software.

E. **Observability of the Framework**: The observability of the framework is done by inviting the potential users, including the local government staff, to move around the local government and observe whether the technology deployed is working as expected or not.

### 7.3.2 Final E-Service Framework Services

There are many services that the E- service framework will offer the citizens, as explained in **section 2.3.3** of the literature review in **Chapter 2**. The citizens will have access to online services such as e-tax, e-payment, e-voting, e-business, e-registration, e-licence, and so on, while browsing through the site as represented in **Figure 74** below. This would enable the citizens to have prompt access to the government information, and promote robust governance including making an impact on the government decision-making process.

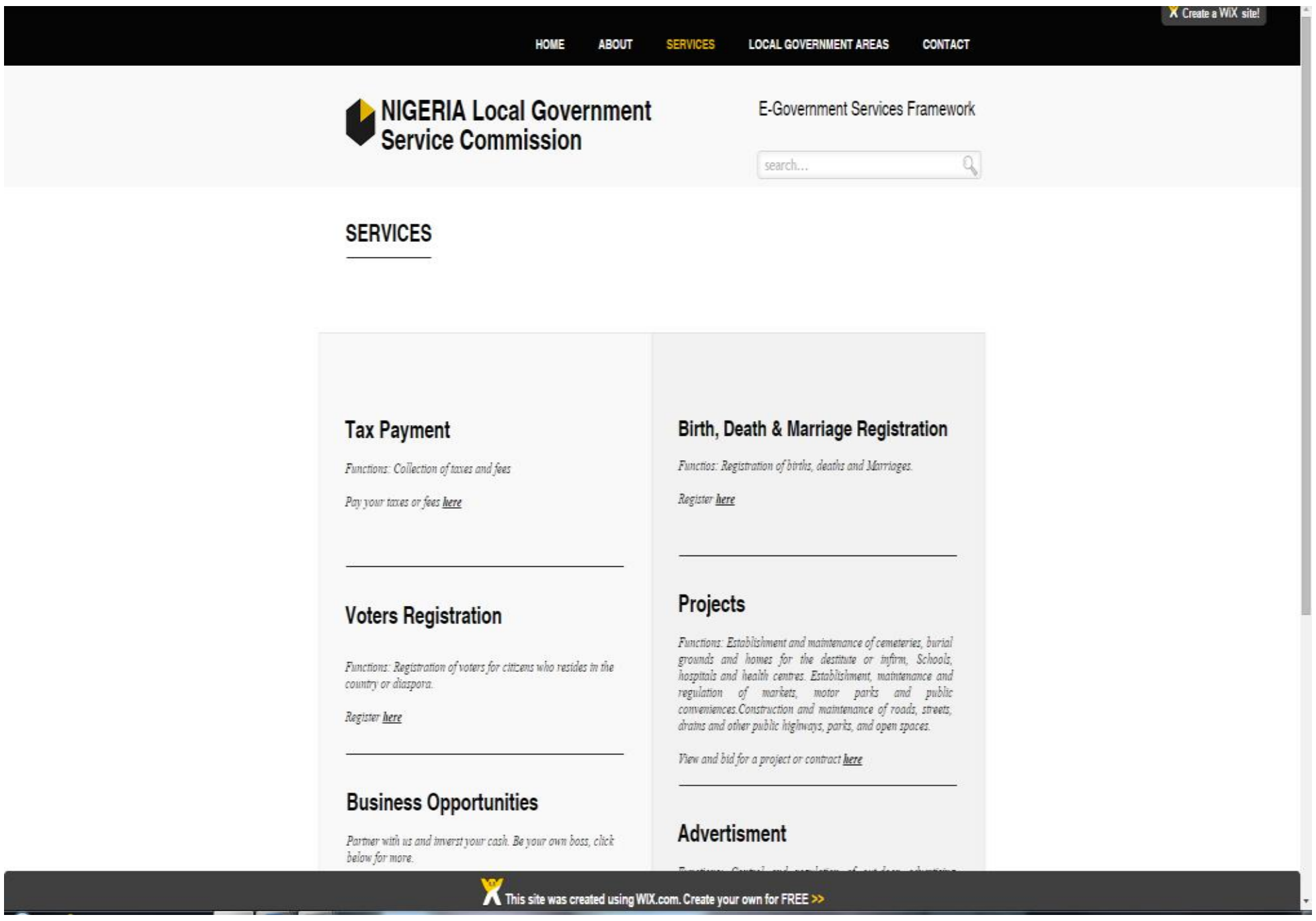

**Figure 74 Public Local E-Service**

Moreover, the government at the local environment level in Nigeria will also benefit from the E-Service framework proposed in this study, as previously discussed in the investigated/identified success factors for the adoption and implementation of E-Services at the local government level in Nigeria. Hence, the E-Service framework will help in the following areas:

- To save costs, and run a cheaper and easier administration by the local governments.
- To encourage transparency and accountability in government.
- To prevent corruption in government (Aizstrauta et al., 2015).
- To eradicate geographical boundaries and the need to travel in order to access local government information, which will now be possible through online access.
- It will prevent ghost workers at the local government level, as workers' salaries will now be paid electronically into their bank account (Aizstrauta et al., 2015).
- It will bridge the digital divide and help with community networking.

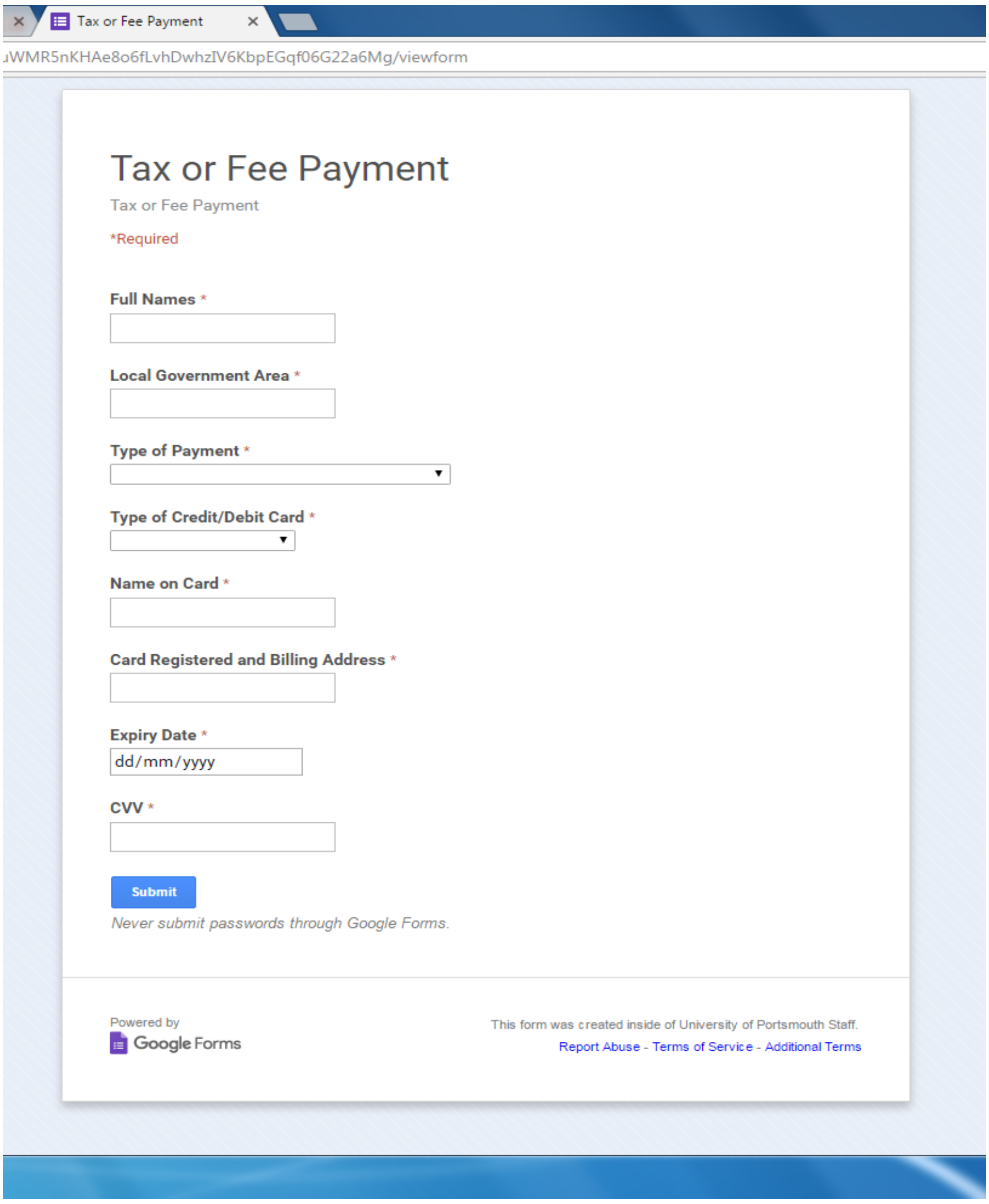

**Figure 75 E-Payment Portal**

**E-Payment**: This enables the citizens to make online payments to the local government, such as the payment of shop licence renewal, radio and television licence fees, birth and death registration fees, council tax, vehicle licence renewal, and so on. This facility, as shown in **Figure 75** above, will help to reduce the stress and time needed to make the payment in person and will prevent double-charging, as is sometimes experienced in Nigeria, where different government officials collect taxes twice, and the money is not remitted to the government account. Therefore, the citizens would make use of an e-payment facility from anywhere in the world at their convenience. An example of an E-Payment process is demonstrated in **Figure 76**.

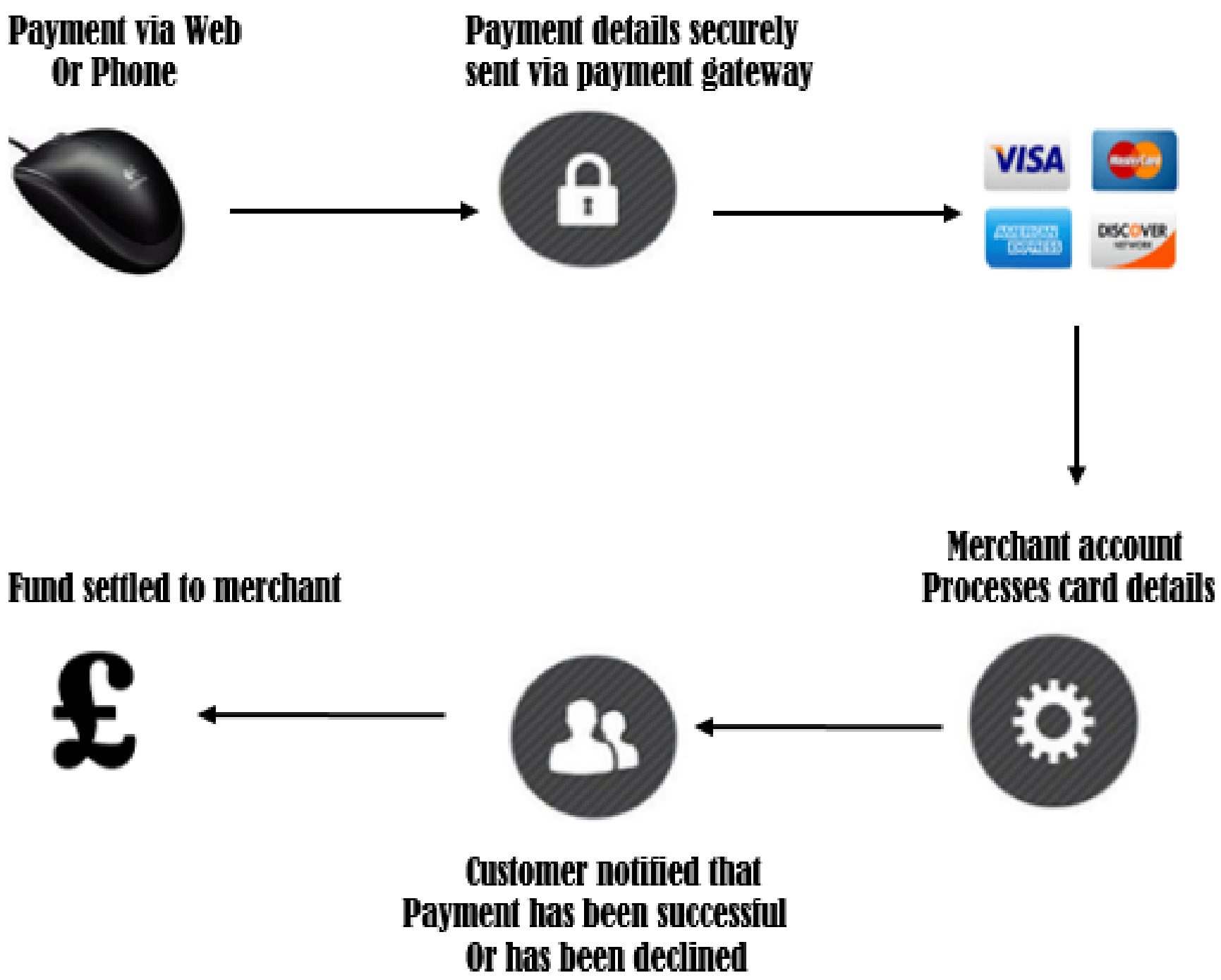

**Figure 76 E-Payment Process**
(Adapted from Aparna et al., 2017)

**E-Voting**: The E-Service framework platform provides the citizens at the local government level in Nigeria the opportunity to register online to vote in the general election. The citizens who reside locally or in the diaspora can register irrespective of their location. They will create a username and a secure password, and the E-Service framework platform has been secured against a security breach as a result of a cyberattack with the authentication and security features. Once they register, the link will automatically direct the users to the national independent electoral website to complete the e-voting registration.

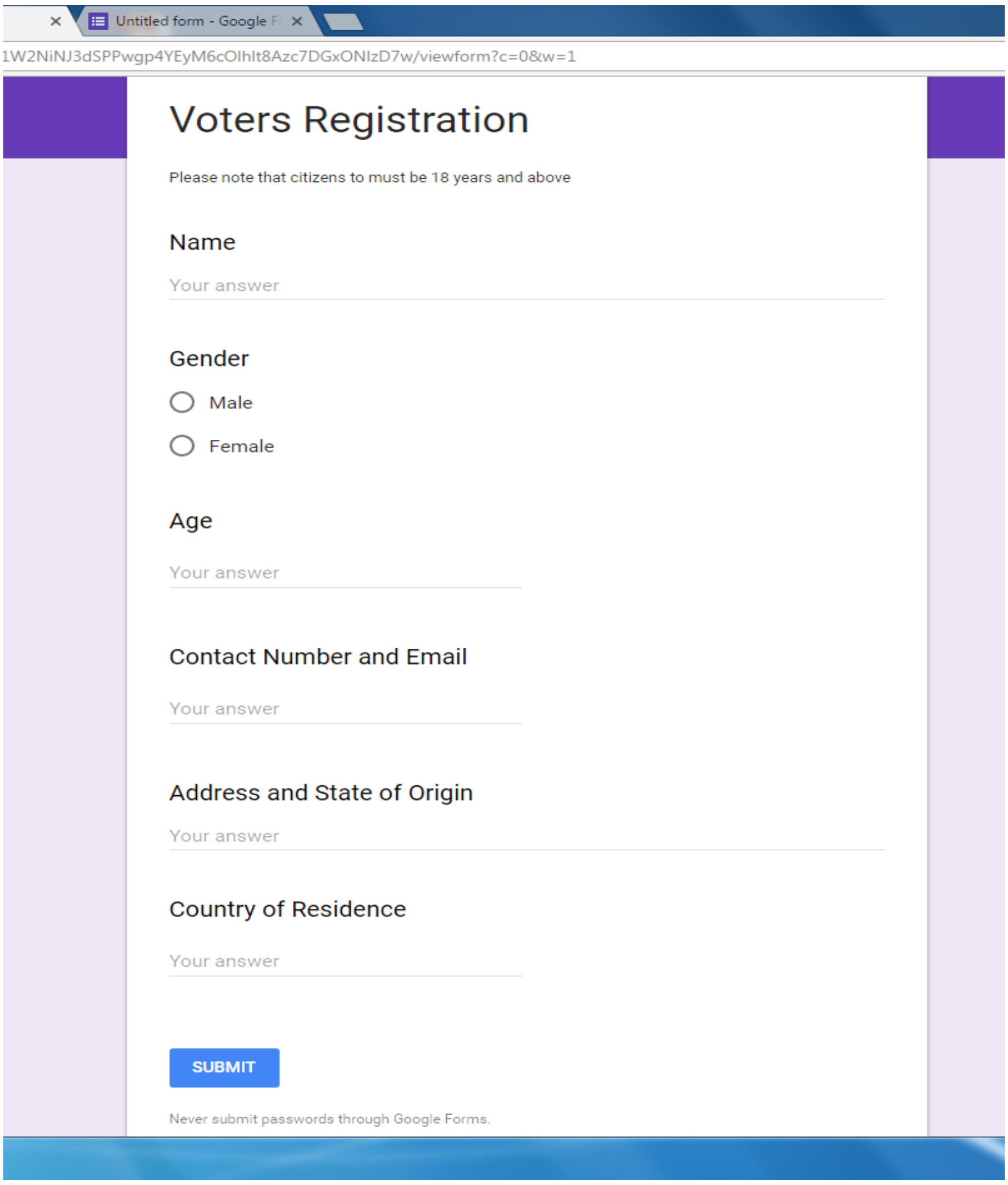

**Figure 77 E-Voting Portal**

## 7.4 Validation of the Identified Barriers Facing the E-Service Adopting and Implementation at the Local Government Level in Nigeria

The identified barriers were also validated by both the interview and focus group participants. **Figure 78** shows that the participants agreed that the identified barriers represent their notion of the barriers facing E-Service adopting and implementation at the local environment level in Nigeria. One interesting finding in the identified barriers for this study is the issue of ***local government autonomy*** which is a new barrier to emerge in E-Government research domain, and this is not part of the barriers that previous studies have suggested. **Figure 78** below shows the identified barriers as emerged from both the interviews and the online survey data for this study.

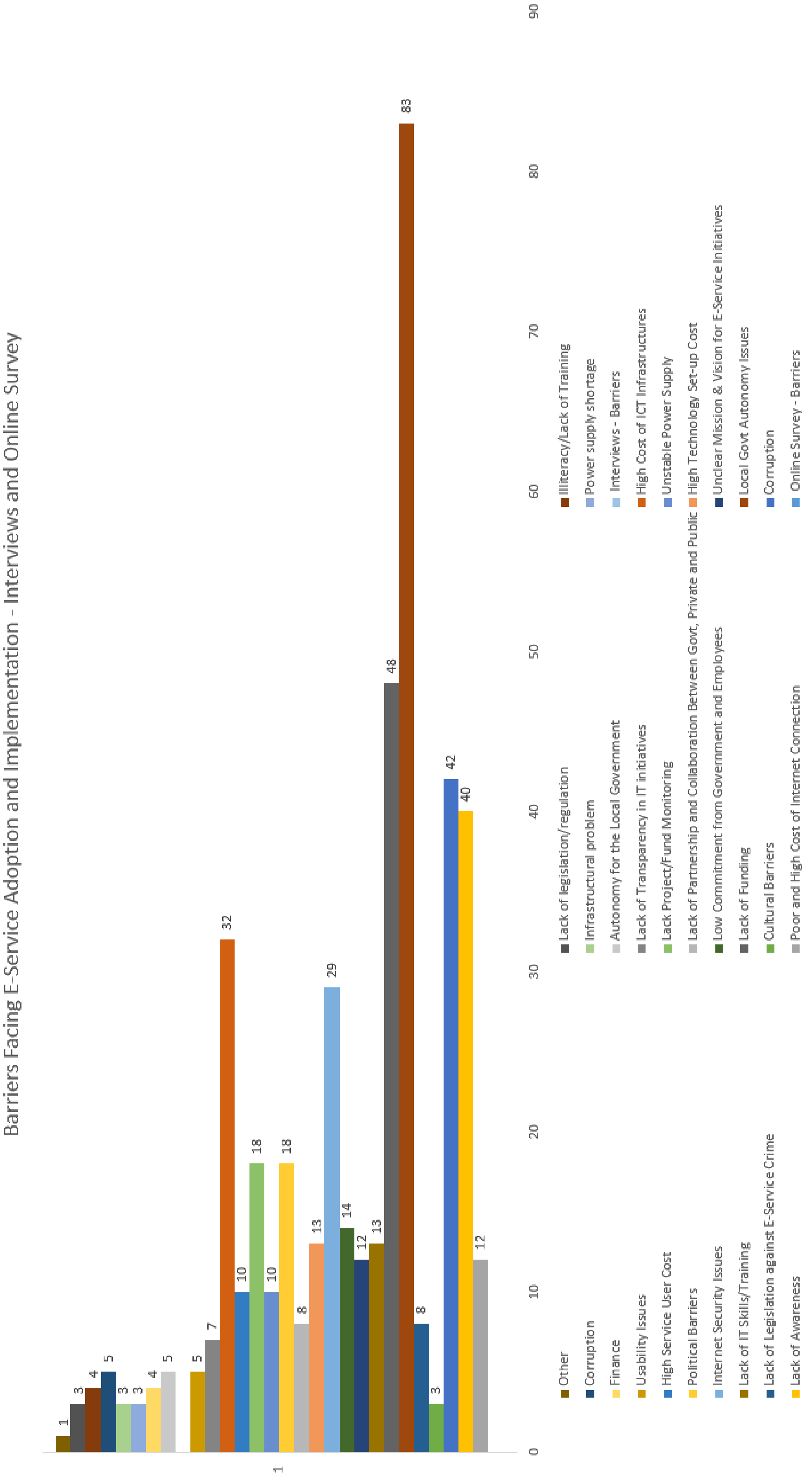

**Figure 78 Identified Barriers Facing E-Service Adopting and Implementation**

## 7.5 General Findings Feedback Sessions

The validation session was carried out with the participants initially at the interview stage and later with the online focus group. The five participants for the online focus group, as shown in **Table 32**, were selected from those who volunteered their contact details during the one-on-one interviews stage of research. Questions were asked relating to the barriers facing the E-Service adoption and implementation, as well as other E-Service initiatives, including the framework. Most of the participants were the senior officers at local government level, and who are also some of the stakeholders involved in E-Service projects. The essence of the validation is to obtain honest comments from the participants regarding the overall research findings after analysing the data collected. Before this, consent and invitation letters were sent out to the participants with full explanations regarding the validation session, and that gave them the opportunity to get themselves prepared. The first comment was made by Participant A, a commissioner at the local government service commission, who felt the findings of this research will be of great help to the E-Service initiatives. However, he was a bit worried about the continuous usage and implementation of the framework due to barriers, such as corruption, that have been identified in this research. The researcher explained that the E-Service framework provides the guidelines that would assist in the E-Service implementation, and maintained that the issue of corruption could be tackled with anti-corruption policies, as explained in the theory of change in **Section 6.4.2**. The anti-corruption model will prevent fraud (e.g. ghost workers).

Furthermore, Participant B, another senior officer, considered the identified barriers, the solutions, and the success factors would be worthy and satisfying, even within a limited time frame for the researcher. Participant B commended the outputs of the framework, which clearly outlined the stages involved in the E-Service project development. Also, the major issue for this participant is how to sustain the relationships created as a result of E-Service project initiatives, as the change in leadership might have an effect on these projects if not well managed. The participant maintained that a transition framework should be put in place in case of sudden leadership change; this is so that the relationships among the E-service managers at the local government levels in Nigeria will forever be maintained to prevent data security breach. The researcher added that the relationships could be retained, even with the change of government if E-Service implementation policies are provided which the framework covered.

Lastly, Participant C, who is a director, believed that the findings and the strategies involved in the framework could easily be used for a successful E-Service initiative implementation. Participant C maintained that the E-Service framework would be of great help in understanding the barriers affecting the adoption and implementation of E-Service. The participant argued whether the framework could work alongside the existing ones at the national level in Nigeria. The researcher explained that the scope of this study is finding the barriers facing the E-Service adoption and to develop an E-Service framework to assist in the E-Service implementation at the local government levels in Nigeria, where the main beneficiaries are the local citizens. The researcher maintained that the E-Service framework could work with any other framework at the national level if the stakeholders are sincere with the E-Service implementation.

## 7.6 Summary

The chapter summary describes the steps that the researcher has taken to validate the E-service framework, including the techniques adopted. As illustrated in **Figure 68** above, this chapter includes the introduction, validation of the final framework, the validation techniques, the validation of the identified barriers, and the general findings feedbacks. The adopted validation methods used are: triangulation, multiple coding, purposive sampling, and respondent validation. A full explanation of the processes involved was given to the participants during the interviews and validation sessions.

Although the research set out to explain the framework evaluation procedures, and some evaluation of the usefulness of the framework was done, it is up to future research to test and thoroughly validate the E-Service framework. The next chapter will be the discussion and conclusion chapter for this thesis.

# Chapter 8 – Discussions and Conclusion

## 8.1 Introduction

The main purpose of this study was to investigate the barriers facing the E-Service adopting and implementation at the local environment level in Nigeria. The main motivation behind this investigation was to assist the local government service commission to develop an E-Service framework that will be suitable for the local government levels in Nigeria as compared to the developed countries. The literature review presented in Chapter Two, together with the review of relevant E-Government models and framework development in Chapter Three, provided by the researcher, there were some research gaps (see **Section 1.4.2**). To achieve the research aims and objectives, and to enable the research to answer the research questions, an appropriate and suitable research method was selected. This was outlined in the research methodology, in Chapter Four. The techniques involved in the data collection and analysis, and the selection of the organisation and participants, were described in Chapter Five. In Chapter Six, the detailed research findings were presented, which include the investigated/identified barriers, lessons and reflections from action research phases, the responses from the respondents, the application of the theory of change and, the E-Service implementation process. The validation of the research findings was conducted in Chapter Seven, and this includes the validation processes for both the final research framework and investigated barriers, and the strategies involved in the validation that confers thoroughness on research findings. The general feedback sessions were also discussed in this chapter.

Hence, the purpose of this chapter is to present and draw the final conclusions on the overall research findings, as emphasised in the thesis. Also, this chapter will provide a summary of the research, the key research contributions to knowledge, the overall research implications, research limitations, and the recommendations for future research. This chapter comprised nine sections. **Section 8.2** provides a discussion of the answers to the study research questions in Chapter One. In **Section 8.3**, the summary of the research findings compared to the findings in the literature were highlighted. **Section 8.4** provides the discussions on the research aims and objectives, as these are summarised against the key findings obtained from this study. The key research contributions to knowledge obtained from this study are presented in **Section 8.5**. The overall research implications are explained in **Section 8.6**, while **Section 8.7** gives the detailed research limitations. **Section 8.8** provides a discussion of the recommendations that emerged from this study, including the recommendation for future research. The chapter concludes with a summary, as shown in **Figure 79** below.

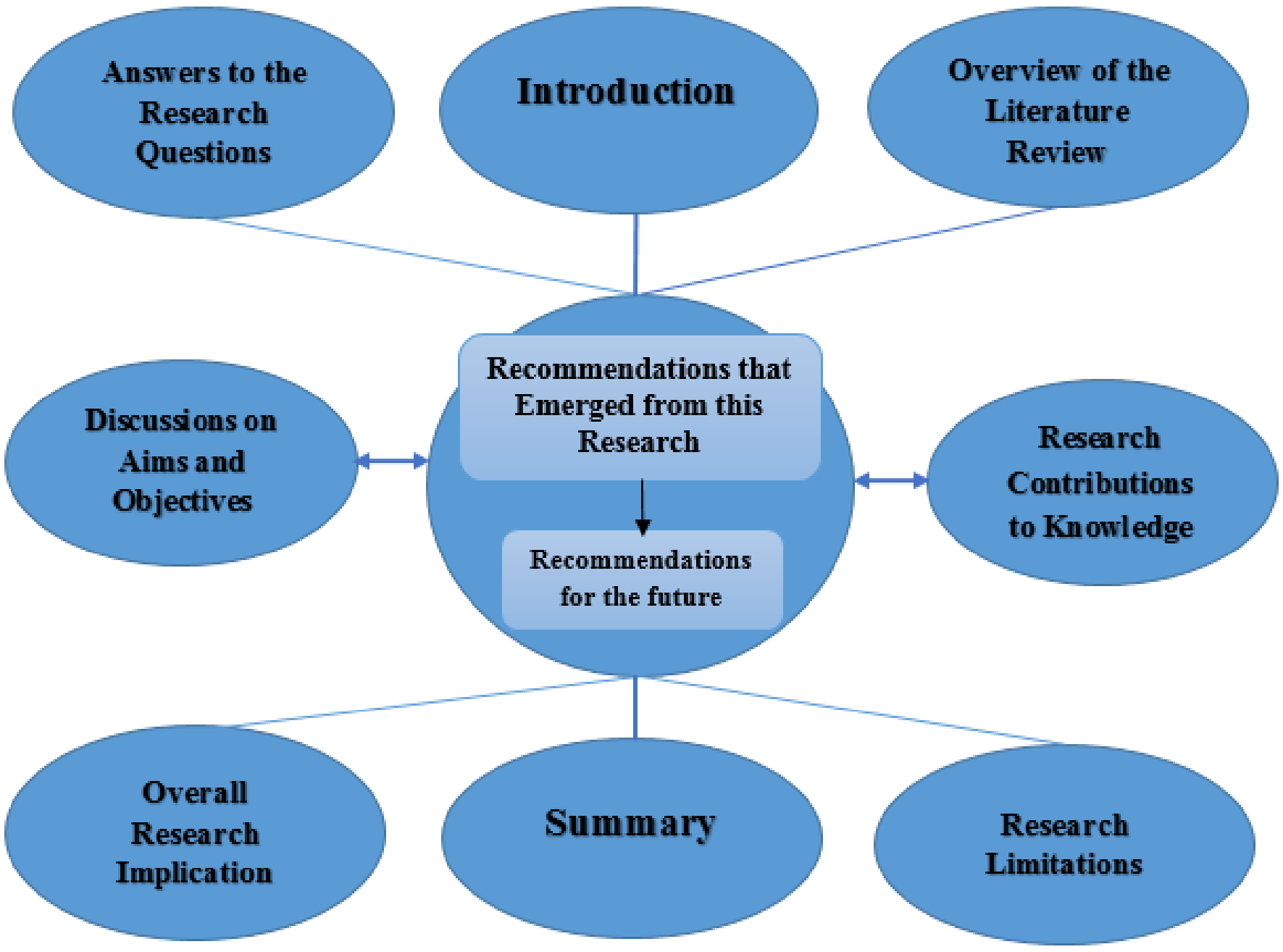

**Figure 79 Chapter Eight Summary**

## 8.2 Answers to the Study Research Questions in Chapter 1

This section provides the summary of the key findings derived from this study that will address the research questions, as presented in **Section 1.5.1** of Chapter One.

In addressing the first research question: **What are the Barriers Facing E-Service Adopting and Implementation at Local Environment Level using Nigeria as a case example**? Firstly, the researcher answered the question by reviewing the various barriers facing E-Service adopting and implementation from previous studies found in the literature. The barriers reviewed range from budget/finance issues, organisational or staff resistance issues, privacy/security and trust issues, the cultural problems. Others are lack of IT skills/literacy/training, lack of ICT infrastructure, lack of partnership and collaborations, lack of strategic planning, lack of leadership and management supports, the high cost of internet connection, and unstable power supplies, especially in many development countries (Oseni and Dingley, 2014). Moreover, there is a lack of maintenance of E-Service systems. Secondly, in addition to the barriers reviewed in the literature, the first research question was also answered by investigations. The methods used to investigate the barriers facing the E-Service adopting and implementation at the local environment level in Nigeria include interviews, online survey, and focus group meetings.

The results obtained and presented in both **Sections 6.2.1** and **6.3.4**, respectively, show that the barriers include poor quality and the high cost of internet connections, lack of awareness of e-service initiatives, and corruption. Other barriers include cultural barriers, lack of legislation against e-service crime, local government autonomy issue, lack of funding, lack of IT skills/training, unclear mission and vision for e-service initiatives, and low commitment from the government and employees. Also, human factors such as internet security issues, high technology set-up costs, lack of partnership and collaboration between the government, private and public, and political barriers. The barriers include infrastructure factors such unstable power supplies, high service user costs, and a high cost of ICT infrastructures. Lack of project/fund monitoring, lack of transparency in IT initiatives and, finally, usability issues are part of the barriers found. These barriers are also categorised into various groups according to Alshehri and Drew (2010); Munda and Musa (2010); Hassan (2011); Alshehri et al. (2012); Khan et al. (2012); Alateyah et al. (2013); Al-Shboul et al. (2014); Abdelkader (2015). The barrier groups are Administrative/Organisational, Economic, Political, Legislative, Technological, and Cultural Barriers.

In answering the second research question: **Is it possible to develop an E-Service Framework that will be suitable for the Local Environment Level in Nigeria as compared to the Developed Countries**? The answer is Yes. Having reviewed various E-Service models, as shown in Chapter Three, and from the proposed E-Service framework (**Figure 71**), it is possible to develop an E-Service framework that will be suitable for local environment level in Nigeria, as opposed to a framework needed for developed countries. The features of some reviewed E-Service frameworks and the implementation in the developed countries were useful in the E-Service framework developed for this study.

## 8.3 Literature Review Overview

The essence of this section is to compare the overall research findings against the literature. An argument has been made that the barriers reviewed in the literature do not totally portray all the barriers facing the E-Service adopting at local environment level in Nigeria. For example, if the infrastructural barrier for the E-Service implementation in some countries is due to environmental issues such as earthquakes, tsunamis, landslides or volcanic activities, this is not applicable in Nigeria, as such environmental hazards are not common in Nigeria. The infrastructural barriers affecting E-Service adopting and implementation in Nigeria will be the non-availability or high cost of ICT amenities. However, there are several issues arising between the literature and the research findings from this study.

The first issue is that the literature argued that E-Service as a term is not only about "electronic" and "service", but the true e-service operation may be part, if not all, of the interactions taking place between service providers and customers over the internet. This was substantiated by Surjadjaja et al. (2003). Moreover, Oseni and Dingley (2014) pointed out that E-Service has moved from the usual manual and traditional way of rendering service to electronic service provision for the public, and that there are several reasons for implementing these services. Kumar et al. (2007) suggest that the quality of services provided to businesses and public could be improved drastically with E-Services. An example is the automated tax payment at both the state and national level in Nigeria (Oseni and Dingley, 2015). The E-tax prevents fraud and reduces corruption because all taxes are remitted through a unified automated system, which enhances E-Service acceptance generally (Oseni and Dingley, 2015). The interviews conducted for this study with the participants also confirmed that E-Service is the interaction between the government and the citizens through the internet, with the aim of ensuring the provision of efficient and real-time public services to the people. One senior IT official explained that:

> "The aim of E-Service is to disseminate government services faster via the internet and should be accessible from any part of the world, regardless of the time and distance. In my opinion, even though the state governments in Nigeria have recorded substantial success over the years in E-Service implementation, there is a need to provide dedicated

> ICT centres at the local governments in Nigeria so that the rendering of local government services could spread easily to the public."

In addition, the research findings pointed out that the barriers are similar apart from a few, such as autonomy and corruption issues, which are the barriers specific to Nigeria that emerged from this study, as discussed in **Section 6.3.4**. In Nigeria, the local government's power to make an independent and bold decision on E-Government deployment may run into chaos with the State governments, due to the **State-Local Joint Account** issue (Okafor, 2010). The **State-Local Joint Account** is the account recognised by the Nigeria Constitution for the local government budget allocation to be paid into by the Federal Government. However, the local government cannot access the fund without the consent of the State Governor (Okafor, 2010). The solution is for the legislatures in Nigeria to enact laws that would address the lack of local autonomy at local government levels and separate the joint **State-Local** account. The joint account separation would give the local governments in Nigeria the much-needed independence to make a bold decision towards initiatives such as E-Services.

Moreover, the research findings pointed out how corruption among the government officials in local government in Nigeria has derailed the implementation of previous E-Service initiatives. Nabafu and Maiga (2012) explained that money meant for projects in local government in Nigeria is being diverted into private accounts. Ugaz (2015) suggested that corruption could be stamped out if abuse of power and bribery is stopped by government officials, and that there is a need for the officials to operate a transparent governance and have attitudinal change towards corruption. Many of the interviewees agreed that local government officials' attitudinal change towards corruption would have positive effects on how easily E-Services projects could be implemented in local government in Nigeria. The findings on corruption (staff attitudinal change) give way to a theory known as the **Theory of Change** as applied by the researcher in **Section 6.4.3**. A commissioner at the local government service commission in Nigeria expressed that:

> "Corruption is the major barrier facing the E-Service adoption and implementation at the local government levels in Nigeria. There is a need to institutionalize the government's services, more support needed from the government on political will since E-Services will block leakages in the system against the personal interest of the political leaders."

Lastly, these research findings placed more emphasis on the corruption of the government officials as a barrier facing the E-Service adopting and implementation at the local environment levels in Nigeria. Hassan (2011), on the other hand, pointed out that the employees' resistance to change, or rejection of technology, is due to their lack of understanding of the benefits of E-Service projects. This, in turn, leads to both intentional and unintentional interruption of the E-Service initiatives (Hassan, 2011).

## 8.4 Discussion of the Research Objectives

This section presents the summary of the key research findings obtained from the research evidence for the purpose of answering the research objectives of this study.

### Objective 1: To Identify the Barriers facing E-Service Adopting and Implementation at Local Environment Level using Nigeria as a case example.

From the research evidence, this objective has been achieved as the research has identified the groups of barriers facing E-Service adopting and implementation at local environment level in Nigeria, as explained in **Section 6.2.1**. The research findings show that the barriers include poor quality and the high cost of internet connections, lack of awareness of e-service initiatives, and corruption. Other barriers include cultural barriers, lack of legislation against e-service crime, local government autonomy issues, lack of funding, lack of IT skills/training, unclear mission and vision for e-service initiatives, and low commitment from the government and employees. Also, human factors such as internet security issues, high technology set-up costs, lack of partnership and collaboration between the government, private and public, and political barriers. The barriers include infrastructure factors such unstable power supplies, high service user costs, and the high costs of ICT infrastructures. Lack of project/fund monitoring, lack of transparency in IT initiatives and, finally, usability issues are part of the barriers found. The research findings also revealed that the nonchalant attitudes of many government officials towards the non-implementation of the E-Service projects were not because of the lack of funds or infrastructures, but mostly due to the massive levels of corruption within the government.

Furthermore, this study discovered that the awareness level among the citizens on E-Service initiatives is very low (Oseni and Dingley, 2014). The achievements recorded so far on E-Service projects at both federal and state levels in Nigeria should be repeated at local government levels, especially as the location of many local councils are very close to the citizens. As discussed in **Section 8.3** above, there is an urgent need for the lawmakers in Nigeria to change laws that will address the lack of local autonomy at local government levels. The research finding shows that many local government functions have been taken over by the state governments, thereby rendering the local governments functionless. Local government finances are no longer independent due to a local-state government joint account. The clause binding the local and state governments together must be removed to enable credible initiatives, such as E-Government services.

Lastly, the research findings also show the success factors for the adopting and implementation of E-Services at the local environment level in Nigeria, as explained in **Section 6.4.2** (Chapter Six) of this thesis. The success factors from the research findings are quick access to information and online resources, accountability, bridging the digital divide, opening business opportunities, and a cashless economy. Others are cheap and easy services, reducing corruption, improving the economy, improving the revenue base, creating employment, boosting record-keeping, and transparent government. Another success of the E-Services adoption and implementation is the stoppage of fund leakages in government. E-Service availability at the local level will promote the timely availability of government information online. The citizens will have easy access to the information, thereby improving the opportunities for them to contribute to the decision-making process at the local government.

Moreover, the adoption of E-Services will help in bridging the digital divide and narrowing the disparity between the spread of internet access in developed and developing nations (Helton, 2012). Also, the full implementation and adopting of the E-Services at the local environment level in Nigeria will help to promote openness, transparency, and accountability and, in the process, reduce corruption. Another success of the E-Service adoption, according to the research findings, is the cashless economy. Presently, the banking sectors in Nigeria are benefiting immensely from this after the introduction of cashless banking by the CBN in 2012 (Ayoola, 2013). Most banking transactions now take place over the internet. According to Ayoola (2013), the introduction of the cashless economy is helping to reduce corruption, and ghost workers are finding it a lot harder to hide in an objective, standardised electronic salary system. Salaries are now being paid directly to the workers' bank accounts after the government conducted physical verification exercises. The era of government officials collecting multiple salaries has stopped, and this has blocked leakages of government revenue. Hence, as a result of E-Service adoption, the cashless payments at the local government level could save citizens time when making payments (e.g., council tax) at the bank. Now, it can easily be done online, anytime and anywhere in the world.

### Objective 2: To Learn from the E-Government Services in Developed and Developing Countries.

This objective has been achieved through studying various E-Government service initiatives in both developed and developing countries, as presented in **Section 2.5** of the literature review. A table of E-Government practices in some countries was also presented in **Figure 13** that explains different stages of the E-Government in these countries, as well as strategies in place to showcase E-Service projects to the citizens. This research study examined the E-Government practices in many countries such as the UK, USA, Taiwan, Singapore, Saudi Arabia, Tunisia, Mauritius, South Korea, Nigeria, and others. However, there is still a gap between the developed and developing countries regarding E-Services projects, especially in the areas of ICT infrastructures, the adoption and implementation level, and practices (Chen et al., 2006; Hassan, 2011). The gap could be closed in many developing countries if their respective governments could collaborate with private companies and individuals in the provision of the necessary infrastructure and funding needed to achieve the full implementation of E-Service projects. Regular IT training is also needed, especially for the local government officials, where sufficient knowledge and skills are required to build and develop effective strategies in developing countries to promote E-Government services.

In reference to some developed countries such as the UK, USA, and Japan, the research finding shows that their governments and economies developed soon after independence. Their economies grew at a constant rate, while productivity increased, with a high standard of living. They have a long history of democracy and more transparent government policy and rule (Hassan, 2011; Nawafleh et al., 2012). There is a high level of internet access and computer literacy in developed countries, but there are still a few digital divide and privacy issues. Developed countries have relatively more experience in the democratic system, and more people actively participate in the governmental policy-making process, which makes the E-Services stronger than we have in developing countries. However, the situation in many developing countries such as Nigeria, Ghana, Georgia, Brazil, Argentina, Romania, and many others, is different from their developed counterpart. The government in developing countries is usually not precisely defined, while the economy is not increasing in productivity.

Furthermore, the economies of many developing countries are not growing and not increasing in productivity, and they are characterised by low standards of living with a relatively short history of democracy and less transparent government policy. Lack of infrastructural facilities and low internet access (Oseni and Dingley, 2014) are also issues in many developing countries. The average citizens in developing countries are reluctant to trust online services because of identity theft. Therefore, they have less active participation in the governmental policy-making process (Hassan, 2011; Nawafleh et al., 2012). In addition to the above, the developing countries do not have local outsourcing abilities and rarely have the financial ability to outsource. However, the current staff may be unable to define specific requirements or build the necessary infrastructure. This is not the case with the developed countries, where the outsourcing abilities and financial resources to outsource are present, and the current staff would be able to define requirements for development.

**Objective 3: To Propose a Diffusion of Innovations (DOI) Theory Based Framework for predicting the Successful E-Service Adopting and Implementation.**

The research gap, as presented in **Section 1.4.2** of Chapter One, called for the development of an E-Service framework, as the literature covered a limited E-Service framework that captures the barriers facing the adopting and implementation of E-Service at the local environment level in Nigeria. Hence, from the research evidence available, the research has developed a DOI-based E-Service framework that explains the main barriers facing E-Service adopting and implementation at local environment level in Nigeria. The solutions to these barriers and the success factors for successful implementation are presented in **Figure 73**. There are stages involved in the development of the E-Service framework. An initial framework was built, based on features of frameworks and models found in the literature. Frameworks developed by Azenabor (2013) for the Federal Government in Nigeria, Mundy and Musa (2010) for the State

Governments in Nigeria (see **Figures 30** and **31**), and the E-Service framework by Hassan (2011) were reviewed. The proposed framework for this study consists of three different developmental stages, and it summarises the barriers facing the adopting and implementation of E-Service based on the literature review in the government context. The stages are **Current State**, the **Transformation State,** and the **Desired End State**.

It is very important to say that this research considered the E-Service development process in the framework, including the transformation from manual (traditional rigid context), as currently experienced at local government level in Nigeria and many developing countries (Hassan, 2011; Azenabor, 2013), to an automated/full public service environment (full public E-Service). Furthermore, the initial framework was modified after the data collection process to add the contributions from the participants and new data that emerged during the interviews and online focus group meetings to the framework, and to effect the required change to improve the system.

### Objective 4: To Evaluate and Validate the proposed E-Service Adoption and Implementation Framework for the Local Environment Level in Nigeria.

From the research evidence, this objective has been achieved by the research. The E-Service framework's evaluation and validation processes were discussed in **section 7.3**. The validation is very important, as it ensures that this E-Service framework will work as expected. The data obtained for this study are from multiple sources (Data Triangulation). Therefore, there is a need for thorough validation. The key findings from this study show that the participants interviewed agreed that the **framework is appropriate** and accepted one for this study. However, there is a need to fully incorporate the research theory (the diffusion of innovation) theory into the validation process. This was achieved through the validation of framework, where the five qualities/attributes of the DOI theory was tested by online focus group participants and the participants earlier interviewed in Nigeria.

In light of the above, the key findings show that the online focus participants also agreed that the framework represents their notion of E-government services. Moreover, the participants concluded that the E-Service framework, if implemented at local government in Nigeria, will boost the revenue base, and reduce corruption.

**Objective 5: To Make Recommendations on the Study Outcomes to the Policy Makers at Local Environment Level in Nigeria especial the Local Government Service Commission.**

This objective has been achieved. From the key research findings and, as explained in **Section 8.8** of this chapter, this study produced an E-Service framework that comprises the group of barriers, the solutions to these barriers, the success factors for the E-Service adopting and implementation at the local environment level in Nigeria. Therefore, part of the recommendations to the local government service commission and the stakeholders is to consider the framework as a tool for a change of management, especially in the area of resistance to change. The framework is a novel contribution that will be very useful to the implementer of changes, scholars and practitioners, and the decision-makers at the local government levels in Nigeria.

To this end, having these recommendations implemented is an important issue for future research, and it is interesting to note that the researcher had full discussions on the research recommendations with the focus group participants, who are senior officers, as shown in **Table 32** below. The participants once again confirmed their interest in the outcomes and recommendations from this study, including the E-Service framework guidelines. The participants maintained that the framework could be used as a pilot for future E-Government services development at a local environment level in Nigeria.

**Table 32 Online Focus Group Participants**

| S/N | Job Role | Years of Experience |
|---|---|---|
| 1 | Commissioner, Local Government Service Commission | 25 |
| 2 | Commissioner, Local Government Service Commission | 22 |
| 3 | Council Manager/Head of Personnel | 25 |
| 4 | Head, Information Technology | 19 |
| 5 | Senior IT Officer | 19 |

## 8.5 Research Contributions to Knowledge

This research attempts to provide original and key contributions to knowledge based on the research findings by focusing on Nigeria local government specific barriers affecting the E-Service adopting and implementation. This study reviewed the literature and looked into various E-Government services initiatives in Nigeria to identify the group of barriers that might affect the E-Service adopting and implementation at the local environment level. The efforts made by this research would help the government at the local environment level in Nigeria and other developing countries to plan and implement E-Services initiatives and increase the adoption rate.

### 8.5.1 Research General Contributions to Knowledge

This section brings together the general research contributions to knowledge as indicated above. They are as follows:

- An early achievement obtained as a result of this research was to explore and identify the barriers facing E-Service adopting and implementation at local environment level, using Nigeria as a case example. This task was accomplished by also investigating the solutions to the identified barriers and the success factors in the adoption and implementation of E-Services at the local environment level in Nigeria.

- The research outcome also guides the Nigeria local government commission with a suitable and appropriate DOI-based E-Service framework for the local government's E-Service development. The framework can be used as a decision-making tool for the successful implementation of local E-Service initiatives.

- The research contributes by making available the research outcomes to the both state and local governments in Nigeria. The research recommendation guides the policy makers at local government level by providing the necessary steps to be taken in the provision and implementation of the E-Government services to the public.

- The research outcomes also produced internationally published journals and papers in the E-Government services domain that might be very useful to academic researchers to further exploit the E-Service areas, both in the developed and developing countries. Hence, more research could be done in this area covering other E-Government communication links, such as Government and Employees (G2E), Government and Business (G2B), and not the Government to Citizen (G2C) that was emphasised in this research.

- The research also contributes to identifying new and novel barriers such as local government autonomy, political barriers, and corruption, which were not fully covered in the literature. These barriers are essential in order to have a full implementation of E-Service at the local environment in Nigeria.

- Empirically, the research contributed to knowledge, as the interview techniques explored the opinions of the senior officials at local environment level in Nigeria responsible for the development and implementation of E-Service initiatives in the government context.

- The research also contributes to developing the Theory of Change model that identified and captured the government official's attitudes towards corruption in the implementation of E-Services at the local environment level in Nigeria. This model suggests rewarding packages and human development such as IT training to boost staff morale, and necessary sanctions and anti-corruption policies to guide against corruption the hindering E-Service development and implementation.

## 8.6 Overall Research Implications

From the research findings, this study highlighted the barriers facing the E-Service adoption and implementation at the local environment level in Nigeria, as well as the solutions and the success factors for the full implementation of E-Service. Furthermore, this research produced an E-Service framework which captured the full context of these barriers and also fills the research gaps identified in **Section 1.4.2** of Chapter One. The implication is the honesty from government at the local environment level in Nigeria in implementing E-Services. The E-Service will promote the availability of government information online for decision-making. There is a need to urgently collaborate with private and public enterprises, as suggested by this research. The partnership will help, as many developing countries don't have the money to outsource and finance IT infrastructures. The E-Service framework produced by this study and other suggestions will be very useful in this regard.

Secondly, corruption is another key research finding from this study. Hence, this is a major barrier affecting the E-Service adopting and implementation at local environment level in Nigeria. The implication here is that the stakeholders, including the staff involved in the E-Service initiatives, should have attitudinal change towards corruption, as suggested during the interviews. A commissioner at the local government service commission in Lagos suggested giving incentives and rewards to boost staff morale and performance. However, necessary sanctions should also be applied to people involved in sharp practices. To this end, the emergence of an appropriate and effective communication structure within the government offices and agencies in local government in Nigeria on anti-corruption will have a major impact on achieving the needed success in E-Service implementation and adoption.

## 8.7 Research Limitations

The main research limitation for this study is having limited one-to-one access to the participants' and organisation's environment as normally practised in the action research methodology. This meant that the research could not accommodate more action research cycles, as this was completed by using an online focus group as suggested and agreed with the participants. The online action research focus group opened a new and novel approach to the data collection process for this study. The ability for the researcher to fully engage and observe action research interactions between the participants was limited, as indicated in **Table 27** (field study) in Chapter Five, due to the limited plenary session.

Another limitation encountered is the limited length of time. This was due to the participants' busy schedules; also, the interviews had to be carried out within a limited period of three months. Travelling between Nigeria and the UK for the purpose of the study needed more time and resources as the researcher was a self-sponsored doctoral candidate. After the initial interviews and early stage validation of results in Nigeria, further validation of the framework and key research findings was carried out online with those focus group members who volunteered from the participants interviewed in Nigeria (see **Section 5.3** of Chapter Five). Hence, arranging online focus group meetings was easier for this researcher and the participants.

Moreover, anonymity and recording issues are another limitation encountered in this research. The issue of the barriers facing the E-Service, using Nigeria as a case example, involved corruption among the government officials, and the participants preferred anonymity and declined their agreement for the interviews to be audio recorded, as explained in **Sections 5.2.6** and **5.4.3**, respectively. Initially, this was seen as a setback; however, the issues were resolved by using questionnaires during the interview meetings and the use of an anonymised online focus group typing in a group session. The use of audio recording would have helped in capturing notes and conversations missed, as the research could go back to play the recorded interactions.

Lastly, from the research findings, and as discussed in **Section 6.4.2**, corruption was a major barrier that emerges from this study. In this regards, the solution to the corruption issues as a factor hindering the full E-Service implementation is for the stakeholders in the E-Services provision at local environment level in Nigeria, which includes the government officials, to have attitudinal change towards this threat. The key findings from the participants could be implemented by putting in place an anti-corruption structure that will repel any form of corruption. However, to complement the anti-corruption drive by the government, this research looked into the theory of change based on the attitudinal change findings from the participants.

The research also looked at scenarios where the theory of change was applied to influence the official attitudes towards corruption, as shown in **Figure 66** (Proposed Theory of Change Model) in Chapter Six. The limitation of this study on the theory of change model is that the researcher was unable to comprehensively validate the theory of change model and the anti-corruption policies explained previously (**Chapter Six**). The busy schedules of many of the participants prevented validation of the theory of change model and to check whether it can achieve the expected long-term goals and outcomes in the necessary change within the Nigeria local government's context. The non-validation of the theory of change model opened the opportunity for further research to be conducted on the theory of change as a model in the anti-corruption drive at the local environment level in Nigeria.

## 8.8 Recommendations that Emerged from this Research

The following are the recommendations that emerged from this research:

- The proposed framework in this research could be used as the basis for more research on E-Service initiatives within the Nigeria local government's context. Future research could investigate the barriers facing the E-Service adoption and implementation at the local government level in Nigeria by covering more local government areas, especially those areas with persistent communal clashes. Presently, some of the local governments in Nigeria are not safe enough to conduct research due to insurgency and attack (Omotosho, 2015).

- The proposed framework consists of a group of barriers, the solutions to these barriers, and the success factors for the E-Service adopting and implementation at the local environment level in Nigeria. Therefore, the framework could serve as a change management tool, especially in the area of resistance to change. The framework is a novel contribution that will be very useful to the implementer of changes, scholars and practitioners, and the decision-makers at the local environment levels in Nigeria.

- From the key research findings, the evidence established that there is a need for further work on why most developing countries, including Nigeria, are still at a low ranking in the E-Readiness in the world due to challenges such as lack of infrastructures, low power supplies, and IT illiteracy.

- Another recommendation that emerges from the research is for the government to improve the provision of ICT infrastructures, and more ICT centres should be established at the local government councils in Nigeria as these infrastructures are very crucial to the country's development. Partnership with private and public companies could provide the necessary resources needed to achieve this.

- The government has a major role to play in the successful implementation of E-Service initiatives in local governments in Nigeria. Therefore, funding should be provided to assist in the procurement of the ICT facilities, legislation and regulations, which are urgently needed to control the E-Service activities. The monthly allocations to the local governments from the Federal Government should be increased. A strong partnership with private and business communities could generate part of the funds needed in the procurement of IT infrastructures.

### 8.8.1 Recommendations for Future Research

The future work of this research will be for researchers to exploit further the action research methodology as it encourages the active participation of those involved. The data capture could be extended to more local government areas in Nigeria, thereby increasing the number of participants involved in the future research, perhaps still using the questions from this research. It is hoped that new findings could emerge that will further strengthen the full implementation of E-Service at local environment level in Nigeria. Furthermore, the future of this research could include the validation of the theory of change model as an agent in the anti-corruption drive towards full public E-Service implementation at the local government level in Nigeria.

## 8.9 Summary

This study has identified the barriers facing the E-Service adopting and implementation at the local environment level in Nigeria, including the solution to these barriers, and the success factors for full E-Service implementation. Also, this research has proposed a suitable framework. Implementing E-Government services in any country requires a well-structured framework. The key findings have exposed the many concerns hindering the full E-Services implementation at local government level, especially local government autonomy and corruption issues. To take this research forward, the researcher also identified factors through suggestions from the participants on how to tackle the threats posed by barriers, and it is also hoped that the recommendations will be adopted to prevent future challenges that E-Service initiatives might encounter.

This chapter critically examined the aims and objectives of this study. The study examined the contributions to knowledge, research limitations, and made recommendations on the E-Service adopting and implementation at the local environment level in Nigeria. The recommendations made were based on the results obtained from the data collected for the study purposed. However, there is always a bias as the researcher did not survey the whole population randomly. The bias emerges from the group of people that the researcher selected. Hence, there are a few limitations to the research which future research can address to improve the research in E-Government services, especially in the developing countries. Part of the research limitations are limited access to the participants, time factor, anonymity, and audio recording issues, which future research may be able to address.

In conclusion, despite the challenges and limitations, this research was able to make the recommendations for both the future study and the policymakers/stakeholders at the local government level in Nigeria. The technology revolution in the world today calls for a robust relationship and communication between the governments and citizens (G2C), and this could only be achieved through full E-Services adopting and implementation.

# References

Abdelkader, Ahmed A.M (2015). A Manifest of Barriers to Successful E-Government: Cases from the Egyptian Programme, *International Journal of Business and Social Science*, Vol 6, Number 1.

Abu-Shanab, E., & Bataineh, L. (2014). Challenges facing e-government projects: How to Avoid Failure? *International Journal of Emerging Sciences*, Volume 4, Issue 4, page 207-217.

Adams, C.A and McNicholas, P. (2007) Making a Difference. Sustainability Reporting, Accountability and Organisational Change. Accounting, Auditing and Accountability Journal, Volume 20, Number 3, pp. 382-402.

Adeyemo, A.B. (2011) E-Government Implementation in Nigeria: An Assessment of Nigeria's Global E-Government Ranking, *Journal of Internet and Information System*, Vol. 2(1).

African Economist (2016). 50 Largest Cities in Africa. Available at http://theafricaneconomist.com/50-largest-cities-in-africa/#.V-uUofmMiUk, accessed on the 28/09/16.

Aghedo, I and Osumah, O (2014) Insurgency in Nigeria: A Comparative Study of Niger Delta and Boko Haram Uprisings. *The Journal of Asian and African Studies*. Sage Publication.

Ahmad, Abdullahi Ayoade (2013) Local Government Autonomy and its Effectiveness in Nigeria, the *Journal of African & Asian Local Government Studies*. Vol. 2, No. 1

Ainabor Augustine. E, Edeh Joseph. N and Onwe Sunday. O (2015) Assessment of Effectiveness of use of ICT Components for Services Delivery in Etsako West Local Government Area of Edo State, Nigeria. *British Journal of Economics, Management & Trade*, 6(3): 197-207, ISSN: 2278-098X.

Aizstrauta, Dace; Ginters, Egils and Eroles, Miquel-Angel Piera (2015) Applying Theory of Diffusion of Innovations to Evaluate Technology Acceptance and Sustainability, Procedia Computer Science, Issue 43, Page 69 – 77.

Ake Gronlund and T.A Horan (2004) "Introducing E-government: History, Definitions and Issues, Communications of the Association for Information Systems, Volume 15.

Aladwani, Adel. M (2016) Corruption as a Source of E-Government Project Failure in Developing Countries: A Theoretical Exposition. *The International Journal of Information Management*, Volume 36, Pages 105-112.

Alateyah, Sulaiman A, Crowder, Richard M and Wills, Gary B (2013) An Exploratory Study of Proposed Factors to Adopt E-Government Services, Saudi Arabia as a Case Study. *International Journal of Advanced Computer Science and Applications*, Vol. 4, No. 11.

Alfarraj, O., Drew, S., & Alghamdi, R. (2011) E-Government Stage Model: Evaluating the Rate of Web Development Progress of Government Websites in Saudi Arabia. *International Journal of Advanced Computer Science and Applications*, Vol. 2, No. 9, pp. 82-90.

Ali, Maged; Weerakkody, Vishanth and El-Haddadeh (2009). The Impact of National Culture on E-Government Implementation: A Comparison Case Study, Proceedings of the Fifteenth Americas Conference on Information Systems, San Francisco, California.

Aliyu, A.A; Bello, M.U; Kasim, Rozilah and Martin, David (2014) Positivist and Non-Positivist Paradigm in Social Science Research: Conflicting Paradigms or Perfect Partners? *Journal of Management and Sustainability*; Vol. 4, No. 3.

Aljazeera News (2014) Nigeria Becomes Africa's Largest Economy, Available at http://www.aljazeera.com/programmes/countingthecost/2014/04/nigeria-truly-strong-economy-2014418134434573256.html. Accessed on the 15th Oct, 2014.

Almarabeh, Tamara and Abu Ali, Amer (2010) A General Framework for E-Government: Definition Maturity Challenge, Opportunities and Success, *European Journal of Scientific Research*, Volume 39, Issue 1, page 29-42.

Alsaghier, H, Ford, M, Nguyen, A, and Hexel, R. (2009) "Conceptualising Citizen's Trust in E-Government: Application of Q Methodology." *Electronic Journal of E-Government*, Volume 7, Issue 4, (pp 295-310).

Al-Shboul, Muhammad; Rababah, Osama; Al-Shboul, M; Ghnemat, Rawan; Al-Saqqa, Samar (2014) Challenges and Factors Affecting the Implementation of E-Government in Jordan. *Journal of Software Engineering and Applications*, 7, Page 1111-1127.

Alshehri, M. and Drew, S. (2011), "E-Government Principles: Implementation, Advantages and Challenges", *International Journal of Electronic Business*, Vol. 9 No. 3, pp. 255-270

Alshehri, M., & Drew, S. (2010). Challenges of E-Government Services Adoption in Saudi Arabia from an E-Ready Citizen Perspective. *International Journal of Social, Behavioural, Educational, Economic, Business and Industrial Engineering*, Volume 4, Issue 6.

Alshehri, M., Drew, S., & Alfarraj, O. (2012). A Comprehensive Analysis of E-Government Services Adoption in Saudi Arabia: Obstacles and Challenges. *International Journal of*

*Advanced Computer Science and Applications*, Vol. 3, No.2.

Al-Sowayegh, Ghada Abdulaziz (2012) Cultural Drivers and Barriers to the Adoption of E-government in the Kingdom of Saudi Arabia, A Thesis submitted to the University of Manchester for the degree of Doctor of Philosophy.

Alter, S. (2006) The Work System Method: Connecting People, Processes, and IT for Business Results, Larkspur, CA: Work System Press.

Anaeto, S. G., Onabanjo, O. S. and Osifeso, J. B. (2008). Models and Theories of Communication. Maryland, USA. Africa Renaissance Books Incorporated.

Anderson, Terry (2005) Design-based Research and its Application to A Call Center Innovation in Distance Education. *Canadian Journal of Learning and Technology*, Volume 31, Number 2, Page 69-84.

Angelopoulos, S; Kitsios, F; Kofakis, P; Papadopoulos, T (2010) Emerging barriers in E-Government implementation, Electronic Government, 6228 (2010), pp. 216–225.

Aparna, M. R., Ostwal, T., Baliga, T., & Sreekumar, N. (2017). Overview of Digital wallets in India. *International Journal of Advanced Research in Computer Science, 6*(8).

Archibald, Mandy M. (2016) Investigator Triangulation: A Collaborative Strategy with Potential for Mixed Methods Research, *Journal of Mixed Methods Research*, Vol. 10(3), page 228–250.

Arjun Neupane, Jeffrey Soar, Kishor Vaidya, and Jianming Yong (2012) Role of Public E-Procurement Technology to Reduce Corruption in Government Procurement. International Public Procurement Conference, August 17-19, Seattle, Washington.

Ashaye, Olusoyi Richard and Irani, Zahir (2013) E-Government Implementation Benefits, Risks and Barriers in Developing Countries: Evidence from of Nigeria, *International Journal of Information Technology & Computer Science (IJITCS)*, Volume 12, Issue 1.

Awortu, Beatrice. E (2015) Boko Haram Insurgency and the Underdevelopment of Nigeria, Research on Humanities and Social Sciences. Vol.5, No.6, ISSN (Paper) 2224-5766, ISSN (Online) 2225-0484

Ayoola T. J (2013) The effect of Cashless Policy of Government on Corruption in Nigeria, The International Review of Management and Business Research, Vol 2, Issue 3.

Azab, N. A., Kamel, S. and Dafoulas, G. (2009) “A Suggested Framework for Assessing

Electronic Government Readiness in Egypt." *Electronic Journal of e-Government*, Volume 7, Issue 1, pp. 11-28, available online at www.ejeg.com

Azenabor C.E (2013) Developing Electronic Government Models for Nigeria: An Analysis. A Thesis Submitted for the award of Doctor of Philosophy (PhD) Degree at the School of Architecture, Computing and Engineering, University of East London, United Kingdom.

Babbie, E. R. (2010) "The Practice of Social Research" Cengage Learning.

Backu, Michiel (2001) E-Governance and Developing Countries, A Research Report.

Bagozzi, Richard P (2007). The Legacy of the Technology Acceptance Model and a Proposal for a Paradigm Shift. *Journal of the Association for Information Systems*, Volume 8, Issue 4, Article 7, pp. 244-254.

Baskerville, R. (1999) Investing Information Systems with Action Research. AIS Volume 2, Issue 19, page 7-17.

Baum, C. and A. Di Maio (2000) Gartner's Four Phases of E-Government Model. Available at http://www.gartner.com, accessed on 06/07/2016.

Bayero, Musa Abdullahi (2015) Effects of Cashless Economy Policy on Financial inclusion in Nigeria: An exploratory study, Procedia – Social and Behavioural Sciences 17 (2), pp 49-56.

Beard, J; Dale, P and Hutchins, J. (2007). The Impact of E-Resources at the Bournemouth University 2004/2006. Performance Measurement and Metrics, Volume 8, No. 1, pp. 7-17.

Bendassolli, Pedro F. (2013). Theory Building in Qualitative Research: Reconsidering the Problem of Induction [50 paragraphs]. Forum Qualitative Sozialforschung/Forum: Qualitative Social Research, 14 (1), Art. 25.

Benson, Tim (2002) E-Services Development Framework Primer, Version 1.0b, Office of the E-Envoy, Cabinet Office, United Kingdom.

Bertot, J; Jaeger, P and Grimes, M (2010) Electronic Government Strategies and Implementation. Government Information Quarterly, 27 (3), Pages 264-271.

Bertot, John Carlo; Estevez, Elsa and Janowski, Tomasz (2016) Digital Public Service Innovation: Framework Proposal, 9th International Conference on Theory and Practice of E-Governance (ICEGOV 2016), Uruguay.

Bertot, John. C; Jaeger, Paul. T and Grimes, Justin. M (2010) Using ICTs to Create a Culture of Transparency: E-Government and Social Media as Openness and Anti-Corruption Tools for Societies. Government Information Quarterly 27, page 264–271.

Bhuiyan, M.S.H (2011) “Public Sector E-Service Development in Bangladesh: Status, Prospects and Challenges” *Electronic Journal of E-Government,* Volume 9, Issue 1, (pp15-29), available online at www.ejeg.com.

Bidgoli, Hossein (2004). The Internet Encyclopaedia, Volume 2, Page 593, John Wiley and Sons, Inc., New Jersey.

Boyd, Paul C (2006) Quirk’s Marketing Research Review, Qualitative Research Issues, Article ID**:** 20061209, page 30.

Bryman A. & Bell, E. (2015) “Business Research Methods” 4th Edition, Oxford University Press, p.27.

Bryman, Alan (2012) Social Research Methods, 4th Edition, Oxford University Press, United Kingdom, ISBN: 978-0-19-958805-3.

Bryman, Alan (2016) Social Research Methods, 5th Edition, Oxford University Press, United Kingdom, ISBN 978-0-19-968945-3.

Buijs, P; Aarnoudse, A; Geers, C; Agbenyadzi, E; Kuma, A; and Doppenberg, A.M. (2012) ‘The Jigsaw of Mayor Clement: Stories and Lessons from an Action Learning Programme in West Africa’ Praxis Note No. 61, Oxford: INTRAC.

Bwalya, Kelvin Joseph and Zulu, Saul (2012) Handbook of Research on E-Government in Emerging Economies: Adoption, E-Participation, and Legal Frameworks. Volume 1, IGI Global, USA.

Calabrese, R.L. (2006) Building Social Capital Through the use of an Appreciative Inquiry Theoretical Perspective in a School and University Partnership, *International Journal of Education Management*, Volume 20, Number 3, pp. 173-182.

Carcary, M. (2008). ‘The Evaluation of ICT Investment Performance in terms of its Functional Deployment’, Ph.D. dissertation, Limerick Institute of Technology Ireland.

Carter, C.W. (1998). An Assessment of the Status of the Diffusion and Adoption of Computer-Based Technology in Appalachian College Association Colleges and Universities (Doctoral Dissertation, Virginia Polytechnic Institute and State University.

Carter, L., & Weerakkody, V. (2008). E-Government Adoption: A cultural comparison. Information Systems Frontiers, 10(4), 473-482.

CFRN (2016) The 1999 Constitution Federal Republic of Nigeria. Available at http://www.nigeria-law.org/ConstitutionOfTheFederalRepublicOfNigeria.htm, accessed on 26/07/16.

Channels Television (2016) Kerry Lauds Buhari's Anti-Corruption Drive. Available at https://www.channelstv.com/2016/01/23/kerry-lauds-buharis-anti-corruption-drive/, retrieved on 04/11/2016.

Chen, Y.N; Chen, H.M; Huang, W; Ching, R.K.H (2006) E-Government Strategies in Developed and Developing Countries: An Implementation Framework and Case Study, *Journal of Global Information Management*, 14(1), 23-46.

Chevalier, J.M. and Buckles, D.J. (2013) Participatory Action Research: Theory and Methods for Engaged Inquiry, Routledge, United Kingdom. ISBN: 978-0415540315.

Christopher M. (1998) "Logistics and Supply Chain Management: Strategies for Reducing Cost and Improving Service" (2nd Edition), ISBN 0 273 63049 0, Pitman Publishing, London.

Chuttur M.Y. (2009) "Overview of the Technology Acceptance Model: Origins, Developments and Future Directions," Indiana University, USA. Sprouts: Working Papers on Information Systems, 9(37).

Coghlan, D. and Brannick, T. (2010). Doing Action Research in Your Own Organization. 3rd Edition. London: Sage.

Cohen, Louis; Manion, Lawrence and Morrison, Keith (2007), Research Methods in Education, 6th Edition, Published by Routledge, London. ISBN 0-203-02905-4.

Collins, H. (2010) "Creative Research: The Theory and Practice of Research for the Creative Industries" AVA Publications, Page 38

Creswell J.W (2009) Research Design: Qualitative, Quantitative and Mixed |Methods Approaches. (Third Edition), SAGE Publication.

Danbatta, Umar (2016) 40 Million Nigerians Lack Network Access, A Publication by the Nigeria Punch Newspapers, available at http://punchng.com/40-million-nigerians-lack-network-access-says-ncc/, accessed on 22/11/2016.

Dawson, Kara; Cavanaugh, Cathy and Ritzhaupt (2013) An Online Tool to Support Teacher

Action Research for Technology Integration. Teacher Education Programs and Online Learning Tools: Innovations in Teacher Preparation.

Dibra, Mirjam (2015) Rogers Theory on Diffusion of Innovation — The Most Appropriate Theoretical Model in the Study of Factors Influencing the Integration of Sustainability in Tourism Businesses, World Conference on Technology, Innovation and Entrepreneurship, Procedia - Social and Behavioural Sciences, Issue 195, Page 1453-1462.

Dick, Bob (1993) You Want To Do an Action Research Thesis? How to Conduct and Report Action Research. Available at http://www.aral.com.au/, accessed on 06/10/16.

Dingley, K. (2001). Internet Video Meditated-Communication in Work and Learning Situations, PhD Thesis Submitted at the University of Portsmouth, United Kingdom.

Dooley, K.E. (1999) Towards a Holistic Model for the Diffusion of Educational Technologies: An Integrative Review of Educational Innovation Studies. Educational Technology & Society, Volume 2(4), 35-45.

Draper, Janet (2004). The Relationship between Research Question and Research Design. In: Crookes, Patrick A. and Davies, Sue eds. Research into Practice: Essential Skills for Reading and Applying Research in Nursing and Health Care, Volume 2nd Ed. Edinburgh: Bailliere Tindall, pp. 69–84.

Easterby-Smith, M, Thorpe, R. & Jackson, P. (2008) “Management Research” 3rd Edition, SAGE Publications Ltd., London.

Easterby-Smith, M; Thorpe, R; and Lowe, A. (2002), Management Research: An Introduction, Second Edition, Sage Publications Ltd, London, UK.

Ebrahim, Zakareya; Irani, Zahir and Al Shawi, Sarmad (2004) A Strategic Framework for E-government Adoption in Public Sector Organisations, Proceedings of the Tenth Americas Conference on Information Systems, New York.

Economist Intelligence Unit (2010) Digital Economy Rankings 2010, Beyond E-Readiness, A Report from the Economist Intelligence Unit.

Elkadi, Hatem (2013) Success and Failure Factors for E-Government Projects: A case from Egypt, Egyptian Informatics Journal, Page 165-173.

El-Sofany, H; Al-Tourki, T; Al-Howimel, H and Al-Sadoon, A. (2012). E-government In Saudi Arabia: Barriers, Challenges and Its Role of Development E-Government in Saudi Arabia.

*International Journal of Computer Applications*, Vol. 48(5), pp. 61-22.

EU (2004) E-Government Research in Europe, European Commission, available at http://europa.eu.int/information_society/programmes/egov_rd/text_en.htm. Accessed on 23/02/2016.

Fajobi O.F. (2010) X-Ray of Local Government Administration in Nigeria, Published by Crest Hill Limited, Basorun, Ibadan, Nigeria.

Fang, Zhiyuan (2002) E-Government in Digital Era: Concept, Practice and Development. *International Journal of the Computer, the Internet and Management*, Vol. 10, No.2, 2002, Pages 1-22

Fath-Allah, A; Cheikhi, L; Al-Qutaish, R.E and Idris, A. (2014) E-Government Maturity Models: A Comparative Study. *International Journal of Software Engineering & Applications (IJSEA),* Vol.5, No.3.

Fatile, Jacob .O. (2012) Electronic Governance: Myth or Opportunity for Nigerian Public Administration? *International Journal of Academic Research in Business and Social Sciences*, Vol. 2, No. 9, ISSN: 2222-6990.

Francis, J; Johnston, M; Robertson, C; Glidewell, L; Entwistle, V; Eccles, M and Grimshaw, J. (2010) 'What is an Adequate Sample Size? Operationalising Data Saturation for Theory-based Interview Studies'. Psychology & Health 25 (10): 1229-1245.

FRN (2016) Federal Republic of Nigeria Website, available at http://www.nigeria.gov.ng/. Accessed on June 1st 2016, 18:26 Hrs.

Gant, J.P (2008) Electronic Government for Developing Countries, Draft of International Telecommunication Union (ITU), ICT Applications and Cybersecurity Division.

Gapp, R. and Fisher, R. (2006), "Achieving Excellence through Innovative Approaches to Student Involvement in Course Evaluation within the Tertiary Education Sector", Quality Assurance in Education, Vol. 14 No. 2, pp. 156-166.

Gbadamosi, Wahab (2007) How E-Collection Contained $92 Million Loss at the Nigeria Inland Revenue Service, FIRS Press Release.

Germonprez, Matt and Mathiassen, Lars (2004). The Role of Conventional Research Methods in Information Systems Action Research, IFIP International Federation for Information Processing, pp. 335-352. ISBN: 978-1-4020-8094-4.

Gil-Garcia, J. Ramon (2013) E-Government Success Factors and Measures: Theories, Concepts, and Methodology, Centro de Investigation Docencia Economicas (CIDE), Mexico.

Gil-Garcıa, J. Ramon and Pardo, Theresa A. (2005) " E-Government Success Factors: Mapping Practical Tools to Theoretical Foundations, Government Information Quarterly 22.

Gill, P; Stewart, K; Treasure, E and Chadwick, B (2008) Methods of Data Collection in Qualitative Research: Interviews and Focus Groups, British Dental Journal 204, 291 – 295.

Goran, Goldkuhl and Erik, Perjons (2014) Focus, Goal and Roles in E-Service Design: Five Ideal Types of the Design Process. E-Service Journal, Volume 9, Issue 2, Page 24-45.

Gravetter, Frederick J and Forzano, Lori-Ann B. (2012) Research Methods for the Behavioural Sciences, 4th Edition. Wadsworth, USA. ISBN-13: 978-1-111-34225-8.

Grefen P., Heiko Ludwig and Samuil Angelov (2002). A Framework for E-Services: A Three-Level Approach towards Process and Data Management, IBM Research Report.

Gregor, S. (2006). “The Nature of Theory in Information Systems,” MIS Quarterly (30:3), pp. 611-642.

Gregory G. Curtin, Michael H. Sommer and Veronika Vis-Sommer (2003) “The World of e-Government” Journal of E-Government, Volume 2, No. 3/4. The Haworth Press Inc. NY, USA.

Griffin, David; Trevorrow, Philippa and Halpin, Edward (2007) Developments in E-Government, A Critical Analysis. Innovation and the Public Sector, Volume 13. ISBN: 978-1-58603-725-3, IOS Press.

Guest, G; Bunce, A. and Johnson, L. (2006) ‘How Many Interviews are Enough? An Experiment with Data Saturation and Variability’. Field Methods 18 (1): 59-82

Hair, J; Black, W; Babin, B; Anderson, R. and Tatham, R. (2006). Multivariate Data Analysis (6th Edition). Uppersaddle River, N.J.: Pearson Prentice Hall.

Hammersley, M. (2016). Reading Ethnographic Research, A Critical Guide. Longman Social Research Series. Routledge, United Kingdom.

Hart, Penny (2013) Investigating Issues Influencing Knowledge Sharing in a Research Organization, using the Appreciative Inquiry Method. Thesis Submitted for the Award of the Degree of Doctor of Philosophy of the University of Portsmouth, UK.

Hashemi, S. M., Razzazi, M., & Teshnehlab, M. (2006). Service Oriented Privacy Modelling in Enterprises with ISRUP E–Service Framework. W3C Workshop on Languages for Privacy Policy Negotiation and Semantics-Driven Enforcement, JRC, European Commission, Italy.

Hassan, H.S.H (2011). An Investigation of E-Service in Developing Countries. The Case Study of E-Government in Egypt. Thesis Submitted to the School of Applied Sciences, Canfield University, United Kingdom.

Hassan H.S, Shehab, E and Peppard, J (2011) "Recent Advances in E-Service in the Public Sector: State-of-the-art and Future Trends", Business Process Management Journal, Vol. 17, Issue 3, pp. 526 – 545

Heeks, Richard and Bailur, Savita (2007) "Analyzing E-Government Research: Perspectives, Philosophies, Theories, Methods, and Practice". Government Information Quarterly, Volume 24, Issue 2, pp 243-265.

Helton, David .A. (2012) Bridging the Digital Divide in Developing Nations Through Mobile Phone Transactions Systems, Peer Reviewed, College of Business, Northern Michigan University, USA.

Huang, Zhenyu, and Peterson O. Bwoma (2003) "An Overview of Critical Issues of E-Government." Issues of Information Systems, Volume 4, Issue 1, Pages 164-170.

Hubbard, William G. and Sandmann, Lorilee R. (2007) Using Diffusion of Innovation Concepts for Improved Program Evaluation, *Journal of Extension*, Volume 45, Number 5.

IMF (2016) The World Economic Outlook Data by International Monetary Fund, available at http://www.imf.org/en/data, accessed on 29/11/2016.

Jiménez, G. and Espadas, J. (2006) Implementation of an E-Services Hub for Small and Medium Enterprises, Proceedings of the Advanced International Conference on Telecommunications and International Conference on Internet and Web Applications and Services.

Johnson Hannes, Johansson Mikael and Andersson Karin (2014) Barriers to Improving Energy Efficiency in Short Sea Shipping: An Action Research Case Study. *Journal of Cleaner Production*, Volume 66, page 317-327.

Johnson, Jesper (2012) Theories of Change in Anti-Corruption Work, A Tool for Programming Design and Evaluation, A Report by the Anti-Corruption Resources Centre.

Jouzbarkand, M; Khodadadi, M and Keivani, F. Sameni (2011) Conceptual Approach to E-

Government, Targets and Barriers facing it. International Conference on Innovation, Management and Service, IPEDR Vol.14

Kakar, Yana Watson; Hausman, Vicky; Thomas, Andria and Denny-Brown, Chris (2012). Women and the Web, Bridging the Internet Gap and Creating New Global Opportunities in Low and Middle-Income Countries. A Report by Intel Corporation.

Kaminski, June (2011) Diffusion of Innovation Theory, *Canadian Journal of Nursing Informatics*, 6(2). Theory in Nursing Informatics Column.

Kapenieks, J. & Salīte, I. (2013). Action Research for Creating Knowledge in an E-Learning Environment. Journal of Teacher Education for Sustainability, 14(2).

Kapenieks, J. (2013). User-friendly E-learning Environment for Educational Action Research. Procedia Computer Science, 26, pages 121-142.

Kelleher, C. and Peppard, J. (2009). The Web Experience – Trends in EService. The Institute of Customer Service.

Keng, Siau and Yuan, Long (2005) Synthesizing E-Government Stage Models: A Meta-Synthesis Based on Meta-Ethnography Approach. *Journal of Industrial Management & Data Systems*, Vol. 105, No. 4, pp. 443-458.

Khadaroo, I., Wong, M. S., & Abdullah, A. (2013) Barriers in Local E–Government Partnership: Evidence from Malaysia. *An International Journal of Electronic Government*, 10(1), 19-33.

Khan Gohar. F; Moon Junghoon; Swar, Bobby; Rho, J.J; Zo, Hangjung (2012) E-Government Service use Intentions in Afghanistan: Technology Adoption and the Digital Divide in a War-Torn Country, Information Development, 28 (4).

Kolsaker, A, and Lee-Kelley, L. (2009) “Singing from the Same Hymnsheet? The Impact of Internal Stakeholders on the Development of e-Democracy.” *Electronic Journal of E-Government*, Volume 7, Issue 2, (pp. 155-162), Available online at www.ejeg.com

Korsgaard, S. (2007) “Social constructionism: and why it should feature in entrepreneurship theory”, CORE Working Paper No. 2007-01, CORE, Aarhus, Denmark.

Kuhn, Thomas (1962). The Structure of Scientific Revolutions, Chicago: University of Chicago Press.

Kumar, V; Mukerji, B; Butt, I and Persaud, A (2007) “Factors for Successful E-Government Adoption: a Conceptual Framework” *The Electronic Journal of E-Government*, Volume 5,

Issue 1, pp 63-76, available online at www.ejeg.com.

Larsen, K. R.; Allen, G; Vance, A and Eargle, D. (Eds.) (2015). Theories Used in IS Research Wiki, available at http://IS.TheorizeIt.org, Retrieved on 01/06/2016.

Lee, S. M., Tan, X., & Trimi, S. (2005). Current Practices of leading E-Government Countries. Communications of the ACM, 48 (10), 99-104.

Leitch, Ruth and Day, Christopher (2000) Action Research and Reflective Practice: Towards a Holistic View, Educational Action Research, Volume 8, Issue 1, Pages 179-193.

Lenk, K. (2002) Electronic Service Delivery – A Driver of Public Sector Modernization, Information Polity, Volume 7, pages 87-96.

Lim, Sanghee; Saldanha, Terence; Malladi, Suresh; and Melville, Nigel P. (2009) "Theories Used in Information Systems Research: Identifying Theory Networks in Leading IS Journals" International Conference on Information Systems (ICIS) Proceedings.

Lindgren, I. (2013) Public E-Service Stakeholders: A Study on Who Matters for the Public E-Service Development and Implementation, Sweden Publication, Ipswich, MA.

LSG (2016) Lagos State Government, Nigeria Website, available at (http://www.lagosstate.gov.ng), accessed on the 30th June, 2016 at 14:26 Hrs.

Macintosh, Ann and Fraser, John (2003). A Framework for E-Government Services, A Governmental Knowledge-based Platform for Public Sector Online Services, IST PROJECT, 2001-35399 SmartGov.

Mack, Lindsay (2010). The Philosophical Underpinnings of Educational Research, Polyglossia, Volume 19.

Mahadeo, J, D. (2009) “Towards an Understanding of the Factors Influencing the Acceptance and Diffusion of e-Government Services.” Electronic Journal of e-Government Vol. 7, Issue 4.

Matavire, R; Chigona, W; Roode, D; Sewchurran, E; Davids, Z; Mukudu, A and Boamah-Abu, C. (2010) “Challenges of E-Government Project Implementation in a South African Context” *The Electronic Journal Information Systems Evaluation*, Volume 13.

Maxwell, Joseph .A. (2013) Qualitative Research Design. An Interactive Approach, 3rd Edition. Sage Publishing.

McCartan, Claire; Schubotz, Dirk & Murphy, Jonathan (2012). The Self-Conscious Researcher:

Post-Modern Perspectives of Participatory Research with Young People. Forum Qualitative Social Forschung/Forum: Qualitative Social Research, 13 (1).

McGill, I., & Beaty, L. (2001). Action Learning: A Guide for Professional, Management & Educational Development. Psychology Press.

McKinley, Jim (2015) Critical Argument and Writer Identity: Social Constructivism as a Theoretical Framework for EFL Academic Writing, Critical Inquiry in Language Studies, Issue 12, Volume 3, pp. 184-207.

Medlin, B.D. (2001). The Factors that may Influence a Faculty Member's Decision to Adopt Electronic Technologies in Instruction (Doctoral Dissertation, Virginia Polytechnic Institute and State University).

Mohammad, Hiba; Almarabeh, Tamara and Abu Ali, Amer (2009), "E-Government in Jordan", European Journal of Scientific Research, Vol. 35 No.2, pp.188-197.

Mundy, D and Musa, B (2010) "Towards a Framework for E-Government Development in Nigeria" *Electronic Journal of E-Government*, Volume 8, Issue 2, pp. 148-161.

Mutula, Stephen M. and Mostert, Janneke (2010) "Challenges and opportunities of e-government in South Africa", The Electronic Library, Vol. 28 Issue: 1, pp.38-53,

Nabafu, R and Maiga, G (2012) A Model of Success Factors for Implementing Local E-government in Uganda, *Electronic Journal of E-Government*, Vol.10, Iss. 1, pp. 31-46.

Nawafleh, S. A; Obiedat, R. F and Harfoushi, O. K (2012) E-Government between Developed and Developing Countries, *International Journal of Advanced Corporate Learning (IJAC)*, Vol. 5.

Neuman, W.L (1997). Social Research Methods. Qualitative and Quantitative Approaches. Boston, London Toronto: Allyn & Bacon.

NBS (2017) National Bureau of Statistics, Nigeria Website. Available at http://nigeria.opendataforafrica.org, accessed on 14/06/2017, 14:04 Hrs.

Nigerian Telecommunication Commission (2014) Mobile Subscriber Data. Available at https://en.wikipedia.org/wiki/List_of_countries_by_number_of_mobile_phones_in_use, retrieved on 15/06/2016.

NIMC (2017) National Identity Management Commission Policies, available at http://www.nimc.gov.ng/policies/, accessed on 14/01/17.

NITDA (2016) National Information Technology Development Agency, E-Library Guidelines, available at http://www.nitda.gov.ng/, accessed on 11/03/2016.

Nkohkwo, Quinta Nven-akeng and Islam, M. Sirajul (2013)" Challenges to the Successful Implementation of E-Government Initiatives in Sub-Saharan Africa: A Literature Review" *Electronic Journal of E-Government*, Volume 11 Issue 2, (pp 253-253).

Nurdin, Nurdin; Stockdale, Rosemary and Scheepers, Helana (2011) Understanding Organizational Barriers Influencing Local Electronic Government Adoption and Implementation: The Electronic Government Implementation Framework, *Journal of Theoretical and Applied Electronic Commerce Research*, Volume 6, Issue 3.

Oates, B. J. (2006) Researching Information Systems and Computing, Sage Publications Ltd, London.

O'Brien, R. (2001). An Overview of the Methodological Approach of Action Research. In Roberto Richardson (Ed.), Theory and Practice of Action Research (English version). Available at http://www.web.ca/~robrien/papers/arfinal.html

Okafor, Jude (2010) Local Government Financial Autonomy in Nigeria: The State Joint Local Government Account, Commonwealth Journal of Local Governance, Issue 6.

Olaiya, T.A (2016) Governance and Constitutional Issues in the Nigerian Local Governments, *Journal of Politics and Law*; Vol. 9, No. 1; ISSN 1913-9047.

Olasupo, F. A (2013). The Scope and Future of Local Government Autonomy in Nigeria, Advances in Applied Sociology. Vol.3, No.5, 207-214.

Olayiwola, Abdul-Rahman .O (2013) Media and Security in Nigeria, *European Journal of Business and Social Sciences*, Vol. 2, No. 9. ISSN: 2235 – 767X.

Omotosho, M. (2015) Dynamics of Religious Fundamentalism: A Survey of Boko Haram Insurgency in Northern Nigeria.

Oseni, K.O (2016) Link between Research Methodology and Theory, Paper Submitted for the PhD Annual Review at the University of Portsmouth, United Kingdom.

Oseni, Kazeem. and Dingley, Kate. (2014) 'Challenges of E-Service Adoption and Implementation in Nigeria: Lessons from Asia' *International Journal of Social, Education, Economics and Management Engineering*, 8(12), 3689-3696.

Oseni, Kazeem Oluwakemi and Dingley, Kate (2015) Roles of E-Service in Economic

Development, Case Study of Nigeria, A Lower-Middle Income Country. *International Journal of Managing Information Technology (IJMIT)*. Volume 7, Number 2. Available in AIRCC Digital Library.

Oseni, Kazeem; Dingley, Kate and Hart, Penny (2015) Barriers Facing E-Service Technology in Developing Countries: A Structured Literature Review with Nigeria as a Case Study, IEEE International Conference on Information Society (i-Society 2015) Proceedings, London. Available in IEEE Xplore Digital Library.

Oseni, Kazeem, Dingley, Kate and Hart, Penny (2017) Instant Messaging and Social Networks – The Advantages in Online Research Methodology, Proceedings of the International Conference on Educational and Information Technology (ICEIT), Cambridge, United Kingdom. Accepted for Publication in the *International Journal of Information and Education Technology (IJIET)*, ISSN: 2010-3689.

Othman, M. K., Yasin, N. M., & Samelan, N. A. (2012) Factors Influencing the Adoption of E-Services in Malaysia, University of Malaya, Malaysia.

Oviasuyi, P. O. and Uwadiae, Jim (2010) The Dilemma of Niger-Delta Region as Oil Producing States of Nigeria, *Journal of Peace, Conflict and Development*, Issue 16.

Parisot, A.H. (1995) Technology and Teaching: The Adoption and Diffusion of Technological Innovations by a Community College Faculty (Doctoral dissertation, Montana State University).

Patton, M. Q. (2001). Qualitative Evaluation and Research Methods (3rd edition). Newbury Park, CA: Sage Publications.

Peersman, Greet (2014) Overview: Data Collection and Analysis Methods in Impact Evaluation, Methodological Briefs: Impact Evaluation 10, UNICEF Office of Research, Florence.

Persson, Anders and Goldkuhl, Göran (2010) "Government Value Paradigms-Bureaucracy, New Public Management, and E-Government," Communications of the Association for Information Systems: Vol. 27, Article 4.

Petersson, J. (2008) "Work System Principles: Towards a Justified Design Theory on the Grounds of Socio-Instrumental Pragmatism," in Proceedings of the 3rd International Conference on the Pragmatic Web (PRAGWEB 2008), 69-76.

Petrescu, Marius; Popescu, Delia; Barbu, Ionuţ and Dinescu, Roxana (2010) Public Management: between the Traditional and New Model. Review of International Comparative Management 408, Volume 11, Issue 3.

Pettigrew, Karen. E and McKechnie, Lynne E.F (2001) The Use of Theory in Information Science Research, *Journal of The American Society For Information Science and Technology*, Volume 52, Issue 1, page 62–73.

Phillips, L.J. and Kristiansen, M. (2012) "Characteristics and Challenges of Collaborative Research: Further Perspective on Reflexive Strategies", in Knowledge and Power in Collaborative Research: A Reflexive Approach. Routledge UK, Ch. 13.

Phythian, Mick; Fairweather, Ben; Howley, Richard (2009) Developing Measures of E-Government Progress Using Action Research, Proceedings of the European Conference on E-Government, p522.

Popplewell R. and Hayman R. (2012) Where, How and Why are Action Research Approaches used by International Development Non-governmental Organisations. Briefing paper No. 32, International NGO Training and Research Centre.

Porta, M. 2008. A Dictionary of Epidemiology. 5th ed. Oxford: Oxford University Press. 320 p.

Préfontaine L. (2002) "New Models of Collaboration for the Implementation of Public Service", Connecting Research and Practice National Conference for Digital Government Research Los Angeles.

Rakhmanov, E. (2009). “The Barriers Affecting E-Government Development in Uzbekistan”, Proceedings of Fourth International Conference on Computer Sciences and Convergence Information Technology, pp. 1474-1480

Ramanathan, R. (2008) “The Role of Organisational Change Management in Offshore Outsourcing of Information Technology Services” Universal Publishers

Rana, Anurag (2014) Scope and Deployment of Strategies of E-Governance in India: A Survey. *International Journal of Interactive Computer Communication*, Volume 03 No. 01 Issue.

Reason, P. and Bradbury, H. (2008). The Sage Handbook of Action Research: Participative Inquiry and Practice. Sage, CA, ISBN: 978-1412920292.

Reason, P., & Bradbury, H. (Eds.) (2001) Handbook of Action Research: Participative Inquiry and Practice, Sage Publisher.

Reeves, Scott; Kuper, Ayelet and Hodges, B.D (2008). Qualitative Research Methodologies: Ethnography, BMJ-2008;337:a1020.

Robinson, L. (2009). A Summary of Diffusion of Innovations. Enabling Change.

Rogers, E.M (1995) Diffusion of Innovations. 4th Edition, New York: Free Press.

Rogers, Everett (2003) Diffusion of Innovations, 5th Edition, Simon and Schuster. ISBN: 978-0-7432-5823-4.

Rogers, Everett M. (1962) Diffusion of Innovations. Glencoe: Free Press. ISBN: 0-612-62843-4. 1st Edition

Rogers, Patricia (2014). Theory of Change, Methodological Briefs: Impact Evaluation 2, UNICEF Office of Research, Florence.

Rokhman, A. (2011), 'E-Government Adoption in Developing Countries; the Case of Indonesia', Journal of Emerging Trends in Computing and Information Sciences, Vol. 2, No. 5, pp. 228-236

Roland T. Rust and P.K Kannan (2003) E-Service: A New Paradigm for Business in the Electronic Environment, Communications of the ACM June 2003, Volume 46, No. 6.

Rose, Susan; Spinks, Nigel and Canhoto, Ana Isabel (2015) Management Research: Applying the Principles of Action Research.

Rowley, J. (2006). An analysis of the E-Service Literature: Towards a Research Agenda. Internet Research, 16(3), 339-359.

Ruyter Ko De, Martin Wetzels and Mirella Kleijnen (2001) "Customer adoption of e-service: an experimental study" *International Journal of Service Management*, Vol.12, No. 2.

Sá, Filipe, Rocha, Álvaro and Cota, Manuel Perez (2015). From the Quality of Traditional Services to the Quality of Local E-Government Online Services: A Literature Review. Government Information Quarterly.

Safeena, R and Kammani, A. (2013) Conceptualization of Electronic Government Adoption, *International Journal of Managing Information Technology (IJMIT)*, Vol.5, No.1.

Saha, Parmita (2008) Government E-Service Delivery: Identification of Success Factors from Citizens' Perspective, Doctoral Thesis, Luleå University of Technology Department of Business Administration and Social Sciences Division of Industrial Marketing, E-Commerce and Logistics.

Saheb, T. (2005). ICT, Education and Digital Divide in Developing Countries. Global Media

Journal, 4(7). Available at http://www.globalmediajournal.com/, retrieved on July 17, 2016.

Sahin, Ismail (2006) Detailed Review of Rogers' Diffusion of Innovations Theory and Educational Technology Related Studies Based on Rogers' Theory. The Turkish Online Journal of Educational Technology – TOJET, ISSN: 1303-6521 volume 5, Issue 2.

Saunders, Mark; Lewis, Philip and Thornhill, Adrian (2016) Research Methods for Business Students, 7th Edition, Published by Pearson Education Limited, United Kingdom. ISBN: 978-1-292-01662-7

Savoldelli, A., Codagnone, C., and Misuraca, G., (2014) Understanding the E-Government Paradox: Learning from Literature and Practice on Barriers to Adoption. Government Information Quarterly, 31, pp 563-571.

Schlæger, J. (2013). E-Government in China: Technology, Power and Local Government Reform (Vol. 34). Routledge.

Schoonenboom, Judith (2015) The Realist Survey: How Respondents' Voices Can Be Used to Test and Revise Correlational Models, *Journal of Mixed Methods Research*, page 1–20.

Schware, Robert and Deane, Arsala (2003) Deploying E-Government Programs: The Strategic Importance of "I" before "E", World Bank Info, Volume 5, Issue 4, available at Emerald site.

Schwester, R (2009) Examining the Barriers to E-Government Adoption, *Electronic Journal of E-Government*, Volume 7, Issue 1. Page 113-122, available online at www.ejeg.com.

Scotland, James (2012) Exploring the Philosophical Underpinnings of Research: Relating Ontology and Epistemology to the Methodology and Methods of the Scientific, Interpretive, and Critical Research Paradigms, English Language Teaching; Vol. 5, No. 9, Published by Canadian Center of Science and Education.

Seifert, J. W. (2008). A Primer on E-Government: Sectors, Stages, Opportunities, and Challenges of Online Governance. In R. B. Ventura (Ed.), E-Government in High Gear. New York: Nova Science Publishers.

Shamoo, A.E., Resnik, B.R. (2003). Responsible Conduct of Research. Oxford University Press.

Shamsuddin, Bolatito and SiddigBalal, Ibrahim (2014) Challenges of Local Government Administration in Nigeria. An Appraisal of Nigerian Experience, *International Journal of Science and Research (IJSR)*, Volume 3 Issue 7, ISSN (Online): 2319-7064.

Sherry, L. & Gibson, D. (2002). The Path to Teacher Leadership in Educational Technology.

Contemporary Issues in Technology and Teacher Education, 2(2), 178-203. Norfolk, VA: Association for the Advancement of Computing in Education (AACE).

Shoniregun C.A. (2007), Synchronizing Internet Protocol Security (SIPSec), Advances in Information Security, Springer-Verlag. ISBN: 0-387-32724-X.

Smith (2005) Philosophy and Methodology of the Social Sciences is a Comprehensive Collection of Readings.

Sukasame, N. (2004) "The Development of E-Service in Thai Government." BU Academic Review 3 (1), page 17-24.

Surendra, S.S. (2001) Acceptance of Web Technology-based Education by Professors and Administrators of a College of Applied Arts and Technology in Ontario (Doctoral Dissertation, University of Toronto).

Surjadjaja, H., Ghosh, S., & Antony, F. (2003) "Determining and Assessing the Determinants of E-Service operations." Managing Service Quality Journal, Volume 13 (1), pages 39-53.

Susanto, T.D and R. Goodwin (2010) "Factors Influencing Citizen Adoption of SMS Based E-Government Services" *Electronic Journal of E-Government*, Volume 8, Issue 1, pages 55-71.

Susman, Gerald I. (1983) "Action Research: A Sociotechnical Systems Perspective." Ed. G. Morgan. London: Sage Publications, pp 95-113.

Taplin, Dana. H and Clark, Helene (2012) Theory of Change Basics. A Primer on Theory of Change. A Report by ActKnowledge, NY, USA.

Tarus, J.K, Gichoya, D and Muumbo, A (2015) Challenge of Implementing E-Learning in Kenya: A case of Kenyan Public Universities. The International Review of Research in Open and Distributed Learning, Volume 16, Issue 1.

Tella, Adeyinka; Adetoro, Niran and Adekunle, P. A. (2009) "A Case Study of the Global System of Mobile Communication in Nigeria". The Spanish CEPIS Society, Vol 5 (3), page 2-7.

Titah, Ryad and Barki, Henri (2006) E-Government Adoption and Acceptance: A Literature Review and Research Framework, *International Journal of Electronic Government Research*, 2 (3), Page 23-57.

TNN (2015) Corruption Can be tackled by E-Governance, The Nation Nigeria Newspapers, available at http://thenationonlineng.net/new/corruption-can-be-tackled-by-e-governance/, accessed on the 15/03/2016.

Torrance, Harry (2012) Triangulation, Respondent Validation, and Democratic Participation in Mixed Methods Research, *Journal of Mixed Methods Research*, 6(2), pages 111–123.

Turnock, Christopher and Gibson, Vanessa (2001) Validity in Action Research: A Discussion on Theoretical and Practice Issues Encountered Whilst Using Observation to Collect Data, *Journal of Advanced Nursing*, 36(3), 471-477.

TVN (2014) Nigeria Need Effective E-Governance for Participatory Democracy, Speech presented at Conference on E-Governance in Nigeria, available at http://www.vanguardngr.com/2014/07/nigeria-needs-effective-e-governance-for-participatory-democracy-dons/, accessed on 16/3/2016.

Ugaz, Jose (2015) Corruption Perceptions Index 2015, Transparency International. Available at https://www.transparency.org/cpi2015/#results-table, accessed on 25/10/16.

Ukoha, Ukiwo (2006) Creation of Local Government Area and Ethnic Conflicts in Nigeria. The Case of Warri, Delta State. Paper Presented at Crise West Africa Workshop, Accra, Ghana.

Umeoji, Emeka (2011). The Impact of Culture and Gender on E-Government Diffusion in a Developing Country: the Case of Nigeria, PhD Thesis, University of Hertfordshire, UK.

United Nations (2016). The United Nations E-Government Ranking Survey, available at https://publicadministration.un.org/egovkb/en-us/reports/un-e-government-survey-2016, accessed on 21/06/17, 15:10 hrs

United Nations Foundation (2004) "A Draft on Governance in Public Private Partnerships for Infrastructure Development", Geneva.

Venkitachalam, R (2015) Validity and Reliability of Questionnaires. A Presentation on Validity and Reliability of Questionnaires in Healthcare, Amrita School of Dentistry. Available at: http://www.slideshare.net/Venkitachalam/validity-and-reliability-of-questionnaires, accessed on 20/09/2016.

Vivek, Agrawal; Vikas, Tripathi and Nitin Seth (2014) A Conceptual Framework on Review of E-Service Quality in Banking Industry World Academy of Science, Engineering and Technology. *International Journal of Social, Education, Economics and Management Engineering,* Vol: 8 No: 12.

Webster, J. and Watson, R. (2002) ‘Analysing the Past to Prepare for the Future; Writing A Literature Review’. MIS Quarterly, 26(2).

Weerakkody, Vishanth; Ramzi El-Haddadeh and Shafi Al-Shafi (2011) "Exploring the Complexity of E-Government Implementation and Diffusion in a Lower Middle Income Country: Some lessons from State of Qatar", *Journal of Enterprise Information Management*, Vol. 24, Issue 2, pp. 172 -196.

Whyte, Grafton and Lamprecht, Adolf (2013) Customer Centricity: A Comparison of Organisational Systems with Social Media Applications, Proceedings of the 4th International Conference on Information Systems Management and Evaluation (ICIME), Vietnam.

World Bank (2017) Open Data, available at http://data.worldbank.org/, accessed on 12/01/2017.

World Bank (2006) The Next Frontier of E-Government: Local Government may hold the Keys to Global Competition, available at http://siteresources.worldbank.org/EXTINFORMATIONANDCOMMUNICATIONANDTECHNOLOGIES/Resources/NextFrontierE_Government.pdf, accessed on 5th April, 2016.

Xu, Jin and Yu, Menggang (2013) Sample Size Determination and Re-estimation for Matched Pair Designs with Multiple Binary Endpoints, Biometrical Journal, Volume 55, Issue 3, pp 271-489.

Yildiz, Mete (2007) "E-government research: Reviewing the Literature, Limitations, and Ways Forward" Government Information Quarterly 24.2, pp 646-665.

Yousafzai, S.Y; Foxall, G.R and Pallister, J.G (2007). Technology Acceptance: A Meta-analysis of the TAM: Part 1, *Journal of Modelling in Management*, Volume 3, and Issue 2.

Zarei, B., Ghapanchi, A., & Sattary, B. (2008). Toward National E-Government Development Models for Developing Countries: A Nine-stage Model. The International Information & Library Review, 40(3), 199-207.

Zhang, Hui; Xiaolin Xu and Jianying Xiao (2014) Diffusion of E-Government: A Literature Review and Directions for Future Directions, Government Information Quarterly, Vol. 31, Issue 4, pp 631

Ziemba, Ewa; Papaj, Tomasz and Żelazny, Rafał (2013) A Model Of Success Factors For E-Government Adoption – The Case Of Poland, Issues in Information Systems, Volume 14, Issue 2, pp.87-100.

Zohrabi, Mohammad (2013) Mixed Method Research: Instruments, Validity, Reliability and Reporting Findings. Theory and Practice in Language Studies, Vol. 3, No. 2, pp. 254-262.

# **Appendix A -** Interview, Online Focus Group and Online Survey Information Sheet, Invitation, Consent form, and Questionnaires

## PARTICIPANT INFORMATION SHEET

**Title of Project**: Barriers Facing E-Service Adopting and Implementation at Local Environment Level in

In Nigeria.

**Name and Contact Details of Researcher**: Kazeem Oluwakemi Oseni, School of Computing, Room 0.15, University of Portsmouth, Buckingham Building, Portsmouth, PO1 3HE. Telephone: 02392 846460.

**Name and Contact Details of Supervisor**: Dr. Kate Dingley, School of Computing, Room 1.21, University of Portsmouth, Buckingham Building, Portsmouth, PO1 3HE. Telephone: 02392 846436.

**FEC Ref No:** 8467-09E4-A95B-634E-19C2-8F74-3458-2931

## Invitation

I would like to invite you to take part in my research study. Joining the study is entirely up to you, but before you decide I would like you to understand why the research is being done and what it would involve for you. I will go through this information sheet with you, to help you decide whether or not you would like to take part and answer any question you may have. There are 20 questions and this interview might take approximately 50 minutes or less. I would like to confirm that all information including names gathered from the interview will be treated with maximum confidentiality.

Please feel free to talk to others about the study if you wish. Do ask if anything is unclear.

I am **Kazeem Oluwakemi OSENI**, a PhD Student at the School of Computing, University of Portsmouth, United Kingdom.

**Study Summary**

This study is aimed at exploring the barriers facing E-Service Adoption and Implementation at the Local Environment Level in Nigeria. This research study is being conducted as a part of PhD Degree at the University of Portsmouth under the supervision of Dr. Kate Dingley, Dr. Penny Hart and Dr. Carl Adams. No personally identifiable information will be associated with your responses and data will be stored securely for the duration of the project after which it may be destroyed. Data summaries will be published in conferences, journals and PhD thesis.

**What is the purpose of the study?**

This study will help the researcher to determine the barriers facing E-Service adopting and implementation at Local Environment Level in Nigeria. Diffusion of Innovations (DOI) theory based framework will emerge through Action Research Methodology. This will also help to provide guidance to the policy makers at the Local Government in Nigeria and developing countries, giving necessary steps in the provision of E-Services to the public and success factors will also be considered. The completion of this research will help the researcher towards gaining a PhD degree in Computing.

**Why have I been invited?**

Participants have been identified and prior notice given earlier.

**Do I have to take part?**

No, taking part in this research is entirely voluntary. It is up to you to decide if you want to volunteer for the study. We will describe the study in this information sheet. If you agree to take part, we will then ask you to sign the attached consent form, dated January 2016, version number 2.0.

**What will happen to me if I take part?**

No risk is involved in taking part in the interview session, it is voluntary and participants can withdraw their participation anytime. Results of this study may be published and/or presented at meetings or academic conferences, but published research would never contain any participant personal information. I would suggest the session might take approximately 50 minutes or less.

**Expenses and payments**

No payments and expenses involved.

**What data will be collected and/or measurements taken?**

The data to be collected prior to the interview session includes previous, current work experience and roles of the participant. Other data to be collected during the interview are E-Government Services knowledge and Usage, barriers facing E-Service adoption and solutions especially from Government perspective, the merit and demerit of E-Service Deployment, E-Service framework, and Success factors for E-Service Adoption. Also, the suggestions and future plans for future planning of e-service at the Local Environment level in Nigeria.

**What are the possible disadvantages, burdens, and risks of taking part?**

No risk is involved in taking part in this study. Though, the interview session may require the participants to devote a little time out of their busy daily schedule to make the session a success.

**What are the possible advantages or benefits of taking part?**

Taking part in the study will help the researcher to determine the barriers facing E-Service adoption and implementation at Local Environment Level in Nigeria. Also, it will help to provide guidance to the policy makers at the Local Environment in Nigeria and other developing countries, giving necessary steps in the provision of E-Services to the public and success factors will also be considered. Lastly, it will help this researcher towards the award of PhD Degree in Computing at the University of Portsmouth, United Kingdom.

**What if there is a problem?**

If you have a query, concern or complaint about any aspect of this study, in the first instance you should contact the researcher. If the researcher is a student, there will also be an academic member of staff listed as the supervisor whom you can contact. If there is a complaint and there is a supervisor listed, please contact the Supervisor with details of the complaint. The contact details for both the researcher and any supervisor are detailed on page 1.

If your concern or complaint is not resolved by the researcher or their supervisor, you should contact the Head of Department:

The Head of Department

Department / School of Computing

University of Portsmouth

Portsmouth

PO1 3HE

Email: nick.savage@port.ac.uk

**If the complaint remains unresolved**, please contact:

The University Complaints Officer

Telephone: 023 9284 3642 Email: complaintsadvice@port.ac.uk

**Who is funding the research?**

This research is self-funded. This researcher will receive no financial reward by conducting this study, other than their normal salary/bursary as an employee/student of the University.

**Who has reviewed the study?**

This study will be reviewed by the Faculty Ethics Committee and if you wish to discuss any ethical issue regarding this research, please contact the chair, faculty of technology ethics committee at **john.williams@port.ac.uk**

**Thank you**

Thank you for taking time to read this information sheet and for considering volunteering for this interview. If you do agree to participate your consent will be sought; please see the accompanying consent form. You will then be given a copy of this information sheet and your signed consent form, to keep.

# Consent Form

## In-depth Interviews

**<u>TOPIC</u>: Barriers Facing e-Service Adopting and Implementation at Local Environment Level in Nigeria**

I agree to participate in this in-depth interview voluntarily, and I understand the purpose of the study. I understand I can withdraw at any time without prejudice.

No personal information will be used. Data collected will only be used for research purpose and may be published in Ph.D. Thesis, journals and conferences.

**……………………..** **…………………….** **………………**

**Name of Participant** **Signature** **Date**

# In-depth Interview Questions

1. Could you please tell me your name and how you like to be addressed?
2. How would you define E-Government and E-Service if asked?
3. In your perception, what should be the main aims of E-Service adoption at Local Government level in Nigeria?
4. (a) Have you ever used any E-Government Service?
   (b) Which E-Government Services have you used?
5. What factors might influence your decision to use E-Service?
6. (a) In your opinion, what are the main barriers facing E-Service adoption?
   (b) Considering political crisis as one of the major barriers facing E-Service adoption and implementation, what is your opinion about various political topics such as; political will, top management support, and commitment, having a clear vision and strategy?
7. Taking account of these barriers facing E-Service adoption and implementation at Local Government level in Nigeria as identified above. In your opinion, will you adopt and use E-Government if deployed?
8. In your opinion, how would you measure the success in the adoption and using E-Government services?
9. In Transparency International 2014 Countries Corruption Index, do you know Nigeria is listed 136 out of 174 countries rated? In your opinion, do you think the deployment of E-Service will be reducing corruption?
10. Can I ask your level of education?
11. In your opinion what do you think Government should do to solve the barriers facing E-Service adoption and implementation?
12. Legislation and regulation have been considered barriers facing E-Service adoption and implementation, what is your opinion for Nigeria?
13. Do you think this government is ready to change the laws that are not appropriate for the introduction of E-Services?
14. Can I ask you how many years of work experience and your current position?
15. Having seen the initial E-Service framework attached? What could be added to produce a final and workable E-Service framework?
16. Do you think it is an appropriate model to be used?
17. Is there anything that might influence your decision using Technology Acceptance Model (TAM) adoption?
18. Would you recommend any framework that might work for E-Service adoption?
19. I will like to contact you again if you don’t mind? May I have your contact details, please?
20. Is there any suggestion that you would like to add in this regard like plans for E-Service adoption at the local government level?

**Invitation**

**Study Title:** Barriers facing E-Service Adopting and Implementation at the Local Environment Level in Nigeria

Dear Potential Participant,

I would like to invite you to participate in **Focus Group Session**

My research is aimed at exploring the barriers facing E-Service Adopting and Implementation at the Local Environment Level in Nigeria. The required time to conduct this focus group will be 40 minutes or less. I would like to confirm that all data collected from the session will be kept confidential and the researcher is responsible for confidentiality. Participants being able to withdraw in focus group session at any point that he or she feels uncomfortable. This has been written in the consent letter and information sheet that will be sent to the participants.

The **Major topics to be discussed during the Focus Group** meetings are as follows:

- The previous and current work experience and roles of the participant.
- E-Government Services knowledge and Usage.
- Barriers facing E-Service adopting and Solutions especially from Government perspective.
- Deployment of E-Service – Merit, and Demerit.
- E-Service framework
- Success factors for E-Service Adoption.
- Suggestions and future plans for future planning of E-service at the Local Government level in Nigeria.

Thank you for taking your time to read this invitation and for considering volunteering for this focus group. If you do agree to participate, your consent will be sought; please see the accompanying consent form. You will then be given a copy of this information sheet and your signed consent form, to keep.

# CONSENT FORM

## Online Focus Group

**Title of Project**: Barriers Facing E-Service Adopting and Implementation at Local Environment Level in Nigeria.

**Name and Contact Details of Researcher**: Kazeem Oluwakemi Oseni, School of Computing, Room 0.15, University of Portsmouth, Buckingham Building, Portsmouth, PO1 3HE. Telephone: 02392 846460.

**Name and Contact Details of Supervisor**: Dr. Kate Dingley, School of Computing, Room 1.21, University of Portsmouth, Buckingham Building, Portsmouth, PO1 3HE. Telephone: 02392 846436.

**FEC Ref No:** KO1

1. I confirm that I have read and understood the information sheet dated January 2016 (version 2.0) for the above study. I have had the opportunity to consider the information, ask questions and have had these answered satisfactorily.

2. I understand that my participation is voluntary and that I am free to withdraw at any time without giving any reason.

3. I understand that data collected during this study could be requested and looked at by regulatory authorities. I give my permission for any authority, with a legal right of access, to view data which might identify me. Any promises of confidentiality provided by the researcher will be respected.

4. I understand that the results of this study may be published and/or presented at meetings or academic conferences, and may be provided to research commissioners. I give my permission for my anonymous data, which does not identify me, to be disseminated in this way.
5. I agree to the data I contribute being retained for any future research that has been approved by a Research Ethics Committee.

6. I agree to take part in the above study.

**Name of Participant:** **Date:** **Signature:**

**Name of Person taking Consent:** **Date:** **Signature:**

***Note**: When completed, one copy to be given to the participant, one copy to be retained in the study file.*

## Additional

I consent for my typed text during the focus group sessions. The typed text will be transcribed and analysed for the research.

**Will my taking part in the study be kept confidential?**

Confidentiality will be provided. Anonymity is being provided by allowing participants to use a pseudonym of their choice. They will use typed text in the session so will see each other's responses but will not know who those others are. They could be sitting in the same room as others in the focus group and not knowing they were participating in the same focus group. The researcher will know the identities, and pseudonyms initially for contact purposes and to ensure that anyone wanting follow-up information can be sent it.

The raw data, which identifies participants, will be kept securely by this researcher using encryption on a password protected desktop and on the University Drive which is password protected. The data will be collected through the typed text only during the online focus group sessions. All data collected will be transcribed where necessary, analysed quickly and original sources will be removed from servers. The data, when made anonymous, may be presented to others at academic conferences, or published as a project report, academic dissertation or in academic journals or book. Anonymous data, which does not identify you, may be used in future research studies approved by an appropriate research ethics committee.

The raw data, which would identify the participant, will not be passed to anyone outside the study team without your express written permission. The exception to this will be any regulatory authority which may have the legal right to access the data for the purposes of conducting an audit or enquiry, in exceptional cases. These agencies treat your personal data in confidence. The Ph.D. Thesis produced by the researcher will be made Open Access by depositing with the University Library and Ethos Library, and it will also be registered with the pure online system in order for them to be eligible for the next Research Excellence Framework (REF) due in 2020.

I consent to verbatim quotes being used in publications; I will not be named, and pseudonyms will be associated with the comments.

I understand that participation will include demanding procedures.

**Limitations to Confidentiality**

I understand that whatever I say in the typed text during the focus group session is confidential unless I tell the researcher that I or someone else is in immediate danger of serious harm, or the researcher sees or is told about something that is likely to cause serious harm. If that happens, the researcher will raise this with me during the focus group session and tell me about what could happen if I continue to talk about it and explore how I would prefer to deal with the situation.

I agree to use the pseudonym of my choice and be referred to accordingly.

## Online Focus Group Questionnaire

1. Have you used E-Government Services before?
2. What type of E-Government Services have you used before?
3. What are the barriers facing E-Service projects success both the adoption and implementation?
4. Will you encourage and use E-Government Services?
5. What are the success factors for adopting E-Government Services?
6. Does the E-Service framework sent to you represent your notion of E-Service?
7. Will you encourage the use of Diffusion of Innovation theory in E-Government services implementation?
8. Do you think the E-Service technology adoption will help in other aspects/areas?
9. As part of the stakeholder in the provision of E-Service at the local government level in Nigeria, will you be ready to receive and implement the recommendations on E-Service from this study?
10. Do you have any suggestion or future plan for E-Government Services?

# Online Survey

**Study Title:** Barriers facing E-Service Adopting and Implementation at the Local Environment Level in Nigeria.

**FEC Ref No:** KO1

**Name of researcher: Mr.** Kazeem Oluwakemi **Oseni**

**Supervisor:** Dr. Kate Dingley

**Contact details:** School of Computing, Room 0.15, University of Portsmouth, Buckingham Building, Portsmouth, PO1 3HE. Telephone: 02392 846460.

**Contact details of Supervisor**: Dr. Kate Dingley, School of Computing, Room 1.21, University of Portsmouth, Buckingham Building, Portsmouth, PO1 3HE. Telephone: 02392 846436.

**Invitation**

Thank you for reading this. I would like to invite you to take part in my research study by completing this questionnaire. It is entirely up to you whether you participate, but your responses would be valued. You have been identified as a potential respondent as a stakeholder in the E-Government development at the local government level in Nigeria. My study will investigate the barriers facing E-Service adopting and implementation at Local Environment Level in Nigeria. I neither need your name nor any identifying details; the questionnaire can be completed anonymously, and all reasonable steps will be taken to ensure confidentiality. Responses from completed questionnaires will be collated for analysis; once this is completed, the original questionnaires might be retained towards the successful completion of my Ph.D. before destroying them. All data collected will be transcribed where necessary, analysed quickly and original sources will be removed from servers.

Up to this stage, completed questionnaires will be stored in a secured and locked filing cabinet. If you wish to learn more about the results of the research, please, feel free to contact this researcher or his supervisor with the contact details above. There is no significant risks or benefits associated with taking path in this survey. Participation is free, and participants can withdraw at any time without prior notice.

# Online Survey - Questionnaire

It will be appreciated if all questions in the survey are completed.

## Part A: Demographic Questionnaires

**1. What is your age group?**

- ☐ 16-25
- ☐ 26-35
- ☐ 36-45
- ☐ 46-55
- ☐ 56 and above

**2. What is your Gender?**

- ☐ Male
- ☐ Female
- ☐ Prefer not to say

**3. Nigeria has six geographical regions. Please, in the box below write the 'Region' in Nigeria you are from?**

**4. What 'State' in Nigeria are you from? Please, write in the box below.**

**5. What is your highest level of education?**

- ☐ PhD or equivalent
- ☐ Masters Level or equivalent
- ☐ Bachelors Level or equivalent
- ☐ Diploma Level or equivalent
- ☐ High School Level or equivalent
- ☐ Other:

**6. Please, state your current Occupation in the box below.**

As discussed above, participants can terminate their involvement in the questionnaire anytime. Please, state your reason in the box below if you wish to discontinue and remember to click submit button OR to go next question to continue.

## Part B: E-Service Adoption and Implementation Questionnaires

**Electronic Service** shortened as ‘E-Service’ which refers to any service that is provided by any electronic means e.g. Internet/website, mobile devices or kiosk (Bhuiyan, 2011). Examples are online bill payment, online tax payment, and online chat.

7. Have you used E-Government Services before?

- Yes
- No
- Not interested in online developments

8. If you answered ‘Yes’ to the previous question. Which E-Government Services from the list below have you used?

- Applying for your passport
- Tax payment
- Applying for a job
- Applying for driving licence or renewal
- Applying for Housing
- Other:

9. If you answered ‘No’ to the previous question. Please, choose from the list below the factors that might influence your decision to use the E-Service?

- Perceived Ease of Use
- Perceived Usefulness of Service
- User Satisfaction (For example, I prefer to talk to someone)
- Attribute of Usability or Accessibility
- Reliability
- Other:

10. **Please, rate your satisfaction level of E-Service you have used (1 - Very Dissatisfied, 2 – Dissatisfied, 3 - Neutral, 4 - Satisfied, 5 - Very Satisfied).***

| | 1 | 2 | 3 | 4 | 5 |
|---|---|---|---|---|---|
| Applying for your passport | ○ | ○ | ○ | ○ | ○ |
| Tax payment | ○ | ○ | ○ | ○ | ○ |
| Applying for a job | ○ | ○ | ○ | ○ | ○ |
| Applying for driving licence or renewal | ○ | ○ | ○ | ○ | ○ |
| Applying for Housing | ○ | ○ | ○ | ○ | ○ |
| Other | ○ | ○ | ○ | ○ | ○ |

11. **In a recent discussion with a focus group, the following have been identified as barriers facing E-Service Adoption and Implementation. On a scale of 1 – 5, how will you rate the potential impact of these barriers facing E-Service adoption and implementation? (1 - Extremely Low, 2 - Low, 3 - Neutral, 4 - High, 5 - Extremely High)***

| | 1 | 2 | 3 | 3 | 5 |
|---|---|---|---|---|---|
| Bad Leadership and Management | ○ | ○ | ○ | ○ | ○ |
| Corruption | ○ | ○ | ○ | ○ | ○ |
| Cultural difference | ○ | ○ | ○ | ○ | ○ |
| Lack of Awareness | ○ | ○ | ○ | ○ | ○ |
| Lack of Funding | ○ | ○ | ○ | ○ | ○ |
| Lack of IT infrastructural | ○ | ○ | ○ | ○ | ○ |
| Lack of IT Skills | ○ | ○ | ○ | ○ | ○ |
| Lack of Partnership and Collaboration | ○ | ○ | ○ | ○ | ○ |
| Lack of Policy and Regulation for E-Services Usage | ○ | ○ | ○ | ○ | ○ |
| Lack of Strategic Plans | ○ | ○ | ○ | ○ | ○ |
| No Regular Training | ○ | ○ | ○ | ○ | ○ |
| Political Crisis | ○ | ○ | ○ | ○ | ○ |
| Resistance to E-Platform Change | ○ | ○ | ○ | ○ | ○ |
| Security | ○ | ○ | ○ | ○ | ○ |
| Trust Issues | ○ | ○ | ○ | ○ | ○ |
| Usability Issues | ○ | ○ | ○ | ○ | ○ |

12. **In Transparency International 2014 Countries Corruption Index, Nigeria is 136 most corrupt country out of 174 countries rated. In your own opinion, how likely do you think the deployment of E-Service will be reducing corruption, and that it will be of great benefits to both government and public?**
Using Likert Scale 1 - 5 (1 - Strongly disagree, 2 - Disagree, 3 - Neutral, 4 - Agree, 5 -Strongly agree)

| 1 | 2 | 3 | 4 | 5 |
|---|---|---|---|---|
| ○ | ○ | ○ | ○ | ○ |

13. In your opinion what do you think Government should do to solve the barriers facing E-Service adoption and implementation at Local Government level in Nigeria?

## Part C - Diffusion of Innovations (DOI) Questionnaires

**Diffusion of Innovations theory** explains how, why and at what rate technology and new ideas are accepted or spread through cultures (Rogers, 2003). In a simpler term, Diffusion is a special type of communication concerned with the spread of messages that are perceived as new ideas. An innovation on the other hand, simply means, "An idea perceived as new by the individual" (Rogers, 1995).

14. Have you heard about Diffusion of Innovations theory (DOI) before?

- Yes
- No
- Not interested in technology deployment

15. If you answered 'No' to the previous question, go to next question and if you answered 'YES', do you think Diffus Innovations theory (DOI) should be used in the provision of electronic (online) public services at Local Government l in Nigeria (1 - Strongly disagree, 2 - Disagree, 3 - Neutral, 4 - Agree, 5 -Strongly agree).

| 1 | 2 | 3 | 4 | 5 |
|---|---|---|---|---|
| ○ | ○ | ○ | ○ | ○ |

16. Despite the barriers facing E-Service adoption and implementation at Local Government level in Nigeria as identi above. In your opinion, will you adopt and use E-Government services like E-Payment at Local Government level in Nigeria if deployed?

- Yes
- No
- Not interested in E-Services at Local Government level.

17. If you answered 'Yes' to the previous question, go to the next question. If you answered 'No' please give your rea below.

18. In your opinion, please state below the success factors for adopting and using E-Government services at Local Government Level in Nigeria?

19. Would you like to be contacted regarding your answers?

- Yes
- No

20. If you answered 'Yes' to the previous question, please, write your name and contact details in the box below. Your contact details will not be disclosed, it will only be useful if there is a need to contact you.

## Use of Data

I understand that anonymised data summaries may be published in conferences, journals and PhD thesis. Thank you very much for taking part in this survey, your contribution is appreciated.

Kind regards,

Kazeem Oluwakemi Oseni
PhD Student
University of Portsmouth
United Kingdom.

If you have any concerns regarding this research please contact me or my supervisor in the first instance. If you are not entirely happy with a response please contact:

The Head of Department

Department / School of Computing

University of Portsmouth

Portsmouth

PO1 3HE

Email: nick.savage@port.ac.uk

**If the complaint remains unresolved, please contact:**

The University Complaints Officer

Telephone: 023 9284 3642 Email: complaintsadvice@port.ac.uk

# **Appendix B -** Participants Responses from the Interviews, Online Focus Group and Online Survey

## A. **Participants Responses from the Interviews**

The researcher used questionnaires for the one-on-one interviews (see, Chapter Five) with the participants to find out the gender, educational qualifications, years of experience, E-Government services knowledge and usage, willingness to adopt E-Government services, the framework for E-Government service adoption and implementation. The reason for these questions is to be able to find answers to the research questions and to help the policy makers at the local government level in Nigeria find out how the citizens feel about the bottlenecks for no availability of the E-Government services at this level. Moreover, it will help the researcher to map out the appropriate strategy and solutions needed to solve the barriers which this study is looking into. IBM SPSS Statistics version 22 was used for this data analysis.

### 1. Gender

Statistically, 53.3% representing 16 participants out of 30 were females, and 46.7% representing 14 participants were male as indicated in **Figure 80** below. Moreover, the interest of the staff towards E-Service initiatives at the local environment level is commendable. Though, many of the staff were unable to discuss openly on the barriers confidently because of the fear of reprisals even with the confidentiality being maintained. Most of the staff interviewed have not used E-Services before and to achieve this when available, most of them need basic IT training, and easy to use websites, when implemented, will help tremendously in E-Services usage.

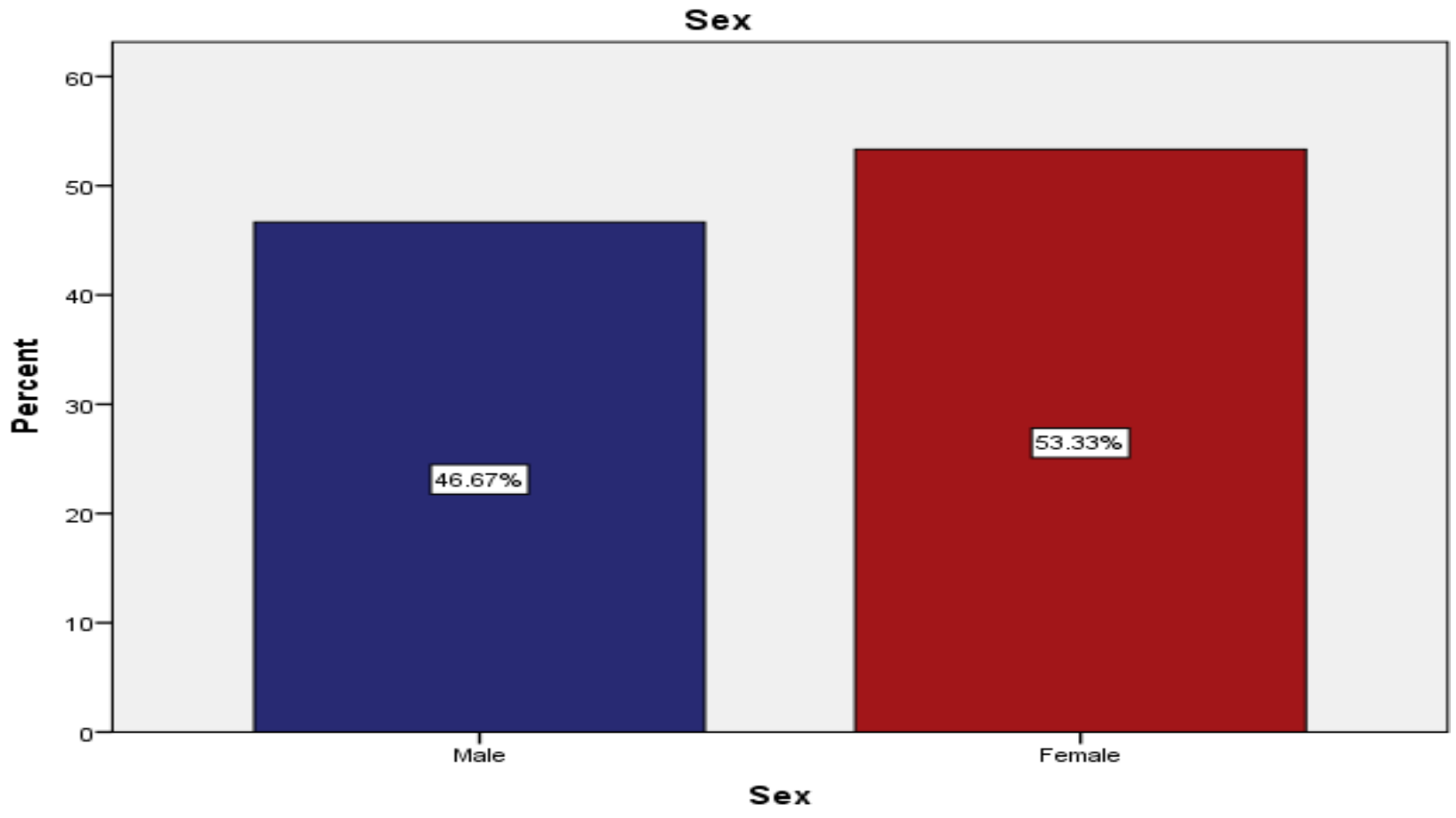

**Figure 80 SPSS Interview Analysis – Sex**

### 2. Definition of E-Government and E-Service

From the results available in **Table 33** below, 53.3% of the participants interviewed were able to give a reasonable definition of both E-Government and E-Service while the remaining 46.7% of the participants indicated that they have no idea. This question was asked to know if the participants have deep knowledge about the E-Services as they have been identified as the stakeholders at the local environment levels in Nigeria for E-Government services development.

**Table 33 Interview Response: E-Government Service Definition**

| | | Frequency | Percent | Valid Percent | Cumulative Percent |
|---|---|---|---|---|---|
| **Valid** | E-Government Service Definition | 16 | 53.3 | 53.3 | 53.3 |
| | No idea | 14 | 46.7 | 46.7 | 100.0 |
| | **Total** | **30** | **100.0** | **100.0** | |

### 3. Aims of E-Service Adopting and Implementation

One striking result that emerged from the interviews conducted with the participants despite the facts that more than half of them have not used E-Government services before the interviews is that twenty-seven (27) out of the thirty (30) total participants were able to identify what they think the aims of E-Service adoption and implementation at the local government level in Nigeria will be. For the researcher, the overwhelming contributions on this might be connected to the high level of education most of the respondent already acquired. This is a welcome development to this study as dealing with participants that understand what the aims and objectives of the research are assisted with the constructive contributions towards obtaining a reach data. Hence, as shown in **Figure 81** below, 7 participants which represent 23.3% mentioned that their aim of E-Service adoption would be for the ***effective provision of online government services***, 4 participants representing 13.3% agreed that their aim to adopt E-Service would be for ***ease collection of revenue***, 3 participants representing 10% decided that their aim of E-Service adoption would be to ***bring the governance closer to the grassroots*** (Citizens at the local government areas in Nigeria). Two participants representing 6.7% agreed that their aim to adopt E-Service would be for ***accountability purposes***, 4 participants which represent 13.3% mentioned that their aim of E-Service adoption would be ***to run a paperless administration***, 6 participants representing 20% decided that their aim of E-Service adoption will be ***to deliver public service online***, 1 respondent representing 3.3% agreed that there are ***other*** aims for adopting E-Service. Finally, 3 participants representing 10 % agreed that they ***have no idea*** on the aims to adopt E-Service and their response might be connected with their unwillingness to adopt and use E-Government services which are not currently available at the local government.

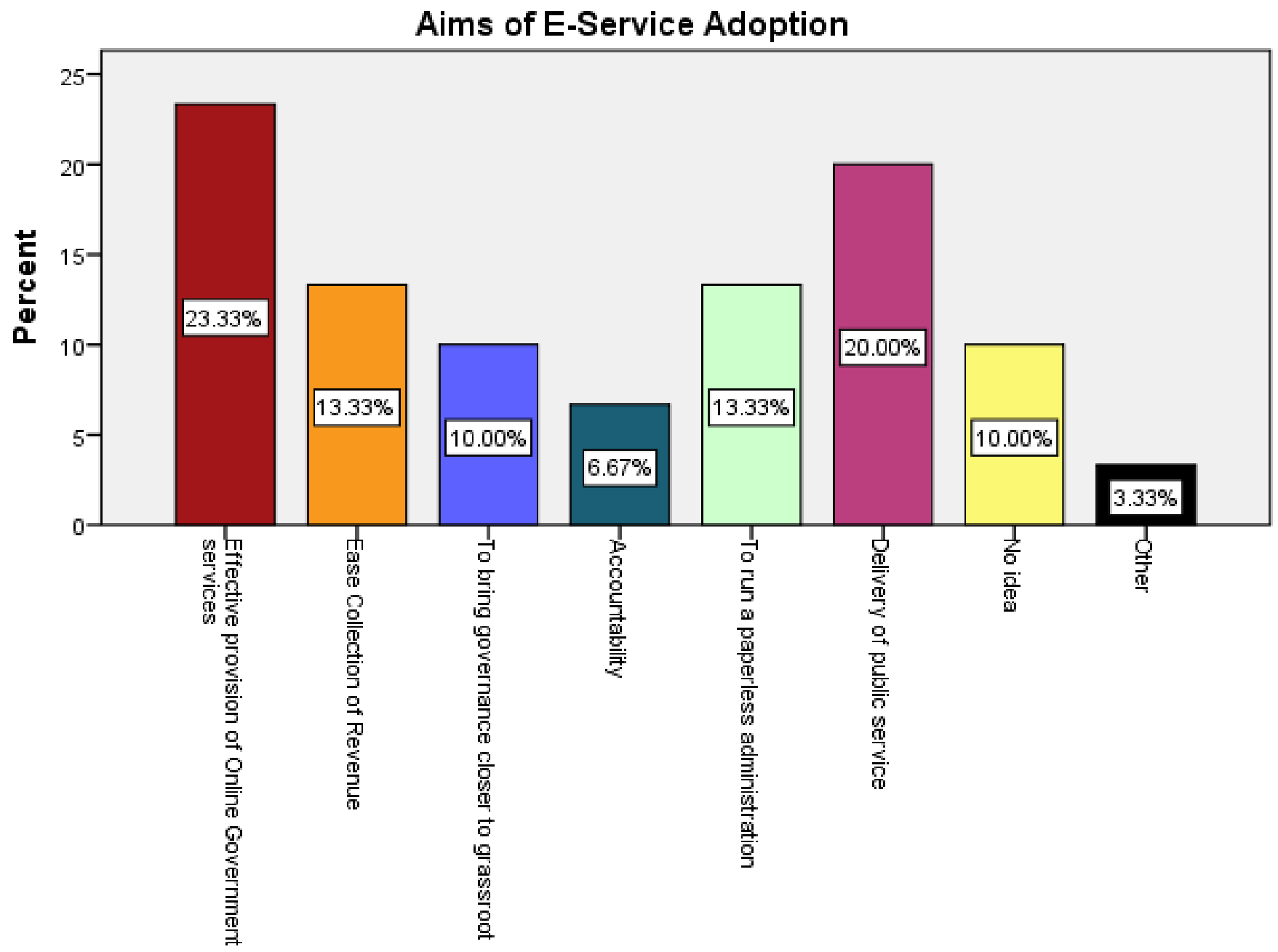

**Figure 81 SPSS Interview Analysis – Aims of E-Service Adoption**

### 4a. E-Government Services Usage

Knowing the level of the E-Government Services usage among the participants who are also part of the stakeholders in the provision and adopting of the E-Government services at the local environment level in Nigeria will assist this researcher in unveiling the barriers facing the initiatives. This researcher asked the participants this question to also find out if the level of usage of the E-Services among the stakeholders has any effect on why the local government unwillingly to implement the initiatives. The result in **Figure 82** below shows that 10 participants representing 33.3% of the total have used E-Government services while the other 20 participants which represent 66.7% have not used any E-Government service before. The single most striking observation to emerge from the data is the result justify the aims of this researcher to investigate the barriers facing the adoption and implementation of the E-Services at the local government level in Nigeria. The facts that most of the staff interviewed have not used E-Services before might be connected to the low enthusiasm towards advising the leadership of the local governments in this direction despite the huge benefits associated with the E-Government services if implemented.

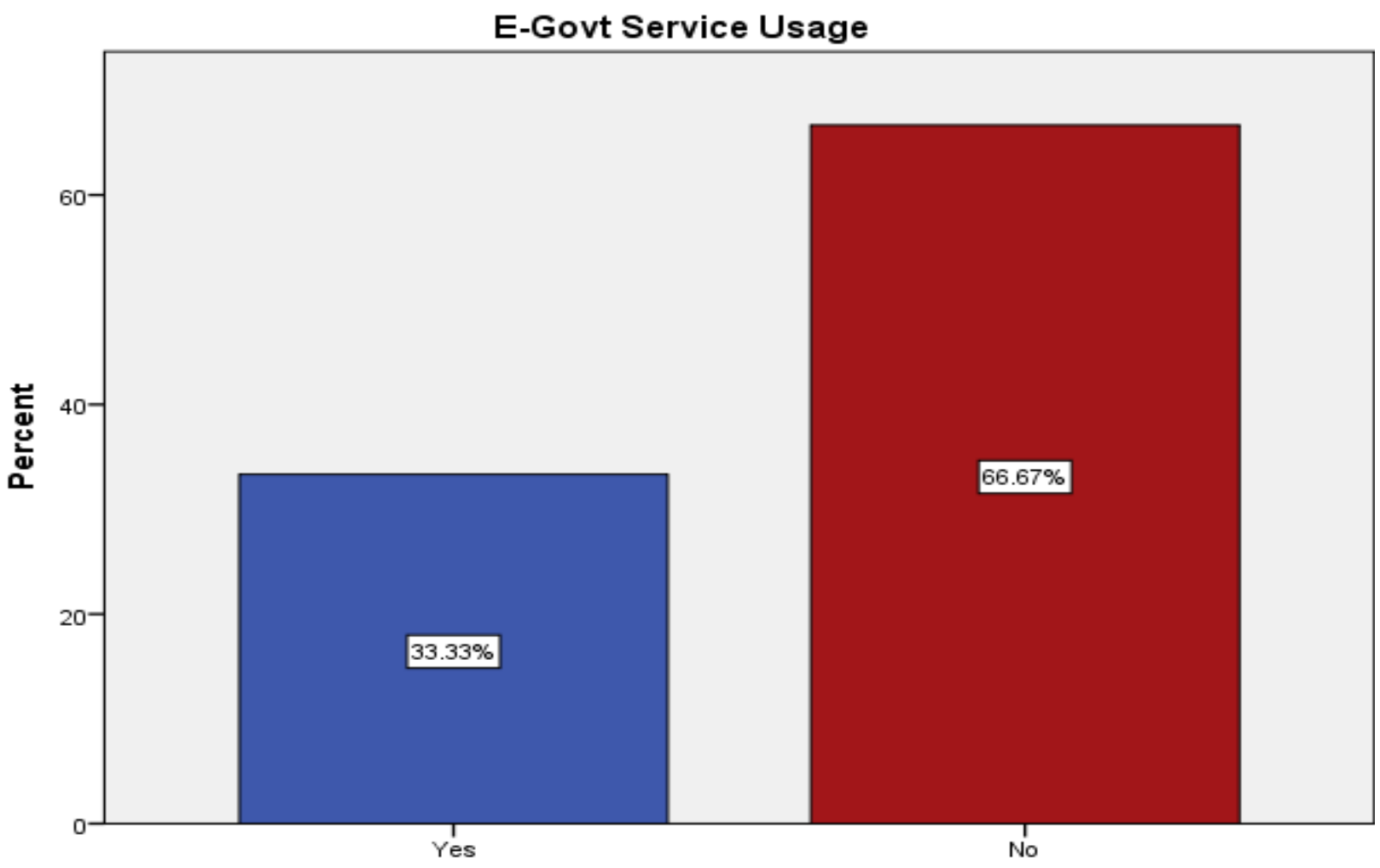

**Figure 82 SPSS Interview Analysis – E-Government Services Usage**

### 4b. Types of E-Government Services Used

As obtained from **Figure 82** above, 10 participants which represent 33.33% of the total participants have used E-Government services before which are either at federal or state government levels. This researcher further asked the participants to find out these E-Government services as it will help to know the services commonly used and the types to be provided at the local environment level. In the analysis as shown in **Figure 83** below, out of the 10 participants that have used E-Government services, 6 participants representing 20% have used E-Payment facilities. Also, one respondent each representing 3.3% have used E-Passport, E-Licence, E-Registration (Auto) respectively. Another respondent representing 3.3% has used another E-Government service. However, 20 participants which represent 66.67% have not used E-Government services before. This result is surprisingly not unexpected as it emerged previously that the same numbers of participants have never used E-Government service.

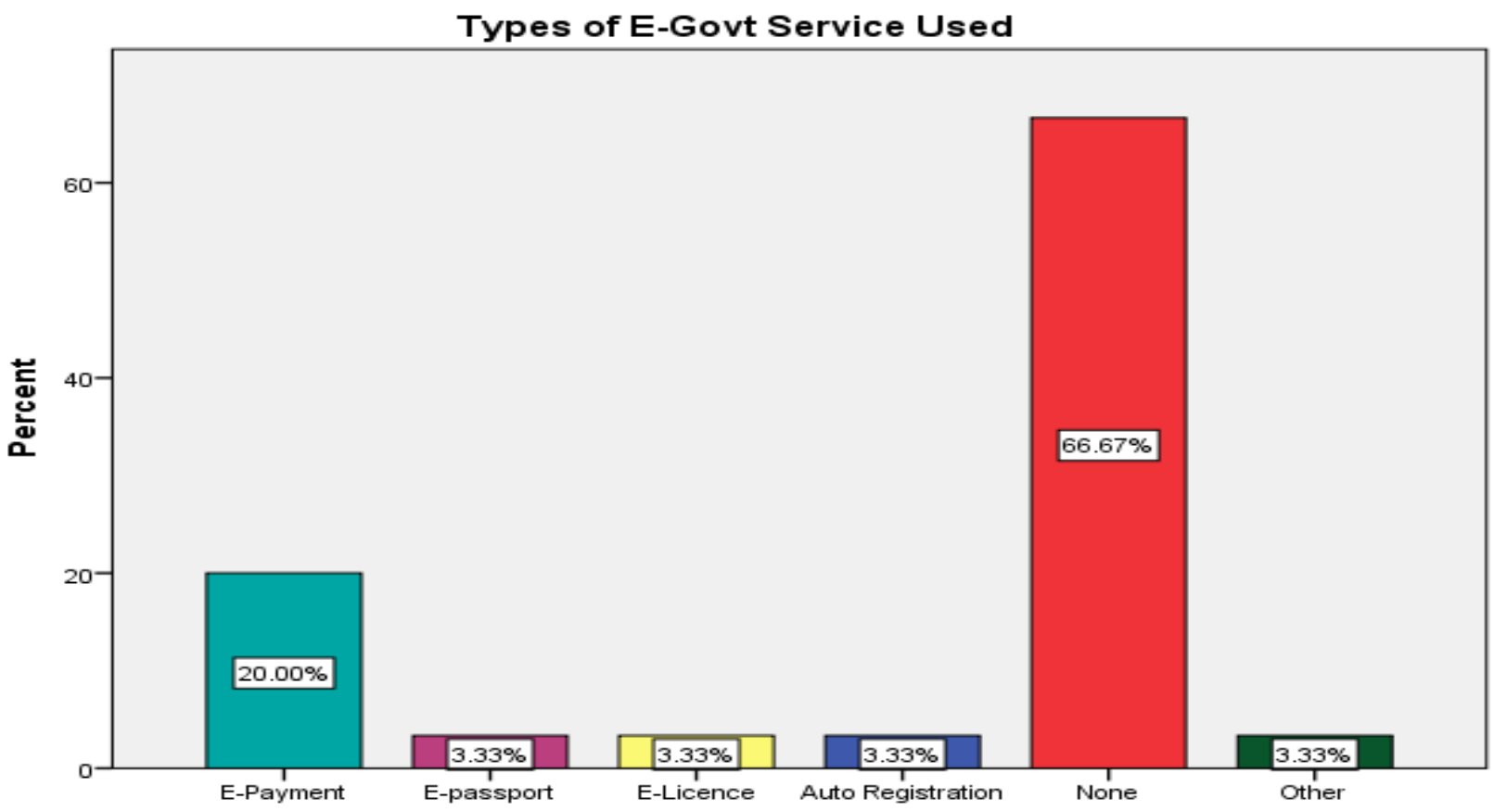

**Figure 83 SPSS Interview Analysis – Types of E-Government Services Used**

### 5. Factors that Might Influence Decision for E-Service Usage

The overall response to this question was very positive. From **Table 34** below, the participants stated the reasons that might influence their decision to adopt E-Service. 26.7% of the total participants agreed that they would adopt E-Service because it is convenient, 6.7% will adopt because it is reliable, and 13.3% of the participants agree to adopt it because it is cost effective. Also, another 13.3% indicated they would adopt E-Service because of the transparency factor.

**Table 34 Interview Response - Factors that Might influence E-Service Usage**

| | | Frequency | Percent | Valid Percent | Cumulative Percent |
|---|---|---|---|---|---|
| Valid | Convenience | 8 | 26.7 | 26.7 | 26.7 |
| | Reliability | 2 | 6.7 | 6.7 | 33.3 |
| | Cost reduction | 4 | 13.3 | 13.3 | 46.7 |
| | Transparency | 4 | 13.3 | 13.3 | 60.0 |
| | Quality public service delivery | 6 | 20.0 | 20.0 | 80.0 |
| | No idea | 4 | 13.3 | 13.3 | 93.3 |
| | Other | 2 | 6.7 | 6.7 | 100.0 |
| | **Total** | **30** | **100.0** | **100.0** | |

Furthermore, 20% of the participants agreed to adopt the E-Service because of quality service delivery while 4 participants which represent 13.3% have no idea about adopting E-Service. Lastly, 6.7% of the participants agreed they would adopt E-Service due to other reasons.

### 6a. Main Barriers Facing E-Service Adopting and Implementation

As discussed in **section 6.3.4** (Chapter Six), there is a strong correlation between the barriers facing E-Service adoption and implementation from the results obtained through the interviews conducted with the participants and the online survey. As shown in both **Figure 54** (Chapter Six) and **Figure 84** below, we can see striking relationships among these barriers, in the results presented in **Figure 84** below, five (5) participants which represent 16.7% agreed that the main barrier facing the E-Service adoption and implementation at the local government level in Nigeria is the ***autonomy*** issues for the local government. This is the highest figure among the participants interviewed. Though, the autonomy issue was discussed briefly in **Section 6.3.4**, it is imperative to say here that it is a major problem at the local environment level in Nigeria as the administrators have no control on their major source of finance, that is, the money shared monthly to the local government areas by the federal government due to joint local-state government account (Okafor, 2010). The funds which are later released to the local government areas by the state governments are so meagre for the administrators at the local government to execute major projects such as E-Service initiatives (Okafor, 2010).

Further analysis showed that 4 participants representing 13.3% agreed that ***finance*** is the main barrier, 3 participants which represent 10% accepted that the main barrier facing E-Service adoption is ***power supply shortage***. Another 3 participants representing 10% acknowledged that the main barrier facing the adoption and implementation at the local government level in Nigeria is the ***infrastructural problem***. Another important finding is that out of the 30 participants who were interviewed, 5 which represent 16.7% agreed that ***corruption*** is the main barrier facing the E-Service adoption and implementation. The result of corruption is joint-highest among the participants interviewed, and this correlates with the results obtained from the online survey. Both local government autonomy issue and corruption are followed closely by the finance problems, and they appeared to the main barriers facing E-Service adoption and implementation at the local government level in Nigeria taking into consideration the results obtained from the online survey and the interviews.

Furthermore, 4 participants representing 13.3% suggested that ***illiteracy/lack of training*** is the main barriers facing E-Service adoption and implementation while another 3 participants which also represent 10% admitted that lack of legislation/regulation is accountable for the obstacles facing E-Service adoption and implementation. ***Leadership problem/political instability*** is another major barrier facing E-Service adoption and implementation at the local government level, 2 participants representing 6.7% agreed to this. Currently, in Nigeria, the level of corruption among the government officials and the political leaders is so high (Oseni and Dingley, 2014). Funds meant for the projects such as E-Service initiatives are being diverted to private pockets, even when the ideas of laudable projects are being conceived, there is no fund available to execute it (Oseni and Dingley, 2014). Political instability is another major issue, the politicians as witnessed today in Nigeria have been identified as sponsors of the Islamic group “Boko Haram” which has caused major political tension in the northerner part of Nigeria, many citizens especially women and children have been killed and displaced from their various home (Oseni and Dingley, 2014). In an unconducive environment of this nature, initiative E-Service project and for citizens to adopt it remain a herculean task. Lastly, another interviewee representing 3.3% of the total alluded to the opinion that there are indeed barriers facing E-Service adoption and implementation at the local government level in Nigeria ***other*** than the ones previously mentioned.

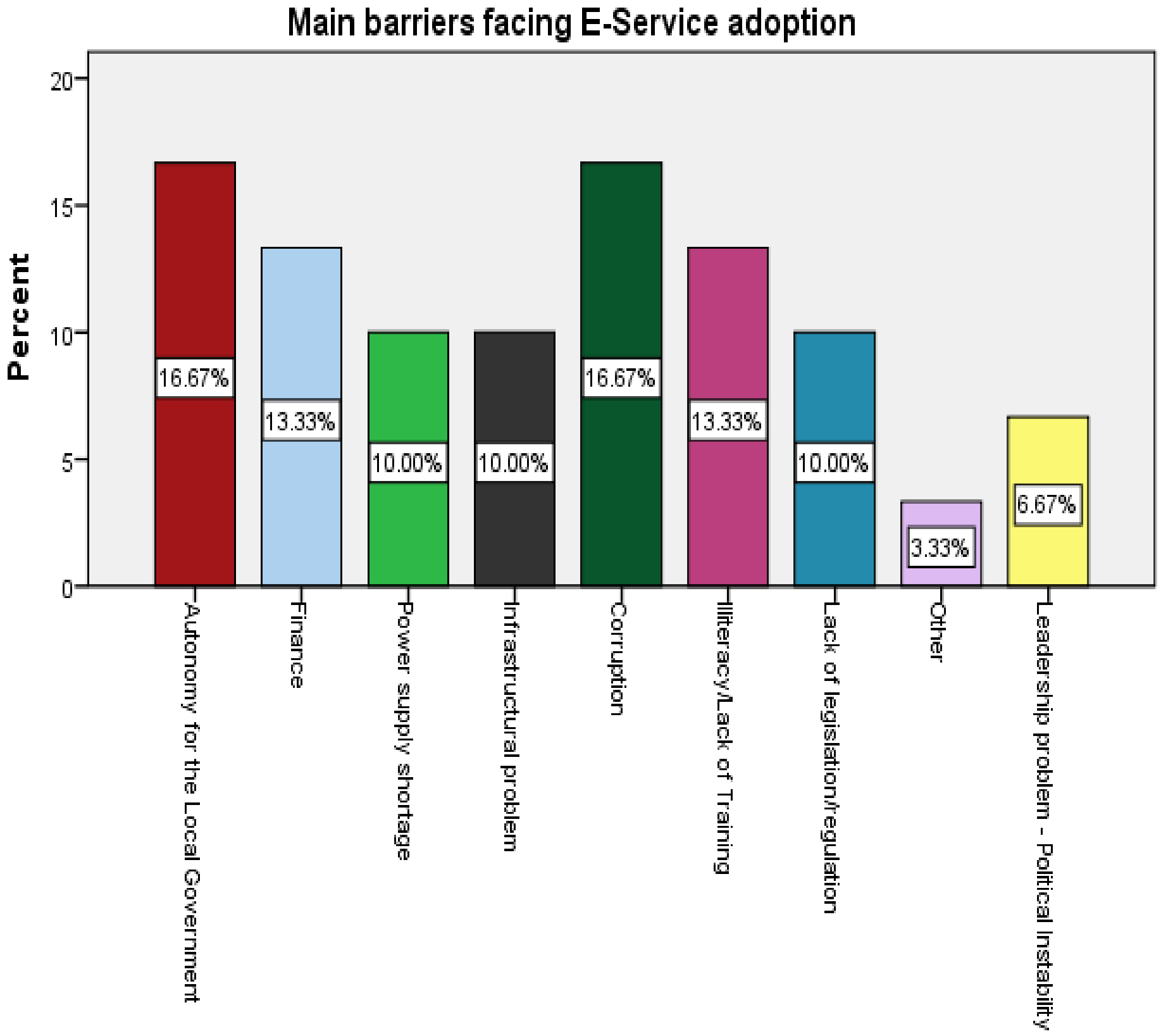

**Figure 84 SPSS Interview Analysis – Main Barriers Facing E-Service Adopting**

### 6b. Opinion on Barriers facing E-Service Adopting e.g. Political Crisis

This question was asked to know the participant's opinion on the barriers facing E-Services especially factors such as political will, top management support, and commitment, having a clear vision and strategy for the E-Service Initiatives. 4 participants (13.3%) of the total agreed that no due process in the management at local government in Nigeria. 9 participants (30%) agreed that political class was not encouraging E-Service, 6.7% of the participants indicated that poor vision and strategy is another issue facing E-Service adoption, 16.7% participants acknowledged that leaders are afraid of change, another 16.7% participants said there is a high level of corruption. Lastly, another 16.7% participants said they have no idea about the opinion on E-Service barriers.

**Table 35 Interview Response - Opinion on Barriers facing E-Service Adopting e.g. Political Crisis**

| | | Frequency | Percent | Valid Percent | Cumulative Percent |
|---|---|---|---|---|---|
| **Valid** | No due management process | 4 | 13.3 | 13.3 | 13.3 |
| | Political class not encouraging E-Service | 9 | 30.0 | 30.0 | 43.3 |
| | Poor vision and strategy | 2 | 6.7 | 6.7 | 50.0 |
| | Leaders afraid of change | 5 | 16.7 | 16.7 | 66.7 |
| | High level of corruption | 5 | 16.7 | 16.7 | 83.3 |
| | No idea | 5 | 16.7 | 16.7 | 100.0 |
| | **Total** | **30** | **100.0** | **100.0** | |

### 7. The Use and Adopting of E-Service if Deployed

The overall response to this question was very positive as 29 participants (96.7%) agreed that they would use and adopt E-Service if deployed at the local environment level in Nigeria as indicated in **Table 36** below. One participant during the interview (3.3%) stated not willing to use E-Service because of the internet security issue.

**Table 36 Interview Response - Will you Adopt E-Service if Deployed**

| | | Frequency | Percent | Valid Percent | Cumulative Percent |
|---|---|---|---|---|---|
| Valid | Yes | 29 | 96.7 | 96.7 | 96.7 |
| | No | 1 | 3.3 | 3.3 | 100.0 |
| | **Total** | **30** | **100.0** | **100.0** | |

### 8. The Success in the Adopting and Using E-Government Services

This question was asked to know how the participants will measure the success of the adoption and using of the E-Service. From the responses shown in **Table 37** below, 13.3% of the participants indicated that the success could be measured because the service is very fast. 6.7% said the success is because it is safer for them to use. Another 6.7% agreed that the success is because the transactions done is very confidential to them. Also, another 6.7% participants said that the success of using E-Service is because the service is reliable while 26.7% of the total participants concluded that the success of using E-Service is because of timely service delivery. Furthermore, another 2 participants which represent 6.7% agreed that the transparent aspect of E-Service is the major success for using it. Almost one-third of the participants (30%) said they have no idea of what the success might be while the last participant (3.3%) agreed that there is another success of using E-Service which is different from others mentioned above.

**Table 37 Interview Response - How Will You Measure E-Service Adopting Success**

| | | Frequency | Percent | Valid Percent | Cumulative Percent |
|---|---|---|---|---|---|
| Valid | Faster | 4 | 13.3 | 13.3 | 13.3 |
| | Safer | 2 | 6.7 | 6.7 | 20.0 |
| | Confidential | 2 | 6.7 | 6.7 | 26.7 |
| | Reliability | 2 | 6.7 | 6.7 | 33.3 |
| | Timely service delivery | 8 | 26.7 | 26.7 | 60.0 |
| | Transparency | 2 | 6.7 | 6.7 | 66.7 |
| | No idea | 9 | 30.0 | 30.0 | 96.7 |
| | Other | 1 | 3.3 | 3.3 | 100.0 |
| | **Total** | **30** | **100.0** | **100.0** | |

### 9. Deployment of E-Service Reducing Corruption

This was asked to know the opinion of the participants towards E-Service reducing the corruption at the local environment level in Nigeria. The overall response to this question was very encouraging as 28 participants (93.3%) agreed that the deployment would reduce corruption.

**Table 38 Interview Response: Will Deployment of E-Service Reduce Corruption**

| | | Frequency | Percent | Valid Percent | Cumulative Percent |
|---|---|---|---|---|---|
| Valid | Yes | 28 | 93.3 | 93.3 | 93.3 |
| | No | 1 | 3.3 | 3.3 | 96.7 |
| | Not sure | 1 | 3.3 | 3.3 | 100.0 |
| | **Total** | **30** | **100.0** | **100.0** | |

Also, one of the participants (3.3%) said deploying it will not reduce corruption but unable to give a reason. Another participant (3.3%) was not sure about the question.

### 10. Level of Education

The level of education for the researcher may be useful to the outcome of this study, and it is worthy to ask the participants. Having a basic education reveals much about the participants other than job experience alone. Education has many benefits as it affects individuals in many ways. For example, having a good education might improve the chances of getting a better job. Hence, as shown in **Figure 85** below, 11 participants (36.7%) has Master's Degree or equivalent. 12 participants (40%) already acquired Bachelor's Degree or equivalent as at the time of this interview. 5 participants representing 16.7% obtained Diploma or equivalent, and lastly, 2 participants out of the 30 staff interviewed already obtained Secondary School Certificate or equivalent, this represents 6.7% of the total.

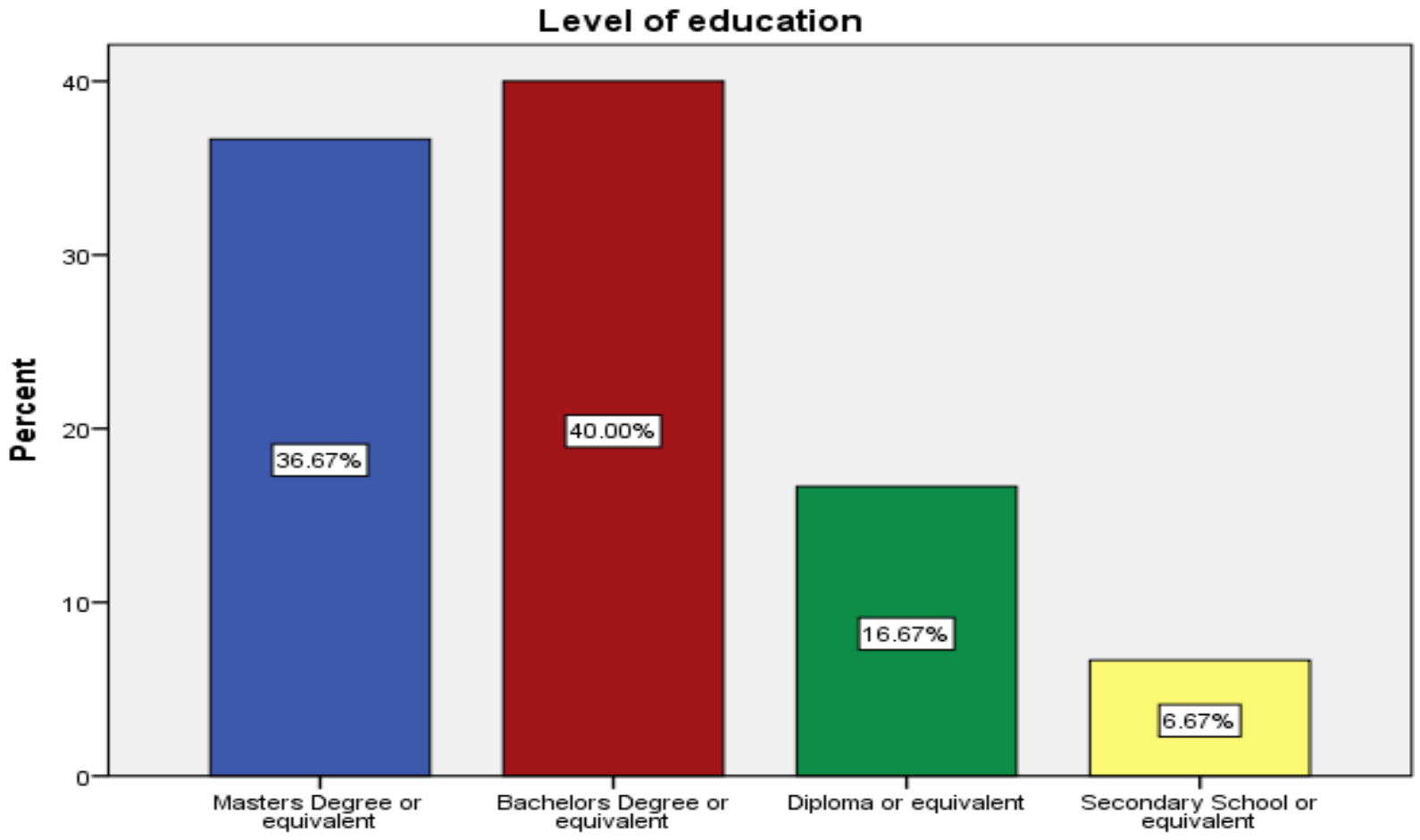

**Figure 85 SPSS Interview Analysis – Level of Education**

### 11. Solutions to the Barriers Facing E-Service Adopting and Implementation

In a real-life scenario, whenever problems arise, there is always solutions to these problems. As many challenges hinder the E-Service adoption and implementation at the local government level in Nigeria, the participants interviewed beliefs there are many solutions to the barriers if the government and the stakeholders could corroborate together in implementing the suggested solutions. From the analysis, as shown in **Figure 86** below, 4 participants which represent 13.3% agree that for the barrier facing E-Service adoption and implementation to be resolved, ***political stability*** in encouraging in Nigeria. Politicians need to behave always like matured aspirants. They need to give peace a chance in the country. Losing an election should not resolve in instigating riot among the citizens, there must be a legal proceeding in the recovery of mandate lost in the election. Participants believe if peace reigns, the conducive environment will help in initiating projects like E-Government services. ***Granting autonomy to the local government*** is another solution suggested by 7 participants which represent 23.3% of the total and the highest. There is a need for the local government finance to be independent, and there must not be a joint account with the state governments in Nigeria (Ahmad, 2013). This will give the administrators at the local government levels more freedom with the availability of funds to initiate a project like E-Government service for the benefit of the citizens.

Additionally, 2 participants representing 6.7% agreed that the provision of ICT ***infrastructural facilities*** would solve the barrier facing E-Service adoption and implementation. The local governments in conjunction with both the federal and state governments should partner with the private sectors in providing suitable and cheaper ICT facilities in the country, and this will aid the implementation of these initiatives. Citizens would embrace the use of E-Services platform if they were able to procure ICT facilities at a cheaper rate in the country. Another important solution to the identified barriers is the issue of ***funding and anti-corruption measures***. The good news for this study is that since the inception of the new government in Nigeria (since May 2015), the anti-corruption drives have been laudable and this has cut the attention of the foreign partners on the seriousness shown by the Muhammad Buhari's led federal government (ChannelsTV, 2016), huge stolen funds now being returned on a daily basis by the officials and this researcher hope that the retrieved money will be invested in major projects like E-Government services which can promote the image of the local governments through the availability of timely government information online. The results as shown in **Figure 86**, indicate that 6 participants which represent 20% and the second highest, explained that another solution is the ***provision of legislation/regulation*** on online content usage and this will protect both electronic document integrity, and proof-of-identity.

Citizens will have no fear in using and adopting the E-Government services as much as their details are secured whenever they are online. Another 2 participants representing 6.7% explained the need to ***improve on security, trust, and privacy*** for online activities. This is another solution which is a bit related to the one earlier mentioned, the provision of legislation/regulation. A secured online transaction will boost citizens and customer's confidence in using online services. The ***Constant power supply*** is another solution to the barriers facing E-Services adoption and implementation as agreed by the participants interviewed. 5 participants representing 16.7% of the total. The epileptic power supply in the country has grounded many businesses as business owners now depend on the private power supply which increased the running costs for their businesses thereby reducing the overall profits if not running the business at lost. The federal government in conjunction with the power holding companies and investors should invest hugely in electricity supply in Nigeria. Above all, 1 participant representing 3.3% of the total supported the opinion that there are certainly solutions to the barriers facing E-Service adoption and implementation at the local government level in Nigeria ***other*** than the ones previously mentioned.

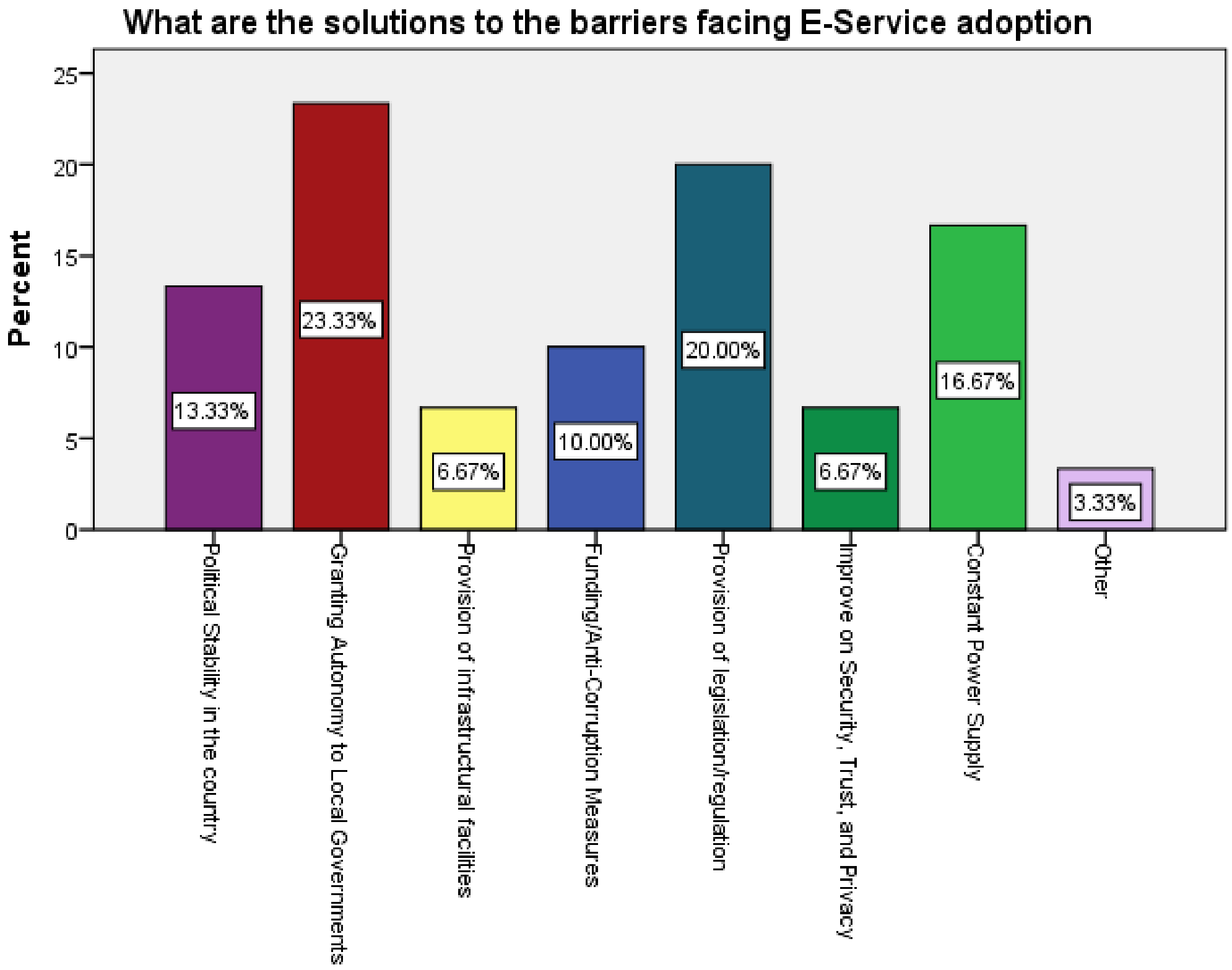

**Figure 86 SPSS Interview Analysis – Solutions to the Barriers Facing E-Service Adopting and Implementation**

**12. Opinion on Legislation and Regulation as a Barrier Facing E-Service Adoption**

The lack of legislation and regulation of E-Service initiatives and the internet usage in Nigeria was identified as a barrier facing the E-Service adopting and implementation. Participants were therefore asked this question to know their opinion on this. Results available in **Table 39** below shows that almost one-third of the participants (30%) agreed that there is a need for the current legislation/law amendment. 13.3% of the participants also agreed that judicial system in the country needs to be improved. Also, 3 participants (10%) indicated that sincerity is necessary from the E-Service stakeholders while another 10% of the participants claimed a stable economy would strengthen the legislation and regulation on E-Service projects. Furthermore, 4 participants (13.3%) said that implementation of the regulations would improve the E-Service adoption at the local government level in Nigeria while another 10% of the participants agreed that political would need to be improved. Moreover, another 10% of the participants said they have no idea while the last participant (3.3%) said another opinion but different from the ones above.

**Table 39 Interview Response - Opinion on Legislation and Regulation Regarding E-Service Adopting**

| | | Frequency | Percent | Valid Percent | Cumulative Percent |
|---|---|---|---|---|---|
| **Valid** | Amendment to existing legislation/law | 9 | 30.0 | 30.0 | 30.0 |
| | Improving judicial system | 4 | 13.3 | 13.3 | 43.3 |
| | Sincerity needed | 3 | 10.0 | 10.0 | 53.3 |
| | Stable economy | 3 | 10.0 | 10.0 | 63.3 |
| | Implementation of regulations | 4 | 13.3 | 13.3 | 76.7 |
| | Improving political will | 3 | 10.0 | 10.0 | 86.7 |
| | No idea | 3 | 10.0 | 10.0 | 96.7 |
| | Other | 1 | 3.3 | 3.3 | 100.0 |
| | **Total** | **30** | **100.0** | **100.0** | |

## 13. Government Readiness to Change Laws not Conforming with E-Service Adopting

The question was asked to find about from the participants who are also the stakeholders at the local government level in Nigeria on projects including the E-Service initiatives. The results available from **Table 40** below indicate that 76.7% of the total participants agreed to the government readiness to change the laws that are not conforming to the E-Service adoption and implementation. One participant (3.3%) said that the government is not ready to change the laws considering personal experience in service. Lastly, 6 participants which represent 20% mentioned that they were not sure about the government readiness to change the laws that are not conforming to the E-Service adopting and implementation in Nigeria.

**Table 40 Interview Response - Government Readiness to Change Laws not Conforming with E-Service Adopting**

| | | Frequency | Percent | Valid Percent | Cumulative Percent |
|---|---|---|---|---|---|
| **Valid** | Yes | 23 | 76.7 | 76.7 | 76.7 |
| | No | 1 | 3.3 | 3.3 | 80.0 |
| | Not sure | 6 | 20.0 | 20.0 | 100.0 |
| | **Total** | **30** | **100.0** | **100.0** | |

**14. Years of Work Experience**

In many job interviews, the years of experience is very important to make a successful application as it will indicate experience in previous work and training. Participants were asked about their years of experience as it will assist in getting more reliable information from old staff interviewed about the current operations within the local government, and their honest opinions on why the E-Government services are still not in place. They could comment on solutions to the barriers facing the e-government services adoption and implementation at the local government level. From the analysis available in **Figure 87** below, 6 participants (20%) with years of work experience between 0 and 5 years, 11 participants representing 36.7% out the total have years of work experience between 5 and 10 years. Another 6 participants which represent 20% have years of work experience between 10 and 15 years. 4 participants representing 13.3% with years of work experience between 15 and 20 years while the last 3 participants out the overall total have years of work experience of 20 years and above, this represent 10%.

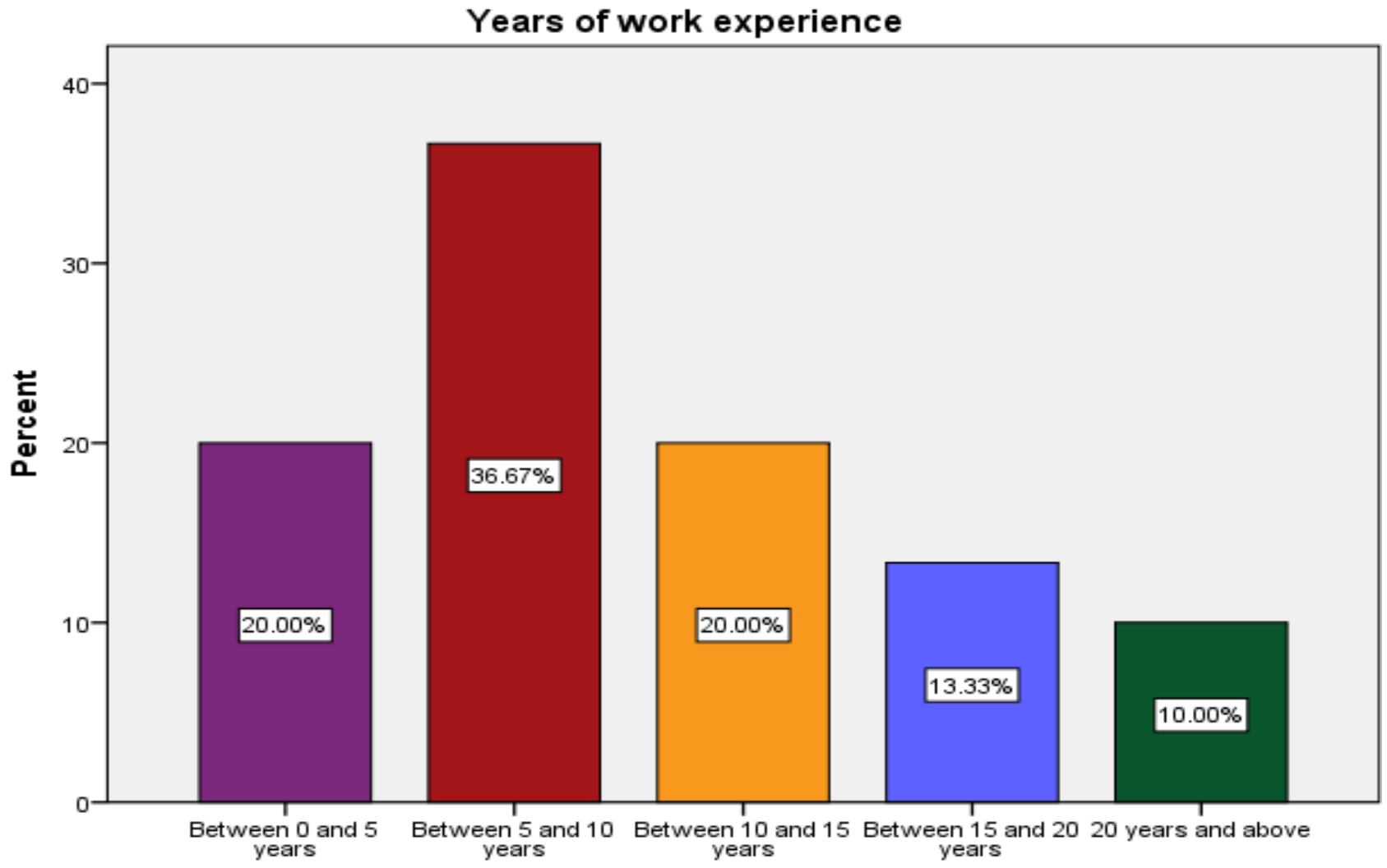

**Figure 87 SPSS Interview Analysis – Years of Work Experience**

**15. New Features to be Added to Initial Framework**

The participants were asked this question to find out if new features could be added to the initial framework given to them by this researcher to produce a final and workable E-Service framework. As shown in **Table 41** below, 5 participants (16.7%) agreed that there are new features to be added to the initial framework. Indeed, most participants questioned the use of Technology Acceptance Model (TAM) as theory choice for this research. Two-thirds of the participants (70%) indicated nothing should be added as the framework represent their notion of the E-Service framework. Lastly, 3 participants (10%) said they have no idea of features to be added.

**Table 41 Interview Response - Features to be added to Initial Framework**

| | | Frequency | Percent | Valid Percent | Cumulative Percent |
|---|---|---|---|---|---|
| **Valid** | Yes | 5 | 16.7 | 16.7 | 16.7 |
| | No | 21 | 70.0 | 70.0 | 86.7 |
| | Not sure | 1 | 3.3 | 3.3 | 90.0 |
| | No idea | 3 | 10.0 | 10.0 | 100.0 |
| | **Total** | **30** | **100.0** | **100.0** | |

**16. E-Service Framework Acceptability and Validity (Appropriate Model)**

The issues about reliability and validity for this study has been discussed in Chapter Four, this researcher deem it necessary to further ask the participants about their notion on E-Service framework proposed for this study. The general acceptability of this proposed E-Service framework by the participants will not only resolve any validity issue, but it will also serve as the method to validate the framework as required for this study. Furthermore, the results that emerged from this analysis will also help this researcher in answering the research question related the proposed framework. From the analysis shown in **Figure 88** below, there was a significant difference between the participants who agreed the framework is appropriate for this study and those participants who are not very sure about their belief. 29 participants out of the total 30 participants representing 96.7% agreed that the ***framework is appropriate*** and accepted one for this study. This is a positive result, and good development for this study as this answered the research question related to the proposed E-Service framework. There may be other possible explanation from the only participant who stated that he was ***not sure*** about the framework. Unfortunately, he was not able to explain to this researcher in detail. This represents 3.3% of the total.

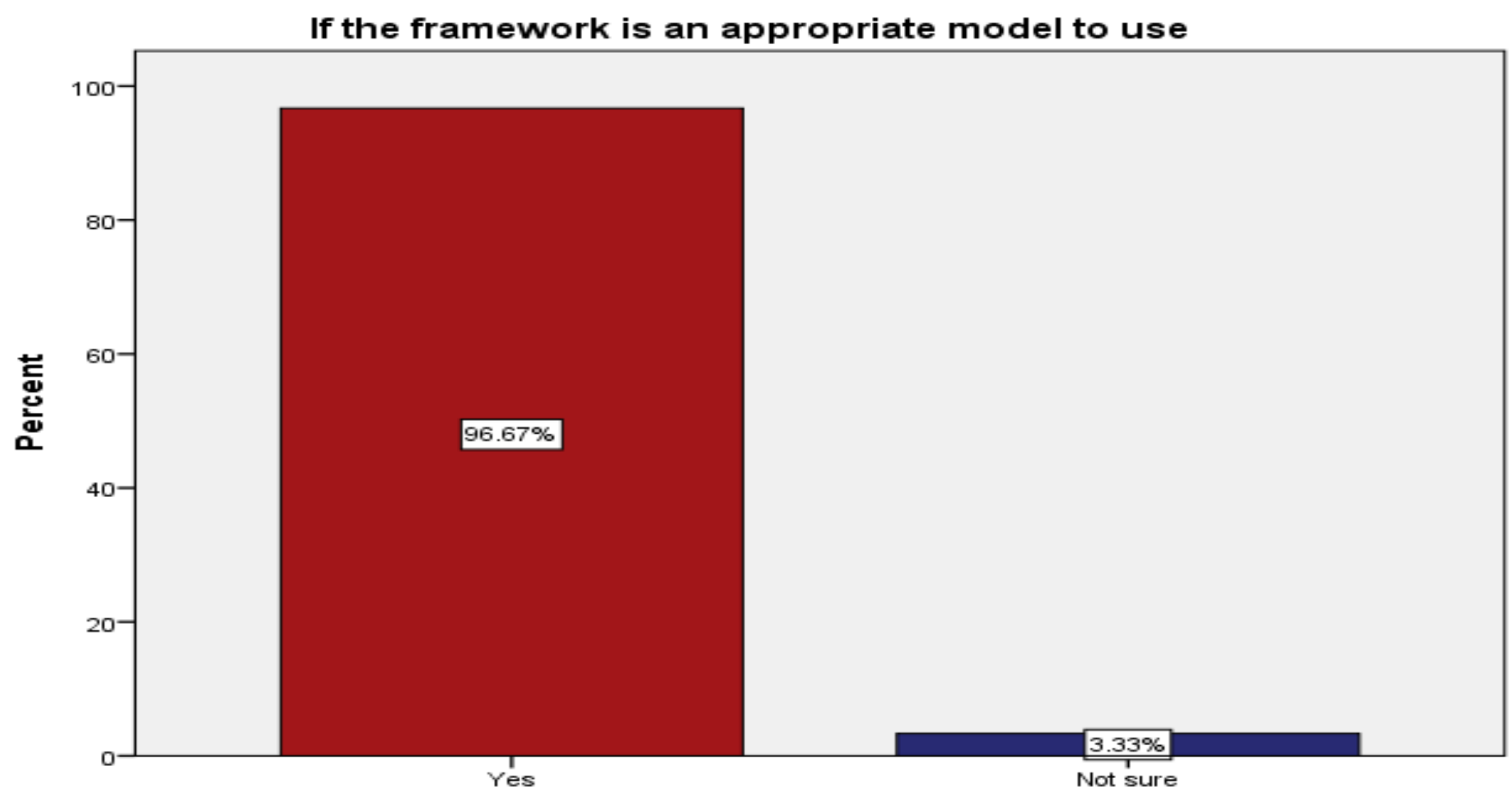

**Figure 88 SPSS Interview Analysis – E-Service Framework Validity**

## 17. Anything that Might Influence Your Decision to Use TAM for E-Service Adoption

This is another turning point for this research as almost all the participants rejected the use of Technology Acceptance Model (TAM) for the E-Service adoption after reviewing the framework. They claimed that the model is not explanatory enough on the role it will play in the E-Service adoption. Hence, as the tradition in action research methodology, the issue of the model have to go through another cycle as the researcher together with the participants needs to examine other possible information system theory that can work well in the proposed E-Service framework.

**Table 42 Interview Response - Is there anything that Might Influence Your Decision to Use TAM for E-Service Adopting**

| | | Frequency | Percent | Valid Percent | Cumulative Percent |
|---|---|---|---|---|---|
| **Valid** | Yes | 1 | 3.3 | 3.3 | 3.3 |
| | No | 23 | 76.7 | 76.7 | 80.0 |
| | Not sure | 1 | 3.3 | 3.3 | 83.3 |
| | No idea | 5 | 16.7 | 16.7 | 100.0 |
| | **Total** | **30** | **100.0** | **100.0** | |

In the light of the above, the result shown in **Table 42** above shows that a participant (3.3%) agreed to the use of Technology Acceptance Model (TAM) for the E-Service proposed framework. More than two-thirds of the participants (76.7%) indicated nothing could influence them and Technology Acceptance Model (TAM) was rejected. Also, another participant (3.3%) was not sure of using the model. Lastly, 5 participants (16.7%) agreed that they have no idea about the Technology Acceptance Model (TAM) for E-Service framework.

### 18. Recommending of Any Other Framework for E-Service Adopting

This question was asked if the participants have any other framework to recommend for E-Service development apart from the proposed E-Service framework by this researcher. The results available in **Table 43** below shows that almost all the participants (96.7%) indicated that they have no other framework to recommend as the one presented by this researcher represent their notion of E-Service framework for this study. The last participant (3.3%) said he was not sure about any other framework.

**Table 43 Interview Response - Would you like to Recommend Any Other Framework for E-Service Adopting**

| | | Frequency | Percent | Valid Percent | Cumulative Percent |
|---|---|---|---|---|---|
| **Valid** | No | 29 | 96.7 | 96.7 | 96.7 |
| | Not sure | 1 | 3.3 | 3.3 | 100.0 |
| | **Total** | **30** | **100.0** | **100.0** | |

### 19. Participants Contact Details for Future Research

This question was asked to find out from the participants if they are willing to give their contact details for the future research. This was important to move this research forward. The focus group participants were approached from the participants who voluntary agreed to leave their contact details.

**Table 44 - Interview Response - Would you like to be Contacted Again**

| | | Frequency | Percent | Valid Percent | Cumulative Percent |
|---|---|---|---|---|---|
| **Valid** | Yes | 25 | 83.3 | 83.3 | 83.3 |
| | No | 2 | 6.7 | 6.7 | 90.0 |
| | Not sure | 1 | 3.3 | 3.3 | 93.3 |
| | Don't want to be contacted | 2 | 6.7 | 6.7 | 100.0 |
| | **Total** | **30** | **100.0** | **100.0** | |

The majority of the participants (83.3%) as shown in **Table 44** above agreed to share their contact details with this researcher as they pointed they have learnt from the interview sessions in E-Government domain. 2 participants (6.7%) said they cannot share their contacts for personal reasons. One participant (3.3%) not sure about sharing her contact while another 2 participants (6.7%) don't want to be contacted at all.

### 20. Any Other Suggestion or Plan for E-Service Adopting

Participants were also asked about suggestion or plan for the E-Service adopting and implementation at local environment levels in Nigeria if any. From the results shown in **Table 45** below, 30% of the total participants suggested that there is an urgent need for the government to establish dedicated ICT centres, 10% agreed that transparency is required for E-service projects, 13.3% of the participants advocated for constant power supply. Also, another 13.3% agreed that awareness should be improved and another 10% of the total participants said a user-friendly E-Service platform is needed. Meanwhile, 4 participants (13.3%) admitted they have no suggestion while 6.7% of the participants agreed they have no idea. Lastly, one participant indicated another suggestion for the E-Service at the local environment level in Nigeria which is different from other suggestions above.

**Table 45 Interview Response - Any Other Suggestion or Plan for E-Service Adopting**

| | | Frequency | Percent | Valid Percent | Cumulative Percent |
|---|---|---|---|---|---|
| **Valid** | Establishment of dedicated ICT centres | 9 | 30.0 | 30.0 | 30.0 |
| | Transparency required for E-service projects | 3 | 10.0 | 10.0 | 40.0 |
| | Constant power supply | 4 | 13.3 | 13.3 | 53.3 |
| | Awareness | 4 | 13.3 | 13.3 | 66.7 |
| | User-friendly platform needed | 3 | 10.0 | 10.0 | 76.7 |
| | No | 4 | 13.3 | 13.3 | 90.0 |
| | No idea | 2 | 6.7 | 6.7 | 96.7 |
| | Other | 1 | 3.3 | 3.3 | 100.0 |
| | **Total** | **30** | **100.0** | **100.0** | |

## B. Participants Responses from the Focus Group (Online)

As explained in Chapter Five, **section 5.4.3**, which includes the use of a focus group for this research and the selection of the focus group participants, this researcher agreed to use the online focus group with the participants in order to have more in-depth discussion where constructive arguments between the participants and the researcher could further be resolved. The online focus group sessions would also be used to validate the framework, further exchanging ideas including the research feedback among other things. An online focus group reduces travelling costs, but more importantly, it made easy, anonymous contact between the participants, if possible (Bryman, 2016; Oseni et al., 2017). **Table 46** below shows the participants responses from the online focus group discussions.

**Table 46 Nvivo Analysis: Participants Responses from the Focus Group (Online)**

| | Gender | E-Government Usage | Will you encourage and use E-Government Services | Do you think the E-Service technology adoption will help in other aspects or areas | Will you encourage the use of Diffusion of Innovation theory in E-Government services implementation | Does the E-Service framework sent to you represent your notion of E-Service | Will you be ready to receive and implement the recommendations on E-Service from this study |
|---|---|---|---|---|---|---|---|
| Participant 1 | Male | No | Yes | Yes | Yes | Yes | Yes |
| Participant 2 | Female | No | Yes | Yes | Yes | Yes | Yes |
| Participant 3 | Male | Yes | Yes | Yes | Yes | Yes | Yes |
| Participant 4 | Male | Yes | Yes | Yes | Yes | Yes | Yes |
| Participant 5 | Male | No | Yes | Yes | Yes | Yes | Yes |

### 1. Gender

The results, as shown in **Table 46** above, indicate that 4 males (80%) and 1 female (20%) participated in the online focus group discussion. Surprisingly, the women's participation here is lower compared to the results obtained from both the interview and the online survey. However, the selection of the participants for the focus group was based on the participants who volunteered their contact details during the one-on-one interviews stage of research as discussed in **section 5.4.3** (Chapter Five) and this produced more males than females.

### 2. E-Government Services Usage

From the results obtained and as shown in **Table 46** above, the participants were asked this question to know their level of E-Government services usage. Perhaps, if the barriers or low rate of adoption of the E-Government services at local government level in Nigeria because, the participants have already used E-Services at other government levels (that is, the state and the federal levels). The most surprising result here is that 3 participants out of the total have not used the E-Government services before while the other 2 participants have used it before. A positive correlation was found between this result and other results obtained from both the interviews and online survey where the majority of the participants and respondents have not used any E-Government service before now. This clearly shows the importance of this research in investigating the barriers facing the E-Service adoption and investigation at the local government level in Nigeria.

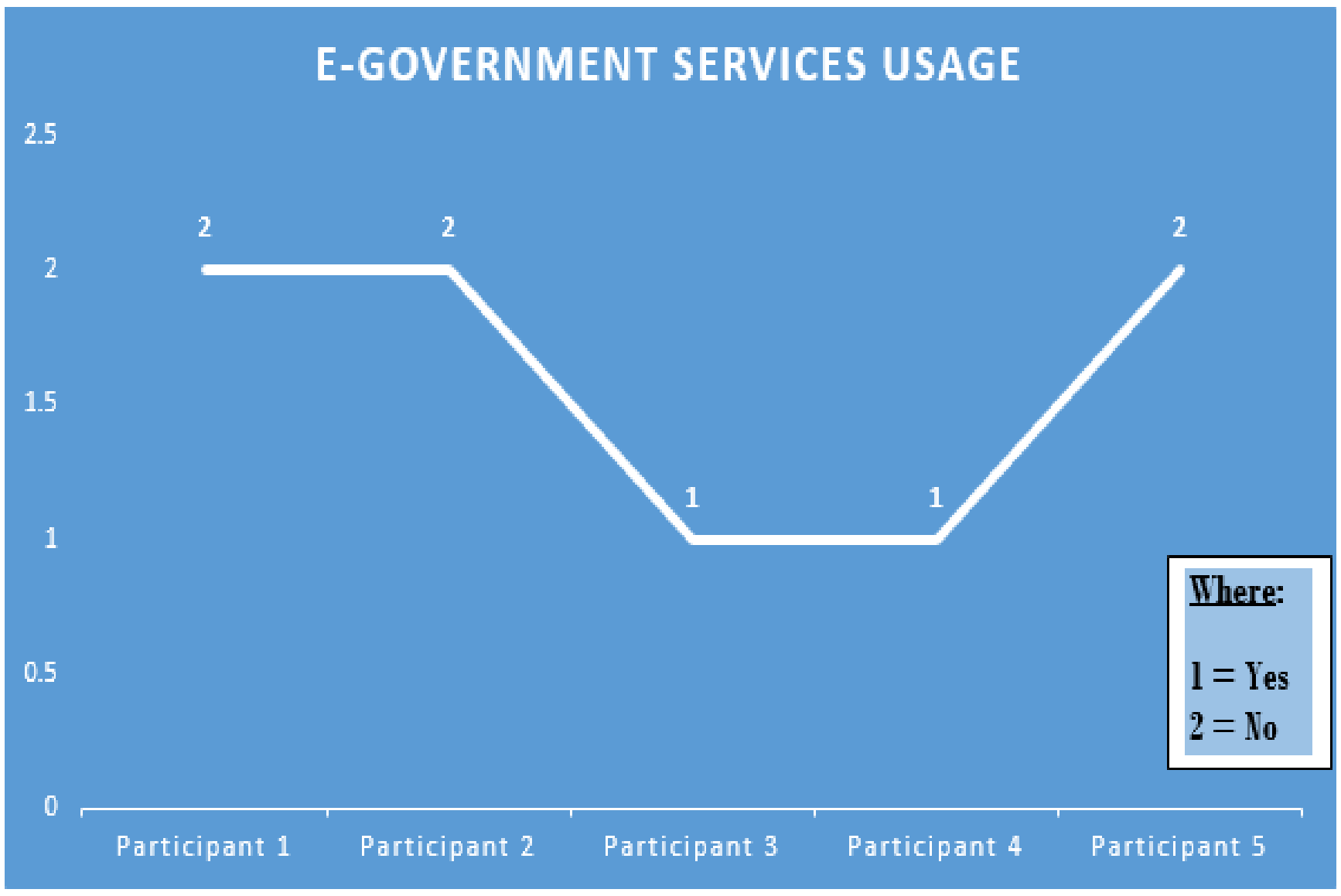

**Figure 89 2-D Line with Marker E-Government Service Usage Chat**

### 3. Encouragement and the Use of E-Government Services

The participants were further asked if they will encourage and use the E-Government services at the local government level in Nigeria. The most interesting aspect of the results, as shown in **Figure 90** below, is that the whole 5 participants representing 100% agreed they would encourage E-Service adoption and also use E-Services if implemented at the local government level in Nigeria. These results provide more important perceptions on the citizen's seal to have E-Government services at the local government level in Nigeria.

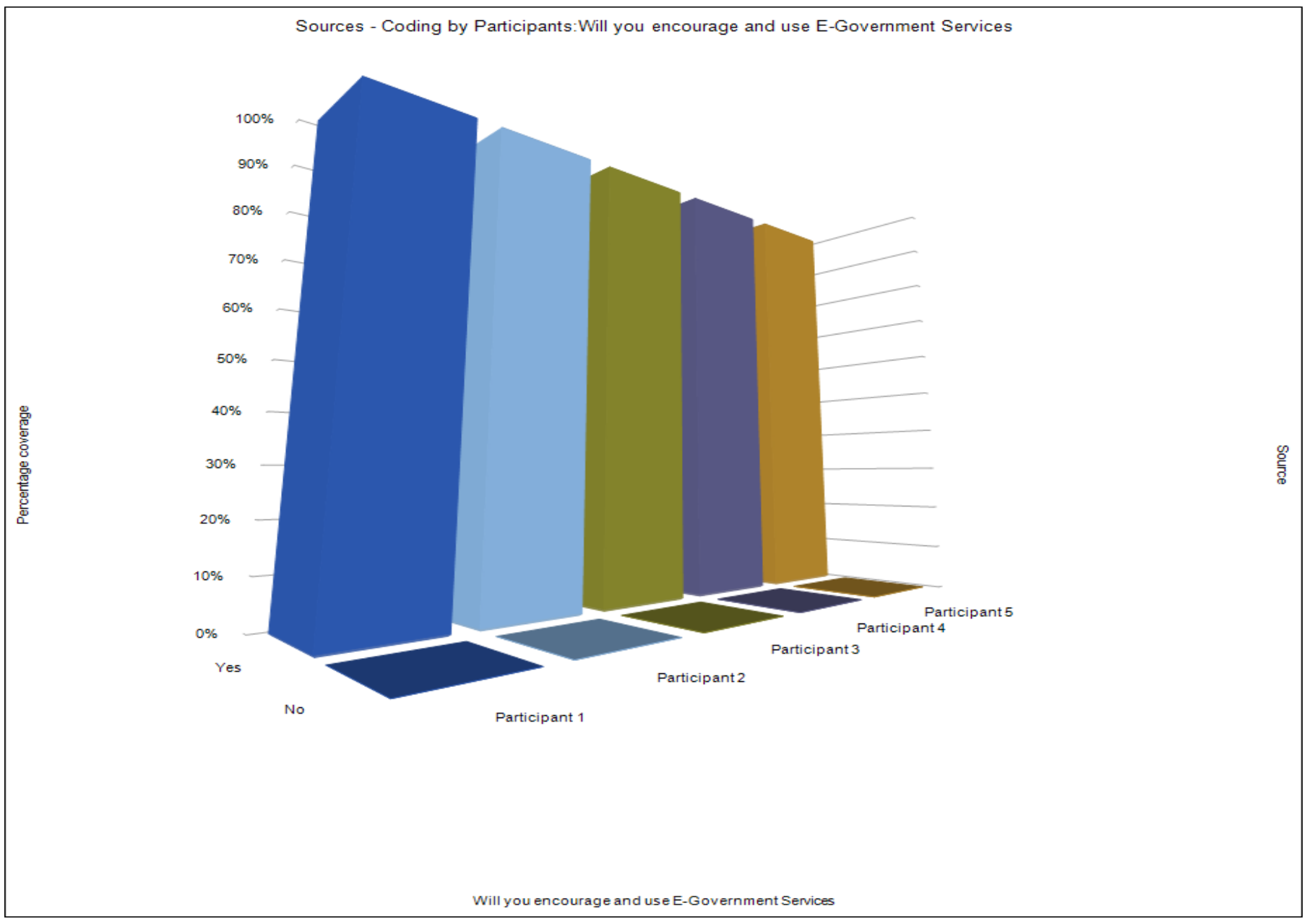

**Figure 90 Nvivo Online Focus Group Analysis - Encouragement and the Use of E-Government Services**

### 4. E-Services Technology Adopting Helpful in Other Aspect or Areas

Participants were also asked if they think E-Service technology adopting will help in other aspects or areas, all the participants (100%) agreed that E-Service technology adoption would be useful in other areas as shown in **Figure 91** below. For this researcher, E-Service technology adoption will help in economic growth as it will boost the e-commerce in the country. For example, the Asia markets have robust e-commerce in the world today where millions are traded daily, and this has helped the Asian economies to be some of the biggest in the world today (IMF, 2016). The achievement from the Asia region however not surprising considering the information available on the world internet users.

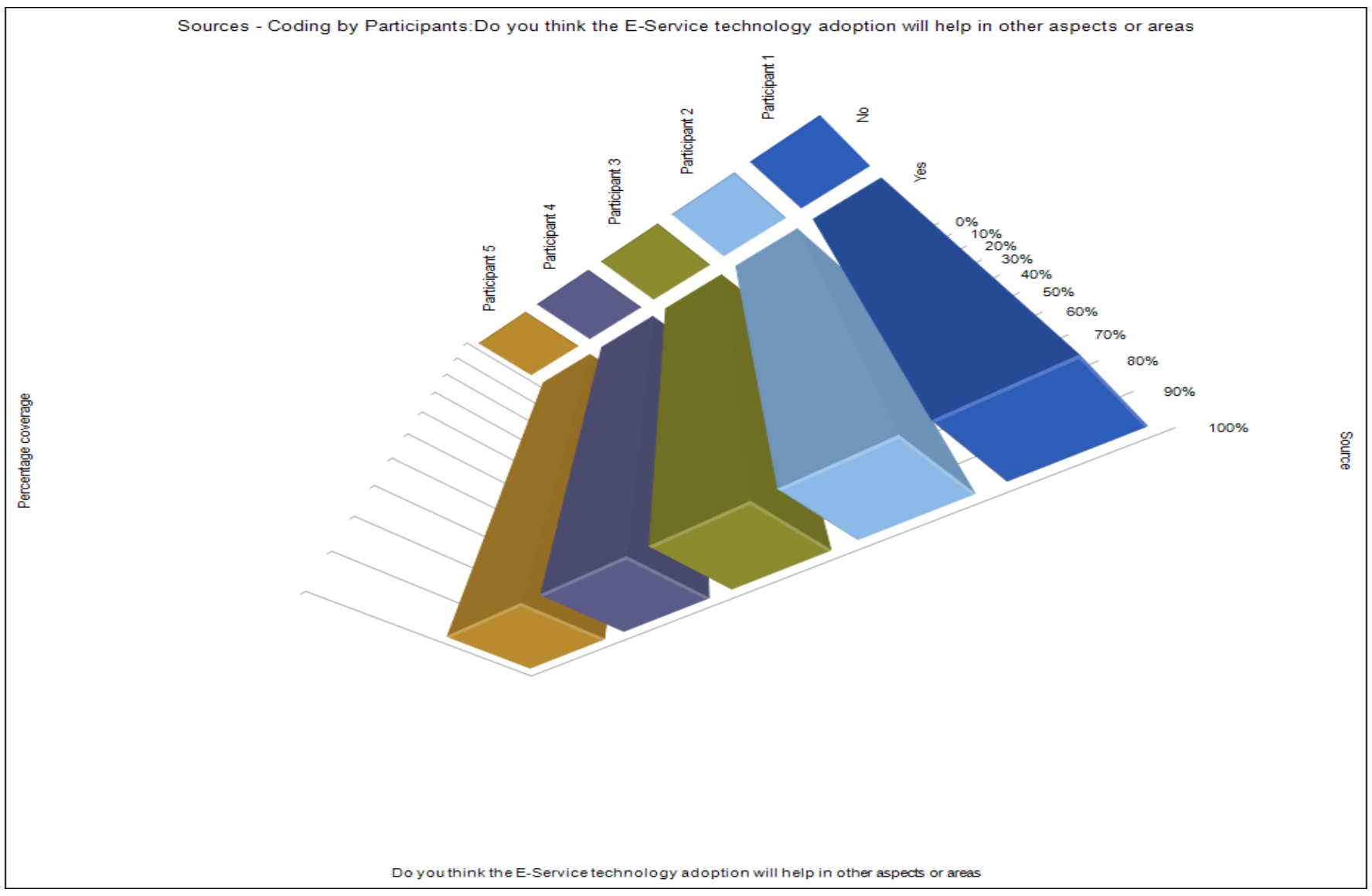

**Figure 91 Nvivo Online Focus Group Analysis: E-Services Technology Adopting Helpful in Other Aspect or Areas**

## 5. The Use of Diffusion of Innovation (DOI) Theory in E-Government Services Implementation

The use of theory to this researcher is a very important tool in ideas generation as it offers the opportunity to integrate the ideas to the existing research knowledge. Bendassolli (2013) pointed out that there is consistency in theories compare to common sense because they are constructed in a systematic way to help with the collection of data. The use of theory will help this researcher to make clear and consistent predictions during the implementation and validation process. Having introduced the participants to the various theories, such as Diffusion of Innovation (DOI) Theory used in information science research during the interview phase, the participants were further asked during the online focus group session if they will encourage the use of Diffusion of Innovation (DOI) Theory in E-Government services implementation. The result here as shown in **Figure 92 below** is inspiring as all the participants (100%) agreed they would encourage the use of Diffusion of Innovation (DOI) Theory in E-Government services implementation.

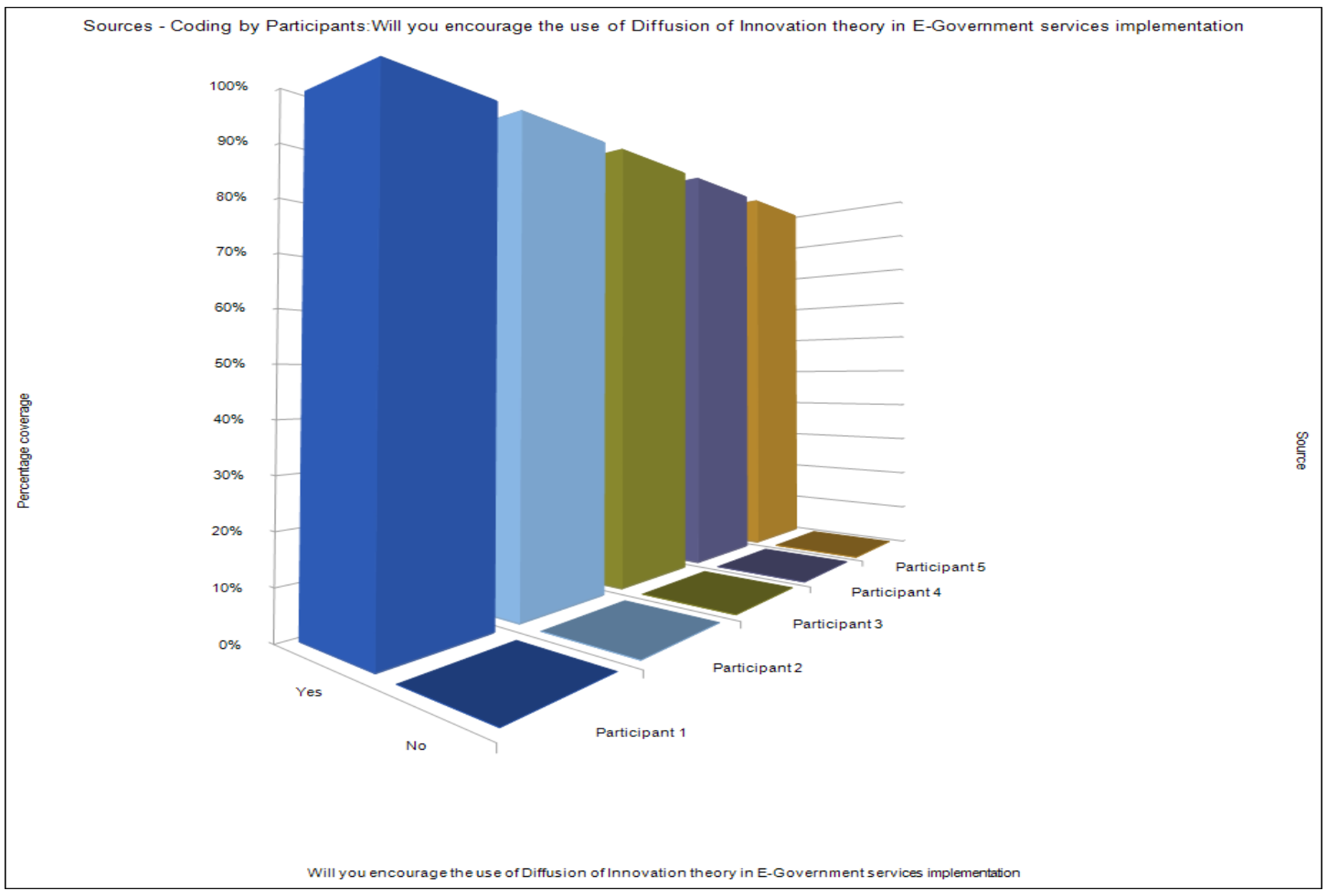

**Figure 92 Nvivo Online Focus Group Analysis - The Use of Diffusion of Innovation (DOI) Theory in E-Government Services Implementation**

**6. Does the E-Service Framework Sent to You Represent Your Notion of E-Service**

The introduction of the various information science theories to the participants during the interview stage opened up many constructive debates on the E-Service framework. After the earlier rejection of the Technology Acceptance Model (TAM) by the participants as the theory for this research, there was a need for this researcher to make changes to the initial E-Service framework following the agreement with the participants to adopt the use of the Diffusion of Innovation (DOI) Theory for this research. Hence, the participants were further asked during the online focus group session if the E-Service framework sent to them represent their notion of E-Service. The overall response to this question was positive as shown in **Figure 93** below, all the participants (100%) agreed they the final and refined E-Service framework sent to them represent their notion of E-Service at the local government level in Nigeria. The result obtained here is very significant to this research especially the fact that the participants here are also part of the stakeholders at the local government level in Nigeria and the E-Service framework produced from this study will be useful in the implementation of E-Government services.

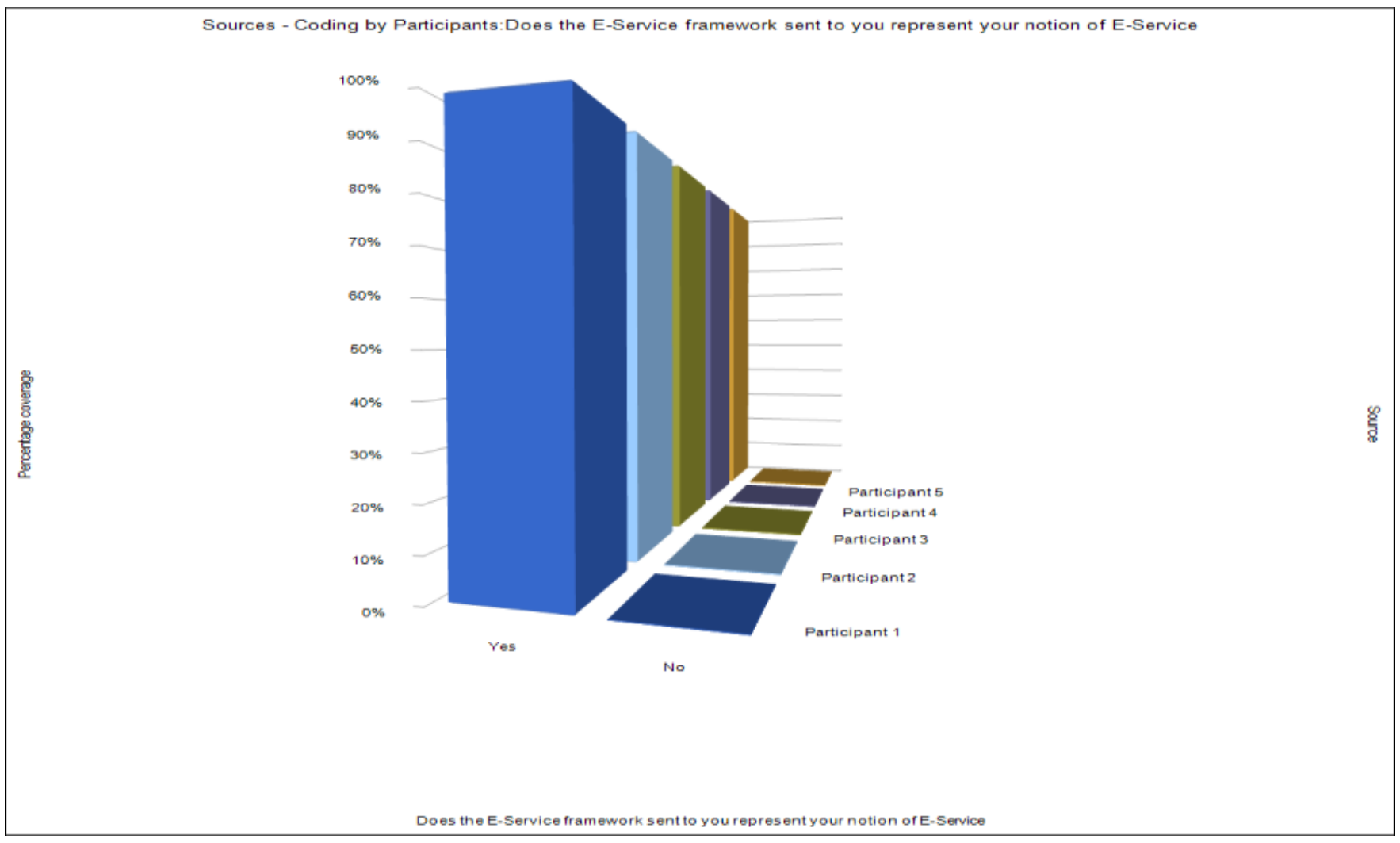

**Figure 93 Nvivo Online Focus Group Analysis – Notion on E-Service framework**

## 7. Will You be Ready To Receive and Implement the Recommendations on E-Service from This Study

For the researcher, asking a question of this nature will not only help in answering the research questions but also adding credibility to the research that has it recommendations being implemented by the stakeholders. Hence, the participants who are also the stakeholders in E-Service implementation at the local government level in Nigeria during the online focus group session were asked if they will be ready to receive and implement the recommendations on E-Service from this study. The responses were anticipated as the whole 5 (100%) participants agreed they would be ready to receive and implement the recommendations from this study on E-Service at the local government level in Nigeria. They are senior officers as indicated in **Table 32** (see, Chapter 8).

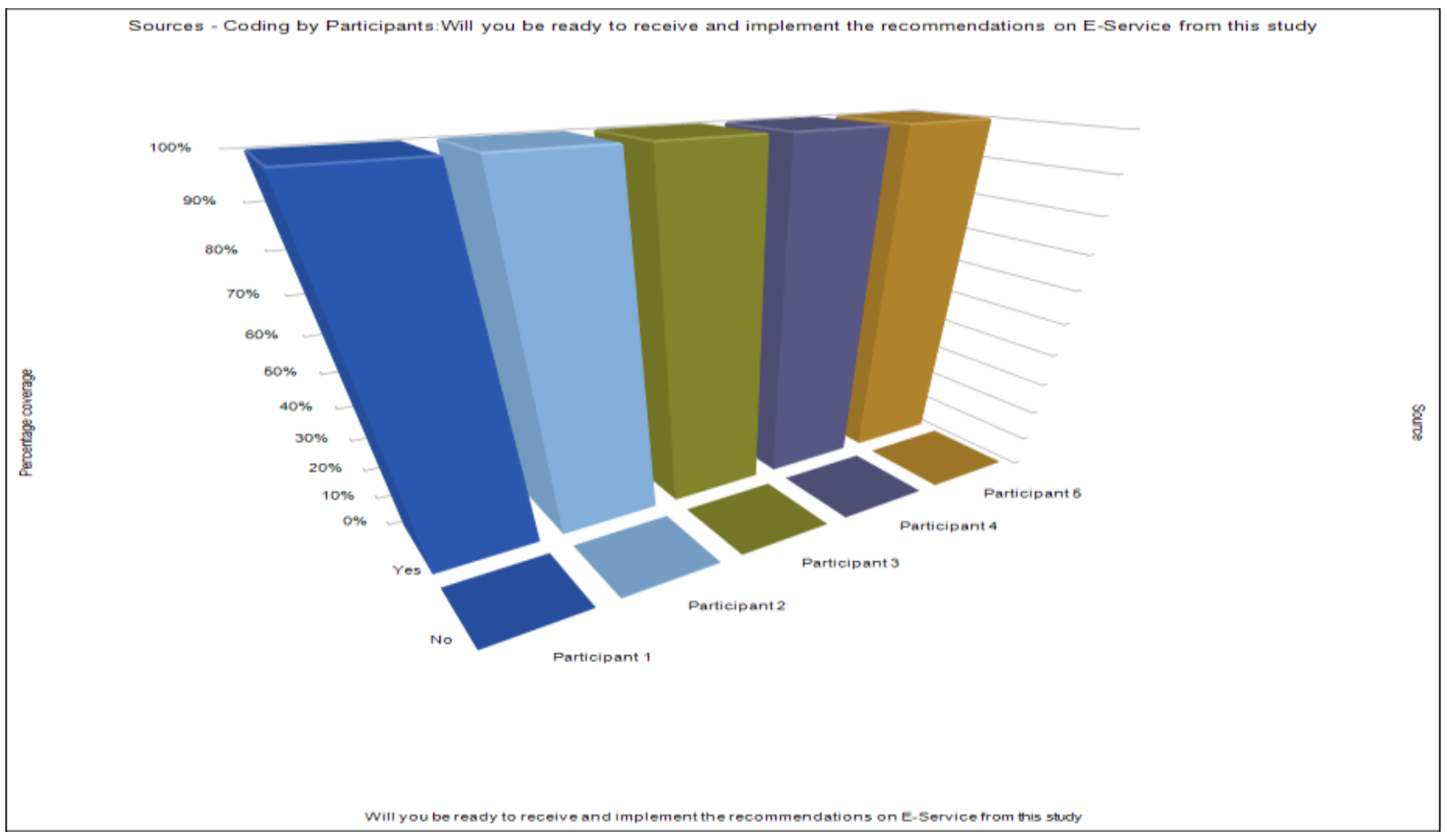

**Figure 94 Nvivo Online Focus Group Analysis – Receiving and Implementing E-Service Recommendations from this Study**

## C. Other Respondents Responses from the Online Survey

This section will explain other responses from the online survey in addition to the responses discussed in **Section 6.4.2**.

### 1. Respondents State of Origin

There are 36 states in Nigeria with the total of 774 local government areas (FRN, 2016). This question was asked to find out the respondents state of origin and to know how spread the E-Service initiatives awareness are all over the states in Nigeria. Respondents from 27 states took part in the online survey which is more than two-thirds of the whole states in Nigeria. Interestingly, the results show that almost half of the whole respondents (43.5%) are from Lagos state. This result is however not surprising as Lagos being the former capital state of Nigeria have many business representatives from the rest of the world. This has indeed improved the infrastructural amenities including ICT facilities and more residents now have access to the internet facilities compared to the other states in Nigeria. However, the surprising and disappointing result as shown in **Table 47** below shows 0.5% of the respondents from Akwa-Ibom state took part in the online survey. Akwa-Ibom is another business and economic state in Nigeria, the low participation recorded from the state might be connected to other unknown reasons.

**Table 47 Online Survey – Respondents State of Origin**

| | | Frequency | Percent | Valid Percent | Cumulative Percent |
|---|---|---|---|---|---|
| Valid | Abia | 5 | 1.2 | 1.2 | 1.2 |
| | Akwa-Ibom | 2 | .5 | .5 | 1.6 |
| | Anambra | 4 | .9 | .9 | 2.6 |
| | Bauchi | 4 | .9 | .9 | 3.5 |
| | Bayelsa | 8 | 1.9 | 1.9 | 5.4 |
| | Benue | 7 | 1.6 | 1.6 | 7.1 |
| | Bornu | 11 | 2.6 | 2.6 | 9.6 |
| | Crossriver | 3 | .7 | .7 | 10.4 |
| | Delta | 7 | 1.6 | 1.6 | 12.0 |
| | Ebonyi | 3 | .7 | .7 | 12.7 |
| | Edo | 12 | 2.8 | 2.8 | 15.5 |
| | Ekiti | 19 | 4.5 | 4.5 | 20.0 |
| | Enugu | 9 | 2.1 | 2.1 | 22.1 |
| | Imo | 15 | 3.5 | 3.5 | 25.6 |
| | Kaduna | 4 | .9 | .9 | 26.6 |
| | Kano | 3 | .7 | .7 | 27.3 |
| | Kebbi | 6 | 1.4 | 1.4 | 28.7 |
| | Kogi | 8 | 1.9 | 1.9 | 30.6 |
| | Kwara | 6 | 1.4 | 1.4 | 32.0 |
| | Lagos | 185 | 43.5 | 43.5 | 75.5 |
| | Ogun | 36 | 8.5 | 8.5 | 84.0 |
| | Ondo | 20 | 4.7 | 4.7 | 88.7 |
| | Osun | 17 | 4.0 | 4.0 | 92.7 |
| | Oyo | 14 | 3.3 | 3.3 | 96.0 |
| | Rivers | 8 | 1.9 | 1.9 | 97.9 |
| | Sokoto | 5 | 1.2 | 1.2 | 99.1 |
| | Zamfara | 4 | .9 | .9 | 100.0 |
| | **Total** | **425** | **100.0** | **100.0** | |

## 2. Respondents Occupation

This question was asked to know the respondent's occupation as the choice of occupation might have an influence on E-Service adoption. It is possible that an accountant might have more interest in using the E-Services than a farmer, though, no research proved this. From the results available in **Table 48** below, almost half of the whole respondents (46.1%) which is the highest are the civil servant, and this is expected as the officials at the local governments are all civil servants, though, they might have areas of specialization depending on the diploma or degree earned. The lowest respondents that took part in the survey with 0.2% each are an architect, an aviation consultant, educational consultant, housing officer, insurance broker, librarian, Offshore Field Engineer, Security Expert, and Surveyor respectively.

**Table 48 Online Survey – Participants Occupation**

| | | Frequency | Percent | Valid Percent | Cumulative Percent |
|---|---|---|---|---|---|
| **Valid** | Accountant | 20 | 4.7 | 4.7 | 4.7 |
| | Administrator | 4 | .9 | .9 | 5.6 |
| | Architect | 1 | .2 | .2 | 5.9 |
| | Auditor | 2 | .5 | .5 | 6.4 |
| | Aviation Consultant | 1 | .2 | .2 | 6.6 |
| | Banker | 3 | .7 | .7 | 7.3 |
| | Business Man | 12 | 2.8 | 2.8 | 10.1 |
| | Business Woman | 11 | 2.6 | 2.6 | 12.7 |
| | Civil Engineer | 3 | .7 | .7 | 13.4 |
| | Civil Servant | 196 | 46.1 | 46.1 | 59.5 |
| | Contractor | 2 | .5 | .5 | 60.0 |
| | Customer Care Officer | 4 | .9 | .9 | 60.9 |
| | Educational Consultant | 1 | .2 | .2 | 61.2 |
| | Farmer | 2 | .5 | .5 | 61.6 |
| | Health Care Worker | 3 | .7 | .7 | 62.4 |
| | Housing Officer | 1 | .2 | .2 | 62.6 |
| | Human Resources Expert | 2 | .5 | .5 | 63.1 |
| | Insurance Broker | 1 | .2 | .2 | 63.3 |
| | IT Professional | 22 | 5.2 | 5.2 | 68.5 |
| | Lawyer | 13 | 3.1 | 3.1 | 71.5 |
| | Lecturer/Teacher | 13 | 3.1 | 3.1 | 74.6 |
| | Librarian | 1 | .2 | .2 | 74.8 |
| | Marketing Consultant | 4 | .9 | .9 | 75.8 |
| | Mechanical Engineer | 5 | 1.2 | 1.2 | 76.9 |
| | Medical Doctor | 2 | .5 | .5 | 77.4 |
| | Nurse | 7 | 1.6 | 1.6 | 79.1 |
| | Offshore Field Engineer | 1 | .2 | .2 | 79.3 |
| | Pension Administrator | 2 | .5 | .5 | 79.8 |
| | Politician | 1 | .2 | .2 | 80.0 |
| | Retired - Civil Servant | 13 | 3.1 | 3.1 | 83.1 |
| | Security Expert | 1 | .2 | .2 | 83.3 |
| | Self Employed | 5 | 1.2 | 1.2 | 84.5 |
| | Student | 11 | 2.6 | 2.6 | 87.1 |
| | Surveyor | 1 | .2 | .2 | 87.3 |
| | Trader | 54 | 12.7 | 12.7 | 100.0 |
| | **Total** | **425** | **100.0** | **100.0** | |

### 3. Adoption and the Use of E-Government Services

The results here not different from other results obtained from the interviews and the focus group sessions where nearly all the participants and the respondents agreed to the use and adoption of E-Services. Indeed, there is a correlation between these results. As shown in **Table 49** below, more than two-thirds of the respondents (89.4%) agreed to use and adopt E-Service while 2.4% indicated they would not adopt and use E-Service. A small number of the respondents (2.4%) said they have no interest in the E-Services at the Local Government levels in Nigeria.

**Table 49 Online Survey – Will you adopt and use E-Government Services**

| | | Frequency | Percent | Valid Percent | Cumulative Percent |
|---|---|---|---|---|---|
| **Valid** | Yes | 380 | 89.4 | 89.4 | 89.4 |
| | No | 35 | 8.2 | 8.2 | 97.6 |
| | Not interested in E-Services at Local Government level | 10 | 2.4 | 2.4 | 100.0 |
| | **Total** | **425** | **100.0** | **100.0** | |

### 4. Deployment of E-Service will be Reducing Corruption

The results as shown in **Table 50** below are in agreement with those obtained from both the interviews and the online focus group sessions. It seems to be consistent as more than two-thirds of the respondents (85.4%) strongly agreed that the deployment of E-Service would be reducing corruption. 11.8% respondents agreed, 2.1% respondents are neutral on the choice of the answer while 3 respondents (0.7%) indicated that they disagreed the fact that deploying E-Service will reduce corruption at the local government level in Nigeria.

**Table 50 Online Survey – Deployment of E-Service will be Reducing Corruption**

| | | Frequency | Percent | Valid Percent | Cumulative Percent |
|---|---|---|---|---|---|
| **Valid** | Disagree | 3 | .7 | .7 | .7 |
| | Neutral | 9 | 2.1 | 2.1 | 2.8 |
| | Agree | 50 | 11.8 | 11.8 | 14.6 |
| | Strongly Agree | 363 | 85.4 | 85.4 | 100.0 |
| | **Total** | **425** | **100.0** | **100.0** | |

### 5. Reason(s) for not Adopting E-Government Services

As shown in **Table 49** above, 2.4% respondents indicated they would not adopt and use E-Service while a small number of the respondents (2.4%) said they have no interest in the E-Services at the Local Government levels in Nigeria. Hence, this researcher asked the respondents as a follow-up question to know their reasons for not interested in E-Service adoption. The results in **Table 51** below shows that 11.8% of the total respondents agreed not to adopt E-Service due to corruption issue, 7.3% high level of illiteracy, 14.1% identity theft while almost one-third of the respondents (30.6%) said they are not adopting

E-Services because of internet security issues. Also, one respondent (0.2%) have no idea, 1.4% no user-friendly policy on E-Service, 25.4% of the total respondents indicated they would not adopt E-Service because of the internet trust issue. Lastly, 9.2% of the respondents agreed not to adopt E-Service due to unskilled IT personnel.

**Table 51 Online Survey - Reason(s) for not Adopting E-Government Services**

| | | Frequency | Percent | Valid Percent | Cumulative Percent |
|---|---|---|---|---|---|
| **Valid** | Corruption | 50 | 11.8 | 11.8 | 11.8 |
| | High Level of Illiteracy | 31 | 7.3 | 7.3 | 19.1 |
| | Identity Theft | 60 | 14.1 | 14.1 | 33.2 |
| | Internet Security Issues | 130 | 30.6 | 30.6 | 63.8 |
| | No Idea | 1 | .2 | .2 | 64.0 |
| | No User-Friendly Policy | 6 | 1.4 | 1.4 | 65.4 |
| | Trust Issues | 108 | 25.4 | 25.4 | 90.8 |
| | Unskilled IT Personnel | 39 | 9.2 | 9.2 | 100.0 |
| | **Total** | **425** | **100.0** | **100.0** | |

### 6. Would You like to be Contacted Regarding Your Answers

This question was also asked if the respondents are willing to give their contact details for future communication regarding their answers. From the results available in **Table 52** below, more than two-thirds of the respondents (86.8%) agreed while a small number of the respondents (13.2%) declined their contact details.

**Table 52 Online Survey - Would you like to be Contacted Regarding Your Answers**

| | | Frequency | Percent | Valid Percent | Cumulative Percent |
|---|---|---|---|---|---|
| **Valid** | Yes | 369 | 86.8 | 86.8 | 86.8 |
| | No | 56 | 13.2 | 13.2 | 100.0 |
| | **Total** | **425** | **100.0** | **100.0** | |

| ACTIVITIES | Year 1 | Year 2 | | | | | | | | | | | | Year 3 |
|---|---|---|---|---|---|---|---|---|---|---|---|---|---|---|
| | | M1 | M2 | M3 | M4 | M5 | M6 | M7 | M8 | M9 | M10 | M11 | M12 | |
| **1. Literature Review** | | | | | | | | | | | | | | |
| 1.1. E-Government/E-Service | | | | | | | | | | | | | | |
| 1.2. Previous framework/Model of E-Service | | | | | | | | | | | | | | |
| 1.3. Barriers/Successful factors of E-Service | | | | | | | | | | | | | | |
| 1.4. Diffusion of Innovation Theory (DOI) | | | | | | | | | | | | | | |
| 1.5. Writing Journals/Papers, attending seminars/conferences. | | | | | | | | | | | | | | |
| **2. Development of the Initial E-Service Framework** | | | | | | | | | | | | | | |
| 2.1 More work on Literature review, writing Journals/Papers, attending seminars. | | | | | | | | | | | | | | |
| **3. Action Research Study – Data Collection** | | | | | | | | | | | | | | |
| 3.1. Case Study - Lagos | | | | | | | | | | | | | | |
| a. In-depth Interview – Stakeholders, local government staff, public. | | | | | | | | | | | | | | |
| b. Survey – Questionnaires, participant observation, recording. | | | | | | | | | | | | | | |
| c. Initial framework – Discussion, refining with Stakeholders. | | | | | | | | | | | | | | |
| **4. Data Processing** | | | | | | | | | | | | | | |
| 4.1 Data Analysis and Processing. | | | | | | | | | | | | | | |
| 4.2 Writing Journals/Papers, attending seminars/conferences. | | | | | | | | | | | | | | |
| **5. Analysis and PhD Thesis Writing** | | | | | | | | | | | | | | |
| 5.1 Final Refined Framework of E-Service/ Writing of thesis/Submission. | | | | | | | | | | | | | | |

**Y**-Year, **M**-Month

**Figure 95 Timetable for the Research Study**

| Required Sample Size | | | | | | | | |
|---|---|---|---|---|---|---|---|---|
| | Confidence = 95% | | | | Confidence = 99% | | | |
| Population Size | Margin of error | | | | Margin of Error | | | |
| | 5.0% | 3.5% | 2.5% | 1.0% | 5.0% | 3.5% | 2.5% | 1.0% |
| 10 | 10 | 10 | 10 | 10 | 10 | 10 | 10 | 10 |
| 20 | 19 | 20 | 20 | 20 | 19 | 20 | 20 | 20 |
| 30 | 28 | 29 | 29 | 30 | 29 | 29 | 30 | 30 |
| 50 | 44 | 47 | 48 | 50 | 47 | 48 | 49 | 50 |
| 75 | 63 | 69 | 72 | 74 | 67 | 71 | 73 | 75 |
| 100 | 80 | 89 | 94 | 99 | 87 | 93 | 96 | 99 |
| 150 | 108 | 126 | 137 | 148 | 122 | 135 | 142 | 149 |
| 200 | 132 | 160 | 177 | 196 | 154 | 174 | 186 | 198 |
| 250 | 152 | 190 | 215 | 244 | 182 | 211 | 229 | 246 |
| 300 | 169 | 217 | 251 | 291 | 207 | 246 | 270 | 295 |
| 400 | 146 | 265 | 318 | 384 | 250 | 309 | 348 | 391 |
| 500 | 217 | 306 | 377 | 475 | 285 | 365 | 421 | 485 |
| 600 | 234 | 340 | 432 | 565 | 315 | 416 | 490 | 579 |
| 700 | 248 | 370 | 481 | 653 | 341 | 462 | 554 | 672 |
| 800 | 260 | 396 | 526 | 739 | 363 | 503 | 615 | 763 |
| 1,000 | 278 | 440 | 606 | 906 | 399 | 575 | 727 | 943 |
| 1,200 | 291 | 474 | 674 | 1,067 | 427 | 636 | 827 | 1,119 |
| 1,500 | 306 | 515 | 759 | 1,297 | 460 | 712 | 959 | 1,376 |
| 2,000 | 322 | 563 | 869 | 1,655 | 498 | 808 | 1,141 | 1,785 |
| 2,500 | 333 | 597 | 952 | 1,984 | 524 | 879 | 1,288 | 2,173 |
| 3,500 | 346 | 641 | 1,068 | 2,565 | 558 | 977 | 1,510 | 2,890 |
| 5,000 | 357 | 678 | 1,176 | 3,288 | 586 | 1,066 | 1,734 | 3,842 |
| 7,500 | 365 | 710 | 1,275 | 4,211 | 610 | 1,147 | 1,960 | 5,165 |
| 10,000 | 370 | 727 | 1,332 | 4,899 | 622 | 1,193 | 2,098 | 6,239 |
| 25,000 | 378 | 760 | 1,448 | 6,939 | 646 | 1,285 | 2,399 | 9,972 |
| 50,000 | 381 | 772 | 1,491 | 8,056 | 655 | 1,318 | 2,520 | 12,455 |
| 75,000 | 382 | 776 | 1,506 | 8,514 | 658 | 1,330 | 2,563 | 13,583 |
| 100,000 | 383 | 778 | 1,513 | 8,762 | 659 | 1,336 | 2,585 | 14,227 |
| 250,000 | 384 | 782 | 1,527 | 9,248 | 662 | 1,347 | 2,626 | 15,555 |
| 500,000 | 384 | 783 | 1,532 | 9,423 | 663 | 1,350 | 2,640 | 16,055 |
| 1,000,000 | 384 | 783 | 1,534 | 9,512 | 663 | 1,352 | 2,647 | 16,317 |
| 2,500,000 | 384 | 783 | 1,536 | 9,567 | 663 | 1,353 | 2,651 | 16,478 |
| 10,000,000 | 384 | 784 | 1,536 | 9,594 | 663 | 1,354 | 2,653 | 16,560 |
| 100,000,000 | 384 | 784 | 1,537 | 9,603 | 663 | 1,354 | 2,654 | 16,584 |
| 300,000,000 | 384 | 784 | 1,537 | 9,603 | 663 | 1,354 | 2,654 | 16,586 |

**Figure 96 Required Sample Size**
**(Boyd, 2006)**

**Appendix C -** Nvivo Coding Summary Report by Source (Focus Group)

22/10/2016 17:27

# Coding Summary By Source

# Barriers facing E-Service Adoption and Implementation

# 22/10/2016 17:27

| Classification | Aggregate | Coverage | Number Of Coding | Reference Number | Coded By Initials | Modified On |
|---|---|---|---|---|---|---|

## Document

## Internals\\Participant 1

### Node

#### Nodes\\Online Focus Group Interview Questions\Q.1. E-Government Services Usage.

| Classification | Aggregate | Coverage | Number Of Coding | Reference Number | Coded By Initials | Modified On |
|---|---|---|---|---|---|---|
| | No | 0.0560 | 1 | | | |
| | | | | 1 | K.O | 20/10/2016 14:50 |

Q.1. E-Government Services Usage.

Researcher

Have you used E-Government Services before?

#### Nodes\\Online Focus Group Interview Questions\Q.1. E-Government Services Usage.\Have you used E-Government Service before

| Classification | Aggregate | Coverage | Number Of Coding | Reference Number | Coded By Initials | Modified On |
|---|---|---|---|---|---|---|
| | No | 0.0186 | 1 | | | |
| | | | | 1 | K.O | 22/10/2016 12:57 |

Have you used E-Government Services before

#### Nodes\\Online Focus Group Interview Questions\Q.1. E-Government Services Usage.\Have you used E-Government Service before\Participant 1

| Classification | Aggregate | Coverage | Number Of Coding | Reference Number | Coded By Initials | Modified On |
|---|---|---|---|---|---|---|
| | No | 0.0124 | 1 | | | |
| | | | | 1 | K.O | 22/10/2016 12:58 |

Participant 1

Answer - No

| Classification | Aggregate | Coverage | Number Of Coding References | Reference Number | Coded By Initials | Modified On |
|---|---|---|---|---|---|---|

## Nodes\\Online Focus Group Interview Questions\Q.10. E-Government Services future

| | No | 0.1120 | 1 | | | |
|---|---|---|---|---|---|---|
| | | | | 1 | K.O | 20/10/2016 14:50 |

Q.10. E-Government Services future plan.

Researcher

Do you have any suggestion or future plan for E-Government Services?

Participant 1

Answer - Government should establish public ICT centres for the citizens at the local government level.

## Nodes\\Online Focus Group Interview Questions\Q.10. E-Government Services future plan.\Do you have any suggestion or future plan for E-Government services

| | No | 0.0297 | 1 | | | |
|---|---|---|---|---|---|---|
| | | | | 1 | K.O | 22/10/2016 13:47 |

Do you have any suggestion or future plan for E-Government Services

## Nodes\\Online Focus Group Interview Questions\Q.10. E-Government Services future plan.\Do you have any suggestion or future plan for E-Government services\Participant 1

| | No | 0.0533 | 1 | | | |
|---|---|---|---|---|---|---|
| | | | | 1 | K.O | 22/10/2016 13:47 |

Participant 1

Answer - Government should establish public ICT centres for the citizens at the local government level.

## Nodes\\Online Focus Group Interview Questions\Q.2. Types of E-Government Services.

| | No | 0.0640 | 1 | | | |
|---|---|---|---|---|---|---|
| | | | | 1 | K.O | 20/10/2016 14:50 |

Q.2. Types of E-Government Services.

Researcher

What type of E-Government Services have you used before?

| Classification | Aggregate | Coverage | Number Of Coding References | Reference Number | Coded By Initials | Modified On |
|---|---|---|---|---|---|---|

## Nodes\\Online Focus Group Interview Questions\Q.2. Types of E-Government Services.\E-Government service types used before

| | No | 0.0244 | 1 | | | |
|---|---|---|---|---|---|---|
| | | | | 1 | K.O | 22/10/2016 13:06 |

What type of E-Government Services have you used before

## Nodes\\Online Focus Group Interview Questions\Q.2. Types of E-Government Services.\E-Government service types used before\Participant 1

| | No | 0.0133 | 1 | | | |
|---|---|---|---|---|---|---|
| | | | | 1 | K.O | 22/10/2016 13:11 |

Participant 1

Answer – None

## Nodes\\Online Focus Group Interview Questions\Q.3. Barriers facing E-Government

| | No | 0.1102 | 1 | | | |
|---|---|---|---|---|---|---|
| | | | | 1 | K.O | 20/10/2016 14:50 |

Q.3. Barriers facing E-Government Services.

Researcher

What are the barriers facing E-Service projects success both implementation and adoption?

## Nodes\\Online Focus Group Interview Questions\Q.3. Barriers facing E-Government Services.\What are the barriers facing E-Service initiatives

| | No | 0.0248 | 1 | | | |
|---|---|---|---|---|---|---|
| | | | | 1 | K.O | 22/10/2016 13:16 |

What are the barriers facing E-Service projects success

## Nodes\\Online Focus Group Interview Questions\Q.3. Barriers facing E-Government Services.\What are the barriers facing E-Service initiatives\Participant 1

| | No | 0.0422 | 1 |
|---|---|---|---|

| | | | | 1 | K.O | 22/10/2016 13:16 |
|---|---|---|---|---|---|---|

Participant 1

Answer – Finance, Lack of Qualified personnel, Corruption, IT Infrastructures.

| Classification | Aggregate | Coverage | Number Of Coding References | Reference Number | Coded By Initials | Modified On |
|---|---|---|---|---|---|---|

## Nodes\\Online Focus Group Interview Questions\Q.4. Adopting E-Government Services.

| | No | 0.0586 | 1 | | | |
|---|---|---|---|---|---|---|
| | | | | 1 | K.O | 20/10/2016 14:50 |

Q.4. Adopting E-Government Services.

Researcher

Will you adopt and use E-Government Services?

## Nodes\\Online Focus Group Interview Questions\Q.4. Adopting E-Government Services.\Will you adopt and use E-Government services\Participant 1

| | No | 0.0128 | 1 | | | |
|---|---|---|---|---|---|---|
| | | | | 1 | K.O | 22/10/2016 13:22 |

Participant 1

Answer - Yes

## Nodes\\Online Focus Group Interview Questions\Q.5. Success Factors for adopting E-Government Services technology.

| | No | 0.1195 | 1 | | | |
|---|---|---|---|---|---|---|
| | | | | 1 | K.O | 20/10/2016 14:50 |

Q.5. Success Factors for adopting E-Government Services technology.

Researcher

What are the success factors for adopting E-Government Services technology?

## Nodes\\Online Focus Group Interview Questions\Q.5. Success Factors for adopting E-Government Services technology.\Success factors for adopting E-Government Services

| | No | 0.0271 | 1 | | | |
|---|---|---|---|---|---|---|

| | | | | 1 | K.O | 22/10/2016 13:23 |
|---|---|---|---|---|---|---|

success factors for adopting E-Government Services technology

| Classification | Aggregate | Coverage | Number Of Coding References | Reference Number | Coded By Initials | Modified On |
|---|---|---|---|---|---|---|

## Nodes\\Online Focus Group Interview Questions\Q.5. Success Factors for adopting E-Government Services technology.\Success factors for adopting E-Government Services

| | No | 0.0457 | 1 | | | |
|---|---|---|---|---|---|---|
| | | | | 1 | K.O | 22/10/2016 13:23 |

Participant 1

Answer - Quick access to information, transparency in government, citizen participation.

## Nodes\\Online Focus Group Interview Questions\Q.6. E-Government Services

| | No | 0.0782 | 1 | | | |
|---|---|---|---|---|---|---|
| | | | | 1 | K.O | 20/10/2016 14:50 |

Q.6. E-Government Services Framework.

Researcher

Does the E-Government Services Framework sent to you represent your notion of E-Service?

## Nodes\\Online Focus Group Interview Questions\Q.6. E-Government Services Framework.\Does the E-Government Services Framework sent to you represent your

| | No | 0.0386 | 1 | | | |
|---|---|---|---|---|---|---|
| | | | | 1 | K.O | 22/10/2016 13:29 |

Does the E-Government Services Framework sent to you represent your notion of E-Service

## Nodes\\Online Focus Group Interview Questions\Q.6. E-Government Services Framework.\Does the E-Government Services Framework sent to you represent your

| | No | 0.0128 | 1 | | | |
|---|---|---|---|---|---|---|
| | | | | 1 | K.O | 22/10/2016 13:29 |

Participant 1

Answer - Yes

| Classification | Aggregate | Coverage | Number Of Coding References | Reference Number | Coded By Initials | Modified On |
|---|---|---|---|---|---|---|

**Nodes\\Online Focus Group Interview Questions\Q.7. The use of DOI theory in E-Government Services.**

| | | | | | | |
|---|---|---|---|---|---|---|
| | No | 0.0884 | 1 | | | |
| | | | | 1 | K.O | 20/10/2016 14:50 |

Q.7. The use of DOI theory in E-Government Services.

Researcher

Will you encourage the use of Diffusion of Innovation Theory (DOI) in E-Service implementation?

**Nodes\\Online Focus Group Interview Questions\Q.7. The use of DOI theory in E-Government Services.\Will you encourage the use of DOI theory in E-Service Implementation and Adoption\Participant 1**

| | | | | | | |
|---|---|---|---|---|---|---|
| | No | 0.0124 | 1 | | | |
| | | | | 1 | K.O | 22/10/2016 13:35 |

Participant 1

Answer - Yes

**Nodes\\Online Focus Group Interview Questions\Q.8. E-Government Services technology adoption diffusion to other aspects or areas.**

| | | | | | | |
|---|---|---|---|---|---|---|
| | No | 0.0964 | 1 | | | |
| | | | | 1 | K.O | 20/10/2016 14:50 |

Q.8. E-Government Services technology adoption diffusion to other aspects or areas.

Researcher

Do you think the E-Service technology adoption will help in other aspects or areas?

**Nodes\\Online Focus Group Interview Questions\Q.8. E-Government Services technology adoption diffusion to other aspects or areas.\Do you think the E-Service technology adoption will help in other aspects or areas**

| | No | 0.0364 | 1 | | | |
|---|---|---|---|---|---|---|
| | | | | 1 | K.O | 22/10/2016 13:37 |

Do you think the E-Service technology adoption will help in other aspects or areas

| Classification | Aggregate | Coverage | Number Of Coding References | Reference Number | Coded By Initials | Modified On |
|---|---|---|---|---|---|---|

**Nodes\\Online Focus Group Interview Questions\Q.8. E-Government Services technology adoption diffusion to other aspects or areas.\Do you think the E-Service technology adoption**

| | No | 0.0120 | 1 | | | |
|---|---|---|---|---|---|---|
| | | | | 1 | K.O | 22/10/2016 13:40 |

Participant 1

Answer - Yes

**Nodes\\Online Focus Group Interview Questions\Q.9. Implementing the E-Government Services recommendations from this study.**

| | No | 0.1408 | 1 | | | |
|---|---|---|---|---|---|---|
| | | | | 1 | K.O | 20/10/2016 14:50 |

Q.9. Implementing the E-Government Services recommendations from this study.

Researcher

As part of the stakeholder in the provision of E-Service at the local government level in Nigeria, will you be ready to receive and implement the recommendations on E-Service from this study?

**Nodes\\Online Focus Group Interview Questions\Q.9. Implementing the E-Government Services recommendations from this study.\Ready to receive and implement the rec**

| | No | 0.0120 | 1 | | | |
|---|---|---|---|---|---|---|
| | | | | 1 | K.O | 22/10/2016 13:46 |

Participant 1

Answer - Yes

| Classification | Aggregate | Coverage | Number Of Coding References | Reference Number | Coded By Initials | Modified On |
|---|---|---|---|---|---|---|

# Internals\\Participant 2

## Node

### Nodes\\Online Focus Group Interview Questions\Q.1. E-Government Services Usage.

| Classification | Aggregate | Coverage | Number Of Coding References | Reference Number | Coded By Initials | Modified On |
|---|---|---|---|---|---|---|
| | No | 0.0525 | 1 | | | |
| | | | | 1 | K.O | 20/10/2016 14:50 |

Q.1. E-Government Services Usage.

Researcher

Have you used E-Government Services before?

### Nodes\\Online Focus Group Interview Questions\Q.1. E-Government Services Usage.\Have you used E-Government Service before\Participant 2

| Classification | Aggregate | Coverage | Number Of Coding References | Reference Number | Coded By Initials | Modified On |
|---|---|---|---|---|---|---|
| | No | 0.0108 | 1 | | | |
| | | | | 1 | K.O | 22/10/2016 12:58 |

Participant 2

Answer - No

### Nodes\\Online Focus Group Interview Questions\Q.10. E-Government Services future

| Classification | Aggregate | Coverage | Number Of Coding References | Reference Number | Coded By Initials | Modified On |
|---|---|---|---|---|---|---|
| | No | 0.1138 | 1 | | | |
| | | | | 1 | K.O | 20/10/2016 14:50 |

Q.10. E-Government Services future plan.

Researcher

Do you have any suggestion or future plan for E-Government Services?

## Nodes\\Online Focus Group Interview Questions\Q.10. E-Government Services future plan.\Do you have any suggestion or future plan for E-Government services\Participant 2

| | No | 0.0604 | 1 | | | |
|---|---|---|---|---|---|---|
| | | | | 1 | K.O | 22/10/2016 13:48 |

Participant 2

Answer - Government should endeavours to partner with private sectors for easier E-Service deployment at Local government level.

| Classification | Aggregate | Coverage | Number Of Coding References | Reference Number | Coded By Initials | Modified On |
|---|---|---|---|---|---|---|

## Nodes\\Online Focus Group Interview Questions\Q.2. Types of E-Government Services.

| | No | 0.0600 | 1 | | | |
|---|---|---|---|---|---|---|
| | | | | 1 | K.O | 20/10/2016 14:50 |

Q.2. Types of E-Government Services.

Researcher

What type of E-Government Services have you used before?

## Nodes\\Online Focus Group Interview Questions\Q.2. Types of E-Government Services.\E-Government service types used before\Participant 2

| | No | 0.0125 | 1 | | | |
|---|---|---|---|---|---|---|
| | | | | 1 | K.O | 22/10/2016 13:12 |

Participant 2

Answer – None

## Nodes\\Online Focus Group Interview Questions\Q.3. Barriers facing E-Government

| | No | 0.1330 | 1 | | | |
|---|---|---|---|---|---|---|
| | | | | 1 | K.O | 20/10/2016 14:50 |

Q.3. Barriers facing E-Government Services.

Researcher

What are the barriers facing E-Service projects success both implementation and adoption?

Participant 2

**Nodes\\Online Focus Group Interview Questions\Q.3. Barriers facing E-Government Services.\What are the barriers facing E-Service initiatives\Participant 2**

| Classification | Aggregate | Coverage | Number Of Coding References | Reference Number | Coded By Initials | Modified On |
|---|---|---|---|---|---|---|
| | No | 0.0683 | 1 | | | |
| | | | | 1 | K.O | 22/10/2016 13:17 |

**Nodes\\Online Focus Group Interview Questions\Q.4. Adopting E-Government Services.**

| Classification | Aggregate | Coverage | Number Of Coding References | Reference Number | Coded By Initials | Modified On |
|---|---|---|---|---|---|---|
| | No | 0.0550 | 1 | | | |
| | | | | 1 | K.O | 20/10/2016 14:50 |

Q.4. Adopting E-Government Services.

Researcher

Will you adopt and use E-Government Services?

**Nodes\\Online Focus Group Interview Questions\Q.4. Adopting E-Government Services.\Will you adopt and use E-Government services\Participant 2**

| Classification | Aggregate | Coverage | Number Of Coding References | Reference Number | Coded By Initials | Modified On |
|---|---|---|---|---|---|---|
| | No | 0.0120 | 1 | | | |
| | | | | 1 | K.O | 22/10/2016 13:21 |

Participant 2

Answer - Yes

**Nodes\\Online Focus Group Interview Questions\Q.5. Success Factors for adopting E-Government Services technology.**

| Classification | Aggregate | Coverage | Number Of Coding References | Reference Number | Coded By Initials | Modified On |
|---|---|---|---|---|---|---|
| | No | 0.1359 | 1 | | | |
| | | | | 1 | K.O | 20/10/2016 14:50 |

Q.5. Success Factors for adopting E-Government Services technology.

Researcher

What are the success factors for adopting E-Government Services technology?

Participant 2

### Nodes\\Online Focus Group Interview Questions\Q.5. Success Factors for adopting E-Government Services technology.\Success factors for adopting E-Government Services

| Classification | Aggregate | Coverage | Number Of Coding References | Reference Number | Coded By Initials | Modified On |
|---|---|---|---|---|---|---|
| | No | 0.0667 | 1 | | | |
| | | | | 1 | K.O | 22/10/2016 13:24 |

Participant 2

Answer - Economic Growth, Stop leakages, Quick service delivery, access to information, improve government business, transparency in government.

| Classification | Aggregate | Coverage | Number Of Coding References | Reference Number | Coded By Initials | Modified On |
|---|---|---|---|---|---|---|

### Nodes\\Online Focus Group Interview Questions\Q.6. E-Government Services

| Classification | Aggregate | Coverage | Number Of Coding References | Reference Number | Coded By Initials | Modified On |
|---|---|---|---|---|---|---|
| | No | 0.0733 | 1 | | | |
| | | | | 1 | K.O | 20/10/2016 14:50 |

Q.6. E-Government Services Framework.

Researcher

Does the E-Government Services Framework sent to you represent your notion of E-Service?

### Nodes\\Online Focus Group Interview Questions\Q.6. E-Government Services Framework.\Does the E-Government Services Framework sent to you represent your

| Classification | Aggregate | Coverage | Number Of Coding References | Reference Number | Coded By Initials | Modified On |
|---|---|---|---|---|---|---|
| | No | 0.0112 | 1 | | | |
| | | | | 1 | K.O | 22/10/2016 13:30 |

Participant 2

Answer - Yes

### Nodes\\Online Focus Group Interview Questions\Q.7. The use of DOI theory in E-Government Services.

| Classification | Aggregate | Coverage | Number Of Coding References | Reference Number | Coded By Initials | Modified On |
|---|---|---|---|---|---|---|
| | No | 0.0825 | 1 | | | |
| | | | | 1 | K.O | 20/10/2016 14:50 |

Q.7. The use of DOI theory in E-Government Services.

Researcher

Will you encourage the use of Diffusion of Innovation Theory (DOI) in E-Service implementation?

## Nodes\\Online Focus Group Interview Questions\Q.7. The use of DOI theory in E-Government Services.\Will you encourage the use of DOI theory in E-Service Implementation and adoption

| | | | | | | |
|---|---|---|---|---|---|---|
| | No | 0.0112 | 1 | | | |
| | | | | 1 | K.O | 22/10/2016 13:34 |

Participant 2

Answer - Yes

| Classification | Aggregate | Coverage | Number Of Coding References | Reference Number | Coded By Initials | Modified On |
|---|---|---|---|---|---|---|

## Nodes\\Online Focus Group Interview Questions\Q.8. E-Government Services technology adoption diffusion to other aspects or areas.

| | | | | | | |
|---|---|---|---|---|---|---|
| | No | 0.0904 | 1 | | | |
| | | | | 1 | K.O | 20/10/2016 14:50 |

Q.8. E-Government Services technology adoption diffusion to other aspects or areas.

Researcher

Do you think the E-Service technology adoption will help in other aspects or areas?

## Nodes\\Online Focus Group Interview Questions\Q.8. E-Government Services technology adoption diffusion to other aspects or areas.\Do you think the E-Service technology adoption will diffuse to other aspects or areas.

| | | | | | | |
|---|---|---|---|---|---|---|
| | No | 0.0112 | 1 | | | |
| | | | | 1 | K.O | 22/10/2016 13:41 |

Participant 2

Answer - Yes

**Nodes\\Online Focus Group Interview Questions\Q.9. Implementing the E-Government Services recommendations from this study.**

| | | | | | | |
|---|---|---|---|---|---|---|
| | No | 0.1321 | 1 | | | |
| | | | | 1 | K.O | 20/10/2016 14:50 |

Q.9. Implementing the E-Government Services recommendations from this study.

Researcher

As part of the stakeholder in the provision of E-Service at the local government level in Nigeria, will you be ready to receive and implement the recommendations on E-Service from this study?

**Nodes\\Online Focus Group Interview Questions\Q.9. Implementing the E-Government Services recommendations from this study.\Ready to receive and implement the recommendations on E-Service from this study\Participant 2**

| | | | | | | |
|---|---|---|---|---|---|---|
| | No | 0.0120 | 1 | | | |
| | | | | 1 | K.O | 22/10/2016 13:45 |

Participant 2

Answer - Yes

| Classification | Aggregate | Coverage | Number Of Coding References | Reference Number | Coded By Initials | Modified On |
|---|---|---|---|---|---|---|

# Internals\\Participant 3

## Node

**Nodes\\Online Focus Group Interview Questions\Q.1. E-Government Services Usage.**

| | | | | | | |
|---|---|---|---|---|---|---|
| | No | 0.0588 | 1 | | | |
| | | | | 1 | K.O | 20/10/2016 14:50 |

Q.1. E-Government Services Usage.

Researcher

Have you used E-Government Services before?

**Nodes\\Online Focus Group Interview Questions\Q.1. E-Government Services Usage.\Have you used E-Government Service before\Participant 3**

| | | | |
|---|---|---|---|
| | No | 0.0125 | 1 |

| | | | | | | |
|---|---|---|---|---|---|---|
| | | | | 1 | K.O | 22/10/2016 12:59 |

Participant 3

Answer - Yes

## Nodes\\Online Focus Group Interview Questions\Q.10. E-Government Services future

| | | | | | | |
|---|---|---|---|---|---|---|
| | No | 0.1029 | 1 | | | |
| | | | | 1 | K.O | 20/10/2016 14:50 |

Q.10. E-Government Services future plan.

Researcher

Do you have any suggestion or future plan for E-Government Services?

## Nodes\\Online Focus Group Interview Questions\Q.10. E-Government Services future plan.\Do you have any suggestion or future plan for E-Government services\Participant 3

| | | | | | | |
|---|---|---|---|---|---|---|
| | No | 0.0435 | 1 | | | |
| | | | | 1 | K.O | 22/10/2016 13:48 |

Participant 3

Answer - Government should step up awareness on these services when deployed.

| Classification | Aggregate | Coverage | Number Of Coding References | Reference Number | Coded By Initials | Modified On |
|---|---|---|---|---|---|---|

## Nodes\\Online Focus Group Interview Questions\Q.2. Types of E-Government Services.

| | | | | | | |
|---|---|---|---|---|---|---|
| | No | 0.0746 | 1 | | | |
| | | | | 1 | K.O | 20/10/2016 14:50 |

Q.2. Types of E-Government Services.

Researcher

What type of E-Government Services have you used before?

## Nodes\\Online Focus Group Interview Questions\Q.2. Types of E-Government Services.\E-Government service types used before\Participant 3

| | | | |
|---|---|---|---|
| | No | 0.0208 | 1 |

| | | | | 1 | K.O | 22/10/2016 13:12 |
|---|---|---|---|---|---|---|

Participant 3

Answer – E-payment, E-passport

## Nodes\\Online Focus Group Interview Questions\Q.3. Barriers facing E-Government

| | No | 0.1015 | 1 | | | |
|---|---|---|---|---|---|---|
| | | | | 1 | K.O | 20/10/2016 14:50 |

Q.3. Barriers facing E-Government Services.

Researcher

What are the barriers facing E-Service projects success both implementation and adoption?

## Nodes\\Online Focus Group Interview Questions\Q.3. Barriers facing E-Government Services.\What are the barriers facing E-Service initiatives\Participant 3

| | No | 0.0296 | 1 | | | |
|---|---|---|---|---|---|---|
| | | | | 1 | K.O | 22/10/2016 13:18 |

Participant 3

Answer - Corruption, Bad Leadership, and Finance.

| Classification | Aggregate | Coverage | Number Of Coding References | Reference Number | Coded By Initials | Modified On |
|---|---|---|---|---|---|---|

## Nodes\\Online Focus Group Interview Questions\Q.4. Adopting E-Government Services.

| | No | 0.0611 | 1 | | | |
|---|---|---|---|---|---|---|
| | | | | 1 | K.O | 20/10/2016 14:50 |

Q.4. Adopting E-Government Services.

Researcher

Will you adopt and use E-Government Services?

## Nodes\\Online Focus Group Interview Questions\Q.4. Adopting E-Government Services.\Will you adopt and use E-Government services\Participant 3

| | No | 0.0125 | 1 | | | |
|---|---|---|---|---|---|---|
| | | | | 1 | K.O | 22/10/2016 13:21 |

Participant 3

Answer - Yes

## Nodes\\Online Focus Group Interview Questions\Q.5. Success Factors for adopting E-Government Services technology.

| | No | 0.1010 | 1 | | | |
|---|---|---|---|---|---|---|
| | | | | 1 | K.O | 20/10/2016 14:50 |

Q.5. Success Factors for adopting E-Government Services technology.

Researcher

What are the success factors for adopting E-Government Services technology?

## Nodes\\Online Focus Group Interview Questions\Q.5. Success Factors for adopting E-Government Services technology.\Success factors for adopting E-Government Services

| | No | 0.0241 | 1 | | | |
|---|---|---|---|---|---|---|
| | | | | 1 | K.O | 22/10/2016 13:28 |

Participant 3

Answer - Transparency in Government.

| Classification | Aggregate | Coverage | Number Of Coding References | Reference Number | Coded By Initials | Modified On |
|---|---|---|---|---|---|---|

## Nodes\\Online Focus Group Interview Questions\Q.6. E-Government Services

| | No | 0.0815 | 1 | | | |
|---|---|---|---|---|---|---|
| | | | | 1 | K.O | 20/10/2016 14:50 |

Q.6. E-Government Services Framework.

Researcher

Does the E-Government Services Framework sent to you represent your notion of E-Service?

**Nodes\\Online Focus Group Interview Questions\Q.6. E-Government Services Framework.\Does the E-Government Services Framework sent to you represent your**

| | | | | | | |
|---|---|---|---|---|---|---|
| | No | 0.0125 | 1 | | | |
| | | | | 1 | K.O | 22/10/2016 13:30 |

Participant 3

Answer - Yes

**Nodes\\Online Focus Group Interview Questions\Q.7. The use of DOI theory in E-Government Services.**

| | | | | | | |
|---|---|---|---|---|---|---|
| | No | 0.0917 | 1 | | | |
| | | | | 1 | K.O | 20/10/2016 14:50 |

Q.7. The use of DOI theory in E-Government Services.

Researcher

Will you encourage the use of Diffusion of Innovation Theory (DOI) in E-Service implementation?

**Nodes\\Online Focus Group Interview Questions\Q.7. The use of DOI theory in E-Government Services.\Will you encourage the use of DOI theory in E-Service Implementation and Adoption\Participant 3**

| | | | | | | |
|---|---|---|---|---|---|---|
| | No | 0.0125 | 1 | | | |
| | | | | 1 | K.O | 22/10/2016 13:34 |

Participant 3

Answer - Yes

| Classification | Aggregate | Coverage | Number Of Coding References | Reference Number | Coded By Initials | Modified On |
|---|---|---|---|---|---|---|

**Nodes\\Online Focus Group Interview Questions\Q.8. E-Government Services technology adoption diffusion to other aspects or areas.**

| | | | | | | |
|---|---|---|---|---|---|---|
| | No | 0.1006 | 1 | | | |
| | | | | 1 | K.O | 20/10/2016 14:50 |

Q.8. E-Government Services technology adoption diffusion to other aspects or areas.

Researcher

Do you think the E-Service technology adoption will help in other aspects or areas?

## Nodes\\Online Focus Group Interview Questions\Q.8. E-Government Services technology adoption diffusion to other aspects or areas.\Do you think the E-Service technology adoption will help in other aspects or areas\Participant 3

| | | | | | | |
|---|---|---|---|---|---|---|
| | No | 0.0125 | 1 | 1 | K.O | 22/10/2016 13:41 |

Participant 3

Answer - Yes

## Nodes\\Online Focus Group Interview Questions\Q.9. Implementing the E-Government Services recommendations from this study.

| | | | | | | |
|---|---|---|---|---|---|---|
| | No | 0.1469 | 1 | 1 | K.O | 20/10/2016 14:50 |

Q.9. Implementing the E-Government Services recommendations from this study.

Researcher

As part of the stakeholder in the provision of E-Service at the local government level in Nigeria, will you be ready to receive and implement the recommendations on E-Service from this study?

## Nodes\\Online Focus Group Interview Questions\Q.9. Implementing the E-Government Services recommendations from this study.\Ready to receive and implement the recommendations on E-Service from this study\Participant 3

| | | | | | | |
|---|---|---|---|---|---|---|
| | No | 0.0134 | 1 | 1 | K.O | 22/10/2016 13:45 |

Participant 3

| Classification | Aggregate | Coverage | Number Of Coding References | Reference Number | Coded By Initials | Modified On |
|---|---|---|---|---|---|---|

# Internals\\Participant 4

### Node

## Nodes\\Online Focus Group Interview Questions\Q.1. E-Government Services Usage.

| | No | 0.0506 | 1 |
|---|---|---|---|

1 K.O 20/10/2016 14:50

Q.1. E-Government Services Usage.

Researcher

Have you used E-Government Services before?

## Nodes\\Online Focus Group Interview Questions\Q.1. E-Government Services Usage.\Have you used E-Government Service before\Participant 4

| | No | 0.0115 | 1 |
|---|---|---|---|

1 K.O 22/10/2016 12:59

Participant 4

Answer – Yes

## Nodes\\Online Focus Group Interview Questions\Q.10. E-Government Services future

| | No | 0.1149 | 1 |
|---|---|---|---|

1 K.O 20/10/2016 14:50

Q.10. E-Government Services future plan.

Researcher

Do you have any suggestion or future plan for E-Government Services?

Participant 4

Answer - Government to make fund available to E-Service projects, proper monitoring should be encouraged during the

## Nodes\\Online Focus Group Interview Questions\Q.10. E-Government Services future plan.\Do you have any suggestion or future plan for E-Government services\Participant 4

| | No | 0.0630 | 1 |
|---|---|---|---|

1 K.O 22/10/2016 13:49

Participant 4

Answer - Government to make fund available to E-Service projects, proper monitoring should be encouraged during the

| Classification | Aggregate | Coverage | Number Of Coding References | Reference Number | Coded By Initials | Modified On |
|---|---|---|---|---|---|---|

## Nodes\\Online Focus Group Interview Questions\Q.2. Types of E-Government Services.

| | | | | | | |
|---|---|---|---|---|---|---|
| | No | 0.0770 | 1 | | | |
| | | | | 1 | K.O | 20/10/2016 14:50 |

Q.2. Types of E-Government Services.

Researcher

What type of E-Government Services have you used before?

## Nodes\\Online Focus Group Interview Questions\Q.2. Types of E-Government Services.\E-Government service types used before\Participant 4

| | | | | | | |
|---|---|---|---|---|---|---|
| | No | 0.0315 | 1 | | | |
| | | | | 1 | K.O | 22/10/2016 13:14 |

Participant 4

Answer – E-passport, E-payment, E-Tax, E-Licence, E-Assessment

## Nodes\\Online Focus Group Interview Questions\Q.3. Barriers facing E-Government

| | | | | | | |
|---|---|---|---|---|---|---|
| | No | 0.1560 | 1 | | | |
| | | | | 1 | K.O | 20/10/2016 14:50 |

Q.3. Barriers facing E-Government Services.

Researcher

What are the barriers facing E-Service projects success both implementation and adoption?

Participant 4

Answer - Lack of strategic plans, Finance, Lack of Qualified personnel, Resistance to Change, Lack of Partnership, Usability

## Nodes\\Online Focus Group Interview Questions\Q.3. Barriers facing E-Government Services.\What are the barriers facing E-Service initiatives\Participant 4

| | | | | | | |
|---|---|---|---|---|---|---|
| | No | 0.0949 | 1 | | | |
| | | | | 1 | K.O | 22/10/2016 13:18 |

Participant 4

Answer - Lack of strategic plans, Finance, Lack of Qualified personnel, Resistance to Change, Lack of Partnership, Usability

| Classification | Aggregate | Coverage | Number Of Coding References | Reference Number | Coded By Initials | Modified On |
|---|---|---|---|---|---|---|

## Nodes\\Online Focus Group Interview Questions\Q.4. Adopting E-Government Services.

| | No | 0.0526 | 1 | | | |
|---|---|---|---|---|---|---|
| | | | | 1 | K.O | 20/10/2016 14:50 |

Q.4. Adopting E-Government Services.

Researcher

Will you adopt and use E-Government Services?

## Nodes\\Online Focus Group Interview Questions\Q.4. Adopting E-Government Services.\Will you adopt and use E-Government services\Participant 4

| | No | 0.0115 | 1 | | | |
|---|---|---|---|---|---|---|
| | | | | 1 | K.O | 22/10/2016 13:20 |

Participant 4

Answer - Yes

## Nodes\\Online Focus Group Interview Questions\Q.5. Success Factors for adopting E-Government Services technology.

| | No | 0.1185 | 1 | | | |
|---|---|---|---|---|---|---|
| | | | | 1 | K.O | 20/10/2016 14:50 |

Q.5. Success Factors for adopting E-Government Services technology.

Researcher

What are the success factors for adopting E-Government Services technology?

## Nodes\\Online Focus Group Interview Questions\Q.5. Success Factors for adopting E-Government Services technology.\Success factors for adopting E-Government Services

| | No | 0.0522 | 1 | | | |
|---|---|---|---|---|---|---|
| | | | | 1 | K.O | 22/10/2016 13:28 |

Participant 4

Answer - Quick service delivery, access to information, transparency, robust government, improved record management.

| Classification | Aggregate | Coverage | Number Of Coding References | Reference Number | Coded By Initials | Modified On |
|---|---|---|---|---|---|---|

## Nodes\\Online Focus Group Interview Questions\Q.6. E-Government Services

| | | | | | | |
|---|---|---|---|---|---|---|
| | No | 0.0702 | 1 | | | |
| | | | | 1 | K.O | 20/10/2016 14:50 |

Q.6. E-Government Services Framework.

Researcher

Does the E-Government Services Framework sent to you represent your notion of E-Service?

## Nodes\\Online Focus Group Interview Questions\Q.6. E-Government Services Framework.\Does the E-Government Services Framework sent to you represent your

| | | | | | | |
|---|---|---|---|---|---|---|
| | No | 0.0115 | 1 | | | |
| | | | | 1 | K.O | 22/10/2016 13:30 |

Participant 4

Answer - Yes

## Nodes\\Online Focus Group Interview Questions\Q.7. The use of DOI theory in E-Government Services.

| | | | | | | |
|---|---|---|---|---|---|---|
| | No | 0.0790 | 1 | | | |
| | | | | 1 | K.O | 20/10/2016 14:50 |

Q.7. The use of DOI theory in E-Government Services.

Researcher

Will you encourage the use of Diffusion of Innovation Theory (DOI) in E-Service implementation?

## Nodes\\Online Focus Group Interview Questions\Q.7. The use of DOI theory in E-Government Services.\Will you encourage the use of DOI theory in E-Service Implementation and Adoption\Participant 4

| | | | | | | |
|---|---|---|---|---|---|---|
| | No | 0.0107 | 1 | | | |

| | | | | 1 | K.O | 22/10/2016 13:33 |
|---|---|---|---|---|---|---|

Participant 4

Answer - Yes

| Classification | Aggregate | Coverage | Number Of Coding References | Reference Number | Coded By Initials | Modified On |
|---|---|---|---|---|---|---|

**Nodes\\Online Focus Group Interview Questions\Q.8. E-Government Services technology adoption diffusion to other aspects or areas.**

| | | | | | | |
|---|---|---|---|---|---|---|
| | No | 0.0865 | 1 | | | |
| | | | | 1 | K.O | 20/10/2016 14:50 |

Q.8. E-Government Services technology adoption diffusion to other aspects or areas.

Researcher

Do you think the E-Service technology adoption will help in other aspects or areas?

**Nodes\\Online Focus Group Interview Questions\Q.8. E-Government Services technology adoption diffusion to other aspects or areas.\Do you think the E-Service technology adoption will help in other aspects or areas\Participant 4**

| | | | | | | |
|---|---|---|---|---|---|---|
| | No | 0.0107 | 1 | | | |
| | | | | 1 | K.O | 22/10/2016 13:42 |

Participant 4

Answer - Yes

**Nodes\\Online Focus Group Interview Questions\Q.9. Implementing the E-Government Services recommendations from this study.**

| | | | | | | |
|---|---|---|---|---|---|---|
| | No | 0.1264 | 1 | | | |
| | | | | 1 | K.O | 20/10/2016 14:50 |

Q.9. Implementing the E-Government Services recommendations from this study.

Researcher

As part of the stakeholder in the provision of E-Service at the local government level in Nigeria, will you be ready to receive and implement the recommendations on E-Service from this study?

**Nodes\\Online Focus Group Interview Questions\Q.9. Implementing the E-Government Services recommendations from this study.\Ready to receive and implement the recommendations on E-Service from this study\Participant 4**

| | | | | | | |
|---|---|---|---|---|---|---|
| | No | 0.0107 | 1 | | | |
| | | | | 1 | K.O | 22/10/2016 13:44 |

Participant 4

Answer - Yes

| Classification | Aggregate | Coverage | Number Of Coding References | Reference Number | Coded By Initials | Modified On |
|---|---|---|---|---|---|---|

# Internals\\Participant 5

## Node

**Nodes\\Online Focus Group Interview Questions\Q.1. E-Government Services Usage.**

| | | | | | | |
|---|---|---|---|---|---|---|
| | No | 0.0535 | 1 | | | |
| | | | | 1 | K.O | 20/10/2016 14:50 |

Q.1. E-Government Services Usage.

Researcher

Have you used E-Government Services before?

**Nodes\\Online Focus Group Interview Questions\Q.1. E-Government Services Usage.\Have you used E-Government Service before\Participant 5**

| | | | | | | |
|---|---|---|---|---|---|---|
| | No | 0.0110 | 1 | | | |
| | | | | 1 | K.O | 22/10/2016 13:00 |

Participant 5

Answer - No

**Nodes\\Online Focus Group Interview Questions\Q.10. E-Government Services future**

| | | | | | | |
|---|---|---|---|---|---|---|
| | No | 0.1070 | 1 | | | |
| | | | | 1 | K.O | 20/10/2016 14:50 |

Q.10. E-Government Services future plan.

Researcher

Do you have any suggestion or future plan for E-Government Services?

Participant 5

Answer - Improve awareness, provision of IT infrastructures and cheap internet access for the citizens.

## Nodes\\Online Focus Group Interview Questions\Q.10. E-Government Services future plan.\Do you have any suggestion or future plan for E-Government services\Participant 5

| | | | | | | |
|---|---|---|---|---|---|---|
| | No | 0.0509 | 1 | | | |
| | | | | 1 | K.O | 22/10/2016 13:49 |

Participant 5

Answer - Improve awareness, provision of IT infrastructures and cheap internet access for the citizens.

| Classification | Aggregate | Coverage | Number Of Coding References | Reference Number | Coded By Initials | Modified On |
|---|---|---|---|---|---|---|

## Nodes\\Online Focus Group Interview Questions\Q.2. Types of E-Government Services.

| | | | | | | |
|---|---|---|---|---|---|---|
| | No | 0.0611 | 1 | | | |
| | | | | 1 | K.O | 20/10/2016 14:50 |

Q.2. Types of E-Government Services.

Researcher

What type of E-Government Services have you used before?

## Nodes\\Online Focus Group Interview Questions\Q.2. Types of E-Government Services.\E-Government service types used before\Participant 5

| | | | | | | |
|---|---|---|---|---|---|---|
| | No | 0.0127 | 1 | | | |
| | | | | 1 | K.O | 22/10/2016 13:13 |

Participant 5

Answer – None

## Nodes\\Online Focus Group Interview Questions\Q.3. Barriers facing E-Government

| | | | |
|---|---|---|---|
| | No | 0.1354 | 1 |

| | | | | | | |
|---|---|---|---|---|---|---|
| | | | | 1 | K.O | 20/10/2016 14:50 |

Q.3. Barriers facing E-Government Services.

Researcher

What are the barriers facing E-Service projects success both implementation and adoption?

Participant 5

Answer – Corruption, No Regular ICT Training, Online Security/Trust, Cultural Problems, Political instability, Lack of ICT regulations and policies.

## Nodes\\Online Focus Group Interview Questions\Q.3. Barriers facing E-Government Services.\What are the barriers facing E-Service initiatives\Participant 5

| | | | | | | |
|---|---|---|---|---|---|---|
| | No | 0.0696 | 1 | | | |
| | | | | 1 | K.O | 22/10/2016 13:18 |

Participant 5

Answer – Corruption, No Regular ICT Training, Online Security/Trust, Cultural Problems, Political instability, Lack of ICT regulations and policies.

| Classification | Aggregate | Coverage | Number Of Coding References | Reference Number | Coded By Initials | Modified On |
|---|---|---|---|---|---|---|

## Nodes\\Online Focus Group Interview Questions\Q.4. Adopting E-Government Services.

| | | | | | | |
|---|---|---|---|---|---|---|
| | No | 0.0560 | 1 | | | |
| | | | | 1 | K.O | 20/10/2016 14:50 |

Q.4. Adopting E-Government Services.

Researcher

Will you adopt and use E-Government Services?

## Nodes\\Online Focus Group Interview Questions\Q.4. Adopting E-Government Services.\Will you adopt and use E-Government services

| | | | | | | |
|---|---|---|---|---|---|---|
| | No | 0.0186 | 1 | | | |
| | | | | 1 | K.O | 22/10/2016 13:19 |

Will you adopt and use E-Government Services

## Nodes\\Online Focus Group Interview Questions\Q.4. Adopting E-Government Services.\Will you adopt and use E-Government services\Participant 5

| | | | | | | |
|---|---|---|---|---|---|---|
| | No | 0.0123 | 1 | | | |
| | | | | 1 | K.O | 22/10/2016 13:20 |

Participant 5

Answer - Yes

## Nodes\\Online Focus Group Interview Questions\Q.5. Success Factors for adopting E-Government Services technology.

| | | | | | | |
|---|---|---|---|---|---|---|
| | No | 0.1290 | 1 | | | |
| | | | | 1 | K.O | 20/10/2016 14:50 |

Q.5. Success Factors for adopting E-Government Services technology.

Researcher

What are the success factors for adopting E-Government Services technology?

| Classification | Aggregate | Coverage | Number Of Coding References | Reference Number | Coded By Initials | Modified On |
|---|---|---|---|---|---|---|

## Nodes\\Online Focus Group Interview Questions\Q.5. Success Factors for adopting E-Government Services technology.\Success factors for adopting E-Government Services

| | | | | | | |
|---|---|---|---|---|---|---|
| | No | 0.0585 | 1 | | | |
| | | | | 1 | K.O | 22/10/2016 13:28 |

Participant 5

Answer - Transparency in government, citizen participation, access to information, economic growth, and robust government.

## Nodes\\Online Focus Group Interview Questions\Q.6. E-Government Services

| | | | | | | |
|---|---|---|---|---|---|---|
| | No | 0.0747 | 1 | | | |
| | | | | 1 | K.O | 20/10/2016 14:50 |

Q.6. E-Government Services Framework.

Researcher

Does the E-Government Services Framework sent to you represent your notion of E-Service?

## Nodes\\Online Focus Group Interview Questions\Q.6. E-Government Services Framework.\Does the E-Government Services Framework sent to you represent your

| | | | | | | |
|---|---|---|---|---|---|---|
| | No | 0.0114 | 1 | | | |
| | | | | 1 | K.O | 22/10/2016 13:31 |

Participant 5

Answer - Yes

## Nodes\\Online Focus Group Interview Questions\Q.7. The use of DOI theory in E-Government Services.

| | | | | | | |
|---|---|---|---|---|---|---|
| | No | 0.0840 | 1 | | | |
| | | | | 1 | K.O | 20/10/2016 14:50 |

Q.7. The use of DOI theory in E-Government Services.

Researcher

Will you encourage the use of Diffusion of Innovation Theory (DOI) in E-Service implementation?

| Classification | Aggregate | Coverage | Number Of Coding References | Reference Number | Coded By Initials | Modified On |
|---|---|---|---|---|---|---|

## Nodes\\Online Focus Group Interview Questions\Q.7. The use of DOI theory in E-Government Services.\Will you encourage the use of DOI theory in E-Service

| | | | | | | |
|---|---|---|---|---|---|---|
| | No | 0.0407 | 1 | | | |
| | | | | 1 | K.O | 22/10/2016 13:32 |

Will use encourage the use of Diffusion of Innovation Theory (DOI) in E-Service implementation?

## Nodes\\Online Focus Group Interview Questions\Q.7. The use of DOI theory in E-Government Services.\Will you encourage the use of DOI theory in E-Service Implementation and Adoption\Participant 5

| | | | | | | |
|---|---|---|---|---|---|---|
| | No | 0.0114 | 1 | | | |
| | | | | 1 | K.O | 22/10/2016 13:33 |

Participant 5

Answer - Yes

## Nodes\\Online Focus Group Interview Questions\Q.8. E-Government Services technology adoption diffusion to other aspects or areas.

| | | | | | | |
|---|---|---|---|---|---|---|
| | No | 0.0921 | 1 | | | |
| | | | | 1 | K.O | 20/10/2016 14:50 |

Q.8. E-Government Services technology adoption diffusion to other aspects or areas.

Researcher

Do you think the E-Service technology adoption will help in other aspects or areas?

## Nodes\\Online Focus Group Interview Questions\Q.8. E-Government Services technology adoption diffusion to other aspects or areas.\Do you think the E-Service technology adoption will help in other aspects or areas\Participant 5

| | | | | | | |
|---|---|---|---|---|---|---|
| | No | 0.0114 | 1 | | | |
| | | | | 1 | K.O | 22/10/2016 13:42 |

Participant 5

Answer - Yes

| Classification | Aggregate | Coverage | Number Of Coding References | Reference Number | Coded By Initials | Modified On |
|---|---|---|---|---|---|---|

## Nodes\\Online Focus Group Interview Questions\Q.9. Implementing the E-Government Services recommendations from this study.

| Classification | Aggregate | Coverage | Number Of Coding References | Reference Number | Coded By Initials | Modified On |
|---|---|---|---|---|---|---|
| | No | 0.1346 | 1 | | | |
| | | | | 1 | K.O | 20/10/2016 14:50 |

Q.9. Implementing the E-Government Services recommendations from this study.

Researcher

As part of the stakeholder in the provision of E-Service at the local government level in Nigeria, will you be ready to receive and implement the recommendations on E-Service from this study?

**Nodes\\Online Focus Group Interview Questions\Q.9. Implementing the E-Government Services recommendations from this study.\Ready to receive and implement the recommendations on E-Service from this study**

| | | | | | | |
|---|---|---|---|---|---|---|
| | No | 0.0819 | 1 | | | |
| | | | | 1 | K.O | 22/10/2016 13:43 |

As part of the stakeholder in the provision of E-Service at the local government level in Nigeria, will you be ready to receive and implement the recommendations on E-Service from this study?

**Nodes\\Online Focus Group Interview Questions\Q.9. Implementing the E-Government Services recommendations from this study.\Ready to receive and implement the recommendations on E-Service from this study\Participant 5**

| | | | | | | |
|---|---|---|---|---|---|---|
| | No | 0.0114 | 1 | | | |
| | | | | 1 | K.O | 22/10/2016 13:43 |

Participant 5

Answer - Yes

| Classification | Aggregate | Coverage | Number Of Coding References | Reference Number | Coded By Initials | Modified On |
|---|---|---|---|---|---|---|

# Memo

## Memos\\Participant 1

### Node

**Nodes\\Online Focus Group Interview Questions\Q.1. E-Government Services Usage.**

| | | | | | | |
|---|---|---|---|---|---|---|
| | No | 0.0560 | 1 | | | |
| | | | | 1 | K.O | 22/10/2016 12:36 |

Q.1. E-Government Services Usage.

Researcher

Have you used E-Government Services before?

## Nodes\\Online Focus Group Interview Questions\Q.10. E-Government Services future

| Classification | Aggregate | Coverage | Number Of Coding References | Reference Number | Coded By Initials | Modified On |
|---|---|---|---|---|---|---|
| | No | 0.1120 | 1 | | | |
| | | | | 1 | K.O | 22/10/2016 12:36 |

Q.10. E-Government Services future plan.

Researcher

Do you have any suggestion or future plan for E-Government Services?

Participant 1

Answer - Government should establish public ICT centres for the citizens at the local government level.

## Nodes\\Online Focus Group Interview Questions\Q.2. Types of E-Government Services.

| Classification | Aggregate | Coverage | Number Of Coding References | Reference Number | Coded By Initials | Modified On |
|---|---|---|---|---|---|---|
| | No | 0.0640 | 1 | | | |
| | | | | 1 | K.O | 22/10/2016 12:36 |

Q.2. Types of E-Government Services.

Researcher

What type of E-Government Services have you used before?

## Nodes\\Online Focus Group Interview Questions\Q.3. Barriers facing E-Government

| Classification | Aggregate | Coverage | Number Of Coding References | Reference Number | Coded By Initials | Modified On |
|---|---|---|---|---|---|---|
| | No | 0.1102 | 1 | | | |
| | | | | 1 | K.O | 22/10/2016 12:36 |

Q.3. Barriers facing E-Government Services.

Researcher

What are the barriers facing E-Service projects success both implementation and adoption?

## Nodes\\Online Focus Group Interview Questions\Q.4. Adopting E-Government Services.

| Classification | Aggregate | Coverage | Number Of Coding References | Reference Number | Coded By Initials | Modified On |
|---|---|---|---|---|---|---|
| | No | 0.0586 | 1 | | | |
| | | | | 1 | K.O | 22/10/2016 12:36 |

Q.4. Adopting E-Government Services.

Researcher

Will you adopt and use E-Government Services?

## Nodes\\Online Focus Group Interview Questions\Q.5. Success Factors for adopting E-Government Services technology.

| | | | | | | |
|---|---|---|---|---|---|---|
| | No | 0.1195 | 1 | | | |
| | | | | 1 | K.O | 22/10/2016 12:36 |

Q.5. Success Factors for adopting E-Government Services technology.

Researcher

What are the success factors for adopting E-Government Services technology?

## Nodes\\Online Focus Group Interview Questions\Q.6. E-Government Services

| | | | | | | |
|---|---|---|---|---|---|---|
| | No | 0.0782 | 1 | | | |
| | | | | 1 | K.O | 22/10/2016 12:36 |

Q.6. E-Government Services Framework.

Researcher

Does the E-Government Services Framework sent to you represent your notion of E-Service?

| Classification | Aggregate | Coverage | Number Of Coding References | Reference Number | Coded By Initials | Modified On |
|---|---|---|---|---|---|---|

## Nodes\\Online Focus Group Interview Questions\Q.7. The use of DOI theory in E-Government Services.

| | | | | | | |
|---|---|---|---|---|---|---|
| | No | 0.0884 | 1 | | | |
| | | | | 1 | K.O | 22/10/2016 12:36 |

Q.7. The use of DOI theory in E-Government Services.

Researcher

Will use encourage the use of Diffusion of Innovation Theory (DOI) in E-Service implementation?

## Nodes\\Online Focus Group Interview Questions\Q.8. E-Government Services technology adoption diffusion to other aspects or areas.

| | No | 0.0964 | 1 | | | |
|---|---|---|---|---|---|---|
| | | | | 1 | K.O | 22/10/2016 12:36 |

Q.8. E-Government Services technology adoption diffusion to other aspects or areas.

Researcher

Do you think the E-Service technology adoption will help in other aspects or areas?

### Nodes\\Online Focus Group Interview Questions\Q.9. Implementing the E-Government Services recommendations from this study.

| | No | 0.1408 | 1 | | | |
|---|---|---|---|---|---|---|
| | | | | 1 | K.O | 22/10/2016 12:36 |

Q.9. Implementing the E-Government Services recommendations from this study.

Researcher

As part of the stakeholder in the provision of E-Service at the local government level in Nigeria, will you be ready to receive and implement the recommendations on E-Service from this study?

| Classification | Aggregate | Coverage | Number Of Coding References | Reference Number | Coded By Initials | Modified On |
|---|---|---|---|---|---|---|

## Memos\\Participant 2

### Node

### Nodes\\Online Focus Group Interview Questions\Q.1. E-Government Services Usage.

| | No | 0.0525 | 1 | | | |
|---|---|---|---|---|---|---|
| | | | | 1 | K.O | 22/10/2016 12:37 |

Q.1. E-Government Services Usage.

Researcher

Have you used E-Government Services before?

### Nodes\\Online Focus Group Interview Questions\Q.10. E-Government Services future

| | No | 0.1138 | 1 |
|---|---|---|---|

| | | | | 1 | K.O | 22/10/2016 12:37 |
|---|---|---|---|---|---|---|

Q.10. E-Government Services future plan.

Researcher

Do you have any suggestion or future plan for E-Government Services?

## Nodes\\Online Focus Group Interview Questions\Q.2. Types of E-Government Services.

| | | | | | | |
|---|---|---|---|---|---|---|
| | No | 0.0600 | 1 | | | |
| | | | | 1 | K.O | 22/10/2016 12:37 |

Q.2. Types of E-Government Services.

Researcher

What type of E-Government Services have you used before?

## Nodes\\Online Focus Group Interview Questions\Q.3. Barriers facing E-Government

| | | | | | | |
|---|---|---|---|---|---|---|
| | No | 0.1330 | 1 | | | |
| | | | | 1 | K.O | 22/10/2016 12:37 |

Q.3. Barriers facing E-Government Services.

Researcher

What are the barriers facing E-Service projects success both implementation and adoption?

Participant 2

| Classification | Aggregate | Coverage | Number Of Coding References | Reference Number | Coded By Initials | Modified On |
|---|---|---|---|---|---|---|

## Nodes\\Online Focus Group Interview Questions\Q.4. Adopting E-Government Services.

| | | | | | | |
|---|---|---|---|---|---|---|
| | No | 0.0550 | 1 | | | |
| | | | | 1 | K.O | 22/10/2016 12:37 |

Q.4. Adopting E-Government Services.

Researcher

Will you adopt and use E-Government Services?

## Nodes\\Online Focus Group Interview Questions\Q.5. Success Factors for adopting E-Government Services technology.

| | | | | | | |
|---|---|---|---|---|---|---|
| | No | 0.1359 | 1 | | | |
| | | | | 1 | K.O | 22/10/2016 12:37 |

Q.5. Success Factors for adopting E-Government Services technology.

Researcher

What are the success factors for adopting E-Government Services technology?

Participant 2

Answer - Economic Growth, Stop leakages, Quick service delivery, access to information, improve government business,

## Nodes\\Online Focus Group Interview Questions\Q.6. E-Government Services

| | | | | | | |
|---|---|---|---|---|---|---|
| | No | 0.0733 | 1 | | | |
| | | | | 1 | K.O | 22/10/2016 12:37 |

Q.6. E-Government Services Framework.

Researcher

Does the E-Government Services Framework sent to you represent your notion of E-Service?

## Nodes\\Online Focus Group Interview Questions\Q.7. The use of DOI theory in E-Government Services.

| | | | | | | |
|---|---|---|---|---|---|---|
| | No | 0.0825 | 1 | | | |
| | | | | 1 | K.O | 22/10/2016 12:37 |

Q.7. The use of DOI theory in E-Government Services.

Researcher

Will use encourage the use of Diffusion of Innovation Theory (DOI) in E-Service implementation?

| Classification | Aggregate | Coverage | Number Of Coding References | Reference Number | Coded By Initials | Modified On |
|---|---|---|---|---|---|---|

### Nodes\\Online Focus Group Interview Questions\Q.8. E-Government Services technology adoption diffusion to other aspects or areas.

| | No | 0.0904 | 1 |
|---|---|---|---|

1 K.O 22/10/2016 12:37

Q.8. E-Government Services technology adoption diffusion to other aspects or areas.

Researcher

Do you think the E-Service technology adoption will help in other aspects or areas?

### Nodes\\Online Focus Group Interview Questions\Q.9. Implementing the E-Government Services recommendations from this study.

| | No | 0.1321 | 1 |
|---|---|---|---|

1 K.O 22/10/2016 12:37

Q.9. Implementing the E-Government Services recommendations from this study.

Researcher

As part of the stakeholder in the provision of E-Service at the local government level in Nigeria, will you be ready to receive and implement the recommendations on E-Service from this study?

# Memos\\Participant 3

## Node

### Nodes\\Online Focus Group Interview Questions\Q.1. E-Government Services Usage.

| | No | 0.0588 | 1 |
|---|---|---|---|

1 K.O 22/10/2016 12:37

Q.1. E-Government Services Usage.

Researcher

Have you used E-Government Services before?

| Classification | Aggregate | Coverage | Number Of Coding References | Reference Number | Coded By Initials | Modified On |
|---|---|---|---|---|---|---|

## Nodes\\Online Focus Group Interview Questions\Q.10. E-Government Services future

| | | | | | | |
|---|---|---|---|---|---|---|
| | No | 0.1029 | 1 | | | |
| | | | | 1 | K.O | 22/10/2016 12:37 |

Q.10. E-Government Services future plan.

Researcher

Do you have any suggestion or future plan for E-Government Services?

## Nodes\\Online Focus Group Interview Questions\Q.2. Types of E-Government Services.

| | | | | | | |
|---|---|---|---|---|---|---|
| | No | 0.0746 | 1 | | | |
| | | | | 1 | K.O | 22/10/2016 12:37 |

Q.2. Types of E-Government Services.

Researcher

What type of E-Government Services have you used before?

## Nodes\\Online Focus Group Interview Questions\Q.3. Barriers facing E-Government

| | | | | | | |
|---|---|---|---|---|---|---|
| | No | 0.1015 | 1 | | | |
| | | | | 1 | K.O | 22/10/2016 12:37 |

Q.3. Barriers facing E-Government Services.

Researcher

What are the barriers facing E-Service projects success both implementation and adoption?

## Nodes\\Online Focus Group Interview Questions\Q.4. Adopting E-Government Services.

| | | | | | | |
|---|---|---|---|---|---|---|
| | No | 0.0611 | 1 | | | |
| | | | | 1 | K.O | 22/10/2016 12:37 |

Q.4. Adopting E-Government Services.

Researcher

Will you adopt and use E-Government Services?

| Classification | Aggregate | Coverage | Number Of Coding References | Reference Number | Coded By Initials | Modified On |
|---|---|---|---|---|---|---|

## Nodes\\Online Focus Group Interview Questions\Q.5. Success Factors for adopting E-Government Services technology.

| | | | | | | |
|---|---|---|---|---|---|---|
| | No | 0.1010 | 1 | | | |
| | | | | 1 | K.O | 22/10/2016 12:37 |

Q.5. Success Factors for adopting E-Government Services technology.

Researcher

What are the success factors for adopting E-Government Services technology?

## Nodes\\Online Focus Group Interview Questions\Q.6. E-Government Services

| | | | | | | |
|---|---|---|---|---|---|---|
| | No | 0.0815 | 1 | | | |
| | | | | 1 | K.O | 22/10/2016 12:37 |

Q.6. E-Government Services Framework.

Researcher

Does the E-Government Services Framework sent to you represent your notion of E-Service?

## Nodes\\Online Focus Group Interview Questions\Q.7. The use of DOI theory in E-Government Services.

| | | | | | | |
|---|---|---|---|---|---|---|
| | No | 0.0917 | 1 | | | |
| | | | | 1 | K.O | 22/10/2016 12:37 |

Q.7. The use of DOI theory in E-Government Services.

Researcher

Will use encourage the use of Diffusion of Innovation Theory (DOI) in E-Service implementation?

## Nodes\\Online Focus Group Interview Questions\Q.8. E-Government Services technology adoption diffusion to other aspects or areas.

| | | | | | | |
|---|---|---|---|---|---|---|
| | No | 0.1006 | 1 | | | |
| | | | | 1 | K.O | 22/10/2016 12:37 |

Q.8. E-Government Services technology adoption diffusion to other aspects or areas.

Researcher

Do you think the E-Service technology adoption will help in other aspects or areas?

| Classification | Aggregate | Coverage | Number Of Coding References | Reference Number | Coded By Initials | Modified On |
|---|---|---|---|---|---|---|

### Nodes\\Online Focus Group Interview Questions\Q.9. Implementing the E-Government Services recommendations from this study.

| | | | | | | |
|---|---|---|---|---|---|---|
| | No | 0.1469 | 1 | | | |
| | | | | 1 | K.O | 22/10/2016 12:37 |

Q.9. Implementing the E-Government Services recommendations from this study.

Researcher

As part of the stakeholder in the provision of E-Service at the local government level in Nigeria, will you be ready to receive and implement the recommendations on E-Service from this study?

## Memos\\Participant 4

## Node

### Nodes\\Online Focus Group Interview Questions\Q.1. E-Government Services Usage.

| | | | | | | |
|---|---|---|---|---|---|---|
| | No | 0.0588 | 1 | | | |
| | | | | 1 | K.O | 22/10/2016 12:37 |

Q.1. E-Government Services Usage.

Researcher

Have you used E-Government Services before?

### Nodes\\Online Focus Group Interview Questions\Q.10. E-Government Services future

| | | | | | | |
|---|---|---|---|---|---|---|
| | No | 0.1029 | 1 | | | |
| | | | | 1 | K.O | 22/10/2016 12:37 |

Q.10. E-Government Services future plan.

Researcher

Do you have any suggestion or future plan for E-Government Services?

| Classification | Aggregate | Coverage | Number Of Coding References | Reference Number | Coded By Initials | Modified On |
|---|---|---|---|---|---|---|

## Nodes\\Online Focus Group Interview Questions\Q.2. Types of E-Government Services.

| Classification | Aggregate | Coverage | Number Of Coding References | Reference Number | Coded By Initials | Modified On |
|---|---|---|---|---|---|---|
| | No | 0.0746 | 1 | | | |
| | | | | 1 | K.O | 22/10/2016 12:37 |

Q.2. Types of E-Government Services.

Researcher

What type of E-Government Services have you used before?

## Nodes\\Online Focus Group Interview Questions\Q.3. Barriers facing E-Government

| Classification | Aggregate | Coverage | Number Of Coding References | Reference Number | Coded By Initials | Modified On |
|---|---|---|---|---|---|---|
| | No | 0.1015 | 1 | | | |
| | | | | 1 | K.O | 22/10/2016 12:37 |

Q.3. Barriers facing E-Government Services.

Researcher

What are the barriers facing E-Service projects success both implementation and adoption?

## Nodes\\Online Focus Group Interview Questions\Q.4. Adopting E-Government Services.

| Classification | Aggregate | Coverage | Number Of Coding References | Reference Number | Coded By Initials | Modified On |
|---|---|---|---|---|---|---|
| | No | 0.0611 | 1 | | | |
| | | | | 1 | K.O | 22/10/2016 12:37 |

Q.4. Adopting E-Government Services.

Researcher

Will you adopt and use E-Government Services?

## Nodes\\Online Focus Group Interview Questions\Q.5. Success Factors for adopting E-Government Services technology.

| Classification | Aggregate | Coverage | Number Of Coding References | Reference Number | Coded By Initials | Modified On |
|---|---|---|---|---|---|---|
| | No | 0.1010 | 1 | | | |
| | | | | 1 | K.O | 22/10/2016 12:37 |

Q.5. Success Factors for adopting E-Government Services technology.

Researcher

What are the success factors for adopting E-Government Services technology?

| Classification | Aggregate | Coverage | Number Of Coding References | Reference Number | Coded By Initials | Modified On |
|---|---|---|---|---|---|---|

## Nodes\\Online Focus Group Interview Questions\Q.6. E-Government Services

| Classification | Aggregate | Coverage | Number Of Coding References | Reference Number | Coded By Initials | Modified On |
|---|---|---|---|---|---|---|
| | No | 0.0815 | 1 | | | |
| | | | | 1 | K.O | 22/10/2016 12:37 |

Q.6. E-Government Services Framework.

Researcher

Does the E-Government Services Framework sent to you represent your notion of E-Service?

## Nodes\\Online Focus Group Interview Questions\Q.7. The use of DOI theory in E-Government Services.

| Classification | Aggregate | Coverage | Number Of Coding References | Reference Number | Coded By Initials | Modified On |
|---|---|---|---|---|---|---|
| | No | 0.0917 | 1 | | | |
| | | | | 1 | K.O | 22/10/2016 12:37 |

Q.7. The use of DOI theory in E-Government Services.

Researcher

Will use encourage the use of Diffusion of Innovation Theory (DOI) in E-Service implementation?

## Nodes\\Online Focus Group Interview Questions\Q.8. E-Government Services technology adoption diffusion to other aspects or areas.

| Classification | Aggregate | Coverage | Number Of Coding References | Reference Number | Coded By Initials | Modified On |
|---|---|---|---|---|---|---|
| | No | 0.1006 | 1 | | | |
| | | | | 1 | K.O | 22/10/2016 12:37 |

Q.8. E-Government Services technology adoption diffusion to other aspects or areas.

Researcher

Do you think the E-Service technology adoption will help in other aspects or areas?

| Classification | Aggregate | Coverage | Number Of Coding References | Reference Number | Coded By Initials | Modified On |
|---|---|---|---|---|---|---|

### Nodes\\Online Focus Group Interview Questions\Q.9. Implementing the E-Government Services recommendations from this study.

| | | | | | | |
|---|---|---|---|---|---|---|
| | No | 0.1469 | 1 | | | |
| | | | | 1 | K.O | 22/10/2016 12:37 |

Q.9. Implementing the E-Government Services recommendations from this study.

Researcher

As part of the stakeholder in the provision of E-Service at the local government level in Nigeria, will you be ready to receive and implement the recommendations on E-Service from this study?

## Memos\\Participant 5

### Node

### Nodes\\Online Focus Group Interview Questions\Q.1. E-Government Services Usage.

| | | | | | | |
|---|---|---|---|---|---|---|
| | No | 0.0535 | 1 | | | |
| | | | | 1 | K.O | 22/10/2016 12:37 |

Q.1. E-Government Services Usage.

Researcher

Have you used E-Government Services before?

### Nodes\\Online Focus Group Interview Questions\Q.10. E-Government Services future

| | | | | | | |
|---|---|---|---|---|---|---|
| | No | 0.1070 | 1 | | | |
| | | | | 1 | K.O | 22/10/2016 12:37 |

Q.10. E-Government Services future plan.

Researcher

Do you have any suggestion or future plan for E-Government Services?

Participant 5

Answer - Improve awareness, provision of IT infrastructures and cheap internet access for the citizens.

| Classification | Aggregate | Coverage | Number Of Coding References | Reference Number | Coded By Initials | Modified On |
|---|---|---|---|---|---|---|

## Nodes\\Online Focus Group Interview Questions\Q.2. Types of E-Government Services.

| Classification | Aggregate | Coverage | Number Of Coding References | Reference Number | Coded By Initials | Modified On |
|---|---|---|---|---|---|---|
| | No | 0.0611 | 1 | | | |
| | | | | 1 | K.O | 22/10/2016 12:37 |

Q.2. Types of E-Government Services.

Researcher

What type of E-Government Services have you used before?

## Nodes\\Online Focus Group Interview Questions\Q.3. Barriers facing E-Government

| Classification | Aggregate | Coverage | Number Of Coding References | Reference Number | Coded By Initials | Modified On |
|---|---|---|---|---|---|---|
| | No | 0.1354 | 1 | | | |
| | | | | 1 | K.O | 22/10/2016 12:37 |

Q.3. Barriers facing E-Government Services.

Researcher

What are the barriers facing E-Service projects success both implementation and adoption?

Participant 5

Answer – Corruption, No Regular ICT Training, Online Security/Trust, Cultural Problems, Political instability, Lack of ICT regulations and policies.

## Nodes\\Online Focus Group Interview Questions\Q.4. Adopting E-Government Services.

| Classification | Aggregate | Coverage | Number Of Coding References | Reference Number | Coded By Initials | Modified On |
|---|---|---|---|---|---|---|
| | No | 0.0560 | 1 | | | |
| | | | | 1 | K.O | 22/10/2016 12:37 |

Q.4. Adopting E-Government Services.

Researcher

Will you adopt and use E-Government Services?

## Nodes\\Online Focus Group Interview Questions\Q.5. Success Factors for adopting E-Government Services technology.

| Classification | Aggregate | Coverage | Number Of Coding References | Reference Number | Coded By Initials | Modified On |
|---|---|---|---|---|---|---|
| | No | 0.1290 | 1 | | | |
| | | | | 1 | K.O | 22/10/2016 12:37 |

Q.5. Success Factors for adopting E-Government Services technology.

Researcher

What are the success factors for adopting E-Government Services technology?

## Nodes\\Online Focus Group Interview Questions\Q.6. E-Government Services

| Classification | Aggregate | Coverage | Number Of Coding References | Reference Number | Coded By Initials | Modified On |
|---|---|---|---|---|---|---|
| | No | 0.0747 | 1 | | | |
| | | | | 1 | K.O | 22/10/2016 12:37 |

Q.6. E-Government Services Framework.

Researcher

Does the E-Government Services Framework sent to you represent your notion of E-Service?

## Nodes\\Online Focus Group Interview Questions\Q.7. The use of DOI theory in E-Government Services.

| Classification | Aggregate | Coverage | Number Of Coding References | Reference Number | Coded By Initials | Modified On |
|---|---|---|---|---|---|---|
| | No | 0.0840 | 1 | | | |
| | | | | 1 | K.O | 22/10/2016 12:37 |

Q.7. The use of DOI theory in E-Government Services.

Researcher

Will use encourage the use of Diffusion of Innovation Theory (DOI) in E-Service implementation?

## Nodes\\Online Focus Group Interview Questions\Q.8. E-Government Services technology adoption diffusion to other aspects or areas.

| Classification | Aggregate | Coverage | Number Of Coding References |
|---|---|---|---|
| | No | 0.0921 | 1 |

| | | | | 1 | K.O | 22/10/2016 12:37 |
|---|---|---|---|---|---|---|

Q.8. E-Government Services technology adoption diffusion to other aspects or areas.

Researcher

Do you think the E-Service technology adoption will help in other aspects or areas?

| Classification | Aggregate | Coverage | Number Of Coding References | Reference Number | Coded By Initials | Modified On |
|---|---|---|---|---|---|---|

**Nodes\\Online Focus Group Interview Questions\Q.9. Implementing the E-Government Services recommendations from this study.**

| Classification | Aggregate | Coverage | Number Of Coding References | Reference Number | Coded By Initials | Modified On |
|---|---|---|---|---|---|---|
| | No | 0.1346 | 1 | | | |
| | | | | 1 | K.O | 22/10/2016 12:37 |

Q.9. Implementing the E-Government Services recommendations from this study.

Researcher

As part of the stakeholder in the provision of E-Service at the local government level in Nigeria, will you be ready to receive and implement the recommendations on E-Service from this study?

# **Appendix D** - Ethical Approvals

University of Portsmouth

Faculty of Technology

**Technology Faculty Ethics Committee**

ethics-tech@port.ac.uk

Date 31/5/16

Applicant: Kazeem Oseni

Address: School of Computing, University of Portsmouth

Dear Kazeem Oseni

| | |
|---|---|
| **Study Title:** | Barriers Facing e-Service Adoption and Implementation at Local Government Level in Nigeria, An Analysis. |
| **Ethics Committee reference:** | **KO1** |

Thank you for submitting your documents for ethical review. The Ethics Committee was content to grant a favourable ethical opinion of the above research on the basis described in the application form, protocol and supporting documentation.

- We would request that phrasing is checked throughout and the document proof read, especially in the documents and questionnaires sent to participants.

*Please note that the favourable opinion of the EC does not grant permission or approval to undertake the research. Management permission or approval must be obtained from any host organisation, including University of Portsmouth, prior to the start of the study.*

**Documents reviewed**

| *Document* | *Version* | *Date* |
|---|---|---|
| Application Form | 26/05/2016 | 3.0 |
| Participant Information Sheet(s) (list if necessary) | 26/05/2016 | 2.0 |
| Consent Form(s) (list if necessary) | 26/05/2016 | 2.0 |
| Invitation Letter | 26/05/2016 | 1.0 |
| Supervisor Email Confirming Application | 16/03/2016 | |

Faculty of Technology

**Statement of compliance**

The Committee is constituted in accordance with the Governance Arrangements set out by the University of Portsmouth

Reporting and other requirements

The attached document acts as a reminder that research should be conducted with integrity and gives detailed guidance on reporting requirements for studies with a favourable opinion, including:

- Notifying substantial amendments
- Notification of serious breaches of the protocol
- Progress reports
- Notifying the end of the study

Feedback

You are invited to give your view of the service that you have received from the Faculty Ethics Committee. If you wish to make your views known please contact the administrator, Emma Bain at ethics-tech@port.ac.uk

**Please quote this number on all correspondence: KO1**

Yours sincerely and wishing you every success in your research

**John Williams**

**Chair**

Email: ethics-tech@port.ac.uk

University of Portsmouth

# Certificate of Ethics Review

| | |
|---|---|
| **Project Title:** | Barriers Facing e-Service Adoption and Implementation at Local Government Level in Nigeria, A Lower Middle Income Country: An Analysis. |
| **User ID:** | 715813 |
| **Name:** | Kazeem Oluwakemi Oseni |
| **Application Date:** | 23/02/2015 11:33:40 |

You must download your referral certificate, print a copy and keep it as a record of this review.

The FEC representative for the School of Computing is Carl Adams

It is your responsibility to follow the University Code of Practice on Ethical Standards and any Department/School or professional guidelines in the conduct of your study including relevant guidelines regarding health and safety of researchers including the following:

- University Policy
- Safety on Geological Fieldwork

All projects involving human participants need to offer sufficient information to potential participants to enable them to make a decision. Template participant information sheets are available from the:

- Univeristy's Ethics Site (Participant information template).

It is also your responsibility to follow University guidance on Data Protection Policy:

- General guidance for all data protection issues
- University Data Protection Policy

**SchoolOrDepartment:** SOC
**PrimaryRole:** PostgraduateStudent
**SupervisorName:** Dr Kate Dingley
**HumanParticipants:** Yes
**ParticipationBeyondAnsweringQuestionsOrInterviews:** Yes
**ParticipantInformationSheets:** I hereby confirm that participant's consent will be obtained.
**ParticipantConfidentiality:** I hereby confirm that I will maintain participant anonymity and confidentiality of data collected. Data collected will only be used for research purpose only.
**InvolvesNHSPatientsOrStaff:** No
**NoConsentOrDeception:** No
**CollectingOrAnalysingPersonalInfoWithoutConsent:** No
**InvolvesUninformedOrDependents:** No
**DrugsPlacebosOrOtherSubstances:** No
**BloodOrTissueSamples:** No

**Certificate Code:** 8467-09E4-A95B-634E-19C2-8F74-3458-2931 Page 1

**PainOrMildDiscomfort:** No
**PsychologicalStressOrAnxiety:** No
**ProlongedOrRepetitiveTesting:** No
**FinancialInducements:** No
**PhysicalEcologicalDamage:** No
**HistoricalOrCulturalDamage:** No
**HarmToAnimal:** No
**HarmfulToThirdParties:** No
**OutputsPotentiallyAdaptedAndMisused:** No
**Confirmation-ConsideredDataUse:** Confirmed
**Confirmation-ConsideredImpactAndMitigationOfPontentialMisuse:** Confirmed
**Confirmation-ActingEthicallyAndHonestly:** Confirmed

# **Appendix E -** UPR 16

## FORM UPR16

**Research Ethics Review Checklist**

University of Portsmouth

**Please include this completed form as an appendix to your thesis (see the Postgraduate Research Student Handbook for more information**

| **Postgraduate Research Student (PGRS) Information** | | | **Student ID:** | 715813 |
|---|---|---|---|---|
| **PGRS Name:** | Mr. Kazeem Oluwakemi OSENI | | | |
| **Department:** | Sch of Computing | **First Supervisor:** | Dr. Kate Dingley | |
| **Start Date:** (or progression date for Prof Doc students) | | **February 1st, 2014** | | |
| **Study Mode and Route:** | Part-time ☐ Full-time ☒ | MPhil ☐ PhD ☒ | MD ☐ Professional Doctorate ☐ | |

| **Title of Thesis:** | Barriers Facing E-Service Adopting and Implementation at Local Environment Level in Nigeria |
|---|---|
| **Thesis Word Count:** (excluding ancillary data) | 60,749 |

If you are unsure about any of the following, please contact the local representative on your Faculty Ethics Committee for advice. Please note that it is your responsibility to follow the University's Ethics Policy and any relevant University, academic or professional guidelines in the conduct of your study

Although the Ethics Committee may have given your study a favourable opinion, the final responsibility for the ethical conduct of this work lies with the researcher(s).

**UKRIO Finished Research Checklist:**
(If you would like to know more about the checklist, please see your Faculty or Departmental Ethics Committee rep or see the online version of the full checklist at: http://www.ukrio.org/what-we-do/code-of-practice-for-research/)

| | |
|---|---|
| a) Have all of your research and findings been reported accurately, honestly and within a reasonable time frame? | YES ☒ NO ☐ |
| b) Have all contributions to knowledge been acknowledged? | YES ☒ NO ☐ |
| c) Have you complied with all agreements relating to intellectual property, publication and authorship? | YES ☒ NO ☐ |
| d) Has your research data been retained in a secure and accessible form and will it remain so for the required duration? | YES ☒ NO ☐ |
| e) Does your research comply with all legal, ethical, and contractual requirements? | YES ☒ NO ☐ |

**Candidate Statement:**

| | |
|---|---|
| I have considered the ethical dimensions of the above named research project, and have successfully obtained the necessary ethical approval(s) | |
| **Ethical review number(s) from Faculty Ethics Committee (or from NRES/SCREC):** | KO1/ 8467-09E4-A95B-634E-19C2-8F74-3458-2931 |
| If you have *not* submitted your work for ethical review, and/or you have answered 'No' to one or more of questions a) to e), please explain below why this is so: | |
| N/A | |
| **Signed *(PGRS)*:** | **Date:** 14/02/2017 |